\documentclass[aps4epj,prd,secnumarabic,twocolumn,nofootinbib,superscriptaddress,floatfix,eqsecnum,showpacs]{revtex4epj}
\usepackage[stable]{footmisc}
\usepackage{graphics}
\usepackage{color}
\usepackage{graphicx}
\usepackage{amssymb}
\usepackage{amsmath}
\usepackage{epstopdf}
\usepackage{xspace}
\usepackage{slashed}
\usepackage{cancel}
\usepackage{hyperref}
\usepackage{mathbbol}
\usepackage{wasysym} 
\usepackage{bbm}     
\usepackage{rotating}
\usepackage{dcolumn}   
\usepackage{bm}        
\usepackage{longtable} 
\usepackage{multirow}
\usepackage{enumitem}
\renewcommand\footnoterule{\hrule width 6em\vspace*{12pt}}

\def\figwid{3.30in}

\def\APS{The Americal Physical Society}

\def\figPermX#1#2#3{From \cite{#1}.}

\def\figPerms#1#2#3#4{{#1} from \cite{#2}, respectively.}

\def\AfigPerms#1#2#3{\figPerms{Adapted}{#1}{#2}{#3}}
\def\AfigPermsAPS#1#2{\AfigPerms{#1}{#2}{\APS}}

\def\beq{\begin{equation}}
\def\eeq{\end{equation}}
\def\beqa{\begin{eqnarray}}
\def\eeqa{\end{eqnarray}}

\newcommand{\kev}  {~\ensuremath{{\mathrm{keV}}     }}
\newcommand{\gev}  {~\ensuremath{{\mathrm{GeV}}     }}
\newcommand{\mev}  {~\ensuremath{{\mathrm{MeV}}     }}

\newcommand{\GeV}  {\gev}
\newcommand{\MeV}  {\mev}

\newcommand{\brat}   {\ensuremath{ \mathcal{B}}}

\newcommand{\epem}   {\ensuremath{ e^+e^- }}
\newcommand{\Dze}   {\ensuremath{ D^0 }}
\newcommand{\Dpl}  {\ensuremath{ D^+ }}

\newcommand{\Dstp} {\ensuremath{ D^{*+} }}
\newcommand{\Dstn} {\ensuremath{ D^{*0} }}

\newcommand{\DDbar}      {\ensuremath{ D   \bar D     }}

\newcommand{\DDst}    {\ensuremath{ D   \bar D^* }}
\newcommand{\DstDst}  {\ensuremath{ D^* \bar D^* }}
\newcommand{\DstnDn}  {\ensuremath{\Dstn \bar{\Dze}}}

\def\Ups{\ensuremath{\Upsilon}}

\def\Unx#1#2{\ensuremath{\Ups({#1}{#2})}}
\def\UnS#1{\ensuremath{\Unx{#1}{S}}}
\def\UoneD{\ensuremath{\Unx{1}{D}}}

\newcommand{\jpsi}{\ensuremath{J/\psi}}

\newcommand{\psip}{\ensuremath{\psi(2S)}}

\newcommand{\chicOne}{\ensuremath{\chi_{c1}}}

\usepackage{relsize}
\def\babar{\mbox{\slshape B\kern-0.1em{\smaller A}\kern-0.1em
    B\kern-0.1em{\smaller A\kern-0.2em R}}}

\newcommand{\dipi}{{\ensuremath{{\pi^+\pi^-}}}}

\def\lamQ{\Lambda_{\rm QCD}}

\def\als{{\alpha_{\rm s}}}
\def\simg{{\ \lower-1.2pt\vbox{\hbox{\rlap{$>$}\lower6pt\vbox{\hbox{$\sim$}}}}\ }}

\newcommand{\etab}{\ensuremath{{\eta_b(1S)}}}
\newcommand{\etac}{\ensuremath{{\eta_c(1S)}}}

\newcommand{\etabp}{\ensuremath{{\eta_b(2S)}}}

\newcommand{\pp}{\mbox{$p$+$p$}}
\newcommand{\PbPb}{Pb+Pb}
\newcommand{\dAu}{$d$+Au}

\newcommand{\AuAu}{Au+Au}
\newcommand{\CuCu}{Cu+Cu}
\newcommand{\ppcoll}{$pp$}
\newcommand{\AAcoll}{$AA$}

\newcommand{\RAA}{\mbox{$R_{AA}$}}
\newcommand{\RCP}{\mbox{$R_{CP}$}}

\newcommand{\RdAu}{\mbox{$R_{\rm dAu}$}}

\newcommand{\cpip}{\ensuremath{\pi^{+}}}

\newcommand{\cga}{\ensuremath{\gamma}}

\newcommand{\chb}{\ensuremath{h_b(1P)}}
\newcommand{\chbp}{\ensuremath{h_b(2P)}}
\newcommand{\chbn}{\ensuremath{h_b(nP)}}
\newcommand{\chbm}{\ensuremath{h_b(mP)}}

\newcommand{\czb}{\ensuremath{Z_b}}
\newcommand{\czbo}{\ensuremath{Z_b(10610)}}
\newcommand{\czbt}{\ensuremath{Z_b(10650)}}

\def\cp{CP}
\def\cme{chiral magnetic effect}

\def\be{\begin{equation}}
\def\ee{\end{equation}}
\def\bea{\begin{eqnarray}}
\def\eea{\end{eqnarray}}
\def\pa{\partial}
\def\la{\lambda}

\newcommand{\dndh}{$dN_{\rm ch}/d\eta$}

\newcommand{\Pt}{$p_{T}$}
\newcommand{\pA}{$p$A}

\def\beq {\begin{equation}}
\def \eeq {\end{equation}}
\def \bea{\begin{eqnarray}}
\def \eea{\end{eqnarray}}

\newcommand{\Dstar} {\ensuremath{ D^{*} }}
\newcommand{\D} {\ensuremath{D}}
\newcommand{\pPb}{$p$+Pb}
\newcommand{\RpPb}{\mbox{$R_{p{\rm Pb}}$}}
\newcommand{\W} {\ensuremath{W}}
\newcommand{\Z} {\ensuremath{Z}}

\newcommand{\bra}[1]{\mbox{$\left\langle #1 \right|$}}
\newcommand{\ket}[1]{\mbox{$\left| #1 \right\rangle$}}

\renewcommand{\thetable}{\arabic{table}}

\begin{document}

\title{ \Large 
QCD and strongly coupled gauge theories: challenges and perspectives}

\thispagestyle{empty}

\renewcommand{\thefootnote}{\fnsymbol{footnote}}


\author{N.~Brambilla\footnotemark[1]\footnotemark[2]}
\affiliation{Physik Department, Technische Universit\"at M\"unchen,
             James-Franck-Stra\ss e 1, 85748 Garching, Germany}

\author{S.~Eidelman\footnotemark[2]}
\affiliation{Budker Institute of Nuclear Physics, SB RAS, Novosibirsk 630090, Russia}
\affiliation{Novosibirsk State University, Novosibirsk 630090, Russia}

\author{P.~Foka\footnotemark[2]\footnotemark[3]}
\affiliation{GSI Helmholtzzentrum f\"ur Schwerionenforschung GmbH, Planckstra\ss e 1, 64291 Darmstadt, 
Germany}

\author{S.~Gardner\footnotemark[2]\footnotemark[3]}
\affiliation{Department of Physics and Astronomy, University of Kentucky, Lexington, KY 40506-0055, USA}

\author{A.S.~Kronfeld\footnotemark[2]}
\affiliation{Theoretical Physics Department,
    Fermi National Accelerator Laboratory, P.O.~Box 500,
    Batavia, Illinois 60510-5011, USA}


\author{\linebreak M.G.~Alford\footnotemark[3]}
\affiliation{Department of Physics, Washington University, St Louis, MO,
63130, USA}

\author{R.~Alkofer\footnotemark[3]}
\affiliation{University of Graz, 8010 Graz, Austria}

\author{M.~Butenschoen\footnotemark[3]}
\affiliation{University of Vienna, Faculty of Physics,
Boltzmanngasse 5, 1090 Wien, Austria}

\author{T.D.~Cohen\footnotemark[3]}
\affiliation{Maryland Center for Fundamental Physics and the Department
of Physics, University of Maryland, College Park, MD 20742-4111, USA}

\author{J.~Erdmenger\footnotemark[3]}
\affiliation{Max-Planck-Institute for Physics, F\"ohringer Ring 6, 80805 M\"unchen, Germany}

\author{L.~Fabbietti\footnotemark[3]}
\affiliation{Excellence Cluster ``Origin and Structure of the Universe'',
Technische Universit\"at M\"unchen,
85748 Garching, Germany}

\author{M.~Faber\footnotemark[3]}
\affiliation{Atominstitut, Technische Universit\"{a}t Wien, 1040 Vienna, Austria}

\author{J.L.~Goity\footnotemark[3]}
\affiliation{Hampton University, Hampton, VA 23668, USA}
\affiliation{Jefferson Laboratory, Newport News, VA 23606, USA}

\author{B.~Ketzer\footnotemark[3]\footnotemark[4]}
\affiliation{Physik Department, Technische Universit\"at M\"unchen,
             James-Franck-Stra\ss e 1, 85748 Garching, Germany}

\author{H.W.~Lin\footnotemark[3]}
\affiliation{Department of Physics, University of Washington, Seattle, WA
98195-1560, USA}

\author{F.J.~Llanes-Estrada\footnotemark[3]}
\affiliation{Department Fisica Teorica I, Universidad Complutense de Madrid, 28040 Madrid, Spain}

\author{H.B.~Meyer\footnotemark[3]}
\affiliation{PRISMA Cluster of Excellence,
Institut f\"ur Kernphysik and Helmholtz~Institut~Mainz,
Johannes Gutenberg-Universit\"at Mainz,
55099 Mainz, Germany}

\author{P.~Pakhlov\footnotemark[3]}
\affiliation{Institute of Theoretical and Experimental Physics, Moscow 117218, Russia}
\affiliation{Moscow Institute for Physics and Technology, Dolgoprudny 141700, Russia}

\author{E.~Pallante\footnotemark[3]}
\affiliation{Centre for Theoretical Physics, University of Groningen, 9747 AG
Groningen, Netherlands}

\author{M.I.~Polikarpov\footnotemark[3]}
\affiliation{Institute of Theoretical and Experimental Physics, Moscow 117218, Russia}
\affiliation{Moscow Institute for Physics and Technology, Dolgoprudny 141700, Russia}

 \author{H.~Sazdjian\footnotemark[3]}
\affiliation{Institut de Physique Nucl\'eaire, CNRS/IN2P3,
             Universit\'e Paris-Sud, 91405 Orsay, France}

\author{A.~Schmitt\footnotemark[3]}
\affiliation{Institut f\"{u}r Theoretische Physik, Technische Universit\"{a}t Wien, 1040 Vienna, Austria}

\author{W.M.~Snow\footnotemark[3]}
\affiliation{Center for Exploration of Energy and Matter and
Department of Physics, Indiana University, Bloomington, IN 47408, USA}

\author{A.~Vairo\footnotemark[3]}
\affiliation{Physik Department, Technische Universit\"at M\"unchen,
             James-Franck-Stra\ss e 1, 85748 Garching, Germany}

\author{R.~Vogt\footnotemark[3]}
\affiliation{Physics Division, Lawrence Livermore National Laboratory, Livermore, CA 94551, USA}
\affiliation{Physics Department, University of California, Davis, CA 95616, USA}

\author{A.~Vuorinen\footnotemark[3]}
\affiliation{Department of Physics and Helsinki Institute of Physics, P.O. Box 64, 
00014 University of Helsinki, Finland}

\author{H.~ Wittig\footnotemark[3]}
\affiliation{PRISMA Cluster of Excellence,
Institut f\"ur Kernphysik and Helmholtz~Institut~Mainz,
Johannes Gutenberg-Universit\"at Mainz,
55099 Mainz, Germany}

\author{\linebreak P.~Arnold}
\affiliation{Department of Physics, University of Virginia, 382 McCormick Rd., P.O. Box 400714, Charlottesville, VA 22904-4714, USA}

\author{P.~Christakoglou}
\affiliation{NIKHEF, Science Park 105, 1098 XG Amsterdam, Netherlands}

\author{P.~Di~Nezza}
\affiliation{Istituto Nazionale di Fisica Nucleare (INFN), Via E. Fermi 40, 00044 Frascati, Italy}

\author{Z.~Fodor}
\affiliation{Wuppertal University, 42119 Wuppertal, Germany}
\affiliation{E\"otv\"os University, 1117 Budapest, Hungary}
\affiliation{Forschungszentrum J\"ulich, 52425 J\"ulich, Germany}

\author{X.~Garcia~i~Tormo}
\affiliation{Albert Einstein Center for Fundamental Physics, Institut f\"ur 
Theoretische Physik, Universit\"at Bern, Sidlerstra\ss e 5, 3012 Bern, Switzerland}

\author{R.~H\"ollwieser}
\affiliation{Atominstitut, Technische Universit\"{a}t Wien, 1040 Vienna, Austria}

\author{M.A.~Janik}
\affiliation{Faculty of Physics, Warsaw University of Technology, 00-662 Warsaw, Poland}

\author{A.~Kalweit}
\affiliation{European Organization for Nuclear Research (CERN), Geneva, Switzerland}

\author{D.~Keane}
\affiliation{Kent State University, Department of Physics, Kent, OH 44242, USA}

\author{E.~Kiritsis}
\affiliation{Crete Center for Theoretical Physics,
Department of Physics, University of Crete, 71003 Heraklion, Greece.}
\affiliation{Laboratoire APC, Universit\'e Paris Diderot, Sorbonne Paris-Cit\'e, 
75205 Paris Cedex 13, France}
\affiliation{Theory Group, Physics Department, CERN, 1211, Geneva 23, Switzerland}


\author{A.~Mischke}
\affiliation{Utrecht University, Faculty of Science, Princetonplein 5, 3584 CC Utrecht, The Netherlands}

\author{R.~Mizuk}
\affiliation{Institute of Theoretical and Experimental Physics, Moscow 117218, Russia}
\affiliation{Moscow Physical Engineering Institute, Moscow 115409, Russia}

\author{G.~Odyniec}
\affiliation{Lawrence Berkeley National Laboratory, 1 Cyclotron Rd., Berkeley, CA 94720, USA}

\author{K.~Papadodimas}
\affiliation{Centre for Theoretical Physics, University of Groningen, 9747 AG
Groningen, Netherlands}

\author{A.~Pich}
\affiliation{IFIC, Universitat de Val\`encia -- CSIC, Apt.\ Correus 22085, 46071
Val\`encia, Spain}

\author{R.~Pittau}
\affiliation{Departamento de Fisica Teorica y del Cosmos and CAFPE, Campus Fuentenueva s.~n., 
Universidad de Granada, 18071 Granada, Spain}

\author{J.-W.~Qiu}
\affiliation{Physics Department, Brookhaven National Laboratory, Upton, NY 11973, USA}
\affiliation{C.~N.~Yang Institute for Theoretical Physics and Department of Physics and Astronomy, 
Stony Brook University, Stony Brook, NY 11794, USA}

\author{G.~Ricciardi}
 \affiliation{Dipartimento di Fisica, Universit\'a degli Studi di Napoli
Federico II, 80126 Napoli, Italy}
 \affiliation{INFN, Sezione di Napoli, 80126 Napoli, Italy}

\author{C.A.~Salgado} 
\affiliation{Departamento de Fisica de Particulas y IGFAE,
Universidade de Santiago de Compostela,\\ 15782 Santiago de Compostela, Galicia, Spain}

\author{K.~Schwenzer}
\affiliation{Department of Physics, Washington University, St Louis, MO,
63130, USA}

\author{N.G.~Stefanis}
\affiliation{Institut f\"{u}r Theoretische Physik II,
                Ruhr-Universit\"{a}t Bochum, 44780 Bochum, Germany}

\author{G.M.~von~Hippel}
\affiliation{PRISMA Cluster of Excellence,
Institut f\"ur Kernphysik and Helmholtz~Institut~Mainz,
Johannes Gutenberg-Universit\"at Mainz,
55099 Mainz, Germany}

\author{V.I.~Zakharov}
\affiliation{Max-Planck-Institute for Physics, F\"ohringer Ring 6, 80805 M\"unchen, Germany}
\affiliation{Institute of Theoretical and Experimental Physics, Moscow 117218, Russia}
\affiliation{Moscow Institute for Physics and Technology, Dolgoprudny 141700, Russia}


\begin{abstract}
We highlight the progress, current status, and
open challenges of QCD-driven physics, in theory and in experiment. 
We discuss how the strong interaction is intimately
connected to a broad sweep of physical problems, in settings
ranging from astrophysics and cosmology to strongly-coupled,
complex systems in particle and condensed-matter physics, as well as to
searches for physics beyond the Standard Model. 
We also discuss how success in describing the strong interaction
impacts other fields, and, in turn, how such subjects can impact 
studies of the strong interaction. 
In the course of the work 
we offer a perspective on the many research streams which 
flow into and out of QCD, as well as a vision for future developments. 
\end{abstract}

\date{\today}
\pacs{12.38.-t\vspace*{-6pt}}

\maketitle

\footnotetext[1]{Corresponding author, \href{mailto:nora.brambilla@ph.tum.de}{nora.brambilla@ph.tum.de}}
\footnotetext[2]{Editor}
\footnotetext[3]{Chapter Convener}
\footnotetext[4]{Present address: 
Helmholtz-Institut f\"ur Strahlen- und Kernphysik, Universit\"at Bonn, 53115 Bonn, Germany}

\renewcommand{\thefootnote}{\textcolor{white}{\arabic{footnote}}} 
{\small \tableofcontents}
\renewcommand{\thefootnote}{\textcolor{black}{\arabic{footnote}}}
\setcounter{footnote}{0}

\clearpage

\section[Chapz]{Overview \protect\footnotemark}
\footnotetext{Contributing authors: N.~Brambilla, S.~Eidelman, P.~Foka, S.~Gardner, A.S.~Kronfeld}
\label{sec:chapz}

This document highlights the status and challenges of strong-interaction physics at the beginning of a new
era initiated by the discovery of the Higgs particle at the Large Hadron Collider at CERN.
It has been a concerted undertaking by many contributing authors, with a smaller group of conveners and
editors to coordinate the effort.
Together, we have sought to address a common set of questions:
What are the latest achievements and highlights related to the strong interaction? %
What important open problems remain? %
What are the most promising avenues for further investigation? %
What do experiments need from theory?
What does theory need from experiments? %
In addressing these questions, we aim to cast the challenges in quantum chromodynamics (QCD) and other
strongly-coupled physics in a way that spurs future developments.

A core portion of the scientific work discussed in this document was nurtured in the framework of the
conference series on ``Quark Confinement and the Hadron Spectrum,'' which has served over the years as a
discussion forum for people working in the field.
The starting point of the current enterprise can be traced to its X$^\mathrm{th}$ edition
(\href{http://www.confx.de}{www.confx.de}), held in Munich in October, 2012.
Nearly 400 participants engaged in lively discussions spurred by its seven topical sessions.
These discussions inspired the chapters that follow, and their organization is loosely connected to the
topical sessions of the conference:
Light Quarks; 
Heavy Quarks; 
QCD and New Physics; 
Deconfinement; 
Nuclear and Astroparticle Physics; 
Vacuum Structure and Confinement; 
and 
Strongly Coupled Theories. 
This document is an original, focused work that summarizes the current status of QCD, broadly interpreted,
and provides a vision for future developments and further research.
The document's wide-angle, high-resolution picture of the field is current through March 15, 2014.

\subsection{Readers' guide}
\label{intro:sec:guide}

We expect that this work will attract a broad readership, ranging from practitioners in one or more subfields
of QCD, to particle or nuclear physicists working in fields other than QCD and the Standard Model (SM), to
students starting research in QCD or elsewhere.
We should note that the scope of QCD is so vast that it is impossible to cover absolutely everything.
Any omissions stem from the need to create something useful despite the numerous, and sometimes rapid,
advances in QCD research.
To help the reader navigate the rest of the document, let us begin with a brief guide to the contents of and
rationale for each chapter.
\pagebreak

\textbf{Chapter~\ref{sec:chapzz}} is aimed at all readers and explains the aims of this undertaking in more
detail by focusing on properties and characteristics that render QCD a unique part of the SM.
We also highlight the broad array of problems for which the study of QCD is pertinent before turning to a
description of the experiments and theoretical tools that appear throughout the remaining chapters.
Chapter~\ref{sec:chapzz} concludes with a status report on the determination of the fundamental parameters of
QCD, namely, the gauge coupling $\als$ and the quark masses.

The wish to understand the properties of the lightest hadrons with the quark model, concomitant with the
observation of partons in deep-inelastic electron scattering, sparked the emergence of QCD.
We thus begin in \textbf{Chapter~\ref{sec:chapb}} with this physics, discussing not only the current status
of the parton distribution functions, but also delving into many aspects of the structure and dynamics of
light-quark hadrons at low energies.
Chapter~\ref{sec:chapb} also reviews the hadron spectrum, including exotic states beyond the quark model,
such as glueballs, as well as chiral dynamics, probed through low-energy observables.
Certain new-physics searches for which control over light-quark dynamics is essential are also described.

Heavy-quark systems have played a crucial role in the development of the SM, QCD especially.
Their large mass, compared to the QCD scale, leads to clean experimental signatures and opens up a new
theoretical toolkit.
\textbf{Chapter~\ref{sec:chapc}} surveys these theoretical tools in systems like quarkonium, i.e., bound
states of a heavy quark and a heavy antiquark, and hadrons consisting of a heavy quark bound to light degrees
of freedom.
Highlights of the chapter include an up-to-date presentation of the exotic states $X$, $Y$, $Z$ that have
been discovered in the charmonium and bottomonium regions, the state of the art of lattice-QCD 
calculations, and an extended discussion of the status of our theoretical understanding of quarkonium 
production at hadron and electron colliders.
The latest results for $B$- and $D$-meson semileptonic decays, which are used to determine some SM parameters
and to look for signs of new physics, are also discussed.

Control of QCD for both heavy and light quarks, and for gluons as well, is the key to many searches for
physics beyond the SM.
\textbf{Chapter~\ref{sec:chape}} reviews the possibilities and challenges of the searches realized through
precision measurements, both at high energy through collider experiments and at low energy through
accelerator, reactor, and table-top experiments.
In many searches, a comparably precise theoretical calculation is required to separate SM from non-SM
effects, and these are reviewed as well.
This chapter has an extremely broad scope, ranging from experiments with multi-TeV $pp$ collisions to those
with ultracold neutrons and atoms; ranging from top-quark physics to the determinations of the weak-mixing
angle at low energies; ranging from searches for new phenomena in quark-flavor violation to searches for
permanent electric dipole moments.

In Chapter~\ref{sec:chape}, QCD is a tool to aid the discovery of exotic phenomena external to QCD.
The next three chapters treat a rich array of as-yet unexplored phenomena that emerge from QCD in complex, 
many-hadron systems.
\textbf{Chapter~\ref{sec:chapd}} begins this theme with a discussion of deconfinement in the context of the
quark-gluon plasma and heavy-ion collisions.
We first give a description of this novel kind of matter and of our present knowledge of the QCD phase
diagram, based on the most recent measurements.
We then turn to describing near-equilibrium properties of the quark-gluon plasma and its approach to
equilibrium.
We explain theorists' present understanding, focusing on ideas and techniques that are
directly connected to QCD.
Hard probes like jet quenching and quarkonium suppression as methods to scrutinize the quark-gluon plasma
properties are also discussed.
The chapter ends with an interesting parallel between thermal field theory calculations in QCD and
cosmology and with a note on the chiral magnetic effect.

\textbf{Chapter~\ref{sec:chapf}} covers cold, dense hadronic systems, including nuclear and hypernuclear
physics and also the ultradense hadronic matter found in neutron stars, noting also the new phases that are
expected to appear at even higher densities.
These topics are informed not only by theory and terrestrial experiments but also by astrophysical
observations.

At this point the reader finds \textbf{Chapter~\ref{sec:chapa}}, which focuses on the biggest question in
QCD: the nature of confinement.
No experiment has detected a colored object in isolation, suggesting that colored objects are trapped inside
color-singlet hadrons.
Chapter~\ref{sec:chapa} focuses on theoretical aspects of confinement and the related phenomenon of 
chiral-symmetry breaking, and how they arise in non-Abelian gauge theories.

\newpage

QCD provides a loose prototype of strongly coupled theories, which are reviewed in
\textbf{Chapter~\ref{sec:chapg}}.
Supersymmetry, string theory, and the AdS/CFT correspondence all play a role in this chapter.
These ideas modify the dynamics of gauge theories profoundly.
Non-supersymmetric theories are also described here, though they are most interesting when the fermion
content is such that the dynamics differ markedly from those of QCD, because they then are candidate models
of electroweak symmetry breaking.
Conformal symmetry is also presented here, both as to help understand the phase diagram of non-Abelian gauge
theories, and to develop additional models of new physics.
New exact results in field theories, sometimes inspired by string theory, are put forward, and their
connection to computations of scattering amplitudes in QCD, with many legs or at many loops, is discussed.
Chapter~\ref{sec:chapg} further discusses techniques devised for strongly-coupled particle physics and their
interplay with condensed-matter physics.

\vskip 54pt

Chapters~\ref{sec:chapb}--\ref{sec:chapg} all contain a section on future directions discussing the most
important open problems and challenges, as well as the most interesting avenues for further research.
The Appendix provides a list of acronyms explaining the meaning of abbreviations used throughout the review
for laboratories, accelerators, and scientific collaborations.
Where available, we provide links to web sites with more information.

\clearpage
\section[Chapzs]{The nature of QCD \protect\footnotemark}
\footnotetext{Contributing authors: N.~Brambilla, S.~Eidelman, P.~Foka, S.~Gardner, 
X.~Garcia~i~Tormo, A.S.~Kronfeld, R. Vogt}
\label{sec:chapzz}

QCD is the sector of the Standard Model (SM) of particle physics that describes the strong interactions of
quarks and gluons.
From a modern perspective, both the SM and general relativity are thought to be effective field theories,
describing the low-energy limit of a more fundamental framework emergent at high energies.
To begin, we would like to focus on one specific theoretical aspect, because it shows how QCD plays a
special role in the~SM.

In quantum field theory, couplings are best understood as depending on an energy scale; roughly speaking,
this is the scale at which the quantum field theory --- understood to be an effective field theory --- is
defined.
In some cases, such as that of the hypercharge coupling or the Higgs self-coupling in the SM, this energy
dependence is such that the coupling increases with increasing energy.
This behavior predicts the failure of the theory at the shortest distance scales.
QCD, on the other hand, is asymptotically free, which means the following.
The QCD Lagrangian in the zero-quark-mass limit is scale invariant, and the interactions of the quarks are
determined by the dimensionless parameter $\als$.
The theory at the quantum (loop) level generates a fundamental, dimensionful scale $\Lambda_{\rm QCD}$ which
controls the variation of the coupling constant $\als$ with energy scale.
In QCD (unlike QED), the coupling \emph{decreases} with increasing energy --- as spectacularly confirmed in
the kinematic variation of cross-section measurements from high-precision, deep-inelastic scattering data.
The decrease is just fast enough that QCD retains its self-consistency in all extreme energy regimes: high
center-of-mass scattering energies, of course, but also high temperatures and large baryon chemical
potentials, etc.
In this way, QCD is the paradigm of a complete physical theory.

Asymptotic freedom allows accurate calculations at high energy with perturbation theory.
The success of the technique does not remove the challenge of understanding the nonperturbative aspects of
the theory.
The two aspects are deeply intertwined.
The Lagrangian of QCD is written in terms of quarks and gluons degrees of freedom which become apparent at
large energy but remain hidden inside hadrons in the low-energy regime.
This confinement property is related to the increase of $\alpha_s$ at low energy, but it has never been
demonstrated analytically.
We have clear indications of the confinement of quarks into hadrons from both experiments and lattice QCD.
Computations of the heavy quark-antiquark potential, for example, display a linear behavior in the
quark-antiquark distance, which cannot be obtained in pure perturbation theory.
Indeed the two main characteristics of QCD: confinement and the appearance of nearly massless pseudoscalar
mesons, emergent from the spontaneous breaking of chiral symmetry, are nonperturbative phenomena whose
precise understanding continues to be a target of research.
Even in the simpler case of gluodynamics in the absence of quarks, we do not have a precise understanding of
how a gap in the spectrum is formed and the glueball spectrum is generated.
Glueball states are predictions of QCD, and their mass spectrum can be obtained with lattice-QCD
calculations.
They have not, however, been unambiguously observed; their predicted mass and width can be significantly
modified by $q\bar q$ mixing effects.

The vacuum of QCD is also difficult to characterize.
One possibility is to characterize the vacuum in terms of several nonperturbative objects.
Such a parametrization has been introduced first in the sum rules approach, yielding a separation of short-
and long-distance physics based on techniques derived from the existence of asymptotic freedom in QCD.
These ideas have proven to be of profound importance, though the specifics have been supplanted, broadly
speaking, by effective field theories in QCD, which, as discussed further in Sec.~\ref{intro:sec:theory},
systematically separate the high- and low-energy contributions.

Once a low-energy (nonperturbative), gauge-invariant quantity has been defined, one could use it to
investigate the low-energy degrees of freedom which could characterize it and their relation to the
confinement mechanism.
Even in the absence of quarks, there is a fascinating and complex landscape of different possible
topological objects: monopoles, vortices, calorons, or dyons, which are investigated using different
methods; either lattice-QCD calculations or QCD vacuum models can be used to this end.
Some of the recent research in this sector is addressed in Chapter~\ref{sec:chapa}.

\subsection{Broader themes in QCD}
\label{intro:sec:broader}

Many of the most influential ideas in field theory have emerged while trying to understand QCD.
The renormalization-group methods of Kenneth Wilson, where short-distance degrees of freedom are
systematically removed, or ``integrated out,'' began with attempts to understand the scale invariance of the
strong interaction.
These ideas flourished in critical phenomena and statistical mechanics, before returning to particle physics
after the asymptotic freedom of gauge theories was discovered.
It is this view of renormalization that provides QCD the high-energy self-consistency we have discussed, and
has also led to one of the two key facets of modern effective field theory.
The other key lies in the work of Steven Weinberg, who argued on the grounds of unitarity and analyticity
that the correct effective Lagrangian would consist of all the operators with the desired fields and
symmetries.
This idea is crucial to the analysis of QCD, because it allows the introduction of an effective theory whose
fields differ from the original ones.
For example, the chiral Lagrangian contains pions and, depending on the context, other hadron fields, but
not quarks and gluons.
Certainly, QCD has been at the heart of the development of most of our tools and ideas in the construction
of the Standard Model.

QCD also has a distinguished pedigree as a description of experimental observations.
It is a merger of two insightful ideas, the quark model and the parton model, which were introduced to
explain, respectively, the discovery of the hadron ``zoo'' in the 1960s and then the deep-inelastic
scattering events seen in the early 1970s.
The acceptance of QCD was forced on us by several discoveries, such as the $J/\psi$ and other charmonium
states in 1974, the analogous $\Upsilon$ and bottomonium states in 1977, and the first observation of
three-jet events, evoking the gluon, in 1979.

Some themes in QCD recur often enough that they appear in many of the chapters to follow, so we list them 
here: 

\medskip
QCD gives rise to the visible mass of the Universe, including everyday objects--- %
The confinement scale, $\Lambda_{\rm QCD}$, sets the mass of the proton and the neutron.
Similar dynamics could, conceivably, play a role in generating the mass of other forms of matter.
\emph{Thus, the confinement mechanism pertains to the origin of mass.}

\medskip
QCD controls many parameters of the SM--- %
QCD is needed to determine $\als$, the six masses of the quarks, and the strong CP-violating parameter, as
well as the Cabibbo-Kobayashi-Maskawa (CKM) mixing matrix.
These tally to 12 parameters, out of the 19 of the SM (or 26--28 with neutrino masses and mixing).
The quark masses and CKM paremeters stem from, and the strong-CP parameter is connected to, the poorly
understood Yukawa couplings of quarks to the Higgs boson; furthermore, $\als$ may unify with the other gauge
couplings.
\emph{Thus, quark couplings are essential in the search for a more fundamental theory.}

\medskip
QCD describes the SM background to non-SM physics--- %
In the high-energy regime, where the coupling constant is small and perturbation theory is applicable, QCD
predicts the calculable background to new phenomena precisely.
For example, QCD calculations of the background were instrumental to the Higgs discovery, and, indeed, QCD
is
ubiquitous at hadron colliders where direct contributions of new physics are most actively sought.
\emph{Thus, QCD plays a fundamental role in our investigations at the high-energy frontier.}

\medskip
In the low energy-regime, QCD is often the limiting factor in the indirect search for non-SM physics--- %
This is true in all searches for new physics in hadronic systems, be it in the study of CP violation in $B$
decays, or in permanent electric dipole moment searches in hadrons and nuclei.
In addition, QCD calculations of hadronic effects are also needed to understand the anomalous magnetic
moment of the muon, as well as aspects of neutrino physics.
\emph{ Thus, QCD also plays a fundamental role in searches for new physics at the intensity frontier.}

\pagebreak 
Nuclear matter has a fascinating phase diagram--- %
At nonzero temperature and nonzero chemical potential, QCD exhibits a rich phase diagram, which we continue
to explore.
The QCD equation of state, the possibility of phase transitions and/or crossovers, and the experimental
search for the existence of a critical point are all current topics of research.
In lattice QCD one can also alter the number of fermions and the number of colors in order to study
different scenarios.
In addition to the hadronic phase, different states of QCD matter are predicted, such as the quark-gluon
plasma, quarkyonic matter, and a color superconductor state with diquark matter.
Experiments studying heavy-ion collisions have shown the quark-gluon plasma to be a surprising substance.
For example, it seems to be a strongly-coupled, nearly perfect liquid with a minimal ratio of shear
viscosity to entropy density.
\emph{Thus, QCD matter in extreme conditions exhibits rich and sometimes unexpected behavior.}

\medskip
QCD impacts cosmology--- %
Probing the region of the QCD phase diagrams at large temperature allows us to probe conditions which have
not existed since the beginning of the universe.
The new state of matter formed in heavy-ion collisions existed microseconds after the Big~Bang, before
hadrons emerged as the universe cooled.
\emph{Thus, characterizing the quark-gluon plasma provides information about the early universe.}

\medskip
QCD is needed for astrophysics--- %
The region of the QCD phase diagram at large chemical potential provides information on the system under
conditions of high pressure and large density, as is the case for astrophysical objects such as compact
stars.
These stars could be neutron stars, quark stars, or hybrids somewhere in between these pure limits.
Moreover, one can use astrometric observational data on such objects to help characterize the QCD equation
of state.
\emph{Thus, terrestrial accelerator experiments and astrophysical observations are deeply connected.}

\medskip
QCD is a prototype of strongly-coupled theories--- %
Strongly-coupled gauge theories have been proposed as alternatives to the SM Higgs mechanism.
Strongly-coupled mechanisms may also underlie new sectors of particle physics that could explain the origin
of dark matter.
Furthermore, the relation between gauge theories and string theories could shed light on the unification of
forces.
\emph{Thus, QCD provides a launching pad for new models of particle physics.}

\medskip
QCD inspires new computational techniques for strongly-interacting systems--- %
As the prototype of an extremely rich, strongly-coupled system, the study of QCD requires a variety of
analytical tools and computational techniques, without which progress would halt.
These developments fertilize new work in allied fields; for example, QCD methods have helped elucidate the
universal properties of ultracold atoms.
Conversely, developments in other fields may shed light on QCD itself.
For example, the possibility of designing arrays of cold atoms in optical lattices with the gauge symmetry
and fermion content of QCD is under development.
If successful, this work could yield a kind of quantum computer with QCD as its specific application.
\emph{Thus, the challenge of QCD cross-fertilizes other fields of science.}

\subsection{Experiments addressing QCD}
\label{intro:sec:expt}

In this section, we offer a brief overview of the experimental tools of QCD.
We discuss $e^+ e^-$ colliders, fixed-target machines, hadron colliders, and
relativistic heavy-ion colliders from a QCD perspective.

From the 1960s to 1990s, $e^+e^-$ colliders evolved from low center-of-mass energies $\sqrt{s}\sim1$~GeV
with
modest luminosity to the Large Electron Positron (LEP) collider with $\sqrt{s}$ up to $209$~GeV and a vastly
greater luminosity.
Along the way, the $e^+e^-$ colliders PETRA (at DESY) and PEP (at SLAC) saw the first three-jet events.
A further breakthrough happened at the end of nineties with the advent of the two $B$-factories at KEK and
SLAC and the operation of lower-energy high-intensity colliders in Beijing, Cornell, Frascati, and
Novosibirsk.
Experiments at these machines are particularly good for studies of quarkonium physics and decays of
open charm and bottom mesons, in a way that spurred theoretical developments.
The copious production of $\tau$ leptons at $e^+e^-$ colliders led to a way to measure $\als$ via their
hadronic decays.
Measurements of the hadronic cross section at various energy ranges play a useful role in understanding the
interplay of QCD and QED.

Experiments with electron, muon, neutrino, photon, or also hadron beams impinging on a fixed target have been
a cornerstone of QCD.
Early studies of deep inelastic scattering at SLAC led to the parton model.
This technique and the complementary production of charged lepton pairs (the so-called Drell-Yan production)
have remained an important tool for understanding proton structure.
Later, the Hadron-Elektron Ring Anlage (HERA) continued this theme with $e^-p$ and $e^+p$ colliding beams.
In addition to nucleon structure, fixed-target experiments have made significant contributions to strangeness
and charm physics, as well as to the spectroscopy of light mesons, and HERA searched for non-SM particles
such as leptoquarks.
This line of research continues to this day at Jefferson Lab, J-PARC, Mainz, Fermilab, and CERN;
future, post-HERA $ep$ colliders are under discussion.

The history of hadron colliders started in 1971 with $pp$ collisions at CERN's Intersecting Storage Rings
(ISR), at a center-of-mass energy of 30~GeV.
The ISR ran for more than ten years with $pp$ and $p\bar{p}$ collisions, as well as ion beams: $pd$, $dd$,
$p\alpha$, and $\alpha\alpha$.
During this time, its luminosity increased by three orders of magnitude.
This machine paved the way for the successful operation of proton-antiproton colliders: the S$p\bar{p}$S at
CERN with $\sqrt{s}=630$~GeV in the 1980s, and the $p\bar{p}$ Tevatron at Fermilab with $\sqrt{s}=1.96$~TeV,
which ran until 2011.
Currently, the Large Hadron Collider (LHC) collides $pp$ beams at the highest energies in history, with a
design energy of 14~TeV and luminosity four orders of magnitude higher than the ISR.
Physics at these machines started from studies of jets at ISR and moved to diverse investigations including
proton structure, precise measurements of the $W$ mass, searches for heavy fundamental particles leading to
discoveries of the top quark and Higgs, production of quarkonia, and flavor physics.

At the same time, pioneering experiments with light ions (atomic number, $A$, around~14) at relativistic
energies started in the 1970s at LBNL in the United States and at JINR in Russia.
The program continued in the 1980s with fixed-target programs at the CERN SPS and BNL AGS.
These first experiments employed light ion beams ($A \sim 30$) on heavy targets ($A \sim 200$).
In the 1990s, the search for the quark-gluon plasma continued with truly heavy ion beams ($A \sim 200$).
In this era, the maximum center of mass energy per nucleon was $\sqrt{s_{NN}} \sim 20$~GeV.
With the new millennium the heavy-ion field entered the collider era, first with the Relativistic Heavy Ion
Collider (RHIC) at BNL at $\sqrt{s_{NN}}=200$~GeV and, in 2010, the LHC at CERN, reaching the highest
currently available energy, $\sqrt{s_{NN}}=2.76$~TeV.

The goal of heavy-ion physics is to map out the nuclear-matter phase diagram, analogous to studies of phase
transitions in other fields.
Proton-proton collisions occur at zero temperature and baryon density, while heavy-ion collisions
can quantify the state of matter of bulk macroscopic systems.
The early fixed-target experiments probed moderate values of temperature and baryon density.
The current collider experiments reach the zero baryon density, high-temperature regime, where the 
quark-gluon plasma can be studied under conditions that arose in the early universe.

While the phase structure observed in collider experiments suggests a smooth crossover from hadronic matter
to the quark-gluon plasma, theoretical arguments, augmented by lattice QCD computations, suggest a
first-order phase transition at nonzero baryon density.
The critical point where the line of first-order transitions ends and the crossover regime begins is of great
interest.
To reach the needed temperature and baryon density, two new facilities---FAIR at GSI and NICA at JINR---are
being built.

Work at all these facilities, from $e^+e^-$ to heavy ions, require the development of novel trigger systems
and detector technologies.
The sophisticated detectors used in these experiments, coupled to farms of computers for on-line data
analysis, permit the study of unprecedentedly enormous data samples, thus enabling greater sensitivity in
searches for rare processes.

\subsection{Theoretical tools for QCD}
\label{intro:sec:theory}

The theory toolkit to study QCD matter is quite diverse, as befits the rich set of phenomena it describes.
It includes QCD perturbation theory in the vacuum, semiclassical gauge theory, and techniques derived from
string theory.
Here we provide a brief outline of some of the wider ranging techniques.

\paragraph{Effective Field Theories (EFTs):}

Effective field theories are important tools in modern quantum field theory.
They grew out of the operator-product expansion (OPE) and the formalism of phenomenological Lagrangians and,
thus, provide a standard way to analyze physical systems with many different energy scales.
Such systems are very common from the high-energy domain of particle physics beyond the Standard Model to
the
low-energy domain of nuclear physics.

Crucial to the construction of an EFT is the notion of \emph{factorization}, whereby the effects in a
physical system are separated into a high-energy factor and a low-energy factor, with each factor
susceptible to calculation by different techniques.
The high-energy factor is typically calculated by making use of powerful analytic techniques, such as
weak-coupling perturbation theory and the renormalization group, while the low-energy part may be amenable
to
lattice gauge theory or phenomenological methods.
A key concept in factorization is the principle of \emph{universality}, on the basis of which a low-energy
factor can be determined from one theoretical or phenomenological calculation and can then be applied in a
model-independent way to a number of different processes.
Factorization appeared first in applications of the OPE to QCD, where a classification of operators revealed
a leading (set of) operator(s), whose short-distance coefficients could be calculated in a power series
in~$\als$.

Apart from their theoretical appeal, EFTs are an extremely practical tool.
In many cases they allow one to obtain formally consistent and numerically reliable predictions for physical
processes that are of direct relevance for experiments.
The essential role of factorization was realized early on in the analysis of deep inelastic scattering data
in QCD and is codified in the determination of parton distribution functions from experiment, allowing SM
predictions in new energy regimes.

Several properties of EFTs are important: they have a power counting in a small parameter which permits
rudimentary error assessment for each prediction; they can be more predictive if they have more symmetry;
they admit an appropriate definition of physical quantities and supply a systematic calculational
framework; finally, they permit the resummation of large logarithms in the ratio of physical scales.
For example, an object of great interest, investigated since the inception of QCD, is the heavy
quark-antiquark static energy, which can be properly defined only in an EFT and subsequently calculated with
lattice gauge theory.

The oldest example is chiral EFT for \emph{light-quark systems}, with roots stemming from the development of
current algebra in the 1960s.
Chiral EFT has supplied us with an increasingly accurate description of mesons and baryons, and it is an
essential ingredient in flavor-physics studies.
The EFT description of pion-pion scattering, together with the data on pionium formation, has given us a
precise way to confirm the standard mechanism of spontaneous breaking of chiral symmetry in QCD.
Chiral effective theory has also allowed lattice QCD to make contact with the physical region of light-quark
masses from simulations with computationally less demanding quark masses.
For more details, see Chapters~\ref{sec:chapb} and~\ref{sec:chape}.

In the case of the heavy quark-antiquark bound states known as \emph{quarkonium}, the EFT known as
Nonrelativistic QCD (NRQCD) separates physics at the scale of the heavy-quark mass from 
those related to the dynamics of quarkonium binding.
This separation has solved the problem of uncontrolled infrared divergences in theoretical calculations and
has opened the door to a systematic improvement of theoretical predictions.
It has given us the tools to understand the data on the quarkonium production cross section at high-energy
colliders, such as the Tevatron, the $B$~factories and the LHC.
It has also made it clear that a complete understanding of quarkonium production and decay involves
processes
in which the quark-antiquark pairs are in a color-octet state, as well as process in which the pairs are in
a
color-singlet state.
New, lower-energy EFTs, such as potential NRQCD (pNRQCD) have given greater control over some technical
aspects of theoretical calculations and have provided a detailed description of the spectrum, decays, and
transitions of heavy quarkonia.
These EFTs allow the precise extraction of the Standard Model parameters, which are relevant for new-physics
searches, from the data of current and future experiments.
See Chapter~\ref{sec:chapc} for applications of NRQCD and pNRQCD.

In the case of strong-interaction processes that involve large momentum transfers and energetic, nearly
massless particles, Soft Collinear Effective Field Theory (SCET) has been developed.
It has clarified issues of factorization for high-energy processes and has proved to be a powerful tool for
resumming large logarithms.
SCET has produced applications over a wide range of topics, including heavy-meson decays, deep-inelastic
scattering, exclusive reactions, quarkonium-production processes, jet event shapes and jet quenching.
Recent developments regarding these applications can be found in Chapters~\ref{sec:chapb}, 
\ref{sec:chapc}, and~\ref{sec:chape}.

In \emph{finite-temperature} and \emph{finite-density} physics, EFTs such as {Hard Thermal Loop} (HTL),
Electric QCD, Magnetic QCD, $\textrm{NRQCD}_\textrm{HTL}$, or p$\textrm{NRQCD}_\textrm{HTL}$ have allowed
progress on problems that are not accessible to standard lattice QCD, such as the evolution of heavy
quarkonia in a hot medium, thermodynamical properties of QCD at the very high temperatures thermalization
rate of non-equilibrium configurations generated in heavy-ion collision experiments, and the regime of
asymptotic density.
These developments are discussed in Chapter~\ref{sec:chapd}.

In \emph{nuclear physics}, chiral perturbation theory has been generalized to provide a QCD foundation to
nuclear structure and reactions.
EFTs have allowed, among other things, a model-independent description of hadronic and nuclear interactions
in terms of parameters that will eventually be determined in lattice calculations, new solutions of
few-nucleon systems that show universality and striking similarities to atomic systems near Feshbach
resonances, derivation of consistent currents for nuclear reactions, and new approaches to understanding
heavier nuclei (such as halo systems) and nuclear matter.
Some recent developments are discussed in Chapter~\ref{sec:chapf}.

\paragraph{Lattice gauge theory:} 

In the past decade, numerical lattice QCD has made enormous strides.
Computing power, combined with new algorithms, has allowed a systematic simulation of sea quarks for the
first time.
The most recently generated ensembles of lattice gauge fields now have 2+1+1 flavors of sea quark,
corresponding to the up and down, strange, and charm quarks.
Most of this work uses chiral EFT to guide an extrapolation of the lightest two quark masses to the physical
values.
In some ensembles, however, the (averaged) up and down mass is now as light as in nature, obviating this
step.
Many quantities now have sub-percent uncertainties, so that the next step will require electromagnetism and
isospin breaking (in the sea).

Some of the highlights include baryon masses with errors of 2--4\%;
pion, kaon and $D$-meson matrix elements with total uncertainty of 1--2\%;
$B$-meson matrix elements to within 5--8\%.
The light quark masses are now known directly from QCD (with the chiral extrapolation), with few per cent
errors.
Several of the best determinations of $\als$ combine perturbation theory (lattice or continuum) with
nonperturbatively computed quantities; these are so precise, because the key input from experiment is just
the scale, upon which $\als$ depends logarithmically.
A similar set of analyses yield the charm- and bottom-quark masses with accuracy comparable to perturbative
QCD plus experiment.
Lattice QCD has also yielded a wealth of thermodynamic properties, not least showing that the deconfinement
transition (at small chemical potential) is a crossover, and the crossover temperature has now been found
reproducibly.

Vigorous research, both theoretical and computational, is extending the reach of this tool into more
demanding areas.
The computer calculations take place in a finite spatial box (because computers' memories are finite), and
two-body states require special care.
In the elastic case of $K\to\pi\pi$ transitions, the required extra computing is now manageable, and
long-sought calculations of direct CP violation among neutral kaons, and related decay rates, now appear on
the horizon.
This success has spurred theoretical work on inelastic, multi-body kinematics, which will be required before
long-distance contributions to, say, $D$-meson mixing can be computed.
Nonleptonic $B$ and $D$ decays will also need these advances, and possibly more.
In the realm of QCD thermodynamics, the phase diagram at nonzero chemical potential suffers from a fermion
sign problem, exactly as in many condensed-matter problems.
This problem is difficult, and several new ideas for workarounds and algorithms are being investigated.

\paragraph{Other nonperturbative approaches:}

The theoretical evaluation of a nonperturbative contribution arising in QCD requires nonperturbative
techniques.
In addition to lattice QCD, many models and techniques have been developed to this end.
Among the most used techniques are: the limit of the large number of colors, generalizations of the original
Shifman-Vainshtein-Zakharov sum rules, QCD vacuum models and effective string models, the AdS/CFT
conjecture,
and Schwinger-Dyson equations.
Every chapter reports many results obtained with these alternative techniques.

\subsection{Fundamental parameters of QCD}
\label{intro:sec:alphas}

Precise determinations of the quark masses and of $\als$ are crucial for many of the problems discussed in
the chapters to come.
As fundamental parameters of a physical theory, they require both experimental and theoretical input.
Because experiments detect hadrons, inside which quarks and gluons are confined, the parameters cannot be
directly measured.
Instead, they must be determined from a set of relations of the form
\begin{equation}
    [M_{\rm HAD}(\Lambda_{\rm QCD}, m_q)]^{\rm TH}=[M_{\rm HAD}]^{\rm EXP}.
\end{equation}
One such relation is needed to determine $\Lambda_{\rm QCD}$, the parameter which fixes the value of
$\als(Q^2)$, the running coupling constant, at a squared energy scale $Q^2$; another six are needed for the
(known) quarks --- and yet another for the CP-violating angle~$\bar\theta$.
The quark masses and $\als$ depend on the renormalization scheme and scale, so that care is needed to ensure
that a consistent set of definitions is used.
Some technical aspects of these definitions (such as the one known as the renormalon ambiguity) are
continuing objects of theoretical research and can set practical limitations on our ability to determine the
fundamental parameters of the theory.
In what follows, we have the running coupling and running masses in mind.

Measurements of $\als$ at different energy scales provide a direct quantitative verification of asymptotic
freedom in QCD.
From the high-energy measurement of the hadronic width of the $Z$ boson, one obtains
$\als(M_Z)=0.1197\pm0.0028$~\cite{Beringer:1900zz}.
From the lower-energy measurement of the hadronic branching fraction of the $\tau$ lepton, one obtains,
after
running to the $Z$~mass, $\als(M_Z^2)=0.1197\pm0.0016$~\cite{Beringer:1900zz}.
At intermediate energies, several analyses of quarkonium yield values of $\als$ in agreement with these two;
see Sec.~\ref{sec:secC5}.
The scale of the $\tau$ mass is low enough that the error assigned to the latter value remains under
discussion; see Sec.~\ref{sec:secB5} for details.
Whatever one makes of these issues, the agreement between these two determinations provides an undeniable
experimental verification of the asymptotic freedom property of QCD.

One can combine $\als$ extractions from different systems to try to obtain a precise and reliable
``world-average'' value.
At present most (but not all) individual $\als$ measurements are dominated by systematic uncertainties of
theoretical origin, and, therefore, any such averaging is somewhat subjective.
Several other physical systems, beyond those mentioned above, are suitable to determine~$\als$.
Those involving heavy quarks are discussed in Sec.~\ref{sec:secC5}.
Lattice QCD provides several different $\als$ determinations.
Recent ones include~\cite{McNeile:2010ji,Shintani:2010ph,Aoki:2009tf,Blossier:2012ef}, in addition to those
mentioned in Sec.~\ref{sec:secC5}.
Some of these determinations quote small errors, because the nonperturbative part is handled cleanly.
They therefore may have quite an impact in world-averages, depending on how those are done.
For example, lattice determinations dominate the error of the current PDG world
average~\cite{Beringer:1900zz}.
Fits of parton-distribution functions (PDFs) to collider data also provide a good way to determine $\als$.
Current analyses involve sets of PDFs determined in next-to-next-to-leading order
(NNLO)~\cite{Martin:2009bu,Ball:2011us,Alekhin:2012ig,JimenezDelgado:2008hf}.
Effects from unknown higher-order perturbative corrections in those fits are difficult to assess, however,
and have not been addressed in detail so far.
They are typically estimated to be slightly larger than the assigned uncertainties of the NNLO extractions.
Jet rates and event-shape observables in $e^+e^-$ collisions can also provide good sensitivity to $\als$.
Current state-of-the-art analyses involve NNLO fixed-order
predictions~\cite{GehrmannDeRidder:2009dp,GehrmannDeRidder:2008ug,GehrmannDeRidder:2007hr,GehrmannDeRidder:2007jk,GehrmannDeRidder:2007bj,Weinzierl:2009yz,Weinzierl:2009ms,Weinzierl:2008iv},
combined with the resummation of logarithmically enhanced terms.
Resummation for the event-shape cross sections has been performed both in the traditional diagrammatic
approach~\cite{Monni:2011gb} and within the SCET framework~\cite{Becher:2008cf,Chien:2010kc,Becher:2012qc}.
One complication with those extractions is the precise treatment of hadronization effects.
It is by now clear~\cite{Dissertori:2009ik} that analyses that use Monte Carlo generators to estimate
them~\cite{Dissertori:2009ik,OPAL:2011aa,Bethke:2008hf,Becher:2008cf,Chien:2010kc} tend to obtain larger
values of $\als$ than those that incorporate power corrections
analytically~\cite{Gehrmann:2012sc,Abbate:2010xh,Gehrmann:2009eh,Abbate:2012jh,Davison:2008vx}.
Moreover, it may not be appropriate to use Monte Carlo hadronization with higher-order resummed
predictions~\cite{Gehrmann:2012sc,Abbate:2010xh,Gehrmann:2009eh}.
We also mention that analyses employing jet rates may be less sensitive to hadronization
corrections~\cite{Frederix:2010ne,Dissertori:2009qa,Schieck:2012mp,Schieck:2006tc}.
The SCET-based results of Refs.~\cite{Abbate:2010xh,Abbate:2012jh} quote remarkably small errors; one might
wonder if the systematics of the procedure have been properly assessed, since the extractions are based only
on thrust.
In that sense, we mention analogous analyses that employ heavy-jet mass, the $C$-parameter, and broadening
are within reach and may appear in the near future.
Note that if one were to exclude the event-shape $\als$ extractions that employ Monte Carlo hadronization,
the impact on the PDG average could be quite significant.
Related analyses employing deep-inelastic scattering data can also be performed~\cite{Kang:2013nha}.

Light-quark masses are small enough that they do not have a significant impact on typical hadronic
quantities.
Nevertheless, the observed masses of the light, pseudoscalar mesons, which would vanish in the
zero-quark-mass limit, are sensitive to them.
Moreover, various technical methods are available in which to relate the quark and hadron masses in this
case.
We refer to Secs.~\ref{sec:subsecB3lattice} and \ref{sec:subsecB32} for discussions of the determination of
the light quark masses from lattice QCD and from chiral perturbation theory.
To determine light-quark masses, one can take advantage of chiral perturbation theory, lattice-QCD
computations, and QCD sum rule methods.
Current progress in the light-quark mass determinations is largely driven by improvements in lattice QCD.

Earlier lattice simulations use $n_f=2$ flavors of sea quark (recent results include
Refs.~\cite{Fritzsch:2012wq,Blossier:2010cr}), while present ones use $n_f=2+1$ (recent results include
Refs.~\cite{Arthur:2012opa,Laiho:2011np,Durr:2010vn,Bazavov:2009fk}).
The influence of charmed sea quarks will soon be studied~\cite{Bazavov:2011fh,Baron:2010bv}.
In addition, some ensembles no longer require chiral extrapolations to reach the physical mass values.
The simulations are almost always performed in the isospin limit, $m_u=m_d=:m_{ud}$, $m_{ud}=(m_u+m_d)/2$,
therefore what one can directly obtain from the lattice is $m_s$, the average $m_{ud}$, and their ratio.
We mention that there is a new strategy to determine the light-quark masses which consists in computing the
ratio $m_c/m_s$, combined with a separate calculation for $m_c$, to obtain
$m_s$~\cite{McNeile:2010ji,Durr:2011ed}.
The advantage of this method is that the issue of lattice renormalization is traded for a continuum
renormalization in the determination of $m_c$.
With additional input regarding isospin-breaking effects, from the lattice results in the isospin limit one
can obtain separate values for $m_u$ and $m_d$; see Sec.~\ref{sec:subsecB3lattice} for additional
discussion.
With the present results, one obtains that $m_u\neq0$, so that the strong-CP problem is not solved by having
a massless $u$ quark~\cite{Beringer:1900zz,Colangelo:2010et,Aoki:2013ldr}; see Sec.~\ref{sec:secE8} for
further
discussion of this issue.

In contrast, heavy-quark masses also affect several processes of interest; for instance, the $b$-quark mass
enters in the Higgs decay rate for $H\to b\bar{b}$.
Many studies of Higgs physics do not, however, use the latest, more precise determinations of $m_b$.
The value of the top-quark mass is also necessary for precision electroweak fits.
To study heavy-quark masses, $m_Q$, one can exploit the hierarchy $m_Q\gg\Lambda_{\rm QCD}$ to construct
heavy-quark effective theories that simplify the dynamics; and additionally take advantage of high-order
perturbative calculations that are available for these systems, and of progress in lattice-QCD computations.
One of the best ways to determine the $b$ and $c$ masses is through sum-rule analyses, that compare
theoretical predictions for moments of the cross section for heavy-quark production in $e^+e^-$ collisions
with experimental data (some analyses that appeared in recent years include~\cite{Dehnadi:2011gc,%
Chetyrkin:2009fv,Hoang:2012us,Penin:2014zaa}) or lattice QCD (e.g.,~\cite{McNeile:2010ji}).
In those analyses, for the case of $m_c$, the approach with lattice QCD gives the most precise determination,
and the errors are mainly driven by perturbative uncertainties.
For $m_b$, the approach with $e^+e^-$ data still gives a better determination, but expected lattice-QCD
progress in the next few years may bring the lattice determination to a similar level of precision.
A complementary way to obtain the $c$-quark mass is to exploit DIS charm production measurements in PDF
fits~\cite{Alekhin:2012vu}.
The best measurement of the top-quark mass could be performed at a future $e^+e^-$ collider, but improvements
on the mass determination, with respect to the present precision, from LHC measurements are possible.

\clearpage
\section[Light Quarks]{Light quarks \protect\footnotemark}
\footnotetext{Contributing authors: R.~Alkofer$^{\dagger}$, J.~L.~Goity$^{\dagger}$, B.~Ketzer$^{\dagger}$, 
H.~Sazdjian$^{\dagger}$, H.~Wittig$^{\dagger}$, S.~Eidelman, S.~Gardner, A.~S.~Kronfeld, 
Felipe J.~Llanes-Estrada, A. Pich, J.-W.~Qiu, C.~Salgado, N.~G.~Stefanis.}
\label{sec:chapb}

\subsection{Introduction}
\label{sec:secB2.intro}
The study of light-quark physics is central to the understanding of QCD.
Light quarks represent a particularly sensitive probe of the 
strong interactions, especially
of nonperturbative effects.

In the two extreme regimes of QCD, namely, in the low-energy regime 
where the energies are (much) smaller than a
typical strong interaction scale $\sim m_\rho$, and in
the high-energy regime where the energies are much higher
than that scale, there are well-established 
theoretical methods, namely, Chiral Perturbation Theory (ChPT) 
and perturbative QCD, respectively, 
that allow for a discussion of the physics 
in a manner consistent with the
fundamental theory, and thus permit in this way to define and 
quantify effects in a more or less rigorous way.
The intermediate-energy regime is less developed 
as there are no analytic methods which would allow for a
complete discussion of the physics, thus requiring 
the introduction of methods which require some degree of
modeling.
However, as discussed in this chapter, 
methods based fundamentally on QCD, such as those based on the
framework of Schwinger-Dyson equations, 
have made great advances, and a promising future lies ahead.
Advances in lattice QCD, in which the excited hadron 
spectrum can be analyzed, are opening new
perspectives for understanding the 
intermediate-energy regime of QCD; and one should expect that this will
result in new strategies, methods, and ideas.
Progress on all of the mentioned fronts continues, 
and in this chapter a representative number of the most
exciting developments are discussed.

Never before has 
the study of the strong interactions 
had as many sources of experimental results as today.
Laboratories and experiments 
around the world, ranging from low- to high-energy accelerators, as well as 
in precision
nonaccelerator physics, give unprecedented access 
to the different aspects of QCD, and 
to light-quark physics in particular. 
In this chapter a broad sample of 
experiments and results from these venues will be given. 
 
The objective of this chapter is to present a selection 
of topics in light-quark physics: partonic structure of light hadrons, 
low-energy properties and structure, excited hadrons, 
the role of light-quark physics in 
extracting fundamental QCD parameters, such as $\alpha_s$ at the
GeV scale, and also of theoretical methods, namely, 
ChPT, perturbative QCD,  
Schwinger-Dyson equations, and lattice QCD.

This chapter is organized as follows: Sec.~\ref{sec:lq.struct} is devoted to 
hadron structure and contains the following topics: 
parton distributions 
(also including their transverse momentum dependence), 
hadron form factors, and generalized parton distributions (GPDs), lattice QCD
calculations of form factors and moments 
of the parton distributions, 
along with a discussion of 
the proton radius puzzle; finally, 
the light pseudoscalar meson form factors, the
neutral pion lifetime, 
and the charged pion polarizabilities complete the section.
Section \ref{sec:lq.spec} deals with hadron spectroscopy and summarizes lattice QCD 
and continuum methods and results, 
along with a detailed presentation of experimental results and perspectives.
Section \ref{sec:secB3} addresses chiral dynamics, 
including studies based on ChPT and/or on lattice QCD.
In Sec.~\ref{sec:secB4} the role of light quarks in precision 
tests of the Standard Model is discussed, with 
the hadronic contributions 
to the muon's anomalous magnetic moment as a particular focus.
The running of the electroweak mixing angle, 
as studied through the weak charge of the
proton, and
the determination
of the strong coupling $\als$ from $\tau$ decay are also addressed.
Finally, Sec.~\ref{sec:secB6} presents some thoughts on future directions.


\subsection{Hadron structure}
\label{sec:lq.struct}

\subsubsection{Parton distribution functions in QCD}
\label{sec:lq.struct.PDF-TMD-theory}
The description of hadrons within QCD
faces severe difficulties because the strength of the color forces becomes
large at low energies and the confinement properties of quarks and
gluons cannot be ignored.  The main concepts and techniques for
treating this nonperturbative QCD regime are discussed in
Chapter~\ref{sec:chapa}, which is devoted to infrared QCD.
Here, we focus on those quantities that enter the description of
hadronic processes in which a large momentum scale is involved,
thus enabling the application of factorization theorems.
Factorization theorems provide the possibility (under certain
assumptions) to compute the cross section for high-energy hadron
scattering by separating short-distance from long-distance effects in
a systematic way.  The hard-scattering partonic processes are
described within perturbative QCD, while the distribution of partons
in a particular hadron --- or of hadrons arising from a particular parton
in the case of final-state hadrons --- is encoded in universal parton
distribution functions (PDFs) or parton fragmentation functions
(PFFs), respectively.  These quantities contain the dynamics of
long-distance scales related to nonperturbative physics and thus are
taken from experiment.

To see how factorization works, consider the measured cross section in
deep-inelastic scattering (DIS) for the generic process
lepton + hadron
$A \to {\rm lepton^{\prime}}$ + anything else~$X$:
\begin{equation}
  d\sigma
=
  \frac{d^{3}\bf{k}'}{2s |\bf{k}'|} \frac{1}{(q)^2}
  L_{\mu\nu}(k,q) W^{\mu\nu}(p,q) \, ,
\label{eq:cs-DIS}
\end{equation}
where $k$ and $k'$ are the incoming and outgoing lepton momenta, $p$
is the momentum of the incoming nucleon (or other hadron),
$s=(p+k)^2$,
and $q$ is
the momentum of the exchanged photon.  The leptonic tensor
$L_{\mu\nu}(k,q)$ is known from the electroweak Lagrangian, whereas
the hadronic tensor $W^{\mu\nu}(p,q)$ may be expressed in terms of
matrix elements of the electroweak currents to which the vector bosons
couple, viz., \cite{Collins:1989gx}
\begin{equation}
  W^{\mu\nu}
=
  \frac{1}{4\pi} \int_{}^{}d^4y e^{iq\cdot y}
  \sum_{X} \left\langle A|j^\mu (y)|X\right\rangle
  \left\langle X|j^\nu (0)|A\right\rangle \, .
\label{eq:hadr-tens}
\end{equation}
For $Q^2=-q^2$ large and Bjorken $x_B=Q^2/2p\cdot q$ fixed, $W^{\mu\nu}$ can be written in the form
of a factorization theorem to read
\begin{eqnarray}
  && \!\!\!\!\!\!\!\!\!\! W^{\mu\nu}(q,p)
=
  \sum_{a} \int_{x_B}^{1} \frac{dx}{x}
  f_{a/A}(x, \mu)\times
\nonumber \\
&& 
  H_{a}^{\mu\nu}(q,xp, \mu, \alpha_s(\mu))
  + \mbox{remainder} \, ,
\label{eq:hadr-fac-the}
\end{eqnarray}
where $f_{a/A}(x, \mu)$ is the PDF
for a
parton $a$ (gluon, $u$, $\bar{u}$, $\ldots$) in a hadron $A$ carrying
a fraction $x$ of its momentum and probed at a factorization scale $\mu$,
$H^{\mu\nu}_a$ is the short-distance contribution of partonic scattering
on the parton $a$, and the sum runs over all possible types of parton $a$.
In Eq.~(\ref{eq:hadr-fac-the}), the (process-dependent) remainder is
suppressed by a power of $Q$.

In deep inelastic scattering (DIS) experiments, $lA \to l^{\prime}X$, we learn about the
longitudinal distribution of partons inside hadron $A$, e.g., the nucleon.
The PDF for a quark $q$ in a hadron $A$ can be defined in a
gauge-invariant way (see \cite{Collins:1989gx} and references cited
therein) in terms of the following matrix element
\begin{eqnarray}
  && \!\!\!\!\!\!\!\!\!\! f_{q/A}(x,\mu)
=
  \frac{1}{4\pi} \int_{}^{} dy^- e^{-i x p^+ y^-}
  \langle
  p| \bar{\psi}(0^+,y^-,{\bf{0}}_T) \times
\nonumber \\
&& \ \ \
  \gamma^+ \mathcal{W}(0^-,y^-) \psi(0^+,0^-,{\bf{0}}_T) |p
  \rangle \, ,
\label{eq:quark-pdf}
\end{eqnarray}
where the light-cone notation, $v^{\pm}=(v^0\pm v^3)/\sqrt{2}$
for any vector $v^\mu$, was used.
Here, $\mathcal{W}$ is the Wilson line operator in the fundamental
representation of $SU(3)_c$,
\begin{equation}
  \mathcal{W}(0^-,y^-)
  =
  {\cal P} \exp \left[
                      ig \int_{0^-}^{y^-} dz^- A_{a}^{+}(0^+, z^-, {\bf{0}}_T)t_a
                \right]
  \label{eq:Wilson-line}
\end{equation}
along a lightlike contour from $0^-$ to $y^-$ with a gluon
field $A_a^\mu$ and the generators $t_a$ for $a=1,2,\dots,8$.
Here, $g$ is the gauge coupling, such that $\als=g^2/4\pi$.
Analogous definitions hold for the antiquark and the gluon --- the latter in
the adjoint representation. These collinear PDFs (and also the fragmentation
functions) represent lightcone correlators of leading twist in which
gauge invariance is ensured by lightlike Wilson lines (gauge links).
The factorization scale $\mu$ dependence of PDFs is controlled by the
DGLAP (Dokshitzer-Gribov-Lipatov-Altarelli-Parisi)
\cite{Altarelli:1977zs,Gribov:1972ri,Dokshitzer:1977sg} evolution
equation \cite{Collins:1981va,Collins:1981uw}.
The PDFs represent the universal part in the factorized cross section
of a collinear process such as Eq.~(\ref{eq:hadr-fac-the}).
They are independent of any specific process in which they are measured.
It is just this universality of the PDFs that ensures the predictive
power of the factorization theorem. For example, the PDFs for the
Drell-Yan (DY) process \cite{Drell:1970yt} are the same as in DIS,
so that one can measure them in a DIS experiment and then use them to
predict the DY cross section \cite{Collins:1989gx,Collins:QCD11}.

The predictive power of the QCD factorization theorem also relies
on our ability to calculate the short-distance, process-specific
partonic scattering part, such as $H_{a}^{\mu\nu}$ in
Eq.~(\ref{eq:hadr-fac-the}), in addition to the universality of the
PDFs.
Since the short-distance partonic scattering part is insensitive
to the long-distance hadron properties, the factorization formalism
for scattering off a hadron in Eq.~(\ref{eq:hadr-fac-the})
should also be valid for scattering off a partonic state.  Applying the
factorization formalism to various partonic states $a$, instead
of the hadron $A$, the short-distance partonic part, $H^{\mu\nu}_a$
in Eq.~(\ref{eq:hadr-fac-the}), can be systematically extracted
by calculating the partonic scattering cross section on the left
and the PDFs of a parton on the right of Eq.~(\ref{eq:hadr-fac-the}),
order-by-order in powers of $\alpha_s$ in perturbative QCD.
The validity of the collinear factorization formalism ensures that any
perturbative collinear divergence of the partonic scattering cross section
on the left is completely absorbed into the PDFs of partons on the right.
The Feynman rules for
calculating PDFs and fragmentation functions have been derived in
\cite{Collins:1981va,Collins:1981uw} having recourse to the concept of
eikonal lines and vertices.  Proofs of factorization of DIS and the DY
process can be found in \cite{Collins:1989gx} and the original works
cited therein.

One of the most intriguing aspects of QCD is the relation between
its fundamental degrees of freedom, quarks and gluons, and
the observable hadrons, such as the proton.  The PDFs
are the most prominent nonperturbative quantities describing the relation
between a hadron and the quarks and gluons within it.
The collinear PDFs, $f(x,\mu)$, give the number density of partons
with longitudinal momentum fraction $x$ in a fast-moving hadron,
probed at the factorization scale $\mu$.
Although they are not direct physical observables, such as the cross
sections of leptons and hadrons, they are well-defined in QCD and
can be systematically extracted from data of cross sections, if
the factorization formulas of the cross sections with perturbatively
calculated short-distance partonic parts are used.
Our knowledge of PDFs has been much improved throughout the years
by many surprises and discoveries from
measurements at low-energy, fixed-target experiments to those at the LHC
--- the highest energy hadron collider in the world. The excellent agreement
between the theory and data on the factorization scale
$\mu$-dependence of the PDFs has provided one of the most stringent
tests for QCD as the theory of strong interaction.
Many sets of PDFs have been extracted from the QCD
global analysis
of existing data, and a detailed discussion on the extraction of PDFs
will be given in the next subsection.

\begin{figure}[b]
  \includegraphics*[height=0.77\columnwidth,
  angle=90]{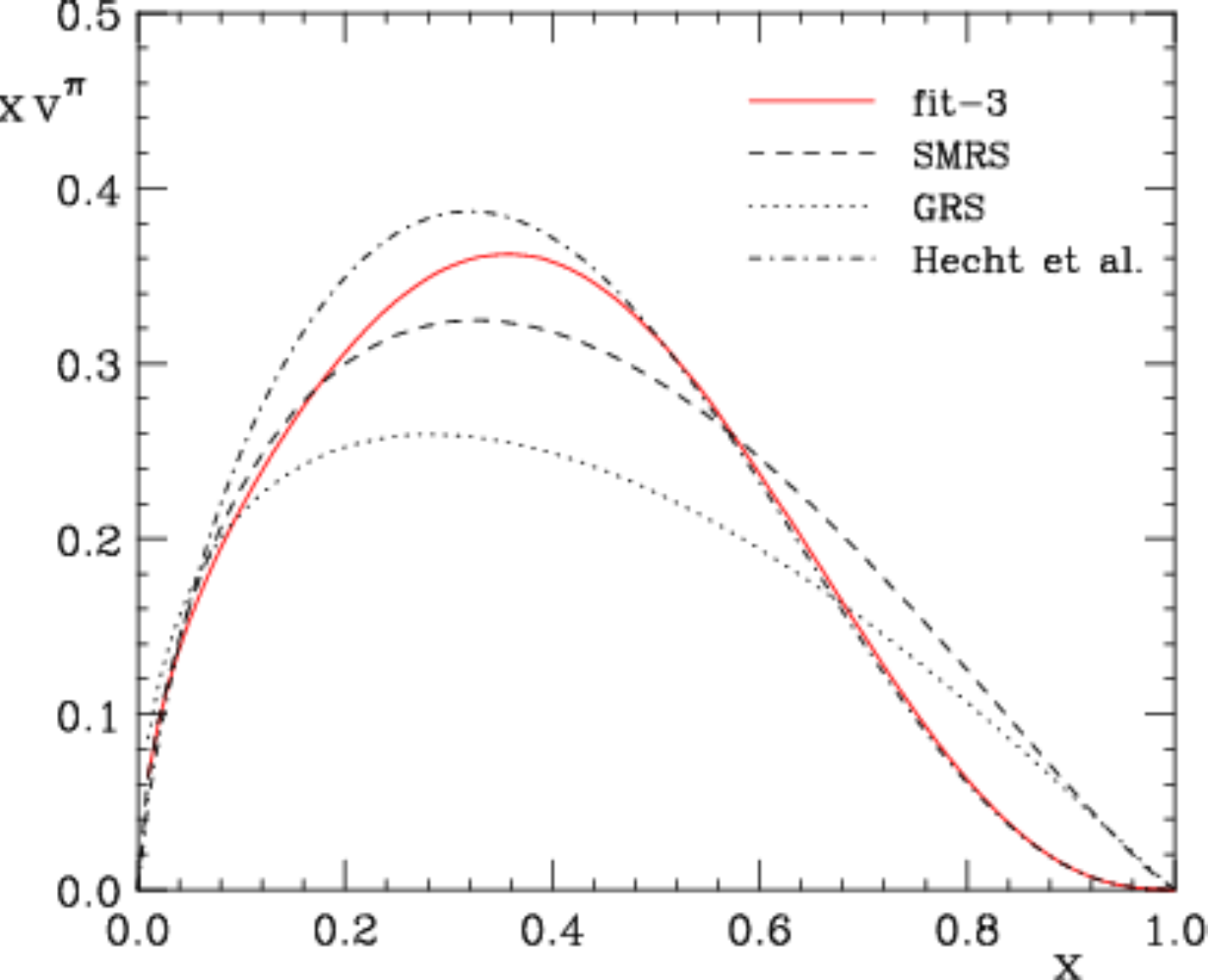}
\caption{Valence distribution of the pion obtained in \cite{Aicher:2010cb}
from a fit to the E615 Drell-Yan data \cite{Conway:1989fs} at $Q=4$~GeV,
compared to the NLO parameterizations of \cite{Sutton:1991ay}
Sutton-Martin-Roberts-Stirling (SMRS) and \cite{Gluck:1999xe}
Gl\"{u}ck-Reya-Schienbein (GRS) and to the distribution obtained from
Dyson-Schwinger equations by Hecht et al. \cite{Hecht:2000xa}.
From \cite{Aicher:2010cb}.
\label{fig:pion-pdf}}
\end{figure}

Understanding the characteristics and physics content of the
extracted PDFs, such as the shape and the flavor dependence of the
distributions, is a necessary step in searching for answers to the
ultimate question in QCD: of how quarks and gluons are confined into
hadrons. However, the extraction of PDFs depends on how well we can
control the accuracy of the perturbatively calculated short-distance
partonic parts.
As an example, consider the pion PDF.  Quite recently, Aicher,
Sch\"{a}fer, and Vogelsang \cite{Aicher:2010cb} addressed the impact
of threshold resummation effects on the pion's valence distribution
$v^\pi\equiv u_v^{\pi^+}= \bar{d}_v^{\pi^+}= d_v^{\pi^-}=
\bar{u}_v^{\pi^-}$ using a fit to the pion-nucleon E615 DY data
\cite{Conway:1989fs}.  They found a falloff much softer than linear,
which is compatible with a valence distribution behaving as
$xv^{\pi}=(1-x)^{2.34}$ (see Fig.\ \ref{fig:pion-pdf}).  This softer
behavior of the pion's valence PDF is due to the resummation of large
logarithmic higher-order corrections --- ``threshold logarithms'' ---
that become particularly important in the kinematic regime accessed by
the fixed-target DY data for which the ratio $Q^2/s$ is large.  Here
$Q$ and $\sqrt{s}$ denote the invariant mass of the lepton pair and
the overall hadronic center-of-mass energy, respectively.  Because
threshold logarithms enhance the cross section near threshold, the
falloff of $v^\pi$ becomes softer relative to previous NLO analyses of
the DY data.  This finding is in agreement with predictions from
perturbative QCD \cite{Sutton:1991ay,Gluck:1999xe} in the low-$x$
regime and from Dyson-Schwinger equation approaches \cite{Hecht:2000xa} in
the whole $x$ region.  Moreover, it compares well with the CERN NA10
\cite{Bordalo:1987cr} DY data, which were not included in the fit
shown in Fig.\ \ref{fig:pion-pdf} (see \cite{Aicher:2010cb} for
details).  Resummation effects on the PDFs in the context of DIS have
been studied in \cite{Corcella:2005us}.

Going beyond a purely longitudinal picture of hadron structure,
one may keep the transverse (spacelike) degrees of freedom of the
partons unintegrated and achieve this way a three-dimensional image
of the hadronic structure by means of transverse-momentum ($k_T$)
dependent (TMD) distribution and fragmentation functions; see, e.g.,
\cite{Boer:2011fh} for a recent review.  Such $x$ and $k_{T}$
dependent quantities provide a useful tool to study semi-inclusive
deep inelastic scattering (SIDIS)
$lH^{\uparrow} \rightarrow l^{\prime} h X$ (HERMES, COMPASS,
JLab at 12~GeV experiments), the Drell-Yan (DY) process
$H_{1}^{(\uparrow)} H_{2}^{\uparrow} \rightarrow l^{+} l^{-} X$
(COMPASS, PAX, GSI, RHIC experiments),
or lepton-lepton annihilation to two almost back-to-back hadrons
$e^{+}e^{-} \rightarrow h_{1} h_{2} X$ (Belle, BaBar experiments),
in which events naturally have two very different momentum scales:
$Q \gg q_T$, where $Q$ is the invariant mass of the exchanged vector
boson, e.g., $\gamma^*$ or $Z^0$, and $q_T$ is the transverse momentum
of the observed hadron in SIDIS or the lepton-pair in DY, or the momentum
imbalance of the two observed hadrons in $e^+e^-$ collisions. It is the
two-scale nature of these scattering processes and corresponding TMD
factorization formalisms \cite{Collins:QCD11,Ji:2004wu,Ji:2004xq}
that enable us to explore the three-dimensional motion of partons inside
a fast moving hadron. The large scale $Q$ localizes the hard collisions
of partons, while the soft scale $q_T$ provides the needed sensitivity
to access the parton $k_T$. Such a two-scale nature makes these observables
most sensitive to both the soft and collinear regimes of QCD dynamics,
and has led to the development of the soft-collinear effective theory approach
in QCD (see Sec.~\ref{sec:LatQcdNucl} for more details and references).

In contrast to collinear PDFs which are related to collinear leading-twist
correlators and involve only spin-spin densities, TMD PDFs
(or simply, TMDs) parameterize spin-spin and momentum-spin correlations,
and also single-spin and azimuthal asymmetries, such as the Sivers
\cite{Sivers:1989cc} and Collins \cite{Collins:1992kk,Collins:2002kn}
effects in SIDIS.  The first effect originates from the correlation of
the distribution of unpolarized quarks in a nucleon with the
transverse polarization vector $S_T$. The second one stems from the
similar correlation between $k_T$ and $S_T$ in the fragmentation
function related to the quark polarization.  The important point is
that the Sivers asymmetry in the DY process flips sign relative to the
SIDIS situation owing to the fact that the corresponding Wilson lines
point in opposite time directions as a consequence of time-reversal
invariance.
This directional (path) dependence breaks the universality
of the distribution functions in SIDIS, DY production,
$e^+ e^-$ annihilation \cite{Collins:2004nx}, and other hadronic processes
that contain more complicated Wilson lines \cite{Bomhof:2006dp}, and lead
to a breakdown of the TMD factorization
\cite{Collins:2007nk,Vogelsang:2007jk,Collins:2007jp,Rogers:2010dm}.
On the other hand, the Collins function seems to possess universal
properties in the SIDIS and $e^+ e^-$ processes \cite{Seidl:2008xc}.
Both asymmetries have been measured experimentally in the SIDIS experiments
at HERMES, COMPASS, and JLab Hall A
\cite{Airapetian:2009ae,Airapetian:2010ds,Alekseev:2008aa,Adolph:2012sn,Adolph:2012sp}.
The experimental test of the breakdown of universality, i.e., a signal of
process dependence, in terms of these asymmetries and their evolution effects
is one of the top-priority tasks in present-day hadronic physics and is
pursued by several collaborations.

\begin{figure}[b]
  \includegraphics*[height=0.77\columnwidth]{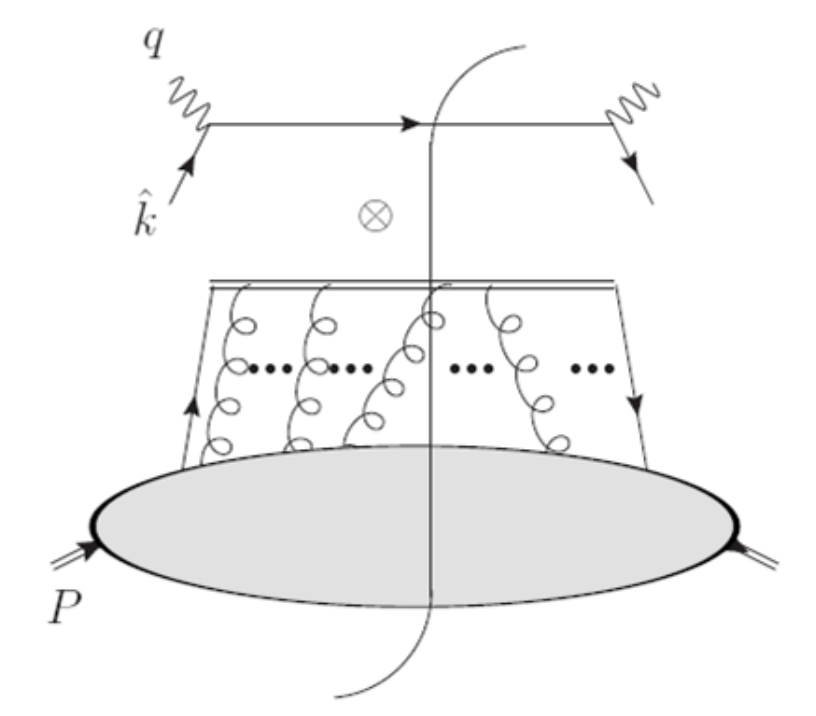}
\caption{Factorization for SIDIS of extra gluons into gauge links
(double lines).Figure  from \cite{Boer:2011fh}.
\label{fig:TMD-SIDIS}}
\end{figure}

Theoretically, the effects described above arise because the TMD field
correlators have a more complicated singularity structure than PDFs,
which is related to the lightlike and transverse gauge links entering
their operator definition
\cite{Ji:2002aa,Belitsky:2002sm,Boer:2003cm}:
\begin{eqnarray}
  \Phi_{ij}^{q[C]}(x, {\bf{k}}_{T};n)
&=& 
  \int\frac{d(y\cdot p) \, d^2 \bm{y}_{T}}{(2\pi)^3}
  e^{-ik \cdot y} \times
  \label{eq:TMD-definition} \\
&  &
      \left\langle
      p| \bar{\psi}_{j}(y)\mathcal{W}(0,y|C)\psi_{i}(0)
      |p\right\rangle_{y\cdot n=0} \, , \nonumber
\end{eqnarray}
where the contour $C$ in the Wilson line $\mathcal{W}(0,y|C)$ has to
be taken along the color flow in each particular process.  For
instance, in the SIDIS case (see Fig.\ \ref{fig:TMD-SIDIS} for an
illustration), the correlator contains a Wilson line at $\infty^-$
that does not reduce to the unity operator by imposing the light-cone
gauge $A^{+}=0$.  This arises because in order to have a closed Wilson
line, one needs in addition to the two eikonal attachments pointing in
the minus direction on either side of the cut in Fig.\
\ref{fig:TMD-SIDIS}, an additional detour in the transverse direction.
This detour is related to the boundary terms that are needed as
subtractions to make higher-twist contributions gauge invariant, see
\cite{Boer:2011fh} for a discussion and references. Hence, the sign
reversal between the SIDIS situation and the DY process is due to the
change of a future-pointing Wilson line into a past-pointing Wilson
line as a consequence of CP invariance (noting CPT is conserved in
QCD) \cite{Collins:2002kn}. In terms of Feynman diagrams this means
that the soft gluons from the Wilson line have ``cross-talk'' with the
quark spectator (or the target remnant) after (before) the hard scattering
took place, which emphasizes the importance of the color flow through the
network of the eikonal lines and vertices.
The contribution of the twist-three fragmentation function to the single
transverse spin asymmetry in SIDIS within the framework of the $k_T$
factorization is another open problem that deserves attention.

The imposition of the light-cone gauge $A^{+}=0$ in combination with
different boundary conditions on the gluon propagator makes the proof
of the TMD factorization difficult --- already at the one-loop order
--- and demands the introduction of a soft renormalization factor to
remove unphysical singularities
\cite{Collins:1999dz,Collins:2000gd,Hautmann:2007uw}.  One may
classify the emerging divergences into three main categories: (i)
ultraviolet (UV) poles stemming from large loop momenta that can be
removed by dimensional regularization and minimal subtraction, (ii)
rapidity divergences that can be resummed by means of the
Collins-Soper-Sterman (CSS) \cite{Collins:1984kg} evolution equation
in impact-parameter space, and (iii) overlapping UV and rapidity
divergences that demand a generalized renormalization procedure to
obtain a proper operator definition of the TMD PDFs.  Rapidity
divergences correspond to gluons moving with infinite rapidity in the
opposite direction of their parent hadron and can persist even when
infrared gluon mass regulators are included, in contrast to the
collinear case in which rapidity divergences cancel in the sum of
graphs.  Their subtraction demands additional regularization
parameters, beyond the usual renormalization scale $\mu$ of the
modified-minimal-subtraction
($\overline{\rm MS}$) scheme.

Different theoretical schemes have been developed to deal with these
problems and derive well-defined expressions for the TMD PDFs.
Starting from the factorization formula for the semi-inclusive
hadronic tensor, Collins \cite{Collins:QCD11} recently proposed a
definition of the quark TMD PDF which absorbs all soft renormalization
factors into the distribution and fragmentation functions, expressing
them in the impact-parameter $b_T$ space.  Taking the limit
$b_T\rightarrow 0$, these semi-integrated PDFs reduce to the collinear
case.

However, this framework has been formulated in the covariant Feynman
gauge in which the transverse gauge links vanish so that it is not
clear how to treat T-odd effects in axial gauges within this
framework.  Moreover, the CSS $b_T$-space approach \cite{Collins:1984kg}
to the evolution of the TMD PDFs requires an extrapolation to the
nonperturbative large-$b_T$ region in order to complete the Fourier
transform in $b_T$ and derive the TMDs in  $k_T$-space.   Different
treatments or approximations of the nonperturbative extrapolation could
lead to uncertainties in the derived TMDs \cite{Qiu:2000hf}.
For example, the TMDs based on Collins' definition predicts
\cite{Aybat:2011zv,Aybat:2011ge,Aybat:2011ta}
asymmetries for DY processes that are a bit too small, while
a more recent analysis \cite{Sun:2013dya,Sun:2013hua}, which derives
from the earlier work in \cite{Ji:2004wu,Ji:2004xq,Idilbi:2004vb}
employing a different treatment on the extrapolation to the large $b_T$
region, seems to describe the evolution of the TMD PDF for both the
SIDIS and the DY process in the range $2$--$100$~GeV$^2$ reasonably
well.

An alternative approach
\cite{Cherednikov:2007tw,Cherednikov:2008ua,Cherednikov:2009wk} to
eliminate the overlapping UV-rapidity divergences employs the
renormalization-group properties of the TMD PDFs to derive an
appropriate soft renormalization factor composed of Wilson lines
venturing off the lightcone in the transverse direction along cusped
contours.  The soft factor encodes contributions from soft gluons with
nearly zero center-of-mass rapidity.  The presence of the soft factor
in the approach of
\cite{Cherednikov:2007tw,Cherednikov:2008ua,Cherednikov:2009wk},
entailed by cusp singularities in the Wilson lines, obscures the
derivation of a correct factorization because it is not clear how to
split and absorb it into the definition of the TMD PDFs to resemble
the collinear factorization theorem.
An extension of this approach, relevant for spin observables beyond
leading twist, was given in \cite{Cherednikov:2010uy}].

Several different schemes to study TMD PDFs and their evolution have also
been proposed \cite{Mantry:2009qz,Mantry:2010mk,Mantry:2010bi},
\cite{Becher:2010tm,Becher:2011xn,Becher:2012yn}
\cite{GarciaEchevarria:2011rb,Echevarria:2012js,Echevarria:2012pw},
\cite{Chiu:2012ir}, \cite{Becher:2011dz,Gehrmann:2012ze}, which are
based on soft collinear effective theory (SCET).  One such framework
\cite{GarciaEchevarria:2011rb,Echevarria:2012js,Echevarria:2012pw} was
shown in \cite{Collins:2012uy} to yield equivalent results to those
obtained by Collins in \cite{Collins:QCD11}.  A detailed comparison of
the Ji-Ma-Yuan scheme \cite{Idilbi:2004vb,Ji:2004xq} with that of
Collins \cite{Collins:QCD11} was given in \cite{Sun:2013hua}.  The
universality of quark and gluon TMDs has been studied in a recent work
by Mulders and collaborators \cite{Buffing:2013kca} in which it was
pointed out that the whole process (i.e., the gauge link) dependence
can be isolated in gluonic pole factors that multiply the universal
TMDs of definite rank in the impact-parameter space.  An analysis of
nonperturbative contributions at the
next-to-next-to-leading-logarithmic
(NNLL) level to the transverse-momentum
distribution of $Z/\gamma^*$ bosons, produced at hadron colliders, has been
presented in \cite{Guzzi:2013aja}.

Last but not least, Sudakov resummation within $k_T$ factorization
of single and double logarithms is an important tool not only
for Higgs boson production in $pA$
collisions, but also for heavy-quark pair production in DIS, used
in the theoretical study of saturation phenomena that
can be accessed experimentally at RHIC and the LHC (see,
\cite{Mueller:2013wwa} for a recent comprehensive analysis).  All
these achievements notwithstanding, the TMD factorization formalism
and the theoretical framework for calculating the evolution of TMD
PDFs and radiative corrections to short-distance dynamics
beyond one loop order have not been fully developed.  Complementary to these
studies, exploratory calculations of TMD nucleon observables in
dynamical lattice QCD have also been performed, which employ nonlocal
operators with ``staple-shaped,'' process-dependent Wilson lines ---
see, for instance, \cite{Musch:2011er}.

\subsubsection{PDFs in the DGLAP approach}
\label{sec:lq.struct.DGLAP}
The PDFs are essential objects in the phenomenology of hadronic
colliders and the study of the hadron structure. In the collinear
factorization framework, the PDFs are extracted from fits to
experimental data for different processes --- they are so-called
global fits. The typical
problem that a global fit solves is to find the set of parameters
$\{p_i\}$ that determine the functional form of the PDFs at a given
initial scale $Q_0^2$, $f_i(x,Q^2_0,\{p_i\})$ so that they minimize a
quality criterion in comparison with the data, normally defined by the
best $\chi^2$. The calculation of the different observables involve i)
the evolution of the PDFs to larger scales $Q^2>Q^2_0$ by means of the
DGLAP evolution equations and ii) the computation of this observable by
the factorized hard cross section at a given order in QCD.
Several
observables are known at next-to-next-to-leading-order (NNLO)
at present, and this order is needed for
precision analyses. This conceptually simple procedure has been
tremendously improved during the last years to cope with the stringent
requirements of more and more precise analyses of the data in the
search of either Standard Model or Beyond the Standard Model
physics. For recent reviews on the topic we refer the readers to \cite{Rojo:2013fta,Forte:2013wc,Perez:2012um,DeRoeck:2011na}.

A standard choice of the initial parametrization, motivated by Regge
theory, is
\begin{equation}
  f_i(x,Q^2_0)=x^{\alpha_i}(1-x)^{\beta_i}g_i(x),
  \label{eq:salgado.IC}
\end{equation}
where $g_i(x)$ is a function whose actual form differs from group to
group. Typical modern sets involve of the order of 30 free parameters
and the released results include not only the best fit (the central value
PDFs) but also the set of {\it error} PDFs to be used to compute
uncertainty bands. These uncertainties are based on Hessian error
analyses which provide eigenvectors of the covariance matrix (ideally)
determined by the one-sigma confidence level criterion or
$\chi^2=\chi^2_{\rm min}+\Delta\chi^2$, with $\Delta\chi^2=1$. Notice,
however, that when applied to a large set of experimental data from
different sources it has long been realized that a more realistic
treatment of the uncertainties requires the inclusion of a {\it
  tolerance} factor $T$ so that $\Delta\chi^2=T^2$
\cite{Pumplin:2002vw,Martin:2002aw}.

An alternative approach which naturally includes the study of the
uncertainties is based on Monte Carlo \cite{Forte:2002fg}, usually by
constructing replicas of the experimental data which encode their
covariance matrix. This approach is employed by the NNPDF Collaboration
\cite{Forte:2002fg,Ball:2010de}, which also makes use of neural
networks for the parametrizations of Eq.~(\ref{eq:salgado.IC}). In this case,
the neural networks provide an unbiased set of basis functions in the
functional space of the PDFs. The Monte Carlo procedure provides a
number of PDF replicas $N_{\rm rep}$ and any observable is computed by
averaging over these $N_{\rm rep}$ sets of PDFs. The main advantage of
this method is that it does not require assumptions on the form of the
probability distribution in parameter space (assumed to be a
multi-dimensional Gaussian in the procedure explained in the previous
paragraph). As a bonus, the method also provides a natural way of
including new sets of data, or check the compatibility of new sets of
data, without repeating the tedious and time-consuming procedure of a
whole global fit. Indeed, in this approach, including a new set of
data would change the relative weights of each of the $N_{\it rep}$
sets of PDFs, so that a new observable can be computed by averaging
over the $N_{\rm rep}$ sets now each one with a different weight
\cite{Giele:1998gw,Ball:2010gb,Ball:2011gg}. This {\it Bayesian
  reweighing} procedure has also been adapted to the Hessian errors
PDFs, where a Monte Carlo representation is possible by simply
generating the PDF sets through a multi-Gaussian distribution in the
parameter space \cite{Watt:2012tq}.

\begin{figure}
  \begin{center}
    \includegraphics[width=0.47\textwidth]{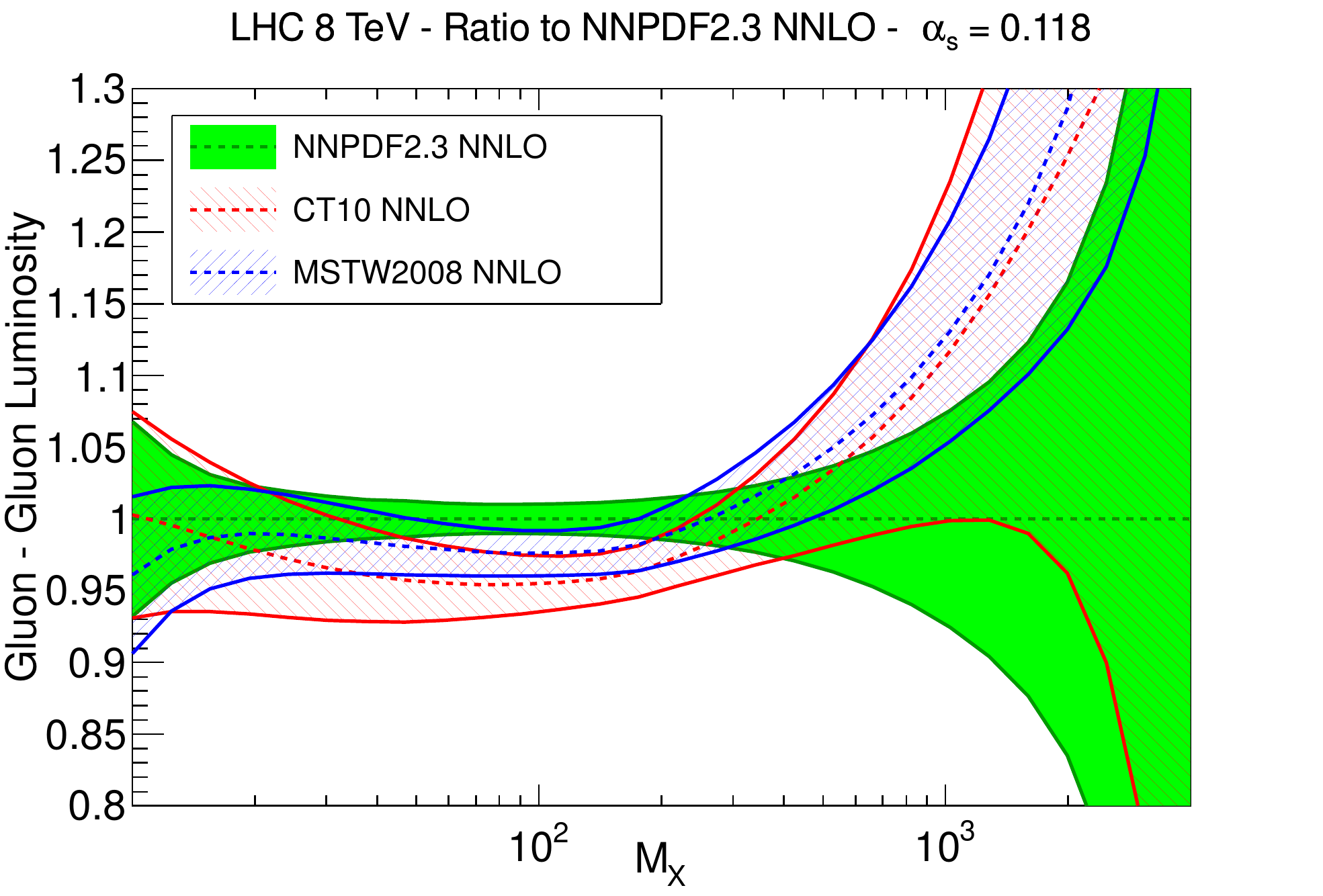}
    \includegraphics[width=0.47\textwidth]{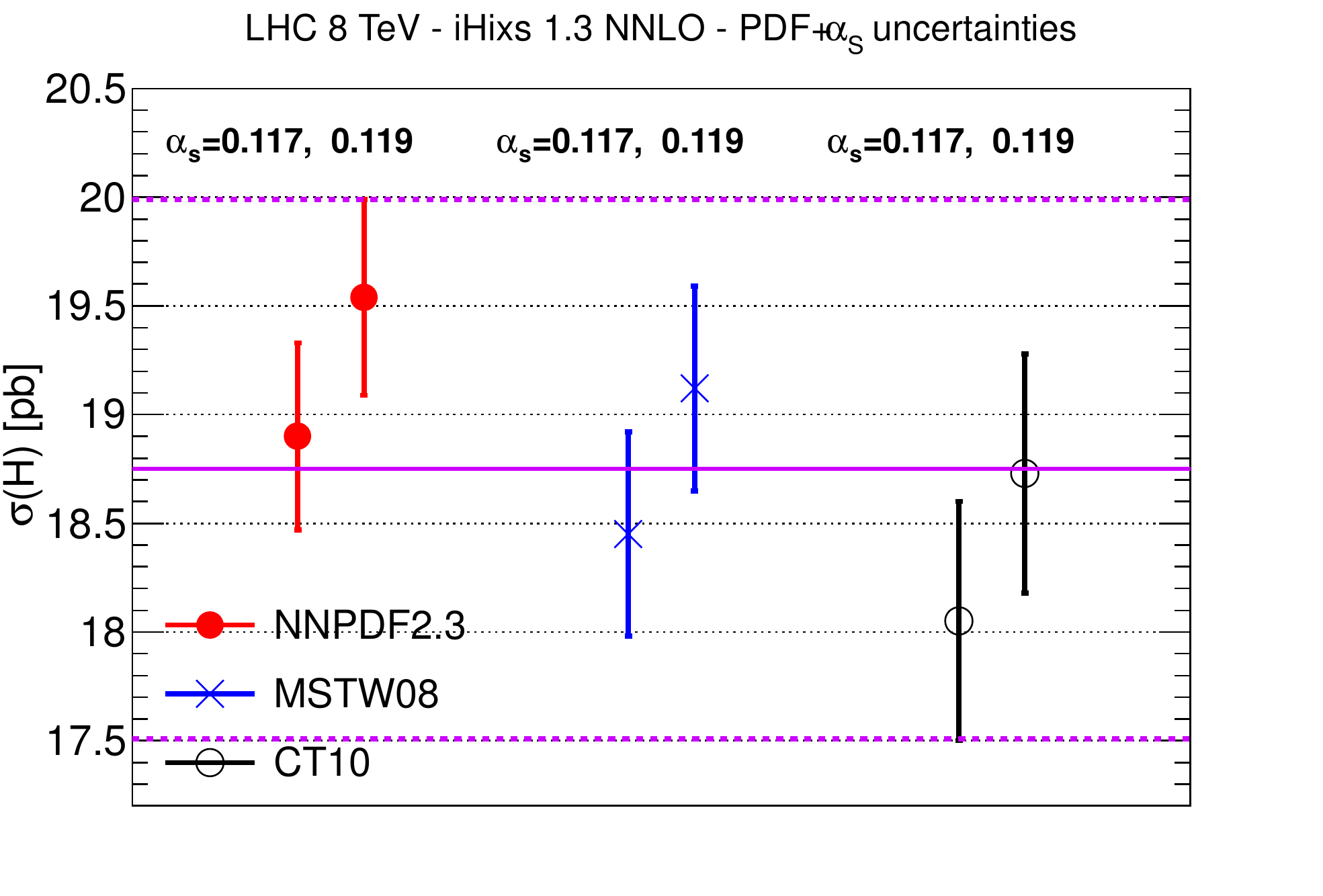}
  \end{center}
  \caption{(Upper figure) Gluon-gluon luminosity to produce a
    resonance of mass $M_X$ for different PDFs normalized to that of
    NNPDF2.3; (Lower figure) The corresponding uncertainties in the
    Higgs cross section from PDFs and $\alpha_s(M_Z)$. Figures
    from~\protect\cite{Ball:2012wy}.}.
  \label{fig:salgado-pdfs}
\end{figure}

Modern sets of unpolarized PDFs for the proton include MSTW08
\cite{Martin:2009iq}, CT10 \cite{Gao:2013xoa}, NNPDF2.3
\cite{Ball:2012cx}, HERAPDF \cite{Aaron:2009aa}, ABM11
\cite{Alekhin:2012ig}, and CJ12 \cite{Owens:2012bv}. Comparison of some of
these sets can be found in Fig.~\ref{fig:salgado-pdfs} as well as of their
corresponding impact on the computation of the Higgs cross section at
NNLO \cite{Ball:2012wy}. Following similar procedures, nuclear PDFs
are also available, that is, nCTEQ \cite{Kovarik:2013sya}, DSSZ
\cite{deFlorian:2011fp}, EPS09 \cite{Eskola:2009uj}, and HKN07
\cite{Hirai:2007sx}, as are polarized PDFs
\cite{Hirai:2008aj,deFlorian:2009vb,Blumlein:2010rn,Leader:2010rb,Ball:2013lla}.

\subsubsection{PDFs and nonlinear evolution equations}
\label{sec:lq.struct.CGC}
Linear evolution equations such as the DGLAP or the
Balitsky-Fadin-Kuraev-Lipatov (BFKL) equations assume a branching
process in which each parton in the hadronic wave function splits into
two lower-energy ones. The divergence of this process in the infrared
makes the distributions more and more populated in the small-$x$
region of the wave function. In this situation it was proposed long
ago that a phenomenon of saturation of partonic densities should
appear at small enough values of the fraction of momentum $x$
\cite{Gribov:1984tu}, or otherwise the unitarity of the scattering
amplitudes would be violated. This idea has been further developed
into a complete and coherent formalism known as the Color Glass
Condensate (CGC, see, e.g.,  \cite{Albacete:2013tpa} for a recent
review).

The CGC formalism is usually formulated in terms of correlators of
Wilson lines on the light-cone in a color singlet state. The simplest
one contains two Wilson lines and can be related to the dipole cross
section; higher-order correlators can sometimes be simplified to the
product of two-point correlators, especially in the large-$N_c$ limit
\cite{Dominguez:2012ad}. The nonlinear evolution equation of the
dipole amplitudes is known in the large-$N_c$ limit with NLO accuracy
\cite{Balitsky:2008zza,Balitsky:2006wa,Kovchegov:2006vj,Albacete:2007yr},
and the LO version of it is termed the Balitsky-Kovchegov equation
\cite{Balitsky:1995ub,Kovchegov:1999ua}. The evolution equations at
finite-$N_c$ are known as the B-JIMWLK equations (using the acronyms
of the authors in
\cite{Balitsky:1995ub,JalilianMarian:1997gr,JalilianMarian:1997dw,Kovner:2000pt,Iancu:2000hn,Ferreiro:2001qy})
and can be written as an infinite hierarchy of coupled nonlinear
differential equations in the rapidity variable, $Y=\log(1/x)$, of the
n-point correlators of the Wilson lines. These equations are very
difficult to solve numerically. However, it has been checked that in the
large-$N_c$ approximation, the BK equations, provides very accurate
results \cite{Rummukainen:2003ns}. The NLO BK equations (or rather
their leading NLO contributions) provide a good description of the
HERA and other small-$x$ physics data with a reduced number of free
parameters \cite{Albacete:2010sy}.

\begin{figure}
  \begin{center}
    \includegraphics[width=0.47\textwidth]{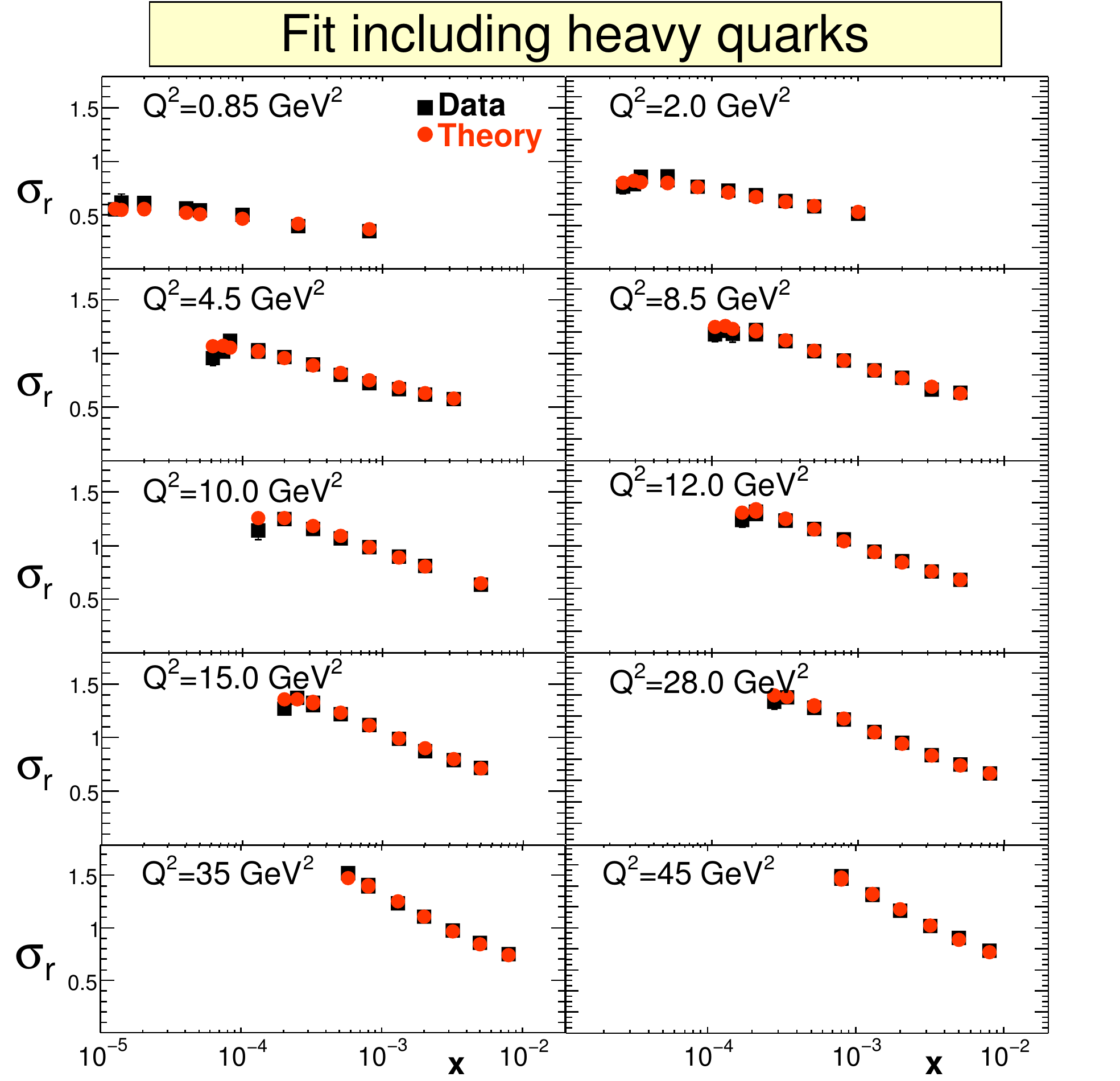}
  \end{center}
  \caption{Fit using the NLO BK nonlinear evolution equations of the
    combined H1/ZEUS HERA data. Figure
    from~\protect\cite{Albacete:2010sy}.}
  \label{fig:nlo-bk}
\end{figure}

One of the main interests of the CGC formalism is that it provides a
general framework in which to address some of the fundamental questions in the
theory of high-energy nucleus-nucleus collisions, in particular, with
respect to the initial stages in the formation of a hot and dense QCD
medium and how local thermal equilibrium is reached (see, e.g.,
\cite{Gelis:2013rba} and references therein). The phenomenological
analyses of different sets of data in such collisions deal with the
multiplicities \cite{ALbacete:2010ad}, the ridge structure in the
two-particle correlations in proton-nucleus collisions, which indicate
very long-range rapidity correlations \cite{Bzdak:2013zma}, or the
coupling of the CGC-initial conditions with a subsequent
hydrodynamical evolution \cite{Gale:2012rq}. These are just examples
of the potentialities of the formalism to provide a complete
description of such complicated systems.

\subsubsection{GPDs and tomography of the nucleon}
\label{sec:lq.gpds}

Quarks and gluons carry color charge, and 
it is very natural to ask how color is distributed
inside a bound and color neutral hadron.  Knowing the color
distribution in space might shed some light on how color is confined in QCD.
Unlike the distribution of electromagnetic charge, which
is given by the Fourier transform of the nucleon's electromagnetic form factors (see the next subsection),
it is very unlikely,
if not impossible, to measure the spatial distribution of color
in terms of scattering cross sections of color-neutral leptons and hadrons.
This is because the gluon carries color, so that 
the nucleon cannot rebound back into a nucleon after absorbing a gluon.
In other words, there is 
no elastic nucleon color form factor.
Fortunately, in the last twenty years, remarkable progress
has been made in both theory and experiment to make it possible to obtain
spatial distributions of quarks and gluons
inside the nucleons.
These distributions, which are also known as tomographic images, are encoded in generalized parton 
distribution functions (GPDs) \cite{Burkardt:2000za,Ralston:2001xs}.

GPDs are defined in terms of generalized parton form factors
\cite{Mueller:2014hsa}, e.g., for quarks,
\begin{eqnarray}
F_{q}(x,\xi,t)
 {\hskip -0.06in} & = & {\hskip -0.06in}
\int \frac{dy^-}{2\pi} e^{-i x p^+ y^-}
  \langle p'| \bar{\psi}({\textstyle\frac{1}{2}}y^-){\textstyle\frac{1}{2}}\gamma^+
  \psi(-{\textstyle\frac{1}{2}}y^-) |p \rangle
  \nonumber\\
&\equiv &  {\hskip -0.05in}
H_q(x,\xi,t) \left[
\overline{\cal U}(p')\gamma^\mu{\cal U}(p)\right]
\frac{n_\mu}{p\cdot n}
\label{eq:quark-gpd}\\
&+ &  {\hskip -0.06in}
E_q(x,\xi,t) \left[
\overline{\cal U}(p') \frac{i\sigma^{\mu\nu}(p'-p)_{\nu}}{2M} {\cal U}(p) \right]
\frac{n_\mu}{p\cdot n} \, ,
\nonumber
\end{eqnarray}
where the gauge link between two quark field operators and
the factorization scale dependence are suppressed,
${\cal U}$'s are hadron spinors,  $\xi=(p'-p)\cdot n/2$ is the skewness, and
$t=(p'-p)^2$ is the squared hadron momentum transfer.  In Eq.~(\ref{eq:quark-gpd}),
the factors $H_q(x,\xi,t)$ and $E_q(x,\xi,t)$ are the quark GPDs.
Unlike PDFs and TMDs, which are defined in terms of forward
hadronic matrix elements of quark and gluon correlators,
like those in Eqs.~(\ref{eq:quark-pdf}) and (\ref{eq:TMD-definition}),
GPDs are defined in terms of non-forward hadronic matrix elements,
$p'\neq p$.
Replacing the $\gamma^\mu$ by $\gamma^\mu\gamma_5$ in
Eq.~(\ref{eq:quark-gpd}) then defines two additional quark GPDs,
$\widetilde{H}_q(x,\xi,t)$ and $\widetilde{E}_q(x,\xi,t)$.
Similarly, gluon GPDs are defined in terms of nonforward hadronic matrix elements
of gluon correlators.

Taking the skewness $\xi\to 0$, the squared hadron momentum transfer 
$t$ becomes $-\vec{\Delta}_{\perp}^2$.  Performing a Fourier transform of GPDs
with respect to $\vec{\Delta}_\perp$ gives the joint distributions of quarks
and gluons in their longitudinal momentum fraction $x$ and
transverse position $b_\perp$, $f_a(x,b_\perp)$ with $a=q,\bar{q},g$,
which are effectively equal to the tomographic images of quarks and gluons
inside the hadron.  Combining the GPDs and TMDs, one could obtain a comprehensive three-dimensional view of
the hadron's quark and gluon structure.

Taking the moments of GPDs, $\int dx\, x^{n-1} H_a(x,\xi,t)$ with $a=q,\bar{q},g$,
gives generalized form factors for a large set of local operators
that can be computed with lattice QCD, as discussed in the next subsection,
although they cannot be directly measured in experiments.  This connects
the hadron structure to lattice QCD  --- one of the main tools
for calculations in the non-perturbative sector of QCD.  For example,
the first moment of the quark GPD, $H(x, 0, t)$ with an appropriate
sum over quark flavors, is equal to the electromagnetic Dirac form factor $F_1(t)$,
which played a major historical role in exploring the internal structure of the proton.

GPDs also play a critical role in addressing the outstanding question of
how the total spin of the proton is built up from the polarization and
the orbital angular momentum of quarks, antiquarks, and gluons.
After decades of theoretical and experimental effort,
following the European Muon Collaboration's discovery \cite{Ashman:1987hv},
it has been established that the polarization of all quarks and antiquarks taken
together can only account for about 30\% of the proton's spin, while about 15\% of 
proton's spin likely stems from gluons, as indicated by RHIC spin  
data~\cite{RHICwp:2012}.  Thus, after all existing measurements, about one half of the 
proton's spin is still not explained, which is a puzzle.  Other 
possible additional contributions from the polarization of quarks and gluons in 
unmeasured kinematic regions, related to the orbital momentum of quarks and 
gluons, could be the major source of the missing portion of the proton's spin.  
In fact, some GPDs are intimately connected with the
orbital angular momentum carried by quarks and gluons \cite{Burkardt:2005km}.
Ji's sum rule is one of the examples that quantify this connection \cite{Ji:1996ek},
\begin{eqnarray}
    J_q = \frac{1}{2} \lim_{t\to 0} \int_0^1 dx\, x \left[H_q(x,\xi,t) + E_q(x,\xi,t) \right]\, ,
    \label{eq:jissumrule}
\end{eqnarray}
which represents the total angular momentum $J_q$ (including both helicity and orbital
contributions) carried by quarks and antiquarks of flavor $q$.  A similar relation holds
for gluons.  The $J_q$ in Eq.~(\ref{eq:jissumrule}) is a generalized form factor at $t=0$
and could be computed in lattice QCD \cite{Hagler:2009ni}.

GPDs have been introduced independently in connection with the
partonic description of deeply virtual Compton scattering (DVCS)
by M\"{u}ller \emph{et al.}~\cite{Mueller:1998fv}, Ji \cite{Ji:1996nm},
and Radyushkin \cite{Radyushkin:1997ki}.
They have also been used to describe deeply virtual meson production
(DVMP) \cite{Radyushkin:1996ru,Collins:1996fb}, and more recently
timelike Compton scattering (TCS) \cite{Berger:2001xd}.
Unlike PDFs and TMDs, GPDs are defined in terms of correlators
of quarks and gluons at the amplitude level.
This allows one to interpret them as an overlap of light-cone wave
functions \cite{Diehl:1998kh,Diehl:2000xz,Brodsky:2000xy}.
Like PDFs and TMDs, GPDs are not direct physical observables.
Their extraction from experimental data relies upon QCD factorization,
which has been derived at the leading twist-two level
for transversely polarized photons in DVCS \cite{Collins:1996fb}
and for longitudinally polarized photons in DVMP \cite{Collins:1998be}.
The NLO corrections to the quark and gluon contributions to
the coefficient functions of the DVCS amplitude were first computed
by Belitsky and M\"uller \cite{Belitsky:1997rh}.
The NLO corrections to the crossed process, namely TCS, have been
derived by Pire \emph{et al.}~\cite{Pire:2011st}.

Initial experimental efforts to measure DVCS and DVMP have been carried
out in recent years by collaborations at HERA and its fixed target experiment HERMES,
as well as by collaborations at JLab and
COMPASS experiment at CERN.
To help extract GPDs from cross-section data for exclusive processes,
such as DVCS and DVMP, various functional forms or representations
of GPDs have been proposed and used for comparing with existing data.
Radyushkin's double distribution ansatz (RDDA)
\cite{Radyushkin:1997ki,Radyushkin:1998es} has been employed
in the Goloskokov-Kroll model
\cite{Goloskokov:2005sd,Goloskokov:2006hr,Goloskokov:2008ib} to
investigate the consistency between the theoretical predictions and
the data from DVMP measurements.  More discussions and references on
various representations of GPDs can be found in a recent article by M\"uller
\cite{Mueller:2014hsa}.

\begin{figure}[b]
\includegraphics*[width=0.9\columnwidth]{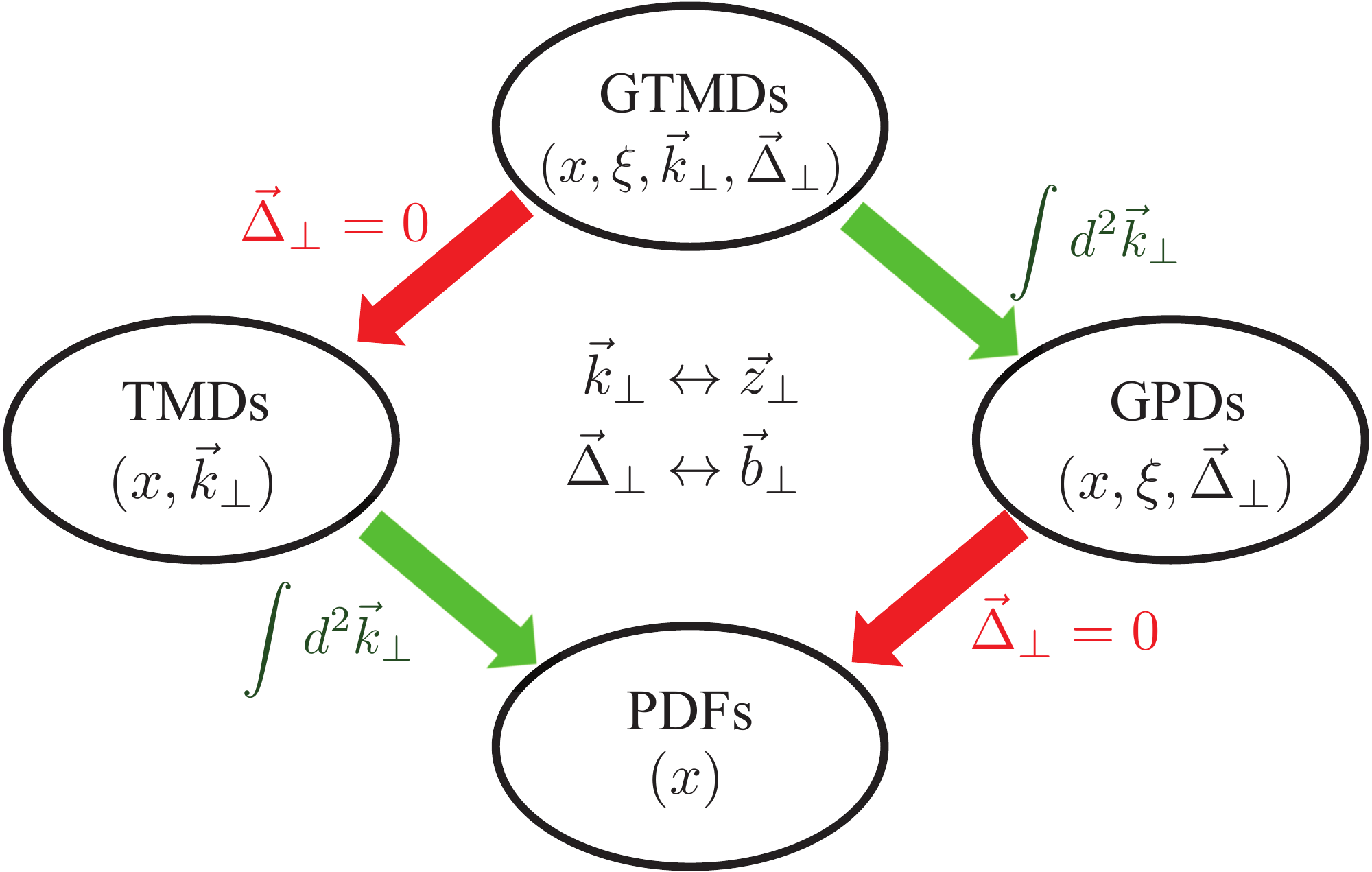}
\caption{Connections among various partonic amplitudes in QCD.
The abbreviations are explained in the text.
\label{fig:GTMD}}
\end{figure}

The PDFs, TMDs, and GPDs represent various aspects of the same hadron's
quark and gluon structure probed in high energy scattering.  They are
not completely independent and, actually, they are encoded in the so-called
{\it mother distributions}, or the generalized TMDs (GTMDs),
which are defined as TMDs with non-forward hadronic matrix elements
\cite{Meissner:2008ay,Meissner:2009ww}.  In addition to the momentum
variables of the TMDs, $x$ and $\vec{k}_\perp$, GTMDs also depend on
variables of GPDs, the skewness $\xi$ and the hadron momentum transfer
$\Delta^\mu =(p'-p)^\mu$ with $t=\Delta^2$.
The Fourier transform of GTMDs can be considered as
Wigner distributions \cite{Belitsky:2003nz},
the quantum-mechanical analogue of classical phase-space distributions.
The interrelationships between GTMDs and the PDFs, TMDs, and GPDs is
illustrated in Fig.~\ref{fig:GTMD}.

Comprehensive and dedicated reviews on the derivation and phenomenology
of GPDs can be found in Refs.~\cite{Goeke:2001tz,Diehl:2003ny,Belitsky:2005qn,Boffi:2007yc,Guidal:2013rya,Mueller:2014hsa}.
More specific and recent reviews of the GPD phenomenology and global
analysis of available data can be found in Ref.~\cite{Kroll:2012sm}
for both the DVCS and DVMP processes, and in Ref.~\cite{Kumericki:2013br}
for DVCS asymmetry measurements of different collaborations pertaining
to the decomposition of the nucleon spin.

With its unprecedented luminosity, the updated 12 GeV program
at JLab will provide good measurements of both DVCS and
DVMP, which will be an excellent source of information on quark
GPDs in the valence region.  It is the future Electron-Ion Collider (EIC)
that will provide the ultimate information on both quark and gluon GPDs,
and the tomographic images of quarks and gluons inside a proton
with its spin either polarized or unpolarized \cite{Accardi:2012qut}.

\subsubsection{Hadron form factors}
\label{sec:lq.struct.form-factors}
The internal structure of hadrons --- most prominently of the nucleon ---
has been the subject of intense experimental and theoretical
activities for decades. Many different experimental facilities have
accumulated a wealth of data, mainly via electron-proton ($ep$)
scattering. Electromagnetic form factors of the nucleon have been
measured with high accuracy, e.g., at MAMI or MIT-Bates. These
quantities encode information on the distribution of electric and
magnetic charge inside the nucleon and also serve to determine the
proton's charge radius. The HERA experiments have significantly
increased the kinematical range over which structure functions of the
nucleon could be determined accurately. Polarized $ep$ and $\mu p/d$
scattering at HERMES, COMPASS and JLab, provide the experimental basis
for attempting to unravel the spin structure of the
nucleon. Furthermore, a large experimental program is planned at
future facilities (COMPASS-II, JLab at 12\,GeV, PANDA@FAIR),
designed to extract quantities such as GPDs, which
provide rich information on the spatial distributions of quarks and gluons
inside the hadrons.  This extensive experimental program requires
equally intense theoretical activities, in order to gain a
quantitative understanding of nucleon structure.


\paragraph{Lattice-QCD calculations}
\label{sec:lq.struct.form-factors.lqcd}
Simulations of QCD on a space-time lattice are becoming increasingly
important for the investigation of hadron structure. Form factors and
structure functions of the nucleon have been the subject of lattice
calculations for many years (see the recent reviews
\cite{Renner:2010ks, Alexandrou:2010cm, Lin:2012ev, Syritsyn:lat13}),
and more complex quantities such as GPDs have also been tackled
recently \cite{Hagler:2007xi, Bratt:2010jn, Alexandrou:2011nr,
  Sternbeck:2012rw, Alexandrou:2013joa, Alexandrou:2013wka}, as
reviewed in \cite{Renner:2012yh,Alexandrou:2012hi}). Furthermore,
several groups have reported lattice results on the strangeness
content of the nucleon \cite{Ohki:2008ff, Toussaint:2009pz,
  Takeda:2010cw, Engelhardt:2010zr, Durr:2011mp, Horsley:2011wr,
  Bali:2011ks,Alexandrou:2013nda, Young:2009zb, Shanahan:2012wh}, as
well as the strangeness contribution to the nucleon spin
\cite{Doi:2009sq,Babich:2010at,QCDSF:2011aa, Liu:2012nz,
  Engelhardt:2012gd, Leinweber:2004tc, Leinweber:2006ug}. Although
calculations of the latter quantities have not yet reached the same
level of maturity concerning the overall accuracy compared to, say,
electromagnetic form factors, they help to interpret experimental data
from many experiments.

Lattice-QCD calculations of baryonic observables are technically more
difficult than for the corresponding quantities in the mesonic
sector. This is largely due to the increased statistical noise which
is intrinsic to baryonic correlation functions, and which scales as
$\exp(m_N-\frac{3}{2}m_\pi)$, where $m_N$ and $m_\pi$ denote the
nucleon and pion masses, respectively. As a consequence, statistically
accurate lattice calculations are quite expensive. It is therefore
more difficult to control the systematic effects relating to lattice
artifacts, finite-volume effects, and chiral extrapolations to the
physical pion mass in these calculations. Statistical limitations may
also be responsible for a systematic bias due to insufficient
suppression of the contributions from higher excited states
\cite{Wittig:2012np}.

Many observables also require the evaluation of so-called
``quark-disconnected'' diagrams, which contain single quark
propagators forming a loop. The evaluation of such diagrams in lattice
QCD suffers from large statistical fluctuations, and specific methods
must be employed to compute them with acceptable accuracy. In a
lattice simulation, one typically considers isovector combinations of
form factors and other quantities, for which the above-mentioned
quark-disconnected diagrams cancel. It should be noted that hadronic
matrix elements describing the ${\pi}N$ sigma term or the strangeness
contribution to the nucleon are entirely based on quark-disconnected
diagrams. With these complications in mind, it should not come as a
surprise that lattice calculations of structural properties of baryons
have often failed to reproduce some well-known experimental results.

In the following we summarize the current status of lattice
investigations of structural properties of the nucleon. The Dirac and
Pauli form factors, $F_1$ and $F_2$, are related to the hadronic
matrix element of the electromagnetic current $V_\mu$ via
\bea \left\langle N(p^\prime,s^\prime)| V_\mu(x) | N(p,s)\right\rangle
= \bar{u}(p^\prime,s^\prime) \left(\gamma_\mu F_1(Q^2) - \right.
\nonumber\\
\left. \sigma_{\mu\nu}\frac{Q_\nu}{2m_N}\, F_2(Q^2) \right) u(p,s),
\eea
where $p,s$ and $p^\prime,s^\prime$ denote the momenta and spins of
the initial and final state nucleons, respectively, and $Q^2=-q^2$ is
the negative squared momentum transfer. The Sachs electric and
magnetic form factors, $G_E$ and $G_M$, which are related to the
electron-proton scattering cross section via the Rosenbluth formula,
are obtained from suitable linear combinations of $F_1$ and $F_2$,
i.e.,
\begin{eqnarray}
  & &G_E(Q^2) = F_1(Q^2) + \frac{Q^2}{(2m_N)^2}F_2(Q^2),\nonumber \\
  & &G_M(Q^2)=F_1(Q^2)+F_2(Q^2).
\end{eqnarray}
The charge radii associated with the form factors are then derived
from
\begin{equation}
  \left\langle r_i^2 \right\rangle = -6\left.
    \frac{d F_i(Q^2)}{d Q^2}\right|_{Q^2=0},\quad i=1,2 \,.
\label{eq:ch_radii}
\end{equation}
Analogous relations hold for the electric and magnetic radii,
$\langle{r_E^2}\rangle$ and $\langle{r_M^2}\rangle$.

Currently there is a large deviation between experimental
determinations of $\langle r_E^2 \rangle$ using muonic hydrogen and
electronic systems that is called the ``proton radius puzzle,'' see
Sec.~\ref{sec:lq.struct.proton-radius} for further discussion.

\begin{figure}[b]
  \includegraphics*[width=\figwid]{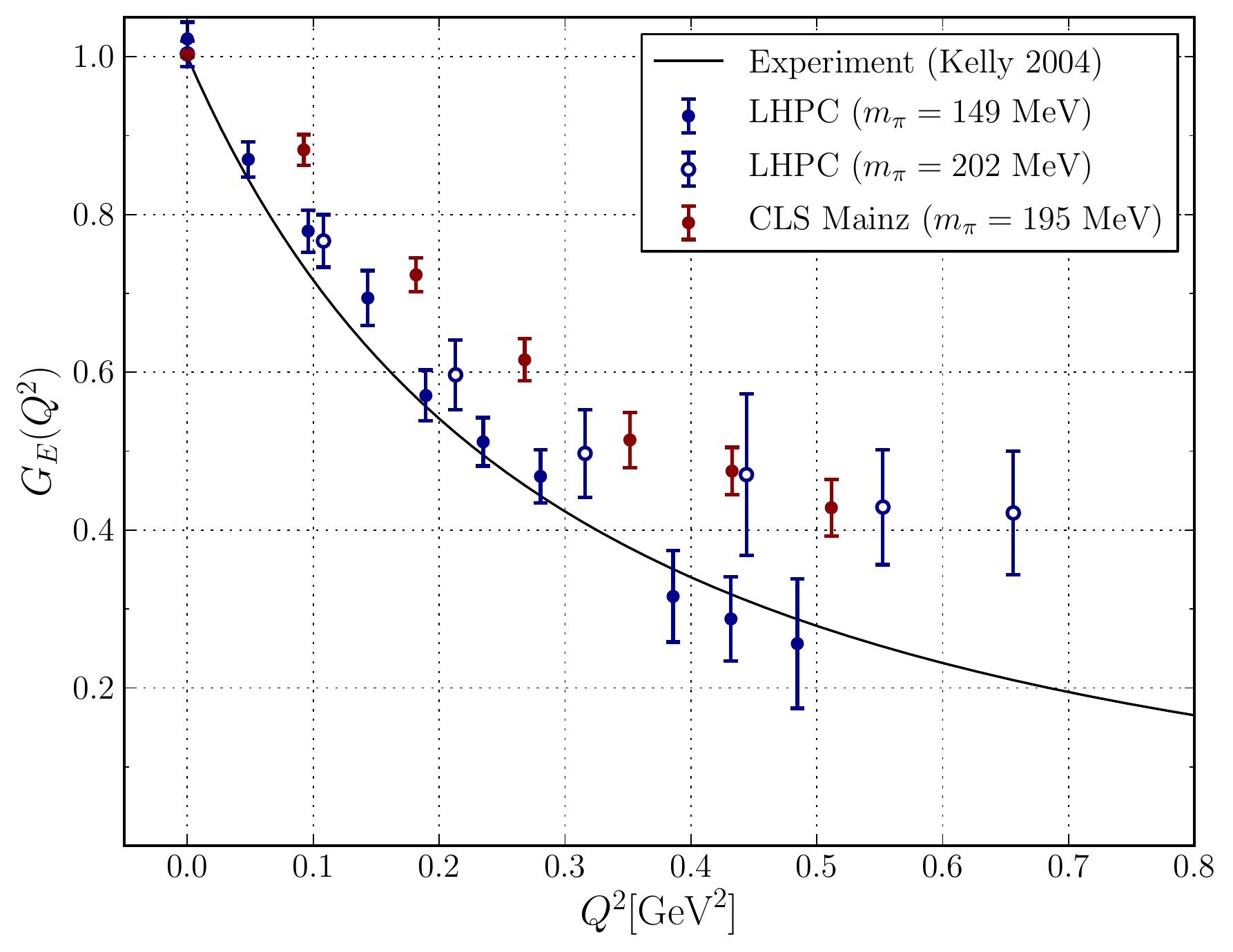}
\caption{The dependence of the nucleon's isovector electric form
factor $G_E$ on the Euclidean four-momentum transfer $Q^2=-q^2$ for
near-physical pion masses, as reported by the LHP Collaboration
\cite{Green:2013hja} and the Mainz group\,\cite{Jager:2013kha}. The
phenomenological parameterization of experimental data is from \cite{Kelly:2004hm}.
\label{fig:GElat}}
\end{figure}

\begin{figure}[b]
  \includegraphics*[width=\figwid]{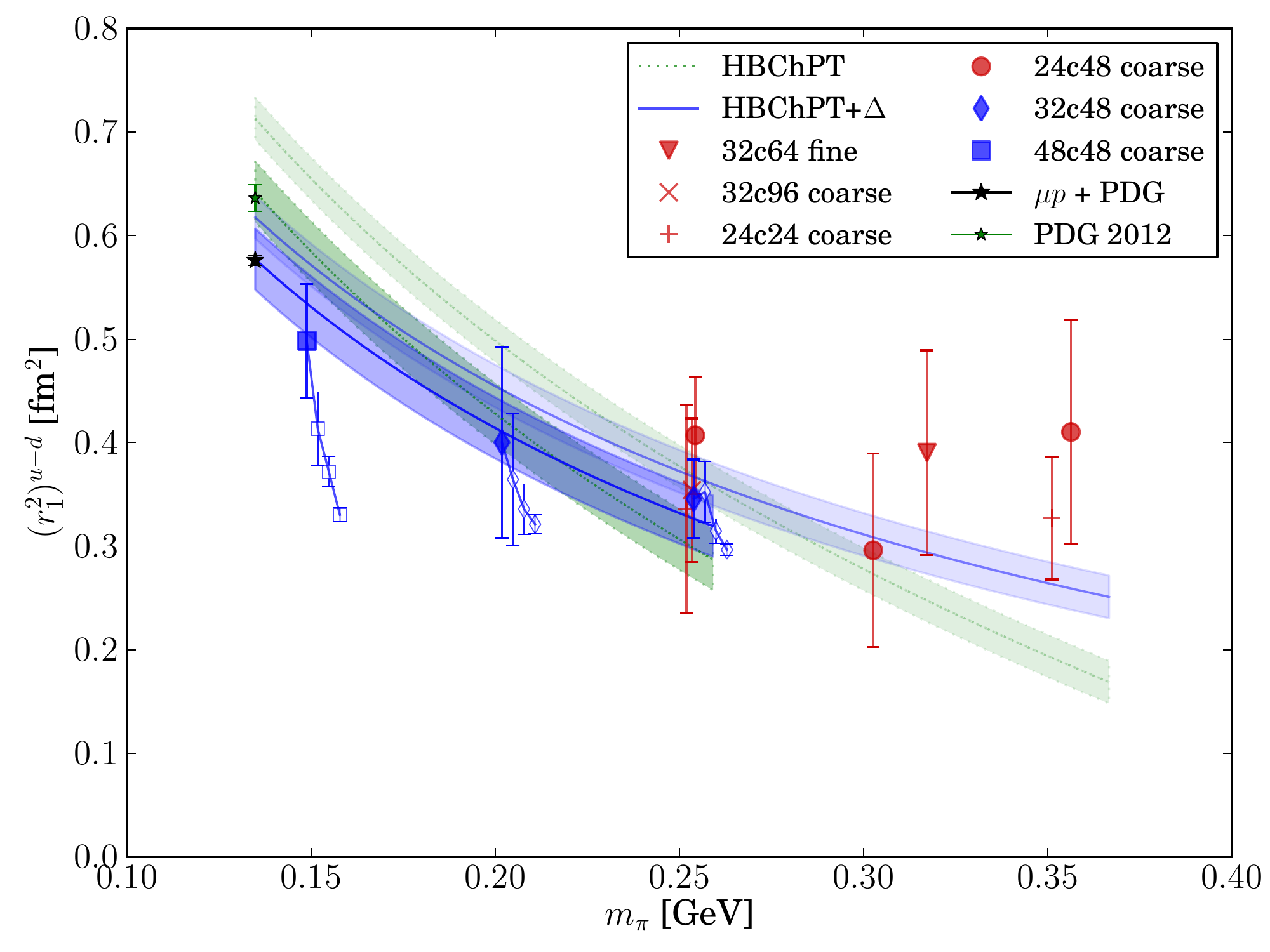}
\caption{The dependence of the isovector Dirac radius $\langle
r_1^2\rangle$ on the pion mass from \cite{Green:2012ud}. Filled
blue symbols denote results based on summed operator insertions,
designed to suppress excited-state contamination.
\label{fig:r1LHPC}}
\end{figure}

There are many cases in which lattice QCD calculations of observables
that describe structural properties of the nucleon compare poorly to
experiment. For instance, the dependence of nucleon form factors on
$Q^2$ computed on the lattice is typically much flatter compared to
phenomenological parameterizations of the experimental data, at least
when the pion mass (i.e., the smallest mass in the pseudoscalar
channel) is larger than about 250~MeV. It is then clear that the
values of the associated charge radii are underestimated compared to
experiment\,\cite{Alexandrou:2006ru,Lin:2008uz,Yamazaki:2009zq,
  Syritsyn:2009mx,Gockeler:2009pe,Bratt:2010jn,Pleiter:2011gw,
  Alexandrou:2011db,Gockeler:2011ze,Collins:2011mk}. The situation
improved substantially after results from simulations with
substantially smaller pion masses became available, combined with
techniques designed to reduce or eliminate excited-state
contamination. The data of \cite{Green:2013hja} and\,\cite{Jager:2013kha},
show a clear trend towards the $Q^2$-behavior seen in a fit of the
experimental results as the pion mass is decreased from around 200~MeV to
almost its physical value (see Fig.~\ref{fig:GElat}). Since different lattice
actions are employed in the two calculations, the results are largely
independent on the details of the fermionic discretization. A key
ingredient in more recent calculations is the technique of summed
operator
insertions\,\cite{Maiani:1987by,Gusken:1989ad,Bulava:2011yz,Capitani:2012gj},
for which excited state contributions are parametrically
suppressed. Alternatively one can employ multi-exponential fits
including the first excited
state\,\cite{Bhattacharya:2013ehc,Green:2013hja} and solve the
generalized eigenvalue problem for a matrix correlation
function\,\cite{Owen:2012ts}, or study the dependence of nucleon matrix
elements for a wide range of source-sink
separations\,\cite{Dinter:2011sg}. Results for the pion mass
dependence of the Dirac radius, $\langle r_1^2\rangle$, from
\cite{Green:2012ud} are shown in Fig.~\ref{fig:r1LHPC}, demonstrating
that good agreement with the PDG value\,\cite{Beringer:2012zz} can be
achieved. Similar observations also apply to the Pauli radius and the
anomalous magnetic moment.

The axial charge of the nucleon, $g_A$, and the lowest moment of the
isovector parton distribution function, $\langle x\rangle_{u-d}$ are
both related to hadronic matrix elements with simple kinematics, since
the initial and final nucleons are at rest. Furthermore, no
quark-disconnected diagrams must be evaluated. If it can be
demonstrated that lattice simulations accurately reproduce the
experimental determinations of these quantities within the quoted
statistical and systematic uncertainties, this would constitute a
stringent test of lattice methods. In this sense $g_A$ and $\langle
x\rangle_{u-d}$ may be considered benchmark observables for lattice
QCD.

\begin{figure}[b]
  \includegraphics*[width=\figwid]{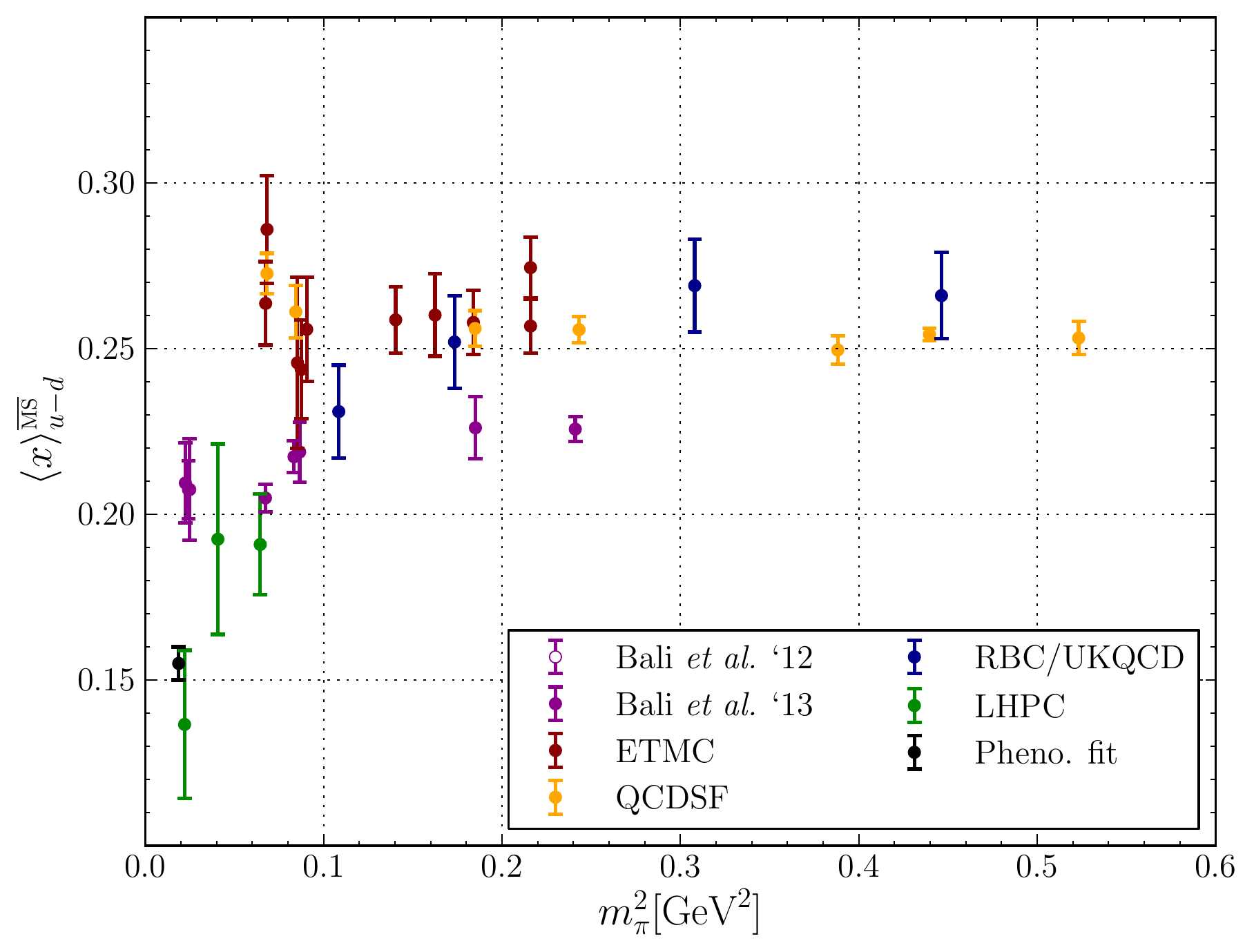}
  \caption{The dependence of the first moment of the isovector PDF
    plotted versus the pion mass. Lattice results are compiled from
    \cite{Aoki:2010xg,Pleiter:2011gw, Alexandrou:2011nr,
      Green:2012ud, Bali:2012av, Bali:2013nla}.\label{fig:avx}}
\end{figure}

Calculations based on relatively heavy pion masses have typically
overestimated $\langle x\rangle_{u-d}$ \cite{Gockeler:2009pe,
  Bratt:2010jn, Aoki:2010xg, Pleiter:2011gw, Alexandrou:2011nr,
  Sternbeck:2012rw} by about 20\%. Moreover, it was found that
$\langle x\rangle_{u-d}$ stays largely constant as a function of the
pion mass (see Fig.\,\ref{fig:avx}). Lower values have been observed
in\,\cite{Bali:2012av,Bali:2013nla}, but given that the overall pion
mass dependence in that calculation is quite weak, it is still
difficult to make contact with the phenomenological estimate.  Other
systematic errors, such as lattice artifacts or insufficient knowledge
of renormalization factors may well be relevant for this
quantity. Recent calculations employing physical pion masses, as well
as methods to suppress excited state
contamination\,\cite{Green:2012ud,Alexandrou:2013jsa}, have reported a
strong decrease of $\langle x\rangle_{u-d}$ near the physical value of
$m_\pi$. Although the accuracy of the most recent estimates does not
match the experimental precision, there are hints that lattice results
for $\langle x\rangle_{u-d}$ can be reconciled with the
phenomenological estimate.

\begin{figure}[b]
  \includegraphics*[width=0.33\textwidth]{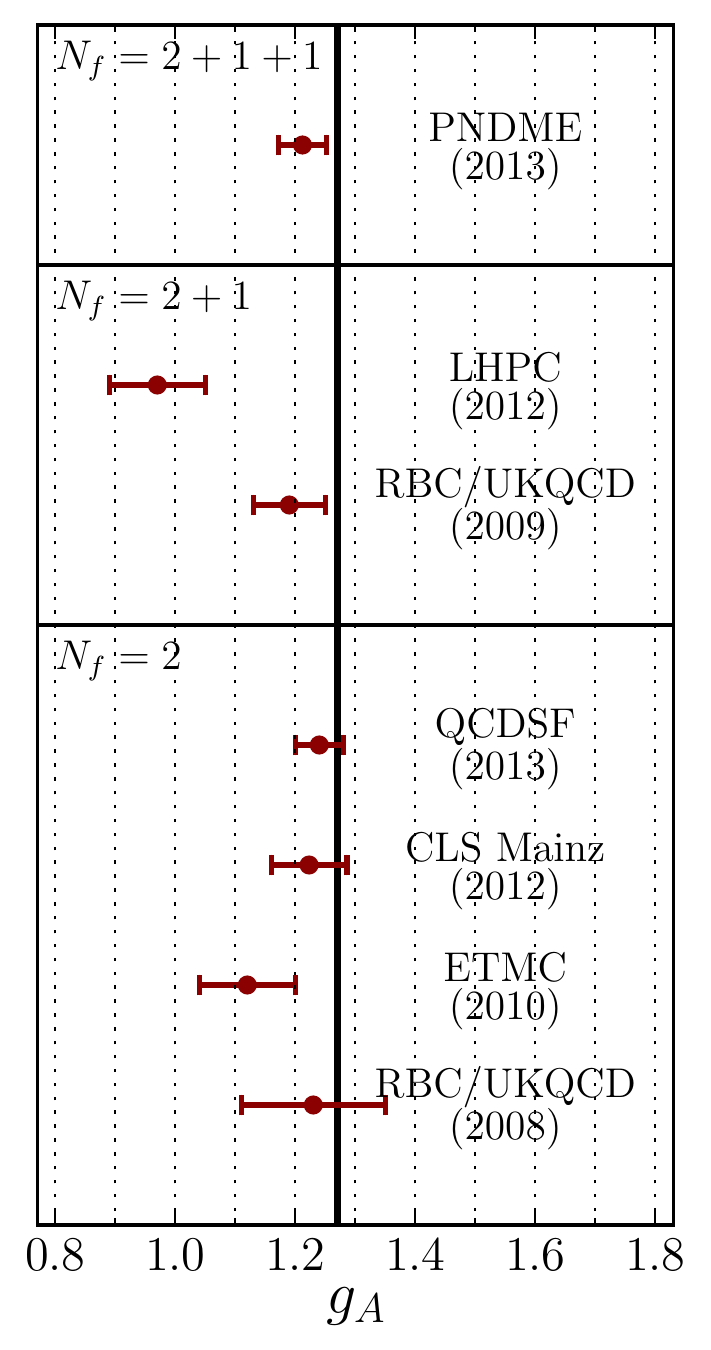}
  \caption{Compilation of recent published results for the axial
    charge in QCD with $N_f=2+1+1$ dynamical
    quarks\,\cite{Bhattacharya:2013ehc} (upper panel),
    $N_f=2+1$\,\cite{Green:2012ud,Yamazaki:2009zq} (middle panel), as
    well as two-flavor
    QCD\,\cite{Horsley:2013ayv,Capitani:2012gj,Alexandrou:2010hf,Lin:2008uz}
    (lower panel).\label{fig:gAcmp}}
\end{figure}

The strategy of controlling the bias from excited states and going
towards the physical pion mass has also helped to make progress on
$g_A$, which, compared to $\langle x\rangle_{u-d}$, is a simpler
quantity. It is the matrix element of the axial current, i.e., a
quark bilinear without derivatives, whose normalization factor is
known with very good accuracy. Lattice simulations using pion masses
$m_\pi > 250$\,MeV typically underestimate $g_A$ by
$10-15\%$\,\cite{Dolgov:2002zm,Ohta:2004mg, Edwards:2005ym, Khan:2006de,
  Yamazaki:2008py, Lin:2008uz,Yamazaki:2009zq,
  Engel:2009nh,Gockeler:2009pe,Bratt:2010jn,Alexandrou:2010hf,
  Pleiter:2011gw,Gockeler:2011ze}. Even more worrisome is the
observation that the data from these simulations show little or no
tendency to approach the physical value as the pion mass is
decreased. However, although some of the most recent calculations
using near-physical pion masses and addressing excited state
contamination\,\cite{Capitani:2012gj,Horsley:2013ayv,Bhattacharya:2013ehc}
produce estimates which agree with experiment (see
Fig.~\ref{fig:gAcmp}), there are notable exceptions: the authors of~\cite{Green:2012ud} still find a very low result, despite using
summed insertions which may be attributed to a particularly strong
finite-size effect in $g_A$. The effects of finite volume have also
been blamed for the low estimates reported
in\,\cite{Lin:2012nv,Ohta:2013qda}.

The current status of lattice-QCD calculations of structural
properties of the nucleon can be summarized by noting that various
sources of systematic effects are now under much better control, which
leads to a favorable comparison with experiment in many
cases. Simulations employing near-physical pion masses and techniques
designed to eliminate the bias from excited-state contributions have
been crucial for this development. Further corroboration of these
findings via additional simulations that are subject to different
systematics is required. Also, the statistical accuracy in the
baryonic sector must be improved.

\paragraph{Poincar{\'e}-covariant Faddeev approach}
\label{sec:lq.struct.form-factors.faddeev}
The nucleons' electromagnetic \cite{Eichmann:2011vu} as well as the
axial and pseudoscalar \cite{Eichmann:2011pv} form factors have been
calculated in the Poincar{\'e}-covariant Faddeev framework based on
Landau-gauge QCD Green's functions. The latter are determined in a
self-consistent manner from functional methods and, if available,
compared to lattice results. Over the last decade, especially the
results for corresponding propagators and some selected vertex
functions have been established to an accuracy that they can serve as
precise input to phenomenological calculations, see also the
discussion in Sec.~\ref{sec:secA2}.

The main idea of the Poincar{\'e}-covariant Faddeev approach is to
exploit the fact that baryons will appear as poles in the six-quark
correlation function.  Expanding around the pole one obtains (in a
similar way as for the Bethe-Salpeter equation) a fully relativistic
bound state equation. The needed inputs for the latter equation are
(i) the tensor structures of the bound state amplitudes, which rest
solely on Poincar{\'e} covariance and parity invariance and provide a
partial-wave decomposition in the rest frame, see, e.g.,
\cite{Eichmann:2009qa,Eichmann:2009zx} and references therein for details;
(ii) the fully dressed quark propagators for {\em complex} arguments; and
(iii) the two- and three-particle irreducible interaction kernels. In
case the three-particle kernel is neglected, the bound-state equation
is then named the Poincar{\'e}-covariant Faddeev equation. The
two-particle-irreducible interaction kernel is
usually modeled within this approach,
and mesons and baryons are then both considered in
the so-called rainbow-ladder truncation, which is the simplest
truncation that fully respects chiral symmetry and leads to a
massless pion in the chiral limit.

In~\cite{Eichmann:2011vu,Eichmann:2011pv} the general
expression for the baryon's electroweak currents in terms of three
interacting dressed quarks has been derived. It turns out that in the
rainbow-ladder truncation the only additional input needed is the
fully dressed quark-photon vertex which is then also calculated in a
consistent way. It is important to note that this vertex then contains
the $\rho$-meson pole, a property which appears essential to obtaining the
correct physics.

In the actual calculations a rainbow-ladder gluon-exchange kernel for
the quark-quark interaction, which successfully reproduces properties
of pseudoscalar and vector mesons, is employed. Then the nucleons'
Faddeev amplitudes and form factors are computed without any further
truncations or model assumptions.  Nevertheless, the resulting
quark-quark interaction is flavor blind\footnote{The quark masses
  introduce a flavor dependence into the quark-quark
  interaction. Furthermore, flavor independence of this interaction is
  in disagreement with experimental facts.}, and by assumption it is a
vector-vector interaction and thus in contradiction to our current
understanding of heavy-quark scalar confinement, cf.\ Sec.~\ref{sec:secA2}.
Reference~\cite{Brodsky:2013ar,deTeramond:2013it} lays out an alternative description of the phenomenology
of confinement, based on the interconnections of light-front QCD, holography, and conformal invariance,
with wide-ranging  implications for the description of hadron structure and dynamics.

Therefore the challenge posed to the Poincar\'e-covariant Faddeev approach is to extend in a
systematically controlled way beyond the rainbow-ladder and the
Faddeev truncations. Given the fact that nonperturbative calculations
of the full quark-gluon vertex and three-gluon vertex have been
published recently and are currently improved, this will become
feasible in the near future. Nevertheless, already the available
results provide valuable insight, and, as can be inferred from the
results presented below, in many observables the effects beyond
rainbow-ladder seem to be on the one hand surprisingly small and on the
other hand in its physical nature clearly identifiable.

\begin{figure}[b]
  \includegraphics*[width=0.33\textwidth]{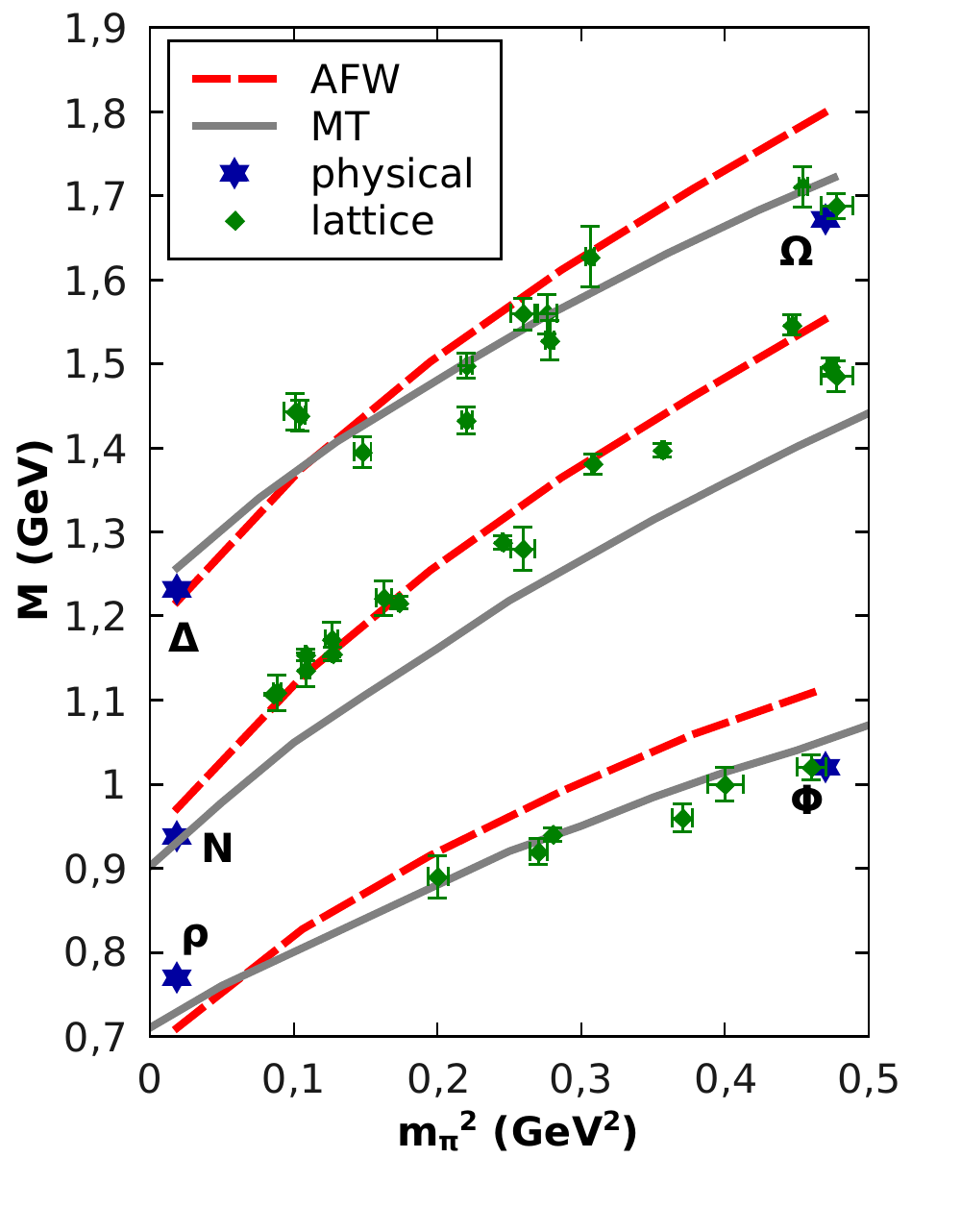}
  \caption{The vector meson, nucleon, and $\Delta$/$\Omega$ masses as a
    function of the pion mass squared in the Poincar{\'e}-covariant
    Faddeev approach (adapted from
    \cite{SanchisAlepuz:2012ej}). \label{fig:massesFaddeev}}
\end{figure}

Figure~\ref{fig:massesFaddeev} shows the results for some selected
hadron masses using two different interaction models, see
\cite{Maris:1999nt} for the MT and \cite{Alkofer:2008et} for the AFW
model. (The main phenomenological difference between these two models
is that the AFW model reproduces the $\eta^\prime$ mass via the
Kogut-Susskind mechanism beyond rainbow-ladder whereas the (older) MT
model does not take this issue into account.)  As one can see, both
model calculations compare favorably with lattice results
\cite{AliKhan:2003cu,Alexandrou:2006ru,Alexandrou:2009hs,Syritsyn:2009mx,Yamazaki:2009zq,Bratt:2010jn,Lin:2010fv,Engel:2010my,Gattringer:2008vj}.
Given the fact that the baryon masses are predictions (with parameters
fixed from the meson sector) and that a rainbow-ladder model kernel
has been used instead of a calculated one, the agreement is even
somewhat better than expected.

\begin{figure}[b]
  \includegraphics*[width=0.49\textwidth]{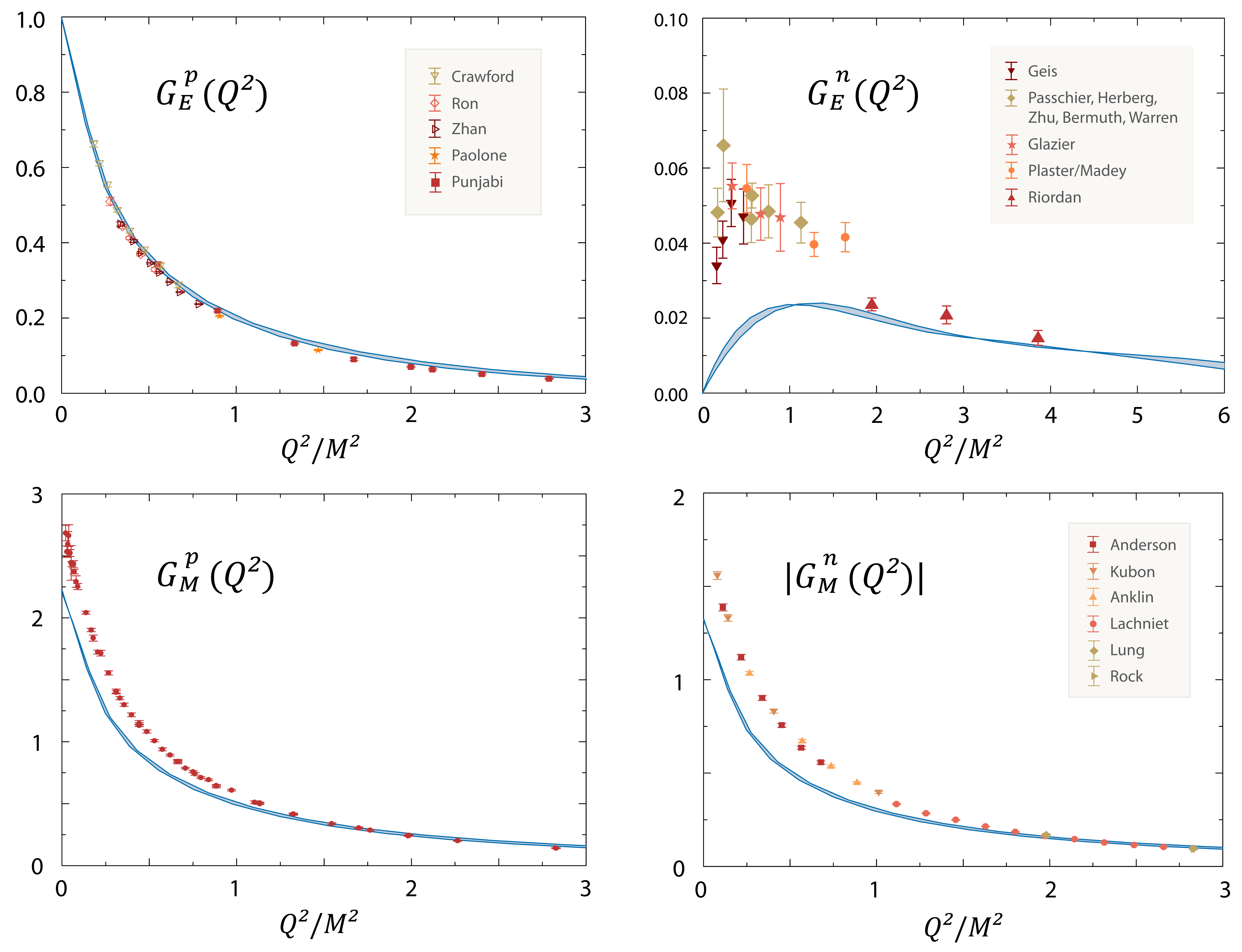}
  \caption{The nucleons' electromagnetic form factors in the
    Poincar{\'e}-covariant Faddeev approach (adapted from
    \cite{Eichmann:2011vu}). \label{fig:NffFaddeev}}
\end{figure}
\begin{figure}[b]
  \includegraphics*[width=0.44\textwidth]{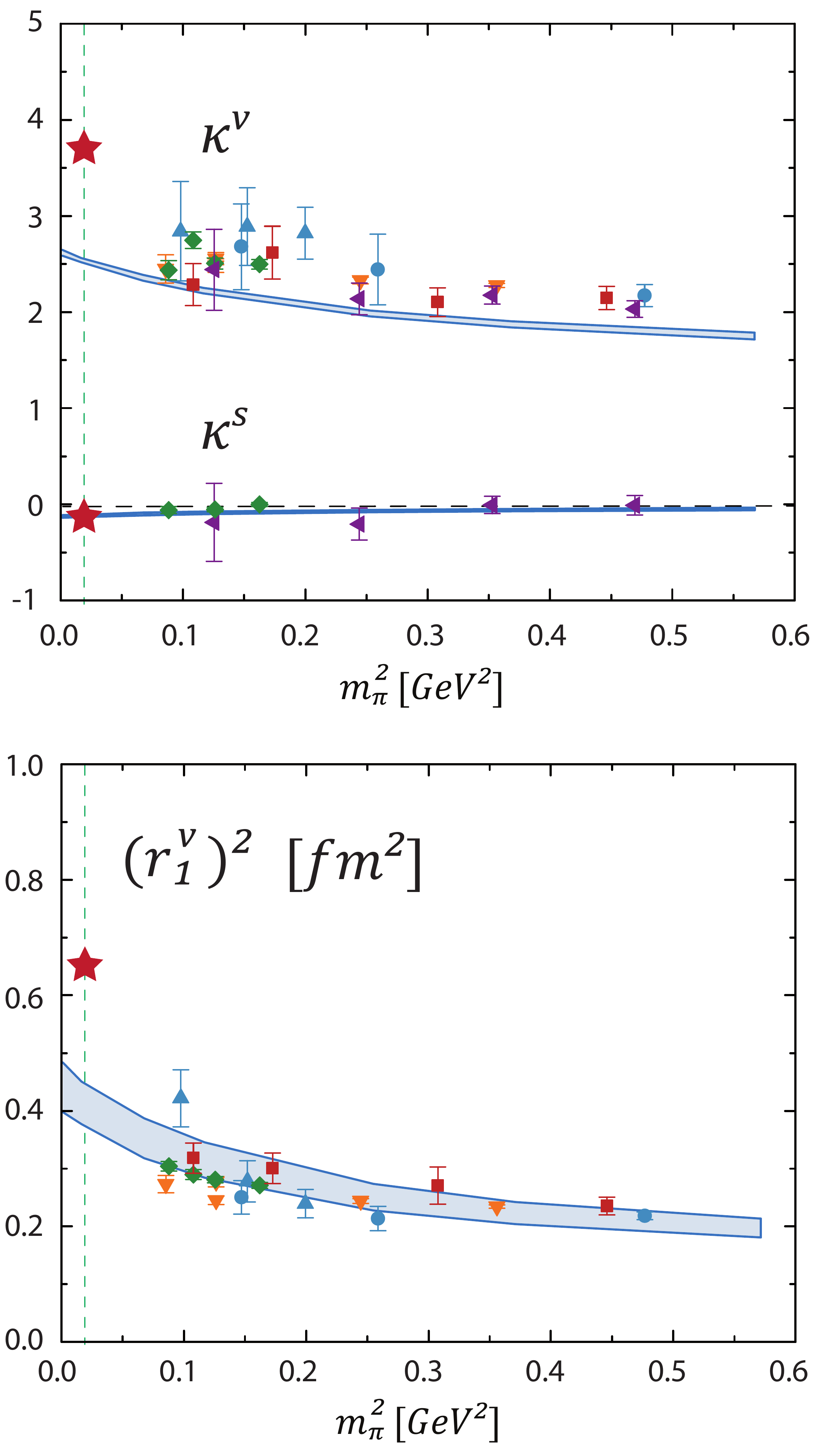}
  \caption{Results for the nucleon's isoscalar and isovector anomalous
    magnetic moments and isovector Dirac radius in the
    Poincar{\'e}-covariant Faddeev approach as compared to lattice QCD
    results and experiment (stars) (adapted from
    \cite{Eichmann:2011vu}). \label{fig:NstaticFaddeev}}
\end{figure}

In Fig.\ \ref{fig:NffFaddeev} the results for the electromagnetic form
factors of the proton and neutron are shown. It is immediately visible
that the agreement with the experimental data at large $Q^2$ is
good. In addition, there is also good agreement with lattice data at
large quark masses. These two observations lead to the expectation
that the difference of the calculated results with respect
to the observed data is due
to missing pion cloud contributions in the region of small explicit
chiral symmetry breaking. This is corroborated by the observation that
the pion-loop corrections of
ChPT are compatible
with the discrepancies appearing in Fig.~\ref{fig:NffFaddeev}.  This
can be deduced in a qualitative way from Fig.~\ref{fig:NstaticFaddeev}.
The results of the Faddeev approach are,
like the lattice results, only weakly dependent on the current quark
mass (viz., the pion mass squared). Whereas lattice results are not
(yet) available at small masses, the Faddeev calculation can be
performed also in the chiral limit. However, pion loop (or pion cloud)
effects are not (yet) contained in this type of
calculations. Correspondingly there are deviations at the physical
pion mass. To this end it is important to note that in the isoscalar
combination of the anomalous magnetic moment leading-order pion
effects are vanishing. As a matter of fact, the Faddeev approach gives
the correct answer within the error margin of the calculation. Details
can be found in \cite{Eichmann:2011vu}.

\begin{figure}[b]
  \includegraphics*[width=0.45\textwidth]{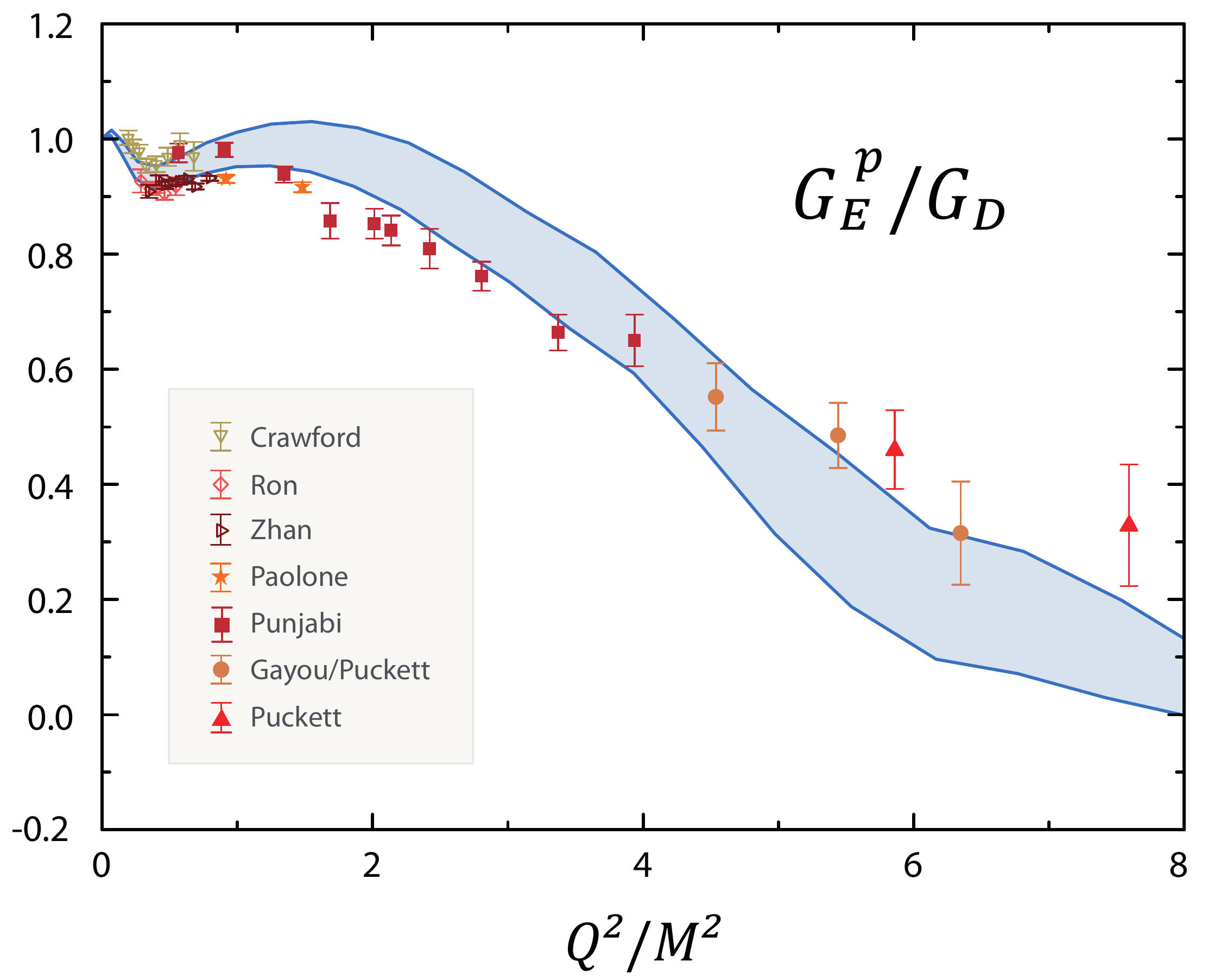}
  \caption{ $Q^2-$evolution of the ratio of the proton's electric form
    factor to a dipole form factor in the Poincar{\'e}-covariant
    Faddeev approach as compared to experimental data (adapted from
    \cite{Eichmann:2011vu}). \label{fig:NratiosFaddeev}}
\end{figure}
Last but not least, the $Q^2-$evolution of the proton's electric form
factor in the multi-GeV region is a topic which has attracted a lot of
interest in the last decade. Contrary to some expectations (raised by
experimental data relying on the Rosenbluth separation) data from
polarization experiments have shown a very strong decrease of the ratio
of the proton's electric to magnetic form factor. Even the possibility
that the proton's electric form factor possesses a zero at $Q^2
\approx 9$ GeV$^2$ is in agreement with the data. However, more details
will be known only after the 12 GeV upgrade of JLab is fully
operational.  In this respect it is interesting to note that the quite
complex Dirac-Lorentz structure of the proton's Faddeev amplitude
quite naturally leads to a strong decrease for $Q^2>2~$GeV$^2$ as shown
in Fig.\ \ref{fig:NratiosFaddeev}.  Several authors attribute the
difference of the data relying on Rosenbluth separation to
polarized-target data to two-photon processes, see, e.g.,
~\cite{Arrington:2006zm}. This has initiated a study of two-photon
processes in the Faddeev approach and an extension to study Compton
scattering has made first but important progress
\cite{Eichmann:2012mp}.

In \cite{Eichmann:2011pv} the axial and pseudoscalar form factors of
the nucleon have been calculated in this approach. It is reassuring
that the Goldberger-Treiman relation is fulfilled for the results of
these calculations for all values of the current quark mass. On the
other hand, the result for the axial charge is underestimated by
approximately 20\%, yielding $g_A\approx 1$ in the chiral limit, which is again
attributed to missing pion effects. This is corroborated by the
finding that the axial and pseudoscalar form factors agree with
phenomenological and lattice results in the range $Q^2>1\ldots
2$~GeV$^2$. In any case, the weak current-mass dependence of $g_A$ in
the Faddeev approach deserves further investigation.

Decuplet, i.e., spin-3/2, baryons possess four electromagnetic form
factors. These have been calculated in the Poincar{\'e}-covariant
Faddeev approach for the $\Delta$ and the $\Omega$
\cite{Sanchis-Alepuz:2013iia}, and the comments made above for the
electric monopole and magnetic dipole form factors for the nucleon
also apply here.  The electric quadrupole (E2) form factor is in good
agreement with the lattice QCD data and provides further evidence for the
deformation of the electric charge contribution from sphericity. The
magnetic octupole form factor measures the deviation from sphericity
of the magnetic dipole distribution, and the Faddeev approach predicts
nonvanishing but small values for this quantity.

Summarizing, the current status of results within the
Poincar{\'e}-covariant Faddeev approach is quite promising. The main
missing contributions beyond rainbow-ladder seem to be pionic effects,
and it will be interesting to see whether future calculations
employing only input from first-principle calculations will verify a
picture of a quark core (whose rich structure is mostly determined by
Poincar{\'e} and parity covariance) plus a pion cloud.

\subsubsection{The proton radius puzzle}
\label{sec:lq.struct.proton-radius}
\begin{figure}
\includegraphics*[width=1.05\linewidth]{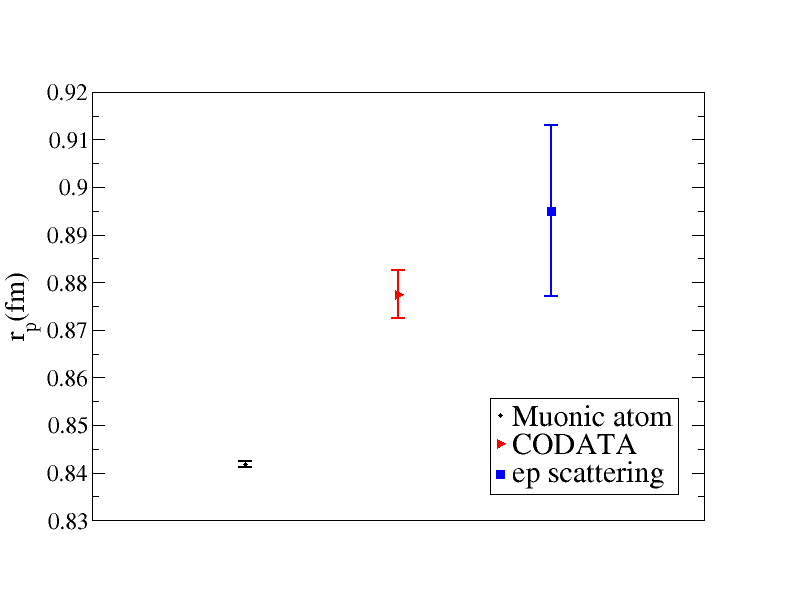}
\caption{\label{fig:muonicpradius}
Proton radius determinations from (i) the muonic-hydrogen Lamb shift (left),
(ii) electron-proton scattering (right), and (iii) the
CODATA-2010 combination of the latter with ordinary hydrogen
spectroscopy (center). Data taken from~\cite{Antognini:2013rsa}.}
\end{figure}

The so-called proton radius puzzle began as a disagreement at the
5$\sigma$ level between its extraction from
a precise measurement of the Lamb shift in
muonic hydrogen~\cite{Pohl:2010zza} and its CODATA value~\cite{Mohr:2008fa},
compiled from proton-radius determinations
from measurements of the Lamb shift in ordinary hydrogen and
of electron-proton scattering.
A recent refinement of the muonic hydrogen Lamb shift measurement
has sharpened the discrepancy with respect to the
CODATA-2010~\cite{Mohr:2012tt}
value to more than 7$\sigma$~\cite{Antognini:1900ns}.
The CODATA values are driven by
the Lamb-shift measurements in ordinary hydrogen, and
a snapshot of the situation
is shown in Fig.~\ref{fig:muonicpradius}, revealing that tensions exist
between all the determinations at varying levels of significance.
The measured Lamb shift in muonic hydrogen is
$202.3706\pm 0.0023$ meV~\cite{Antognini:1900ns}, and
theory~\cite{Pachucki:1996zz,Pachucki:1999zza,Eides:2000xc,Jentschura:2010ej}
yields a value of
$206.0336 \pm 0.0015 - (5.2275 \pm 0.0010)r_E^2 + \Delta E_{\rm TPE}$
in meV~\cite{Antognini:2013rsa}, where $r_E$ is the
proton charge radius and
$\Delta E_{\rm TPE}$ reflects the possibility of two-photon
exchange between the electron and proton. The first
number is the prediction from QED theory and experiment.
%
The proton-radius disagreement amounts to about a 300 $\mu$eV change
in the prediction of the Lamb shift.
Considered broadly the topic shows explicitly how a precise, low-energy
experiment interplays with highly accurate theory (QED)
to reveal
potentially new phenomena. We now turn to a discussion of
possible resolutions, noting the review of \cite{Pohl:2013yb}.

%
%
%

Since the QED calculations are believed to be
well-understood and indeed would have to be grossly wrong to explain
the discrepancy~\cite{Jentschura:2010ej} (and a recently
suggested nonperturbative
QED effect does not exist~\cite{Eides:2014swa,Melnikov:2014lwa}),
a lot of attention has focused on the
hadronic contribution arising from the proton's structure, to which the muonic
atom, given its smaller Bohr radius $a_0(\mu)\simeq (m_e/m_\mu) a_0
(e)$, is much more sensitive. If the disagreement is assigned to an
error in the proton-radius determination, then, as we have noted,
the disagreement between the muonic-atom
determination~\cite{Antognini:1900ns} ($r_E^{(\mu)}$)
and the
CODATA-2010~\cite{Mohr:2012tt}
value (based on hydrogen spectroscopy as well as elastic
electron-proton scattering data) ($r_E^{(e)}$) is very large, namely,
\begin{eqnarray}
  r_{E}^{(\mu)} &=& 0.84087 \pm 0.00039 \,{\rm fm} \,,\nonumber \\
  r_E^{(e)} &=& 0.8775 \pm 0.0051 \, {\rm fm}\,.
\label{eq:pradius}
\end{eqnarray}
It has been argued~\cite{Robson:2013nwa} that atomic physicists
measure the rest-frame proton radius, but electron-scattering data,
parametrized in terms of the Rosenbluth form factors, yields the
Breit-frame proton radius, and these do not coincide.
A resolution by definition might be convenient, but
it is not true: precisely
the same definition, namely, that of Eq.~(\ref{eq:ch_radii}), is used
in both contexts~\cite{Eides:2000xc,Eides:2014swa}.
The value of $r_E^{(e)}$ from hydrogen spectroscopy does rely,
though, on the value of the Rydberg constant $R_\infty$~\cite{Parthey:2011},
and new experiments plan to improve the determination of this
important quantity~\cite{Antognini:1900ns}.

The precision of the experimental extraction of the
vector form factor
from $ep$ scattering, from which the proton radius is extracted as per
Eq.~(\ref{eq:ch_radii})~\cite{Bernauer:2010wm}, has also been
questioned~\cite{Hill:2010yb,Lorenz:2012tm}.
In particular,
it has been noted that
the low-energy Coulomb correction from $ep$ final-state interactions
is sizeable, and this ameliorates the discrepancy between
the charge radii determined from hydrogen spectroscopy and its
determination in $ep$ scattering~\cite{Lee:2014uia}.

Higher-order hadronic corrections
involving two-photon processes have also been considered as a way of
resolving the puzzle~\cite{Carlson:2011zd,Hill:2011wy}.
Revised, precise dispersive
reevaluations of the proton's two-photon
kernel~\cite{Gorchtein:2013yga} based on experimental input (photo-
and electro-production of resonances off the nucleon and high-energy
pomeron-dominated cross-section) yield a contribution of $40\pm 5\
\mu$eV to the muonic hydrogen Lamb shift.
The small uncertainty which remains
is controlled with the ``J=0'' fixed pole of Compton scattering,
i.e., the local coupling of two photons to the proton, and which is
phenomenologically known only for real photons.
This result is in tension with
the value $\Delta E_{\rm TPE}=33.2 \pm 2.0\,\mu{\rm eV}$
used in \cite{Antognini:1900ns}, but it remains
an order of magnitude too small
to explain the discrepancy in the Lamb shift.
The appearance of different energy scales in the analysis of
muonic hydrogen makes it a natural candidate for the application
of effective field theory techniques~\cite{Pineda:2004mx,Hill:2011wy}.
Limitations in the ability to assess
the low-energy constants would seem to make such analyses
inconclusive. Nevertheless, a systematic treatment under the combined use
heavy-baryon effective theory, and (potential) 
non-relativistic QED~\cite{Pineda:2004mx,Peset:2014yha} has recently been
employed to determine a proton radius of $0.8433\pm 0.0017\,{\rm fm}$
from the muonic hydrogen data\cite{Antognini:1900ns},  
assuming that the underlying 
power counting determines the numerical size of the neglected terms. 
This result remains $6.4\sigma$ away from the CODATA-2010 result. 

To summarize, hopes that hadronic contributions to the two-photon
exchange between the muon and the proton would resolve the issue
quickly are starting to fade away because the correction needed to
explain the discrepancy is unnaturally
large~\cite{Miller:2012rba}. 
Therefore, 
it might be useful to test ideas of physics beyond the
Standard Model, i.e., a different interaction of muons and electrons,
in the context of 
the proton radius puzzle, see Sec.~\ref{sec:secE6} for a corresponding
discussion.

\subsubsection{The pion and other pseudoscalar mesons}
\label{sec:structure.pion}

The lightest hadron, the pion, is one of the most important strongly
interacting particles and serves as a ``laboratory'' to test various
methods within QCD, both on the perturbative and the nonperturbative
side.  The electromagnetic form factor at spacelike momenta has been
treated by many authors over the last decades using various techniques
based 
on collinear factorization \cite{Lepage:1980fj,Efremov:1978rn,Chernyak:1983ej} 
with calculations up to the NLO order of perturbation theory, see, e.g.,
\cite{Melic:1998qr,Bakulev:2004cu}.  A novel method was recently
presented in \cite{Chang:2013nia} which uses the Dyson-Schwinger
equation framework in QCD (see \cite{Cloet:2013jya} for a review).
This analysis shows the prevalence of the leading-twist perturbative
QCD result (i.e., the hard contribution) for $Q^2F_{\pi}(Q^2)$ beyond
$Q^2 \gtrsim 8$~GeV$^2$ in agreement with the earlier results of
\cite{Bakulev:2004cu}.  Furthermore, it reflects via the dressed quark
propagator the scale of dynamical chiral symmetry breaking (D$\chi$SB)
which is of paramount importance and still on the wish list of hadron
physics, because a detailed microscopic understanding of this
mechanism is still lacking.  Moreover, our current understanding of the pion's
electromagnetic form factor in the timelike region is still marginal
\cite{Bakulev:2000uh}.

Nevertheless, the 
dual nature of the pion --- being on the one hand
the would-be Goldstone boson of D$\chi$SB and on the other hand a
superposition of highly relativistic bound states of quark-antiquark
pairs in quantum field theory --- is basically understood and generally
accepted. Furthermore, as discussed in
Sec.~\ref{sec:lq.struct.PDF-TMD-theory}, its valence parton distribution
function has been recently determined with a higher precision using
threshold resummation techniques \cite{Aicher:2010cb}. Finally, the quark
distribution amplitude for the pion, which universally describes its
strong interactions in exclusive reactions, has been reconstructed
from the world data on the pion-photon transition form factor as we
will see below and is found to be wider than the asymptotic
one \cite{Bakulev:2012nh}.

\paragraph{Form factors of pseudoscalar mesons}
\label{sec:lq.struct.pion.form-factors}
The two-photon processes $\gamma^{*}(q_{1}^2)\gamma(q_{2}^{2})
\rightarrow P$ with $q_{1}^{2}=-Q^2$ and $q_{2}^{2}=-q^2\sim 0$ of
pseudoscalar mesons $P=\pi^0, \eta, \eta'$ in the high-$Q^2$ region
have been studied extensively within QCD (see
\cite{Stefanis:2012yw,Bakulev:2011rp,Cloet:2013jya} for analysis and
references). This theoretical interest stems from the fact that in
leading order such processes are purely electromagnetic with all
strong-interaction (binding) effects factorized out into the
distribution amplitude of the pseudoscalar meson in question by virtue
of collinear factorization.  This implies that for $Q^2$ sufficiently
large, the transition form factor for such a process can be formulated
as the convolution of a hard-scattering amplitude $T(Q^2,
q^2\rightarrow 0, x) = Q^{-2}(1/x + \mathcal{O}(\alpha_s))$,
describing the elementary process $\gamma^*\gamma \longrightarrow
q\bar{q}$, with the twist-two meson distribution amplitude
\cite{BL81-2phot}.  Therefore, this process constitutes a valuable
tool to test models of the distribution amplitudes of these mesons.

Several experimental collaborations have measured the cross section
for $Q^{2}F^{\gamma^*\gamma\pi^0}(Q^2,q^2\to 0)$ and
$Q^2F^{\gamma^*\gamma\eta(\eta')}(Q^2,q^2\to 0)$ in the two-photon
processes $e^+e^- \to e^+e^- \gamma^*\gamma \to e^+e^- X$, where
$X=\pi^{0}$ \cite{Behrend:1990sr,Gronberg:1997fj,Aubert:2009mc},
$\eta$ and $\eta'$ \cite{Gronberg:1997fj,BABAR:2011ad}, through the
so-called single-tag mode in which one of the final electrons is
detected.  From the measurement of the cross section the meson-photon
transition form factor is extracted as a function of $Q^2$.  The
spacelike $Q^2$ range probed varies from $0.7$--$2.2$~GeV$^2$
\cite{Behrend:1990sr}(CELLO), to $1.64$--$7.90$~GeV$^2$
\cite{Gronberg:1997fj} (CLEO), to $4$--$40$~GeV$^2$
\cite{Aubert:2009mc} (BaBar) and \cite{Uehara:2012ag} (Belle).  A
statistical analysis and classification of all available experimental
data versus various theoretical approaches can be found in
\cite{Bakulev:2012nh}.  The BaBar Collaboration extended substantially
the range of the spacelike $Q^2$, which had been studied before by
CELLO~\cite{Behrend:1990sr} and CLEO~\cite{Gronberg:1997fj} below 9
GeV$^2$ to $Q^2 < 40$~GeV$^2$.  While at low momentum transfers the
results of BaBar agree with those of CLEO and have significantly
higher accuracy, above 9 GeV$^2$ the form factor shows rapid growth
and from $\sim 10$ GeV$^2$ it exceeds the asymptotic limit predicted
by perturbative QCD~\cite{Lepage:1980fj}.  The most recent results
reported by Belle \cite{Uehara:2012ag} for the wide kinematical region
$4 \lesssim Q^2 \lesssim 40$~GeV$^2$ have provided important evidence
in favor of the collinear factorization scheme of QCD.  The rise of
the measured form factor $Q^{2}F^{\gamma^*\gamma\pi^0}$, observed
earlier by the BaBar Collaboration \cite{Aubert:2009mc} in the
high-$Q^2$ region, has not been confirmed.  This continued rise of the
form factor would indicate that the asymptotic value of the form
factor predicted by QCD would be approached from above and at much
higher $Q^2$ than currently accessible, casting serious doubts on the
validity of the QCD factorization approach and fueling intensive
theoretical investigations in order to explain it (see, for example,
\cite{Radyushkin:2009zg}).  The results of the Belle measurement are
closer to the standard theoretical expectations \cite{Lepage:1980fj}
and do not hint to a flat-like pion distribution amplitude as proposed
in \cite{Radyushkin:2009zg}.  Further support for this comes from the
data reported by the BaBar Collaboration \cite{BABAR:2011ad} for the
$\eta(\eta')$-photon transition form factor that also complies with
the QCD theoretical expectations of form-factor scaling at higher
$Q^2$.  A new experiment by KLOE-2 at Frascati will provide
information on the $\pi-\gamma$ transition form factor in the
low-$Q^2$ domain, while the BES-III experiment at Beijing will measure
this form factor below $5$~GeV$^2$ with high statistics.

\paragraph{Neutral pion lifetime} \label{sec:lqstructpionlifetime}
In the low-energy regime, the two-photon process $\pi^0\rightarrow
\gamma\gamma$ is also important because one can test at once the
Goldstone boson nature of the $\pi^0$ and the chiral Adler-Bell-Jackiw
anomaly \cite{Bell:1969ts,Adler:1969gk}.  While the level of
accuracy achieved long ago makes existing tests satisfactory,
deviations due to the nonvanishing quark masses should become
observable at some point.  The key quark-mass effect is due to the
isospin-breaking-induced mixing: $\pi^0$-$\eta$ and $\pi^0$-$\eta'$,
with the mixing being driven by $m_d-m_u$.  The full ChPT correction
has been evaluated by several authors and an enhancement of the decay
width of about $4.5\pm 1.0$\% has been found
\cite{Goity:2002nn,Ananthanarayan:2002kj,Kampf:2009tk}, leading to the
prediction $\Gamma_{\pi^0\to 2 \gamma}= 8.10$~eV.

The most recent measurement was carried out by the PRIMEX
collaboration at JLab with an experiment based on the
Primakoff effect \cite{Larin:2010kq}, providing the result
$\Gamma(\pi^0\rightarrow \gamma\gamma)=7.82$ eV with a global
uncertainty of 2.8\%, which is by far the most precise result to date.
Taking into account the uncertainties, it is marginally compatible
with the ChPT predictions.  With the aim of reducing the error down to
2\%, a second PRIMEX experiment has been completed and results of the
analysis should appear soon.  A measurement of
$\Gamma_{\pi^0\rightarrow 2\gamma}$ at the per cent level is also
planned in the study of two-photon collisions with the KLOE-2 detector
\cite{Babusci:2011bg}.  A recent review of the subject can be found in
\cite{Bernstein:2011bx}.

\begin{figure}[tbp]
  \includegraphics[width=\columnwidth]{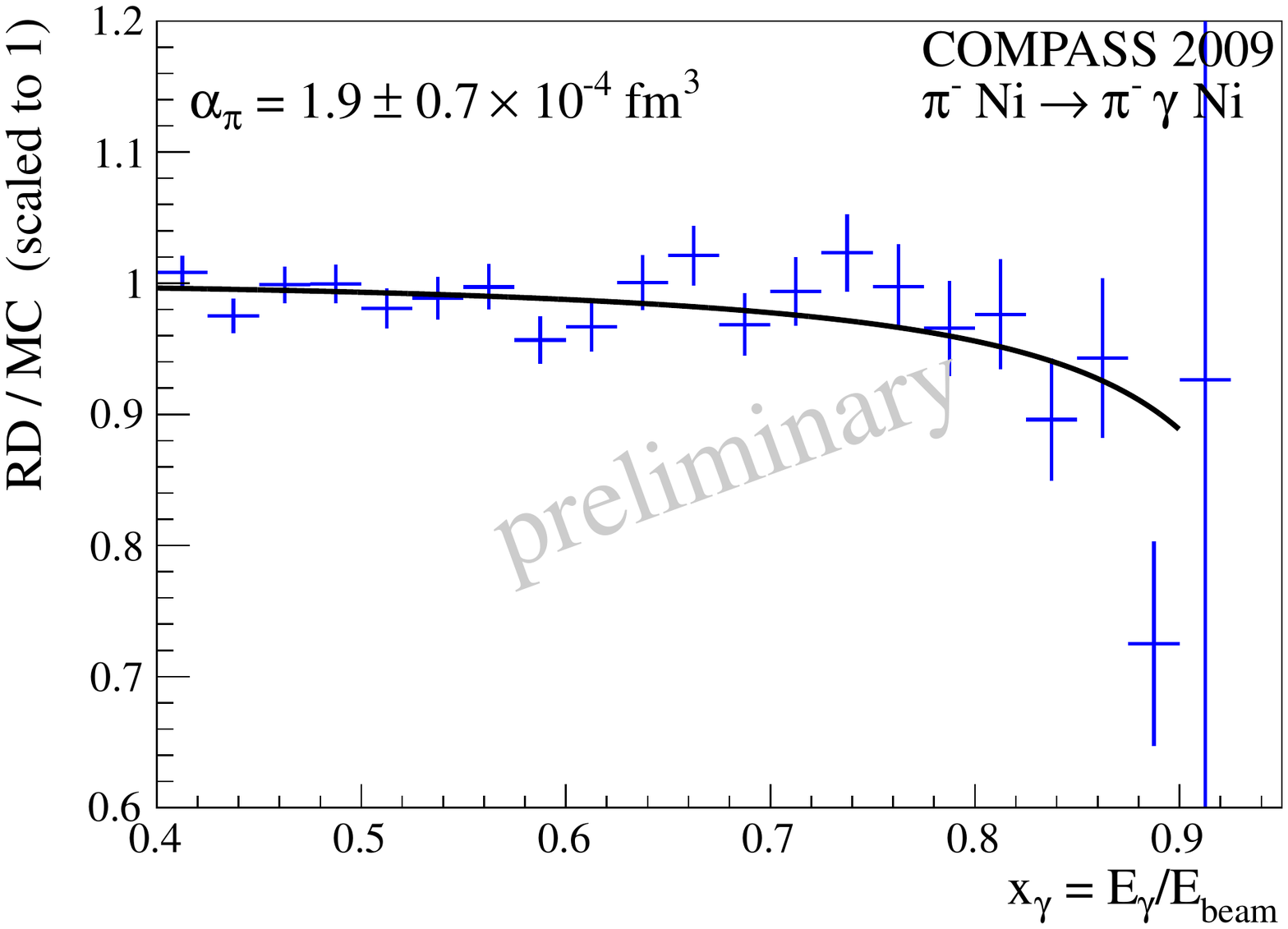}
  \caption{Determination of the pion polarizability at COMPASS through
    the process $\pi^- \mathrm{Ni}\rightarrow \pi^-\gamma
    \mathrm{Ni}$ \cite{Friedrich:2012ara}.}
  \label{fig:secB1.pion_polarizability}
\end{figure}

\paragraph{Pion polarizabilities}
\label{sec:lq.struct.pion.polarizability}
Further fundamental low-energy properties of the pion are its electric
and magnetic polarizabilities $\alpha_\pi$ and $\beta_\pi$. While firm
theoretical predictions exist based on 
ChPT~\cite{Gasser:2006qa,Kaiser:2011zz}, the experimental determination of
these quantities from pion-photon interactions using the Primakoff
effect \cite{Antipov:1982kz}, radiative pion photoproduction
\cite{Ahrens:2004mg}, and $\gamma\gamma\rightarrow \pi\pi$
\cite{Fil'kov:2005ss}, resulted in largely scattered and inconsistent
results. The COMPASS experiment at CERN has performed a first
measurement of the pion polarizability in pion-Compton scattering with
$190\,\GeV/c$ pions off a Ni target via the Primakoff effect.  The
preliminary result, extracted from a fit to the ratio of measured
cross section and the one expected for a point-like boson shown in
Fig.~\ref{fig:secB1.pion_polarizability}, is $\alpha_\pi=(1.9 \pm
0.7_\mathrm{stat}\pm 0.8_\mathrm{sys})\cdot 10^{-4}\,\mathrm{fm}^3$,
where the relation between electric and magnetic polarizability
$\alpha_\pi=-\beta_\pi$ has been assumed \cite{Friedrich:2012ara}.
This result is in tension with previous experimental results, but is
in good agreement with the expectation from 
ChPT~\cite{Gasser:2006qa}. New data taken with the COMPASS spectrometer in
2012 are expected to decrease the statistical and systematic error,
determined at COMPASS by a control measurement with muons in the same
kinematic region, by a factor of about three.  The data will for the
first time allow an independent determination of $\alpha_\pi$ and
$\beta_\pi$, as well as a first glimpse on the polarizability of the
kaon. Studies of the charged pion polarizability have been proposed
and approved at JLab, where the photon beam delivered to Hall
D will be used for the Primakoff production of $\pi^+\pi^-$ of a
nuclear target. A similar study of the $\pi^0$ polarizability will also
be possible.

\subsection{Hadron spectroscopy}
\label{sec:lq.spec}




In contrast to physical systems bound by electromagnetic interactions, 
the masses of light hadrons are not dominated by the masses of their
elementary building blocks but are 
to a very large extent generated dynamically by the strong
force. 
The coupling of the light quarks to the Higgs field is only responsible for
$\sim 1\%$ of the visible mass of our present-day universe, the rest
is a consequence of the interactions between quarks and gluons.   
While at high energies 
the interactions between partons 
become asymptotically free, allowing systematic calculations in QCD
using perturbation theory, the average energies and momenta of partons
inside hadrons are below the scale where perturbative methods are
justified. 
As a consequence, the fundamental degrees of
freedom of the underlying theory of QCD do not directly manifest
themselves in the physical spectrum of hadrons, which, rather, are
complex, colorless, many-body systems.  
One of the main goals of the physics of strong interactions for many 
years has been the determination and the understanding of the
excitation spectrum of these strongly bound states. 
In the past, phenomenological models have been developed, which quite
successfully describe certain aspects of the
properties of hadrons in terms of effective degrees of freedom, e.g.,   
the quark model \cite{GellMann:1964nj,Zweig:1981pd}, the bag model 
\cite{Chodos:1974je,DeGrand:1975cf}, the flux-tube model
\cite{Isgur:1984bm}, or   
QCD sum rules \cite{Shifman:1978bx}.   
A full understanding of 
the hadron spectrum from the underlying theory of QCD, however,
is still missing. Nowadays, QCD solved numerically on a discrete
spacetime lattice \cite{Wilson:1974sk} is one of the most promising
routes towards this goal. 

On the experimental side, 
significant advances in the light-quark sector have been 
made in the last few years. Data with
unprecedented statistical accuracy have become available from
experiments at both electron and hadron machines, often coupled with
new observables related to polarization or precise determination of
the initial and final state properties. 
In the light-meson sector, 
the unambiguous identification and systematic study of bound states
beyond the constituent 
quark degrees of freedom, e.g., multiquark states or states with 
gluonic degrees of freedom (hybrids, glueballs), allowed by QCD due to
its non-Abelian 
structure, 
is within reach of present and future generations of
experiments.
For a recent review, see e.g.\ \cite{Klempt:2007cp}. For the light
baryons, photoproduction experiments shed new light on the
long-standing puzzle of missing resonances. Here, the recent progress 
is summarized in\ \cite{Crede:2013kia}. 

On the theoretical side, hadron spectroscopy has received a huge boost
from lattice QCD. Simulations with dynamical up, down, and 
strange quarks are now routinely performed, and in many cases the need
for chiral extrapolations is becoming obsolete thanks to the ability
to simulate at or near the physical values of the up and down quark
masses~\cite{Aoki:2009ix,Durr:2010aw}. This concerns, in particular,
lattice calculations of the masses of the lightest mesons and
baryons~\cite{Aoki:2008sm,Durr:2008zz,Alexandrou:2009qu}, which show
excellent agreement with experiment. 
Lattice-QCD calculations for the masses of higher-lying mesons, baryons, as
well as possible 
glueball and hybrid states can provide guidance for experiments to
establish a complete understanding of the hadron spectrum. 
Other theoretical tools, such as dispersion relations, provide a way to
extract physically relevant quantities such as pole positions and
residues of amplitudes.

\subsubsection{Lattice QCD}
\label{sec:lq.spec.lqcd}
The long-sought objective of studying hadron resonances with 
lattice QCD is finally becoming a reality. 
The discrete energy spectrum of hadrons  
can be determined by computing  
correlation functions between creation and annihilation of an 
interpolating operator 
$\mathcal{O}$ at Euclidean times
$0$ and $t$, respectively,
\begin{equation}
  \label{eq:lq.spec.corr}
  C(t) = \bra{0}\mathcal{O}(t)\mathcal{O}^\dagger(0)\ket{0}. 
\end{equation}
Inserting a complete set of eigenfunctions $\ket{n}$ of the
Hamiltonian $\hat{H}$ which satisfy
$\hat{H}\ket{n}=E_k\ket{n}$, the correlation function can be written
as a sum of contributions from all states in the spectrum with the
same quantum numbers,
\begin{equation}
  \label{eq:lq.spec.corr2}
  C(t) = \sum_n{\left|\bra{0}\mathcal{O}\ket{n}\right|^2
  e^{-E_n t}}.
\end{equation}
For large times, the ground state dominates, while the excited states
are subleading contributions. To measure the energies of excited
states, it is thus important to construct
operators which have a large overlap with a given state. The technique
of smearing the quark-field creation operators is well-established to
improve operator overlap
\cite{Albanese:1987ds,Teper:1987wt,Gusken:1989ad,Allton:1993wc}. A
breakthrough for the study of excited 
states was the introduction of the distillation technique
\cite{Peardon:2009gh}, where the smearing function is replaced by a
cost-effective low-rank approximation.  
The interpolating operators are usually constructed from a sum of 
basis operators $\mathcal{O}_i$ for a given channel, 
\begin{equation}
  \label{eq:lq.spec.basis}
  \mathcal{O}=\sum_i{v_i}\mathcal{O}_i, 
\end{equation}
and a variational method \cite{Michael:1982gb} is then employed to
extract the best linear combination of operators within a finite
basis for each state which maximizes $C(t)/C(t_0)$. This 
requires the determination of all elements of the correlation matrix
\begin{equation}
  \label{eq:lq.spec.cm}
  C_{ij}(t) = \bra{0}\mathcal{O}_i(t)\mathcal{O}_j^\dagger(0)\ket{0},
\end{equation} 
and the solution of the generalized eigenvalue problem \cite{Luscher:1990ck,Blossier:2009kd}
\begin{equation}
  \label{eq:lq.spec.gevp}
  \mathbf{C}(t)\bm{v}_n = \lambda_n \mathbf{C}(t_0)\bm{v}_n.
\end{equation}
The procedure requires a good basis set of operators that resembles
the states of interest. 

\begin{figure*}
    \centering
    \includegraphics[width=0.80\textwidth]{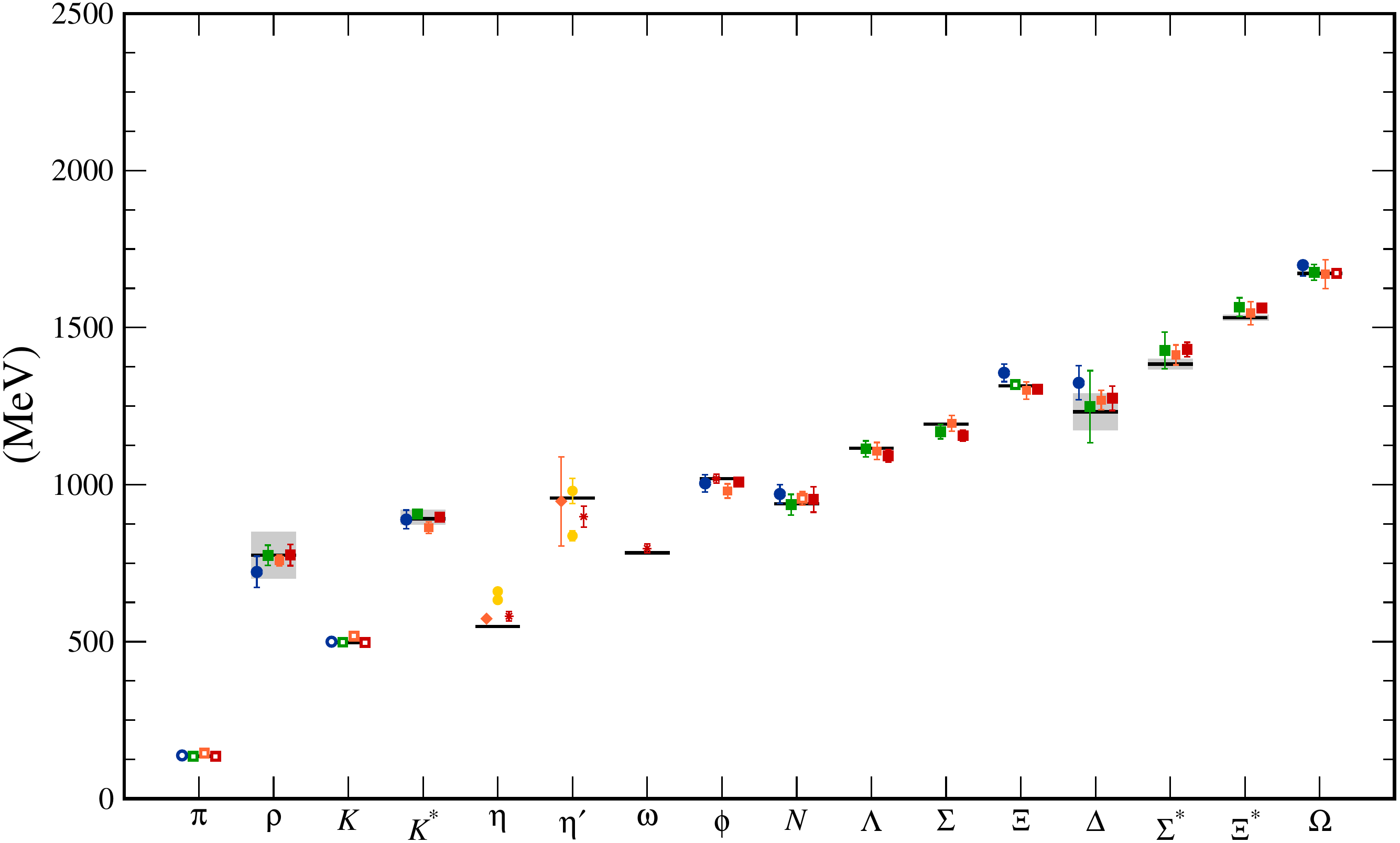}
    \caption[fig:spectrum]{Hadron spectrum from lattice QCD.
        Wide-ranging results are from 
        MILC~\cite{Aubin:2004wf,Bazavov:2009bb},
        PACS-CS~\cite{Aoki:2008sm},
        BMW~\cite{Durr:2008zz}, and
        QCDSF~\cite{Bietenholz:2011qq}.
        Results for $\eta$ and $\eta'$ are from 
        RBC \& UKQCD~\cite{Christ:2010dd},
        Hadron Spectrum~\cite{Dudek:2011tt} (also the only $\omega$ mass), and
        UKQCD~\cite{Gregory:2011sg}.
        Symbol shape denotes the formulation used for sea quarks. 
        Asterisks represent anisotropic lattices.
        Open symbols denote the masses used to fix parameters. 
        Filled symbols (and asterisks) denote results.
        Red, orange, yellow, green, and blue stand for increasing numbers of ensembles 
        (i.e., lattice spacing and sea quark mass).
        Horizontal bars (gray boxes) denote experimentally measured masses (widths).
        Adapted from \cite{Kronfeld:2012uk}.}
    \label{fig:lq.spec.lqcd_light_hadron_spectrum}
\end{figure*}

Thanks to algorithmic and computational advances in recent years, lattice-QCD calculations of the
lowest-lying mesons and baryons with given quantum numbers and quark content have been performed with full
control of the systematics due to lattice artifacts (see the review in \cite{Fodor:2012gf}).
Figure~\ref{fig:lq.spec.lqcd_light_hadron_spectrum}   
shows a 2012 compilation of lattice-QCD calculations of the 
light-hadron spectrum \cite{Kronfeld:2012uk}.
The pion and kaon masses have been used to fix the masses of light and strange 
quarks, and (in each case) another observable is used to set the overall mass scale.
The experimentally observed spectrum of the baryon octet and decuplet 
states, as well as the masses of some
light vector mesons, are well reproduced within a few percent of accuracy.
Except for the isosinglet mesons, 
the calculations shown use several lattice spacings and a wide range of pion 
masses. 
They also all incorporate $2+1$ flavors into the sea, but the chosen
discretization of the QCD action differs.
The consistency across all calculations suggests that the systematics, which are different for different 
calculations, are well controlled.
This body of work is a major achievement for lattice QCD, and the precision will improve while the methods 
are applied to more challenging problems.

Also for simulations of excited mesons and baryons huge progress has been made,
although the control of the systematics is still much less advanced
than in the case of the ground states \cite{Bulava:2009jb, Dudek:2010wm,
  Bulava:2010yg, Engel:2010my, Mahbub:2010rm, Dudek:2011tt, Edwards:2011jj,
  Menadue:2011pd, Engel:2011aa, Dudek:2012ag,
  Mahbub:2012ri, Edwards:2012fx, Engel:2013ig, Michael:2013gka}. These
calculations are typically 
performed for relatively few fairly coarse lattice spacings, and no
continuum extrapolation is attempted. A systematic study of
finite-volume effects, as well as the extrapolation to physical
quark masses, have not yet been performed. 
The goal of these calculations is to
establish a general excitation pattern rather than to perform precision
calculations. The focus at the moment is therefore on identifying a 
good operator basis, on disentangling various excitations in a given
channel, and on separating resonances from multihadron states. 

\paragraph{Light mesons}
\label{sec:lq.spec.lqcd.mesons}
As an example of recent progress, the work
of the Hadron Spectrum Collaboration  
\cite{Dudek:2010wm,Dudek:2011tt,Edwards:2011jj} is   
highlighted here, which recently performed a 
fully dynamical (unquenched) lattice-QCD calculation of
the complete light-quark spectrum of mesons and baryons.   
The simulations are carried out on anisotropic lattices 
with lattice spacings $a_\mathrm{s}\sim 0.12\,\mathrm{fm}$ and
$a_t^{-1}\sim 5.6\,\mathrm{GeV}$ in the spatial and temporal
directions, respectively, and 
with spatial volumes of $L^3 \sim(2.0\,\mathrm{fm})^3$ and 
$(2.5\,\mathrm{fm})^3$. 
They are performed with three flavors of order-$a$ improved Wilson quarks,
i.e., a mass-degenerate light-quark doublet, corresponding to a pion mass 
down to $396\,\MeV$, and a 
heavier quark whose mass is tuned to that of the strange quark. 
A large basis of smeared operators for single mesons was built
using fermion bilinears projected onto zero meson momentum,
including up to three gauge-covariant derivatives. No operators
corresponding to multiparticle states, however, were used. The
distillation method was used to 
optimize coupling to low-lying excited states. 
The correlators are analyzed by a variational method, which gives the
best estimate for masses and overlaps. 
The spins of states are determined by
projection of angular momentum eigenstates onto the irreducible
representations of the hypercubic group. 

\begin{figure*}[tbp]
  \includegraphics[width=\textwidth]{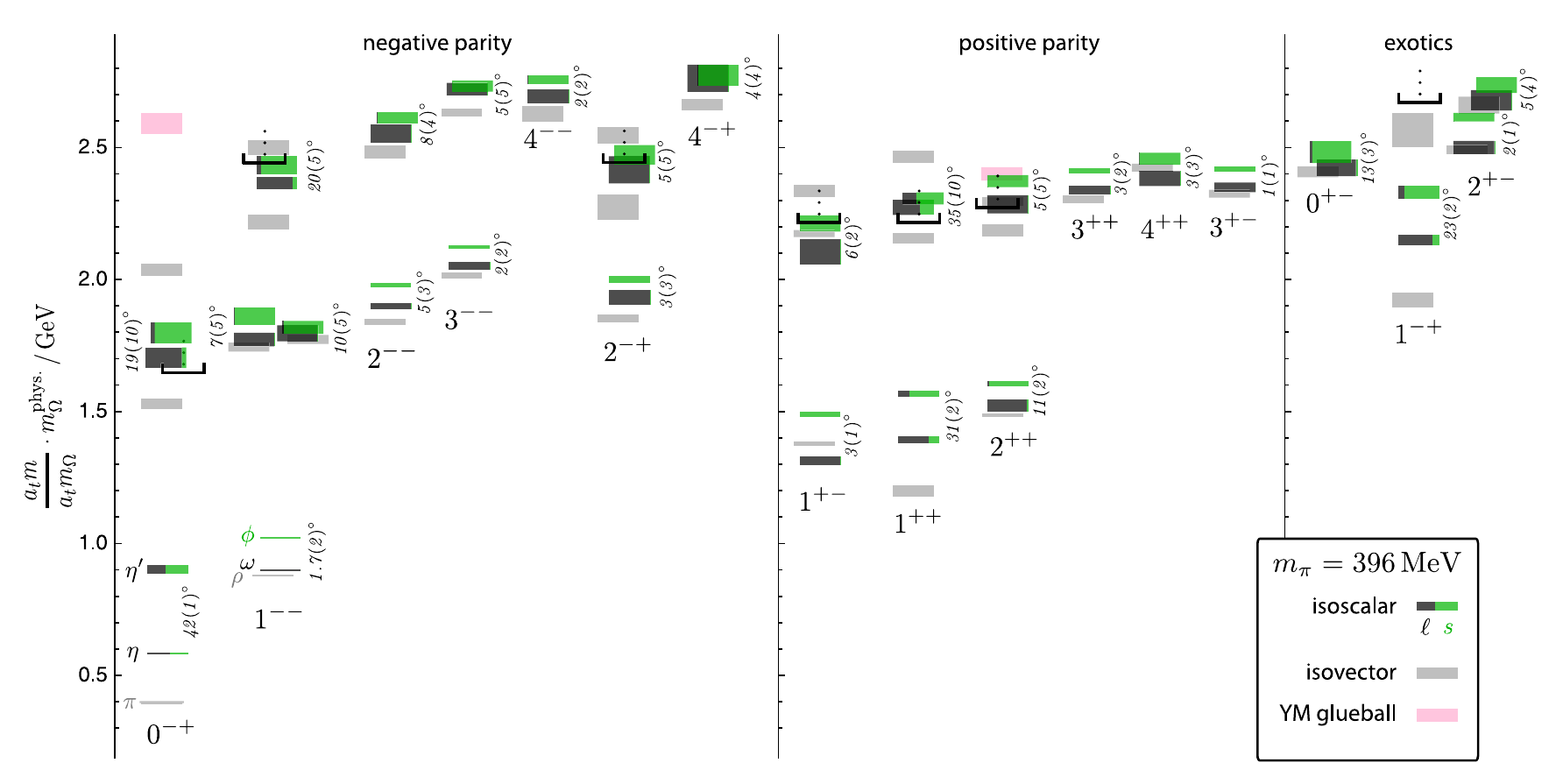}
  \caption{Light-quark meson spectrum resulting from
    lattice QCD \cite{Dudek:2011tt}, sorted by the quantum
    numbers $J^{PC}$. Note that these results have been obtained with
    an unphysical pion mass, $m_\pi=396\,\MeV$.}
  \label{fig:secB2.lqcd_meson_spectrum}
\end{figure*}

The resulting isoscalar and isovector 
meson spectrum is shown in Fig.~\ref{fig:secB2.lqcd_meson_spectrum}
\cite{Dudek:2011tt}.  
Quantum numbers and the quark-gluon structure of a meson state
$n$ with a given
mass $m_{n}$ are extracted by studying matrix elements 
$\langle n|\mathcal{O}_i|0\rangle$, which encode the extent
to which operator $\mathcal{O}_i$ overlaps with state $n$. States
with high spins, up to~$4$, are 
resolved. The
resulting spectrum, as well as the strange-nonstrange mixing of
isoscalar mesons, compares well with the currently 
known states \cite{Beringer:2012zz}.   
The calculated masses come out about
$15\%$ too high, 
probably owing to the unphysical pion mass, $m_\pi=396\,\MeV$.  
The lattice-QCD simulations also predict a number of 
extra states, that are not yet well established experimentally. These
include a series of exotic states with quantum numbers which cannot be
produced by pairing a quark and an antiquark, like
$J^{PC}=0^{+-},1^{-+},2^{+-},\ldots$, which have been previously postulated to
exist also in various models. For some states, a significant overlap
with operators containing the gluon field strength tensor has been
found, making them candidates for hybrids. It is interesting to note
that the quantum numbers and the degeneracy pattern predicted by lattice QCD 
for hybrid mesons are 
quite different from those of most
models. Lattice QCD predicts four low-mass hybrid multiplets at masses around
$2\,\GeV$ with quantum
numbers  
$1^{-+},0^{-+},1^{--},2^{-+}$, in agreement with 
the bag model \cite{Chanowitz:1982qj,Barnes:1982tx}, but at variance
with the flux-tube model \cite{Isgur:1984bm,Barnes:1995hc}, which  
predicts eight nearly degenerate hybrid multiplets. At masses larger
than $2.4\,\GeV$, lattice QCD predicts a group of ten hybrid multiplets,
in disagreement with bag and flux-tube model predictions. 
The pattern emerging from lattice QCD, i.e., of four low-mass and ten
higher-mass multiplets, can be 
reproduced by a $q\overline{q}'$ pair in an $S$- or $P$-wave 
coupled to a $1^{+-}$ chromomagnetic gluonic excitation, 
which can be modeled by a quasi-gluon in a $P$-wave with respect to the
$q\overline{q}'$ pair \cite{Guo:2007sm}.   

The spectrum of glueballs has first been calculated on a lattice in
pure SU(3) Yang-Mills theory, i.e.\ in the quenched
approximation to QCD~\cite{Bali:1993fb,Morningstar:1999rf,Chen:2005mg} at a
lattice spacing of $a\sim 0.1-0.2\,\mathrm{fm}$. A
full spectrum  
of states is predicted with the lightest one having scalar quantum
numbers, $0^{++}$, and a mass between $1.5\,\GeV$ and $1.7\,\GeV$. 
Also the next-higher glueball 
states have nonexotic quantum numbers,  
$2^{++}$ (mass $2.3$--$2.4\,\GeV$) 
and $0^{-+}$ (mass $2.3$--$2.6\,\GeV$), and hence will be 
difficult to identify experimentally. In a simple constituent gluon
picture, these three states correspond to two-gluon systems in
relative $S$~wave, with different combinations of
helicities. Table~\ref{tab:lq.spec.lqcd.glueballs} summarizes the
quenched lattice results for the masses of the lightest glueballs. 
\begin{table}[tpb]
  \caption{Continuum-limit glueball masses (in MeV) from
    quenched lattice 
    QCD. The first parentheses contain the statistical errors, while
    the second, where present, include the scale uncertainty.}
  \label{tab:lq.spec.lqcd.glueballs}
  \centering
  \begin{tabular}{l*{3}{@{\quad\;}l}}
    \hline\hline
    $J^{PC}$ & Bali~\cite{Bali:1993fb} &
    Morningstar~\cite{Morningstar:1999rf} & Chen~\cite{Chen:2005mg} \\ 
    \hline
    $0^{++}$ & $1550(50)$ & $1730(50)(80)$ & $1710(50)(80)$ \\
    $2^{++}$ & $2270(100)$ & $2400(25)(120)$ & $2390(30)(120)$ \\
    $0^{-+}$ & $2330(270)$ & $2590(40)(130)$ & $2560(35)(120)$ \\
    \hline\hline
  \end{tabular}
\end{table}

\begin{figure}[tbp]
  \includegraphics[width=0.5\textwidth]{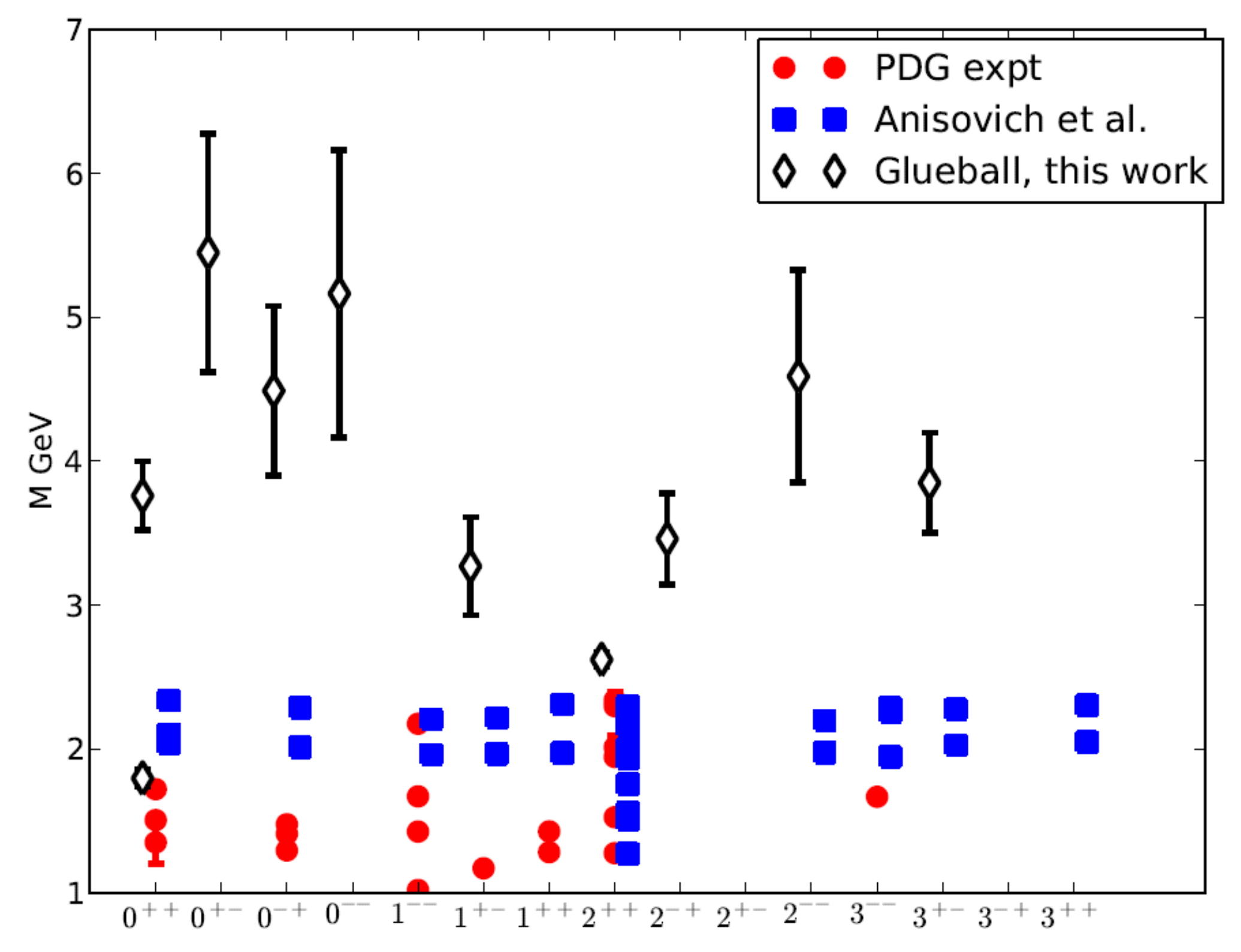}
  \caption{Glueball masses resulting from unquenched 
    lattice QCD \cite{Gregory:2012hu}, compared with
    experimental meson masses \cite{Beringer:2012zz,Anisovich:2011in}. 
  From \cite{Gregory:2012hu}.}
  \label{fig:secB2.lqcd_glueball_spectrum}
\end{figure}

While the glueball spectrum in 
pure SU(3) Yang-Mills theory is theoretically
well defined, because the glueball operators do not
mix with fermionic operators, unquenched lattice calculations are more 
difficult. The dynamical sea quarks will cause the glueball and flavor
singlet fermionic $0^{++}$ interpolating operators to couple to
the same physical states. In addition, decays of the $0^{++}$ states 
into two mesons are allowed for sufficiently light quark masses, and
may thus play an important role and dynamically modify the properties
of the glueball state.  
Hence, lattice-QCD calculations of the glueball spectrum with dynamical 
$q\overline{q}$ contributions are
still at a relatively early stage
\cite{Hart:2001fp,Hart:2006ps,Gregory:2012hu}. One particular problem
is the unfavorable signal-to-noise ratio of the relevant correlation
functions, which requires large statistics. 
The authors of~\cite{Gregory:2012hu}, using 2+1 flavors of ASQTAD
improved staggered fermions and a variational
technique which includes glueball scattering states, 
found no evidence for large effects from including dynamical sea
quarks. 
Their mass for the
$0^{++}$ glueball, $1795(60)\,\MeV$, is only slightly higher compared
to the quenched result of~\cite{Chen:2005mg}. 
Figure~\ref{fig:secB2.lqcd_glueball_spectrum} shows the glueball
masses calculated in~\cite{Gregory:2012hu}, compared to some 
experimental meson masses. 
No extrapolation
to the continuum, however, was performed, and no fermionic scattering
states were included. Much higher statistics will be needed for precise
unquenched calculations of 
flavor singlet sector 
on the lattice,   
with a 
$200\,\MeV$  
resolution needed, e.g., to distinguish the three isoscalar mesons in
the $1.5\,\GeV$ mass range. 
A technique designed to overcome the problem of an exponentially increasing
noise-to-signal ratio in glueball calculations has been proposed and tested in
the quenched approximation \cite{DellaMorte:2010yp}. However, it is not known
whether it can be generalized to full QCD.

\paragraph{Light baryons}
\label{sec:lq.spec.lqcd.baryons}
In addition to the meson spectrum, also the spectrum of baryons containing the 
light quarks $u, d$, and~$s$ has been calculated recently by different groups 
\cite{Bulava:2009jb, Bulava:2010yg, Engel:2010my, Mahbub:2010rm,
  Edwards:2011jj, Menadue:2011pd, Dudek:2012ag, Alexandrou:2012xk,
  Mahbub:2012ri, Edwards:2012fx, Engel:2013ig}. While the focus lies mostly on
establishing the spectral pattern of baryon resonances, the possibility of the
existence of hybrid baryons has also been addressed. For instance,
the Hadron Spectrum Collaboration has obtained 
spectra for $N$ 
and $\Delta$ baryons with $J\leq\frac{7}{2}$ and masses up to
$\sim 1.9M_\Omega$ \cite{Edwards:2011jj}. The well-known pattern
of organizing the states in multiplets of $SU(6)\times O(3)$, 
where the
first is the spin-flavor group, clearly emerges when checking the
overlap of the states with the different source/sink operators. The
multiplicity of states observed is similar to that of the
nonrelativistic quark model. 
The first excited positive-parity state, however, is found to have
significantly higher mass than its 
negative spin-parity partner, in contrast to the experimental ordering of
the $N(1440)\frac{1}{2}^+$ and the $N(1535)\frac{1}{2}^-$. The chiral behavior
of the observed level structure in these calculations has been analyzed in
detail in \cite{Engel:2010my,Mahbub:2010rm,Mahbub:2012ri,Engel:2013ig}.
Furthermore, no obvious pattern of degenerate levels with opposite
parity (parity doubling) emerges from the simulations for higher masses,
in contrast to indications from experiments \cite{Anisovich:2011ye}.  
Lattice-QCD calculations have been extended to include excited hyperons
\cite{Menadue:2011pd,Edwards:2012fx,Engel:2013ig}. In particular, the nature of the
$\Lambda(1405)$ has been the subject of the study in \cite{Menadue:2011pd}.
Moreover, lattice QCD presents the possibility of testing for the presence of
excited glue in baryons (hybrids) for the first time, and it has been
carried out in~\cite{Dudek:2012ag}.  
In contrast to the meson sector, however, 
all possible $J^P$ values for baryons can be built up from states
consisting of three quarks with nonvanishing orbital angular momentum
between them, so that there is no spin-exotic signature of hybrid baryons.  
It is found that their multiplet
structure is compatible with a color-octet chromomagnetic excitation
with quantum numbers $J^P=1^{+-}$, coupling to three quarks in a
color-octet state and forming a color-neutral object, as in the case
of hybrid mesons. Also the mass 
splitting between 
the $qqq$ states and the hybrid states is the same as observed for the
meson sector, indicating a common bound-state structure for hybrid
mesons and baryons. 

\paragraph{Future directions}
\label{sec:lq.spec.lqcd.future}
These very exciting developments still lack two aspects: First, there is the
issue of controlling systematic effects such as lattice artifacts,
finite-volume effects, and long chiral extrapolations owing to the use of
unphysical quark masses. Second, the fact that hadron resonances have a
nonzero width is largely ignored in the calculations described above,
i.e., resonances are treated as stable particles. While the first issue will be
dealt with once gauge ensembles with finer lattice spacing and smaller pion
masses are used for spectrum calculations, the second problem requires a
different conceptual approach. The position and width of a resonance are
usually determined from the scattering amplitude. However, as noted in
\cite{Maiani:1990ca}, the latter cannot be determined directly from
correlation functions computed in Euclidean space-time. L\"uscher pointed out
in his seminal work \cite{Luscher:1985dn, Luscher:1986pf, Luscher:1990ux,
  Luscher:1991cf} that the phase shift of the scattering amplitude in the
elastic region can be determined from the discrete spectrum of multi-particle
states in a finite volume. When plotted as a function of ${m_\pi}L$,
resonances can be identified via the typical avoided level crossing.

The L\"uscher formalism, which was originally derived for the center-of-mass
frame of mass-degenerate hadrons, has since been generalized to different
kinematical situations \cite{Rummukainen:1995vs, Kim:2005gf, Feng:2011ah,
  Fu:2011xz, Leskovec:2012gb, Doring:2012eu, Gockeler:2012yj}. Numerical
applications of the method are computationally quite demanding, since they
require precise calculations for a wide range of spatial volumes, as well as
the inclusion of multi-hadron interpolating operators. Most studies have
therefore focused on the simplest case, i.e., the $\rho$-meson
\cite{Aoki:2007rd, Frison:2010ws, Feng:2010es, Aoki:2011yj, Lang:2011mn,
  Pelissier:2012pi, Dudek:2012xn}. As reviewed in \cite{Mohler:2012nh}, other
mesonic channels such as $K\pi$, $D\pi$, and $D^{\ast}\pi$, as well as the
$N\pi$ system (i.e., the $\Delta$-resonance) have also been considered. While
the feasibility of extracting scattering phase shifts via the L\"uscher method
has been demonstrated, lattice calculations of resonance properties are still
at an early stage. In spite of the technical challenges involved in its
implementation, the L\"uscher method has been extended to the
phenomenologically more interesting cases of multi-channel scattering
\cite{He:2005ey, Lage:2009zv, Bernard:2010fp} and three-particle intermediate
states \cite{Polejaeva:2012ut,Kreuzer:2012sr}.


\subsubsection{Continuum methods}
\label{sec:lq.spec.models}
Although lattice calculations will provide answers to many questions
in strong QCD, the development of reliable analytical continuum
methods is a necessity to develop an intuitive understanding of QCD
from first principles, to construct advanced
phenomenological models,
and to address computationally challenging tasks 
like the extrapolation to physical quark masses or large hadronic
systems. The tools at our disposal include effective theories such as 
ChPT, Dyson-Schwinger methods, fixed gauge
Hamiltonian QCD approaches, and QCD sum rule methods.  

A study of baryon resonances with various models has been carried out
since time immemorial. In recent times, dynamical models based on 
meson-baryon degrees of freedom have received much attention
\cite{Meissner:1999vr,Oller:2000fj,Jido:2003cb,Hyodo:2011ur}, in
particular in the case of $S$~wave resonances, such as the
$\Lambda(1405)$. These models use effective Lagrangians to couple
light mesons to the ground state baryons, and in this way generate
resonances dynamically. Since baryons couple strongly to the
continuum, it is known that meson-baryon dynamics plays an important
role; one would like to eventually understand how to better quantify
that role by using improved models. One can speculate that this can
also be an interesting topic of exploration in the framework of
lattice QCD, where the possibility of varying the quark masses can
illuminate how excited baryon properties change with the pion mass.
Models of baryons based on the Schwinger-Dyson equations have also
been studied \cite{Eichmann:2008ef,Nicmorus:2008eh}, and are being
developed into important tools to study excited baryons with a
framework anchored in the principles of QCD.
  
In the spirit of effective theories, one approach based on the $1/N_c$
expansion has been developed
\cite{Goity:1996hk,Pirjol:1997sr,Pirjol:1997bp,Goity:2004pw}. In the
limit of large $N_c$, a spin-flavor dynamical symmetry emerges in the
baryon sector, which is broken at subleading order in $1/N_c$ and thus
provides a starting point for
the description of baryon observables in a power
series in $1/N_c$. As in every effective theory, it is necessary to
give inputs, namely baryon observables determined phenomenologically,
and the $1/N_c$ expansion serves to organize and relate them at each
order in the expansion. The framework is presented as an expansion in
composite operators, where quantities or observables are expanded on a
basis of operators at a given order in $1/N_c$, and the coefficients
of the expansion, which encode the QCD dynamics, are determined by
fitting to the observables.  It has been applied to baryon masses
\cite{Carlson:1998gw,Carlson:1998vx,Schat:2001xr,Goity:2002pu,Goity:2003ab,Matagne:2006zf,Goity:2007sc},
partial decay widths \cite{Goity:2004ss,Goity:2009wq,Jayalath:2011uc}, 
and photocouplings \cite{Scoccola:2007sn}. Through those analyses it
is observed that the different effects, which are classified by their
$SU(2 N_f)\times O(3)$ structure ($N_f$ is the number of light flavors) and by their power in $1/N_c$, seem
to follow the natural order of the $1/N_c$ expansion, that is, they
have natural magnitude. An interesting challenge is the implementation
of the $1/N_c$ expansion constraints in models, in order to have a
more detailed understanding of the dynamics. One such nice and illustrative
example 
has been given in \cite{Pirjol:2008gd}.



\subsubsection{Experiments}
\label{sec:lq.spec.exp}
The fundamental difficulty in studying the light-hadron spectrum is
that in most cases resonances do not appear as isolated, narrow
peaks. Instead, states have rather large widths of several hundred
$\MeV$ and consequently overlap. Peaks observed in a spectrum may be
related to thresholds opening up or interference effects rather than
to genuine resonances, not to speak of kinematic reflections or
experimental acceptance effects. In addition, nonresonant
contributions and final-state effects may also affect the measured
cross section. Partial wave or
amplitude analysis (PWA) techniques are the state-of-the-art way 
to disentangle 
contributions from  individual, and even small, resonances and to determine
their quantum numbers. 
Multiparticle decays are
usually modeled using the phenomenological approach of the isobar
model, which describes 
multiparticle final states by sequential
two-body decays into intermediate resonances (isobars),
that eventually decay into the final state observed in the
experiment. 
Event-based fits allow one to take into account the
full correlation between final-state particles. Coupled-channel
analyses are needed to reliably extract resonance parameters from
different reactions or final states. 

One notoriously difficult problem is the
parameterization of the dynamical properties of resonances. Very
often, masses and
widths of resonances are  
determined from Breit-Wigner parameterizations, although this approach
is strictly only valid for isolated, narrow states with a single decay
channel. For two-body processes, e.g., the K-matrix formalism provides a
way to ensure that the 
amplitudes fulfill the unitarity condition also in the case of
overlapping resonances. 
The rigorous definition of a resonance is by means of a pole in the
second (unphysical) Riemann sheet of the complex energy plane.  
For poles deep in the complex plane, however, none of the above
approaches yield 
reliable results, although they might describe the data well. The
correct analytical properties of the amplitude are 
essential for an extrapolation from the experimental data (real
axis) into the complex plane in order to determine the pole
positions. Dispersion relations provide a rigorous way to do this by
relating the amplitude at any point in the complex plane to an
integral over the (imaginary part of the) amplitude evaluated on the real
axis (i.e., the data) making use of Cauchy's theorem. 

\paragraph{Scalar mesons and glueballs}
\label{sec:lq.spec.exp.scalars}
The identification and classification of scalar mesons with masses
below $2.5\,\GeV$ is a long-standing puzzle. Some of them have large
decay widths and couple strongly to the two-pseudoscalar 
continuum. 
The opening of nearby thresholds such as 
$K\overline{K}$ and $\eta\eta$ strongly distort the resonance
shapes. In addition, non-$q\overline{q}'$ scalar objects like 
glueballs and multi-quark states are expected in the mass range below
$2\,\GeV$, which will mix with the states composed of $q\overline{q}'$. 
The Particle Data Group (PDG) currently lists
the following light scalars \cite{Beringer:2012zz}, sorted according
to their isospin, ($I=0$)  
$f_0(500)$, $f_0(980)$, $f_0(1370)$, $f_0(1500)$, $f_0(1710)$, ($I=1/2$)
$K_0^\ast(800)$ (listed as still requiring confirmation),
$K_0^\ast(1430)$, 
($I=1$) $a_0(980)$, $a_0(1450)$. 
One possible interpretation is that the scalars with masses below
$1\,\GeV$ form a new nonet with an inverted mass hierarchy, with the
wide, isoscalar $f_0(500)$ as the lightest member, the $K_0^\ast(800)$ 
(neutral and 
charged), and the 
isospin-triplet $a_0(980)$, which does not have any $s$-quark content
in the quark model, and its isospin-singlet counterpart
$f_0(980)$ as the heaviest members. The high masses of the
$a_0(980)$ and the $f_0(980)$ and their large coupling to $K\bar{K}$
could be explained by interpreting them as tightly bound tetraquark
states \cite{Jaffe:1976ig} or $K\bar{K}$ molecule-like objects
\cite{Weinstein:1983gd}. The scalar mesons above $1\,\GeV$ would form
another nonet, with one supernumerary isoscalar state, indicating the
presence 
of a glueball in the $1.5\,\GeV$ mass region mixing with the
$q\bar{q}$ states\ \cite{Amsler:1995tu}.    
Other interpretations favor an ordinary
$q\bar{q}$ nonet consisting of $f_0(980)$, $a_0(980)$,
$K_0^\ast(1430)$, and $f_0(1500)$\
\cite{Minkowski:1998mf,Ochs:2013gi}. The $f_0(1370)$ is interpreted as
an interference effect. The $K_0^\ast(800)$ is not required in this
model, and the supernumerary broad $f_0(500)$ would then have a large
admixture of a light glueball.  
In view of these different interpretations it is important to clarify
the properties of scalar mesons. 
An updated review on the
topic can be found, e.g., in the 
PDG's ``Note on Scalar
Mesons below $2\,\GeV$'' \cite{Beringer:2012zz}. 

When it comes to the lightest scalar mesons, the $f_0(500)$, 
huge progress has been made in recent years towards a confirmation of its
resonant nature and the determination of its pole position. 
Although omitted from the PDG's
compilation 
for many years, its existence has been verified in several
phenomenological analyses of $\pi$-$\pi$ scattering data.  As for other
scalar particles, the $f_0(500)$, also known as $\sigma$, is produced in, 
{e.g.}, $\pi$-$N$-scattering or $\bar pp$-annihilation, and data is, 
in particular, obtained from $\pi$-$\pi$, $K$-$\bar K$, $\eta$-$\eta$, 
and 4$\pi$ systems in the $S$-wave channel. The analyses of several
processes require four poles, the $f_0(500)$ and three other scalars,
in the region from the $\pi$-$\pi$ threshold to $1600\,\MeV$. Hereby the
missing distinct resonance structure below $900\,\MeV$ in $\bar p
p$-annihilation was somehow controversial. However, by now it is
accepted that also this data is described well with the standard
solution requiring the existence of the broad $f_0(500)$.

The pole position, {i.e.}, the pole mass and related width, is also
accurately determined. The combined analysis with ChPT and dispersion
theory of $\pi$-$\pi$ scattering \cite{Caprini:2005zr} has led to a
particularly accurate determination of those parameters. The PDG
quotes a pole position of $M-i\Gamma/2\simeq\sqrt{s_\sigma} =
(400$--$550)-i(200$--$350)\,\MeV$, whereas averaging over the most advanced
dispersive analyses gives a much more restricted value of
$\sqrt{s_\sigma} = (446\pm 6)-i(276\pm 5)\,\MeV$.  Especially relevant
for the precise determination of the $f_0(500)$ pole were recent data
from the NA48/2 experiment at CERN's Super Proton Synchrotron (SPS) on
$K^\pm\rightarrow \pi^+\pi^- e^\pm \nu$ (K$_{e4}$) decays
\cite{Batley:2010zza}, which have a much smaller systematic
uncertainty than the older data from $\pi N\rightarrow \pi\pi N$
scattering due to the absence of other hadrons in the final state.
NA48/2 has collected $1.13$ million K$_{e4}$ events using simultaneous
$K^+$ and $K^-$ beams with a momentum of $60\,\GeV$.

As mentioned above, however, 
there exist many, 
partly mutually-excluding interpretations of the $f_0(500)$:
a quark-antiquark bound state, 
$\pi$-$\pi$ molecule, tetraquark, QCD dilaton, to name the most
prominent ones. In addition, it will certainly also mix with the
lightest glueball.
From the phenomenological side it is evident that the large
$\pi$-$\pi$ decay width is the largest obstacle in gaining more
accurate information. However, it is exactly the pattern of D$\chi$SB
which makes this width quite naturally so large. An $f_0(500)$ with a
small width could only occur if there is substantial explicit breaking
of chiral symmetry (because, e.g., a large current mass would lead to
$m_\sigma< 2 m_\pi $) or if by some other mechanism the scalar mass
would be reduced.

Here a look to the electroweak sector of the
Standard Model is quite enlightening.  The scalar particle claimed
last year by CMS and ATLAS with mass $125$--$126\,\GeV$ is consistent
with 
the Standard Model (SM) Higgs boson, cf.\ 
Sec.~\ref{subsubsec:secE2Higgstop}.  It
appears to be very narrow as its width-to-mass ratio is small. Though
the mass is a free parameter of the SM, one natural explanation of its
lightness relative to its ``natural'' mass of about $300$--$400\,\GeV$ is
fermion-loop mass renormalization, strongest by the top quark
loop. This is one clear contribution that makes the Higgs
light\footnote{Other possibilities exist, such as making the Higgs an
  additional Goldstone boson of new physics, e.g., a dilaton.}. In any
case, the accident $m_H < 2 m_W$ prevents the decay $h\to WW$. Since
the longitudinal $W$ components are the Goldstone bosons of
electroweak symmetry breaking, the analogy to $\sigma\pi\pi$ in QCD is
evident. If the top quark were much lighter, or if it would be less
strongly coupled (such as the nucleon to the sigma), the Higgs mass
could naturally be higher by some hundreds of GeV, the decay channel
to $WW$ would open, and the Higgs would have a width comparable
in magnitude to its mass.  This comparison makes it plain that the
$f_0(500)$, for which no fermion that 
strongly couples to
it is similar in mass, is naturally so broad because of the existence of
pions as light would-be Goldstone bosons and its strong coupling of
the two-pion channel. Unfortunately, this also implies that the nature
of the $f_0(500)$ can be only revealed by yet unknown nonperturbative
methods. It has to be emphasized that the lack of understanding of the
ground state in the scalar meson channel is an unresolved but
important question of hadron physics.

In recent dispersive analyses
\cite{GarciaMartin:2011jx,Moussallam:2011zg} of $\pi\pi$ 
scattering data and the very recent $K_{\ell 4}$ experimental results
\cite{Batley:2010zza}, the pole  
positions of the $f_0(500)$ and $f_0(980)$ were  
determined simultaneously, and the results, summarized in
Table~\ref{tab:lq.spec.f0}, are in excellent agreement with each other. 
\begin{table}[tbp]
  \centering
    \caption{Positions of the complex poles of the $f_0(500)$ and
    $f_0(980)$, determined in dispersive analyses
    \cite{GarciaMartin:2011jx,Moussallam:2011zg} 
    of $\pi\pi$ scattering data and $K_{\ell 4}$ decays.}
    \begin{tabular*}{\columnwidth}{c@{\extracolsep{\fill}}c@{\extracolsep{\fill}}c} \hline\hline
         & \multicolumn{2}{c}{$\sqrt{s_0}$ ($\MeV$)} \\ 
    Ref. & $f_0(500)$ & $f_0(980)$ \\ \hline
    \cite{GarciaMartin:2011jx} & 
    $\left(457^{+14}_{-13}\right)-i\left(279^{+11}_{-7}\right)$ &
    $\left(996\pm 7\right)-i\left(25^{+10}_{-6}\right)$ \\
    \cite{Moussallam:2011zg} &
    $\left(442^{+5}_{-8}\right)-i\left(274^{+6}_{-5}\right)$ &
    $\left(996^{+4}_{-14}\right)-i\left(24^{+11}_{-3}\right)$ \\ \hline\hline
  \end{tabular*}
  \label{tab:lq.spec.f0}
\end{table}

The situation with the lightest strange scalar, $K_0^\ast(800)$ or
$\kappa$, is more complicated. A dispersive analysis of $\pi
K\rightarrow \pi K$ scattering data gives a pole position of the
$K_0^\ast(800)$ of $\left(658\pm 13\right)-i/2\left(557 \pm
  24\right)\,\MeV$ \cite{DescotesGenon:2006uk}, while recent
measurements by BESII in $J/\psi\rightarrow K_S K_S \pi^+\pi^-$ decays
\cite{Ablikim:2010ab} give a slightly higher value for the pole
position of $\left(764\pm 63^{+71}_{-54}\right)-i\left(306\pm
  149^{+143}_{-85}\right)\,\MeV$. Similar results from dispersive
analyses are expected for the $a_0(980)$. A broad scalar with mass
close to that above is also needed for the interpretation of the
$K\pi$ invariant mass spectrum observed by Belle in $\tau^- \to
K^0_S\pi^-\nu_\tau$ decay~\cite{Epifanov:2007rf}.  Numerous
measurements of invariant mass spectra in hadronic decays of $D$ and
$B$ mesons are hardly conclusive because of the large number of
interfering resonances involved in parameterizations and different
models used in the analyses.

\begin{figure*}[tbp]
  \begin{center}
    \includegraphics[width=0.9\textwidth]{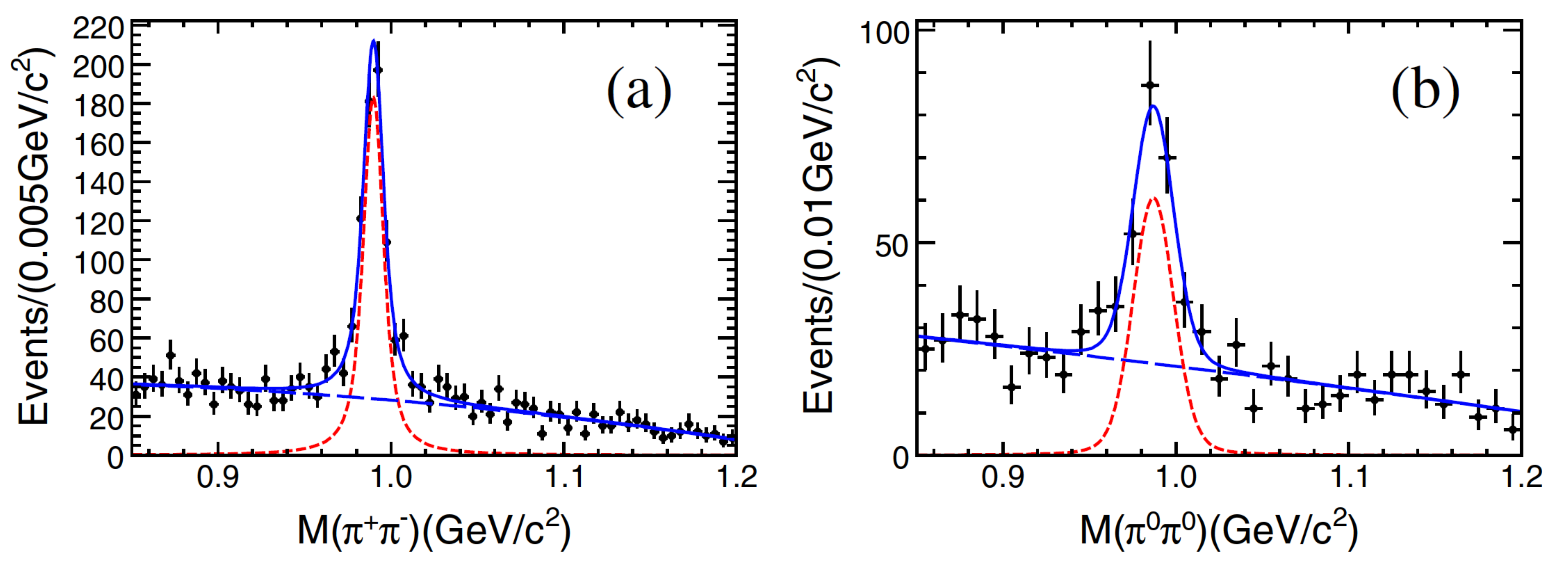}
    \caption{Invariant mass of $\pi^+\pi^-$ and $\pi^0\pi^0$ with the 
      $\pi^+\pi^-\pi^0$ ($3\pi^0$) mass in the $\eta(1405)$ mass
      region, measured at BES~III \cite{BESIII:2012aa}.} 
    \label{fig:f0.2pi}
  \end{center}
\end{figure*}

New data are being collected by BES~III at the
recently upgraded  
BEPCII $e^+e^-$ collider in Beijing in the
$\tau$-charm mass region at a luminosity of
$10^{33}\,\mathrm{cm}^{-2}\,\mathrm{s}^{-1}$ (at a center-of-mass (CM)
energy of
$2\times 1.89\,\GeV$), 
with a maximum CM energy of $4.6\,\GeV$ \cite{Ablikim:2009aa}.   
In the last three years, the experiment has collected the
world's largest data samples of $J/\psi$, $\psi(2S)$, and $\psi(3770)$  
decays. These data are also being used to make a
variety of studies in light-hadron 
spectroscopy, especially in the scalar meson sector. Recently, BES~III
reported the first observation of the isospin-violating decay $\eta(1405)
\rightarrow\pi^0 
f_0(980)$ in $J/\psi\rightarrow\gamma 3\pi$
\cite{BESIII:2012aa}, together with an anomalous  
lineshape of the $f_0(980)$ in the $2\pi$ invariant mass spectra, as
shown in Fig.~\ref{fig:f0.2pi}.
The $f_0$ mass, deduced from a Breit-Wigner
fit to the mass spectra, is slightly shifted compared to its nominal
value, with a width of $<11.8\,\MeV$ ($90\%$ C.L.), much smaller than its
nominal value. The observed isospin violation is $(17.9\pm 4.2)\%$,
too large 
to be explained by $f_0(980)$-$a_0(980)$ mixing, also observed
recently by BES~III at the $3.4\sigma$ level \cite{Ablikim:2010aa}. 
Wu et al.\ \cite{Wu:2011yx} suggest that a $K$ triangle anomaly could be
large enough to account for the data. 

\begin{figure}[tbp]
  \begin{center}
    \includegraphics[width=0.9\columnwidth]{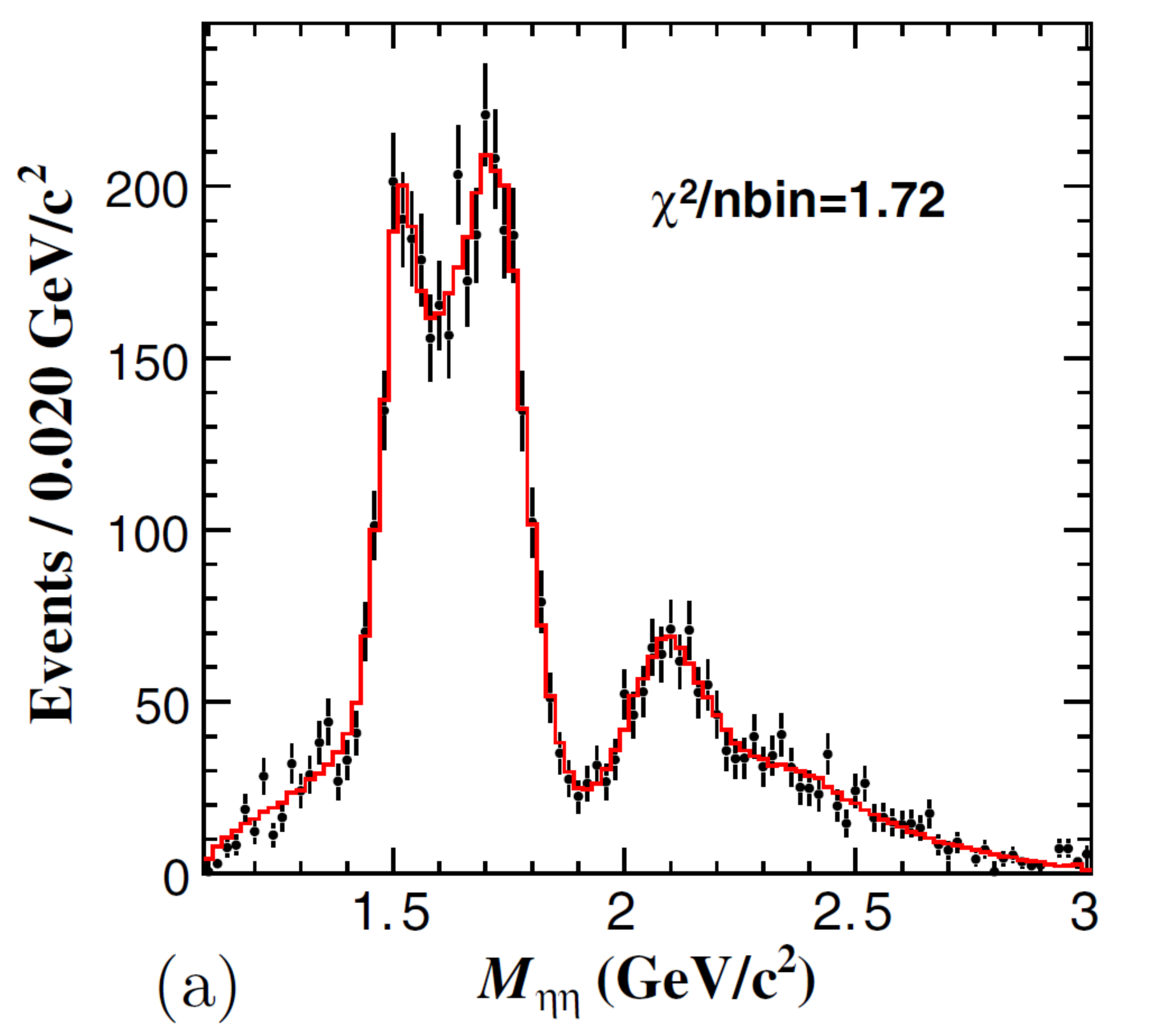}
    \caption{Invariant mass distribution of $\eta\eta$ from
      $J/\psi\rightarrow \gamma\eta\eta$, and the projection of the
      PWA fit from BES~III \cite{Ablikim:2013hq}.} 
    \label{fig:etaeta.pwa}
  \end{center}
\end{figure}

BES~III has recently performed a full PWA of
$5460$ radiative
$J/\psi$ decays 
to two 
pseudoscalar 
mesons, $J/\psi\rightarrow\gamma\eta\eta$, commonly regarded as an
ideal system to look for scalar and tensor glueballs. In its baseline
solution, the fit 
contains six 
scalar and tensor resonances
\cite{Ablikim:2013hq},   
$f_0(1500)$, $f_0(1710)$, $f_0(2100)$, $f_2'(1525)$, $f_2(1810)$, and
$f_2(2340)$, as well as $0^{++}$ phase space and $J/\psi\rightarrow
\phi\eta$.  
The scalars $f_0(1710)$, $f_0(2100)$, and $f_0(1500)$ are found to be
the dominant 
contributions, with the production rate for the latter being about
one order of magnitude smaller than for the first two. 
No evident contributions from $f_0(1370)$ or
other scalar mesons are seen. The well-known tensor resonance
$f_2'(1525)$ is clearly observed, but 
several $2^{++}$ tensor components are also needed in the
mass range between $1.8$--$2.5\,\GeV$. The statistical precision of the
data, however, is not yet sufficient to distinguish the contributions. 
Figure~\ref{fig:etaeta.pwa} shows the resulting PWA fit result of the
$\eta\eta$ 
invariant mass spectrum.

In conclusion, the situation in the scalar meson sector is
still unresolved. The lightest glueball is predicted to have scalar
quantum numbers and is 
therefore expected to    
mix with nearby isoscalar scalar $q\bar{q}$ $P$-wave states.
For recent reviews
on glueballs, see \cite{Crede:2008vw,Mathieu:2008me,Ochs:2013gi}.   
On the experimental side, further new
results from BES~III,  
from Belle on 
two-photon production of meson pairs \cite{Uehara:2013mbo,Liu:2012eb},
and from COMPASS on 
central production \cite{Austregesilo:2013pba} may help to 
resolve some of the questions in the scalar sector in the future.  

\paragraph{Hybrid mesons}
\label{sec:lq.spec.exp.hybrids}
Experimental evidence for the existence of hybrid mesons can come from
two sources. The observation of an overpopulation of states with
$q\overline{q}'$ quantum numbers may indicate the existence of 
states beyond the quark model, i.e., hybrids, glueballs, or multi-quark
states. The densely populated spectrum of light mesons in the mass
region between $1$ and $2\,\GeV/c^2$, and the broad nature of the
states involved, however, makes this approach difficult. It requires the
unambiguous identification of all quark-model states of a given
$J^{PC}$ nonet, a 
task which has been achieved only for the ground state nonets so
far. The identification of a resonant state with exotic, i.e., 
non-$q\overline{q}'$ states, however, is considered a ``smoking gun''
for the existence of such states. 
Table~\ref{tab:lq.spec.exp.hybrids} lists experimental candidates for
hybrid mesons and their main properties \footnote{Here we restrict
  ourselves to states used by the PDG in their averages\
  \cite{Beringer:2012zz}, together with recent data from COMPASS not yet listed in
  the summaries.}. 
\begin{table*}[tbp]
  \caption{Experimental properties of low-mass hybrid candidate states 
    with quantum numbers $J^{PC}=1^{-+}$, $0^{-+}$, $2^{-+}$,
    $1^{--}$.
  } 
  \label{tab:lq.spec.exp.hybrids}
  \centering{\small
  \begin{tabular}{llllllrl} \hline\hline
    State & $J^{PC}$ & Final state & Decay mode(s) & Mass
    ($\MeV$) & Width ($\MeV$) & Events & Reference \\ \hline
    $\pi_1(1400)$ & $1^{-+}$ & $\pi^+\pi^-\pi^0\pi^0$ & $\eta\pi^0$ & $1257\pm  20\pm 25$ & 
    $354\pm 64\pm 60$ & $24$k  & E852\ \cite{Adams:2006sa} \\
    & & $2\pi^+2\pi^-$ & $\rho\pi$ & $1384\pm  20\pm 35$ & 
    $378\pm 58$ & $90$k  & OBELIX\ \cite{Salvini:2004gz} \\
    & & $\pi^0\pi^0\eta(2\gamma)$ & $\eta\pi^0$ & $1360\pm  25$ & 
    $220\pm 90$ & $270$k  & CB\ \cite{Abele:1999tf} \\
    & & $\pi^-\pi^0\eta(2\gamma)$ & $\eta\pi$ & $1400\pm  20\pm 20$ & 
    $310\pm 50^{+50}_{-30}$ & $53$k  & CB\ \cite{Abele:1998gn} \\
    & & $\pi^-\eta(2\gamma)$ & $\eta\pi^-$ & $1370\pm  16^{+50}_{-30}$ & 
    $385\pm 40^{+65}_{-105}$ & $47$k  & E852\ \cite{Thompson:1997bs} \\
    \hline
    $\pi_1(1600)$ & $1^{-+}$ & $\pi^+\pi^-\pi^-$ & $\rho\pi^-$ & $1660\pm  10^{+0}_{-64}$ & 
    $269\pm 21^{+42}_{-64}$ & $420$k  & COMPASS\ \cite{Alekseev:2009xt} \\
    & & $\pi^-\pi^0\omega(\pi^+\pi^-\pi^0)$ & $b_1(1235)\pi^-$ & $1664\pm  8\pm 10$ & 
    $185\pm 25\pm 28$ & $145$k  & E852\ \cite{Lu:2004yn} \\
    & & $\pi^-\pi^-\pi^+\eta(\gamma\gamma)$ & $f_1(1285)\pi^-$ & $1709\pm  24\pm 41$ & 
    $403\pm 80\pm 115$ & $69$k  & E852\ \cite{Kuhn:2004en} \\
    & & $\pi^-\pi^-\pi^+\eta(\gamma\gamma)$ & $\eta'\pi^-$ & $1597\pm  10^{+45}_{-10}$ & 
    $340\pm 40\pm 50$ & $6$k  & E852\ \cite{Ivanov:2001rv} \\ 
    \hline
    $\pi_1(2015)$ & $1^{-+}$ & $\pi^-\pi^0\omega(\pi^+\pi^-\pi^0)$ &
    $b_1(1235)\pi^-$ & $2014\pm  20\pm 16$ &  
    $230\pm 32\pm 73$ & $145$k  & E852\ \cite{Lu:2004yn} \\
    & & $\pi^-\pi^-\pi^+\eta(\gamma\gamma)$ & $f_1(1285)\pi^-$ & $2001\pm  30\pm 92$ & 
    $333\pm 52\pm 49$ & $69$k  & E852\ \cite{Kuhn:2004en} \\
    \hline
    $\pi(1800)$ & $0^{-+}$ & $3\pi^-2\pi^+$ & $f_0(1500)\pi^-$ & 
    $1781\pm 5^{+1}_{-6}$ & $168\pm 9^{+5}_{-14}$ & $200$k & COMPASS\ \cite{Neubert:2010zz} \\
    & & $\pi^+\pi^-\pi^-$ & $f_0(980)\pi^-$ & $1785\pm  9^{+12}_{-6}$ & 
    $208\pm 22^{+21}_{-37}$ & $420$k  & COMPASS\ \cite{Alekseev:2009xt} \\ 
    & & $\eta(\gamma\gamma)\eta(\pi^+\pi^-\pi^0)\pi^-$ &
    $a_0(980)\eta$, $f_0(1500)\pi^-$ &
    $1876\pm  18\pm 16$ & 
    $221\pm 26\pm 38$ & $4$k  & E852\ \cite{Eugenio:2008zza} \\ 
    & & $\pi^+\pi^-\pi^-$ & $f_0(980)\pi^-$ & $1774\pm  18\pm 20$ &  
    $223\pm 48\pm 50$ & $250$k  & E852\ \cite{Chung:2002pu} \\ 
    & & $\pi^+\pi^-\pi^-$ & $(\pi\pi)_S\pi^-$ & $1863\pm  9\pm 10$ & 
    $191\pm 21\pm 20$ & $250$k  & E852\ \cite{Chung:2002pu} \\ 
    & & $\eta(\pi^+\pi^-\pi^0)\eta(\gamma\gamma)\pi^-$ &
    $a_0(980)\eta$ & $1840\pm  10\pm 10$ &  
    $210\pm 30\pm 30$ & $1$k  & VES\ \cite{Amelin:1996zz} \\ 
    & & $\pi^+\pi^-\pi^-$ & $f_0(980)\pi^-$, $(\pi\pi)_S\pi^-$ & $1775\pm  7\pm 10$ & 
    $190\pm 15\pm 15$ & $2000$k  & VES\ \cite{Amelin:1995zz} \\ 
    & & $K^+K^-\pi^-$ & $f_0(980)\pi^-$, $K_0^\ast(800) K^-$ & $1790\pm  14$ & 
    $210\pm 70$ & $145$k  & VES\ \cite{Berdnikov:1994zz} \\ 
    & & $\eta'(\pi^+\pi^-\eta(\gamma\gamma),\rho^0\gamma)\eta(\gamma\gamma)\pi^-$ &
    $\eta'\eta\pi^-$ & $1873\pm 33\pm 20$ &  
    $225\pm 35\pm 20$ & $1.9$k  & VES\ \cite{Bityukov:1992zz} \\ 
    & & $\eta(\pi^+\pi^-\pi^0)\eta(\gamma\gamma)\pi^-$ &
    $\eta\eta\pi^-$ & $1814\pm  10\pm 23$ & 
    $205\pm 18\pm 32$ & $0.4$k  & VES\ \cite{Bityukov:1991zz} \\ 
    & & $\pi^+\pi^-\pi^-$ & $(\pi\pi)_S\pi^-$ & $1770\pm  30$ & 
    $310\pm 50$ & $120$k  & SERP\ \cite{Bellini:1982zz} \\ 
    \hline
    $\pi_2(1880)$ & $2^{-+}$ & $3\pi^-2\pi^+$ & $f_2(1270)\pi^-$, $a_1(1260)\rho$, & \\
     &  &  & \quad $a_2(1320)\rho$ & $1854\pm 6^{+6}_{-4}$ 
    & $259\pm 13^{+7}_{-17}$ & $200$k & COMPASS\ \cite{Neubert:2010zz} \\
    & & $\eta(\gamma\gamma)\eta(\pi^+\pi^-\pi^0)\pi^-$ & $a_2(1320)\eta$ &
    $1929\pm  24\pm 18$ & 
    $323\pm 87\pm 43$ & $4$k  & E852\ \cite{Eugenio:2008zza} \\ 
    & & $\pi^-\pi^0\omega(\pi^+\pi^-\pi^0)$ &
    $\omega\rho^-$ & $1876\pm  11\pm 67$ &  
    $146\pm 17\pm 62$ & $145$k  & E852\ \cite{Lu:2004yn} \\ 
    & & $\pi^-\pi^-\pi^+\eta(\gamma\gamma)$ & $f_1(1285)\pi^-$,
    $a_2(1320)\eta$ & $2003\pm  88\pm 148$ & 
    $306\pm 132\pm 121$ & $69$k  & E852\ \cite{Kuhn:2004en} \\ 
    & & $\pi^0\pi^0\eta(\gamma\gamma)\eta(\gamma\gamma)$ & 
    $a_2(1320)\eta$ & $1880\pm  20$ & 
    $255\pm 45$ & $15$k  & CB\ \cite{Anisovich:2001zz} \\ 
    \hline
    $\eta_2(1870)$ & $2^{-+}$ &
    $\eta(\gamma\gamma,\pi^+\pi^-\pi^0)\pi^+\pi^-$ & $a_2(1320)\pi$, 
    $a_0(980)\pi$ & $1835\pm 12$ 
    & $235\pm 23$ & & WA102\ \cite{Barberis:2000zz}\\ 
    & & $2\pi^+2\pi^-$, $\pi^+\pi^-\pi^0\pi^0$ & $a_2(1320)\pi$ & $1844\pm 13$ 
    & $228\pm 23$ & $1500$k & WA102\ \cite{Barberis:1999wn}\\ 
    & & $2\pi^+2\pi^-$ & $a_2(1320)\pi$ & $1840\pm 25$ 
    & $200\pm 40$ & $1200$k & WA102\ \cite{Barberis:1997ve}\\ 
    & & $\eta(\gamma\gamma)3\pi^0$ & $f_2(1270)\eta$ & $1875\pm 20\pm 35$ 
    & $200\pm 25\pm 45$ & $5$k & CB\ \cite{Adomeit:1996nr}\\ 
    & & $\eta(\gamma\gamma)\pi^0\pi^0$ & $a_2(1320)\pi$, $a_0(980)\pi$
    & $1881\pm 32\pm 40$ & $221\pm 92\pm 44$ & $1.2$k & CBall\
    \cite{Karch:1991sm}\\ 
    \hline
    $\rho(1450)$ & $1^{--}$ & $\pi\pi$, $4\pi$, $e^+e^-$ & & $1465\pm
    25$ & $400\pm 60$ & & PDG est.\ \cite{Beringer:2012zz} \\ 
    \hline
    $\rho(1570)$ & $1^{--}$ & $K^+K^-\pi^0$ & $\phi\pi^0$ & 
    $1570\pm 36\pm 62$ & $144\pm 75 \pm 43$ & $54$ & BABAR\
    \cite{Aubert:2007ym}\\ 
    \hline\hline 
  \end{tabular}
}
\end{table*}

Models, as well as lattice QCD, consistently predict a light hybrid
multiplet with spin-exotic quantum numbers $J^{PC}=1^{-+}$. 
Currently, there are 
three 
experimental candidates for a light $1^{-+}$ hybrid 
\cite{Beringer:2012zz} (for recent reviews, see
\cite{Meyer:2010ku,Ketzer:2012vn}): the 
$\pi_1(1400)$ and the $\pi_1(1600)$, observed in diffractive reactions
and $\overline{p}N$ annihilation, and the $\pi_1(2015)$, seen only in
diffraction. 
The $\pi_1(1400)$ has only been observed in the $\pi\eta$ final state, and
is generally considered too light to be a hybrid meson. 
In
addition, a hybrid should not decay into a $P$-wave $\eta\pi$ system from SU(3) symmetry
arguments\ \cite{Chung:2002fz}. 
There are
a number of studies that suggest it is a nonresonant effect,
possibly related to cusp effects due to two-meson thresholds. 
The $\pi_1(1600)$ has been seen decaying into
$\rho\pi$, $\eta'\pi$, 
$f_1(1285)\pi$, and
$b_1(1235)\pi$.  
New data on the $1^{-+}$ wave have recently been
provided by COMPASS, CLEO-c, and CLAS and will be reviewed in the
following.  

\begin{figure*}[tbp]
  \begin{center}
    \includegraphics[width=0.45\textwidth]{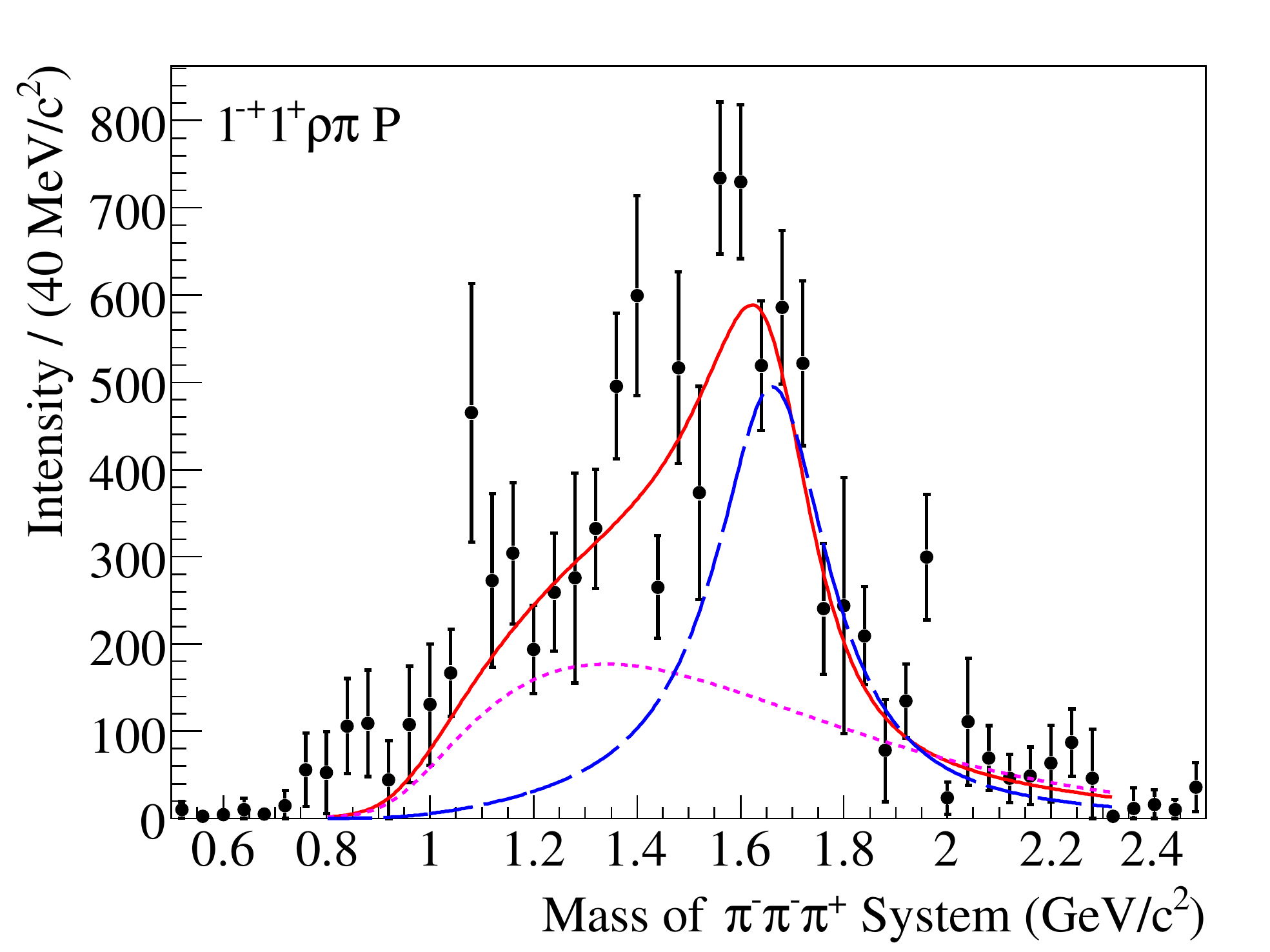}
    \hfill
    \includegraphics[width=0.45\textwidth]{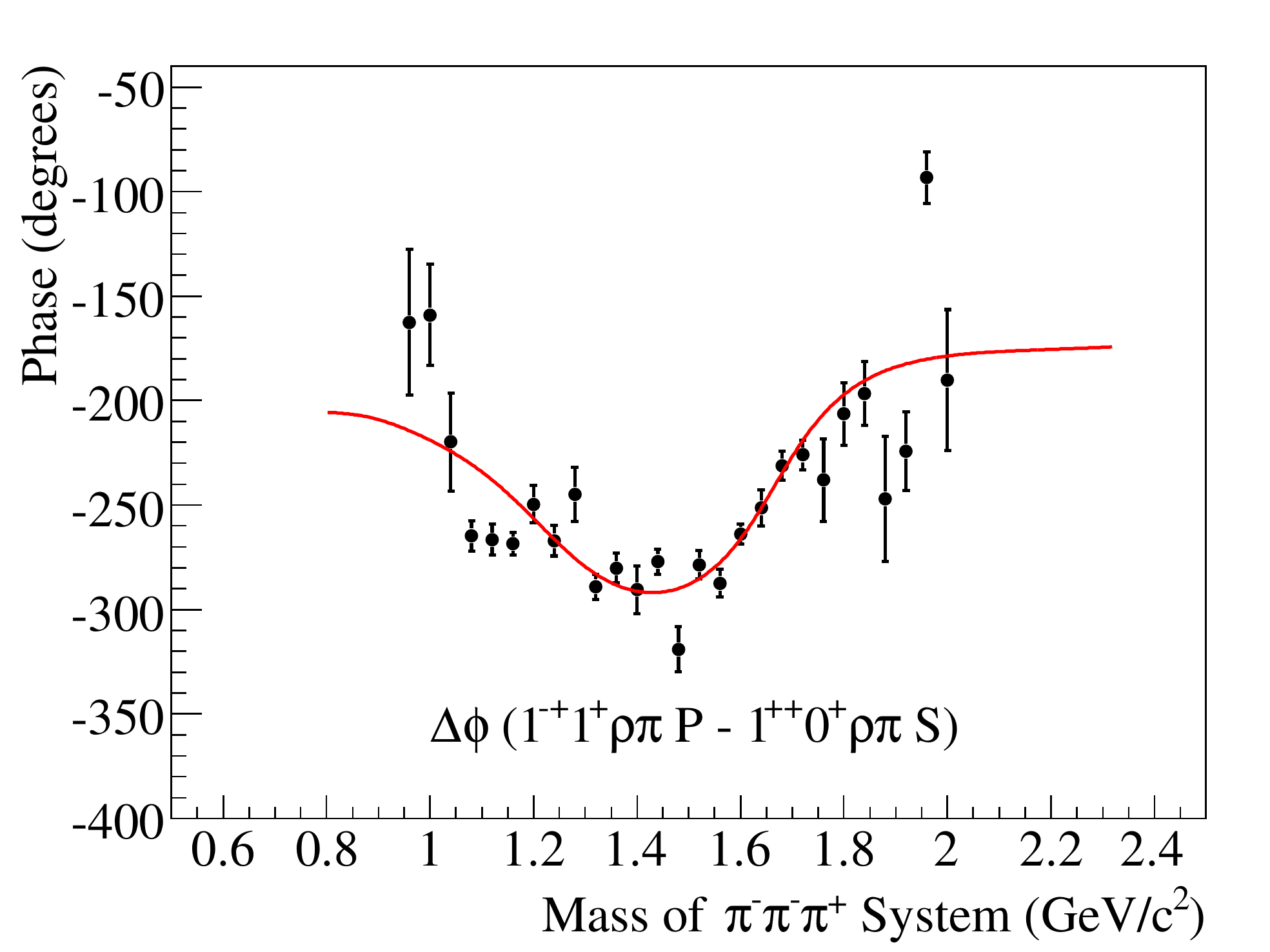}
    \caption{Exotic $1^{-+}1^+\,\rho\pi\,P$ wave observed at the COMPASS
      experiment~\cite{Alekseev:2009xt} for 4-momentum transfer between
      $0.1$ and $1.0\,\GeV^2$ on a Pb
      target and $\pi^-\pi^-\pi^+$ final state. (Left) intensity, 
      (right) phase difference to the
      $1^{++}0^+\,\rho\pi\,S$ wave as a function of the $3\pi$
      invariant mass. The data points
      represent the result of the fit in mass bins, the lines are the
      result of the mass-dependent fit.} 
    \label{fig:3pi.1-+.intensity}
  \end{center}
\end{figure*}

The COMPASS experiment \cite{Abbon:2007pq} at CERN's Super Proton
Synchrotron (SPS) is investigating 
diffractive and Coulomb
production reactions of hadronic beam particles into
final states 
containing charged and neutral particles. 
In a first analysis of the 
$\pi^-\pi^-\pi^+$ final state from
scattering of 
$190\,\GeV$ $\pi^-$
on a Pb target,  
a clear signal in
intensity and phase motion in the
$1^{-+}1^+\,\rho\pi\,P$ partial wave has been
observed \cite{Alekseev:2009xt}, as shown in 
Fig.~\ref{fig:3pi.1-+.intensity}.  
A much bigger data set was
taken by the same experiment with a liquid hydrogen target, surpassing
the existing world data set  
by about one order of magnitude\ \cite{Haas:2011rj,Nerling:2012ei}.  
For both the Pb and the H targets a large
broad nonresonant contribution at lower masses is needed to describe
the mass dependence of the spin-density matrix.  
First studies suggest that the background can be reasonably
well described by Deck-like processes \cite{Deck:1964hm} which proceed
through 1-pion exchange. A more refined analysis in bins of $3\pi$
mass and $t$ is being performed on the larger data set and is expected
to shed more light on the relative contribution of resonant and
nonresonant processes in this and other waves. 

\begin{figure}[tbp]
  \begin{center}
    \includegraphics[width=0.95\columnwidth]{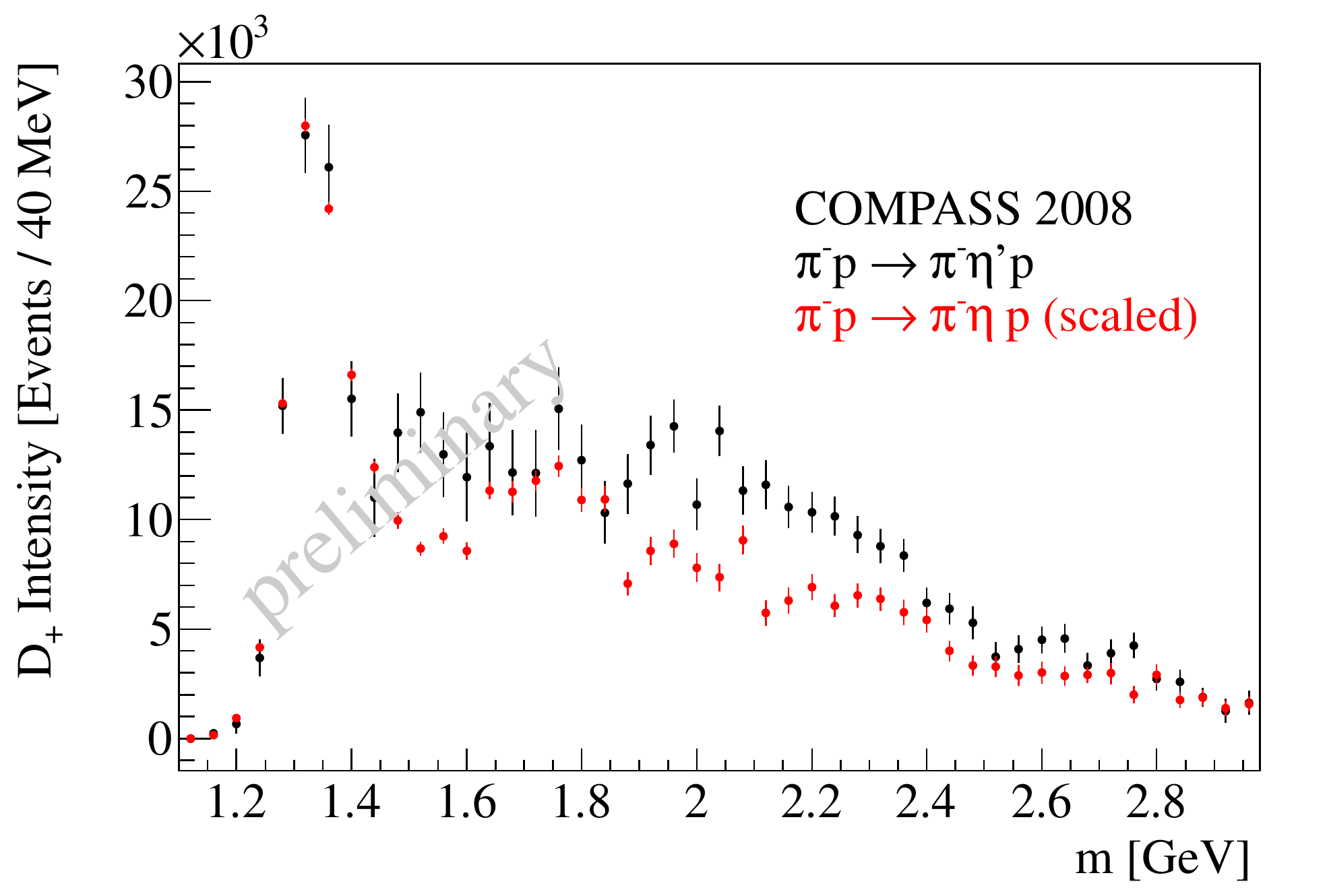}    
    \includegraphics[width=0.95\columnwidth]{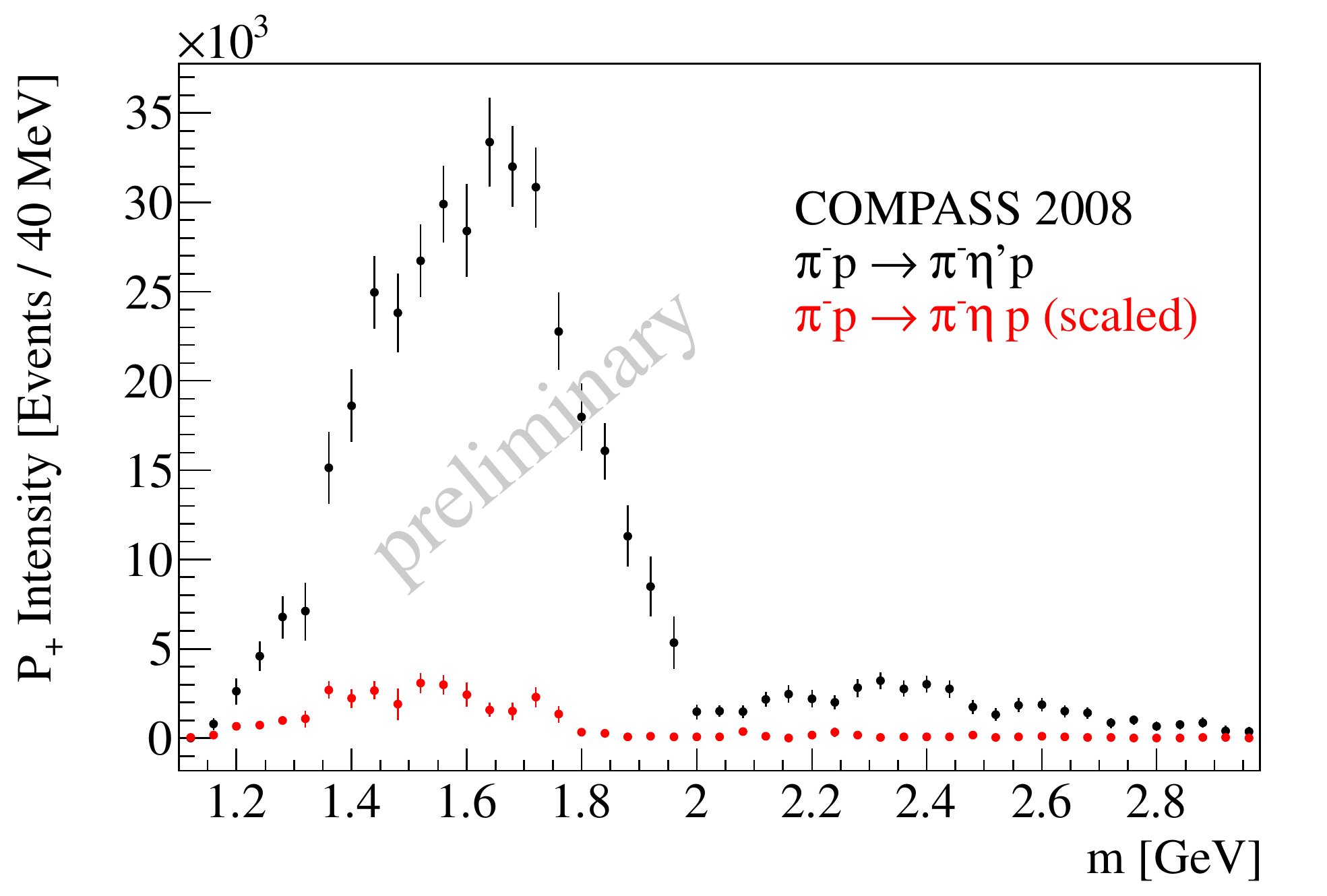}    
    \caption{Comparison of waves for $\eta\pi$ (red data points) and
      $\eta'\pi$ (black data points) final  
      states. (Top) Intensity of the $J^{PC}=2^{++}$ $D$ wave,
      (bottom)
      intensity of the spin-exotic $1^{-+}$ $P$ wave from COMPASS
      \cite{Schluter:2012re}.}  
    \label{fig:H.etapi}
  \end{center}
\end{figure}

COMPASS has also presented data for $\eta\pi$
($\eta\rightarrow\pi^+\pi^-\pi^0$) and $\eta'\pi$
($\eta'\rightarrow\pi^+\pi^-\eta$, $\eta\rightarrow\gamma\gamma$) 
final states from diffractive scattering of
$\pi^-$ off the H target \cite{Schluter:2012re}, which 
exceed the statistics of previous 
experiments by more than a factor of $5$. 
Figure~\ref{fig:H.etapi} shows the intensities in the
(top panel) $2^{++}1^+$ and 
(bottom panel) $1^{-+}1^+$ waves for the $\eta'\pi$ (black data points) and the
$\eta\pi$ final state (red data points), respectively, where the data
points for the latter final state have been scaled by a phase-space
factor. While the intensities in
the $D$ wave are
remarkably similar in intensity and shape in both final states after
normalization, the $P$ 
wave intensities appear to be very different. For $\eta\pi$, the $P$
wave is 
strongly suppressed, while for $\eta'\pi$ it is the dominant wave. The
phase differences between the $2^{++}1^+$ and the $1^{-+}1^+$ waves
agree for
the two final states for masses below $1.4\,\GeV$, showing a
rising behavior due to the resonating 
$D$ wave, while they evolve quite
differently at masses larger than $1.4\,\GeV$, suggesting a different resonant contribution in the two
final states. 
As for the $3\pi$ final
states, resonant, as well as nonresonant, contributions to the exotic
wave have to be included 
in a fit to the spin-density matrix 
in order to describe both intensities and
phase shifts \cite{Schluter:2012re}. 
Regardless of this, the spin-exotic contribution to the
total intensity is found to be much larger for the $\eta'\pi$ final
state than for the $\eta\pi$ final state, as expected for a hybrid
candidate. 

\begin{figure*}[tbp]
  \begin{center}
    \includegraphics[width=0.7\textwidth]{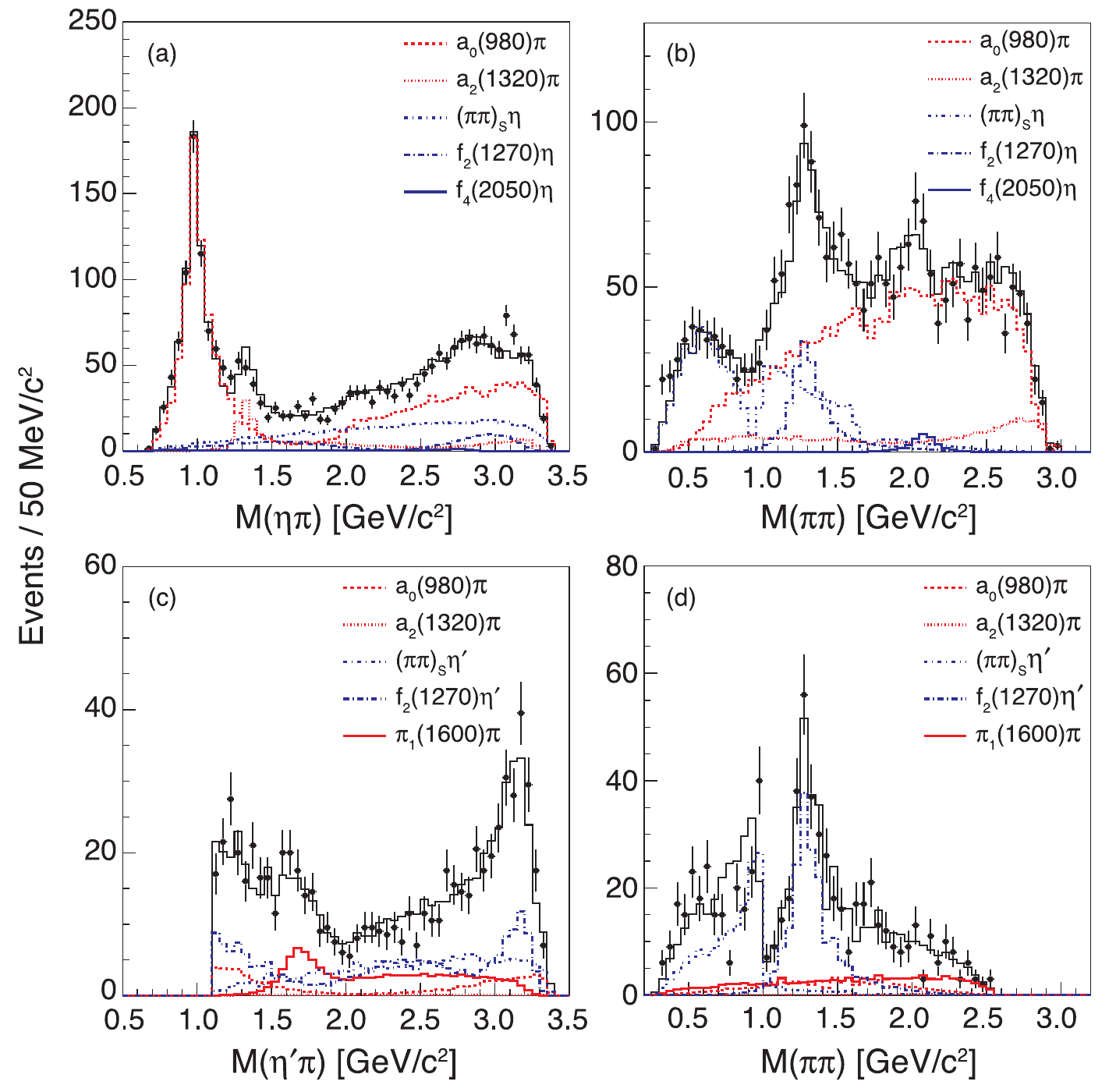}    
    \caption{Invariant mass projections from the analyses of
      (a,b) $\chi_{c1}\rightarrow\eta\pi^+\pi^-$, and
      (c,d) $\chi_{c1}\rightarrow\eta'\pi^+\pi^-$ measured by CLEO-c 
      \cite{Adams:2011sq}. The contributions of the individual fitted
      decay modes are indicated by lines, the data points with full
      points.} 
    \label{fig:chic1.etapipi}
  \end{center}
\end{figure*}

The CLEO-c detector \cite{briere2001cleo} at the Cornell Electron
Storage Ring studied 
charmed mesons at high luminosities until 2008. 
The advantage of using charmonium states as a
source for light-quark states is a clearly defined initial state,
which allows one to limit the available decay modes and to select the
quantum numbers through which the final state is reached. 
Using the full CLEO-c  
data sample of $25.9\times 10^{6}$ $\psi(2S)$ decays, an amplitude analysis 
of the decay chains $\psi(2S)\rightarrow\gamma\chi_{c1}$, with 
$\chi_{c1}\rightarrow\eta\pi^+\pi^-$ or
$\chi_{c1}\rightarrow\eta'\pi^+\pi^-$ has been performed 
\cite{Adams:2011sq}. For these final states, the only allowed $S$-wave
decay of the $\chi_{c1}$ goes through the spin-exotic $1^{-+}$ wave,
which then decays to $\eta(')\pi$. There was no need to
include a spin-exotic wave for the $\eta\pi^+\pi^-$ final state, for which
2498 events had been observed. In
the $\eta'\pi^+\pi^-$ channel with 698 events, a significant
contribution of an exotic $\pi_1$ state decaying to $\eta'\pi$ is
required in order to describe the data, as can be seen from
Fig.~\ref{fig:chic1.etapipi}. This is consistent with the 
COMPASS observation of a strong exotic $1^{-+}$ wave in the same final
state in diffractive production, and is the first evidence of a light-quark
meson with exotic quantum numbers in charmonium decays. 

\begin{figure*}[tbp]
  \begin{center}
    \includegraphics[width=0.45\textwidth]{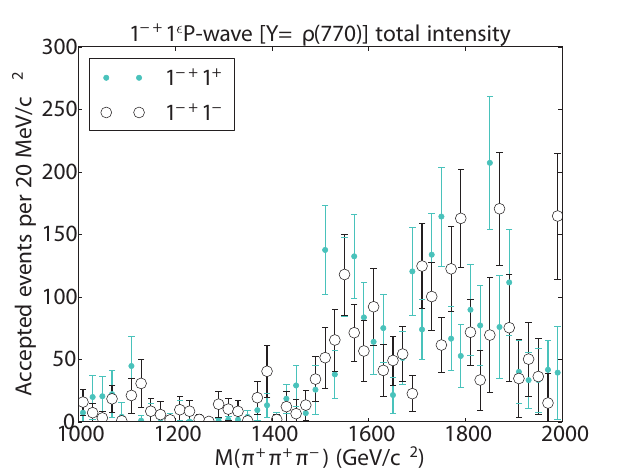}    
    \includegraphics[width=0.45\textwidth]{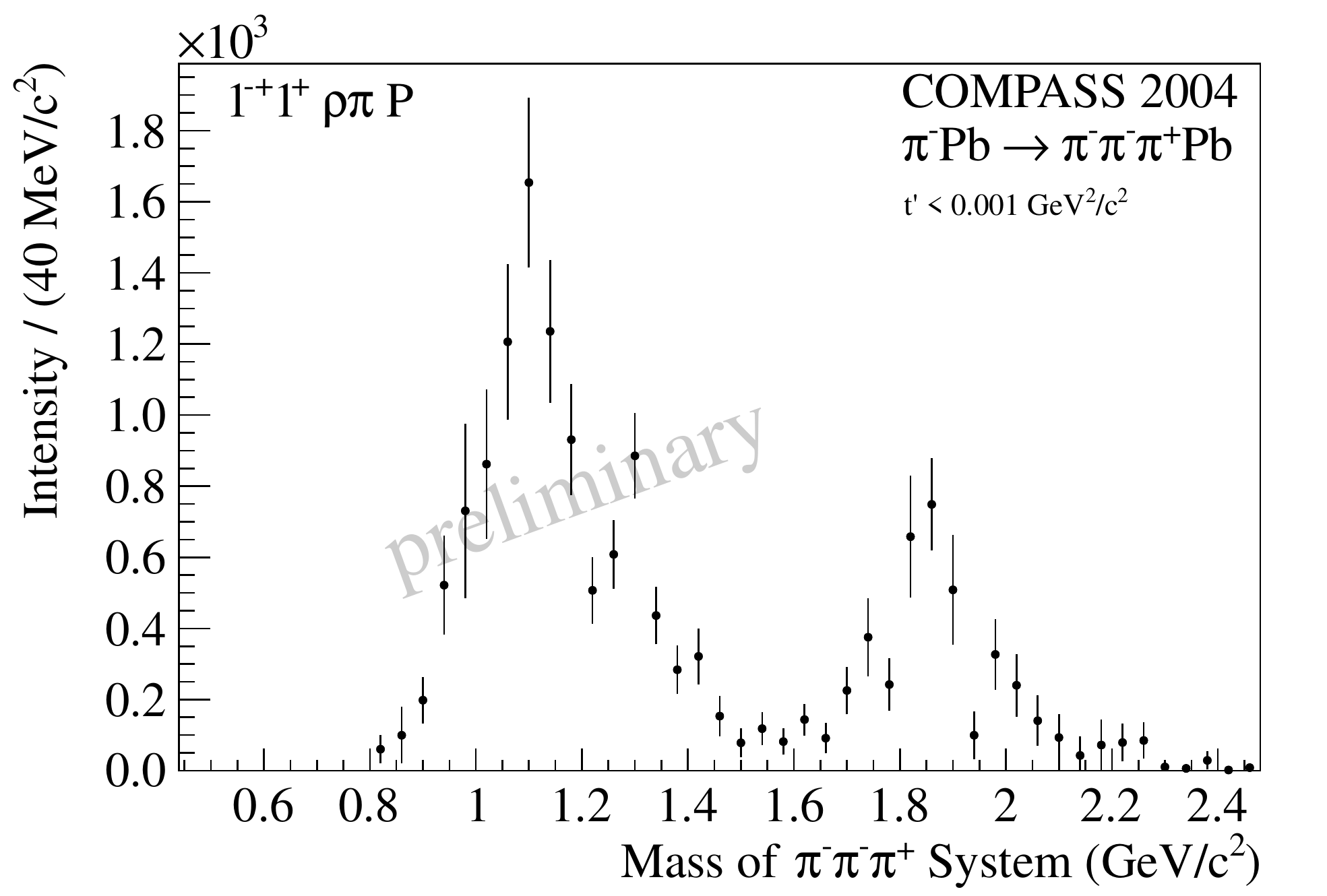}    
    \caption{Intensity of the $1^{-+}1^\pm\,\rho\pi\,P$ waves from
      photoproduction 
      at (left) CLAS \cite{Bookwalter:2011cu} and (right) COMPASS
      \cite{Ketzer:2012vn} as a  
      function of $3\pi$ invariant mass.}
    \label{fig:photo.3pi.1-+}
  \end{center}
\end{figure*}

The CEBAF Large Acceptance Spectrometer (CLAS) \cite{Mecking:2003zu}
at Hall B of JLab
is studying photo- and electro-induced hadronic reactions
by detecting final states containing charged and neutral
particles. Since the coverage for photon detection is limited in CLAS,
undetected neutral particles are inferred mostly via energy-momentum
conservation from the precisely measured
4-momenta of the charged particles. 
CLAS investigated the reaction $\gamma p\rightarrow
\Delta^{++}\eta\pi^-$ in order to search for an exotic $\pi_1$ meson
decaying to the $\eta\pi$ final state \cite{Schott:2012wqa}. 
They found the $J^{PC}=2^{++}$
wave to be dominant, with Breit-Wigner parameters consistent with the
$a_2(1320)$. No structure or clear phase motion was observed for the
$1^{-+}$ wave. 
Two CLAS experimental campaigns in 2001 and 2008 
were dedicated to a search for exotic mesons photoproduced in the
charge exchange reaction $\gamma p\rightarrow\pi^+\pi^+\pi^- (n)$. 
The intensity of the exotic $1^{-+}1^\pm\,\rho\pi\,P$ wave, shown in
Fig.~\ref{fig:photo.3pi.1-+} (left) as a function of the $3\pi$
invariant mass, does not exhibit any evidence for structures
around $1.7\,\GeV$. 
Also its phase difference relative to the
$2^{-+}1^\pm\,f_2\pi\,S$ wave does not suggest any resonant behavior of the
$1^{-+}$ wave in this mass region. 
The conclusion from the CLAS experiments is that there is no evidence
for an exotic $1^{-+}$ wave in photoproduction.  
This is in contradiction to some models
\cite{Close:1994pr,Afanasev:1997fp,Szczepaniak:2001qz}, according to
which photoproduction of 
mesons with 
exotic quantum numbers was expected to occur with a strength comparable to
$a_2(1320)$ production.
   
The COMPASS experiment studied pion-induced reactions on a Pb target
at very low values of 4-momentum transfer,  
which proceed via 
the exchange of quasireal photons from the Coulomb field of the
heavy nucleus. 
A partial wave analysis of
this data set does not 
show any sign of a resonance in the exotic  $1^{-+}1^\pm\,\rho\pi\,P$ wave
at a mass of 
$1.7\,\GeV$ (see Fig.~\ref{fig:photo.3pi.1-+}), consistent with
the CLAS observation.  

While there is some evidence for an isovector member of a light
$1^{-+}$ exotic 
nonet, as detailed in the previous paragraphs, members
of nonexotic hybrid multiplets will be more difficult to
identify. Most of 
the light meson resonances observed until now are in fact compatible
with a $q\overline{q}'$ interpretation.  
Taking the lattice-QCD predictions as guidance, the lowest
isovector hybrids with ordinary quantum numbers should have
$J^{PC}=0^{-+}$, $1^{--}$, and $2^{-+}$ (see
Sec.~\ref{sec:lq.spec.lqcd.mesons}). In the following paragraphs,
recent experimental results for states with these quantum numbers are
summarized. 

There is clear experimental evidence for the $\pi(1800)$ 
\cite{Beringer:2012zz}. The latest measurements of this state come from the
COMPASS experiment which observes it in 
the $3\pi$ and $5\pi$ final states,    
using a $190\,\GeV$ $\pi^-$ beam impinging on a
Pb target. 
Table~\ref{tab:lq.spec.exp.hybrids} includes
the masses and 
widths obtained by fitting Breit-Wigner functions to
the spin density matrix. 
More statistics and advanced coupled-channel analyses
are certainly needed to clarify the decay pattern and
thus the hybrid or $3S$ $q\overline{q}'$ interpretation of this state. 

There is growing experimental evidence for the existence of the $\pi_2(1880)$. 
The latest high-statistics measurements of this state again
come from COMPASS. 
For both Pb and H 
targets a clear
peak is observed in the 
intensity of the $2^{-+}0^+\,f_2\pi\,D$ wave of the $3\pi$ final state
\cite{Haas:2011rj}, which is 
shifted in mass with respect to the $\pi_2(1670)$, and also exhibits a
phase motion relative to the latter in the $f_2\pi\,S$ wave. This
observation, however, was also explained 
differently, including, e.g.,   
the interference of the $f_2\pi\,S$ wave with a Deck-like amplitude,
which shifts the true $\pi_2$ peak to lower masses \cite{Dudek:2006ud}.  
For $5\pi$ final states \cite{Neubert:2010zz}, a total of three
resonances are needed to 
describe the $2^{-+}$ sector, the $\pi_2(1670)$, the $\pi_2(1880)$,
and a high-mass $\pi_2(2200)$. 
The resulting mass and width deduced from this fit for the $\pi_2(1880)$ are also 
included in Table~\ref{tab:lq.spec.exp.hybrids}.
A possible isoscalar partner of the $\pi_2(1880)$, the $\eta_2(1870)$
has also been reported \cite{Beringer:2012zz}, but needs confirmation. 

The PDG lists two $\rho$-like excited states, the
$\rho(1450)$ and the $\rho(1700)$, observed in $e^+e^-$ annihilation,
photoproduction, antiproton annihilation and $\tau$ decays
\cite{Beringer:2012zz}. Their masses are  
consistent with the $2^3S_1$ and $1^3D_1$ $q\overline{q}'$ states,
respectively, but their decay
patterns do not follow the $^3P_0$ rule \cite{Barnes:1996ff}. The
existence of a light vector hybrid state, mixing with the
$q\overline{q}'$ states, was proposed to solve
these discrepancies \cite{Donnachie:1999re}. 
Recently, BaBar has reported the observation of a $1^{--}$ state
decaying to $\phi\pi^0$ \cite{Aubert:2007ym}, the $\rho(1570)$,  
which might be identical
to an earlier observation in Serpukhov \cite{Bityukov:1986yd}. 
Interpretations of this signal include a new state, a
threshold effect, and an OZI-suppressed decay of the $\rho(1700)$. 
A very broad vector state with pole position
$M=(1576^{+49+98}_{-55-91}+\frac{i}{2}
818^{+22+64}_{-23-133})\,\MeV$ has been reported by BES
\cite{Ablikim:2006hp} and is listed as $X(1575)$ by the PDG\
\cite{Beringer:2012zz}. It has been interpreted to be due to
interference 
effects in final state interactions, and in tetraquark scenarios. 
In conclusion, there is no clear evidence for a hybrid state with
vector quantum numbers. A clarification of the nature of the
$\rho$-like states, especially above $1.6\,\GeV$, requires
more data than those obtained in previous ISR measurements at BaBar and
Belle, which will hopefully be reached in current $e^+e^-$ experiments
(CMD-3 and SND at the VEPP-2000 collider, BES~III at BEPCII) as well as
with ISR at the future Belle II detector.

The final test for
the hybrid hypothesis of these candidate states will, of course, be the
identification of the isoscalar and strange members of a multiplet. 
Identification of some reasonable subset of these
states is needed to experimentally confirm what we now expect from
lattice QCD.
New experiments with higher statistical significance and better acceptance, allowing
for more elaborate analysis techniques, are
needed in order to shed new light on these questions. 

\paragraph{Light baryons}
\label{sec:lq.spec.exp.baryons}
Light baryon resonances represent one of the key areas for studying
the strong QCD dynamics.  
Despite large efforts, the fundamental degrees of freedom underlying 
the baryon spectrum are not yet fully understood. 
The determinations of baryon resonance 
parameters, namely quantum numbers, masses and partial widths and
their structure such as electromagnetic (EM) 
helicity amplitudes are currently among the
most active areas in hadron physics, with a convergence of
experimental programs, and analysis and theoretical activities. An
appraisal of the present status of the field can be found in \cite{Crede:2013kia}.  Many important questions and open problems
motivate those concerted efforts. Most important among them is the
problem of missing resonances: 
in quark models based on approximate flavor SU(3) symmetry it is expected
that resonances form multiplets; many excited 
states are predicted which have not been observed (for a review see
\cite{Capstick:2000qj}), with certain
configurations seemingly not realized in nature at all
\cite{Anisovich:2011sv}.  
More recently lattice-QCD calculations
(at relatively large quark masses) \cite{Edwards:2011jj} also predict
a similar proliferation of states. Do (some of) those predicted states
exist, and if so, is it possible to identify them in the experimental
data?  
In addition to $N$ and $\Delta$ baryons made of $u$ and $d$ quarks,
the search for hyperon resonances remains an 
important challenge. Efforts in that direction are ongoing at current
facilities, in particular at JLab (CLAS), where studies of
$S=-1$ excited hyperons,  
e.g., in photoproduction of $\Lambda(1405)$ 
\cite{Moriya:2013eb,Moriya:2013hwg}, have been completed. 
A program to study hyperons with 
$S=-1$, $-2$, and even $-3$ is part of the CLAS12
upgrade. 

Another important task is quantifying and understanding the
structure of resonances, which still is in its early
stages. Experimentally, one important access to structure is provided
by measurements at resonance electro-production, as exemplified by
recent work \cite{Aznauryan:2011qj,Aznauryan:2012ba} where the EM
helicity amplitudes $A_{1/2}(Q^2)$ (electro-couplings) of the Roper
and $N$(1520) resonances have been determined from measurements at 
CLAS, an effort that will continue with the CLAS12
program. An additional tool is provided by meson transition couplings
which can be obtained from single meson EM production. Both
experimental and theoretical studies of resonance structure are key to
further progress.
 
Since 
most of the information on light-quark baryon resonances listed in
\cite{Beringer:2012zz} comes from partial
wave analyses of $\pi N$ scattering, one possible reason why many
predicted resonances  
were not observed may be due to small couplings to $\pi N$. 
Additional information may come from the observation of 
other final states like $\eta N$, $\eta' N$, $KY$, $\omega N$, or 
$2\pi N$.  
A significant number of the current and future experimental efforts
are in electro- and photoproduction experiments, namely JLab
(CLAS and 
CLAS12), Mainz (MAMI-C), Bonn (ELSA) and Osaka (SPring8-LEPS).
Experiments with proton beams are being carried out at CERN (COMPASS),
J-PARC (Japan; also K beam), COSY and GSI (Germany), and at the proton synchrotron at
ITEP (Russia).
Resonance production in charmonium decays (BES~III and
CLEO-c) is also an important source of new excited baryon data. 

As in the light-meson sector, the broad and overlapping nature of
baryon resonances in the mass region below $2.5\,\GeV$ requires the
application of sophisticated 
amplitude or partial-wave analyses in order to disentangle the
properties of the 
contributing states. 
Partial wave analysis is currently a very active area, with several
important groups employing different methods and models.  Among the
groups are SAID (George Washington Univ.), MAID (Mainz), EBAC
(Jefferson Lab), Bonn-Gatchina (BnGa), Bonn-J\"ulich, Valencia,
Gie{\ss}en,  
and others. While at present the analyses are based to the largest  
extent on $\pi N$ and $K N$ data, the large data sets already
accumulated and to be acquired in the near future in photo- and
electroproduction 
are expected to have a big impact in future
analyses.

The extraction of amplitudes from the measured differential cross
sections suffers from ambiguities, as the latter are bilinear products
of amplitudes. These ambiguities can be resolved or at least minimized
by imposing physical
constraints on the amplitudes, 
or by 
measuring a well-chosen set of 
single and double 
polarization observables 
which further constrain the problem.  
For photoproduction experiments, a ``complete experiment'' to
extract 
the full scattering amplitude 
unambiguously 
\cite{Chiang:1996em} 
requires a
combination of 
linearly and circularly polarized photon 
beams, 
longitudinally and transversely polarized targets (protons and
neutrons), or the polarization of the recoil nucleon, measured for each
energy. These amplitudes are then expanded in terms of partial waves,
which are usually truncated at some values of angular momentum.   
Such measurements are one of the
main objectives for the near future, which will give unprecedented
detailed access to established baryon resonances, means to
confirm or reject less established ones and also possibly lead to the
discovery of new resonances.

\begin{figure*}[tbp]
  \begin{center}
    \includegraphics[width=0.51\textwidth]{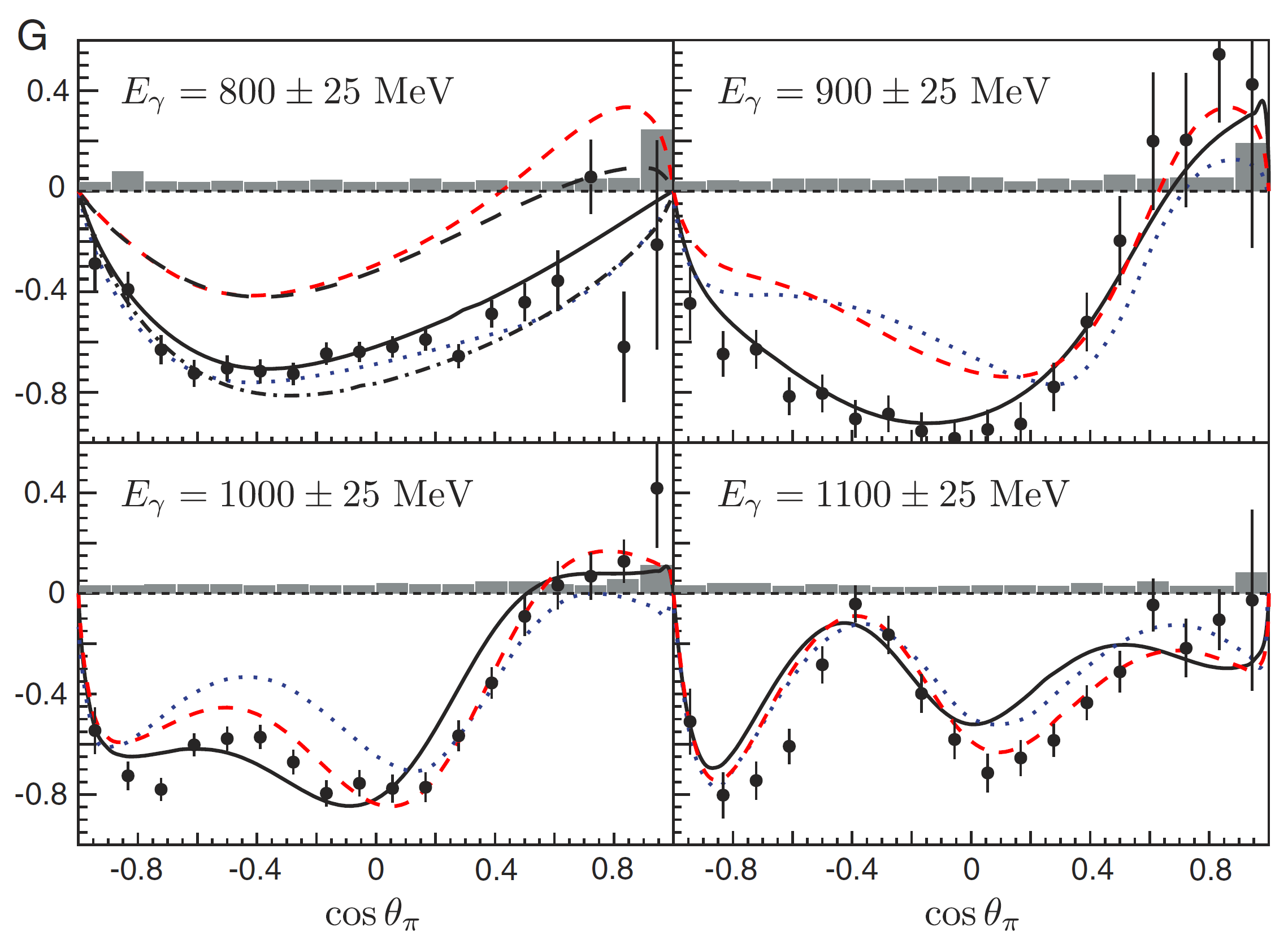}    
    \includegraphics[width=0.39\textwidth]{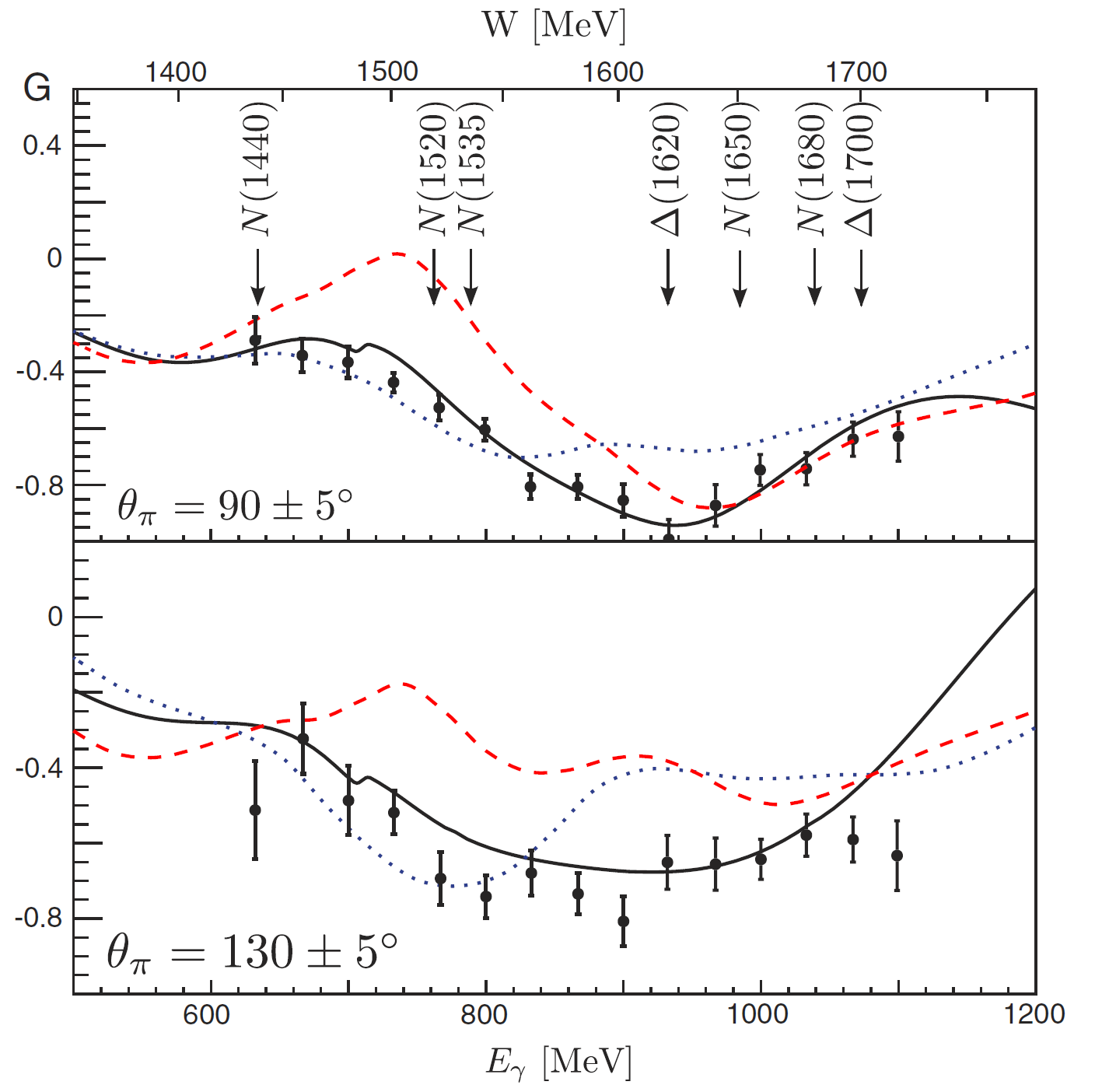}    
    \caption{Double-polarization observable $G$ measured at
      CBELSA~\cite{Thiel:2012yj}, 
      (left) as a function of 
      $\cos{\theta_\pi}$ for four different photon energies, (right)
      as a function of 
      photon energy for two different pion polar angles $\theta_\pi$, 
      compared to predictions by different PWA
      formalisms, (blue) SAID, (red) BnGa, (black) MAID.}
    \label{fig:baryons.cbelsa.g}
  \end{center}
\end{figure*}

Even for the simplest photoproduction reaction, $\gamma p\rightarrow
\pi^0 p$, recently investigated in a double-polarization experiment at
CBELSA/TAPS (Bonn) using linearly polarized photons hitting longitudinally
polarized protons \cite{Thiel:2012yj}, discrepancies between the
latest  
PWA predictions and the data were found at rather low energies in
the region of the four-star
resonances $N(1440)$, $N(1535)$, and $N(1520)$. 
Figure~\ref{fig:baryons.cbelsa.g} (left) shows the observable $G$ as a
function of $\cos{\theta_\pi}$ for four different photon energies,
where $\theta_\pi$ is the polar angle of the outgoing pion, compared
to predictions by several PWA formalisms. $G$ is 
the amplitude of a $\sin{2\phi_\pi}$ modulation of the 
cross section in a double-polarization experiment, where $\phi_\pi$ is
the azimuthal angle of the produced pion.   
Figure~\ref{fig:baryons.cbelsa.g} (right) shows $G$ as a function of the 
photon energy $E_\gamma$ for two selected bins in pion polar angle
$\theta_\pi$. The differences in the theory predictions 
arise from different descriptions of two multipoles, $E_{0^+}$ and
$E_{2^-}$, in the three analyses, which are related to the properties
of the $N(1520)$ $J^P=\frac{3}{2}^-$ and $N(1535)\frac{1}{2}^-$  resonances, respectively. 

\begin{figure*}[tbp]
  \begin{center}
    \includegraphics[width=0.9\textwidth]{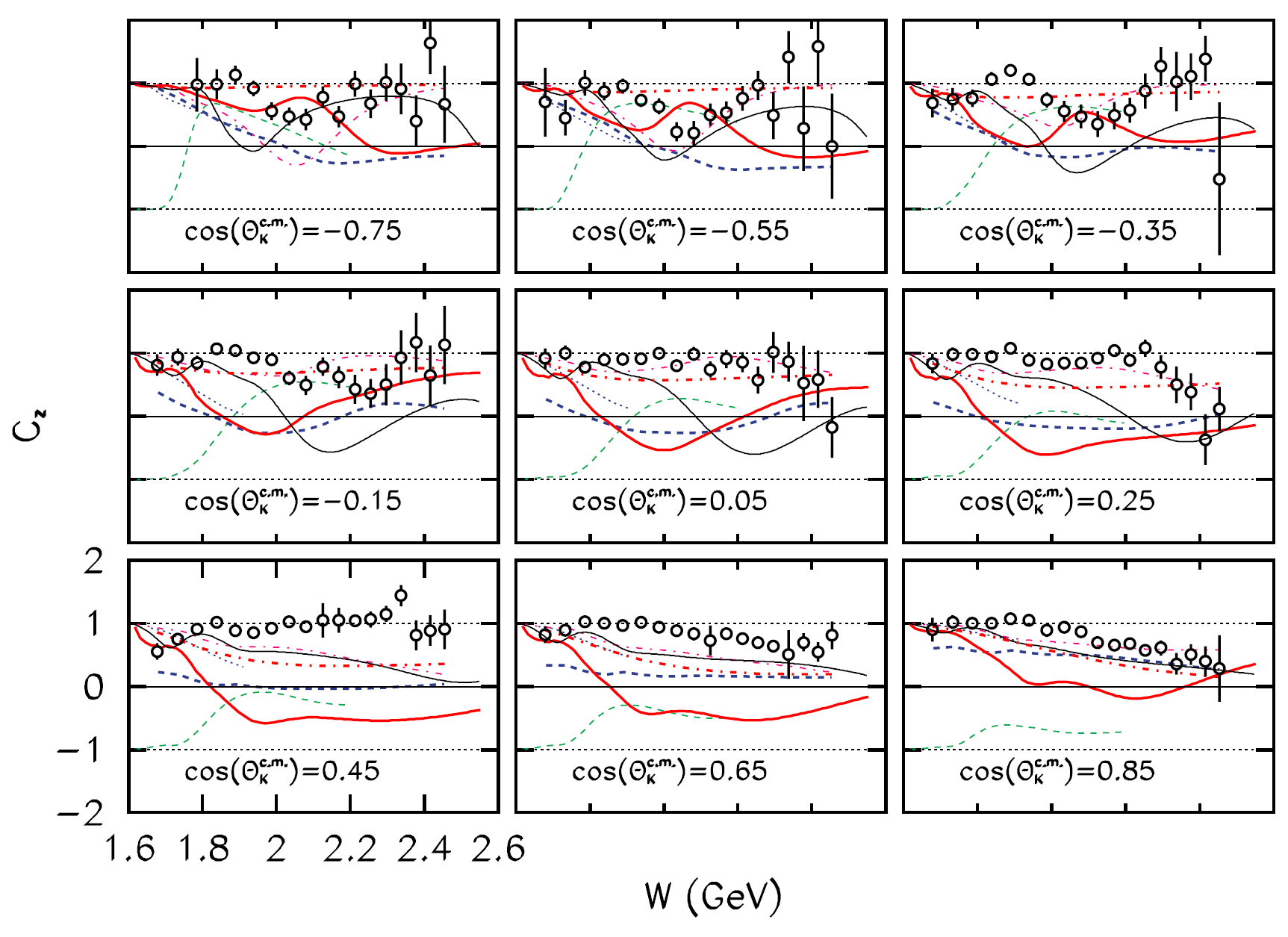}    
    \caption{Beam-recoil observable $C_z$ for circularly polarized
      photons in the reaction $\gamma p\rightarrow K^+\Lambda$ 
      as a function of $\gamma$-$p$ CM energy for 
      different kaon polar angles $\theta_K^{\mathrm{CM}}$
      measured by CLAS~\cite{Bradford:2006ba}. The data points are
      compared to different models (see \cite{Bradford:2006ba} for
      details).}
    \label{fig:baryons.clas.c_z}
  \end{center}
\end{figure*}

Photoproduction of strangeness, where a hyperon is produced in
association with a strange meson, e.g., $\gamma
p\rightarrow K Y$ ($Y=\Lambda,\Sigma$), provides complementary access
to nonstrange baryon resonances that may couple only weakly to
single-pion final 
states.  
In addition, the self-analyzing weak decay of hyperons
offers a convenient way to 
access double polarization observables, as has been recently exploited
at CLAS and GRAAL.  
Using a beam of circularly polarized 
photons, the polarization transfer to the recoiling hyperon along
orthogonal axes in the production plane is characterized by $C_x$ and
$C_z$. The CLAS collaboration \cite{Bradford:2006ba} reported that for
the case of $\Lambda$ 
photoproduction the polarization transfer along the photon momentum
axis 
$C_z\sim +1$ over a wide kinematic range (see
Fig.~\ref{fig:baryons.clas.c_z}), and the corresponding 
transverse polarization transfer $C_x\sim C_z-1$. The magnitude of
the total 
$\Lambda$ polarization vector $\sqrt{P^2+C_x^2+C_z^2}$, including the
induced polarization $P$, is consistent with unity at all measured
energies and angles for a fully polarized photon beam, an observation
which still lacks a proper understanding. Consistent results were
obtained by GRAAL \cite{Lleres:2008em} for the double polarization
observables $O_{x,z}$ using linearly polarized photons.     

Decays to vector mesons provide additional polarization information
by a measurement of the spin-density matrix,  
which constrains the PWA of the reaction. Additionally, the
photoproduction of $\omega$ mesons, like that of $\eta$, serves as an
isospin filter for $N^\ast$ resonances. 
A PWA based on a recent high-statistics CLAS measurement of the
unpolarized cross 
section of the reaction $\gamma
p\rightarrow \omega p$ at CM energies up to $2.4\,\GeV$ \cite{Williams:2009aa} required
contributions from at least two $\frac{5}{2}^+$ resonances, identified
as the $N(1680)\frac{5}{2}^+$ and $N(2000)\frac{5}{2}^+$, and a 
heavier $N(2190)\frac{7}{2}^-$ resonance. The latter had previously
only been observed in $\pi N$ scattering, and was confirmed more
recently by CBELSA/TAPS in $\pi^0$ photoproduction
\cite{Crede:2011dc}. 

As a consequence of the recent high-statistics data sets from
photoproduction, in particular for the reaction $\gamma p\rightarrow
K^+\Lambda$, several baryon resonances, some of which had 
previously been only weakly observed 
in $\pi N$ scattering, have 
been newly proposed in a recent multichannel analysis of the
Bonn-Gatchina PWA group 
\cite{Anisovich:2011fc} 
and are now listed in the 2012 PDG review 
\cite{Beringer:2012zz}. Table~\ref{tab:baryons.states} shows the new 
states in bold letters. 
\begin{table}[tbp]
  \caption{Summary of new light-quark baryon resonances (in bold)
    proposed in \cite{Anisovich:2011fc} and listed in the 
    2012 review of particle physics \cite{Beringer:2012zz}.} 
  \label{tab:baryons.states}
  \centering
  \begin{tabular}{lllll} \hline\hline
    $J^P$ & \multicolumn{4}{c}{Resonance region} \\ \hline 
    $1/2^+$ & $N(1440)$**** & $N(1710)$*** & {\boldmath$N(1880)$}{\bf **} & 
    {\boldmath$N(2100)$}{\bf *} \\
    $1/2^-$ & $N(1535)$**** & $N(1650)$**** & {\boldmath$N(1895)$}{\bf
      **} &  \\
    $3/2^+$ & & $N(1720)$**** & $N(1900)$**{\bf *} &  
    {\boldmath$N(2040)$}{\bf *}\\
    $3/2^-$ & $N(1520)$**** & $N(1700)$*** & {\boldmath$N(1875)$}{\bf
      ***} & 
    {\boldmath$N(2120)$}{\bf **} \\
    $5/2^+$ & & $N(1680)$**** & {\boldmath$N(1860)$}{\bf **} &
    $N(2000)$** \\ 
    $5/2^-$ & & $N(1675)$**** & & {\boldmath$N(2060)$}{\bf **} \\ \hline
    $3/2^-$ & & $\Delta(1700)$*** & $\Delta(1940)$*{\bf *} & \\
    \hline\hline
  \end{tabular}
\end{table}

Some solutions of the partial wave analyses of the world data seem to indicate the existence of parity
doublets at higher masses \cite{Anisovich:2011ye,Anisovich:2011sv}, i.e., two approximately degenerate states
with the same spin but opposite parity (see also Table~\ref{tab:baryons.states}).
This is consistent with predictions based on the effective restoration of chiral symmetry at high baryon
masses \cite{Glozman:2007ek,Glozman:2012fj}.
Similar patterns, however, are also predicted in models which do not make explicit reference to chiral
symmetry \cite{Capstick:1986bm,Loring:2001kx}.
In contrast, the most recent lattice-QCD calculations of excited, higher-spin baryon masses
\cite{Edwards:2011jj} uncover no evidence for the existence of parity doublets.
Thus, the question of whether or not chiral doublets exist in the upper reaches of the baryon spectrum
remains unanswered.

\paragraph{Future directions}
Spectroscopy of light hadrons will remain an active field of research
in the future. 
In order to arrive at a full 
understanding of the excitation spectrum of QCD, a departure from simplistic
Breit-Wigner resonance descriptions towards a full specification of
the pole positions of the amplitude in the complex plane, including
dynamical effects, thresholds, cusps, is required. 
As masses increase, multiparticle channels open up, leading to broad
and overlapping resonances. 
Partial-wave 
analysis models have to be extended to fully respect unitarity,
analyticity, and crossing symmetry, 
in order to extract fundamental,
process-independent quantities. 
The rigorous way of determining the
poles and residues of the amplitude from experiment, which has been
performed at physical values of $s$ and $t$, is by means of dispersion
relations, which provide the correct analytic extension of the
amplitudes to the complex plane. If and how these can be incorporated
into fit models for multiparticle final states remains an open question.
A clear separation of resonant and nonresonant contributions, 
a recurring question for  many of the observed signals in the
light meson sector, e.g., 
requires coupled-channel analyses of different final states, but also
studies in different reactions and kinematics in order to clarify the
underlying production mechanisms.  

New results from running experiments are to be expected in the near
future. The extraction of polarization observables for baryon
resonances in electromagnetically induced reactions will continue at ELSA and MAMI,
which in turn will provide input to multichannel PWA.   
COMPASS, whose data set with hadron beams ($\pi$,
$p$, $K$) is currently being analyzed,  
will continue to take 
data for a couple of 
years with muon and pion beams ~\cite{COMPASS:2010}. 
New experiments are on the horizon or have already started to take
data, which are expected to considerably advance our 
understanding of the excitation spectrum of QCD. Key features
of these experiments 
will be large data sets requiring highest possible luminosities and
sensitivity to production cross sections in the sub-nanobarn
region. This can only be achieved by hermetic detectors with excellent
resolution and particle 
identification capabilities, providing a very high acceptance for charged
and neutral particles. 

Although not their primary goal, $e^+ e^-$ machines, operating at
charmonium or bottomonium center-of-mass energies, have
initiated a renaissance of hadron spectroscopy in the past few years
by discovering many new and yet unexplained states containing charm
and bottom quarks. 
In $e^+ e^-$ collisions states with
photon quantum numbers are directly formed. Other states including
exotics can be accessed  
via hadronic or radiative decays of heavy mesons, or are produced
recoiling against other particles. 
Hadronic decays of heavy-quark states may serve as a source for
light-quark states, with a clearly defined initial state facilitating
the partial wave analysis.  
BES~III at the
BEPCII collider in Beijing has already started to take data in the
$\tau$-charm region with a luminosity of
$10^{33}\,\mathrm{cm}^{-2}\,\mathrm{s}^{-1}$ at a CM energy of
$2\times 1.89\,\GeV$, and will continue to do so over the next
years. 
The Belle II experiment at SuperKEKB \cite{Abe:2010sj}, aiming at a
40-fold luminosity 
increase to values of 
$8\times 10^{35}\,\mathrm{cm}^{-2}\,\mathrm{s}^{-1}$, is expected to 
increase  
the sensitivity for new states in the charm and bottom sector
dramatically, but will also feed the light-quark sector.   
Experiments at the LHC, especially LHCb with its excellent resolution,
are also expected to deliver 
high-statistics data on the meson spectrum. 

GlueX \cite{Eugenio:2012cra} is a new experiment which will study
photoproduction of mesons 
with masses below $3\,\GeV$ at the $12\,\GeV$ upgrade of CEBAF at 
JLab. An important advantage of the experiment will be
the use of polarized photons, which narrows down the possible initial
states and gives direct information on the production process. 
Hadron spectroscopy in
Hall B of JLab will be extended to a new domain of
higher mass resonances and 
the range of higher transferred momentum using electron
beams up to $11\,\GeV$ and the upgraded CLAS12 detector
\cite{Stepanyan:2010kx}. In addition 
to studying GPDs,
CLAS12 will perform
hadron spectroscopy using photoproduction of high-mass baryon and meson
resonances, either by electron scattering via quasireal photons or by
high-energy real photon beams. The detector will consist of a 
forward detector, making use of partly existing equipment with new
superconducting torus coils, and a central
detector with a new $5\,\mathrm{T}$ solenoid magnet and a barrel tracker, 
providing nearly $4\pi$ solid angle coverage for hadronic final
states. 

PANDA, a new experiment at the FAIR antiproton storage ring HESR, is
designed for high-precision studies of the 
hadron spectrum in the charmonium mass range \cite{Lutz:2009ff}. 
In $\overline{p}p$
annihilations, all states with 
nonexotic quantum numbers can be formed directly. Consequently, the
mass resolution 
for these states is only limited by the beam momentum
resolution.
Spin-exotic states can be obtained in production experiments. 
PANDA is expected to run at center-of-mass energies between $2.3$ and
$5.5\,\GeV$ 
with a maximum luminosity of $2\times
10^{32}\,\mathrm{cm}^{-2}\,\mathrm{s}^{-1}$.  
As for the $e^+e^-$ machines, hadronic decays of heavy hadrons may
also serve as a well-defined source for light mesons. 
The study of multistrange hyperons in proton-antiproton annihilation
is also foreseen in the PANDA experiment. 

\subsection{Chiral dynamics}
\label{sec:secB3}

The low-energy regime of light hadron physics plays a key role in 
tests of the nonperturbative phenomena of QCD. In particular, the 
approximate chiral $SU_L(3)\times SU_R(3)$ symmetry and its 
spontaneous breaking sets  the stage for low-energy QCD. The rigorous 
description of low-energy QCD in terms of effective theories, namely 
Chiral Perturbation Theory (ChPT) in its various versions, the availability 
of fundamental experiments, 
and most recently the advent of lattice-QCD calculations with small 
quark masses, are signs of progress that continues unabated, leading to 
very accurate tests of QCD's chiral dynamics. 
\par
ChPT is a low energy effective field theory
of QCD, in which the degrees of freedom are the eight Goldstone
bosons of the hadronic world, corresponding to the $\pi$, $K$, and $\eta$
mesons, and resulting from the spontaneous breakdown of the chiral 
$SU(3)_L\times SU(3)_R$ symmetry  in the limit of massless $u$, $d$, $s$ 
quarks \cite{Gasser:1983yg,Gasser:1984gg}.  ChPT can be readily extended 
to include the low energy physics of ground state light baryons, as well 
as that of heavy mesons and baryons.
\par
We review here the
most salient experimental and theoretical developments that have been 
accomplished recently in the areas of meson-meson and meson-nucleon 
dynamics, along with an outlook for the future.

\subsubsection{$\pi\pi$ and $\pi K$ scattering lengths}
\label{sec:subsecB31}

Measurements of the $S$-wave $\pi\pi$ scattering lengths 
represent one of the most precise tests of the $SU(2)_L^{}\times SU(2)_R^{}$ 
sector of chiral dynamics. 
The NA48/2 experiment at the CERN SPS \cite{Batley:2010zza} 
has analyzed, on the basis of more than one million events, the  $K_{e4}$  
decay $K^{\pm}\rightarrow \pi^+\pi^-e^{\pm}\nu$. The analysis of the 
corresponding form factors, and through them of the $\pi\pi$ final-state 
interactions, has led to the currently most accurate determination of 
the $S$-wave isospin-0 and isospin-2 scattering lengths $a_0^0$ and $a_0^2$,
where $a_\ell^I$ denotes the channel with orbital angular momentum~$\ell$ and isospin~$I$. 
In this analysis, a crucial role  is played by 
isospin breaking effects \cite{Colangelo:2008sm}. An additional 
improvement has been attained by combining the latter results with 
those of the experiment NA48/2 on the nonleptonic decay 
$K^{\pm}\rightarrow \pi^{\pm}\pi^0\pi^0$, with more than 60 million 
events, and the impact of the cusp properties at $\pi^0\pi^0$ 
threshold, due to the mass difference between charged and neutral 
pions. The current results are summarized by:
\begin{eqnarray}
& &\lefteqn{\hspace{-0.5 cm} m_{\pi}a_0^0=0.2210\pm 0.0047_{\mathrm{stat}}
\pm 0.0040_{\mathrm{syst}},} \nonumber\\
& &{\hspace{-0.5 cm} m_{\pi}a_0^2=-0.0429\pm 0.0044_{\mathrm{stat}}
\pm 0.0018_{\mathrm{syst}},} \nonumber\\  
& &{\hspace{-0.5 cm} m_{\pi}(a_0^0-a_0^2)=0.2639\pm 0.0020_{\mathrm{stat}}
\pm 0.0015_{\mathrm{syst}},}
\label{eq:a0Exp}
\end{eqnarray}
where $m_{\pi}$ is the charged pion mass. The agreement with 
the ChPT  result at two-loop order \cite{Colangelo:2001df} is striking:
\begin{eqnarray}
& &m_{\pi}a_0^0=0.220\pm 0.005, \nonumber\\
& &m_{\pi}a_0^2=-0.0444\pm 0.0010, \nonumber\\   
& &m_{\pi}(a_0^0-a_0^2)=0.265\pm 0.004.
\label{eq:a0Th}
\end{eqnarray}
\par
The $\pi\pi$ scattering amplitude is usually analyzed with the aid of
the so-called Roy equations \cite{Roy:1971tc}, which are fixed-$t$
dispersion relations based on analyticity, crossing symmetry and 
unitarity. The corresponding representation has been used in
\cite{Colangelo:2001df} to check the consistency of the chiral 
representation and of the corresponding values of the scattering
lengths and to restrict as much as possible the resulting uncertainties.
Dispersion relations and Roy equations have also been used in    
\cite{GarciaMartin:2011cn}, without the input of ChPT, to analyze   
the $\pi\pi$ scattering amplitude; using high-energy data and the 
$K_{e4}$ decay measurements, results in agreement with those of 
\cite{Colangelo:2001df} have been found.  
\par
Recently, the NA48/2 collaboration also measured the branching ratio of 
$K_{e4}$ decay \cite{Batley:2012rf}, which permits the determination of 
the normalization of the corresponding form factors. This in turn can be
used for additional tests of ChPT predictions.
\par
On the other hand, the measurement of the $K_{\mu 4}$ decay 
\cite{Madigozhin:2013fda} will give access to the $R$ form factor,
which is not detectable in $K_{e4}$ decay, since it contributes 
to the differential decay rate with a multiplicative factor proportional 
to the lepton mass squared. $R$ is one of the three form factors associated 
with the matrix element of the axial vector current; it is mostly
sensitive to the matrix element of the divergence of the axial vector 
current and hence brings information about the chiral symmetry breaking 
parameters. 
\par
Distinct access to the $\pi\pi$ scattering lengths is provided through
the DIRAC experiment at CERN, which measures   the lifetime
of the pionium atom.  The atom, because of the mass difference between the 
charged and neutral pions, decays mainly into two $\pi^0$'s. The decay 
width is proportional, at leading nonrelativistic order, to $(a_0^0-a_0^2)^2$ 
\cite{Deser:1954vq}. Corrections coming from relativistic effects, 
photon radiative corrections, and isospin breaking must be taken
into account to render the connection between the lifetime and the
strong interaction scattering lengths more accurate: these amount to
a 6\% effect \cite{Gasser:2007zt} (and references therein). 
The DIRAC experiment, which started
almost ten years ago, reached last year the objective of measuring
the pionium lifetime with an error smaller  than 10\%. From a sample
of $21000$ pionium atoms a 4\% measurement of the difference
of the $\pi\pi$ scattering lengths has been obtained \cite{Adeva:2011tc}:
\begin{equation}
m_{\pi}|a_0^0-a_0^2|=0.2533^{+0.0080}_{-0.0078}|_
{\mathrm{stat}}{}^{+0.0078}_{-0.0073}|_{\mathrm{syst}},
\end{equation}
a result which is in agreement with those of Eqs. (\ref{eq:a0Th}) and
(\ref{eq:a0Exp}), taking into account the relatively large uncertainty.
\par
In the future, the DIRAC Collaboration also aims to measure the
$2s-2p$ energy splitting, which would allow for the separate 
measurements of the two $S$-wave scattering lengths. 
Another project of the collaboration is the study of the properties
of the $\pi K$ atom, in analogy with the pionium case, thus providing
the $S$-wave $\pi K$ scattering lengths 
\cite{Schweizer:2004qe,Jallouli:2006an}. Preliminary tests of the
experiment at CERN have already begun \cite{Adeva:2009zz}. 
\par
A review of the status of several scattering processes which are 
sensitive to the spontaneous and explicit chiral symmetry breaking 
of QCD can be found in \cite{Meissner:2012ku}. 
\par
The analysis of the $\pi K$ scattering process is a particularly 
representative computation in ChPT in the presence of a strange quark. 
Calculations, similar to those of the $\pi\pi$ scattering amplitude, have 
been carried out. The elastic scattering amplitude has been
evaluated in one- and two-loop order 
\cite{Bernard:1990kw,Bijnens:2004bu}. One finds a slow but  reasonable
convergence of the results at each step of the evaluation. The $S$-wave 
isospin 1/2 and 3/2 scattering lengths are found at the two-loop order:
\begin{equation}
m_{\pi}a_0^{1/2}=+0.220,\ \ \ \ \ \ 
m_{\pi}a_0^{3/2}=-0.047.
\end{equation}
The uncertainties, not quoted explicitly, depend on the variations of 
the parameters that enter in the modeling of the $O(p^6)$ low energy
constants.
\par
The experimental values of the scattering lengths are obtained by
using Roy-Steiner equations \cite{Roy:1971tc,Steiner:1971ms}, which 
generalize the Roy equations to the $\pi K$ system, and high-energy data 
for $\pi K$ scattering \cite{Buettiker:2003pp}, leading to:
\begin{eqnarray}
m_{\pi}a_0^{1/2}&=&+0.224\pm 0.022,\nonumber\\ 
m_{\pi}a_0^{3/2}&=&-0.0448\pm 0.0077.
\end{eqnarray}
The agreement between the ChPT evaluation and the experimental output seems satisfactory, with, however,
larger uncertainties than in the $\pi\pi$ case.

Efforts are also being made to extract the $\pi K$ phase shifts from the nonleptonic decays of $D$ and $B$
mesons \cite{Poluektov:2004mf,Aitala:2005yh,Aubert:2008bd,Link:2009ng}.
The results are not yet sufficiently precise to allow for quantitative comparisons with previous work.

In recent years, the lattice-QCD determination of the $\pi\pi$ and $\pi K$ scattering lengths is providing
increasingly accurate results in full QCD
\cite{Beane:2005rj,Beane:2007xs,Nagata:2008wk,Feng:2009ij,Sasaki:2010zz,Fu:2011wc,Lang:2012sv}.
This work is still maturing, as can be seen in the wide range of both central values and error estimates
(some of which are not yet complete).
A comparative summary of lattice-QCD results can be found in \cite{Lang:2012sv}.
Once all source of uncertainty are controlled, however, one can foresee the time when lattice QCD will
compete with and even supersede the experimental extraction of scattering lengths.

\subsubsection{Lattice QCD calculations: quark masses and effective couplings}
\label{sec:subsecB3lattice}

While the determination of scattering lengths in lattice QCD is still at an
early stage, other quantities, such as quark masses or low-energy constants
(LECs) of mesonic ChPT, have been obtained with high overall precision and
controlled systematic uncertainties. The ``Flavour Averaging Group'' (FLAG)
has set itself the task of collecting and compiling the available lattice
results for phenomenologically relevant quantities
\cite{Colangelo:2010et,Aoki:2013ldr}. Furthermore, FLAG provides a critical
assessment of individual calculations regarding control over systematic
effects. Results which satisfy a set of quality criteria are then combined to
form global estimates. Here we briefly summarize the results and discussions
in \cite{Aoki:2013ldr}, relating to determinations of the light quark masses
and LECs. We focus on QCD with $2+1$ dynamical quarks, which corresponds to a
degenerate doublet of $u,d$ quarks, supplemented by the heavier strange quark.

The FLAG estimates for the strange quark mass, $m_s$, and the average light
quark mass, $\hat{m}\equiv\frac{1}{2}(m_u+m_d)$, were obtained by combining
the results of \cite{Bazavov:2009fk,Durr:2010vn,Arthur:2012opa}, with
\cite{McNeile:2010ji} as an important cross check. In the $\rm \overline{MS}$
scheme at a scale 2\,GeV one finds
\begin{eqnarray}
& & \hat{m} = 3.42 \pm 0.06_{\rm stat}\pm0.07_{\rm sys}\,{\rm MeV},
\label{eq:mhatres} \\
& & {m_s} = 93.8 \pm 1.5_{\rm stat}\pm1.9_{\rm sys}\,{\rm MeV}.
\label{eq:msres}
\end{eqnarray}
The FLAG estimate for the scheme- and scale-independent ratio $m_s/\hat{m}$,
in which some systematic effects cancel, reads
\begin{equation}
   m_s/\hat{m} = 27.46\pm0.15_{\rm stat}\pm0.41_{\rm sys}.
\end{equation}
In order to provide separate estimates for the masses of the up and down
quarks, one has to account for isospin breaking effects, stemming from both
the strong and electromagnetic interactions. Current lattice estimates of
$m_u$ and $m_d$ are mostly based on additional input from phenomenology
\cite{Bazavov:2009fk,Durr:2010vn,deDivitiis:2011eh}. In some cases,
electromagnetic effects (i.e., corrections to Dashen's theorem
\cite{Dashen:1969eg}) have been determined via the inclusion of a quenched
electromagnetic field \cite{Blum:2007cy,Blum:2010ym}. The FLAG results for
$m_u, m_d$ are obtained by combining the global lattice estimate for $\hat{m}$
with the ChPT estimate for the ratio $m_u/m_d$ and phenomenological estimates
of electromagnetic self-energies. In the $\overline{\rm MS}$ scheme at
2~GeV this yields 
\begin{eqnarray}
& & {m_u} = 2.16 \pm 0.09_{\rm stat+sys}\pm0.07_{\rm e.m.}\,{\rm MeV},
\label{eq:mures} \\
& & {m_d} = 4.68 \pm 0.14_{\rm stat+sys}\pm0.07_{\rm e.m.}\,{\rm MeV},
\label{eq:mdres} \\
& & m_u/m_d = 0.46\pm0.02_{\rm stat+sys}\pm0.02_{\rm e.m.}.
\label{eq:mumdrat}
\end{eqnarray}
For a detailed discussion we refer the reader to the FLAG report
\cite{Aoki:2013ldr}. The quark mass
ratio $Q$, defined by 
\begin{equation}
Q^2=(m_s^2-\hat{m}^2)/(m_d^2-m_u^2),
\label{eq:Qdef}
\end{equation}
is a measure of isospin-breaking effects. By combining
Eqs.~(\ref{eq:mhatres}), (\ref{eq:msres}), (\ref{eq:mures}), and
(\ref{eq:mdres}) one arrives at the lattice estimate
\begin{equation}
  Q=22.6\pm0.7_{\rm stat+sys}\pm0.6_{\rm e.m.}.
\end{equation}

In addition to providing accurate values for the light quark masses, lattice
QCD has also made significant progress in determining the effective couplings
of ChPT. This concerns not only the LECs that arise at order~$p^2$ in the
chiral expansion, i.e., the chiral condensate $\Sigma$ and the pion decay
constant in the chiral limit, $F$, but also the LECs that appear at
$O(p^4)$. Moreover, lattice QCD can be used to test the convergence properties
of ChPT, since the bare quark masses are freely tunable parameters, except for
the technical limitation that simulations become less affordable near the
physical pion mass.

The recent FLAG averages for the leading-order LECs for QCD with $2+1$ flavors
read \cite{Aoki:2013ldr}
\begin{equation}
\Sigma=(265\pm17)^3\,{\rm MeV}^3, \quad F_\pi/F=1.0620\pm0.0034,
\end{equation}
where $F_\pi/F$ denotes the ratio of the physical pion decay constant over its
value in the chiral limit. As discussed in detail in Sec.~5 of
\cite{Aoki:2013ldr} there are many different quantities and methods which
allow for the determination of the 
LECs of either SU(2) or SU(3) ChPT. The overall
picture that emerges is quite coherent, as one observes broad consistency
among the results, independent of the details of their extraction. For
specific estimates of the $O(p^4)$ LECs we again refer to the FLAG
report. Despite the fact that the LECs can be determined consistently using a
variety of methods, some collaborations \cite{Allton:2008pn,Aoki:2008sm} have
reported difficulties in fitting their data to SU(3) ChPT for pion masses
above 400\,MeV. Whether this is due to the employed ``partially quenched''
setting, in which 
the sea and valence quark masses are allowed to differ, remains
to be clarified.

\subsubsection{$SU(3)_L\times SU(3)_R$ global fits}
\label{sec:subsecB32}

Due to the relatively large value of the strange quark mass with
respect to the masses of the nonstrange quarks, the matter of the 
convergence and accuracy of $SU(3)_L^{}\times SU(3)_R^{}$ ChPT 
becomes of great importance. In the meson sector, this has been investigated 
over a long period of time by Bijnens and collaborators \cite{Amoros:2001cp} 
at NNLO in the chiral expansion. Taking into account new experimental data, 
mainly on the $K_{e4}$ and $K_{\ell 3}$
form factors, a new global analysis has been  carried out up to $O(p^6)$
effects \cite{Bijnens:2011tb}. The difficulty of the task comes from
the fact that the number of 
LECs at two-loop
order, $C_i^r$, is huge and no unambiguous predictions of them are possible. 
One is left here with educated guesses based on naive dimensional analysis
or model calculations. Several methods of estimate have been 
used and compared with each other. It turns out that the most consistent
estimate of the LECs $C_i^r$ comes from their evaluation with the 
resonance saturation scheme. One is then able to extract from the various
experimental data the values of the $O(p^4)$ LECs $L_i^r$. 
$SU(3)_L\times SU(3)_R$ ChPT seems now to satisfy improved convergence 
properties concerning the expressions of the meson masses and decay 
couplings, a feature which was not evident in the past evaluations. 
Nevertheless, the new global fit still suffers from several drawbacks, 
mainly related to a bad verification of the expected
large-$N_c$ properties of some OZI-rule violating quantities. 
Another drawback is related to the difficulty of reproducing the 
curvature of one of the form factors of the $K_{e4}$ decay. 
Incorporation of latest lattice-QCD results is expected to improve 
the precision of the analysis.
\par 
The question of the convergence of $SU(3)_L\times SU(3)_R$ ChPT has also
led some authors to adopt a different line of approach. It has been 
noticed that, because of the proximity of the strange quark mass value 
to the QCD scale parameter $\Lambda_{QCD}$, vacuum fluctuations of 
strange quark loops may be enhanced in OZI-rule violating scalar 
sectors and hence may cause instabilities invalidating the 
conventional counting rules of ChPT \cite{DescotesGenon:2003cg} 
in that context. 
To cure that difficulty, it has been proposed to treat the quantities that 
may be impacted by such instabilities with resummation techniques. Analyses,
supported by some lattice-QCD calculations \cite{Allton:2008pn,Aoki:2008sm},
seem to provide a consistent picture of three-flavor ChPT 
\cite{Bernard:2010ex}, at least for pion masses below about 400\,MeV.  
\par
The problem of including strangeness in Baryon Chiral Perturbation 
Theory (BChPT) is, on the other hand, 
still a wide open question. It is particularly striking in the quark mass 
expansion of the baryon masses, where very large nonanalytic terms 
proportional to $M_K^3$ indicate a failure of the chiral expansion, and this 
happens in every known version of BChPT. In other observables, such as the 
axial couplings,  magnetic moments, and meson-baryon scattering, certain 
versions of BChPT, namely, those including the decuplet baryons as explicit 
degrees of freedom, lead to important improvements in its convergence. As 
discussed later, these latter versions are motivated by the $1/N_c$ 
expansion, and they provide several such improvements which lend a strong 
support to their use.

\subsubsection{$\eta\rightarrow 3\pi$ and the nonstrange quark masses}
\label{sec:subsecB33}

A process of particular interest in ChPT is $\eta\rightarrow 3\pi$
decay. This process is due to the breaking of isospin symmetry and
therefore should allow for measurements of the nonstrange quark-mass
difference. Nevertheless, attempts to evaluate the decay 
through the Dalitz plot analysis, at one-loop order \cite{Gasser:1984pr}, 
as well as at two-loop order \cite{Bijnens:2007pr}, do not seem 
successful. One of the difficulties is related to the fit to the 
neutral-channel Dalitz plot slope parameter  $\alpha$, whose experimental 
value is negative, while ChPT calculations yield a positive  value. 
To remedy difficulties inherent to higher-order effects, a dispersive 
approach has been suggested, in which $\pi\pi$ rescattering effects are 
taken into account in a more systematic way \cite{Anisovich:1996tx}. 
\par
Including new experimental data accumulated during recent
years (Crystal Barrel \cite{Abele:1998yi}, Crystal Ball
\cite{Tippens:2001fm,Unverzagt:2008ny,Prakhov:2008ff}, 
WASA \cite{Bashkanov:2007aa,Adolph:2008vn}, KLOE \cite{Ambrosino:2010mj}), 
several groups have reanalyzed the
$\eta\rightarrow 3\pi$ problem. Reference \cite{Schneider:2010hs}, 
using a modified nonrelativistic effective field theory approach,
shows that the failure to reproduce $\alpha$ in ChPT can be traced back 
to the neglect of $\pi\pi$ rescattering effects.
References \cite{Kampf:2011wr} and \cite{Colangelo:2011zz} tackle this
problem using the dispersive method, which takes into account higher-order
rescattering effects. The two groups use similar methods of approach
and the same data, but differ in the imposed normalization conditions. The sign of the parameter $\alpha$ is found to be negative
in both works, but it is slightly greater in magnitude than the experimental 
value. The parameter that measures the isospin-breaking effect is $Q$, 
defined in terms of quark masses, see Eq.~(\ref{eq:Qdef}).
The value found for $Q$ in \cite{Kampf:2011wr} is 
$Q=23.1\pm 0.7$, to be compared with the lattice-QCD evaluation
$Q=22.6\pm 0.9$ \cite{Aoki:2013ldr}. (Results of \cite{Colangelo:2011zz} 
are still preliminary and will not be quoted.) 
\par
It is possible to obtain the values of the nonstrange quark masses 
$m_u$ and $m_d$ from the value of $Q$, provided one has additional
information about the the strange quark mass $m_s$ and about $\hat m$.
Using the lattice-QCD results $m_s=(93.8\pm 2.4)$ MeV and $\hat m=(3.42\pm 0.09)$ MeV
\cite{Aoki:2013ldr}, calculated in the $\overline{\rm MS}$ scheme at the 
running scale $\mu=2$ GeV, one finds \cite{Kampf:2011wr} 
\begin{equation}
m_u=(2.23\pm 0.14)\ \mathrm{MeV},\ \ \ \ 
m_d=(4.63\pm 0.14)\ \mathrm{MeV},
\end{equation}
which are in good agreement with the lattice-QCD results \cite{Aoki:2013ldr}
quoted in Eqs.~(\ref{eq:mures}) and~(\ref{eq:mdres}).
\par
Some qualitative differences exist between \cite{Kampf:2011wr}
and \cite{Colangelo:2011zz}. The key point concerns the Adler zeros 
\cite{Adler:1964um} for the $\eta\rightarrow 3\pi$ decay amplitude, whose 
existence is derived here as a consequence of a $SU(2)_L^{}\times SU(2)_R^{}$ 
low-energy theorem \cite{Leutwyler:2013wna}, therefore not using the expansion 
in terms of the strange quark mass. While the two solutions are close
to each other in the physical region, they differ in the unphysical region
where the Adler zeros exist. The solution obtained in 
\cite{Kampf:2011wr} does not seem to display, for small nonzero values of 
the nonstrange quark masses, any nearby Adler zeros. However, the 
authors of \cite{Kampf:2011wr} point out that the quadratic slopes of the 
amplitude are not protected by the above mentioned symmetry and might 
find larger
corrections than expected.
\par  
If the difference between the results of the above two approaches persists 
in the future, it might be an indication that the detailed properties of 
the $\eta\rightarrow 3\pi$ decays are not yet fully under control. 
A continuous effort seems still to be needed to reach a final 
satisfactory answer. For the most recent appraisal of the theoretical 
status, see~\cite{Lanz:2013ku}.
\par
On the experimental side, the $\eta\rightarrow 3\pi$ width is  
determined through the branching ratio from the measurement of the 
$\eta\rightarrow \gamma\gamma$ width. For a long time, 
measurements of the latter using the reaction 
$e^+e^-\rightarrow e^+e^-\eta$ consistently gave a significantly higher 
value~\cite{Beringer:2012zz} than that of the old determination via the 
Primakoff effect~\cite{Browman:1974sj}.
However, a reanalysis of this result based on a new calculation of the 
inelastic background, due to incoherent $\eta$ photoproduction, brought 
the Primakoff measurement in line with those at $e^+e^-$ 
colliders~\cite{Rodrigues:2008zza}.
A new  Primakoff measurement has been proposed by the PRIMEX Collaboration
at JLab, using the 11~GeV tagged photon beam to be delivered 
to Hall D, with the aim of a width measurement with an error of 3\%
or less. Also, the large data base collected by Hall B at JLab 
contains a large sample of $\eta\rightarrow\pi^+\pi^-\pi^0$, of the order 
of $2\times 10^6$ events, which will significantly improve the 
knowledge of its Dalitz distribution. A recent precise measurement  of 
$\Gamma_{\eta \rightarrow \gamma\gamma}$ by KLOE~\cite{Babusci:2012ik} 
shows a high promise of the new measurement planned with KLOE-2.
\par
Isospin-breaking effects are also being investigated with lattice-QCD
calculations, as recently reviewed in \cite{Tantalo:2013maa}.
The effects of the quark-mass difference $m_d-m_u$ on kaon 
masses, as well as on nucleon masses, have recently been studied in 
\cite{deDivitiis:2011eh}, in which earlier references can also be found. 
In addition, QED effects have also been included 
\cite{Aoki:2012st,deDivitiis:2013xla,Basak:2013iw,Borsanyi:2013lga}.
These concern mainly the evaluation of the corrections 
to Dashen's theorem \cite{Dashen:1969eg}, which establishes, 
in the $SU(3)_L\times SU(3)_R$ chiral limit, relationships between the 
electromagnetic mass differences of hadrons belonging to the same 
$SU(3)_V$ multiplet. A summary of results regarding the latter
subject, as well as about the ratio of the nonstrange
quark masses, can be found in \cite{Portelli:2013jla}. 
The issue of the ChPT LECs in the presence of electromagnetism
and isospin breaking through lattice-QCD calculations is also
considered in \cite{Ishikawa:2012ix}. 
\par

\subsubsection{Other tests with electromagnetic probes}
\label{sec:subsecB34p}

One of the classic low-energy processes is 
$\pi^0\rightarrow \gamma\gamma$ decay, which tests 
the Goldstone boson nature of the $\pi^0$ and the chiral 
Adler-Bell-Jackiw anomaly \cite{Bell:1969ts,Adler:1969gk}. This 
subject is considered at the end of 
Sec.~\ref{sec:lqstructpionlifetime}
to which the reader is referred.
\par
One important  test remaining to be improved is that of the process
$\gamma \pi\to 2\pi$,  whose amplitude $F_{3\pi}$ is fixed in the
chiral and low-energy limit by the chiral box anomaly. The two
existing results for   $F_{3\pi}$, namely the Primakoff one
\cite{Antipov:1986tp} from Serpukhov and the recent analysis
\cite{Giller:2005uy} of the $e^-\pi^-\to e^-\pi^-\pi^0$ data 
\cite{Amendolia:1985bs}, disagree with each other and with leading order 
ChPT. Currently, data from COMPASS using the   Primakoff  effect for 
measuring  $\gamma \pi\to 2\pi$ are under analysis (for early results on 
the $2\pi$ invariant mass spectrum see the COMPASS-II proposal 
\cite{Gautheron:2010wva}), and this result is expected to have an important 
impact for   experimentally establishing  this important process. Recently, 
and motivated by the COMPASS measurement,  a new theoretical analysis of 
$\gamma \pi\to 2\pi$ has been carried out \cite{Hoferichter:2012pm}, in 
which  the whole kinematic  domain of this  process is taken into account 
using  ChPT supplemented with dispersion relations. In particular, this 
analysis gives also information that can be used to describe more accurately 
the amplitude $\pi^0\gamma\gamma^*$, important in the analysis of  
light-by-light scattering and the muon's $g-2$. Theoretical works on the  
corrections to the contributions of the anomaly to  $\gamma \pi\to 2\pi$ 
have been addressed in ChPT to higher orders  in 
\cite{Bijnens:1989ff,Bijnens:2012hf,Hannah:2001ee}, and in the vector meson 
dominance model \cite{Holstein:1995qj}.   $F_{3\pi}$  has also been 
calculated from the pion's Bethe-Salpeter amplitude, see 
\cite{Cotanch:2003xv} and references therein. In these and related studies  
 three key constraints are met:   the quark propagator and the pion amplitude 
respect the  axial-vector Ward identity, the full quark-photon vertex  
fulfills an electromagnetic Ward identity, and  a complete set of 
ladder diagrams beyond the impulse approximation are taken into account. 
The three conditions are necessary to reproduce the  low-energy theorem for 
the anomalous form factor. Results at large momentum and nonvanishing pion 
mass agree with the limited   data and exhibit the same resonance behavior 
as the phenomenological vector meson dominance model. The latter property 
signals    that a dynamically calculated quark-photon vertex contains the 
$\rho$ meson pole, and that in the relevant kinematical regions the vector 
mesons are the key physical ingredient in this QCD-based calculation.
 It seems that the time for an accurate test of   $\gamma \pi\to 2\pi$ has 
arrived.
A recent additional test of the box anomaly contributions is the decay 
$\eta\to\pi^+\pi^-\gamma$, which is currently being investigated in 
measurements  at COSY (WASA) \cite{Lersch:2012qca}.

Another test of ChPT is provided by the measurement at COMPASS of 
$\pi^- \gamma\to \pi^-\pi^-\pi^+$ at $\sqrt{s}\leq 5 M_\pi$ with an 
uncertainty in the cross section of 20\%. The results have been published 
in~\cite{Adolph:2011it}, along with a discussion of 
the good agreement with the leading-order ChPT result~\cite{Kaiser:2008ss}.


\subsubsection{Hard pion ChPT}
\label{sec:subsecB35}

ChPT also describes situations in which pions are emitted by heavy mesons
($K$, $D$, $B$, etc.). In such decays, there are regions of phase space 
where the pion is hard and where chiral counting rules can no longer be 
applied. It has been, however, argued that chiral logarithms, calculated in 
regions with soft pions, might still survive in hard pion regimes and
therefore might enlarge, under certain conditions, the domain of validity
of the ChPT analysis \cite{Flynn:2008tg,Bijnens:2010ws}. This line of 
approach has been called ``Hard pion ChPT'' and assumes that the chiral 
logarithms factorize with respect to the energy dependence in the chiral 
limit. Such factorization properties have been carefully analyzed in
\cite{Colangelo:2012ew} using dispersion relations and explicitly shown to
be violated for pion form factors by the inelastic contributions, 
starting at three loops. The study in
\cite{Colangelo:2012ew} is presently being extended to heavy-light form
factors. This will help clarify to what degree of approximation and in
what regions of phase space Hard pion ChPT might have practical
applications in the analysis of heavy meson decays. 
\par
   
\subsubsection{Baryon chiral dynamics}
\label{sec:subsecB36}

Baryon chiral dynamics still represents a challenge, but very exciting 
progress is being made on three fronts:   experimental, 
theoretical, and lattice QCD. Here we highlight some of them.
\par
Combining data from pionic hydrogen and deuterium 
\cite{Gotta:2008zza,Strauch:2010vu}, the extraction of the $\pi N$ 
scattering lengths have been evaluated with the so-called Deser formula
\cite{Deser:1954vq,Gasser:2007zt}, leading to \cite{Baru:2012ir}:
$m_{\pi}a_0^-=(86.1\pm 0.1)\times 10^{-3}$ and
$m_{\pi}a_0^+=(7.6\pm 3.1)\times 10^{-3}$, to be compared with the 
leading-order predictions \cite{Weinberg:1966kf}:   
$m_{\pi}a_0^-\simeq 80\times 10^{-3}$ and $m_{\pi}a_0^+=0$. 
($a_0^+$ and $a_0^-$ are the $S$-wave isospin even and isospin odd scattering
lengths, respectively. They are related to the isospin $1/2$ and $3/2$
scattering lengths through the formulas $a_0^+=(a_0^{1/2}+2a_0^{3/2})/3$
and $a_0^-=(a_0^{1/2}-a_0^{3/2})/3$.) In the same spirit, 
kaon-nucleon scattering lengths have been extracted
from the combined data coming from kaonic hydrogen X-ray emissions 
\cite{Meissner:2004jr} and kaon deuterium scattering \cite{doring:2011xc}.
The latter analysis uses data coming from the recent SIDDHARTA
experiment at the DA$\Phi$NE electron-positron collider 
\cite{Bazzi:2012eq}.
\par
In spite of existing huge data sets on pion-nucleon scattering, the 
low-energy scattering amplitudes are still not known with great precision. 
And yet this is the region in which low-energy theorems and ChPT 
predictions exist. To remedy this deficiency, a systematic construction of
$\pi N$ scattering amplitudes has been undertaken in~\cite{Ditsche:2012fv} 
using the Roy-Steiner equations, based on a
partial wave decomposition, crossing symmetry, analyticity, and 
dispersion relations. This approach parallels the one undertaken for
$\pi K$ scattering \cite{Buettiker:2003pp}, although in the present 
case the spin degrees of freedom of the nucleon considerably increase 
the number of Lorentz invariant amplitudes. It is hoped that  a 
self-consistent iterative procedure between solutions obtained in 
different channels will yield a precise description of the low-energy 
$\pi N$ scattering amplitude.
\par
Another long-standing problem in $\pi N$ physics is the 
evaluation of the pion-nucleon sigma term. In general, sigma terms 
are defined as forward matrix elements of quark mass operators  
between single hadronic states and are denoted, with appropriate
indices, by $\sigma$. More generally, the sigma terms are related
to the scalar form factors of the hadrons, denoted by $\sigma(t)$,
where $t$ is the momentum transfer squared, with $\sigma(0)$
corresponding to the conventional sigma term. The interest in the 
sigma terms resides in their property of being related to the mass
spectrum of the hadrons and to the scattering amplitudes
through Ward identities. Concerning the pion-nucleon sigma term, 
in spite of an existing low-energy theorem \cite{Cheng:1970mx}, its full 
evaluation necessitates an extrapolation of the low-energy $\pi N$ 
scattering amplitude to an unphysical region \cite{Gasser:1990ce}. The 
result depends crucially on the way the data are analyzed. Several 
contradictory results have been obtained in the past, and this has 
given rise to much debate. Recent evaluations of the sigma term 
continue to raise the same questions. In~\cite{Alarcon:2011zs}, 
a relatively large value of the sigma term is found, $\sigma=(59\pm 7)$ 
MeV, while in~\cite{Stahov:2012ca}, the relatively low value of~\cite{Gasser:1990ce} is confirmed, $\sigma=(43.1\pm 12.0)$ MeV; 
the two evaluations remain, however, marginally compatible. 
One application of the equations of~\cite{Ditsche:2012fv} concerns 
a dispersive analysis of the scalar form factor of the nucleon 
\cite{Hoferichter:2012wf}. This has allowed the evaluation of the correction 
$\Delta_{\sigma}=\sigma(2m_{\pi}^2)-\sigma(0)$ of the scalar form factor
of the nucleon, needed for the extraction of the $\pi N$ sigma term from 
$\pi N$ scattering. Using updated phase shift inputs, the value 
$\Delta_{\sigma}=(15.2\pm 0.4)$ MeV has been found, confirming the earlier 
estimate of~\cite{Gasser:1990ap}.
\par 
A complementary access to the sigma term is 
becoming possible thanks to lattice-QCD calculations of the nucleon mass 
at varying values of the quark masses \cite{Alvarez-Ruso:2013fza}. The 
current limitations reside in the relatively large quark masses used, 
and also in the still significant error bars from calculations which employ 
the lowest possible quark masses. It is however feasible that in the near 
future results competitive in accuracy to the ones obtained from $\pi N$ 
analyses will be available from lattice QCD. 
\par
One issue that has been open for a long time is the precise  value of the 
$g_{\pi NN}$ coupling. A new extraction by an analysis in~\cite{Baru:2011bw} based on the Gell-Mann-Oakes-Renner (GMO) sum rule 
gives $g_{\pi NN}^2/(4\pi)=13.69(12)(15)$. This value agrees 
with those of analyses favoring smaller values of the coupling. It, in 
particular, supports the argument based on the 
naturalness of the Goldberger-Treiman discrepancy  when  extended to  
$SU(3)$ \cite{Goity:1999by}.
\par   
A theoretical development in BChPT which has been taking place over many 
years  is the development of effective theories with explicit spin 3/2 baryons 
degrees of freedom.  It has been known for a long time \cite{Jenkins:1991ne} 
that the inclusion of the spin 3/2 decuplet improved the convergence of the 
chiral expansion for certain key quantities. The theoretical foundation for  
it is found in the $1/N_c$ expansion \cite{Dashen:1993ac,Dashen:1993as}, 
the key player 
being the (contracted) spin-flavor symmetry of baryons in large $N_c$.  
This has led to formulating BChPT in conjunction with the $1/N_c$ expansion 
\cite{Jenkins:1995gc,FloresMendieta:1998ii,FloresMendieta:2006ei,
CalleCordon:2012xz}, a framework which has yet to be applied extensively 
but which has already shown its advantages. Evidence of this is provided 
by several works on baryon semileptonic decays \cite{FloresMendieta:1998ii,
FloresMendieta:2006ei,FloresMendieta:2012dn}, and 
in particular in the analysis of the nucleon's axial coupling 
\cite{CalleCordon:2012xz}, where the cancellations between the contributions 
from the nucleon and $\Delta$ to one-loop chiral corrections are crucial for 
describing the near independence of $g_A$ with respect to the quark masses 
as obtained from lattice-QCD calculations 
\cite{Edwards:2005ym,Bratt:2010jn,Alexandrou:2010hf,
Yamazaki:2008py,Yamazaki:2009zq,Lin:2008uz}.  We expect that many further 
applications of the BChPT$\otimes 1/N_c$ framework will take place in the 
near future, and it will be interesting to see what its impact becomes in 
some of the most difficult problems such as   baryon polarizabilities, 
spin-polarizabilities, $\pi N$ scattering, etc. Further afield, and 
addressed elsewhere in this review, are the applications to few-nucleon 
effective theories, of which the effective theory in the one-nucleon sector 
is a part. 
An interesting recent development in baryon lattice QCD is the calculation 
of masses at varying $N_c$ \cite{DeGrand:2012hd}. 
Although at this point the calculations are limited to quenched QCD, they 
represent a new tool for understanding the validity of $N_c$ counting 
arguments in the real world, which will be further improved by calculations 
in full QCD and at lower quark masses. 
For an analysis of the results in \cite{DeGrand:2012hd} in the light of  
BChPT$\otimes 1/N_c$ framework, see~\cite{CC-DeG-G}.
\par
A new direction worth mentioning is the application of BChPT to the study 
of the nucleon  partonic structure at large transverse distances 
\cite{Strikman:2009bd}, which offers an example of  the possible 
applications of effective theories to the soft structures accompanying hard 
processes in QCD.
\par 

\subsubsection{Other topics}
\label{sec:subsecB37}

Many other subjects are in the domains of interest and expertise
of ChPT and are being studied actively. We merely quote some of them:
pion and eta photoproduction off protons 
\cite{Mai:2012wy,Ruic:2011wf,FernandezRamirez:2009su,FernandezRamirez:2009jb,Hornidge:2012ca,FernandezRamirez:2012nw}, 
pion polarizabilities \cite{Gasser:2006qa,Kaiser:2011zz} 
(see also Sec.~\ref{sec:lq.struct.pion.polarizability}) and two-pion 
production in $\gamma\gamma$ collisions \cite{Hoferichter:2011wk}, the decay 
$\eta'\rightarrow \eta\pi\pi$ \cite{Escribano:2010wt}, the 
electromagnetic rare decays $\eta'\rightarrow \pi^0\gamma\gamma$ and 
$\eta'\rightarrow \eta\gamma\gamma$ \cite{Escribano:2012dk},
$K$ meson rare decays \cite{D'Ambrosio:1996sw,Goudzovski:2012gh}, 
hadronic light-by-light scattering \cite{Engel:2012xb}, etc. The 
incorporation of the $\eta'$ meson into the ChPT calculations is usually
done in association with the $1/N_c$ expansion \cite{Kaiser:2000gs}, since 
for finite $N_c$, the $\eta'$ is not a Goldstone boson in the chiral limit. 
\par
The above processes enlarge the field  of investigation of ChPT, by
allowing for the determination of new LECs and tests of nontrivial predictions.
Some of the amplitudes of these processes do not receive contributions 
at tree level and have as leading terms $O(p^4)$ or $O(p^6)$ loop 
contributions. Therefore they offer more sensitive tests of higher-order
terms of the chiral expansion.
\par
An important  area of applications of ChPT is to weak decays, which 
unfortunately cannot be covered in this succinct review. Of particular 
current interest are the inputs to nonleptonic kaon decays, where 
lattice-QCD calculations have  been steadily progressing and are 
making headway in 
understanding old, difficult problems such as the $|\Delta I|=1/2$ rule 
\cite{Boyle:2012ys}. We refer to \cite{Colangelo:2010et} for a review of 
the current status of kaon nonleptonic decays vis-\`a-vis lattice QCD.    
Many topics in baryon physics have also not been touched upon, among them 
the study of low-energy aspects of the EM properties of baryons such as the 
study of polarizabilities, in particular, the spin polarizabilities and 
generalized polarizabilities as studied with electron scattering 
\cite{Chen:2010qc,Fonvieille:2012cd}.
\par

\subsubsection{Outlook and remarks}
\label{sec:subsecB38}

As a low-energy effective field theory of QCD, ChPT offers a solid and 
reliable framework for a systematic evaluation of various dynamical
contributions, where the unknown parts are encoded within a certain 
number of low energy constants (LECs). Two-flavor ChPT is well 
established, lying on a firm ground. The main challenge now concerns the
convergence properties of three-flavor ChPT, where a definite progress 
in our understanding of the role of the strange quark is still missing.
Efforts are being continued in this domain, and it 
is hoped that new results coming from lattice-QCD 
calculations will help clarify the situation. Another 
specific challenge concerns the understanding of isospin breaking, 
including the evaluation of electromagnetic effects, in the decay 
$\eta\rightarrow 3\pi$. New domains of interest, such as the probe of 
hard-pion regions in heavy-particle decays, $\eta'$ physics, and rare kaon 
decays, are being explored. This, together with data provided 
by high-precision experimental projects, gives confidence in the progress 
that should be accomplished in the near future.
\par
In baryons, the present progress in lattice QCD is leading to an important 
understanding of the behavior of the chiral expansion thanks to the 
possibility of studying the quark-mass dependence of key observables.  
Although issues remain, such as the problem in confronting with the 
empirical value of $g_A$, it is clear that lattice 
QCD will have a fundamental impact in our understanding of the chiral 
expansion in baryons. Further, the union of BChPT and the $1/N_c$ expansion 
represents a very promising framework  for further advances in the 
low-energy effective theory for baryons. 
\par

\subsection{Low-energy precision observables and tests of the Standard Model}
\label{sec:secB4}



\subsubsection{The muon's anomalous magnetic moment}

The muon's anomalous magnetic moment, $a_\mu=(g-2)_\mu/2$, is one of the most precisely measured quantities
in particle physics, reaching a precision of 0.54\,ppm.
The most recent experimental measurement, BNL 821~\cite{Bennett:2006fi}, is
\begin{equation}
    10^{10}a_\mu = 11\,659\,208.9(6.3).
    \label{eq:secB4:BNL821}
\end{equation}
This result should be compared with the theoretical calculation within the Standard Model
(a topic worthy of a review in itself):
\begin{eqnarray}
    10^{10}a_\mu &=&  11\,658\,471.81 \hphantom{(0.10)} \quad\text{4-loop QED} \nonumber \\
      &+&  \hphantom{11\,658\,4}19.48 \hphantom{(0.20)} \quad\text{1-loop EW} \nonumber \\
      &-&  \hphantom{11\,658\,40}3.91 \pm 0.10             \quad\text{2-loop EW} \nonumber \\
      &+&  \hphantom{11\,658}\,692.3~\,\pm 4.2\hphantom{5} \quad\text{LO~HVP} \nonumber \\
      &-&  \hphantom{11\,658\,49}9.79 \pm 0.09              \quad\text{NLO~HVP} \nonumber \\
      &+&  \hphantom{11\,658\,4}10.5~\,\pm 2.6\hphantom{5} \quad\text{HLbL \cite{Prades:2009tw}}\nonumber\\[-4pt]
      & &                                                 \rule{16em}{1pt} \nonumber \\
      &=&            11\,659\,180.4 \pm 4.2 \pm 2.6             \quad\text{Total}
    \label{eq:secB4:SMg-2}
\end{eqnarray}
using the compilation in~\cite{Hoecker:2010qn}.
Here the leading-order (LO) hadronic vacuum polarization is taken from measurements of
$R(e^+e^-\to\text{hadrons})$, and the electroweak (EW) corrections have been adjusted slightly to account for
the (since measured) Higgs mass $M_H=125~\text{GeV}$.
While QED and electroweak contributions account for more than 99.9999\% of the value $a_\mu$, the dominant
errors in Eq.~(\ref{eq:secB4:SMg-2}) 
stem from the hadronic vacuum polarization (HVP) and hadronic light-by-light
(HLbL) scattering --- they stem from QCD.

The difference between the values in Eqs.~(\ref{eq:secB4:BNL821}) and~(\ref{eq:secB4:SMg-2}) is
$28.5\pm 6.3_\text{expt}\pm 4.9_\text{SM}$, which is both large --- 
larger than the EW contributions
$19.5-3.9=15.6$ --- and significant --- around $3.5\sigma$.
This deviation has persisted for many years and, if corroborated, would provide a strong hint for physics
beyond the Standard Model.
This situation has motivated two new experiments with a target precision of 0.14\,ppm, FNAL~989 
\cite{Carey:2009zzb}, and J-PARC~P34 
\cite{Aoki:2009p34}.
The new experiments have, in turn, triggered novel theoretical efforts with the objective to obtain a
substantial improvement of the theoretical values of the QCD corrections to the muon anomaly.
In this section, we address HVP and HLbL in turn, discussing approaches 
(such as $R(e^+e^-)$) involving other 
experiments, lattice QCD, and for HLbL also models of QCD.

The principal phenomenological approach to computing the HVP contribution $a_\mu^{\rm had;VP}$ is based
on the optical theorem and proceeds by evaluating a dispersion integral, using the experimentally measured
cross section for $e^{+}e^{-}\to\rm hadrons$.
Evaluations of various authors are using the same data sets and basically agree, slightly differing in the computational methods and final uncertainties depending on the conservatism of the authors~\cite{Davier:2010nc,Jegerlehner:2011ti,Teubner:2012qb}.
Note that recent measurements of $\sigma(e^+e^- \to \pi^+\pi^-)$, the process
dominating the LO HVP contribution, performed by initial-state radiation at BaBar~\cite{Lees:2012cj} and KLOE~\cite{Ambrosino:2010bv,Babusci:2012rp} do not show complete agreement
between each other and with the previous measurements based on direct scans~\cite{Achasov:2006vp,Akhmetshin:2006bx}. Determination of
the cross section in all these experiments, in particular those using
initial state radiation, crucially depends on the rather complicated
radiative corrections.

An alternative phenomenological approach is to use $\tau$ decay to hadrons to estimate the HVP.
This approach is very sensitive to the way isospin-breaking corrections
are evaluated. While a model-dependent method trying to take into account
various effects due to $m_d \neq m_u$ still shows notable deviation from the
$e^+e^-$ based estimate~\cite{Davier:2009ag}, the authors of \cite{Jegerlehner:2011ti} claim that after correcting the $\tau$ data for the 
missing $\rho-\omega$ mixing contribution, 
in addition to the other known isospin-symmetry-violating 
corrections, $e^+e^-$ and $\tau$-based
calculations give fully compatible results.

\begin{figure}[b]
\includegraphics*[width=\figwid]{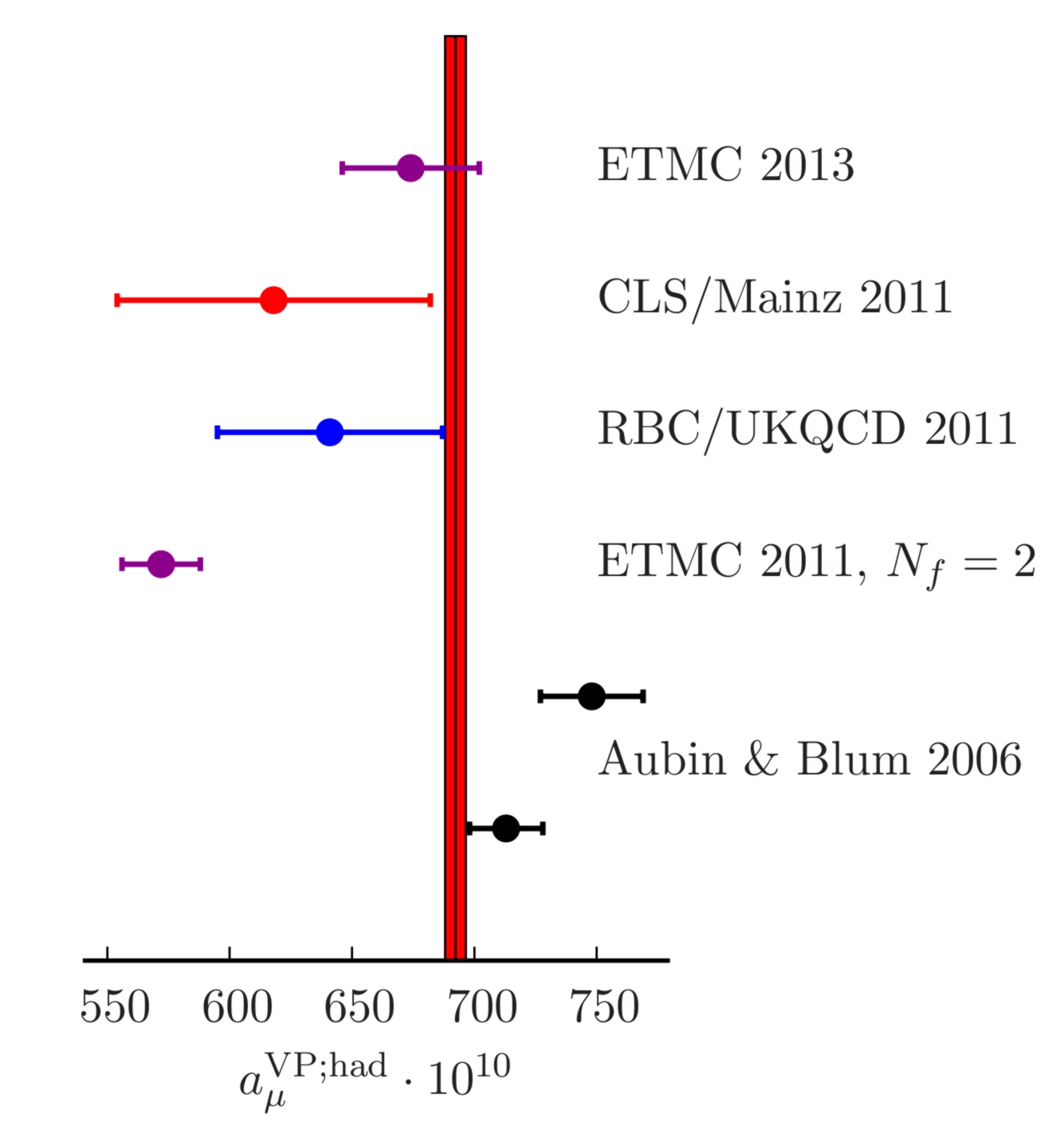}
\caption{Compilation of recently published lattice QCD results for the
  leading hadronic vacuum polarization contribution to the muon's
  anomalous magnetic moment. Displayed is $10^{10}a_\mu^{\rm VP;had}$,
  from ETM\,\cite{Feng:2011zk,Burger:2013jya},
  CLS/Mainz\,\cite{DellaMorte:2011aa}, RBC/UKQCD\,\cite{Boyle:2011hu}
  and Aubin et al.\,\cite{Aubin:2006xv}. The position and width of the
  red vertical line denote the phenomenological result from dispersion
  theory and its uncertainty, respectively.}
  \label{fig:amuhad}
\end{figure}

To complement the phenomenological approach, it is desirable to determine the contributions due to HVP from
first principles.
Lattice QCD is usually restricted to space-like momenta, and in\,\cite{Lautrup:1971jf,Blum:2002ii} it was
shown that $a_\mu^{\rm had;VP}$ can be expressed in terms of a 
convolution integral, i.e., 
\begin{equation}
  a_\mu^{\rm VP;had} = 4\pi^2\left(\frac{\alpha}{\pi}\right)^2
  \int_0^\infty {\rm d}Q^2\, f(Q^2)\left\{ \Pi(Q^2)-\Pi(0) \right\},
\label{eq:amuhad}
\end{equation}
where the vacuum polarization amplitude, $\Pi(Q^2)$, is determined by computing the correlation function of
the vector current.
Recent calculations based on this approach appeared in\,\cite{Aubin:2006xv,Feng:2011zk,Boyle:2011hu,
DellaMorte:2011aa, DellaMorte:2012cf,Burger:2013jya,Gregory:2013taa}, and a compilation of published results
is shown in Fig.\,\ref{fig:amuhad}.

The evaluation of the correlation function of the electromagnetic current involves quark-disconnected
diagrams, which are also encountered in isoscalar form factors of the nucleon discussed earlier in this
section.
Given that a statistically precise evaluation is very costly, such contributions have been largely neglected
so far.
Another major difficulty arises from the fact that the known convolution function $f(Q^2)$ in
Eq.\,(\ref{eq:amuhad}) is peaked at momenta around the muon mass, which is a lot smaller than the typical
nonzero momentum that can be achieved on current lattices.
Therefore, it appears that lattice estimates of $a_\mu^{\rm VP;had}$ are afflicted with considerable
systematic uncertainties relating to the low-$Q^2$ region.
In \cite{DellaMorte:2011aa} it was therefore proposed to apply partially twisted boundary
conditions\,\cite{deDivitiis:2004kq,Sachrajda:2004mi} to compute the quark-connected part of the correlator.
In this way, it is possible to obtain a very high density of data points, which penetrate the region where
the convolution integral receives its dominant contribution.

Recently, there have been proposals which are designed to overcome this problem.
In\,\cite{Feng:2013xsa,Francis:2013fzp} the subtracted vacuum polarization amplitude, $\Pi(Q^2)-\Pi(0)$, is
expressed as an integral of a partially summed vector-vector correlator, which is easily evaluated on the
lattice for any given value of the $Q^2$.
Furthermore, a method designed to compute the additive renormalization $\Pi(0)$ directly, i.e., 
without the need for an extrapolation to vanishing $Q^2$, has been proposed\,\cite{deDivitiis:2012vs}.

The compilation of recent lattice results for the leading hadronic vacuum polarization contributions and
their comparison to the standard dispersive approach in Fig.\,\ref{fig:amuhad} shows that the accuracy of
current lattice estimates of $a_\mu^{\rm{VP;had}}$ is not yet competitive.
In particular, statistical uncertainties will have to be considerably reduced, before lattice results can
challenge the accuracy of dispersion theory.
One step in this direction was undertaken in\,\cite{Blum:2012uh}, which advocates the use of efficient noise
reduction techniques, dubbed ``all-mode-averaging.''
Other recent activities study the systematic effects relating to the use of twisted boundary
conditions\,\cite{Aubin:2013daa} and the Ansatz used to extrapolate $\Pi(Q^2)$ to vanishing
$Q^2$\,\cite{Aubin:2012me}.

For HLbL, a direct experimental determination analogous to those discussed for 
HVP is not directly available.
HLbL enters in ${\cal O}(\alpha^3)$, just as the NLO HVP does. The latter, however, 
is assessed in a dispersion relation framework~\cite{Hagiwara:2006jt}, 
similar to that of the LO piece --- the piece 
associated with the electromagnetic dressing of the HVP is part of the 
final-state radiation correction 
to the LO HVP term~\cite{Hagiwara:2003da}. 
As for the HLbL term it must be calculated; we 
refer to \cite{Prades:2009tw,Jegerlehner:2009ry} for reviews. 
 The diagrammatic contributions to it 
can be organized in a simultaneous expansion in momentum $p$ and number of colors
$N_c$~\cite{deRafael:1993za}; the leading contribution in $N_c$ is a $\pi^0$
exchange graph. The computation of HLbL requires integration over three of the
four photon momenta. Detailed analysis reveals that the bulk of the integral
does not come from small, virtual momenta, 
making ChPT
of little use. Consequently heavier meson
exchanges should be included as well; this makes the uncertainties in the
HLbL computation more challenging to assess. 
We have reported the HLbL result determined by 
the consensus of different groups~\cite{Prades:2009tw}. 
Recently there has been discussion of the charged-pion loop graph 
(which enters as a subleading effect) in chiral perturbation
theory, arguing that existing model calculations of HLbL are inconsistent with
the low-energy structure of QCD~\cite{Engel:2012xb}. Including the omitted
low-energy constants in the usual framework 
does modify the HLbL prediction at the 10\% level~\cite{Bijnens:2012an}. 
The upshot is that the uncertainties can be better controlled 
through measurement of the pion polarizability 
(or generally of processes 
involving a $\pi^+\pi^- \gamma^*\gamma^*$ vertex such 
as $e^+ e^- \to e^+ e^- \pi^+\pi^0$), which is possible
at JLab~\cite{Engel:2013kda}. As long recognized, 
data on $\pi^0\to \gamma\gamma^*$, 
$\pi^0\to \gamma^*\gamma^*$, as well as $\pi^0 \to e^+ e^- (\gamma)$, 
should also help in constraining the primary $\pi^0$ exchange contribution. 
Recently, a dispersive framework for the analysis of the $\pi$ and 
$2\pi$ intermediate states (and generalizable to other mesons)
to HLbL has been 
developed~\cite{Colangelo:2014dfa}; we are hopeful in regards to its future
prospects.

Unfortunately, lattice-QCD calculations of HLbL are still at a very early stage. 
A survey of recent ideas with a status report is given in\,\cite{Blum:2013qu}.
%
%
Here, we comment briefly on only two approaches: The extended Nambu--Jona-Lasinio (ENJL) model (see, e.g.,
\cite{Bijnens:1995xf}), and a functional approach based on calculations of Landau-gauge-QCD Green's functions
(see, e.g., \cite{Goecke:2012qm,Goecke:2013fpa}).
The latter is based on a model interaction (cf.\ the remarks on the 
Faddeev approach to nucleon observables
in Sec.~\ref{sec:lq.struct.form-factors.faddeev}).
However, such a calculation based on input determined from first-principle calculations is highly desirable.

In the ENJL model one has a nonrenormalizable contact interaction, and consequently a momentum-independent
quark mass and no quark wave-function renormalization.
The quark-photon vertex is modeled as a sum of the tree-level term and a purely transverse term containing
the vector meson pole.
On the other hand, the Green's function approach is based on an interaction according to the ultraviolet
behavior of QCD and is therefore renormalizable.
The resulting quark propagator is characterized by a momentum-dependent quark mass and a momentum-dependent
quark wave-function renormalization.
The quark-photon vertex is consistently calculated and contains a dynamical vector meson pole.
Although the different momentum dependencies cancel each other partly (which is understandable when
considering the related Ward identities), remarkable differences in these calculations remain.
Based on a detailed comparison the authors of \cite{Goecke:2012qm} argue that the suppression of the
quark-loop reported in the ENJL model is an artifact of the momentum-independent quark mass and the momentum
restriction within the quark-photon vertex, which, in turn, are natural consequences of the contact
interaction employed there.
Regardless of whether one concludes from these arguments that the standard value for the hadronic
light-by-light scattering contribution may be too small, one almost inevitably needs to conclude that the
given comparison provides evidence that the systematic error attributed to the ENJL calculation is largely
underestimated.

As it is obvious that an increased theoretical error leads to a different conclusion on the size of the
discrepancy of the value for $a_\mu$ between the theoretical and experimental values, an increased effort on
the QCD theory side is needed.
One important aspect of future lattice calculations of the hadronic light-by-light contribution is to employ
them in a complementary way together with other methods.
For instance, an identification of the relevant kinematics of the hadronic contribution to the photon
four-point function through the cross-fertilization of different approaches might already pave the way for
much more accurate computations.
The forthcoming direct measurement of $a_\mu$ at FNAL is expected to reduce the overall error by a factor of 
five.
Therefore, a significant improvement of the theoretical uncertainty for the hadronic light-by-light
scattering contribution down to the level of 10\% is required.
Hereby the systematic comparison of different approaches such as effective models, functional methods, and
lattice gauge theory may be needed to achieve this goal.

\subsubsection{The electroweak mixing angle}

The observed deviation between direct measurements and theoretical predictions
of the muon anomalous magnetic moment --- if corroborated in the future ---
may be taken as a strong hint for physics beyond the Standard Model. Another
quantity which provides a stringent test of the Standard Model is the
electroweak mixing angle, $\sin^2\theta_W$. There is, however, a three-sigma
difference bet\-ween the most precise experimental determinations of
$\sin^2\theta_W(M_Z)_{\rm\overline{MS}}$ at SLD\,\cite{Abe:2000dq}, measuring
the left-right asymmetry in polarized $e^{+}e^{-}$ annihilation, and
LEP\,\cite{ALEPH:2005ab}, which is based on the forward-backward asymmetry in
$Z\to b\bar{b}$. The origin of the tension between these two results has never
been resolved. While an existing measurement at the
Tevatron\,\cite{Abazov:2011ws,Han:2011vw} is not accurate enough to decide the
issue, it will be interesting to see whether the LHC experiments can improve
the situation.

The value of $\sin^2\theta_W$ can be translated into a value of the Higgs
mass, given several other SM parameters as input, including the strong
coupling constant $\alpha_s$, the running of the fine-structure constant
$\Delta\alpha$, and the mass of the top quark, $m_t$. The two conflicting
measurements at the $Z$-pole lead to very different predictions for the Higgs
mass $m_H$\,\cite{Marciano:2006zu}, which can be confronted with the direct
Higgs mass measurement at the LHC. In order to decide whether any observed
discrepancy could be a signal for physics beyond the Standard Model, further
experimental efforts to pin down the value of
$\sin^2\theta_W(M_Z)_{\rm\overline{MS}}$ are required.

\begin{figure}[b]
\includegraphics*[width=\figwid]{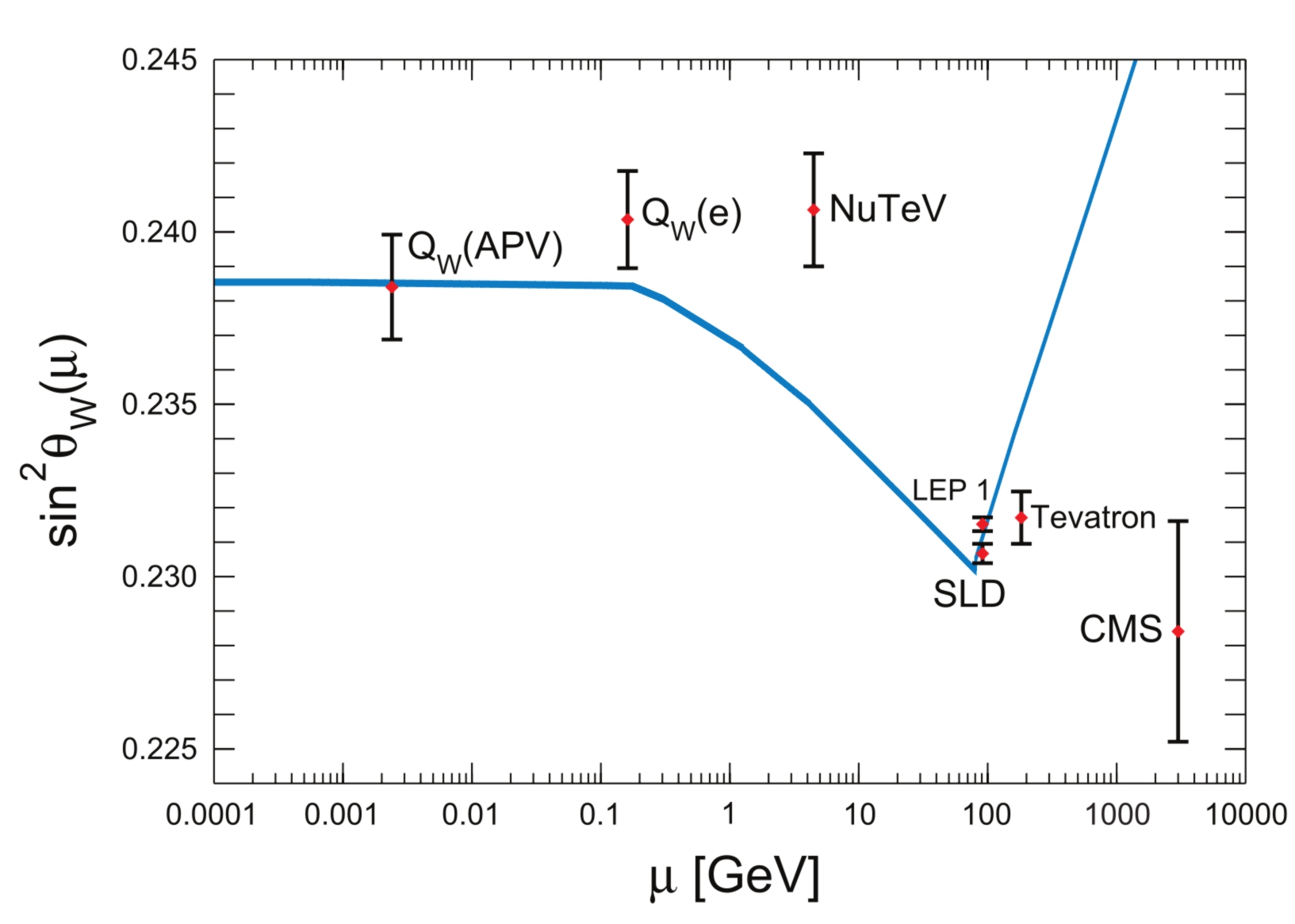}
\caption{The scale dependence of the electroweak mixing angle in
  the $\rm\overline{MS}$ scheme. The blue band is the theoretical
  prediction, while its width denotes the theoretical uncertainty from
  strong interaction effects. From \cite{Beringer:1900zz}).
\label{fig:weinberg}}
\end{figure}

In addition to the activities at high-energy colliders, there are also
new (QWEAK \cite{Armstrong:2012ps}) and planned experiments
(MOLLER\,\cite{Mammei:2012ph}, P2@MESA), designed to measure the
electroweak mixing angle with high precision at low energies, by
measuring the weak charge of the proton. These efforts extend earlier
measurements of the parity-violating asymmetry in M{\o}ller scattering
\cite{Anthony:2005pm} and complement other low-energy determinations,
based on atomic parity violation (APV) and neutrino-DIS (NuTeV). The
collection of measurements across the entire accessible energy range
can be used to test whether the running of $\sin^2\theta_W$ is
correctly predicted by the SM, i.e., by checking that the different
determinations can be consistently translated into a common value at
the $Z$-pole. The current status is depicted in
Fig.\,\ref{fig:weinberg}.

We will now discuss the particular importance of low-energy hadronic
determinations of the electroweak mixing angle, and the role of new
experiments (for an in-depth treatment, see\,\cite{Kumar:2013yoa}). These are
based on measuring the weak charge of the proton, $Q_W^p$, which is accessible
by measuring the helicity-dependent cross section in polarized $ep$
scattering. For a precise determination of the electroweak mixing angle, one
must augment the tree-level relation $Q_W^p=1-4\sin^2\theta_W$ by radiative
corrections \cite{Erler:2003yk}. It then turns out that the dominant
theoretical uncertainty is associated with hadronic effects, whose evaluation
involves some degree of modeling\,\cite{Erler:2004in}. Radiative corrections
arising from ${\gamma}Z$ box graphs play a particularly important role, and
their contributions have been evaluated
in\,\cite{Gorchtein:2008px,Gorchtein:2011mz,Sibirtsev:2010zg,Rislow:2010vi,Blunden:2011rd}.
An important feature is that they are strongly suppressed at low energies. It
is therefore advantageous to measure the weak charge in low-energy $ep$
scattering, since the dominant theoretical uncertainties in the relation
between $Q_W^p$ and $\sin^2\theta_W$ are suppressed.

\subsubsection{$\alpha_s$ from inclusive hadronic $\tau$ decay}
\label{sec:secB5}



As remarked several times in this review,
the precise determination of $\alpha_s$
at different scales, and hereby especially the impressive agreement between
experimental determinations and theoretical predictions, provides an important
test of asymptotic freedom and plays a significant role in establishing QCD as
the correct fundamental theory of the Strong Interaction.
Hadronic $\tau$ decays allow for a determination of $\alpha_s$ at quite low momentum 
scales~\cite{Braaten:1991qm}.
The decisive experimental observable is the inclusive ratio of $\tau$ decay widths,
\begin{equation}
R_\tau \equiv \frac{\Gamma [\tau^- \to\nu_\tau {\mathrm {hadrons} \, (\gamma)]}}
{\Gamma [\tau^- \to\nu_\tau e^-\bar \nu_e (\gamma)]}  ,
\label{eq:Rtau}
\end{equation}
which can be rigorously analyzed with the short-distance operator product expansion.
Since nonperturbative corrections are heavily suppressed by six powers of the $\tau$ mass, the theoretical prediction is dominated by the perturbative contribution, which is already known to $O(\alpha_s^4)$
and amounts to a 20\% increase of the naive parton-model result $R_\tau = N_C = 3$.
Thus, $R_\tau$ turns out to be very sensitive to the value of the strong coupling at the $\tau$ mass scale; see, e.g., \cite{Pich:2013lsa} and references therein.

From the current $\tau$ decay data, one obtains \cite{Pich:2013lsa}
\begin{equation}\label{eq:alphas_tau}
\alpha_s(m_\tau^2) = 0.331\pm 0.013\, .
\end{equation}
The recent Belle measurement of the $\tau$ lifetime \cite{Belous:2013dba} has slightly increased the central value by $+0.002$, with respect to the previous result \cite{Pich:2013sqa}.
After evolution to the scale $M_Z$, the strong coupling decreases to
\begin{equation}
\alpha_s(M_Z^2) = 0.1200\pm 0.0015\, ,
\end{equation}
in excellent agreement with the direct measurement at the $Z$ peak,
$\alpha_s(M_Z^2) = 0.1197\pm 0.0028$ \cite{Beringer:1900zz}. Owing to the QCD running, the error on $\alpha_s$ decreases by one order of magnitude from $\mu = m_\tau$ to $\mu=M_Z$.

The largest source of uncertainty has a purely perturbative origin.
The $R_\tau$ calculation involves a closed contour integration in the complex $s$-plane, along the circle
$|s| = m_\tau^2$.
The long running of $\alpha_s(-s)$ generates powers of large logarithms, $\log{(-s/m_\tau^2)}=i \phi$,
$\phi\in [-\pi,\pi]$, which need to be resummed using the renormalization group.
One gets in this way an improved perturbative series, known as contour-improved perturbation theory (CIPT)
\cite{Le_Diberder:1992te}, which shows quite good convergence properties and a mild dependence on the
renormalization scale.
A naive expansion in powers of $\alpha_s(m_\tau^2)$ (fixed-order perturbation theory, FOPT), without
resumming those large logarithms, gives instead a badly-behaved series which suffers from a large
renormalization-scale dependence.
A careful study of the contour integral shows that, even at $O(\alpha_s^4)$, FOPT overestimates the total
perturbative correction by about 11\%; therefore, it leads to a smaller fitted value for $\alpha_s$.
Using CIPT one obtains $\alpha_s(m_\tau^2) = 0.341\pm 0.013$, while FOPT results in $\alpha_s(m_\tau^2) =
0.319\pm 0.014$ \cite{Pich:2013lsa}.

The asymptotic nature  of the perturbative QCD series has been argued to play an important role even at low orders in the coupling expansion. Assuming that the fourth-order series is already governed by the lowest ultraviolet and infrared renormalons, and fitting the known expansion coefficients to ad-hoc renormalon models, one predicts a positive correction from the unknown higher orders, which results in a total perturbative contribution to $R_\tau$ close to the naive FOPT result \cite{Beneke:2008ad}. This conclusion is however model dependent \cite{Pich:2013lsa}. In the absence of a better understanding of higher-order corrections, the CIPT and FOPT determinations have been averaged in Eq.~(\ref{eq:alphas_tau}), but keeping conservatively the smallest error.

A precise extraction of $\alpha_s$ at such low scale necessitates also a thorough understanding of the small
nonperturbative condensate contributions to $R_\tau$.
Fortunately, the numerical size of nonperturbative effects can be determined from the measured
invariant-mass distribution of the final hadrons in $\tau$ decay \cite{Le_Diberder:1992fr}.
With good data, one could also analyze the possible role of corrections beyond the operator product expansion.
The latter are called duality violations (because they signal the breakdown of quark-hadron duality
underlying the operator product expansion), and there is (yet) no first-principle theoretical description
available.
These effects are negligible for $R_\tau$, because the operator-product-expansion uncertainties near the real
axis are kinematically suppressed in the relevant contour integral; however, they could be more relevant for
other moments of the hadronic distribution.

The presently most complete and precise experimental analysis, performed with the ALEPH data, obtains a total
nonperturbative correction to $R_\tau$, $\delta_{\mathrm{NP}} = -(0.59\pm 0.14)\% $ \cite{Davier:2008sk}, in
good agreement with the theoretical expectations and previous experimental determinations by ALEPH, CLEO, and
OPAL \cite{Pich:2013lsa}.
This correction has been taken into account in the $\alpha_s$ determination in Eq.~(\ref{eq:alphas_tau}).
A more recent fit to rescaled OPAL data (adjusted to reflect current values of exclusive hadronic
$\tau$-decay branching ratios), with moments chosen to maximize duality violations, finds
$\delta_{\mathrm{NP}} = -(0.3\pm 1.2)\% $ \cite{Boito:2012cr}, in agreement with the ALEPH result but less
precise because of the much larger errors of the OPAL data.

A substantial improvement of the $\alpha_s(m_\tau^2)$ determination requires more accurate $\tau$ spectral-function data, which should be available in the near future, and a better theoretical control of higher-order perturbative contributions, i.e., an improved understanding of the asymptotic nature of the QCD perturbative series.

Experimental knowledge on $\alpha_s$ at even lower scales ($s<m^2_\tau$), at the
borderline of the perturbative to nonperturbative regime of QCD, could profit
from lattice simulations of appropriately chosen observables. Last but not
least, it should be noted that in the nonperturbative domain, i.e., at scales below
1 GeV, an unambiguous definition of the strong coupling is missing; for a
corresponding discussion see, e.g.,~\cite{Shirkov:2002gw}.

\subsection{Future Directions}
\label{sec:secB6}

In a broad sense, the physics of light quarks  remains  a key for understanding  strong QCD dynamics, from its more fundamental nonperturbative effects to the varied dynamical effects which manifest themselves in the different properties of hadrons. Recent progress in the theoretical and experimental fronts is indeed remarkable.

Numerous experimental results keep flowing from different facilities employing hadron (J-PARC, COSY, COMPASS, VES) or electron beams (CLAS, MAMI-C, ELSA, SPring-8, CLEO-c, BESIII, KLOE-2, and  CMD-3 and SND at VEPP-2000). The experiments aim at investigating the full hadron spectrum, searching e.g. for exotic and hybrid mesons or missing baryon resonances, as well as at determining dynamical properties of those excited states such as helicity amplitudes and form factors. New facilities are planned to come into operation in the next few years, which are expected to deliver data with extremely high statistical accuracy. The upcoming 12 GeV upgrade of JLab with the new Hall D is one of the key new additions to that line of research. Also the upgraded CLAS12 detector at JLab is expected to contribute to hadron spectroscopy. Hadronic decays of heavy-quark states produced at future $e^+e^-$ (Belle II) or $p\bar{p}$ machines (PANDA) will serve as abundant source of light-quark states with clearly defined initial states. In addition,   the particularly clean   access to light hadron states via  direct production    in $e^+e^-$ annihilation with initial-state radiation as well as via $\gamma \gamma$ fusion is possible at Belle II.   The anticipated data from these next-generation experiments should in principle allow us to clarify the existence and nature of hadronic resonances beyond the quark model and 
to 
determine resonance parameters reliably for states where this has not been possible in the past because of pole positions far in the complex plane, overlapping resonances, or weak couplings to experimentally accessible channels. A model- and reaction-independent characterization of resonance parameters in terms of pole positions and residues, however, also requires advances on the analysis side to develop models which respect the theoretical constraints of unitarity and analyticity.

Experiments on the ground-state mesons and baryons will continue at the intermediate- and high-energy facilities, which can have an impact in and beyond QCD. Examples include  the elucidation of the spin structure of the nucleon at the partonic level, which is one of the motivations for the work currently underway  designing an Electron Ion Collider, precision photo-production on the nucleon and of light mesons, and experiments that impact the Standard Model, such as those necessary for improving the calculation of the hadronic  contributions  to the muon's anomalous magnetic moment, and the measurements of the weak charge of the nucleon that impacts the knowledge of the EW angle at lower energies. Naturally, most of the topics discussed in this review are part of the broad experimental programs in place today and planned for the near future.

\newpage

On the theoretical front, LQCD is opening new perspectives. Full QCD calculations with light quark masses nearing the physical limit are becoming standard. This is allowing for unprecedented insights into the quark mass dependencies of meson and baryon observables, which especially influence the determination of numerous LECs in EFT which are poorly known from phenomenology, and also in the knowledge of form factors and moments of structure functions. The study of excited light hadrons in LQCD is one of the most important developments in recent years, with the promise of illuminating the present rather sparse knowledge of those excited states as well as possibly leading to the ``discovery'' of new states which are of difficult experimental access.
It is  clear that the  progress in LQCD will continue, turning it into  a key exploratory/discovery as well as a precision tool for light quark physics.

Progress also continues with analytic methods, in particular with methods rooted in QCD, such as Schwinger-Dyson equations, ChPT, dispersion theory combined with ChPT, SCET, various approaches in perturbative QCD, $1/N_c$ expansion, AdS/QCD, etc.  Most analytic methods rely on experimental and/or lattice QCD information, which is currently fueling theoretical progress thanks to the abundance and quality of that information.

\clearpage
\section[Chapc]{Heavy quarks \protect\footnotemark}
\footnotetext{Contributing authors: M.~Butenschoen$^{\dagger}$, P.~Pakhlov$^{\dagger}$, 
A.~Vairo$^{\dagger}$, N.~Brambilla, X.~Garcia~i~Tormo, G.~M. von~Hippel, R.~Mizuk, J.-W.~Qiu, G.~Ricciardi.}  

\label{sec:chapc}

Heavy quarks have played a crucial role in the establishing and 
development of QCD in particular, and the Standard Model of particle 
physics in general. Experimentally this is related to a clean signature 
of many observables even in the presence of only few rare events, which 
allows the study of both new emergent phenomena in the realm of QCD 
and new physics beyond the Standard Model.
Theoretically, the clean signature may be traced back to the fact that 
\begin{equation}
m_Q \gg \lamQ\,,
\label{eq:heavy:hierarchy}
\end{equation}
which implies that processes happening at the scale of the heavy-quark mass 
$m_Q$ can be described by perturbative QCD and that nonperturbative 
effects, including the formation of background low-energy 
light hadrons,  are suppressed by powers of $\lamQ/m_Q$.
The hierarchy \eqref{eq:heavy:hierarchy} gets complicated by lower energy 
scales if more than one heavy quark is involved in the physical process, 
but the basic fact that high-energy physics at the scale $m_Q$ 
can be factorized from low-energy nonperturbative physics at the 
hadronic scale $\lamQ$ is at the core of the dynamics of any system
involving a heavy quark. 

The hierarchy \eqref{eq:heavy:hierarchy} is usually exploited to replace 
QCD with equivalent Effective Field Theories (EFTs) that make manifest 
at the Lagrangian level the factorization of the high-energy modes from 
the low-energy ones. Examples are the Heavy Quark Effective Theory (HQET)
\cite{Isgur:1989vq,Isgur:1989ed,Eichten:1989zv,Neubert:1993mb} suitable 
to describe systems made of one heavy quark, and EFTs like Nonrelativistic 
QCD (NRQCD) \cite{Caswell:1985ui,Bodwin:1994jh} or potential Nonrelativistic QCD (pNRQCD) 
\cite{Pineda:1997bj,Brambilla:1999xf},
suitable to describe systems made of two or more heavy quarks.
Nonrelativistic EFTs \cite{Brambilla:2004jw} have been systematically used both in analytical 
and in numerical (lattice) calculations involving heavy quarks.
Concerning lattice studies, nowadays the standard approach 
is to resort to EFTs when bottom quarks are involved, 
and to rely on full lattice QCD calculations 
when studying systems made of charm quarks.

The chapter aims at highlighting some of the most relevant progress 
made in the last few 
years in the heavy-quark sector of QCD both from the methodological 
and phenomenological point of view. There is no aim of completeness.
It is organized in the following way. 
In Sec.~\ref{sec:secC1} we discuss methodological novelties in the formulation of nonrelativistic EFTs 
and in lattice QCD, whereas the following sections are devoted to more phenomenological 
aspects. These are divided in phenomenology of heavy-light mesons, discussed in  Sec.~\ref{sec:secC3} and in phenomenology of heavy quarkonia.
In Sec.~\ref{sec:secC2} we present recent progress in quarkonium spectroscopy with 
particular emphasis on the quarkonium-like states at and above the open flavor threshold.
Section~\ref{sec:secC5} provides an updated list of $\als$ extractions from quarkonium observables. 
Section~\ref{sec:secC6} summarizes our current understanding of quarkonium production.
Finally, Sec.~\ref{sec:secC7} outlines future directions.

\subsection{Methods}

\label{sec:secC1}

\subsubsection{Nonrelativistic effective field theories}

\label{sec:subsecC11}

The nonrelativistic EFT of QCD suited to describe a heavy quark 
bound into a heavy-light meson is HQET \cite{Eichten:1989zv,Eichten:1990vp}
(see \cite{Neubert:1993mb} for an early review). Heavy-light mesons are characterized 
by only two energy scales: the heavy quark mass $m_Q$ and the hadronic scale $\lamQ$.
Hence the HQET Lagrangian is organized as an expansion in $1/m_Q$ and physical observables 
as an expansion in $\lamQ/m_Q$ (and $\als$ encoded in the Wilson coefficients).  
In the limit where  $1/m_Q$ corrections are neglected, the HQET Lagrangian is 
independent of the flavor and spin of the heavy quark. This symmetry is called 
the heavy quark symmetry. Some of its phenomenological consequences for $B$ and $D$ 
decays will be discussed in Sec.~\ref{sec:secC3}.

In the case of two or more heavy quarks, the system is characterized by more energy scales.
We will focus on systems made of a quark and an antiquark, i.e. quarkonia, although EFTs have 
been also developed for baryons made of  three quarks 
\cite{Brambilla:2005yk,Brambilla:2009cd,Brambilla:2013vx}.
For quarkonia, one has to consider at least the scale of the typical momentum transfer between the quarks, 
which is also proportional to the inverse of the typical distance, and the scale of the 
binding energy. In a nonrelativistic bound state, the first goes parametrically like $m_Qv$ 
and the second like $m_Qv^2$, where $v$ is the velocity of the heavy quark in the center-of-mass frame. An EFT suited to describe heavy quarkonia at a scale lower than 
$m_Q$ but larger than $m_Qv$ and $\lamQ$ is NRQCD \cite{Caswell:1985ui,Bodwin:1994jh} (whose lattice version
was formulated in \cite{Thacker:1990bm,Lepage:1992tx}). Also the NRQCD Lagrangian is organized 
as an expansion in $1/m_Q$ and physical observables as an expansion in $v$ 
(and $\als$ encoded in the Wilson coefficients). 
In the heavy-quark bilinear sector the Lagrangian coincides with the one of HQET 
(see also \cite{Manohar:1997qy}), but the Lagrangian contains also four-quark operators.
These are necessary to describe heavy-quarkonium annihilation and production, which are processes  
happening at the scale $m_Q$. The NRQCD factorization for heavy quarkonium annihilation processes 
has long been rigorously proved  \cite{Bodwin:1994jh}, while  this is not the case 
for heavy quarkonium production. Due to its relevance, we devote the entire 
Sec.~\ref{sec:subsecC13} to the most recent progress towards a proof of factorization for heavy 
quarkonium production. The state of the art of our understanding of heavy quarkonium 
production in the framework of NRQCD is presented in Sec.~\ref{sec:secC6}.

The power counting of NRQCD is not unique because the low energy matrix elements 
depend on more than one residual energy scale. These residual scales are $m_Qv$, $m_Q v^2$, $\lamQ$ 
and possibly other lower energy scales. The ambiguity in the power counting is reduced and 
in some dynamical regimes solved by integrating out modes associated to the scale $m_Qv$ 
and by replacing NRQCD by pNRQCD, an EFT suited to describe quarkonium physics 
at the scale $m_Qv^2$ \cite{Pineda:1997bj,Brambilla:1999xf}.
The pNRQCD Lagrangian is organized as an expansion in $1/m_Q$, inherited from NRQCD, 
and an expansion in powers of the distance between the heavy quarks.
This second expansion reflects the expansion in the scale $m_Qv^2$ relative to the scale $m_Qv$
specific to pNRQCD. Like in NRQCD, contributions to physical observables are counted in powers of $v$ 
(and $\als$ encoded in the high energy Wilson coefficients). The degrees of freedom of pNRQCD depend 
on the specific hierarchy between $m_Qv^2$ and $\lamQ$ for the system under examination. 

The charmonium ground state and the lowest bottomonium states may have a sufficiently small 
radius to satisfy the condition $m_Qv^2 \simg \lamQ$. If this is the case, the degrees of freedom 
of pNRQCD are quark--antiquark states and gluons. The system can be studied in perturbative QCD, nonperturbative 
contributions are small and in general one may expect precise theoretical determinations 
once potentially large logarithms have been resummed by solving renormalization group equations 
and renormalon-like singularities have been suitably subtracted. For early applications we refer to 
\cite{Kniehl:1999ud,Brambilla:1999xj,Brambilla:2001fw,Pineda:2001zq,Kniehl:2003ap,Penin:2004ay}, for a dedicated review see \cite{Pineda:2011dg}.
As an example of the quality of these determinations, we mention 
the  determination of the $\eta_b$ mass in \cite{Kniehl:2003ap}. 
This was precise and solid enough to challenge early experimental measurements, 
while being closer to the most recent ones. We will come back to this and 
other determinations in Sec.~\ref{sec:secC2}.

Excited bottomonium and charmonium states are likely strongly bound, which implies that 
$\lamQ \simg m_Qv^2$. The degrees of freedom of pNRQCD are colorless and made of 
color-singlet quark-antiquark and light quark states
\cite{Brambilla:2000gk,Pineda:2000sz,Brambilla:2001xy,Brambilla:2002nu,Brambilla:2003mu}.
The potentials binding the quark and antiquark have a rigorous expression in terms 
of Wilson loops and can be determined by lattice QCD \cite{Koma:2006si,Koma:2006fw,Koma:2007jq,Koma:2012bc}.
It is important to mention that lattice determinations of the potentials have been performed 
so far in the quenched approximation. Moreover, at order $1/m_Q^2$ not all the necessary 
potentials have been computed (the set is complete only for the spin-dependent potentials).
This implies that the quarkonium dynamics in the strongly coupled regime is not 
yet exactly known beyond leading $1/m_Q$ corrections. 

For states at or above the open flavor threshold, new degrees of freedom may become important 
(heavy-light mesons, tetraquarks, molecules, hadro-quarkonia, hybrids, glueballs, ...).
These states can in principle be described in a very similar framework to the one discussed above 
for states below threshold \cite{Brambilla:2008zz,Vairo:2009tn,Brambilla:2010cs,Braaten:2013boa}. 
However, a general theory does not exist so far and specific EFTs have been built to describe 
specific states (an example is the well-known $X(3872)$ 
\cite{Braaten:2004ai,Braaten:2005jj,Braaten:2005ai,Braaten:2007dw,Fleming:2007rp,Fleming:2008yn}).
This is the reason why many of our expectations for these states still rely on potential models.

In Sec.~\ref{sec:secC2} we will discuss new results concerning the charmonium and bottomonium 
spectroscopy below, at and above threshold, the distinction being dictated by our different understanding 
of these systems. 
For instance, we will see that there has been noteworthy progress in describing 
radiative decays of quarkonium below threshold and that 
the theory is now in the position to provide for many of the transitions competitive and model-independent results.

Finally, on a more theoretical side, since the inception of nonrelativistic EFTs 
there has been an ongoing investigation on how they realize Lorentz invariance.
It has been shown in \cite{Luke:1992cs,Manohar:1997qy} that HQET is 
reparameterization invariant. Reparameterization invariance constrains the form of the Wilson coefficients 
of the theory. In \cite{Brambilla:2001xk,Brambilla:2003nt} it was shown that the same 
constraints follow from imposing the Poincar\'e algebra on the generators of the Poincar\'e group in the EFT.
Hence reparameterization invariance appears as the way in which Lorentz invariance, which is 
manifestly broken by a nonrelativistic EFT, is retained order by order in $1/m_Q$ by the EFT.
This understanding has recently been further substantiated in \cite{Heinonen:2012km}, where 
the consequences of reparameterization and Poincar\'e invariance have been studied 
to an unprecedented level of accuracy.

\subsubsection{The progress on NRQCD factorization}

\label{sec:subsecC13}

The NRQCD factorization approach to heavy quarkonium production, introduced as a conjecture \cite{Bodwin:1994jh}, is phenomenologically successful in describing existing data, although there remain challenges particularly in connection with polarization observations \cite{Brambilla:2010cs}.  In the NRQCD factorization approach, the inclusive cross section for the direct production of a quarkonium state $H$  at large momentum transfer ($p_T$) is written as a sum of ``short-distance'' coefficients times NRQCD long-distance matrix elements (LDMEs), 
\begin{equation}
\sigma^H(p_T,m_Q) = \sum_n \sigma_n(p_T,m_Q,\Lambda) \langle 0| {\cal O}_n^H(\Lambda)|0\rangle \, .
\label{eq:nrqcd-fac}
\end{equation}
Here $\Lambda$ is the ultraviolet cut-off of the NRQCD effective theory.  The {\em short-distance} coefficients $\sigma_n$ are essentially the process-dependent partonic cross sections to produce a $Q\bar{Q}$ pair in various color, spin, and orbital angular momentum states $n$ (convoluted with the parton distributions of incoming hadrons for hadronic collisions), and perturbatively calculated in powers of $\alpha_s$.   The LDMEs are nonperturbative, but universal, representing the probability for a $Q\bar{Q}$ pair in a particular state $n$ to evolve into a heavy quarkonium.  The sum over the $Q\bar{Q}$ states $n$ is organized in terms of powers of the pair's relative velocity $v$, an intrinsic scale of the LDMEs. For charmonia, $v^2\approx0.3$, and for bottomonia, $v^2\approx0.1$. The current successful phenomenology of quarkonium production mainly uses only NRQCD LDMEs through relative order $v^4$, as summarized in Table~\ref{tab:cveloscal}.  The traditional color singlet model is recovered as the $v\to 0$ 
limit. In case of $P$ wave quarkonia and relativistic corrections to $S$ state quarkonia \cite{Bodwin:2012xc}, the color singlet model is incomplete, due to uncanceled infrared singularities.

\begin{table}[t]
\caption{NRQCD velocity scaling of the LDMEs contributing to ${^3}S_1$ quarkonium production up to the order $O(v^4)$ relative to the leading ${^3}S_1$ color singlet contribution \cite{Bodwin:1994jh}. Upper indices~$^{[1]}$ refer to color singlet states and upper indices~$^{[8]}$ to color octet states. The $\langle {\cal O}^{J/\psi}({^3}S_1^{[1]}) \rangle$, $\langle {\cal P}^{J/\psi}({^3}S_1^{[1]}) \rangle$, and $\langle {\cal Q}^{J/\psi}({^3}S_1^{[1]}) \rangle$ LDMEs correspond to the leading order, and the $O(v^2)$ and  $O(v^4)$ relativistic correction contributions to the color singlet model. The contributions involving the $\langle {\cal O}^{J/\psi}({^1}S_0^{[8]}) \rangle$, $\langle {\cal O}^{J/\psi}({^3}S_1^{[8]}) \rangle$ and $\langle {\cal O}^{J/\psi}({^3}P_J^{[8]}) \rangle$ LDMEs are often referred to as {\em the} Color Octet states.\vspace{2mm}}
\label{tab:cveloscal}
\begin{tabular}{c@{\quad}l}
    \hline\hline
Relative scaling & Contributing LDMEs \\ \hline 
1 & $\langle {\cal O}^{H}({^3}S_1^{[1]}) \rangle$ \\
$v^2$ & $\langle {\cal P}^{H}({^3}S_1^{[1]}) \rangle$ \\
$v^3$ & $\langle {\cal O}^{H}({^1}S_0^{[8]}) \rangle$ \\
$v^4$ & $\langle {\cal Q}^{H}({^3}S_1^{[1]}) \rangle$, $\langle {\cal O}^{H}({^3}S_1^{[8]}) \rangle$, 
$\langle {\cal O}^{H}({^3}P_J^{[8]}) \rangle$ \\
\hline\hline
\end{tabular}
\end{table}

Despite the well-documented phenomenological successes, there remain two major challenges for the NRQCD factorization approach to heavy quarkonium production.  One is the validity of the factorization itself, which has not been proved, and the other is the difficulty in explaining the polarization of produced quarkonia in high energy scattering, as will be reviewed in Sec.~\ref{sec:secC6}.  These two major challenges could well be closely connected to each other, and could also be connected to the observed tension in extracting LDMEs from global analyses of all data from different scattering processes \cite{Butenschoen:2011yh}.  A proof of the factorization to all orders in $\alpha_s$ is complicated because gluons can dress the basic factorized production process in ways that apparently violate factorization.  Although there is a clear scale hierarchy for heavy quarkonium, $m_Q \gg m_Q v \gg m_Q v^2$, which is necessary for using an effective field theory approach, a full proof of NRQCD factorization would require a demonstration that all partonic 
diagrams at each order in $\alpha_s$ can be reorganized such that (1) all soft singularities cancel or can be absorbed into NRQCD LDMEs, 
and (2) all collinear singularities and spectator interactions can be either canceled or absorbed into incoming hadrons' parton distributions. 
This has at all orders so far only been established for exclusive production in helicity-non-flip processes 
in $e^+e^-$ annihilation and $B$-meson decay \cite{Bodwin:2008nf,Bodwin:2010fi,Bodwin:2010xb}.

For heavy quarkonium production at collider energies, there is sufficient phase space to produce more than one pair of heavy quarks, and additional observed momentum scales, such as $p_T$.  The NRQCD factorization in Eq.~(\ref{eq:nrqcd-fac}) breaks when there are co-moving heavy quarks \cite{Nayak:2007mb,Nayak:2007zb}.  The short-distance coefficient $\sigma_n(p_T,m_Q,\Lambda)$ in Eq.~(\ref{eq:nrqcd-fac}) for a $Q\bar{Q}(n)$ state can have different power behavior in $p_T$ at different orders in $\alpha_s$.  For example, for $Q\bar{Q}(^3S_1^{[1]})$-channel, the Leading Order (LO) coefficient in $\alpha_s$ is dominated by $1/p_T^8$, and the Next-to-Leading Order (NLO) dominated by $1/p_T^6$, while the Next-to-Next-to-Leading Order (NNLO) coefficient has terms proportional to $1/p_T^4$.  When $p_T$ increases, the logarithmic dependence of $\alpha_s$ on the hard scale cannot compensate the power enhancement in $p_T$ at higher orders, which leads to a unwanted phenomenon that the NLO correction to a given channel 
could be an order of magnitude larger than the LO contribution \cite{Campbell:2007ws,Artoisenet:2008fc}.  Besides the power enhancement at higher orders, the perturbative coefficients at higher orders have higher powers of large $\ln(p_T^2/m_Q^2)$-type logarithms, which should be systematically resummed.  That is, when $p_T\gg m_Q$, a new organization of the short-distance coefficients in Eq.~(\ref{eq:nrqcd-fac}) or a new factorization formalism is necessary. Very significant progress has been made in recent years.

Two new factorization formalisms were derived for heavy quarkonium production at large $p_T$.  One is based on perturbative QCD (pQCD) collinear factorization \cite{Nayak:2005rt,Nayak:2006fm,Kang:2011zza,Kang:2011mg,KMQS:2013,KMQS-hq1,KMQS-hq2}, and the other based on soft collinear effective theory (SCET) \cite{Fleming:2012wy,Fleming:2013qu}.  Both approaches focus on quarkonium production when $p_T\gg m_Q$, and explore potential connections to the NRQCD factorization.   

The pQCD collinear factorization approach, also referred to as the fragmentation function approach \cite{Brambilla:2010cs}, organizes the contributions to the quarkonium production cross section in an expansion in powers of $1/p_T$, and then factorizes the leading power (and the next-to-leading power) contribution in terms of ``short-distance" production of a single-parton of flavor $f$ (and a heavy quark pair $[Q\bar{Q}(\kappa)]$ with $\kappa$ labeling the pair's spin and color) convoluted with a universal fragmentation function for this parton (and the pair) to be evolve into a heavy quarkonium,
\begin{eqnarray}
d\sigma_H(p_T,m_Q)
&\approx &
\sum_{f} d\hat{\sigma}_f(p_T,z)\otimes D_{f\to H}(z,m_Q)\, 
\nonumber\\
&+&
\sum_{[Q\bar{Q}(\kappa)]} 
d\hat{\sigma}_{[Q\bar{Q}(\kappa)]}(p_T,z,u,v)
\label{eq:pqcd-fac}\\
&\ & \hskip 0.4in
\otimes
{\cal D}_{[Q\bar{Q}(\kappa)]\to H}(z,u,v,m_Q)\, ,
\nonumber
\end{eqnarray}
where factorization scale dependence was suppressed, $z,u,v$ are momentum fractions, and $\otimes$ represents the convolution of these momentum fractions.  Both the single parton and heavy quark pair fragmentation functions, $D_f$ and ${\cal D}_{[Q\bar{Q}(\kappa)]}$, are universal, and we can resum large logarithms by solving the corresponding evolution equations \cite{Kang:2011zza,KMQS:2013,KMQS-hq1}. 
The factorization formalism in Eq.~(\ref{eq:pqcd-fac}) holds to all orders in $\alpha_s$ in pQCD up to corrections of ${\cal O}(1/p_T^4)$ (${\cal O}(1/p_T^2)$) with (without) a heavy quark pair, $[Q\bar{Q}(\kappa)]$, being produced \cite{Nayak:2005rt,Kang:2011zza,KMQS-hq1}.

Including the $1/p_T$-type power correction into the factorized production cross section in Eq.~(\ref{eq:pqcd-fac}) necessarily requires modifying the evolution equation of a single parton fragmentation function as \cite{Kang:2011zza,KMQS-hq1},
\begin{eqnarray}
\frac{\partial}{\partial\ln\mu^2}D_{f\to H}(z,\mu^2;m_Q)
&=& \sum_{f'}
\gamma_{f\to f'} \otimes D_{f'\to H}
\label{eq:pqcd-evo} \\
&\ & \hskip -1.0in
+\, \frac{1}{\mu^2} \sum_{[Q\bar{Q}(\kappa')]} 
\gamma_{f\to [Q\bar{Q}(\kappa')]} \otimes {\cal D}_{[Q\bar{Q}(\kappa')]\to H}\, ,
\nonumber
\end{eqnarray}
where $\otimes$ represents the convolution of momentum fractions as those in Eq.~(\ref{eq:pqcd-fac}), and the dependence of momentum fractions in the right-hand-side is suppressed.  The first line in Eq.~(\ref{eq:pqcd-fac}) is effectively equal to the well-known DGLAP evolution equation.  The second term on the right of Eq.~(\ref{eq:pqcd-fac}) is new, and is needed for the single-parton fragmentation functions to absorb the power collinear divergence of partonic cross sections producing a ``massless'' ($m_Q/p_T \sim 0$) heavy quark pair to ensure that the short-distance hard part, $\hat{\sigma}_{[Q\bar{Q}(\kappa)]}(p_T,z,u,v)$ in Eq.~(\ref{eq:pqcd-fac}), is infrared and collinear safe \cite{KMQS-hq2}.  The modified single-parton evolution equation in Eq.~(\ref{eq:pqcd-evo}), together with the evolution equation of heavy quark-pair fragmentation functions
\cite{Kang:2011zza,KMQS-hq1,Fleming:2012wy},
\begin{eqnarray}
\hskip -0.2in
&\ &
\frac{\partial}{\partial\ln\mu^2}{\cal D}_{[Q\bar{Q}(\kappa)]\to H}(z,u,v,\mu^2;m_Q)
\label{eq:pqcd-evo2} \\
\hskip -0.2in
&\ & \hskip 0.5in
=  \sum_{[Q\bar{Q}(\kappa')]} 
\Gamma_{[Q\bar{Q}(\kappa)]\to [Q\bar{Q}(\kappa')]} \otimes 
{\cal D}_{[Q\bar{Q}(\kappa')]\to H}\, ,
\nonumber
\end{eqnarray}
forms a closed set of evolution equations of all fragmentation functions in Eq.~(\ref{eq:pqcd-fac}).  The ${\cal O}(\alpha_s^2)$ evolution kernels for mixing the single-parton and heavy quark-pair fragmentation functions, $\gamma_{f\to [Q\bar{Q}(\kappa')]}$ in Eq.~(\ref{eq:pqcd-evo}), are available \cite{KMQS-hq1}, and the ${\cal O}(\alpha_s)$ evolution kernels of heavy quark-pair fragmentation functions, $\Gamma_{[Q\bar{Q}(\kappa)]\to [Q\bar{Q}(\kappa')]}$ in Eq.~(\ref{eq:pqcd-evo2}), were derived by two groups \cite{Fleming:2013qu,KMQS-hq1}.  

For production of heavy quarkonium, it is necessary to produce a heavy quark pair.  The combination of the QCD factorization formula in Eq.~(\ref{eq:pqcd-fac}) and the evolution equation in Eq.~(\ref{eq:pqcd-evo}) presents a clear picture of how QCD organizes the contributions to the production of heavy quark pairs in terms of distance scale (or time) where (or when) the pair was produced.  The first (the second) term in Eq.~(\ref{eq:pqcd-fac}) describes the production of the heavy quark pairs after (at) the initial hard partonic collision.  The first term in Eq.~(\ref{eq:pqcd-evo}) describes the evolution of a single active parton before the creation of the heavy quark pair, and the power suppressed second term summarizes the leading contribution to the production of a heavy quark pair at any stage during the evolution.  Without the power-suppressed term in Eq.~(\ref{eq:pqcd-evo}), the evolved single-parton fragmentation function is restricted to the situation when the heavy quark pair is only produced {\it 
after} the time corresponding to the input scale of the evolution $\mu_0\gtrsim 2m_Q$.  With perturbatively calculated short-distance hard parts \cite{KMQS-hq2} and evolution kernels \cite{KMQS-hq1,Fleming:2013qu}, the predictive power of the pQCD factorization formalism in Eq.~(\ref{eq:pqcd-fac}) relies on the experimental extraction of the universal fragmentation functions at the input scale $\mu_0$, at which the $\ln(\mu_0^2/(2m_Q)^2)$-type contribution is comparable to $(2m_Q/\mu_0)^2$-type power corrections.  It is these input fragmentation functions at $\mu_0$ that are responsible for the characteristics of producing different heavy quarkonium states,  such as their spin and polarization, since perturbatively calculated short-distance partonic hard parts and evolution kernels of these fragmentation functions are universal for all heavy quarkonium states.

The input fragmentation functions are universal and have a clear scale hierarchy $\mu_0 \gtrsim 2m_Q \gg m_Q v$. It is natural to apply the NRQCD factorization in Eq.~(\ref{eq:nrqcd-fac}) to these input fragmentation functions as \cite{Kang:2011mg,KMQS-hq2}
\begin{eqnarray}
&&
D_{f\to H}(z,m_Q,\mu_0)
=
\sum_n d_{f\to n}(z,m_Q,\mu_0)\langle 0| {\cal O}_n^H |0\rangle
\nonumber\\
&&
{\cal D}_{[Q\bar{Q}(\kappa)]\to H}(z,u,v,m_Q,\mu_0)
\label{eq:nrqcd-fac-d}\\
&& \hskip 0.3in 
= \sum_n d_{Q\bar{Q}(\kappa)]\to n}(z,u,v,m_Q,\mu_0)\langle 0| {\cal O}_n^H | 0 \rangle\, .
\nonumber
\end{eqnarray}
The above NRQCD factorization for single-parton fragmentation functions was verified to NNLO \cite{Nayak:2005rt}, 
and was also found to be valid for heavy-quark pair fragmentation functions at NLO \cite{Swave,Pwave}.
But a proof to all orders in NRQCD is still lacking.  If the factorization in Eq.~(\ref{eq:nrqcd-fac-d}) would be proved to be valid, the pQCD factorization in Eq.~(\ref{eq:pqcd-fac}) is effectively a reorganization of the NRQCD factorization in Eq.~(\ref{eq:nrqcd-fac}) when $p_T\gg m_Q$, which resums the large logarithmic contributions to make the perturbative calculations much more reliable \cite{Kang:2011mg,KMQS-hq2}.  In this case, the experimental extraction of the input fragmentation functions at $\mu_0$ is reduced to the extraction of a few universal NRQCD LDMEs.  

When $p_T\gg m_Q$, the effective theory, NRQCD, does not contain all the relevant degrees of freedom.  In addition to the soft modes absorbed into LDMEs, there are also dangerous collinear modes when $m_Q/p_T \sim 0$.  On the other hand, SCET \cite{Bauer:2000ew,Bauer:2000yr} is an effective field theory coupling soft and collinear degrees of freedom and should be more suited for studying heavy quarkonium production when $p_T\gg m_Q$.  The SCET approach matches QCD onto massive SCET at $\mu\sim p_T$ and expands perturbatively in powers of $\alpha_s(p_T)$ with a power counting parameter $\lambda \sim (2m_Q)/p_T$.  
The approach derives effectively the same factorization formalism for heavy quarkonium production as that in Eq.~(\ref{eq:pqcd-fac}) 
for the first two powers in $\lambda$.  However, the derivation in SCET, due to the way the effective theory was set up, 
does not address the cancellation of Glauber gluon interactions between spectators, and may face further difficulties having to do with infinite 
hierarchies of gluon energy scales, and therefore may be not as complete as in the pQCD approach.  As expected, the new fragmentation functions for a heavy quark pair to fragment into a heavy quarkonium obey the same evolution equations derived in the pQCD collinear factorization approach.  Recently, it was verified that the first order evolution kernels for heavy-quark pair fragmentation functions calculated in both pQCD and SCET approaches are indeed consistent \cite{Fleming:2013qu,KMQS:2013,KMQS-hq1}.  However, it is not clear how to derive the evolution kernels for mixing the single-parton and heavy quark-pair 
fragmentation functions, like $\gamma_{f\to [Q\bar{Q}(\kappa')]}$ in Eq.~(\ref{eq:pqcd-evo}), in SCET \cite{Adam:2013}.

In the SCET approach to heavy quarkonium production, the heavy-quark pair fragmentation functions defined in SCET are matched onto NRQCD after running the fragmentation scale down to the order of $2m_Q$.  It was argued \cite{Fleming:2012wy} that the matching works and NRQCD results can be recovered under the assumption that the LDMEs are universal.  However, the NRQCD factorization in Eq.~(\ref{eq:nrqcd-fac-d}) has not been proved to all orders in pQCD because of the potential for the input fragmentation functions to have light-parton jet(s) of order 
of $m_Q$.  It is encouraging that major progress has been achieved in understanding 
heavy quarkonium production and its factorization recently, but more work is still needed.

\subsubsection{Lattice gauge theory}

\label{sec:subsecC12}

With ensembles at very fine lattice spacings becoming increasingly
available due to the continuous growth of available computer power,
simulations employing relativistic valence charm quarks are now becoming
more and more common. Indeed, the first ensembles incorporating dynamical
(sea) charm quarks
\cite{Bazavov:2010ru,Baron:2010bv,Briceno:2012wt}
are beginning to become available.

The heavy mass of the charm quark means that (since $m_ca\not\ll 1$)
discretization effects cannot be completely neglected and have to be
accounted for properly. This is possible using the Symanzik effective
theory formalism
\cite{Symanzik:1983dc,Symanzik:1983gh}.
For any given lattice action, it is possible to formulate an effective
theory (the Symanzik effective theory) defined in the continuum, which
has the lattice spacing $a$ as its dimensionful expansion parameter and
incorporates all operators compatible with the symmetries of the lattice
action (including Lorentz-violating term with hypercubic symmetry), and
the short-distance coefficients of which are fixed by determining that it
should reproduce the on-shell matrix elements of the lattice theory up to
some given order in $a$. The use of this effective theory in lattice QCD
is twofold
\cite{Kronfeld:2002pi}:
firstly, it provides a means to parameterize the discretization
artifacts as a function of the lattice spacing, thus allowing an extrapolation
to the $a\to 0$ continuum limit from a fit to results obtained at a range
of (sufficiently small) lattice spacings. Secondly, one can take different
lattice actions discretizing the same continuum theory and consider a lattice
action formed from their weighted sum with the weights chosen so as to ensure
that the leading short-distance coefficients of the Symanzik effective action
become zero for the resulting (improved) action. Examples of improved actions
in current use are the Sheikholeslami-Wohlert (clover) action
\cite{Sheikholeslami:1985ij,Wohlert:1987rf},
which removes the O($a$) artifacts of the Wilson quark action, and the
asqtad ($a^2$ tadpole-improved)
\cite{Orginos:1999cr}
and HISQ (Highly Improved Staggered Quark)
\cite{Follana:2006rc}
actions for staggered quarks. Likewise, it is possible to improve the
lattice action for NRQCD
\cite{Lepage:1992tx,Horgan:2009ti}
so as to remove O($a^2$) artifacts. The operators used to measure correlation
functions may be improved in a similar fashion;
cf.~e.g.~\cite{Kurth:2000ki,Grimbach:2008uy,Blossier:2010jk}
for the O($a$) improvement of the static-light axial and vector currents
used in HQET.
Finally, one can use HQET instead of the Symanzik theory to understand the
cutoff effects with heavy quarks
\cite{Kronfeld:2000ck,Harada:2001fi,Harada:2001fj},
which when applied to the Wilson or claver action is known as the Fermilab
method
\cite{ElKhadra:1996mp}.

Since the experimental discovery of the $X(3872)$ resonance by the Belle
collaboration
\cite{Choi:2003ue},
and the subsequent emergence of more and more puzzling charmonium-like
states, the spectroscopy of charmonium has gained increased interest.
Lattice studies of states containing charm quarks are thus of great
importance, as they provide an {\em a priori} approach to charm spectroscopy.
The use of relativistic charm quarks eliminates systematic uncertainties
arising from the use of effective theories, leaving discretization errors
as the leading source of systematic errors, which can in principle be
controlled using improved actions.

A variety of lattice studies with different actions are now available, with
both the HISQ 
\cite{Follana:2006rc}
action
\cite{Donald:2012ga},
and O($a$)-improved Wilson fermions
\cite{Bali:2012ua}
having been used for a fully relativistic treatment of the charm quark.
In addition, anisotropic lattices have been employed to improve the
time resolution of the correlation functions to  allow for better control
of excited states
\cite{Liu:2012ze}.
An important ingredient in all spectroscopy studies is the use of the
variational method
\cite{Michael:1982gb,Luscher:1990ck,Blossier:2009kd}
to resolve excited states.

\begin{figure*}[t]
  \includegraphics*[width=0.8\linewidth]{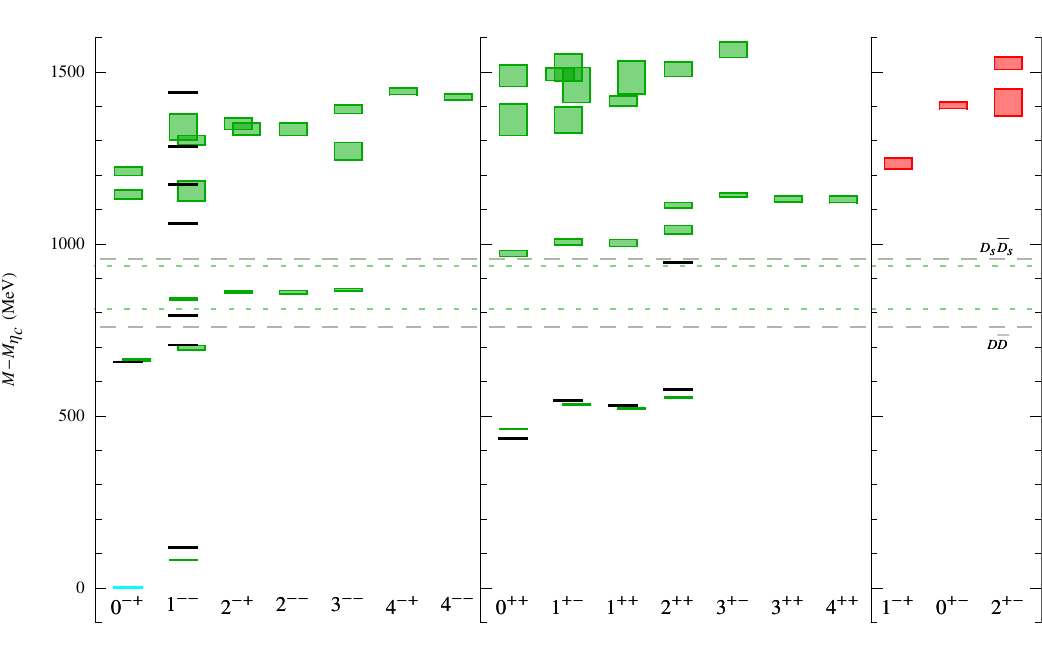}
  \caption{ The charmonium spectrum from lattice simulations of the Hadron
            Spectrum Collaboration using $N_f=2+1$ flavors of dynamical light
            quarks and a relativistic valence charm quark on anisotropic lattices.
            The shaded boxes indicate the $1\sigma$ confidence interval
            from the lattice for the masses relative to the simulated $\eta_c$
            mass, while the corresponding experimental mass differences are
            shown as black lines.
            The $D\overline{D}$ and $D_s\overline{D}_s$ thresholds from lattice
            simulation and experiment are shown as green and grey dashed lines,
            respectively.
            \figPermX{Liu:2012ze}{2012}{SISSA}
            }
\label{fig:charm-spect-lat}
\end{figure*}

As flavor singlets, charmonium states also receive contributions from
quark-disconnected diagrams representing quark--antiquark annihilation
and mixing with glueball and light-quark states
\cite{Levkova:2010ft}.
Using improved stochastic estimators, Bali et al.~%
\cite{Bali:2011rd}
have studied disconnected contributions, finding no resulting energy shift
within the still sizeable statistical errors. The use of the new
``distillation'' method
\cite{Peardon:2009gh,Morningstar:2011ka}
to estimate all-to-all propagators has been found to be helpful in resolving
disconnected diagrams, whose contributions have been found to be small
\cite{Liu:2011rn}.

Besides the mixing with non-$c\bar{c}$ states arising from the disconnected
diagram contributions, quarkonium states above or near the open-charm threshold
may also mix with molecular $D\overline{D}$ and tetraquark states. Studies
incorporating these mixings
\cite{Bali:2009er,Bali:2011rd}
have found evidence for a tightly bound molecular $D\overline{D}^*$ state.
Recently, a study of $DD^*$ scattering on the lattice
\cite{Prelovsek:2013cra}
using L\"uscher's method
\cite{Luscher:1990ux}
found the first evidence of a $X(3872)$ candidate. It was found that the observed
spectrum of states near the threshold depends strongly on the basis of operators
used; in particular, the $X(3872)$ candidate was not observed if only $\bar{c}c$
operators, but no dimeson operators, were included in the basis, nor if the
basis contained only dimeson, but no $\bar{c}c$ operators. This was interpreted
as evidence that the $X(3872)$ might be the consequence of an accidental
interference between $\bar{c}c$ and scattering states. On the other hand, it
could also be seen as rendering the results of this and similar studies doubtful
in so far as it cannot be easily excluded that the inclusion of further operators
might not change the near-threshold spectrum again.
A significant challenge in this area is thus to clarify which operators are
needed to obtain reliable physical results.
Recently, it has been suggested
\cite{Guo:2013nja}
based on large-$N$ arguments that for tetraquark operators the singly
disconnected contraction is of leading order in $1/N$ whenever it contributes.
This would appear to apply also to the tetraquark operators relevant
near the open-charm threshold, making use of all-to-all methods such as
distillation
\cite{Peardon:2009gh,Morningstar:2011ka}
(which was used in \cite{Prelovsek:2013cra})
mandatory for near-threshold studies.

The spectra of the open-charm $D$ and $D_s$ mesons have been studied by
Mohler and Woloshyn
\cite{Mohler:2011ke}
using the Fermilab formalism for the charm quark.
It was found that while the ground state $D$, $D^*$, $D_s$ and $D_s^*$ masses
were reasonably well reproduced, the masses of the $D_J$ and $D_{sJ}$ states
from their simulation strongly disagreed with experiment; possible reasons
include neglected contributions from mixing with multihadron states.

As for $b$ quarks, the currently achievable lattice spacings do not allow the
direct use of relativistic actions. An interesting development in this direction
is the use of highly improved actions (such as HISQ
\cite{Follana:2006rc})
to simulate at a range of quark masses around and above the physical charm quark
mass, but below the physical $b$ quark mass, in order to extrapolate to the
physical $b$ quark mass using Bayesian fits
\cite{Lepage:2001ym}
incorporating the functional form of the expected discretization artifacts and
$1/m_Q$ corrections
\cite{McNeile:2011ng,McNeile:2012qf}.
This method relies on the convergence of the Symanzik expansion up to values of
$m_Qa\sim 1$, and of the heavy-quark expansion in the vicinity of the charm quark
mass; neither assumption can be proven with present methods, but empirical
evidence
\cite{Blossier:2010mk}
suggests that at least for the heavy-quark expansion convergence is much better
than might naively be expected. The removal of as many sources of discretization
errors as possible, including using the $N_f=2+1+1$ HISQ MILC ensembles
\cite{Bazavov:2012xda}
with reduced sea quark discretization effects
\cite{Hart:2008sq}
might be helpful in addressing the question of the convergence of the expansion
in $a$.

Otherwise, simulations of $b$ quarks need to rely on effective field theories,
specifically nonperturbatively matched HQET
\cite{Heitger:2001ch,Sommer:2002en,Heitger:2003nj}
for heavy--light systems, and NRQCD or m(oving)NRQCD
\cite{Lepage:1992tx,Braaten:1996ix,Horgan:2009ti}
for heavy--heavy, as well as heavy--light, systems. An important point to note
in this context is that each discretization choice (such as the use of HYP1
versus HYP2 links
\cite{Hasenfratz:2001hp,DellaMorte:2003mn,DellaMorte:2005yc}
in the static action of HQET, or the use of different values of the stability
parameter in the lattice NRQCD action
\cite{Thacker:1990bm,Gray:2005ur,Hammant:2013sca})
within either approach constitutes a separate theory with its own set of
renormalization constants which must be matched to continuum QCD separately.

The nonperturbative matching of HQET to quenched QCD at order $1/m_b$ has been
accomplished in
\cite{Blossier:2010jk},
and subsequent applications to the spectroscopy
\cite{Blossier:2010vz}
and leptonic decays
\cite{Blossier:2010mk}
of the $B_s$ system have showcased the power of this approach. The extension
to $N_f=2$ is well under way
\cite{Blossier:2011dk,Bernardoni:2012ti},
and future studies at $N_f=2+1$ are to be expected. Beyond the standard
observables such as masses and decay constants, observables featuring in
effective descriptions of strong hadronic interactions, such as the
$B^*B\pi$ coupling in Heavy Meson Chiral Perturbation Theory
\cite{Bulava:2010ej,Bulava:2011yz,Bahr:2012vt}
and the $B^{*'}\to B$ matrix element
\cite{Blossier:2013qma}
have been studied successfully in this approach.

In NRQCD, until recently only tree-level actions were available. In
\cite{Muller:2009af,Dowdall:2011wh},
the one-loop corrections to the coefficients $c_1$, $c_5$ and $c_6$ of the
kinetic terms in an $\mathrm{O}(v^4)$ NRQCD lattice action have been calculated,
and in
\cite{Hammant:2011bt,Hammant:2013sca},
the background field method has been used to calculate also the one-loop
corrections to the coefficients $c_2$ and $c_4$ of the chromomagnetic
$\sigma\cdot{\bf B}$ and chromoelectric Darwin terms for a number of lattice
NRQCD actions. Simulations incorporating these perturbative improvements
\cite{Dowdall:2011wh,Dowdall:2012ab}
have shown a reduced lattice-spacing dependence and improved agreement with
experiment.

Matching the NRQCD action to QCD beyond tree-level has a significant
beneficial effect on lattice determinations of bottomonium spectra,
in particular for the case of the bottomonium $1S$ hyperfine
splitting, which moves from $\Delta M_{\rm HF}(1S) = 61(14)$~MeV without the
perturbative improvements
\cite{Gray:2005ur}
to $\Delta  M_{\rm HF}(1S) =70(9)$~MeV with the perturbative improvements
\cite{Dowdall:2011wh}.
The most recent determination based on lattice NRQCD, including $O(v^6)$
corrections, radiative one-loop corrections to $c_4$, nonperturbative 4-quark
interactions and the effect of $u$, $d$, $s$ and $c$ sea quarks, 
gives  $\Delta M_{\rm HF}(1S) = (62.8 \pm 6.7)\,\mev$
\cite{Dowdall:2013jqa}, 
which is to be compared to the PDG value of
$\Delta  M_{\rm HF}(1S) = 69.3(2.9)$~MeV
\cite{Beringer:1900zz}
excluding the most recent Belle data
\cite{Collaboration:2011chb},
or $\Delta  M_{\rm HF}(1S) = 64.5(3.0)$~MeV 
\cite{Beringer:1900zz}
when including them.

The resulting prediction for the bottomonium 2S hyperfine splitting
of $\Delta M_{\rm HF}(2S) = 35(3)(1)$~MeV
\cite{Dowdall:2011wh}
is in reasonable agreement with the Belle result
$\Delta M_{\rm HF}(2S) =24.3^{+4.0}_{-4.5}$~MeV
\cite{Mizuk:2012pb},
but disagrees with the CLEO result of Dobbs et al., 
$\Delta M_{\rm HF}(2S) = 48.7(2.3)(2.1)$~MeV
\cite{Dobbs:2012zn},
%
%
 see discussion in Sec.~\ref{sec:secC2}.

Another factor with a potentially significant influence on the bottomonium
hyperfine splitting is the lack of, or the inclusion of, spin-dependent
interactions at higher orders in the nonrelativistic expansion. In
\cite{Meinel:2010pv},
it was shown that including the $\mathrm{O}(v^6)$ spin-dependent terms in
the NRQCD action leads to an increase in the 1S hyperfine splitting, moving it
away from the experimental value. The results of
\cite{Hammant:2013sca}
suggest that this effect will at least partially be compensated by the inclusion
of perturbative corrections to the coefficients of the spin-dependent operators.
The $2S$ hyperfine splitting is not similarly affected, and the prediction
$\Delta M_{\rm HF}(2S) = 23.5(4.1)(2.1)(0.8)$~MeV of
\cite{Meinel:2010pv}
is in excellent agreement with the Belle value
\cite{Mizuk:2012pb}.

The $B_c$ system combines the challenges of both the $b$ and charm sectors,
while also allowing for one of the relatively few predictions from QCD that
is \emph{not} to some extent a ``postdiction'' in that it precedes experiment,
{\em viz.} the mass of the as yet undiscovered $B_c^*$ meson, which has been
predicted by the HPQCD collaboration to be $M_{B_c^*}=6330(7)(2)(6)$~MeV
\cite{Gregory:2009hq}
using NRQCD for the $b$ and the HISQ action for the charm quarks. Reproducing
this prediction using another combination of lattice actions might be worthwhile.
For the time being, we note that the lattice prediction compares very well 
with the perturbative calculation of \cite{Penin:2004xi}, which 
gives $M_{B_c^*}=6327(17)^{+15}_{-12}(6)$~MeV to next-to-leading logarithmic accuracy.

\subsection{Heavy semileptonic decays}

\label{sec:secC3}

Semileptonic decays of $B$ and $D$ mesons have been extensively studied in the last years.
They provide information about the CKM matrix elements $ |V_{cb}| $,
$|V_{ub}|$, $|V_{cd}|$ and $|V_{cs}|$ through exclusive and inclusive processes driven by $ b \rightarrow c(u)$ and
$c \rightarrow  s(d)$  decays, respectively (for recent reviews see, e.g., 
Refs.~\cite{Ricciardi:2012pf,Ricciardi:2012dj,Ricciardi:2013cda, Ricciardi:2013xaa,Ricciardi:2014iga}).

The leptonic decays $ B^+ \rightarrow l^+ \nu $ and $ D^+_{(s)}
\rightarrow l^+ \nu $ can also be used for the determination of CKM
matrix elements. The advantages of semileptonic decays are that they
are not helicity suppressed and new physics is not expected to play a
relevant role; so, it is generally, but not always, disregarded.

In  deep inelastic neutrino (or antineutrino)--nucleon scattering,  single charm particles  can be
produced through $dc$ and $sc$ electroweak currents.
Analyses based on neutrino and antineutrino
interactions  give a determination of $|V_{cd}|$ with comparable, and often better,  precision  than the ones obtained
from semileptonic charm decays. Not so for
the determination of $|V_{cs}|$,  which  suffers from the
uncertainty of the s-quark sea content \cite{Beringer:1900zz}. On-shell $W^\pm$
decays  sensitive to $|V_{cs}|$ have also been used \cite{Abreu:1998ap}, but
semileptonic $D$ or leptonic $D_s$ decays provide direct and more precise determinations.

\subsubsection{Exclusive and inclusive $D$ decays}
\label{sect1}

The hadronic matrix element for a generic semileptonic decay $H \to P l \nu$, where $H$ and $P$
denote a heavy and a light pseudoscalar meson respectively, is usually written in terms of two
form factors $f_+(q^2)$ and $f_0(q^2)$
\begin{eqnarray}
\langle P(p_P)| J^\mu | H(p_H) \rangle  \!\! &=&  \!\! f_+(q^2) \left( p_H^\mu+p_P^\mu - \frac{m^2_H-m_P^2}{q^2} q^\mu \right) \nonumber \\  
&&\!\! + f_0(q^2) \frac{m^2_H-m_P^2}{q^2} q^\mu,
\end{eqnarray}
where $ q \equiv p_H - p_P$ is the momentum transferred to the lepton pair, and $J^\mu$ denotes
the heavy-to-light vector current.
In the case of massless leptons, the form factor
$f_0(q^2)$  is absent and the differential decay rate depends on $f_+(q^2) $ only.
The main theoretical challenge is  the nonperturbative evaluation of the form factors.
In this section, we consider $H$ to be a $D_{(q)}$ meson.
 For  simplicity's sake, one can split the nonperturbative evaluation of the form factors into two steps, the  evaluation of their normalization at $q^2=0$ and the determination of their $q^2$ dependence.

 The form factors are expected to decrease at low values of $q^2$, that is at high values of spectator quark recoil.  Indeed,  in the leading  spectator diagram,  the probability of forming a hadron in the final state decreases as the recoil momentum of the spectator quark increases. Moreover, the form factors are expected to be analytic functions everywhere in
the complex $q^2$  plane  outside a cut extending along the positive $q^2$ axis from the mass
of the lowest-lying $c \bar q$ resonance. That implies they can be described by
 dispersion relations, whose exact form is not known a priori, but can be  reasonably assumed  to be  dominated, at high $q^2$,  by the  nearest poles to $q^2_{\mathrm{max}}= (m_{D_{(q)}}-m_P)^2$.  Pole dominance implies current conservation at large $q^2$.
 We expect the form factors to have a singular behavior as $q^2$ approaches the lowest lying poles, without reaching them,  since they are
beyond the kinematic cutoff.
The simplest parametrization of the $q^2$ dependence motivated by this behavior is the simple pole model,
where a single pole dominance is assumed. By restricting to  the form factor $f_+(q^2)$,  we have
\beq
f_+(q^2)= \frac{f_+(0)}{1-\frac{q^2}{m_{\mathrm{pole}}^2}}.
\label{singlepole1}
\eeq
In $D \rightarrow \pi  l \nu $  decays, the pole  for $f_+(q^2)$ corresponds  to the  $c \bar d$ vector meson of lowest mass $D^\star$.
In $D \rightarrow K l \nu $ and   $D_s \rightarrow \eta^{(\prime)} l \nu $ decays, the poles  correspond  to the  $c \bar s$ vector mesons and  the lowest  resonance compatible with $J^P=1^-$  is $D_s^{\ast \pm}$, with mass $M_ {D_s^{\ast}}= 2112.3 \pm 0.5$ MeV. Form factor fits have been performed for $D \rightarrow K(\pi) l \nu$   by the CLEO \cite{Besson:2009uv} and   BESIII Collaborations \cite{Liu:2012bn}, where several models for the $q^2$ shape have been considered.
In the simple pole model,  agreement  with data is only reached when the value of   $m_{\mathrm{pole}}$
is not  fixed  at the  $D^\star_{(s)}$ mass, but  is  a free parameter.
In order to take into account higher poles, while keeping the number of free parameters low, a modified pole model has been proposed \cite{Becirevic:1999kt}, where
\beq
f_+(q^2)= \frac{f_+(0)}{ \left(1-\frac{q^2}{m_{\mathrm{pole}}^2}\right)  \left(1- \alpha\frac{q^2}{m_{\mathrm{pole}}^2}\right)}.
\label{doublepole1}
\eeq
Another  parametrization, known as the series or $z$-expansion, is based on a  transformation that maps the cut in the $q^2$ plane onto a
unit circle in another variable, $z$, and fits the form factor as a power series (in $z$) with improved properties of convergence \cite{Becher:2005bg, Arnesen:2005ez, Bourrely:2008za}.
More in detail,  the first step is to remove poles  by the form factors, that is, for the $B \to K$ decays
\begin{eqnarray}
\tilde f_0^{D \to K } (q^2) &=&  \left ( 1 - \frac{q^2}{ m_{D^\ast_{s 0}}^2 }\right)  f_0^{D \to K } (q^2) \nonumber \\
\tilde f_+^{D \to K } (q^2) &=&  \left ( 1 - \frac{q^2}{ m_{D^\ast_{s}}^2 }\right)  f_+^{D \to K } (q^2)
\end{eqnarray}
The  variable $z$ is defined as
\beq
z(q^2) = \frac{ \sqrt{t_+ - q^2}- \sqrt{t_+ - t_0} } { \sqrt{t_+ - q^2} +  \sqrt{t_+ - t_0}} \qquad t_+ = (m_D+m_K)^2
\eeq
The final step consists in fitting $\tilde f$ as a power series in $z$,
\begin{eqnarray}
\tilde f_0^{D \to K } (q^2)  &=& \sum_{n \ge 0} c_n z^n \nonumber \\
 \tilde f_+^{D \to K } (q^2) &=&  \sum_{n \ge 0} b_n z^n \qquad c_0=b_0
\end{eqnarray}
Employing this parametrization, the shapes of $ f_{0,+}^{D \to K }$ form factors have been very recently estimated by the  HPQCD
 Collaboration \cite{Koponen:2013tua}.

To evaluate the  normalization of the form factors, lattice and QCD sum rules are generally employed.
Lately, high statistics studies on the lattice have become available and
preliminary results for  $ f_{0,+}^{D \to K/\pi }$  have been presented  by
ETMC \cite{DiVita:2011py, Riggio},  HPQCD  \cite{Koponen:2012di} and  Fermilab/MILC \cite{Bailey:2012sa}.

The most recent  published $|V_{cd}|$ estimates are from HPQCD \cite{Na:2012iu}, where
the value of  $|V_{cd}|$   has been evaluated using the Highly Improved Staggered
Quark (HISQ) action for valence charm and light quarks on MILC $N_f=2+1$
 lattices
with experimental
inputs from CLEO  \cite{Eisenstein:2008aa}
and  BESIII \cite{BESIII}.
The value $ |V_{cd}| = 0.223 \pm 0.010_{\mathrm{exp}} \pm  0.004_{\mathrm{lat}}$ \cite{Na:2012iu}, 
with the first error coming from experiment and the second from the lattice computation, is in agreement with the value of $|V_{cd}|$ the same collaboration  has recently 
extracted from leptonic decays. It also agrees, with a competitive error,  with the value  $ |V_{cd}| = 0.230 \pm 0.011$ \cite{Beringer:1900zz} from neutrino scattering.

The same HPQCD collaboration gives also the most recent $|V_{cs}| $ estimate
 by analyzing $D \to K/\pi \,  l \, \nu $,  $D_s \rightarrow \phi/\eta_s \, l \, \nu $  and using experimental  inputs from    CLEO  \cite{Besson:2009uv},   
BaBar \cite{Aubert:2007wg, Aubert:2008rs}, Belle \cite{Widhalm:2006wz}.
and BESIII (preliminary) \cite{Liu:2012bn}.
Their   best value $ |V_{cs}| = 0.963 \pm 0.005_{\mathrm{exp}} \pm  0.014_{\mathrm{lat}}$  is in agreement with
values from indirect fits \cite{Beringer:1900zz}.
The big increase in accuracy with respect to their older determinations,
is due to the larger amount of data employed. Specifically they have used all experimental
$q^2$ bins, rather than just the $q^2 \to 0$ limit or the total rate.
The FLAG $N_f=2+1$ average value  from semileptonic decays gives
$|V_{cs}|= 0.9746 \pm 0.0248 \pm 0.0067$ \cite{Sachrajda:EPS}.
Experiments at BESIII, together with experiments at present and future flavor factories, all have the potential to reduce the
errors on the measured decay branching fractions of $D^+_{(s)}$ and $D^0$ leptonic and semileptonic
decays,  in order to allow  more precise comparison of these CKM matrix elements. In particular, BESIII is actively working
 on semileptonic charm decays; new preliminary results on the branching fractions and form factors in the parameterizations  mentioned above, for the $D \rightarrow K/\pi e \nu$ channels, have  been recently reported  \cite{Dong:EPS}.

Lattice determinations of the decay constant $f_{D_s}$ governing the leptonic
decays $D_s^+\to\mu^+\nu$ and $D_s^+\to\tau^+\nu$ have for several years
exhibited the ``$f_{D_s}$ puzzle'',
an apparent $(3-4)\sigma$ discrepancy between lattice determinations of $f_{D_s}$
\cite{Aubin:2005ar,Follana:2007uv,vonHippel:2008pc,Blossier:2009bx}
and the value of $f_{D_s}$ inferred from experimental measurements of
the branching ratios $B(D_s^+\to\mu^+\nu)$ and $B(D_s^+\to\tau^+\nu)$
\cite{Aubert:2006sd,Pedlar:2007za,Artuso:2007zg,Widhalm:2007ws,Ecklund:2007aa}.
When this discrepancy first appeared, it was immediately discussed as a signal
for new physics \cite{Dobrescu:2008er};
in the meantime, however, careful investigation of all sources of systematic
error, combined with increased statistics, has led to the lattice values
shifting up slightly \cite{Davies:2010ip,Bazavov:2011aa,Dimopoulos:2011gx}
and the experimental values shifting down noticeably \cite{Alexander:2009ux,Onyisi:2009th,Naik:2009tk,Zupanc:2012cd},
thus more or less eliminating the ``puzzle'' \cite{Kronfeld:2009cf}.
However, the most recent determinations still show some tension versus
the FLAG $N_f=2+1$ average value from leptonic decays $|V_{cs}|= 1.018 \pm
0.011 \pm 0.021$ \cite{Sachrajda:EPS}.

It is interesting to observe that, according to  lattice determinations in  \cite{Koponen:2012di},  the form factors
are insensitive to the spectator quark: The $D_s  \rightarrow \eta_s l \nu $ and $D \rightarrow K l \nu $ form
factors are  equal  within 3\%, and the same holds for   $D_s  \rightarrow K l \nu $ and $D \rightarrow \pi l \nu $
 within 5\%.
This result, which  can be tested experimentally,  is expected  by heavy quark symmetry to hold
also for $B$ meson decays so that the $B_s \rightarrow D_s$ and $B \rightarrow  D$ form factors would be equal.

QCD light-cone sum rules have also been employed to extract $|V_{cs}|$ and $|V_{cd}|$ \cite{Khodjamirian:2009ys}, giving  substantial  agreement on the averages and
 higher theoretical error with respect to the previously-quoted  lattice  results.
 By using the same data and a revised version of QCD sum rules, errors on $|V_{cd}|$  have been  reduced, but
 a higher average value has been obtained:
$|V_{cd}|= 0.244 \pm 0.005 \pm 0.003 \pm 0.008 $.  The first and second errors are of an experimental origin and the third is due to the
theoretical uncertainty \cite{Li:2012gr}.

Form factors for semileptonic transitions to a vector or a pseudoscalar meson have also been investigated within a
model which combines heavy quark symmetry and properties of the chiral Lagrangian \cite{Fajfer:2004mv, Fajfer:2005ug, Fajfer:2006uy}.

Exclusive semileptonic $D$ decays play also  a role in better understanding the composition   of the $\eta$ and $\eta^{\prime}$ wave functions,  a long-standing problem.
The transitions $D_s^+ \rightarrow \eta^{(\prime )} l^+ \nu$ and  $D^+\rightarrow \eta^{(\prime )} l^+ \nu$  are driven by
weak interactions at the Cabibbo-allowed and Cabibbo-suppressed levels, and provide us with  complementary information since
they produce the $\eta^{(\prime )}$ via their $s \bar s$ and $d \bar d$ components, respectively.
In addition, $\eta^{(\prime )}$ could be excited via a $gg$  component.
That is important since it would validate, for the first time, an independent role of gluons
 in hadronic spectroscopy, outside their traditional domain of mediating strong interactions.
Also $B$ decays, semileptonic or hadronic, have  been similarly  employed (see e.g.,
Refs.~\cite{Ricciardi:2012pf, DiDonato:2011kr, Fleischer:2011ib}). 
Experimental evidence of glueballs are searched 
for in a variety of processes at  several experiments, e.g.,  BESIII and PANDA. In 2009  the first absolute
measurement of  ${\cal{B}} ( D_s^+ \rightarrow \eta^{(\prime )} e^+ \nu_e)$ \cite{Yelton:2009aa} and
 the first observation of the $ D^+ \rightarrow \eta \, e^+ \nu_e$ decay \cite{Mitchell:2008kb} were reported by CLEO.
Improved  branching
fraction measurements, together with the first observation of the decay mode $ D^+ \rightarrow \eta^\prime e^+ \nu_e  $  and
the first form factor determination for $ D^+ \rightarrow \eta \,  e^+ \nu_e$,   followed in 2011 \cite{Yelton:2010js}.
On the theoretical side,
recent lattice results have  become available for the values of mixing angles \cite{Christ:2010dd, Dudek:2011tt},
 quoting values of the mixing angle $\phi$ between
$  40^\circ$ and $50^\circ$.
The latest  analysis, by ETM, leads to a value of $\phi  = (44 \pm 5)^\circ$ \cite{Ottnad:2012fv},
with a statistical error only.  Systematic uncertainties, difficult to estimate on the lattice, are likely to affect this result.
Preliminary  results by the QCDSF Collaboration \cite{Bali:2011yx, Kanamori:2013rha}  give  a mixing angle $\theta \sim -(7^\circ, 8^\circ)$ in the octet-singlet basis, that is, in the quark-flavor basis,  $\phi = \theta+\arctan \sqrt{2} \sim 47^\circ $.
 Out of chorus is the lower value favored by  the recent UKQCD staggered investigation \cite{Gregory:2011sg},
 $\phi  = (34 \pm 3)^\circ$.
All lattice analyses do not include a gluonic operator, discussing only the
relative quark content.
The agreement with other determinations from semileptonic decays based on different phenomenological approaches and older data   is remarkable (see, e.g., Refs.~\cite{Feldmann:1998vh, Gronau:2010if, Anisovich:1997dz, DiDonato:2011kr}).
Recent experimental and theoretical progress has increased the role of semileptonic $D$ decays with respect to traditional, low energy analyses \cite{Ricciardi:2012xu}.

 In the vector sector, the  $\phi$--$\omega$ mixing is not expected as large as in the pseudoscalar one,   because there is no additional mixing induced  by the axial $U(1)$ anomaly.
In the absence of mixing,   the state $\omega $ has no strange valence quark and  corresponds to  $|u \bar u + d \bar d\rangle/\sqrt 2 $.
Cabibbo-favored semileptonic decays of $D_s$ are expected to lead to final states that
can couple to $|\bar s s \rangle$, in the quark flavor basis. The  decay $ D^+_s \rightarrow \omega e^+ \nu_e $
 occurs through $\phi$--$\omega$ mixing and/or  Weak Annihilation (WA) diagrams, where  the lepton pair couples weakly to the $c \bar s$ vertex.
Experimentally,
only an upper limit is available
on  the  branching fraction  ${\cal{B}}( D^+_s \rightarrow \omega e^+ \nu_e) <0.20\%$,
 at 90\% C.L. \cite{Martin:2011rd}.

Exclusive semileptonic $D$ decays also offer the chance to explore possible exotic states.
An interesting channel is the  $D^+_s \rightarrow f_0(980) \, l^+ \nu $ decay.
The nontrivial nature of the experimentally  well-established $f_0(980)$ state has been discussed
for decades and there are still different interpretations, from the conventional quark-antiquark picture,  to a multiquark  or molecular
bound state.  The channels $D_{(s)}^+ \rightarrow f_0(980) \, l^+ \nu $
  can be used as a probe of  the hadronic structure  of the light scalar resonance; more recent experimental investigation has been made available by CLEO
\cite{Ecklund:2009aa}.
A further handle is given by the possibility  to correlate observables related to the charm  semileptonic
branching ratios with theoretical and experimental analyses of the  hadronic $B_s \rightarrow   J/\psi f_0$ decay
\cite{Stone:2008ak, Ecklund:2009aa,  Fleischer:2011au}.

The most recent  experimental results on inclusive $D^0$ and $D_{(s)}^+$  semileptonic branching fractions  have been derived using the complete CLEO-c data sets \cite{Asner:2009pu}.
Besides being important in their own right, these measurements, due to similarities between the $D$ and $B$ sectors,
can be helpful to  improve understanding of $B$ semileptonic decays, with the hope
 to reduce the theoretical uncertainty  in the determination of the still-debated weak
mixing parameter $|V_{ub}|$. In \cite{Asner:2009pu},
knowledge about exclusive semileptonic modes and form factor models is used to
extrapolate the spectra below the 200 MeV momentum cutoff. The ratios of the
semileptonic decay widths are determined to be
$\Gamma_{D^+}^{\mathrm{SL}}/\Gamma_{D^0}^{\mathrm{SL}}
= 0.985 \pm  0.015 \pm  0.024 $
and $\Gamma_{D^+_s}^{\mathrm{SL}}/\Gamma_{D^0_s}^{\mathrm{SL}}
= 0.828 \pm  0.051 \pm  0.025 $.
The former agrees with isospin symmetry, while
the latter ratio shows an indication of difference.
Significant improvements of the branching ratio measurement ${\cal  B} (D  \rightarrow X \mu^+ \nu_\mu)$   can be expected at BESIII,
because of advantages provided by the  capabilities of the BESIII $\mu$ detection
system \cite{Asner:2008nq}.
The  $D^{0,\pm}$ and $D_s$ inclusive decays are differently affected by the WA diagrams, since they are Cabibbo-suppressed
in the $D^\pm$ case, Cabibbo-favored in  $D_s$ decays,  and completely absent in $D^0$ decays.
The  semileptonic  decays of $D$ and $D_s$ can be helpful in constraining the WA matrix elements that enter the $B \rightarrow X_u \, l \bar \nu$  decay, via heavy quark symmetry.
By comparison of measured total
semileptonic rates or moments in these channels, we can hope to  extract information on the WA contributions.
The ``theoretical background'' to take into account  is the fact  that such contributions compete with
additional ones arising from $SU(3)$ breaking in the matrix elements, and/or from weak annihilation.
 However,
no relevance or clear evidence of  WA effects has  been found considering
 the semileptonic widths  \cite{Ligeti:2010vd} or  the widths  and the lepton--energy moments \cite{Gambino:2010jz}.

\subsubsection{Exclusive $B$ decays}

Most  theoretical approaches exploit
the fact that the mass $m_b$ of the $b$ quark is large compared to
the QCD scale  that determines low-energy hadronic physics in order to build  differential ratios.  Neglecting  the charged lepton and neutrino masses,  we can recast
the differential ratios as
\begin{eqnarray}
\frac{d\Gamma}{d \omega} (\bar{B}\rightarrow D\,l \bar{\nu})  &=&  \frac{G_F^2}{48 \pi^3}\, K_1\,
(\omega^2-1)^{\frac{3}{2}}\,  |V_{cb}|^2 {\cal G}^2(\omega)\nonumber \\
\qquad\frac{d\Gamma}{d \omega}(\bar{B}\rightarrow D^\ast\,l \bar{\nu})
&=&  \frac{G_F^2}{48 \pi^3} K_2   (\omega^2-1)^{\frac{1}{2}} |V_{cb}|^2  {\cal F}^2(\omega)\end{eqnarray}
where $K_1=   (m_B+m_D)^2  m_D^3 $,  $K_2 =  (m_B-m_{D^\ast})^2 m_{D^\ast}^3 \chi (\omega) $ and  $\chi (\omega)$ is a known phase space.
 The semileptonic  decays
$ \bar{B}\rightarrow D \, l \, \bar{\nu}$ and $ \bar{B}\rightarrow D^\ast \, l \, \bar{\nu}$
 depend on the form factors ${\cal G}(\omega)$ and ${\cal F}(\omega)$, respectively, where
$\omega $ is  the product of the heavy quark velocities  $v_B= p_B/m_B$ and $v_{D^{(\ast)}}= p_{D^{(\ast)}}/m_{D^{(\ast)}}$.
The form factors,
in the heavy-quark limit, are both  normalized to unity at the zero recoil point $\omega=1$. Corrections to this limit have been calculated
in the lattice unquenched approximation, giving  $ {\cal G}(1) = 1.074 \pm 0.024 $ \cite{Okamoto:2004xg} and
$ {\cal F}(1) =0.906 \pm 0.004 \pm 0.012 \label{F11}  \label{VcbexpF2} $ \cite{Bailey:2014tva},
including  the enhancement factor 1.007, due to the electroweak corrections to the four-fermion operator mediating
the semileptonic decay.

The lattice calculations have been compared with non-lattice ones (see, e.g., Ref. \cite{UraltsevProc}).
By combining  the heavy-quark expansion with a  ''BPS" expansion \cite{Uraltsev:2003ye}, in which $\mu_\pi^2=\mu^2_G$, the following value is quoted
$ {\cal G}(1) =1.04 \pm  0.02 $. Recently, the value $ {\cal F}(1) = 0.86 \pm 0.02 \label{gmu} $ \cite{Gambino:2010bp, Gambino:2012rd}
has  been  calculated, using zero recoil sum rules, including full $\alpha_s$ and estimated effects up to $1/m_Q^5$.

Since the zero recoil point is not accessible experimentally, due to the kinematical suppression of the differential
decay rates, the $V_{cb}$ estimates rely on the
extrapolation from $\omega \neq 0$ to the zero recoil point.
In Table \ref{phidectab2} we list the results of the $V_{cb}$ determinations obtained from the comparison of the previous form factors at zero recoil with experimental data.
The errors are experimental and theoretical, respectively.
The first three averages are taken by HFAG \cite{Amhis:2012bh}, the fourth one  by PDG \cite{Beringer:1900zz}. The slightly smaller values for the form factors in non-lattice determinations imply slightly higher values of $V_{cb}$.
 In the last line, we quote the result due to
an alternative lattice determination,  currently available only in the quenched approximation, which  consists of calculating  the form factor normalization directly  at values $\omega >1$,
 avoiding the large extrapolation to $\omega=1$ and thus reducing the model dependence \cite{deDivitiis:2007ui}.
This approach, by using 2009  BaBar  data \cite{Aubert:2009ac}, gives a slightly higher value than the unquenched  lattice result.
The errors are  statistical, systematic and due to the theoretical uncertainty in the form factor $ {\cal G}$, respectively.
Calculations of form factors at non-zero recoil have been recently completed
for $B \to D$ semileptonic decays, giving the value
$|V_{cb}|=(38.5 \pm 1.9_{\rm exp+lat} \pm 0.2_{\rm QED}) \times 10^{-3}$ \cite{Qiu:2013ofa}.

\begin{table}[t]
\centering
\vskip 0.1 in
\caption{Comparison of some exclusive  determinations of $|V_{cb}|$.}
\label{phidectab2}
\begin{tabular}{l c} \hline \hline
{\it \bf Theory}  &   $|V_{cb}| \times 10^{3}$ \\
\hline
{\centering{$ \bar{B}\rightarrow D^\ast \, l \, \bar{\nu}$  }} \\
HFAG (Lattice) \cite{Amhis:2012bh, Bernard:2008dn, Bailey:2014tva}   & $ 39.04 \pm 0.49_{\mathrm{exp}} \pm 0.53_{\mathrm{QCD}}$ \\
& \qquad\qquad\qquad\quad $\pm 0.19_{\mathrm{QED}}$\\
HFAG (SR) \cite{Amhis:2012bh, Gambino:2010bp, Gambino:2012rd} & $   41.6\pm 0.6_{\mathrm{exp}}\pm 1.9_{\mathrm{th}} $ \\
$ \bar{B}\rightarrow D \, l \, \bar{\nu}$ &   \\
HFAG (Lattice)  \cite{Amhis:2012bh,Okamoto:2004xg}  & $39.70 \pm 1.42_{\mathrm{exp}} \pm 0.89_{\mathrm{th}} $\\
PDG (BPS)   \cite{Beringer:1900zz, Uraltsev:2003ye} & $ 40.7 \pm 1.5_{\mathrm{exp}} \pm 0.8_{\mathrm{th}} $\\
BaBar (Lattice $\omega \neq 1$) \cite{Aubert:2009ac, deDivitiis:2007ui} & $  41.6 \pm 1.8 \pm 1.4
\pm 0.7_{FF}  \label{lattunq} $\\
\hline\hline
\end{tabular}
\end{table}

Until a few years ago,  only exclusive decays where the final lepton was an electron or a muon had been observed,
since  decays into a $\tau$ lepton are suppressed because of the large $\tau$ mass. Moreover,
these modes are very difficult to measure because of the multiple neutrinos in the final state, the low lepton momenta,
and the large associated background contamination.
Results of semileptonic decays with a $\tau$ in the final state  were limited to inclusive and
semi-inclusive  measurements in LEP experiments.
The first observation of an  exclusive semileptonic $B$ decay was reported  by the Belle Collaboration in 2007. They measured the branching fraction
$ {\cal{B}} (\bar B^0 \rightarrow   D^{\ast +} \tau^-  \bar \nu_\tau)$ \cite{Matyja:2007kt}.
Recently the BaBar Collaboration has published
results of their measurements of $B \to D^{(\ast)} \tau \nu$  branching fractions normalized to the
corresponding $B \to D^{(\ast)} l \nu$ modes (with $l=e , \mu$) by using  the full BaBar data sample \cite{Lees:2012xj}.
Their results are in agreement with measurements by
 Belle using $657  \times 10^6$  $B \bar B$ events \cite{Adachi:2009qg}, and  indicate  an enhancement  of order $(2 \sim 3) \sigma$  above theoretical results within the SM.
It will be interesting to compare   with the
final Belle results on these modes using  the full data
sample of  $772 \times 10^6$  $B \bar B$ pairs together with improved hadronic tagging.
Indeed, a  similar  deviation from the SM has been previously observed also in leptonic decays $B^- \rightarrow \tau^- \bar \nu_\tau$, but
 Belle finds now
a much lower value, in agreement with the SM,  by using the full data set
 of    $B \bar B$ events \cite{Adachi:2012mm}.
 By using Belle data and the FLAG $N_f=2+1$ determination of  $f_B$, one obtains
 the value $|V_{ub}| = (3.35 \pm 0.65 \pm 0.07) \times 10^{-3}$ \cite{Sachrajda:EPS}. The accuracy is not sufficient to make this channel competitive for $
|V_{ub}|$ extraction, but
the intriguing experimental situation has led to  a reconsideration of SM predictions as well as exploring the possibility of new physics contributions,  traditionally not expected in processes driven by the tree level  semileptonic $b$ decay. (For more details see, e.g., Refs.~\cite{Ricciardi:2013jf, Ricciardi:2012dj}.)

The analysis of exclusive charmless semileptonic decays, in particular the  $\bar B \rightarrow \pi l \bar \nu_l$ decay, 
is currently employed to determine the CKM parameter $|V_{ub}|$, which plays a crucial role in the study  of
the unitarity constraints. Also here,  information about hadronic matrix elements is required via form factors.
Recent $|V_{ub}|$ determinations  have been reported by the BaBar collaboration, see Table VII of Ref.  \cite{Lees:2012vv}
(see also \cite{Ricciardi:2012pf}),  all in agreement with each other and with
 the value
$
|V_{ub}| = (3.25 \pm 0.31) \times 10^{-3} \label{exclus} $, determined
from the simultaneous fit to the  experimental data
and the lattice theoretical predictions  \cite{Lees:2012vv}.
They are also in agreement with the Belle results
for $
|V_{ub}| = (3.43 \pm 0.33) \times 10^{-3}  $ extracted from the  $\bar B \rightarrow \pi l \bar \nu_l$ decay channel  \cite{Ha:2010rf} and for $
|V_{ub}| $ from the  $\bar B \rightarrow \rho l \bar \nu_l$ decay channel,  with precision
of twice as good as than the world average \cite{Sibidanov:HEP}.

Finally, we just mention that exclusive $B_s$ decays are attracting a lot of attention due to the avalanche of recent data
and to the expectation of new data. $B_s$ physics has been, and is, the domain of Tevatron and LHCb, but
also present and future $e^+ e^-$ colliders can give their contribution, since  the $\Upsilon \mathrm{(5S)}$ decays in about 20\% of the cases to $B_s^{(\star)}$ meson-antimeson pairs.
The measurement of the semileptonic asymmetry and its analysis are particularly interesting, since  CP violation is   expected to be tiny  in the SM and any  significant enhancement  would be evidence for NP (see also \cite{Ricciardi:2012pf, Ricciardi:2012qm}).

\subsubsection{Inclusive $B$  decays}

In most of the phase space for  inclusive $ B \rightarrow X_q  l  \nu$ decays,  long and short distance dynamics are factorized by means of  the heavy
quark expansion.
However, the  phase space region includes a  region of singularity, also called endpoint or threshold region, plagued by the presence
 of large double (Sudakov-like)  perturbative  logarithms at all orders in the strong coupling \footnote{For theoretical aspects of threshold
resummation in $B$ decays see, e.g., Refs.~\cite{Aglietti:2002ew, Aglietti:2005mb, Aglietti:2005bm, Aglietti:2005eq, Aglietti:2007bp, DiGiustino:2011jn}.}.
For  $b \rightarrow c$ semileptonic decays, the effect of the small region of singularity is not very important; in addition,  corrections are not expected  as singular as in the $ b \rightarrow u$ case, being  cut off by the charm mass.

Recently,  a global fit \cite{Amhis:2012bh} has been  performed to the width and all
available measurements of moments in $ B \rightarrow X_c  l  \nu$ decays, yielding, in the kinetic scheme
  $|V_{cb}| = (41.88 \pm 0.73) \times 10^{-3}$ and in the 1S scheme   $|V_{cb}| = (41.96 \pm 0.45) \times 10^{-3}$.
Each scheme has its own nonperturbative parameters that have been estimated together with the charm and
bottom masses. The inclusive
averages are in good agreement with the values extracted from exclusive decays in Table \ref{phidectab2}, within the errors.

 In principle,  the method of extraction of $|V_{ub}|$  from inclusive $ \bar  B \rightarrow X_u  l \bar \nu_l$  decays  follows in the footsteps of the $|V_{cb}|$ determination from $ \bar  B \rightarrow X_c  l \bar \nu_l$, but the copious background from the
$ \bar B \rightarrow X_c l \bar \nu_l$  process, which has a rate about 50 times higher, limits
the experimental sensitivity to restricted regions of phase space,
where the  background  is  kinematically suppressed. The relative weight of  the threshold  region, where the previous approach fails, increases  and new
theoretical issues need to be addressed.
 Latest results by Belle \cite{Urquijo:2009tp} and BaBar   \cite{Lees:2011fv}
 access about $ 90$\% of the $ \bar B \rightarrow X_u  l \bar \nu_l$ phase space.
On the theoretical side, several approaches have been devised
to analyze data in the threshold region,  with differences
in treatment of perturbative corrections and the
parameterization of nonperturbative effects.
\begin{table}[t]
\centering
\vskip 0.1 in
\caption{ Comparison of  inclusive  determinations of $|V_{ub}|$ \cite{Amhis:2012bh}. }
\label{phidectab3}
\begin{tabular}{l c} \hline \hline
{\it \bf Theory}  &   $|V_{ub}| \times 10^{3}$ \\
\hline
BLNP  & $ 4.40 \pm 0.15^{+0.19}_{-0.21}  $\\
DGE   & $4.45 \pm 0.15^{+ 0.15}_{- 0.16}$\\
ADFR   & $4.03 \pm 0.13^{+ 0.18}_{- 0.12}$\\
GGOU   & $4.39 \pm  0.15^{ + 0.12}_ { -0.20} $\\
\hline
\hline
\end{tabular}
\end{table}
The average values for $|V_{ub}|$ have  been extracted  by HFAG \cite{Amhis:2012bh}  from the partial branching fractions, adopting
a specific theoretical framework and taking into account
 correlations among the various measurements
and theoretical uncertainties.
In Table \ref{phidectab3} we list some determinations, specifically the QCD theoretical calculations taking into account  the whole set of experimental results, or most of it, starting from 2002 CLEO data \cite{Bornheim:2002du}.
They refer to the BLNP approach by Bosch, Lange, Neubert, and Paz \cite{Lange:2005yw}, the
GGOU one  by Gambino, Giordano, Ossola and Uraltsev \cite{Gambino:2007rp}, the DGE one, the
dressed gluon exponentiation, by Andersen and Gardi \cite{Andersen:2005mj, Gardi:2008bb} and  the ADFR approach, by Aglietti, Di Lodovico, Ferrara, and Ricciardi,
\cite{Aglietti:2004fz, Aglietti:2006yb, Aglietti:2007ik}.
The results  listed in  Table \ref{phidectab3}  are consistent within the errors, but  the theoretical uncertainty among determinations can reach 10\%.
Other theoretical approaches  have also been proposed in \cite{Bauer:2001rc, Leibovich:1999xf, Ligeti:2008ac}.
Notwithstanding all the experimental and theoretical efforts,
the values of $|V_{ub}|$ extracted from inclusive decays remain about two $\sigma$ above the values given by exclusive determinations.

\subsubsection{Rare charm  decays}
\label{raredecays}

The decays driven by $ c \rightarrow u l^+ l^-$ are forbidden at tree level in the standard model (SM) and proceed via one-loop diagrams
(box and penguin)  at leading order in the electroweak interactions.
Virtual quarks in the loops are of the down type,  and no
breaking due to the large top mass occurs. The  GIM  mechanism works more effectively  in suppressing flavor (charm) changing neutral currents than  their strangeness and beauty analogues, leading to tiny decay rates,  dominated by
long-distance effects.  On the other side, we expect possible enhancements due to new physics to stand out, once we exclude potentially large long-distance SM
contributions.

In the SM, a very low branching ratio has been estimated  for  inclusive decays, largely dominated
by long-distance contributions $  { \cal B} (D \rightarrow X_u l^+ l^-) = { \cal B}_\mathrm{LD} (D \rightarrow X_u l^+ l^-) \sim  O(10^{-6})$ \cite{Burdman:2001tf}.
Long-distance contributions are assumed  to proceed from intermediate vector resonances such as
 $ D \rightarrow X_u V$, $V \rightarrow l^+l^-$, where $V = \phi$, $\rho$ or $\omega$, which set the scale with  branching fractions of order $10^{-6}$.
Short-distance contributions lay far behind \cite{Burdman:2001tf,Fajfer:2001sa,Paul:2011ar}; the latest estimate gives 
$ { \cal B}_\mathrm{SD} (D \rightarrow X_u e^+ e^-) \sim 4 \,\times \,  10^{-9}$ \cite{Paul:2011ar}.
Handling long-distance dynamics in these
processes becomes equivalent to handling several intermediate charmless resonances, in a larger  number than in the case of $B$ meson analogues.
Their effect can be separated from short-distance contributions by   applying selection criteria on the invariant mass of the leptonic pair.

To consider exclusive decays, let us start from $D_{(s)}^\pm \rightarrow  h^\pm l^+ l^-$, with $h \in (\pi, \rho, K, K^\star)$ and $l  \in (e, \mu)$,
none of which   has  been observed up to now. The best experimental limits on branching fractions are  $O(10^{-6})$ or higher, at 90\% confidence level (CL), coming all  from BaBar \cite{Lees:2011hb, Godang:2013im}, with a few exceptions: very old limits on    $D^+ \rightarrow \rho^+ \mu^+ \mu^-$ and $D_s^+ \rightarrow K^{\ast +}(892) \mu^+ \mu^-$ decays,  given by E653 \cite{Kodama:1995ia}, and the  recent limits on  $D^{+}_{(s)} \rightarrow \pi^{\pm} \mu^\mp \mu^+$ decays, given by LHCb with an integrated luminosity of 1.0 ${\mathrm{fb}}^{-1}$ \cite{Aaij:2013sua}.
The BESIII collaboration will be able to reach a sensitivity of $O(10^{-7})$ for $D^+ \rightarrow K^+/\pi^+ \,  l^+ l^-$ at 90\% CL with a 20 fb$^{-1}$ data sample taken at the $\psi(3770)$ peak \cite{Asner:2008nq}.
The LHCb collaboration can also search for  $D_{(s)}^\pm \rightarrow  h^\pm l^+ l^-$ decays.
The very recent update on the $ D_{(s)}^+ \rightarrow \pi^+ \mu^+ \mu^-$ channel with 3 ${\mathrm{fb}}^{-1}$ full data sample is still orders of magnitudes above the SM prediction; new searches for the $ D_{(s)}^+ \rightarrow K^+ \mu^+ \mu^-$ decays are ongoing~\cite{Kochebina:EPS}.
Also  decays  $D^0 \rightarrow  h^0 l^+ l^-$ have not been observed yet; the best experimental  limits at 90\% CL are of order
$O(10^{-5})$ or higher, and are given by older analyses of  CLEO \cite{Freyberger:1996it}, E653 \cite{Kodama:1995ia} and  E791 \cite{Aitala:2000kk}.
Future Super B factories are expected to reach a sensitivity of $O(10^{-8})$ on a 90\% CL on various rare decays, including $D^+ \rightarrow \pi^+ l^+ l^-$
and  $D^0 \rightarrow \pi^0 l^+ l^-$ \cite{Asner:2007kw}.

A way to disentangle possible new physics is to  choose appropriate observables containing mainly short distance
contributions. Last year, hints of possible new physics (NP) have been advocated in the charm sector to explain the  nonvanishing
direct CP asymmetry in $D^0 \rightarrow K^+ K^- $ and  $D^0 \rightarrow \pi^+ \pi^- $, measured by LHCb \cite{Aaij:2011in}, confirmed by CDF \cite{Aaltonen:2011se} and supported by recent data from Belle \cite{Ko:2012px}. Encouraged by these results,  effects of the same kind of possible NP  have been looked for in other processes, including  rare charm decays.
CP asymmetries can be generated by imaginary parts of Wilson coefficients in the
effective Hamiltonian for $ c \rightarrow u l^+ l^-$ driven decays. They have  been investigated in
$ D^+ \rightarrow \pi^+ \mu^+ \mu^-$ and $D_s^+ \rightarrow K^+ \mu^+ \mu^-$ decays, around the
$\phi$ resonance peak in the spectrum of dilepton invariant mass, concluding that in favorable conditions their value can be as high as 10\% \cite{Fajfer:2012nr}.
Older studies report  investigations of semileptonic decays in the framework of other NP models, such as R-parity violating supersymmetric models,
extra heavy up vector-like quark models  \cite{Fajfer:2007dy}, Little Higgs \cite{Paul:2011ar}, or leptoquark models \cite{Fajfer:2008tm}.
The   parameter space discussed in  older analyses  cannot take into account the constraints given by recent LHC data, most notably  the discovery of the  125 GeV resonance. In several cases,
a reassessment  in the updated framework could be used advantageously.

\subsection{Spectroscopy}

\label{sec:secC2}

The year 2013 marks the tenth anniversary of the observation of the
$X(3872)$ charmonium-like state~\cite{Choi:2003ue} that put an end to
the era when heavy quarkonium was considered as a relatively well
established bound system of a heavy quark and antiquark. Since 2003
every year has been bringing discoveries of new particles with
unexpected properties, not fitting a simple $q\bar{q}$ classification
scheme. The wealth of new results is mainly from B- and c-factories,
Belle, BaBar and BES~III, where data samples with unprecedented
statistics became available.

In this section we first describe experiments that contribute to the
subject, discuss recent developments for low-lying states, then we
move to the open flavor thresholds and beyond. We consider the
charmonium- and bottomonium-(like) states in parallel to stress
similarities between the observed phenomena in the two quarkonium
sectors.

\subsubsection{Experimental tools}
Over the last decade the main suppliers of new information about
quarkonium states have been the $B$-factories, the experiments working at
asymmetric energy \epem\ colliders operated at center-of-mass energies
in the $\Upsilon$-resonance region. Both Belle and BaBar detectors
are general-purpose 4$\pi$ spectrometers with excellent momentum
resolution, vertex positioning and particle identification for charged
tracks, as well as with high-resolution electromagnetic
calorimeters. Although the main purpose of the $B$-factories is to
study CP asymmetries in $B$-decays, these experiments allow for many
other searches apart from the major goal. Charmonium states at
$B$-factories are copiously produced in $B$-decays, two-photon fusion,
charm quark fragmentation in $\epem \to c\bar{c}$ annihilation
(mostly via double $c\bar{c}$ production) and via initial state
radiation, when the energy of \epem\ annihilation is dumped by emission
of photons in the initial state. Both $B$-factories intensively
studied also bottomonium states, taking data at different $\Upsilon$
states that allow to access lower mass bottomonia via hadronic and
radiative transitions. Although both $B$-factories completed their data
taking already long ago (BaBar in 2008 and Belle in 2010), the
analysis of the collected data is still ongoing, and many interesting
results have been obtained recently. The data samples of the two
experiments are summarized in Table~\ref{tab:samples}.

Another class of experiments where charmonium states are extensively
studied are the charm-$\tau$ factories. For the last decade BES~II,
CLEOc, and finally BES~III have successively covered measurements of
$\epem$ annihilation around the charmonium region. The BES~III
experiment started data taking in 2009 after a major upgrade of the
BEPC \epem\ collider and the BES~II spectrometer. The BEPC~II
accelerator operates in the c.m.  energy range of $\sqrt{s} = (2 -
4.6)\gev$ and has already reached a peak luminosity close to the
designed one. Starting late 2012 BES~III has collected data at high
energies to study $Y(4260)$ and other highly excited charmonium-like
states.

Experiments at hadron machines (Tevatron and LHC) can investigate
quarkonium produced promptly in high energy hadronic collisions in
addition to charmonium produced in $B$-decays. The Tevatron
experiments CDF and D0 completed their experimental program in 2010,
after CERN started operating the LHC. Four LHC experiments are
complementary in tasks and design. While LHCb has been optimized for
mainly heavy flavor physics, ATLAS and CMS are contributing to the
field by investigating certain signatures in the central rapidity
range with high statistics. The LHC accelerator performance has
fulfilled and even exceeded expectations. The integrated luminosity
delivered to the general-purpose experiments (ATLAS and CMS) in 2011
was about 6\,fb$^{-1}$, and more than 20\,fb$^{-1}$ in 2012. The
instantaneous luminosity delivered to LHCb is leveled to a constant
rate due to limitations in the LHCb trigger and readout, and to
collect data under relatively clean conditions. The integrated luminosity
delivered to LHCb was 1 fb$^{-1}$ and 2 fb$^{-1}$ in 2011 and 2012,
respectively.

\begin{table}[tb]
\caption{Integrated luminosities (in fb$^{-1}$) collected by the BaBar 
and Belle experiments at different $e^+e^-$ energies.}
\begin{tabular}{ccc}
    \hline\hline
 & \ \ \ BaBar \ \ \ & \ \ \ Belle \ \ \   \\ \hline
\UnS{1} & -- & 5.7  \\
\UnS{2} & 14 & 24.1 \\
\UnS{3} & 30 & 3.0 \\
\UnS{4} & 433 & 711  \\
off-resonance & 54 & 87  \\
\UnS{5} & - & 121   \\
\UnS{5}- \UnS{6} scan & 5 & 27  \\ \hline\hline
\end{tabular}
\label{tab:samples}
\end{table}

The new $B$ factory at KEK, SuperKEKB, will be commissioned in 2015
according to the current planning schedule. It is expected that
the target integrated luminosity, 50 ab$^{-1}$ , will be collected by 2022.

\subsubsection{Heavy quarkonia below open flavor thresholds}
Recently, significant progress has been achieved in the studies of
the spin-singlet bottomonium states. In addition, last year two more
states have been found below their corresponding open flavor thresholds,
the $\psi_2(1D)$ charmonium and the $\chi_b(3P)$ bottomonium (in the
latter case the levels with different $J$ are not resolved),  see
Table \ref{tab:qq_below_thresh}.  All these new data provide important
tests of the theory, which, due to lattice and effective field
theories, is rather solid and predictive below the open flavor
threshold.  The theory verification in this particular region becomes
even more important given the difficulties of the theory for states
near or above the open flavor threshold.

\begin{table*}[tb]
\footnotesize
\caption{Quarkonium states below the corresponding open flavor thresholds.}
\label{tab:qq_below_thresh}
\begin{ruledtabular}
\begin{tabular}{lrcllclc}
State & $M,\,\mev$ & $\Gamma,\,\mev$ & $J^{PC}$ & Process (mode) & Experiment (\#$\sigma$) & Year & Status \\
\hline
      $\psi_2(1D)$ & $3823.1\pm1.9$ & $<24$ & $2^{--}$ &
      $B\to K(\gamma\,\chi_{c1})$ & Belle~\cite{Bhardwaj:2013rmw} (3.8) & 
      2013 & NC! \\
      $\eta_b(1S)$ & $9398.0\pm3.2$ & $11^{+6}_{-4}$ & $0^{-+}$ &
      $\Upsilon(3S)\to\gamma\,(...)$ & BaBar~\cite{Aubert:2008ba} (10), CLEO~\cite{Bonvicini:2009hs} (4.0) & 
      2008 & Ok \\
      & & & &
      $\Upsilon(2S)\to\gamma\,(...)$ & BaBar~\cite{Aubert:2009as} (3.0) & 2009 & NC! \\
      & & & &
      $h_b(1P,2P)\to\gamma\,(...)$ & Belle~\cite{Mizuk:2012pb} (14) & 2012 & NC! \\
      $h_b(1P)$ & $9899.3\pm1.0$ & ? & $1^{+-}$ &
      $\Upsilon(10860)\to\pi^+\pi^-\,(...)$ & Belle~\cite{Adachi:2011ji,Mizuk:2012pb} (5.5) & 
      2011 & NC! \\
      & & & &
      $\Upsilon(3S)\to\pi^0\,(...)$ & BaBar~\cite{Lees:2011zp} (3.0) & 2011 & NC! \\
      $\eta_b(2S)$ & $9999\pm4$ & $<24$ & $0^{-+}$ &
      $h_b(2P)\to\gamma\,(...)$ & Belle~\cite{Mizuk:2012pb} (4.2) & 
      2012 & NC! \\
      $\Upsilon(1D)$ & $10163.7\pm1.4$ & ? & $2^{--}$ &
      $\Upsilon(3S)\to\gamma\gamma\,(\gamma\gamma\,\Upsilon(1S))$ & CLEO~\cite{Bonvicini:2004yj} (10.2) & 
      2004 & NC! \\
      & & & &
      $\Upsilon(3S)\to\gamma\gamma\,(\pi^+\pi^-\Upsilon(1S))$ & BaBar~\cite{delAmoSanchez:2010kz} (5.8) & 
      2010 & NC! \\
      & & & &
      $\Upsilon(10860)\to\pi^+\pi^-(\gamma\gamma\,\Upsilon(1S))$ & Belle~\cite{Krokovny:LaThuile2012} (9) & 
      2012 & NC! \\
      $h_b(2P)$ & $10259.8\pm1.2$ & ? & $1^{+-}$ &
      $\Upsilon(10860)\to\pi^+\pi^-\,(...)$ & Belle~\cite{Adachi:2011ji,Mizuk:2012pb} (11.2) & 
      2011 & NC! \\
      $\chi_{bJ}(3P)$ & $10534\pm9$ & ? & $(1,2)^{++}$ &
      $pp,p\bar{p}\to(\gamma\Upsilon(1S,2S))\,...$ & ATLAS~\cite{Aad:2011ih} ($>$6), D0~\cite{Abazov:2012gh} (5.6) & 
      2011 & Ok \\
\end{tabular}
\end{ruledtabular}
\end{table*}

Spin-singlet bottomonium states do not have production or decay
channels convenient for experimental studies. Therefore their
discovery became possible only with the high statistics of the
$B$-factories. An unexpected source of the spin-singlet states turned
out to be the di-pion transitions from the $\UnS{5}$.  The states are
reconstructed inclusively using the missing mass of the accompanying
particles.  Belle observed the $\chb$ and $\chbp$ states in the
transitions $\UnS{5}\to\dipi\chbn$~\cite{Adachi:2011ji}.  The
hyperfine splittings were measured to be $(+0.8\pm1.1)\,\mev$ for
$n=1$ and $(+0.5\pm1.2)\,\mev$ for $n=2$~\cite{Mizuk:2012pb}.  The
results are consistent with perturbative QCD
expectations \cite{Titard:1993nn,Titard:1994id,Titard:1994ry,Brambilla:2004wu}.
This shows in particular that the spin--spin potential does not have a
sizeable long-range contribution~\cite{Vairo:2006pc}, an observation
supported by direct lattice computations~\cite{Koma:2006fw}.  For
comparison, in the charmonium sector the measured $1P$ hyperfine
splitting of $(-0.11\pm0.17)\,\mev$~\cite{Beringer:1900zz} is also
consistent with zero with even higher accuracy.

The $\eta_b(1S)$ is found in M1 radiative transitions
from \UnS{3}~\cite{Aubert:2008ba,Bonvicini:2009hs}
and \UnS{2}~\cite{Aubert:2009as}. The measured averaged hyperfine
splitting $\Delta M_{\rm HF}(1S)=M_{\UnS{1}}-M_\etab= (69.3\pm
2.8)\mev$~\cite{Beringer:1900zz} was larger than perturbative pNRQCD
$(41 \pm 14)\,\mev$~\cite{Kniehl:2003ap} and lattice $(60 \pm
8)\,\mev$~\cite{Meinel:2010pv} estimates.  In 2012, using a large
sample of \chbm\ from \UnS{5} Belle observed the $\chb\to\etab\cga$
and $\chbp\to\etab\cga$ transitions~\cite{Mizuk:2012pb}. The Belle \etab\
mass measurement is more precise than the PDG2012 average and is
$(11.4 \pm 3.6)\,\mev$ above the central value, which is in better
agreement with the perturbative pNRQCD determination.  The residual
difference of about $17\,\mev$ is consistent with the uncertainty of
the theoretical determination. Also lattice determinations have
improved their analyses (see Sec.~\ref{sec:subsecC12}).  The latest
determination based on lattice NRQCD, which includes spin-dependent
relativistic corrections through $O(v^6)$, radiative corrections to
the leading spin-magnetic coupling, nonperturbative 4-quark
interactions and the effect of $u$, $d$, $s$ and $c$ quark vacuum
polarization, gives $\Delta M_{\rm HF}(1S) = (62.8 \pm
6.7)\,\mev$~\cite{Dowdall:2013jqa}.  Belle measured for the first time
also the \etab\ width, $\Gamma_{\etab} =
(10.8\,^{+4.0}_{-3.7}\,^{+4.5}_{-2.0})\,\mev$, which is consistent
with expectations.

Belle found the first strong evidence for the \etabp\ with a
significance of $4.4\,\sigma$ using the $\chbp\to\cga\etabp$
transition. The hyperfine splitting was measured to be $\Delta M_{\rm
HF}(2S)=(24.3^{+4.0}_{-4.5})\,\mev$.  The ratio $\Delta M_{\rm
HF}(2S)/\Delta M_{\rm HF}(1S)=0.420^{+0.071}_{-0.079}$ is in agreement
with NRQCD lattice
calculations~\cite{Meinel:2010pv,Dowdall:2011wh,Dowdall:2013jqa},
the most recent of which gives $\Delta M_{\rm HF}(2S)/\Delta M_{\rm
HF}(1S)=0.425\pm 0.025$~\cite{Dowdall:2013jqa} (see also Sec.~\ref{sec:subsecC12}).  
The measured branching fractions
$\brat(\chb\to\cga\etab)=(49.2\pm5.7\,^{+5.6}_{-3.3})\%$,
$\brat(\chbp\to\cga\etab)=(22.3\pm3.8\,^{+3.1}_{-3.3})\%$, and
$\brat(\chbp\to\cga\etabp)=(47.5\pm10.5\,^{+6.8}_{-7.7})\%$ are
somewhat higher than the model predictions~\cite{Godfrey:2002rp}.

There is another claim of the \etabp\ signal by the group of K.~Seth
from Northwestern University, that used CLEO data~\cite{Dobbs:2012zn}.
The $\UnS{2}\to\etabp\cga$ production channel is considered and
the \etabp\ is reconstructed in 26 exclusive channels with up to 10
charged tracks in the final state. The measured hyperfine splitting
$\Delta M_{\rm HF}(2S)=(48.7\pm3.1)\,\mev$ is $5\,\sigma$ away from
the Belle value and is in strong disagreement with theoretical
expectations~\cite{Burns:2012pc}. In \cite{Dobbs:2012zn} the
contribution of final state radiation is not considered, therefore the
background model is incomplete and the claimed significance of
$4.6\,\sigma$ is overestimated. Belle repeated the same analysis with
17 times higher statistics and found no
signal~\cite{Sandilya:2013rhy}. The Belle upper limit is an order of
magnitude lower than the central value in \cite{Dobbs:2012zn}. We
conclude that the evidence for the \etabp\ with the anomalous mass
reported in \cite{Dobbs:2012zn} is refuted.

The $n=3$ radial excitation of the $\chi_{bJ}$ system was recently
observed by ATLAS~\cite{Aad:2011ih} and confirmed by
D0~\cite{Abazov:2012gh}. The $\chi_{bJ}(3P)$ states are produced
inclusively in the $pp$ and $p\overline{p}$ collisions and are
reconstructed in the $\cga\Upsilon(1S,2S)$ channels with
$\Upsilon\to\mu^+\mu^-$. Converted photons and photons reconstructed
from energy deposits in the electromagnetic calorimeter are used.  The
mass resolution does not allow to discern individual $\chi_{bJ}(3P)$
states with $J=0$, 1 and 2. A measured barycenter of the triplet
$10534\pm9\,\mev$ is close to the quark model expectations of
typically $10525\,\mev$~\cite{Kwong:1988ae,Motyka:1997di}.

Potential models predict that $D$-wave charmonium levels are situated
between the \DDbar\ and \DDst\ thresholds~\cite{Eichten:2002qv}.
Among them the states $\eta_{c2}$ ($J^{PC}=2^{-+}$) and $\psi_2$
($J^{PC}=2^{--}$) cannot decay to \DDbar\ because of unnatural
spin-parity, and they are the only undiscovered charmonium levels that are
expected to be narrow. Recently Belle reported the first evidence for
the $\psi_2(1D)$ using the $B^+\to K^+\psi_2(1D)[\to\cga\chi_{c1}]$
decays~\cite{Bhardwaj:2013rmw}, with a mass of
$M=(3823.1\pm1.9)\,\mev$ and width consistent with zero,
$\Gamma<24\,\mev$. The full width is likely to be very small, since
the state is observed in the radiative decay and the typical
charmonium radiative decay widths are at the $O(100)\,\kev$ level. The
odd $C$-parity (fixed by decay products) discriminates
between the $\eta_{c2}$ and $\psi_2$ hypotheses. No signal is found in
the $\cga\chi_{c2}$ channel, in agreement with expectations for the
$\psi_2$~\cite{Eichten:2002qv}. Belle measured $\brat(B^+\to
K^+\psi_2)\times\brat(\psi_2\to\cga\chi_{c1})=
(9.7{^{+2.8}_{-2.5}}{^{+1.1}_{-1.0}})\times10^{-6}$. Given that one
expects $\brat(\psi_2\to\cga\chi_{c1})\sim 2/3$~\cite{Eichten:2002qv},
$\brat(B^+\to K^+\psi_2)$ is a factor of 50 smaller than the
corresponding branching fractions for the \jpsi, \psip\ and \chicOne\
due to the factorization suppression~\cite{Suzuki:2002sq,Colangelo:2002mj}.

Many of the above studies and, in particular, many discovery channels
involve radiative decays. For states below threshold, theory has made
in the last few years remarkable progress in the study of these decay
channels.  From the EFT side, pNRQCD provides now an (almost) complete
description of E1 and M1
transitions \cite{Brambilla:2005zw,Brambilla:2012be}, which means that
we have expressions for all these decay channels up to and including
corrections of relative order $v^2$. The only exception are M1
transitions for strongly bound quarkonia that depend at order $v^2$ on
a yet uncalculated Wilson coefficient.  The kind of insight in the QCD dynamics
of quarkonia that one may get from having analytical expressions for
these decay rates can be understood by looking at the transition
$J/\psi \to \eta_c(1S)\gamma$. The PDG average for the width
$\Gamma(J/\psi \to \eta_c(1S)\gamma)$ is $(1.58 \pm 0.37) \,\kev$,
which is clearly lower than the leading order estimate $2.83\,\kev$.
Corrections of relative order $v^2$ are positive in the case of a
confining potential, whereas they are negative in the case of a
Coulomb potential \cite{Brambilla:2005zw}.  Therefore the current PDG average  
favors an interpretation of the $J/\psi$ as a Coulombic bound state.
This interpretation may be challenged by the most recent KEDR analysis
that finds $\Gamma(J/\psi\to\eta_c(1S)\gamma) = (2.98 \pm 0.28)\,{\rm
keV}$~\cite{Eidelman:Hadron2013}. The KEDR result has a better accuracy than
the current world average and is $3.0 \sigma$ above its central value. 

In \cite{Becirevic:2012dc}, a determination of
$\Gamma(J/\psi \to \eta_c(1S)\gamma)$ based on lattice QCD in the
continuum limit with two dynamical quarks, the authors find
$\Gamma(J/\psi \to \eta_c(1S)\gamma) = (2.64\pm 0.11\pm 0.03)\,\kev$.
Earlier lattice determinations of the charmonium radiative transitions in
quenched lattice QCD can be found in \cite{Dudek:2006ej,Dudek:2009kk}.
In \cite{Pineda:2013lta}, a determination of
$\Gamma(J/\psi \to \eta_c(1S)\gamma)$ in perturbative pNRQCD, the
authors find $\Gamma(J/\psi \to \eta_c(1S)\gamma) = (2.12\pm 0.40) \,{\rm keV}$.  
Both theoretical determinations are consistent with each other 
and fall in between the PDG average and the latest KEDR determination 
with the lattice determination favoring a somewhat larger value 
and the perturbative QCD determination a somewhat smaller value 
of the transition width. Part of the tension between data, and between 
data and theoretical determinations may be due to the fact that 
the extraction of the $J/\psi \to \eta_c(1S)\gamma$ branching fraction
from the photon energy line shape in $J/\psi \to X\gamma$ is not free
from uncontrolled uncertainties~\cite{Brambilla:2010ey}.  

Bottomonium M1 transitions have been studied in perturbative pNRQCD in 
\cite{Brambilla:2005zw} and \cite{Pineda:2013lta}. In particular, 
in \cite{Pineda:2013lta}
a class of large perturbative contributions associated with the static
potential has been resummed providing an improved determination of
several M1 transitions:
$\Gamma({\Upsilon(1S) \rightarrow \eta_b(1S)\gamma}) = (15.18 \pm
0.51) \,{\rm eV}$, $\Gamma({h_b(1P) \rightarrow \chi_{b0}(1P)\gamma})
= (0.962\pm 0.035) \,{\rm eV}$,
$\Gamma({h_b(1P) \rightarrow \chi_{b1}(1P)\gamma}) = (8.99\pm
0.55) \times 10^{-3}\,{\rm eV}$, $\Gamma({\chi_{b2}(1P) \rightarrow
h_b(1P)\gamma}) = (0.118\pm 0.006) \,{\rm eV}$ and
$\Gamma({\Upsilon(2S) \rightarrow \eta_b(1S)\gamma}) =6^{+26}_{-6} \,
{\rm eV}$.  The improved determination of
$\Gamma({\Upsilon(2S) \rightarrow \eta_b(1S)\gamma})$ is particularly
noteworthy because it is consistent with the most recent data,
$(12.5\pm 4.9)\,{\rm eV}$ from BaBar \cite{Aubert:2009as}, while the
leading-order determination is off by at least one order of
magnitude. Bottomonium transitions in lattice NRQCD with $2+1$
dynamical quarks have been computed in \cite{Lewis:2011ti,Lewis:2012bh}.

E1 transitions are more difficult to study both on the lattice and
with analytical methods.  The reason is that even at leading order
they involve a nonperturbative matrix element.  A complete theoretical
formulation in the framework of pNRQCD can be found in
\cite{Brambilla:2012be} with a preliminary but promising 
phenomenological analysis in \cite{Pietrulewicz:2013ct}.

The theoretical status of quarkonium hadronic transitions, inclusive
and exclusive hadronic and electromagnetic decays has been summarized
in \cite{Brambilla:2004wf,Brambilla:2010cs,Bodwin:2013nua}.  There has been a limited
use of the pNRQCD factorization for these processes and only
restricted to inclusive hadronic and electromagnetic decays
\cite{Brambilla:2001xy,Brambilla:2002nu,Vairo:2003gh,Pineda:2003be,Penin:2004ay,Pineda:2011dg}, 
while most of the recent work has concentrated on improving the expansion 
in the NRQCD factorization framework to higher orders in $v$ and $\alpha_s$
\cite{Bodwin:2002hg,Brambilla:2006ph,Bodwin:2008vp,Gong:2008ue,Brambilla:2008zg,Jia:2011ah,Guo:2011tz,Sang:2011fw,
Chen:2011ph,Sang:2012yh,Li:2012rn,Feng:2012by,Xu:2012uh}.

\subsubsection{Quarkonium-like states at open flavor thresholds} 
There are several states in both the charmonium and bottomonium
sectors lying very close to the threshold of their decay to a pair of
open flavor mesons, see Table \ref{tab:qq_at_thresh}.  This proximity
suggests a molecular structure for these states.

\begin{table*}[tb]
\footnotesize
\caption{Quarkonium-like states at the open flavor thresholds. 
For charged states, the $C$-parity is given for the neutral members of the corresponding 
isotriplets.}
\label{tab:qq_at_thresh}
\begin{ruledtabular}
\begin{tabular}{lrcllclc}
State & $M,\,\mev$ & $\Gamma,\,\mev$ & $J^{PC}$ & Process (mode) & Experiment (\#$\sigma$) & Year & Status \\
\hline
      $X(3872)$ & $3871.68\pm0.17$ & $<1.2$ & $1^{++}$ &
      $B\to K(\pi^+\pi^-J/\psi)$ & Belle~\cite{Choi:2003ue,Choi:2011fc} ($>$10), BaBar~\cite{Aubert:2008gu} (8.6) & 
      2003 & Ok \\
      & & & &
      $p\bar{p}\to(\pi^+\pi^-J/\psi)\,...$ & CDF~\cite{Abulencia:2006ma,Aaltonen:2009vj} (11.6), D0~\cite{Abazov:2004kp} (5.2) & 2003 & Ok \\
      & & & &
      $pp\to(\pi^+\pi^-J/\psi)\,...$ & LHCb~\cite{Aaij:2011sn,Aaij:2013zoa} (np) & 2012 & Ok \\
      & & & &
      $B\to K(\pi^+\pi^-\pi^0J/\psi)$ & Belle~\cite{Abe:2005ix} (4.3), BaBar~\cite{delAmoSanchez:2010jr} (4.0) & 2005 & Ok \\
      & & & &
      $B\to K(\gamma\, J/\psi)$ & Belle~\cite{Bhardwaj:2011dj} (5.5), BaBar~\cite{Aubert:2008ae} (3.5) & 2005 & Ok \\
      & & & & & LHCb~\cite{Aaij:2014ala}~($>10$) \\
      & & & &
      $B\to K(\gamma\, \psi(2S))$ & BaBar~\cite{Aubert:2008ae} (3.6), Belle~\cite{Bhardwaj:2011dj} (0.2)& 2008 & NC! \\
      & & & & & LHCb~\cite{Aaij:2014ala}~(4.4) \\
      & & & &
      $B\to K(D\bar{D}^*)$ & Belle~\cite{Adachi:2008sua} (6.4), BaBar~\cite{Aubert:2007rva} (4.9) & 2006 & Ok \\
      $Z_c(3885)^+$ & $3883.9\pm4.5$ & $25\pm12$ & $1^{+-}$ &
      $Y(4260)\to\pi^-(D\bar{D}^*)^+$ & BES~III~\cite{Ablikim:2013xfr} (np) & 
      2013 & NC! \\
      $Z_c(3900)^+$ & $3891.2\pm3.3$ & $40\pm8$ & $?^{?-}$ &
      $Y(4260)\to\pi^-(\pi^+J/\psi)$ & BES~III~\cite{Ablikim:2013mio} (8), Belle~\cite{Liu:2013dau} (5.2) & 
      2013 & Ok \\
      & & & & & T.~Xiao {\it et al.}\ [CLEO data]~\cite{Xiao:2013iha} ($>$5) & & \\
      $Z_c(4020)^+$ & $4022.9\pm2.8$ & $7.9\pm3.7$ & $?^{?-}$ &
      $Y(4260,4360)\to\pi^-(\pi^+h_c)$ & BES~III~\cite{Ablikim:2013wzq} (8.9) & 
      2013 & NC! \\
      $Z_c(4025)^+$ & $4026.3\pm4.5$ & $24.8\pm9.5$ & $?^{?-}$ &
      $Y(4260)\to\pi^-(D^*\bar{D}^*)^+$ & BES~III~\cite{Ablikim:2013emm} (10) & 
      2013 & NC! \\
      $Z_b(10610)^+$ & $10607.2\pm2.0$ & $18.4\pm2.4$ & $1^{+-}$ &
      $\Upsilon(10860)\to\pi(\pi\Upsilon(1S,2S,3S))$ & Belle~\cite{Adachi:2011gja,Belle:2011aa,Krokovny:2013mgx} ($>$10) & 
      2011 & Ok \\
      & & & & 
      $\Upsilon(10860)\to\pi^-(\pi^+h_b(1P,2P))$ & Belle~\cite{Belle:2011aa} (16) & 2011 & Ok \\
      & & & & 
      $\Upsilon(10860)\to\pi^-(B\bar{B}^*)^+$ & Belle~\cite{Adachi:2012cx} (8) & 2012 & NC! \\
      $Z_b(10650)^+$ & $10652.2\pm1.5$ & $11.5\pm2.2$ & $1^{+-}$ &
      $\Upsilon(10860)\to\pi^-(\pi^+\Upsilon(1S,2S,3S))$ & Belle~\cite{Adachi:2011gja,Belle:2011aa} ($>$10) & 
      2011 & Ok \\
      & & & & 
      $\Upsilon(10860)\to\pi^-(\pi^+h_b(1P,2P))$ & Belle~\cite{Belle:2011aa} (16) & 2011 & Ok \\
      & & & & 
      $\Upsilon(10860)\to\pi^-(B^*\bar{B}^*)^+$ & Belle~\cite{Adachi:2012cx} (6.8) & 2012 & NC! \\
\end{tabular}
\end{ruledtabular}
\end{table*}

The $X(3872)$ is a state very close to the \DstnDn\ threshold,
$\delta
m_{X(3872)}=m_{X(3872)}-m_{D^{*0}}-m_{D^0}=-0.11\pm0.22\,\mev$~\cite{Beringer:1900zz,Aaij:2013uaa,Lees:2013dja}.
The decays $X(3872)\to\rho\jpsi$ and $X(3872)\to\omega\jpsi$ have
similar branching fractions, $\brat_{\omega}/\brat_{\rho}
=0.8\pm0.3$~\cite{Abe:2005ix,delAmoSanchez:2010jr}; this corresponds
to a strong isospin violation. The favorite $X(3872)$ interpretation is a
mixture of a charmonium state $\chi_{c1}(2P)$ and an $S$-wave
\DstnDn\ molecule~\cite{Voloshin:2003nt,Braaten:2004rw,Braaten:2004fk,Braaten:2004jg,Voloshin:2004mh,Braaten:2004ai,Braaten:2005jj,Braaten:2005ai,Voloshin:2005rt,Braaten:2006sy,Dubynskiy:2006cj,Fleming:2007rp,Voloshin:2007hh,Braaten:2007dw,Dubynskiy:2007tj,Braaten:2007ft,Braaten:2007sh,Voloshin:2007dx,Fleming:2008yn,Artoisenet:2009wk,Braaten:2010mg,Artoisenet:2010va,Mehen:2011ds,Fleming:2011xa,Esposito:2013ada,Braaten:2013poa},
with the molecular component responsible for the isospin violation and
the charmonium component accounting for the production in $B$ decays
and at hadron collisions. For alternative interpretations we refer
to~\cite{Brambilla:2010cs} and references therein.  The molecular
hypothesis is valid only for the spin-parity assignment
$J^{PC}=1^{++}$. Experimentally $1^{++}$ was favored, but $2^{-+}$ was
not excluded for a long
time~\cite{Abe:2005iya,Abulencia:2006ma,Choi:2011fc}.  This intrigue
has been settled recently by LHCb with a clear preference of $1^{++}$
and exclusion of $2^{-+}$ hypothesis~\cite{Aaij:2013zoa}.
Progress towards a lattice understanding of the  $X(3872)$ has been 
discussed in Sec.~\ref{sec:subsecC12}.

The contributions to the $\delta m_{X(3872)}$ uncertainty are
$0.17\,\mev$ from the $X(3872)$ mass, $0.13\,\mev$ from the $D^0$ mass
and $0.07\,\mev$ from the $D^{*0}-D^0$ mass difference. LHCb can
improve the accuracy in $m_{X(3872)}$ and $m_{D^0}$, BES~III and KEDR
can contribute to the $m_{D^0}$ measurement. The $D^{*0}-D^0$ mass
difference was measured 20 years ago by ARGUS and CLEO and also can be
improved.
The radiative $X(3872)\to\cga\jpsi$ decay is established, while there is an experimental controversy
regarding $X(3872)\to\cga\psip$~\cite{Abe:2005ix,Aubert:2008ae,Bhardwaj:2011dj}, with recent LHCb evidence
pointing towards existence of this channel~\cite{Aaij:2014ala}.
The dominant decay mode of the $X(3872)$ is \DstnDn~\cite{Gokhroo:2006bt,Aubert:2007rva,Adachi:2008sua}, as
expected for the molecule, however, the absolute branching fraction is not yet determined.
These questions will have to wait for Belle~II data.

Charged bottomonium-like states $\czbo$ and $\czbt$ are
observed by Belle as intermediate $\UnS{n}\pi^{\pm}$ and
$h_b(mP)\pi^{\pm}$ resonances in the $\UnS{5}\to\pi^+\pi^-\UnS{n}$ and
$\UnS{5}\to\pi^+\pi^- h_b(mP)$ decays~\cite{Belle:2011aa}. The
nonresonant contribution is sizable for the
$\UnS{5}\to\pi^+\pi^-\UnS{n}$ decays and is consistent with zero for
the $\UnS{5}\to\pi^+\pi^-h_b(mP)$ decays. The mass and width of the
$\czb$ states were measured from the amplitude analysis, assuming
a Breit--Wigner form of the $\czb$ amplitude. The parameters are in
agreement among all five decay channels, with the average values
$M_1=(10607.4\pm2.0)\,\mev$, $\Gamma_1=(18.4\pm2.4)\,\mev$ and
$M_2=(10652.2\pm1.5)\,\mev$, $\Gamma_2=(11.5\pm2.2)\,\mev$.
The measured $\czbo$ and $\czbt$ masses coincide within uncertainties with
the $B\bar{B}^*$ and $B^*\bar{B}^*$ thresholds, respectively.

Belle observed the $\czbo\to B\bar{B}^*$ and $\czbt\to B^*\bar{B}^*$
decays with the significances of $8\,\sigma$ and $6.8\,\sigma$,
respectively, using the partially reconstructed
$\UnS{5}\to(B^{(*)}\bar{B}^*)^{\pm}\pi^{\mp}$
transitions~\cite{Adachi:2012cx}. Despite much larger phase space, no
significant signal of the $\czbt\to B\bar{B}^*$ decay was
found. Assuming that the decays observed so far saturate the $Z_b$
decay table, Belle calculated the relative branching fractions of
$\czbo$ and $\czbt$ (Table~\ref{tab:zb_dec}).
\begin{table}[tb]
\centering
\caption{Branching fractions (\brat) of \czb s in per cent.}
\label{tab:zb_dec}
\center
\begin{tabular}{lcc}
\hline\hline
Channel & \brat\ of \czbo & \brat\ of \czbt \\
\hline
$\cpip\UnS{1}$  & $0.32\pm0.09$ & $0.24\pm0.07$ \\
$\cpip\UnS{2}$  & $4.38\pm1.21$ & $2.40\pm0.63$ \\
$\cpip\UnS{3}$ & $2.15\pm0.56$ & $1.64\pm0.40$ \\
$\cpip\chb$  & $2.81\pm1.10$ & $7.43\pm2.70$ \\
$\cpip\chbp$ & $4.34\pm2.07$ & $14.8\pm6.22$ \\
$B^+\bar{B}^{*0}+\bar{B}^0B^{*+}$ & $86.0\pm3.6$ & -- \\
$B^{*+}\bar{B}^{*0}$ & -- & $73.4\pm7.0$ \\
\hline
\hline
\end{tabular}
\end{table}
The $\czbo\to B\bar{B}^*$ and $\czbt\to B^*\bar{B}^*$ decays are
dominant with a branching fraction of about 80\%.  If the $\czbt\to
B\bar{B}^*$ channel is included in the decay table, its branching
fraction is $\brat(\czbt\to B\bar{B}^*)=(25\pm10)\,\%$ and all other
$\czbt$ branching fractions are reduced by a factor of 1.33.

Belle observed the neutral member of the $Z_b(10610)$ isotriplet by
performing a Dalitz analysis of the $\UnS{5}\to\pi^0\pi^0\UnS{n}$
($n=1,2,3$) decays~\cite{Krokovny:2013mgx}. The $Z_b(10610)^0$ significance
combined over the $\pi^0\UnS{2}$ and $\pi^0\UnS{3}$ channels 
is $6.5\,\sigma$. The measured mass value $M_{Z_b(10610)^0}=(10609\pm6)\,\mev$
is in agreement with the mass of the charged $\czbo^{\pm}$. No significant
signal of the $Z_b(10650)^0$ is found; the data are consistent with
the existence of the $Z_b(10650)^0$ state, but the available
statistics are insufficient to observe it.

To determine the spin and parity of the $Z_b$ states, Belle performed
a full 6-dimensional amplitude analysis of the
$\Upsilon(5S)\to\pi^+\pi^-\Upsilon(nS)$ $(n=1,2,3)$
decays~\cite{Garmash:2014dhx}. The $Z_b(10610)$ and
$Z_b(10650)$ are found to have the same spin-parity $J^P=1^+$, while
all other hypotheses with $J\leq2$ are rejected at more than
$10\,\sigma$ level. The highest discriminating power is provided by
the interference pattern between the $Z_b$ signals and the
nonresonant contribution.

Proximity to the $B\bar{B}^*$ and $B^*\bar{B}^*$ thresholds suggests
that the $\czb$ states have molecular structure, i.e., their
wave function at large distances is the same as that of an S-wave meson
pair in the $I^G(J^P)=1^+(1^+)$ state~\cite{Bondar:2011ev}. 
The assumption of the molecular structure can naturally explain all
the properties of the $\czb$ states without specifying their dynamical
model~\cite{Bondar:2011ev}. The decays into constituents
[i.e. $\czbo\to B\bar{B}^*$ and $\czbt\to B^*\bar{B}^*$], if
kinematically allowed, should dominate. The molecular spin function,
once decomposed into $b\bar{b}$ spin eigenstates, appears to be a
mixture of the ortho- and para-bottomonium components. The weights of
the components are equal, therefore the decays into $\pi\Upsilon$ and
$\pi h_b$ have comparable widths. The $B\bar{B}^*$ and $B^*\bar{B}^*$
states differ by a sign between the ortho- and para-bottomonium
components. This sign difference is observed in the interference
pattern between the $\czbo$ and $\czbt$ signals in the $\pi\Upsilon$ and
$\pi h_b$ final states~\cite{Belle:2011aa}.

The question of the dynamical model of the molecules remains
open. Among different possibilities are nonresonant
rescattering~\cite{Chen:2011pv}, multiple rescatterings that result in
a pole in the amplitude, known as a coupled channel
resonance~\cite{Danilkin:2011sh}, and deuteron-like molecule bound by
meson exchanges~\cite{Ohkoda:2011vj}. All these mechanisms are closely
related and a successful phenomenological model should probably account
for all of them. Predictions for the $Z_b$ lineshapes that can be used
to fit data would be useful to discriminate between different
mechanisms.

Alternatively, the $\czb$ states are proposed to have
diquark--antidiquark structure~\cite{Ali:2011ug}. In this model the
$B^{(*)}\bar{B}^*$ channels are not dominant and the lighter (heavier)
state couples predominantly to $B^*\bar{B}^*$ ($B\bar{B}^*$). The data
on the decay pattern of the $\czb$ states strongly disfavor the
diquark-antidiquark hypothesis.

Observation of the charged $Z_b$ states motivated a search for their
partners in the charm sector. Since late 2012 BES~III has been collecting
data at different energies above $4\,\gev$ to study charmonium-like
states.
In the course of 2013 the states $Z_c(3885)^{\pm}\to
(D\bar{D}^*)^{\pm}$, $Z_c(3900)^{\pm}\to\pi^{\pm}J/\psi$,
$Z_c(4020)\to\pi^{\pm}h_c$, $Z_c(4025)\to (D^*\bar{D}^*)^{\pm}$ were
observed (see Table~\ref{tab:qq_at_thresh}). 
The masses and widths of the $Z_c(3885)$/$Z_c(3900)$ and
$Z_c(4020)$/$Z_c(4025)$ pairs agree at about $2\,\sigma$ level. All
current measurements disregard the interference between the $Z_c$
signal and the nonresonant contribution, which is found to be
significant in all channels (including $\pi h_c$, in contrast to the
$\pi h_b$ case). Interference effects could shift the peak position by as
much as half the resonance width. A more accurate measurement of masses
and widths as well as spins and parities using the amplitude analyses
will help to clarify whether the above $Z_c$ pairs could be merged.

The $Z_c(3885)$ and $Z_c(3900)$ [$Z_c(4020)$ and $Z_c(4025)$] states
are close to the $D\bar{D}^*$ [$D^*\bar{D}^*$] threshold. In fact, all
the measured masses are about $10\,\mev$ {\it above} the
thresholds. This is a challenge for a molecular model, but could be an
experimental artifact caused by neglecting the interference.

If the $Z_c(3885)$ and $Z_c(3900)$ states are merged, the properties
of the resulting state agree with the expectations for the
$D\bar{D}^*$ molecular structure. The $D\bar{D}^*$ channel is
dominant~\cite{Ablikim:2013xfr},
\begin{equation}
    \frac{\Gamma[Z_c(3885)^{\pm}\to(D\bar{D}^*)^{\pm}]}%
        {\Gamma[Z_c(3900)^{\pm}\to \pi^{\pm}J/\psi]}=6.2\pm2.9.
\end{equation}
A $2.1\,\sigma$ hint for the $Z_c(3900)\to\pi^{\pm}h_c$
transition~\cite{Ablikim:2013wzq} implies that the state couples to both
ortho- and para-charmonium, with a weaker $\pi h_c$ signal due to 
phase space suppression. Finally, the spin-parity measured for the
$Z_c(3885)$ $J^P=1^+$ corresponds to $D\bar{D}^*$ in S-wave.

Identification of the $Z_c(4020)$ or $Z_c(4025)$ as a
$D^*\bar{D}^*$ molecule is difficult. If the $D\bar{D}^*$ molecule
decays to $\pi^{\pm}J/\psi$, then according to heavy-quark spin
symmetry the $D^*\bar{D}^*$ molecule should also decay to
$\pi^{\pm}J/\psi$. However, no hint of $Z_c(4020)$ or $Z_c(4025)$ is
seen in the $\pi^{\pm}J/\psi$ final state.
It could be that the $D^*\bar{D}^*$ molecule is not produced in the
$Y(4260)$ decays, as would be the case if the $Y(4260)$ is a
$D_1(2420)\bar{D}$ molecule (see next section).

The $Z_c(4020)$ could be a candidate for hadrocharmonium, a
color-neutral quarkonium state in a cloud of light
meson(s)~\cite{Voloshin:2013dpa}. The decay into constituent
charmonium and light meson should dominate, while the decay to another
charmonium is suppressed. The available experimental information on
$Z_c(4020)$ agrees with this picture. The $Z_c(4025)$ is not a
suitable hadrocharmonium candidate since the $D^*\bar{D}^*$ channel
dominates. Hadrocharmonium was proposed to explain the affinity of many
charmonium-like states [$Y(4260)$, $Y(4360)$, $Z(4050$, $Z(4250)$,
$Z(4430)$,\ldots] to some particular channels with charmonium and light
hadrons, as discussed below~\cite{Dubynskiy:2008mq}.

Another configuration proposed for the $Z_c$ states is a
Born-Oppenheimer tetraquark~\cite{Braaten:2013boa}. In such a state a
colored $c\bar{c}$ pair is moving in the adiabatic potential created
by the light degrees of freedom. This approach aims at providing a
general framework for the description of all $XYZ$ states.

To summarize, the properties of the $Z_b(10610)$ and $Z_b(10650)$
states are in good agreement with the assumption that they have
molecular structure. The $Z_c(3885/3900)$ state is a candidate for the
$D\bar{D}^*$ molecule, while the absence of the $Z_c(4020,4025)\to
\pi J/\psi$ signal disfavors the interpretation of $Z_c(4020)$ or
$Z_c(4025)$ as a $D^*\bar{D}^*$ molecule. The peak positions of the
$Z_c$ signals are about $10\,\mev$ above the $D\bar{D}^*$ or
$D^*\bar{D}^*$ thresholds. Unless future amplitude analyses find
values that are closer to the thresholds, this could be a challenge
for the molecular model.
Upcoming BES~III results on the $Z_c$ masses, widths, branching
fractions and spin-parities from the amplitude analyses, and on the
search for other decay channels ($\pi\psi(2S)$ and $\rho\eta_c$), are
crucial for interpreting the $Z_c$ states.

The $Z_c$ and $Z_b$ states provide a very rich testing ground for
phenomenological models and, given intensive experimental and
theoretical efforts, one can expect progress in understanding of the
hadronic systems near the open flavor thresholds.

\subsubsection{Quarkonium and quarkonium-like states above 
open flavor thresholds }
More than ten new charmonium and charmonium-like states well above the
\DDbar\ threshold have been observed in the last decade by the $B$-factories
and other experiments, see Table \ref{tab:qq_above_thresh}. 
We discuss first the states that can be assigned to vacant charmonium levels. 
In 2008 Belle observed the $\chi_{c2}(2P)$ state in $\cga\cga$ collisions, 
later confirmed by BaBar. Almost all of the $\chi_{c2}(2P)$ properties 
(diphoton width, full width, decay mode) are in nice agreement with the theory
expectations, only the mass of the state is $\sim 50\mev$ below 
potential model predictions.  Another two charmonium candidates (for
the third and fourth radial excitations of \etac) are observed by
Belle~\cite{Abe:2007jn,Abe:2007sya} in the double charmonium production
process $\epem \to\jpsi X(3940/4160)$, that decay to \DDst\ and
\DstDst\ channels, respectively. BaBar has not reported any studies of
these processes yet. While production processes and decay modes are
typical of conventional charmonium, the masses of these states are
significantly lower than potential model expectations (e.g.,
$\eta_c(4S)$ is expected to be $\sim 300\mev$ heavier than the observed
$X(4160)$). The assignment can be tested by studying the angular
distribution of the final state at Belle~II.

\begin{table*}[tb]
\footnotesize
\caption{Quarkonium-like states above the corresponding open flavor thresholds.
For charged states, the $C$-parity is given for the neutral members of the corresponding 
isotriplets.}
\label{tab:qq_above_thresh}
\begin{ruledtabular}
\renewcommand*{\arraystretch}{1.225}
\begin{tabular}{lllllclc}
State & $M,\,\mev$ & $\Gamma,\,\mev$ & $J^{PC}$ & Process (mode) & Experiment (\#$\sigma$) & Year & Status \\
\hline
      $Y(3915)$ & $3918.4\pm1.9$ & $20\pm5$ & $0/2^{?+}$ &
      $B\to K(\omega J/\psi)$ & Belle~\cite{Abe:2004zs} (8), BaBar~\cite{Aubert:2007vj,delAmoSanchez:2010jr} (19) &
      2004 & Ok \\
      & & & &
      $e^+e^-\to e^+e^-(\omega J/\psi)$ & Belle~\cite{Uehara:2009tx} (7.7), BaBar~\cite{Lees:2012xs} (7.6)
      & 2009 & Ok \\
      $\chi_{c2}(2P)$ & $3927.2\pm2.6$ & $24\pm6$ & $2^{++}$ &
      $e^+e^-\to e^+e^-(D\bar{D})$ & Belle~\cite{Uehara:2005qd} (5.3), BaBar~\cite{Aubert:2010ab} (5.8) &
      2005 & Ok \\
      $X(3940)$ & $3942^{+9}_{-8}$ & $37^{+27}_{-17}$ & $?^{?+}$ &
      $e^+e^-\to J/\psi\,(D\bar{D}^*)$ & Belle~\cite{Abe:2007jn,Abe:2007sya} (6) &
      2005 & NC! \\
      $Y(4008)$ & $3891\pm42$ & $255\pm42$ & $1^{--}$ &
      $e^+e^-\to (\pi^+\pi^-J/\psi)$ & Belle~\cite{Yuan:2007sj,Liu:2013dau} (7.4) & 
      2007 & NC! \\
      $\psi(4040)$ & $4039\pm1$ & $80\pm10$ & $1^{--}$ &
      $e^+e^-\to (D^{(*)}\bar{D}^{(*)}(\pi))$ & PDG~\cite{Beringer:1900zz} & 
      1978 & Ok \\
      & & & &
      $e^+e^-\to(\eta J/\psi)$ & Belle~\cite{Wang:2012bgc} (6.0) & 
      2013 & NC! \\
      $Z(4050)^+$ & $4051^{+24}_{-43}$ & $82^{+51}_{-55}$ & $?^{?+}$ &
      $\bar{B}^0\to K^-(\pi^+\chi_{c1})$ & Belle~\cite{Mizuk:2008me} (5.0), BaBar~\cite{Lees:2011ik} (1.1) & 
      2008 & NC! \\
      $Y(4140)$ & $4145.8\pm2.6$ & $18\pm8$ & $?^{?+}$ &
      $B^+\to K^+(\phi J/\psi)$ & CDF~\cite{Aaltonen:2011at} (5.0), Belle~\cite{Brodzicka:2010zz} (1.9), & 
      2009 & NC! \\
      & & & & & LHCb~\cite{Aaij:2012pz} (1.4), CMS~\cite{Chatrchyan:2013dma} ($>$5) & & \\
      & & & & & D0~\cite{Abazov:2013xda} (3.1) & & \\
      $\psi(4160)$ & $4153\pm3$ & $103\pm8$ & $1^{--}$ &
      $e^+e^-\to (D^{(*)}\bar{D}^{(*)})$ & PDG~\cite{Beringer:1900zz} & 
      1978 & Ok \\
      & & & &
      $e^+e^-\to(\eta J/\psi)$ & Belle~\cite{Wang:2012bgc} (6.5) & 
      2013 & NC! \\
      $X(4160)$ & $4156^{+29}_{-25}$ & $139^{+113}_{-65}$ & $?^{?+}$ &
      $e^+e^-\to J/\psi\,(D^*\bar{D}^*)$ & Belle~\cite{Abe:2007sya} (5.5) &
      2007 & NC! \\
     $Z(4200)^+$ & $4196^{+35}_{-30}$ & $370^{+99}_{-110}$ & $1^{+-}$ &
     $\bar{B}^0\to K^-(\pi^+J/\psi)$ & Belle~\cite{Chilikin:MoriondQCD2014} (7.2) &
     2014 & NC! \\
      $Z(4250)^+$ & $4248^{+185}_{-45}$ & $177^{+321}_{-72}$ & $?^{?+}$ &
      $\bar{B}^0\to K^-(\pi^+\chi_{c1})$ & Belle~\cite{Mizuk:2008me} (5.0), BaBar~\cite{Lees:2011ik} (2.0) & 
      2008 & NC! \\
      $Y(4260)$ & $4250\pm9$ & $108\pm12$ & $1^{--}$ &
      $e^+e^-\to (\pi\pi J/\psi)$ & BaBar~\cite{Aubert:2005rm,Lees:2012cn} (8), CLEO~\cite{Coan:2006rv,He:2006kg} (11) & 
      2005 & Ok \\
      & & & & & Belle~\cite{Yuan:2007sj,Liu:2013dau} (15), BES~III~\cite{Ablikim:2013mio} (np) & & \\
      & & & &
      $e^+e^-\to(f_0(980)J/\psi)$ & BaBar~\cite{Lees:2012cn} (np), Belle~\cite{Liu:2013dau} (np) & 
      2012 & Ok \\
      & & & &
      $e^+e^-\to(\pi^-Z_c(3900)^+)$ & BES~III~\cite{Ablikim:2013mio} (8), Belle~\cite{Liu:2013dau} (5.2) & 
      2013 & Ok \\
      & & & &
      $e^+e^-\to(\gamma\,X(3872))$ & BES~III~\cite{Ablikim:2013vva} (5.3) & 
      2013 & NC! \\
      $Y(4274)$ & $4293\pm20$ & $35\pm16$ & $?^{?+}$ &
      $B^+\to K^+(\phi J/\psi)$ & CDF~\cite{Aaltonen:2011at} (3.1), LHCb~\cite{Aaij:2012pz} (1.0), & 
      2011 & NC! \\
      & & & & & CMS~\cite{Chatrchyan:2013dma} ($>$3), D0~\cite{Abazov:2013xda} (np) & & \\
      $X(4350)$ & $4350.6^{+4.6}_{-5.1}$ & $13^{+18}_{-10}$ & $0/2^{?+}$ &
      $e^+e^-\to e^+e^-(\phi J/\psi)$ & Belle~\cite{Shen:2009vs} (3.2) & 
      2009 & NC! \\
      $Y(4360)$ & $4354\pm11$ & $78\pm16$ & $1^{--}$ &
      $e^+e^-\to(\pi^+\pi^-\psi(2S))$ & Belle~\cite{Wang:2007ea} (8), BaBar~\cite{Lees:2012pv} (np) & 
      2007 & Ok \\
      $Z(4430)^+$ & $4458\pm15$ & $166^{+37}_{-32}$ & $1^{+-}$ &
      $\bar{B}^0\to K^-(\pi^+\psi(2S))$ & Belle~\cite{Mizuk:2009da,Chilikin:2013tch} (6.4), BaBar~\cite{Aubert:2008aa} (2.4) &
      2007 & Ok \\
      & & & & & LHCb~\cite{Aaij:2014jqa} (13.9) & & \\
      & & & &
      $\bar{B}^0\to K^-(\pi^+J/\psi)$ & Belle~\cite{Chilikin:MoriondQCD2014} (4.0) &
      2014 & NC! \\
      $X(4630)$ & $4634^{+9}_{-11}$ & $92^{+41}_{-32}$ & $1^{--}$ &
      $e^+e^-\to(\Lambda_c^+\bar{\Lambda}_c^-)$ & Belle~\cite{Pakhlova:2008vn} (8.2) & 
      2007 & NC! \\
      $Y(4660)$ & $4665\pm10$ & $53\pm14$ & $1^{--}$ &
      $e^+e^-\to(\pi^+\pi^-\psi(2S))$ & Belle~\cite{Wang:2007ea} (5.8), BaBar~\cite{Lees:2012pv} (5) & 
      2007 & Ok \\
      $\Upsilon(10860)$ & $10876\pm11$ & $55\pm28$ & $1^{--}$ &
      $e^+e^-\to(B_{(s)}^{(*)}\bar{B}_{(s)}^{(*)}(\pi))$ & PDG~\cite{Beringer:1900zz} & 
      1985 & Ok \\
      & & & &
      $e^+e^-\to(\pi\pi\Upsilon(1S,2S,3S))$ & Belle~\cite{Abe:2007tk,Belle:2011aa,Krokovny:2013mgx} ($>$10) & 2007 & Ok \\
      & & & &
      $e^+e^-\to(f_0(980)\Upsilon(1S))$ & Belle~\cite{Belle:2011aa,Krokovny:2013mgx} ($>$5) & 2011 & Ok \\
      & & & &
      $e^+e^-\to(\pi Z_b(10610,10650))$ & Belle~\cite{Belle:2011aa,Krokovny:2013mgx} ($>$10) & 2011 & Ok \\
      & & & &
      $e^+e^-\to(\eta\Upsilon(1S,2S))$ & Belle~\cite{Krokovny:LaThuile2012} (10) & 2012 & Ok \\
      & & & &
      $e^+e^-\to(\pi^+\pi^-\Upsilon(1D))$ & Belle~\cite{Krokovny:LaThuile2012} (9) & 2012 & Ok \\
      $Y_b(10888)$ & $10888.4\pm3.0$ & $30.7^{+8.9}_{-7.7}$ & $1^{--}$ &
      $e^+e^-\to(\pi^+\pi^-\Upsilon(nS))$ & Belle~\cite{Chen:2008xia} (2.3) & 
      2008 & NC! \\
\end{tabular}
\end{ruledtabular}
\end{table*}

For the majority of the other new particles, their assignments to the
ordinary charmonium states are not well recognized.  Contrary to expectations,
most of the new states above the open charm threshold, the so-called 
"$XYZ$ states", decay into final states containing
charmonium, but do not decay into open charm pairs with a detectable
rate. This is the main reason why they are discussed as candidates for
exotic states. An extended discussion on the different interpretations
of these states can be found in \cite{Brambilla:2010cs} and references
therein. In the following we discuss recent results on the states
above open heavy flavor thresholds.

BaBar confirmed the observation of the process $\gamma\gamma\to
Y(3915)\to\omega J/\psi$ that was observed by Belle in
2009~\cite{Lees:2012xs}. From angular analyses BaBar determined the
$Y(3915)$ spin-parity to be $J^P=0^+$~\cite{Lees:2012xs}. In this
analysis it is assumed that in the alternative hypothesis of $J=2$ it
is produced in the helicity-two state, analogous to the production of
$\chi_{c2}(1P)$. Given the unknown nature of $Y(3915)$, this assumption
could be unjustified. The $J^P=0^+$ state can decay to $D\bar{D}$ in
S-wave. Since $Y(3915)$ is $200\,\mev$ above the $D\bar{D}$ threshold,
its width of $20\,\mev$ looks extremely narrow and points to its
exotic nature. In addition, the mass difference relative to
$\chi_{c2}(2P)$ of $9\,\mev$ is too small~\cite{Guo:2012tv} to
interpret the $Y(3915)$ as $\chi_{c0}(2P)$.

CMS and D0 studied the $B^+\to K^+\phi J/\psi$
decays~\cite{Chatrchyan:2013dma,Abazov:2013xda} and confirmed the
$Y(4140)$ state near the $\phi J/\psi$ threshold that was observed in
2008 by CDF~\cite{Aaltonen:2011at}. The experiments also see a second
structure, the $Y(4274)$, though the mass measurements agree only at
about $3\,\sigma$ level. The background under the $Y(4274)$ can be
distorted by reflections from the $K^{*+}\to\phi K^+$ decays, which
makes an estimate of the $Y(4274)$ significance
difficult~\cite{Chatrchyan:2013dma}. The $Y(4140)$ and $Y(4274)$
states were not seen in $B$ decays by Belle~\cite{Brodzicka:2010zz}
and LHCb~\cite{Aaij:2012pz} and in $\gamma\gamma$ collisions by
Belle~\cite{Shen:2009vs}. Amplitude analyses with increased statistics
at the LHC could settle the controversy.

BaBar updated the $e^+e^-\to\pi^+\pi^-\psi(2S)$ study using ISR
photons and confirmed the $Y(4660)$ that was earlier observed by
Belle~\cite{Lees:2012pv}. Both BaBar and Belle updated the
$e^+e^-\to\pi^+\pi^- J/\psi$
analyses~\cite{Lees:2012cn,Liu:2013dau}. Belle confirms the $Y(4008)$
using an increased data sample. However, the mass becomes smaller,
$M=3891\pm42\,\mev$. BaBar sees events in the same mass region,
but they are attributed to a contribution with an exponential
shape. BES~III data taken in this region will help to clarify the
existence of the $Y(4008)$ resonance.

BES~III measured the $e^+e^-\to\pi^+\pi^-h_c$ cross-section at several
energies above $4\,\gev$~\cite{Ablikim:2013wzq}. Unlike the
$e^+e^-\to\pi^+\pi^-h_b$ reaction, the final three-body state is
mainly nonresonant. The shape of the cross section looks different
from that of the $\pi^+\pi^-J/\psi$ final state and possibly exhibits
structures distinct from known $Y$ states~\cite{Yuan:2013lma}. Since
hybrids contain a $c\bar{c}$ pair in the spin-singlet state, such
structures could be candidates for hybrids. A more detailed scan by
BES~III is underway.

Belle performed the full amplitude analysis of the $B^0\to
K^+\pi^-\psi(2S)$ decays to determine the spin-parity of the 
$Z(4430)^{\pm}$~\cite{Chilikin:2013tch}, which is the first charged
quarkonium-like state observed by Belle in
2007~\cite{Choi:2007wga,Mizuk:2008me}. The $J^P=1^+$ hypothesis is
favored over the $0^-$, $1^-$ and $2^-$ and $2^+$ hypotheses at the
levels of $3.4\,\sigma$, $3.7\,\sigma$, $4.7\,\sigma$ and
$5.1\,\sigma$, respectively. The width of the $Z(4430)^{\pm}$ became
broader, $\Gamma=200^{+49}_{-58}\,\mev$. This state and two more
states $Z(4050)^{\pm}$ and $Z(4250)^{\pm}$ in the $\pi^{\pm}\chi_{c1}$
channel are not confirmed by BaBar~\cite{Aubert:2008aa,Lees:2011ik}.
The long-standing question of the $Z(4430)^\pm$'s existence has finally
been settled by the LHCb experiment, which confirmed both the state itself
and its spin-parity assignment of $1^+$~\cite{Aaij:2014jqa}.
For the first time, LHCb has demonstrated resonant behavior of the
$Z(4430)^\pm$ amplitude.
Belle has performed a full amplitude analysis of the $\bar{B}^0\to K^-\pi^+J/\psi$
decays and observed a new charged charmonium-like state $Z(4200)^+$ and 
evidence for the $Z(4430)^+$ decay to $\pi^+J/\psi$~\cite{Chilikin:MoriondQCD2014}. 
This decay is within the reach of LHCb. 
Further studies of $Z(4050)^{\pm}$ and $Z(4250)^{\pm}$ could be more
difficult at LHCb because of soft photons in the final state and might
have to wait for Belle~II to run.

Given that the signals of $Y(4260)$, $Y(4360)$ and $Y(4660)$ are not
seen in the $e^+e^-\to\mathrm{hadrons}$ cross section ($R_c$ scan), one
can set the limit
$\Gamma[Y\to\pi^+\pi^-\psi]\gtrsim1\,\mev$~\cite{Mo:2006ss}. This is
at least one order of magnitude higher than that of $\psi(2S)$ and
$\psi(3770)$. Recently Belle found that  $\psi$ states seen as
prominent peaks in the $R_c$ scan, can also have anomalous hadronic transitions
to lower charmonia. Belle observed $\psi(4040)$ and $\psi(4160)$
signals in the $\epem\to\eta\jpsi$ cross-section measured using
ISR~\cite{Wang:2012bg}. The partial widths are measured to be
$\Gamma[\psi(4040,4160)\to\eta\jpsi]\sim1\,\mev$. Thus it seems to be
a general feature of all charmonium(-like) states above the open charm
thresholds to have intense hadronic transitions to lower charmonia. A
similar phenomenon is found in the bottomonium sector: In 2008 Belle
observed anomalously large rates of the $\UnS{5}\to\dipi\UnS{n}$
($n=1,~2,~3$) transitions with partial widths of
$300-400\,\kev$~\cite{Abe:2007tk}. Recently Belle reported preliminary
results on the observation of $\UnS{5}\to\eta\Ups(1S,2S)$ and $\UnS{5}
\to\dipi\UoneD$ with anomalously large rates~\cite{Krokovny:LaThuile2012}. 
It is proposed that these anomalies are due to
rescatterings~\cite{Simonov:2007cj,Meng:2007tk}. The large branching
fraction of the $\UnS{4}\to\UnS{1}\eta$ decay observed in 2010 by
BaBar could have a similar origin~\cite{Voloshin:2011hw}.
The mechanism can be considered either as a rescattering of
the \DDbar\ or $B\bar{B}$ mesons, or as a contribution of the
molecular component to the quarkonium wave function.
The model in which $Y(4260)$ is a $D_1(2420)\bar{D}$ molecule
naturally explains the high probability of the intermediate molecular
resonance in the $Y(4260)\to\pi^+\pi^-J/\psi$
transitions~\cite{Wang:2013cya,Cleven:2013mka} and predicts the
$Y(4260)\to\gamma X(3872)$ transitions with high
rates~\cite{Guo:2013zbw}. Such transitions have recently been observed
by BES~III, with~\cite{Ablikim:2013vva}
\begin{equation}
\frac{\sigma[e^+e^-\to\gamma X(3872)]}{\sigma[e^+e^-\to\pi^+\pi^-J/\psi]}\sim11\%.    
\end{equation}
\pagebreak

Despite striking similarities between the observations in the
charmonium and bottomonium sectors, there are also clear
differences. In the charmonium sector, each of the $Y(3915)$,
$\psi(4040)$, $\psi(4160)$, $Y(4260)$, $Y(4360)$ and $Y(4660)$ decays
to only one particular final state with charmonium [$\omega\jpsi$,
$\eta\jpsi$, $\dipi\jpsi$ or $\dipi\psip$]. In the bottomonium sector,
there is one state with anomalous properties, the $\UnS{5}$, and it
decays to different channels with similar rates [$\dipi\UnS{n}$,
$\dipi\chbm$, $\dipi\UoneD$, $\eta\UnS{n}$]. There is no general
model describing these peculiarities. To explain the affinity of the
charmonium-like states to some particular channels, the notion of
``hadrocharmonium'' was proposed in~\cite{Dubynskiy:2008mq}. It is a
heavy quarkonium embedded into a cloud of light hadron(s), thus the
fall-apart decay is dominant. Hadrocharmonium could also provide an
explanation for the charged charmonium-like states $Z(4430)^+$,
$Z(4050)^+$ and $Z(4250)^+$.

\subsubsection{Summary}
Quarkonium spectroscopy enjoys an intensive flood of new results.  The
number of spin-singlet bottomonium states has increased from one to
four over the last two years, including a more precise measurement of
the \etab\ mass, $11\,\mev$ away from the PDG2012 average.  
There is evidence for one of the two still missing narrow
charmonium states expected in the region between the \DDbar\
and \DDst\ thresholds.  Observations and detailed studies of the {\it
charged} bottomonium-like states
\czbo\ and \czbt\ and first results on the charged charmonium-like
states $Z_c$ open a rich phenomenological field to study exotic states
near open flavor thresholds. There is also significant progress and a
more clear experimental situation for the highly excited heavy
quarkonium states above open flavor thresholds. Recent highlights
include confirmation of the $Y(4140)$ state by CMS and D0, observation
of the decays $\psi(4040,4160)\to\eta J/\psi$ by Belle, measurement
of the energy dependence of the $e^+e^-\to\pi^+\pi^-h_c$ cross section
by BES~III, observation of the $Y(4260)\to\gamma X(3872)$ by BES~III and
determination of the $Z(4430)$ spin-parity from full amplitude
analysis by Belle. A general feature of highly excited states is their
large decay rate to lower quarkonia with the emission of light
hadrons. Rescattering is important for understanding their properties, 
however, there is no general model explaining their decay patterns. 
The remaining experimental open questions or controversies are within 
the reach of the LHC or will have to wait for the next generation $B$-factory.

From the theoretical point of view, low quarkonium excitations are in
agreement with lattice QCD and effective field theories calculations,
which are quite accurate and able to challenge the accuracy of the
data. Higher quarkonium excitations show some unexpected
properties. Specific effective field theories have been developed for
some of these excitations. Lattice studies provide a qualitative
guide, but in most cases theoretical expectations still rely on models
and a quantitative general theory is still missing.

\subsection{Strong coupling $\alpha_s$}

\label{sec:secC5}

There are several heavy-quark systems that are suitable for a precise
determination of $\als$, mainly involving quarkonium, or
quarkonium-like, configurations, which are basically governed by the
strong interactions. One can typically take advantage of
nonrelativistic effective theories, high-order perturbative calculations that are available for these
systems, and of progress in lattice computations. 

Using moments of
heavy-quark correlators calculated on the lattice, and the continuum
perturbation theory results for them~\cite{Allison:2008xk}, the HPQCD collaboration has obtained
$\als(M_Z)=0.1183\pm0.0007$~\cite{McNeile:2010ji}. This result is very close, both in the 
central value and error, to the one
obtained from measuring several quantities related to short-distance
Wilson loops by the same collaboration~\cite{McNeile:2010ji}. The energy between
two static sources in the fundamental representation, 
as a function of its separation, is also suitable for a precise $\als$
extraction. The perturbative computation has now reached
a three-loop level~\cite{Anzai:2009tm,Smirnov:2009fh,Brambilla:2006wp,Brambilla:2009bi,%
Brambilla:2010pp,Tormo:2013tha}, and lattice-QCD results with $n_f=2+1$ sea quarks
are available~\cite{Bazavov:2011nk}. A comparison of the two gives
$\als(M_Z)=0.1156^{+0.0021}_{-0.0022}$~\cite{Bazavov:2012ka}. New lattice data for the
static energy, including points at shorter distances, will be
available in the near future, and an update of the result for $\als$
can be expected, in principle with reduced errors.

Quarkonium decays, or more precisely ratios of their widths (used to reduce the
sensitivity to long-distance effects), were readily
identified as a good place for $\als$ extractions. One complication is
the dependence on color-octet configurations. The best ratio for
$\als$ extractions, in the sense that the sensitivity to color-octet
matrix elements and relativistic effects is most reduced, turns out to
be $R_{\gamma}:=\Gamma(\Upsilon\to\gamma X)/\Gamma(\Upsilon\to X)$,
from which one obtains
$\als(M_Z)=0.119^{+0.006}_{-0.005}$~\cite{Brambilla:2007cz}. The main
uncertainty in this result comes from the systematic errors of the
experimental measurement of $R_{\gamma}$~\cite{Besson:2005jv}. Belle
could be able to produce an improved measurement of $R_{\gamma}$,
which may translate into a better $\als$ determination.

Very recently the CMS collaboration has presented a determination of
$\als$ from the measurement of the inclusive cross section for
$t\bar{t}$ production, by comparing it with the NNLO QCD
prediction. The analysis is performed with different NNLO PDF sets,
and the result from the NNPDF set is used as the main
result. Employing $m_t=173.2\pm1.4$~GeV,
$\als(M_Z)=0.1151^{+0.0033}_{-0.0032}$ is obtained~\cite{Chatrchyan:2013haa}, the first $\als$ determination from top-quark production.

\subsection{Heavy quarkonium production}

\label{sec:secC6}

Forty years after the discovery of the $\jpsi$, the mechanism underlying quarkonium production has still not been clarified. Until the mid-90s mostly the traditional color singlet model was used in perturbative cross section calculations. The dramatic failure to describe $\jpsi$ production at the Tevatron 
led, however, to a search for alternative explanations. The NRQCD factorization conjecture has by now received most acceptance, although not yet being fully established.

\subsubsection{Summary of recent experimental progress}
\label{subsec:summaryrep}
The past couple of years have seen incredible progress in measurements of quarkonium production observables, which was mainly, but not solely, due to the operation of the different LHC experiments. Here we will give an overview of the most remarkable results of the past years.

The production rates of a heavy quarkonium $H$ are split into direct, prompt, and nonprompt contributions. Direct production refers to the production of $H$ directly at the interaction point of the initial particles, while prompt production also includes production via radiative decays of higher quarkonium states, called {\em feed-down} contributions. Nonprompt production refers to all other production mechanisms, mainly the production of charmonia from decaying $B$ mesons, which can be identified by a second decay vertex displaced from the interaction point.

\paragraph{$\jpsi$ production in $pp$ collisions}

The 2004 CDF transverse momentum $p_T$ distribution measurement of the $\jpsi$ production cross section \cite{Acosta:2004yw} is still among the most precise heavy quarkonium production measurements. But since theory errors in all models for heavy quarkonium production are still much larger than today's experimental errors, it is in general not higher precision which is needed from the theory side, but rather new and more diverse production observables. And this is where the LHC experiments have provided very important input. As for the $\jpsi$ hadroproduction cross section, they have extended the CDF measurement \cite{Acosta:2004yw} into new kinematic regions: Obviously, the measurements have been performed at much higher center-of-mass energies than before, namely at $\sqrt{s}=2.76$, 7, and 8~TeV. But more important for testing quarkonium production models is the fact that there are measurements which exceed the previously 
measured $p_T$ range both at high $p_T$, as by ATLAS \cite{Aad:2011sp} and CMS \cite{Chatrchyan:2011kc}, and at low $p_T$, as in the earlier CMS measurement \cite{Khachatryan:2010yr}, but also in the recent measurement by the PHENIX collaboration at RHIC \cite{Adare:2011vq}. We note that this list is not complete, and that there have been many more $\jpsi$ hadroproduction measurements recently than those cited here.

\paragraph{$\psi(2S)$ and $\chi_c$ production in $pp$ collisions}

$\jpsi$ is the quarkonium which is easiest to be measured due to the large branching ratio of its leptonic decay modes, but in recent years, high precision measurements have been also performed for the $\psi(2S)$, namely by the CDF \cite{Aaltonen:2009dm}, the CMS \cite{Chatrchyan:2011kc}, and the LHCb \cite{Aaij:2012ag} collaborations. Also the $\chi_c$ production cross sections were measured via their decays into $\jpsi$ by LHCb \cite{LHCb:2012af}, the first time since the CDF measurement \cite{Affolder:2001ij} in 2001.
The $\chi_{c2}$ to $\chi_{c1}$ production ratio was measured at LHCb \cite{LHCb:2012ac}, CMS \cite{Chatrchyan:2012ub} and previously by CDF \cite{Abulencia:2007bra}.
These measurements are of great importance for the theory side since they allow fits of NRQCD LDMEs for these charmonia and  determine direct $\jpsi$ production data, which can in turn be compared to direct production theory predictions.

\paragraph{$\Upsilon$ production in $pp$ collisions}

$\Upsilon(1S)$, $\Upsilon(2S)$, and $\Upsilon(3S)$ production cross sections were measured at the LHC by ATLAS \cite{:2012iq} and LHCb \cite{LHCb:2012aa,Aaij:2013yaa}. Previously, $\Upsilon$ was produced only at the Tevatron \cite{Acosta:2001gv,Abazov:2005yc}.

\paragraph{Polarization measurements in $pp$ collisions}
\label{par:cpolmeasurements}

\begin{figure*}[p]
\begin{minipage}[c]{0.495\linewidth}
   \includegraphics*[width=0.8\linewidth]{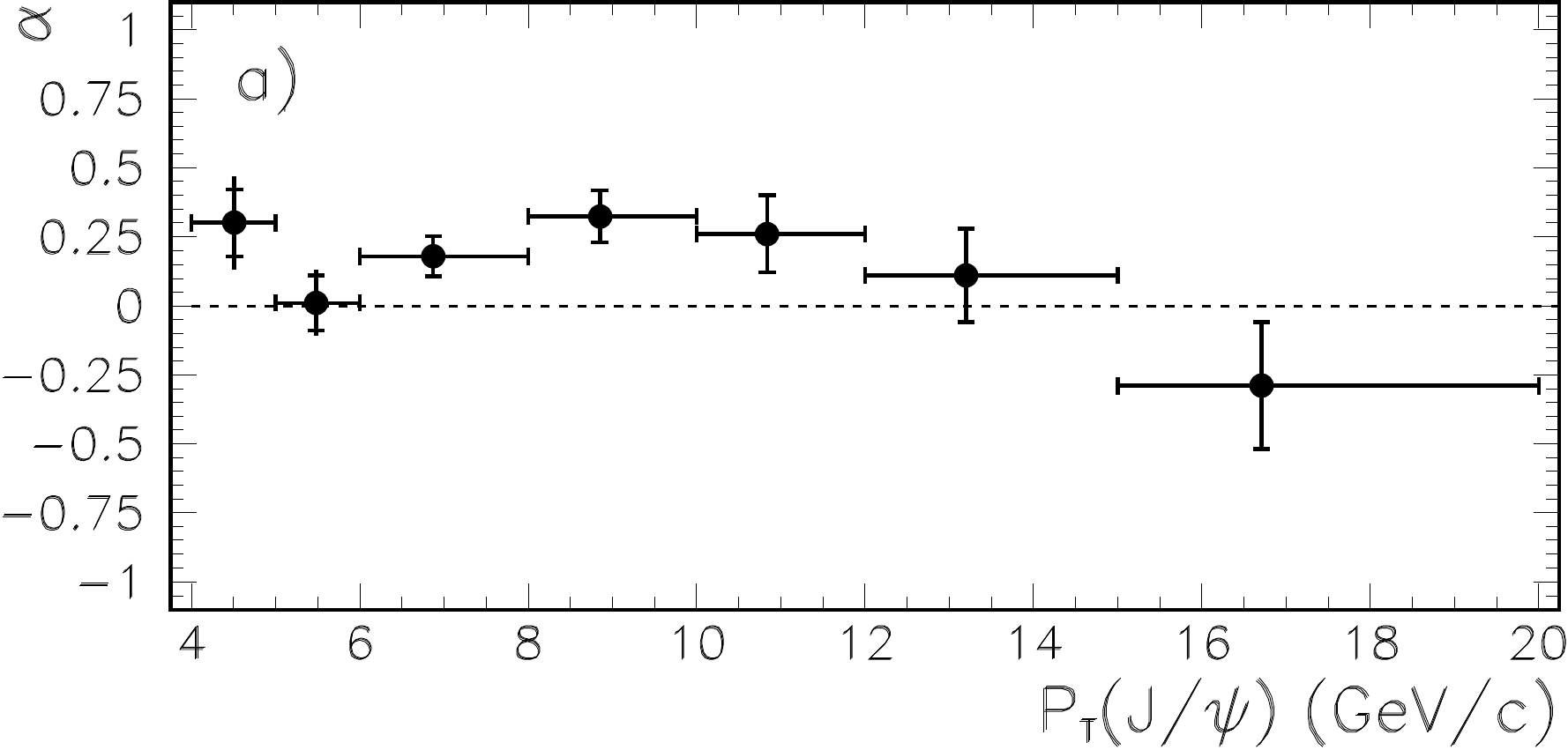}
   \includegraphics*[width=0.65\linewidth]{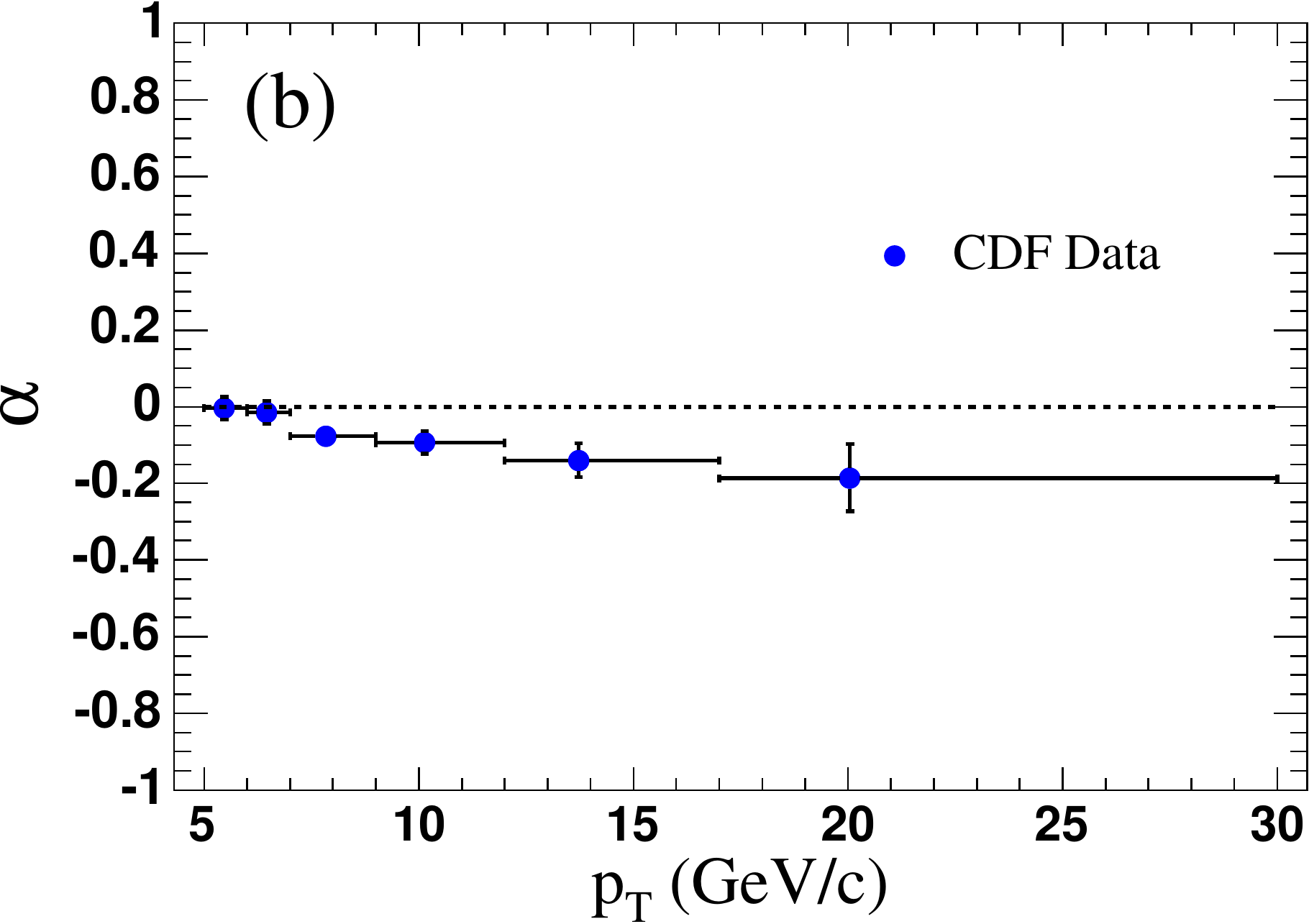}
\end{minipage}
\begin{minipage}[c]{0.495\linewidth}
  \setlength{\unitlength}{\linewidth}
   \includegraphics*[width=\linewidth]{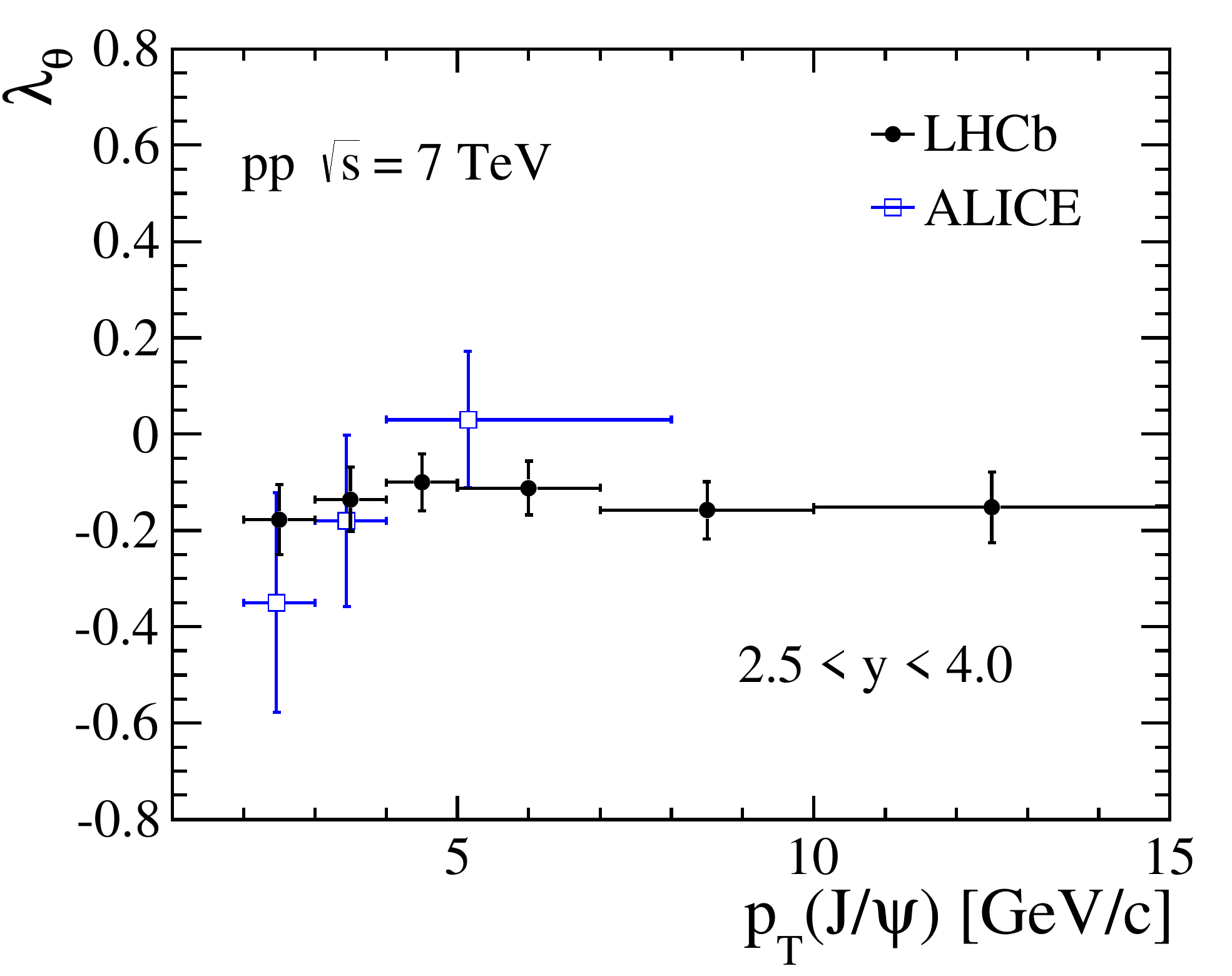}  
\end{minipage}

  \caption{The $\jpsi$ polarization parameter $\alpha\equiv\lambda_\theta$ in the helicity frame as measured by CDF in Tevatron run I \cite{Affolder:2000nn} (a), run II  \cite{Abulencia:2007us} (b), and by ALICE \cite{Abelev:2011md} and LHCb \cite{Aaij:2013nlm} at the LHC (right).
            \AfigPermsAPS{Affolder:2000nn,Abulencia:2007us,Aaij:2013nlm}{2000, 2007, 2013}
            }
\label{fig:cJPsiPolA}
\end{figure*}

\begin{figure*}[p]
\includegraphics*[width=0.8\linewidth]{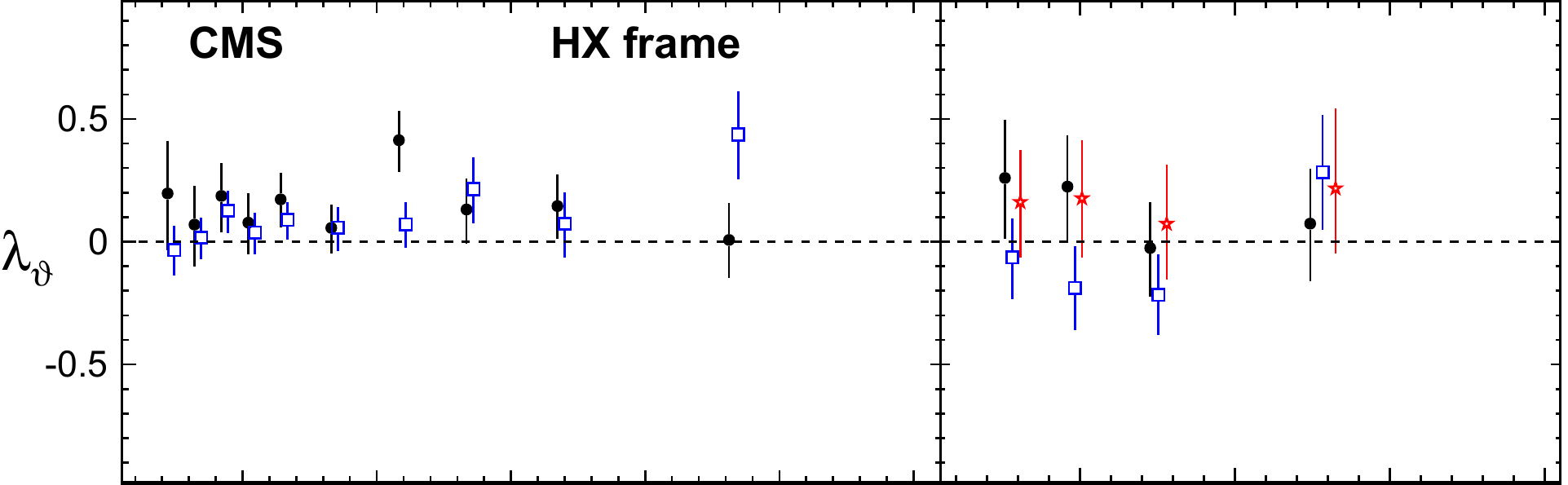}
\\
\vspace{-0.095\linewidth}
\hspace{-0.05\linewidth}
\includegraphics*[width=0.15\linewidth]{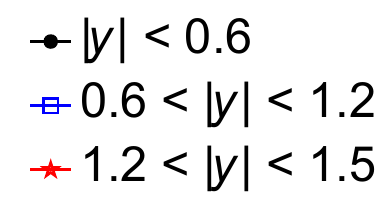}
\\
\vspace{0.02\linewidth}
\includegraphics*[width=0.8\linewidth]{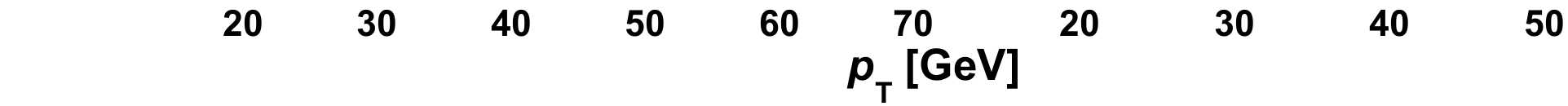}
\caption{The polarization parameter $\lambda_\theta$ in the helicity frame for $J/\psi$ (left) and $\psi(2S)$ (right) production as measured by CMS \cite{Chatrchyan:2013cla}. Adapted from \cite{Chatrchyan:2013cla}.}
\label{fig:cJPsiPolB}
\end{figure*}

\begin{figure*}[p]
\hspace{-20pt}
\begin{minipage}[c]{0.49\linewidth}
   \includegraphics*[width=0.95\linewidth]{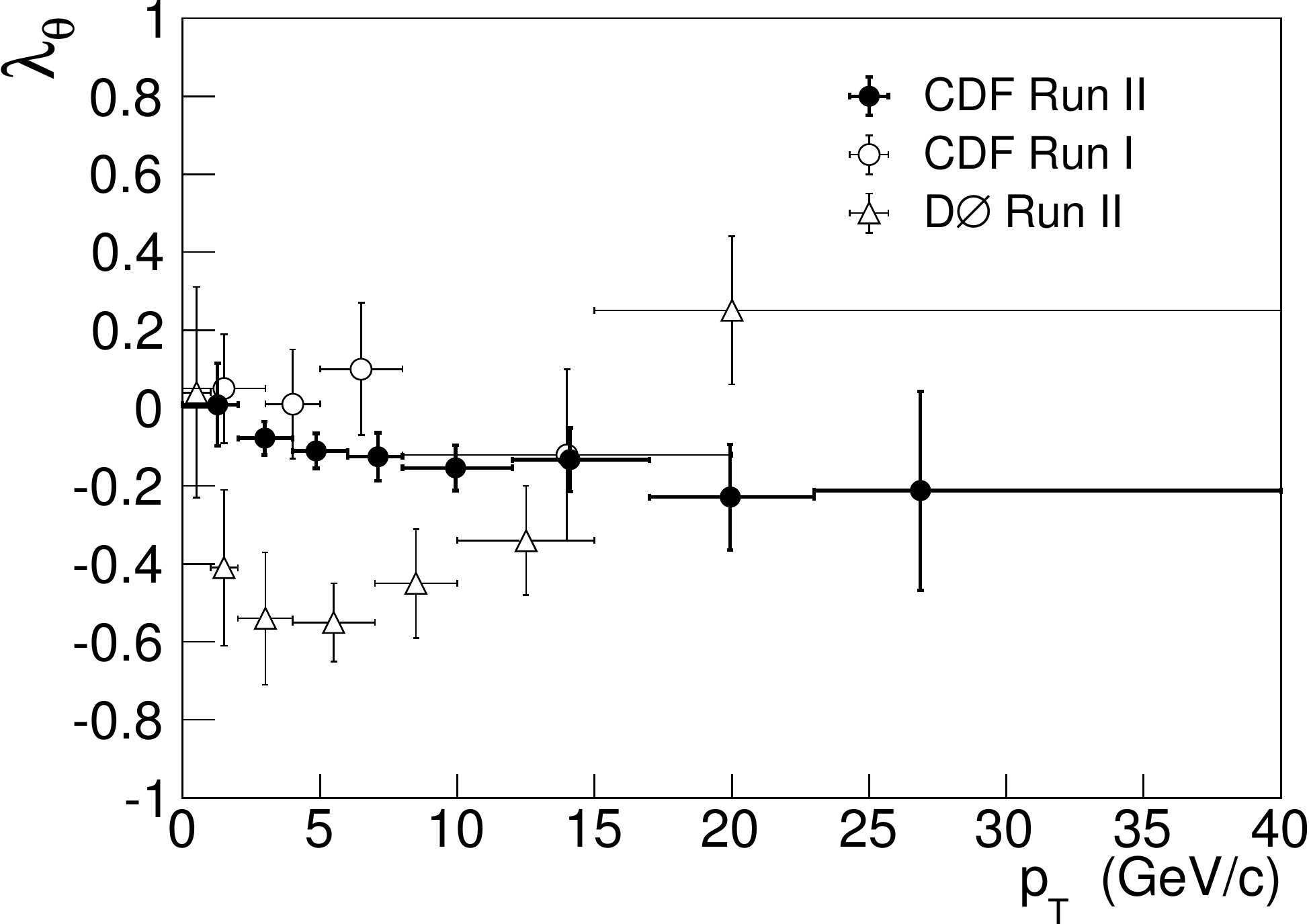}
\end{minipage}
\begin{minipage}[c]{0.49\linewidth}
   \vspace{5pt}
   \includegraphics*[width=0.98\linewidth]{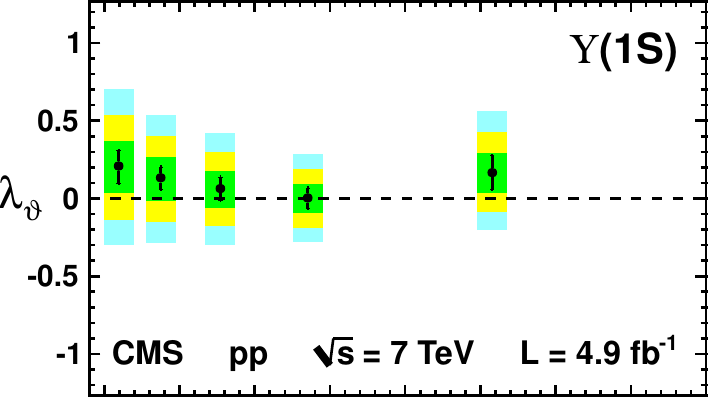}
   \includegraphics*[width=0.98\linewidth]{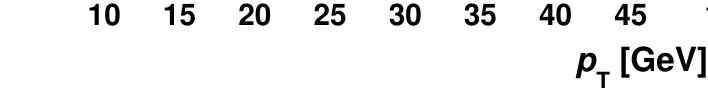}
\end{minipage}
  \caption{The $\Upsilon(1S)$ polarization parameter $\lambda_\theta$ in the helicity frame as measured by CDF \cite{Acosta:2001gv,CDF:2011ag}, D0 \cite{Abazov:2008aa} and CMS \cite{:2012pia}.
            \AfigPermsAPS{CDF:2011ag,:2012pia}{2012, 2012}
            }
\label{fig:cUpsilonPol}
\end{figure*}

The measurements of the angular distributions of the quarkonium decay leptons are among the most challenging experimental tasks in quarkonium physics, because much more statistics and a much better understanding of the detector acceptances than in cross section measurements is needed. These angular distributions $W(\theta,\phi)$ are directly described by the polarization parameters $\lambda_\theta$, $\lambda_\phi$, and $\lambda_{\theta\phi}$ via
\begin{eqnarray}
W(\theta,\phi)&\propto&1+\lambda_\theta\cos^2\theta
+\lambda_\phi\sin^2\theta\cos(2\phi)
\nonumber\\
&&{}+\lambda_{\theta\phi}\sin(2\theta)\cos\phi,
\end{eqnarray}
where $\theta$ and $\phi$ are,  respectively, the polar and azimuthal angles of the positively charged decay lepton in the quarkonium rest frame.
These polarization measurements pose highly nontrivial tests for quarkonium production models, and have therefore probably been the most anticipated LHC results on quarkonium. Previous Tevatron measurements tended to give ambiguous results: The CDF measurements of $\jpsi$ polarization in Tevatron run I \cite{Affolder:2000nn} and II \cite{Abulencia:2007us} have been in partial disagreement, see Fig.~\ref{fig:cJPsiPolA}, similar to 
the $\Upsilon(1S)$ polarization measured by D0 \cite{Abazov:2008aa} and by CDF in Tevatron run I \cite{Acosta:2001gv} and II \cite{CDF:2011ag}, see Fig.~\ref{fig:cUpsilonPol}. At RHIC, $\jpsi$ polarization has been measured by PHENIX \cite{Adare:2009js} and STAR \cite{Adamczyk:2013vjy}. At the LHC, $J/\psi$ polarization has so far been measured by ALICE \cite{Abelev:2011md}, LHCb \cite{Aaij:2013nlm}, and CMS \cite{Chatrchyan:2013cla}, see Figs.~\ref{fig:cJPsiPolA} and \ref{fig:cJPsiPolB}. Furthermore, CMS has measured the $\psi(2S)$ \cite{Chatrchyan:2013cla} and $\Upsilon$ \cite{:2012pia} polarization, see Figs.~\ref{fig:cJPsiPolB} and \ref{fig:cUpsilonPol}. None of the CDF Tevatron run II and the LHC measurements have found a strong and significant transverse or longitudinal polarization for any quarkonium. CDF at Tevatron run II and LHCb do, however, seem to prefer slight longitudinal polarizations in their helicity frame quarkonium polarization measurements, whereas in the CMS measurements there seems to 
be a tendency for slight transverse polarizations in the helicity frame,
see Figs.~\ref{fig:cJPsiPolA},~\ref{fig:cJPsiPolB}~and \ref{fig:cUpsilonPol}.

\paragraph{Recent $ep$ collision results}

For testing theory predictions, in particular the universality of NRQCD long distance matrix elements, we need to consider experimental data from a variety of different production mechanisms. Very important charmonium production data has thereby in the past come from inelastic photoproduction at the $ep$ collider HERA, which came in distributions in the transverse charmonium momentum $p_T$, the photon-proton invariant mass $W$ and the inelasticity variable $z$. The latest update on inclusive $\jpsi$ production cross sections was in 2012 by the ZEUS collaboration \cite{Abramowicz:2012dh}. This publication also presented values for $\sigma(\psi(2S))/\sigma(\jpsi)$ with error bars reduced by about two thirds relative to the previous ZEUS measurement \cite{Chekanov:2002at} at HERA 1. The $\jpsi$ polarization measurements by the ZEUS \cite{Chekanov:2009ad} and H1 \cite{Aaron:2010gz} collaborations were, however, still associated with such large errors that no unambiguous picture of the $\jpsi$ polarization in 
photoproduction 
emerged. Furthermore, no $\Upsilon$ photoproduction could be observed at HERA. Therefore, from the theory side, a new $ep$ collider at much higher energies and luminosities than HERA, like possibly an LHeC, would be highly desired. On the other hand, there is still no NLO calculation for $\jpsi$ production in deep inelastic scattering available, as, for example, measured most recently by H1 \cite{Aaron:2010gz}.

\paragraph{Further production observables} 
The LHCb experiment with its especially rich quarkonium program has also measured completely new observables which still need to be exploited fully in theory tests: For the first time in $pp$ collisions the double $\jpsi$ production cross section was measured \cite{Aaij:2011yc}, as well as the production of $\jpsi$ in association with charmed mesons \cite{Aaij:2012dz}. Like double charmonium production, $\jpsi+c\overline{c}$ was previously only measured at the $B$ factories, latest in the Belle analysis \cite{Pakhlov:2009nj}, which was crucial for testing $\jpsi$ production mechanisms in $e^+e^-$ production. $J/\psi$ production in association with $W$ bosons has for the first time been measured by the ATLAS collaboration \cite{Aad:2014rua}. Exclusive charmonium hadroproduction has been observed recently by CDF \cite{Aaltonen:2009kg} and LHCb \cite{Aaij:2013jxj,Aaij:2014iea}.
Exclusive production had previously been a domain of $ep$ experiments, see \cite{Alexa:2013xxa} for a recent update by the H1 collaboration. Another observable for which theory predictions exist is the $\jpsi$ production rate in $\gamma\gamma$ scattering. This observable has previously been measured at LEP by DELPHI \cite{Abdallah:2003du} with very large uncertainties and could possibly be remeasured at an ILC.

\subsubsection{NLO tests of NRQCD LDME universality}

\begin{table*}[p]
\caption{Overview of different NLO fits of the CO LDMEs. Analysis \cite{Butenschoen:2011yh} is a global fit to inclusive $\jpsi$ yield data from 10 different $pp$, $\gamma p$, $ee$, and $\gamma\gamma$ experiments. In \cite{Gong:2012ug}, fits to $pp$ yields from CDF \cite{Acosta:2004yw,Aaltonen:2009dm} and LHCb \cite{Aaij:2011jh,Aaij:2012ag,LHCb:2012af} were made. In \cite{Chao:2012iv}, three values for their combined fit to CDF $\jpsi$ yield and polarization \cite{Affolder:2000nn,Abulencia:2007us} data are given: A default set, and two alternative sets. Analysis \cite{Ma:2010vd} is a fit to the $\chi_{c2}/\chi_{c1}$ production ratio measured by CDF \cite{Abulencia:2007bra}. The analyses \cite{Butenschoen:2011yh} and \cite{Chao:2012iv} refer only to direct $\jpsi$ production, and in the analyses \cite{Gong:2012ug} and \cite{Chao:2012iv} $p_T<7$~GeV data was not considered. The color singlet LDMEs for the ${^3}S_1^{[1]}$ and ${^3}P_0^{[1]}$ states were not fitted. The values of the LDMEs given in the 
second through sixth column (referring to \cite{Butenschoen:2011yh}, \cite{Gong:2012ug}, and \cite{Chao:2012iv}) were used for the plots of Fig.~\ref{fig:ccomparegraphs}.}
\label{tab:ccompareldmes}
\begin{tabular}{ccccccc}
\hline \hline
& Butenschoen, & Gong, Wan, & \multicolumn{3}{c} {Chao, Ma, Shao, Wang, Zhang \cite{Chao:2012iv}:} & Ma, Wang, \\
& Kniehl \cite{Butenschoen:2011yh}: & Wang, Zhang \cite{Gong:2012ug}: & (default set) & (set 2) & (set 3) &  Chao \cite{Ma:2010vd}:\\
\hline
$\langle {\cal O}^{\jpsi}({^3}S_1^{[1]}) \rangle /\mbox{GeV}^3 $ &
$1.32$ &
$1.16$ &
$1.16$ &
$1.16$ &
$1.16$
\\
$\langle {\cal O}^{\jpsi}({^1}S_0^{[8]}) \rangle /\mbox{GeV}^3$ &
$0.0497\pm0.0044$ &
$0.097\pm0.009$ &
$0.089\pm0.0098$ &
$0$ &
$0.11$
\\
$\langle {\cal O}^{\jpsi}({^3}S_1^{[8]}) \rangle /\mbox{GeV}^3$ &
$0.0022\pm0.0006$ &
$-0.0046\pm0.0013$ &
$0.0030\pm0.012$ &
$0.014$ &
$0$
\\
$\langle {\cal O}^{\jpsi}({^3}P_0^{[8]}) \rangle /\mbox{GeV}^5$ &
$-0.0161\pm0.0020$ &
$-0.0214\pm0.0056$ &
$0.0126\pm0.0047$ &
$0.054$ &
$0$
\\
$\langle {\cal O}^{\psi(2S)}({^3}S_1^{[1]}) \rangle /\mbox{GeV}^3$ & &
$0.758$ &\phantom{(default set)} &\phantom{(default set)} & \phantom{(default set)}&
\\
$\langle {\cal O}^{\psi(2S)}({^1}S_0^{[8]}) \rangle /\mbox{GeV}^3$ & &
$-0.0001\pm0.0087$ & & & &
\\
$\langle {\cal O}^{\psi(2S)}({^3}S_1^{[8]}) \rangle /\mbox{GeV}^3$ & &
$0.0034\pm0.0012$ & & & &
\\
$\langle {\cal O}^{\psi(2S)}({^3}P_0^{[8]}) \rangle /\mbox{GeV}^5$ & &
$0.0095\pm0.0054$ & & & &
\\
$\langle {\cal O}^{\chi_0}({^3}P_0^{[1]}) \rangle /\mbox{GeV}^5$ & &
$0.107$ & & & &
$0.107$
\\
$\langle {\cal O}^{\chi_0}({^3}S_1^{[8]}) \rangle /\mbox{GeV}^3$ & &
$0.0022\pm0.0005$ & & & &
$0.0021\pm0.0005$\\
\hline \hline
\end{tabular}
\end{table*}

\begin{figure*}[p]

\centering
\includegraphics[width=0.22\linewidth]{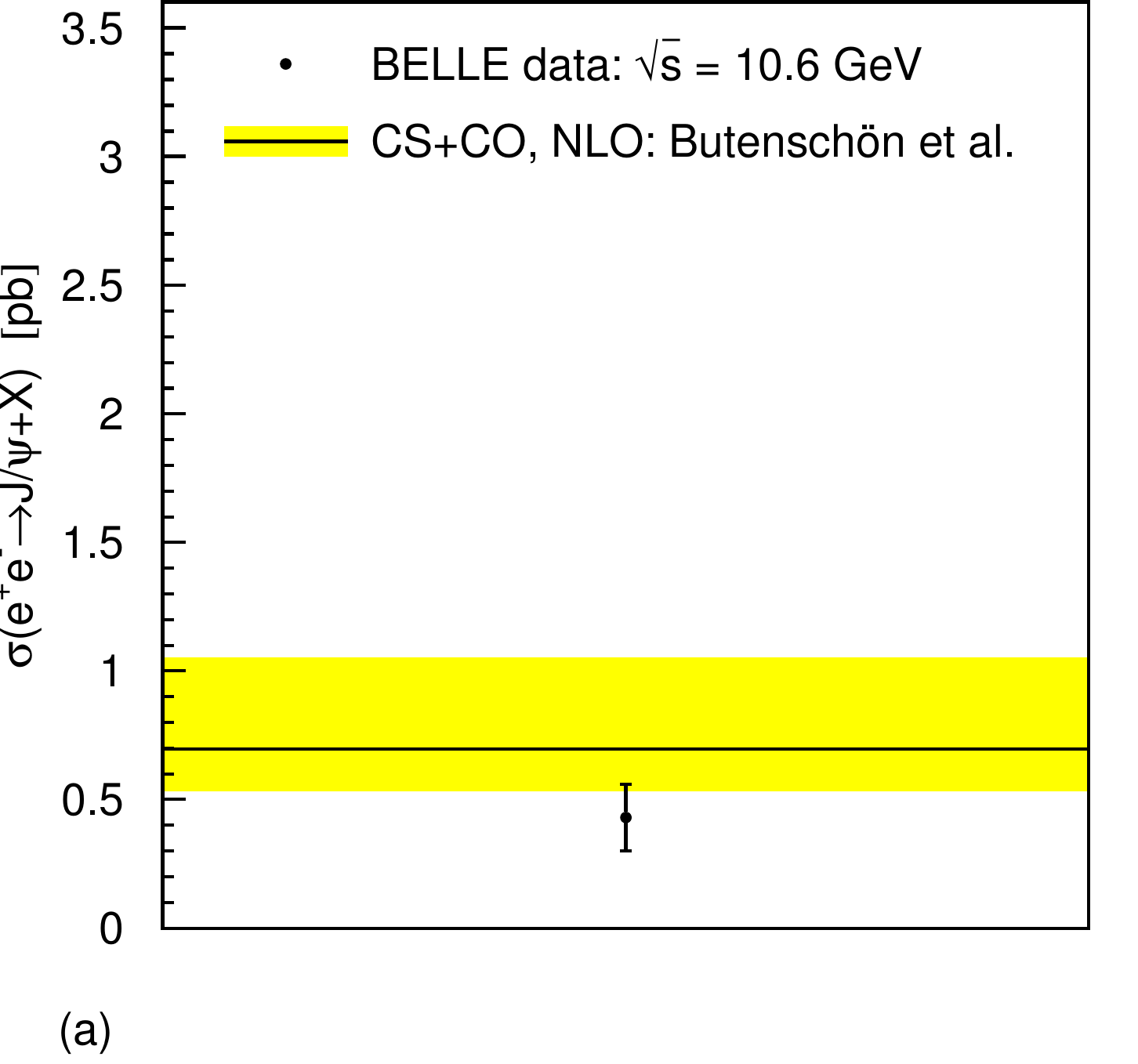}\hspace{2pt}
\includegraphics[width=0.22\linewidth]{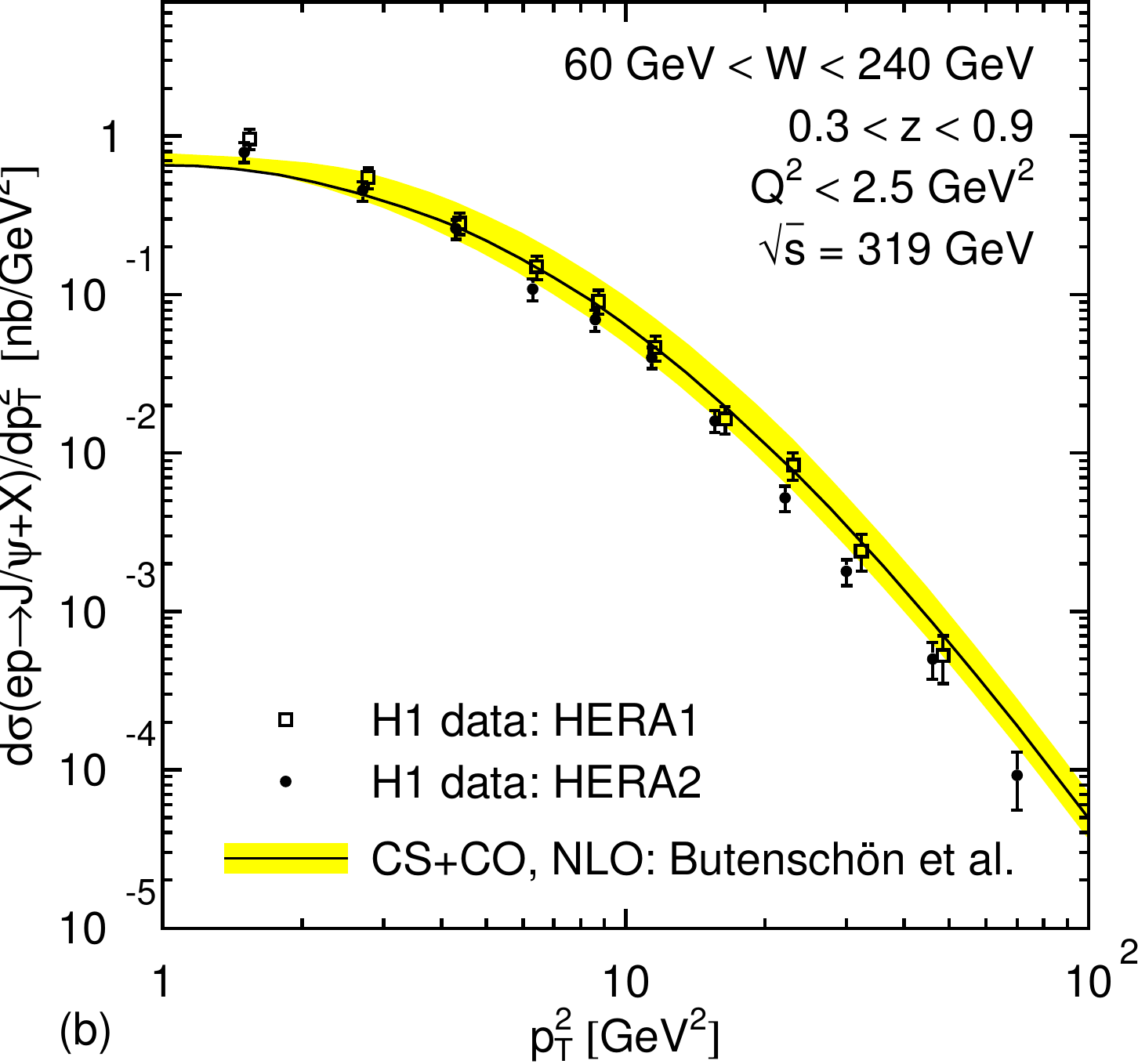}\hspace{2pt}
\includegraphics[width=0.22\linewidth]{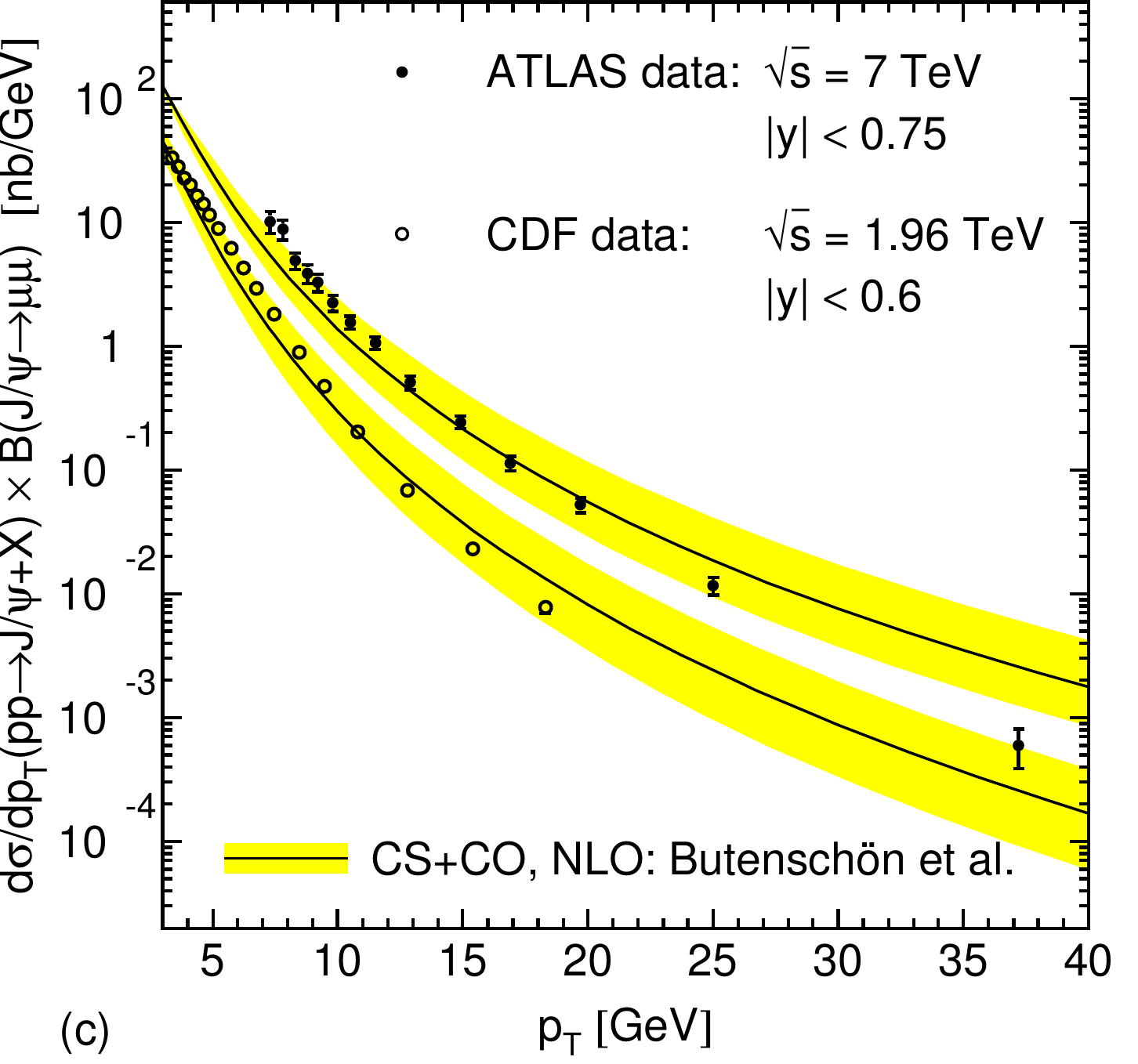}\hspace{2pt}
\includegraphics[width=0.22\linewidth]{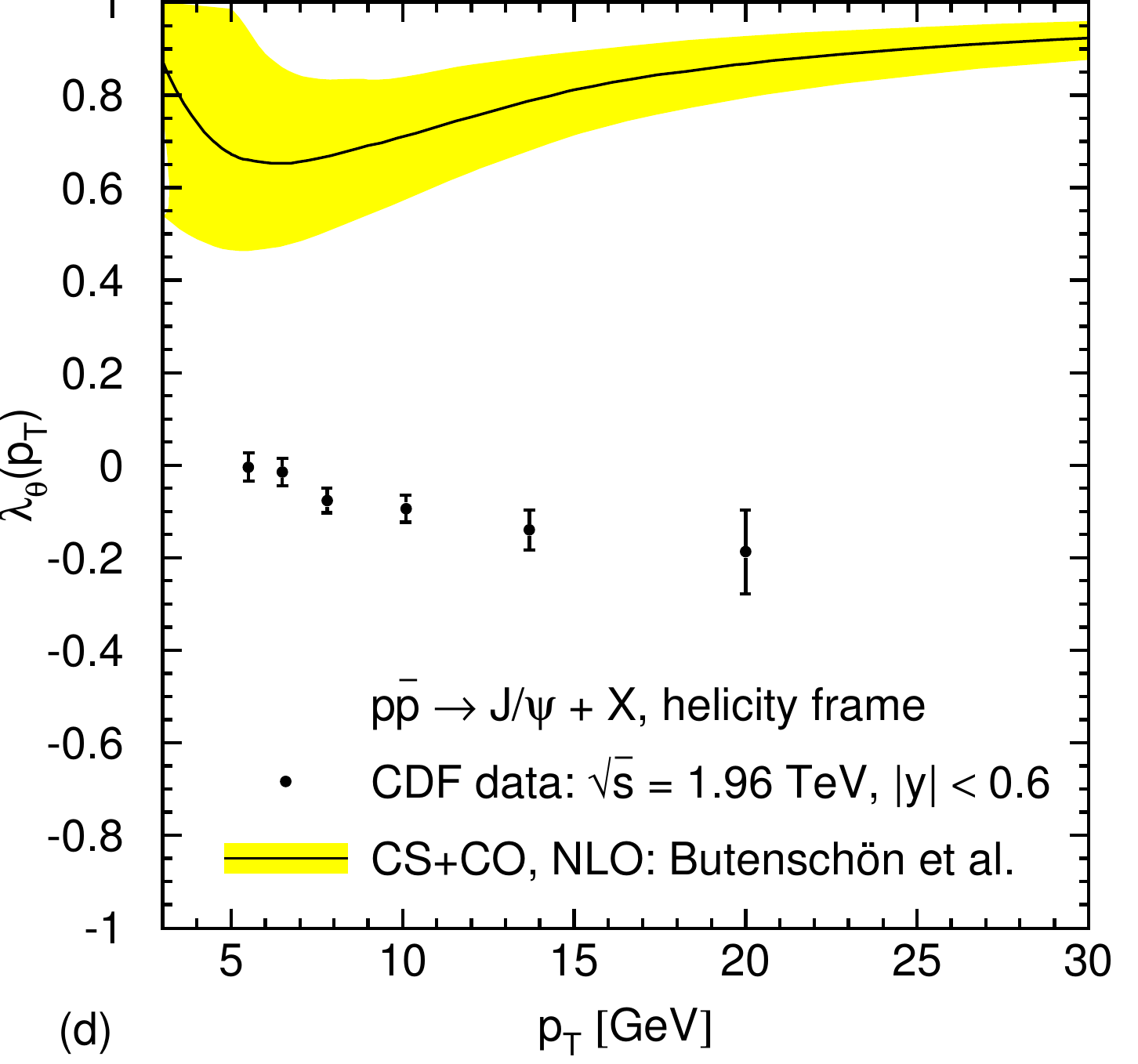}

\vspace{5pt}

\includegraphics[width=0.22\linewidth]{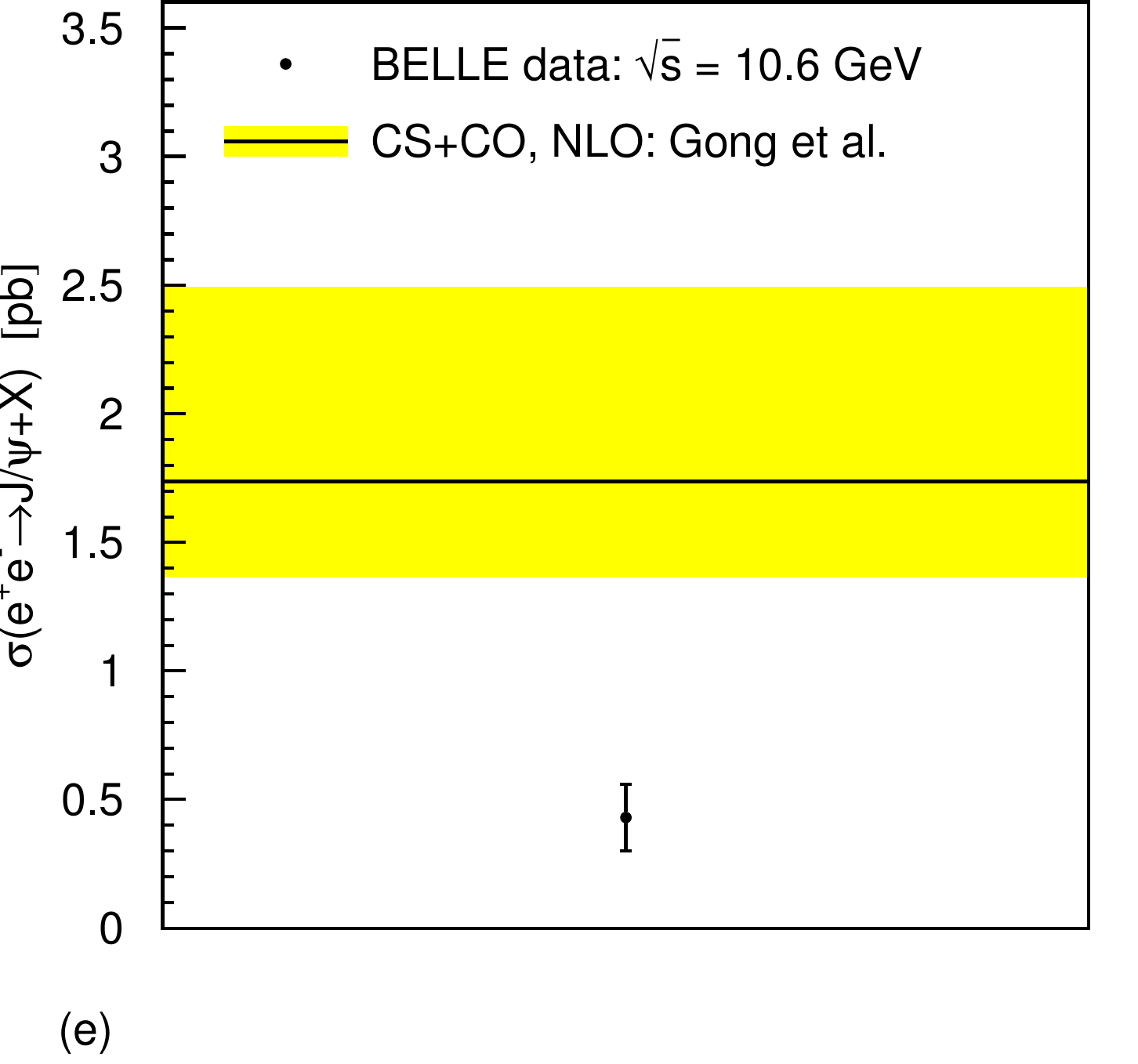}\hspace{2pt}
\includegraphics[width=0.22\linewidth]{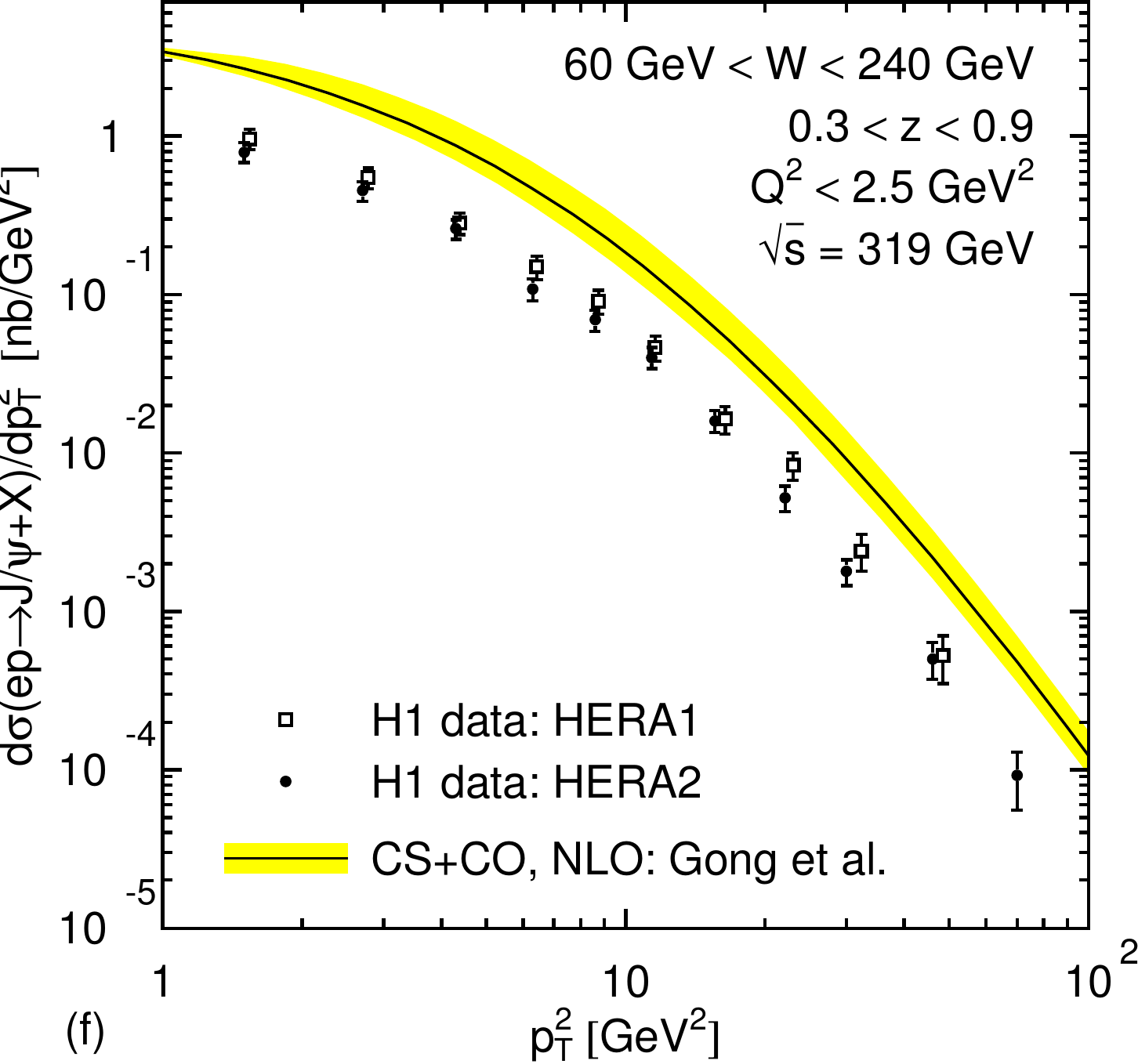}\hspace{2pt}
\includegraphics[width=0.22\linewidth]{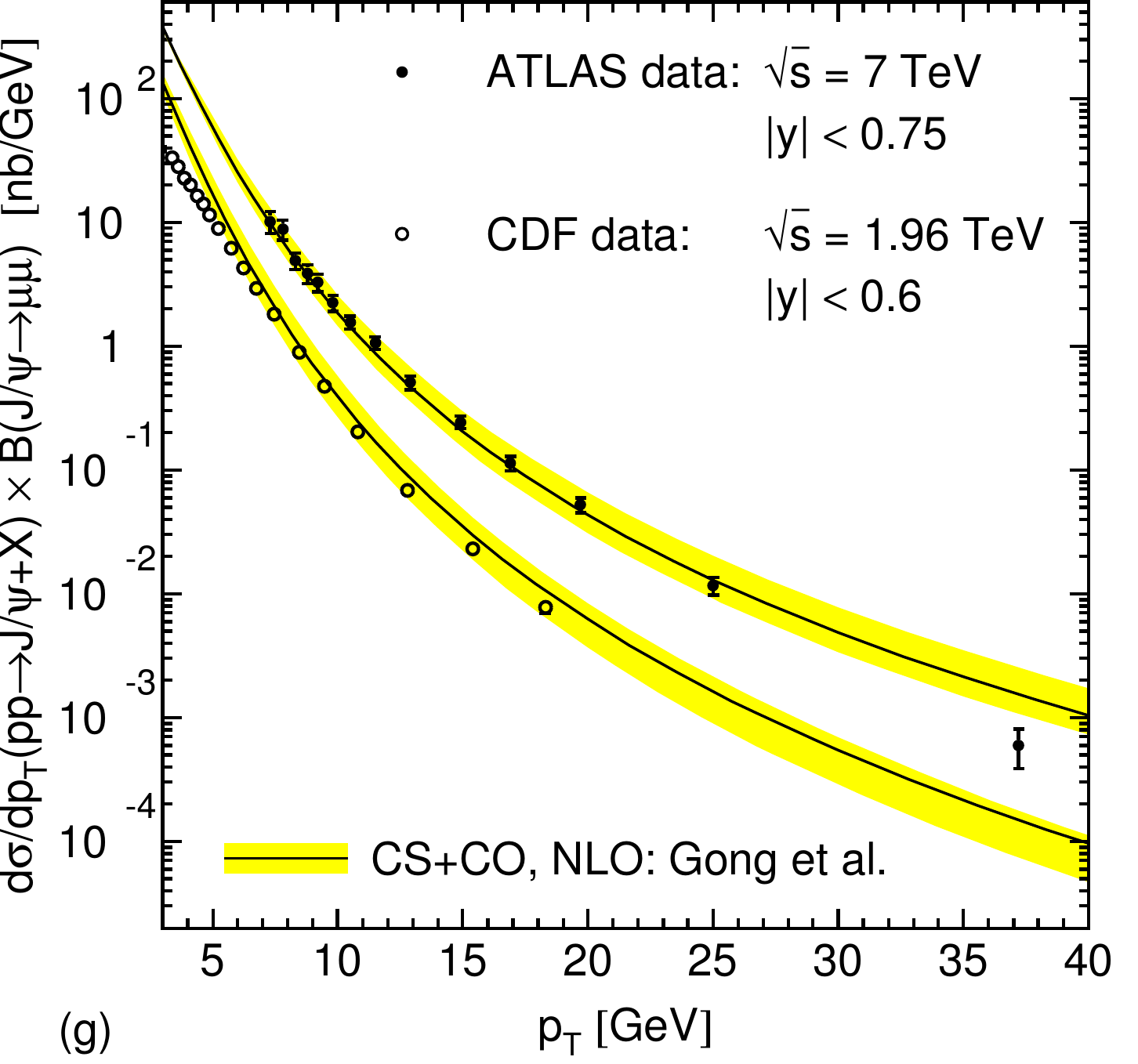}\hspace{2pt}
\includegraphics[width=0.22\linewidth]{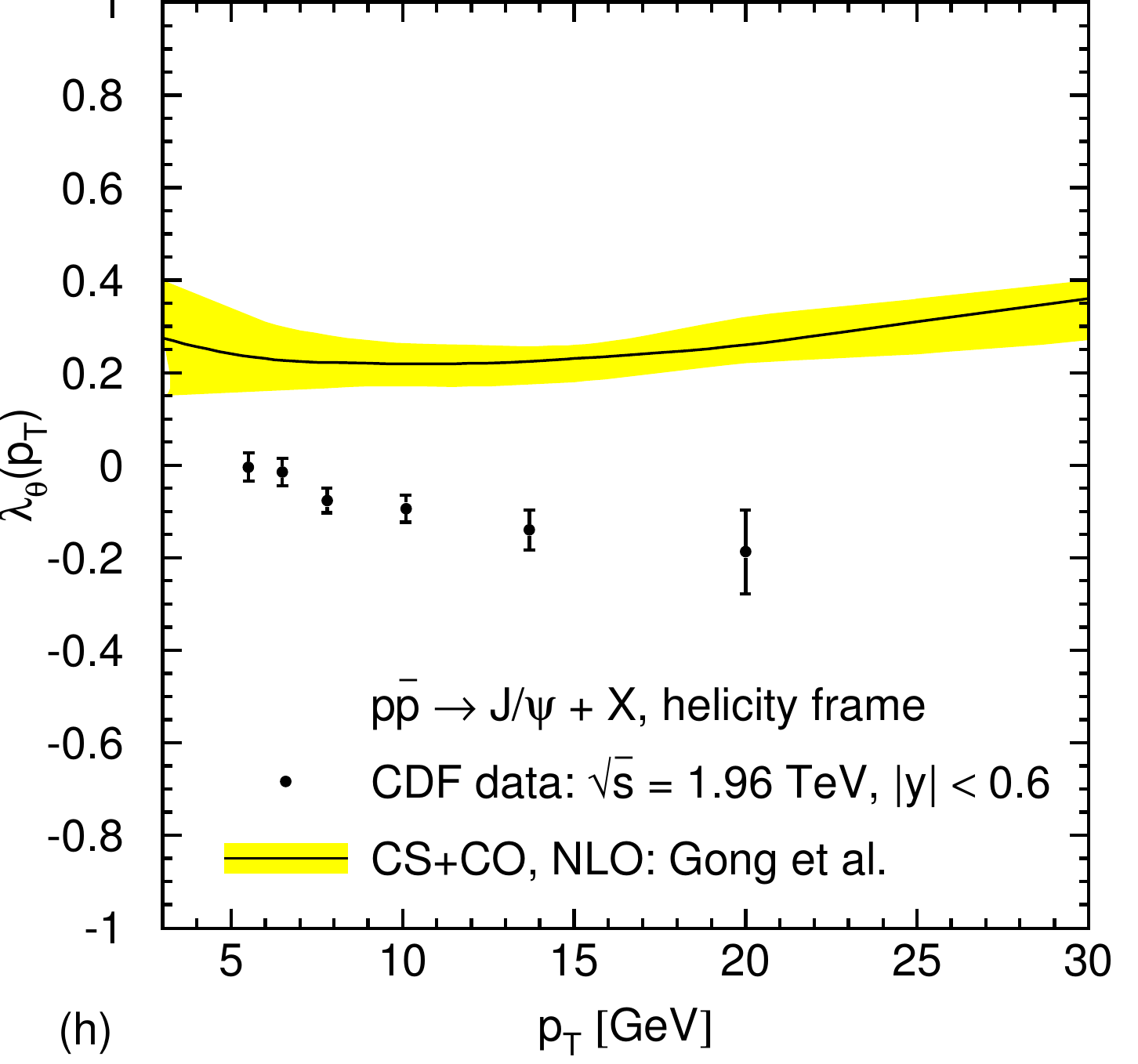}

\vspace{5pt}

\includegraphics[width=0.22\linewidth]{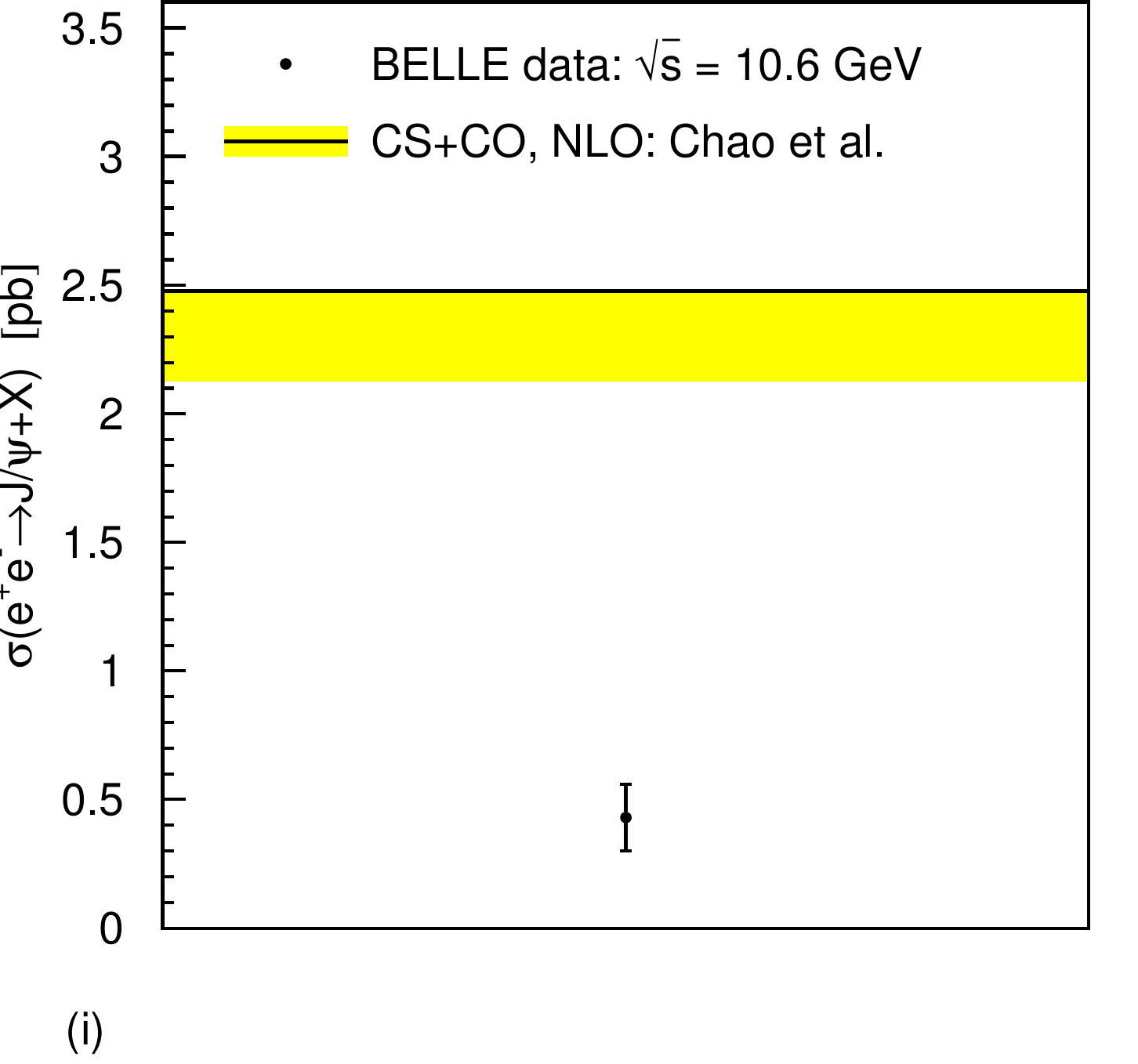}\hspace{2pt}
\includegraphics[width=0.22\linewidth]{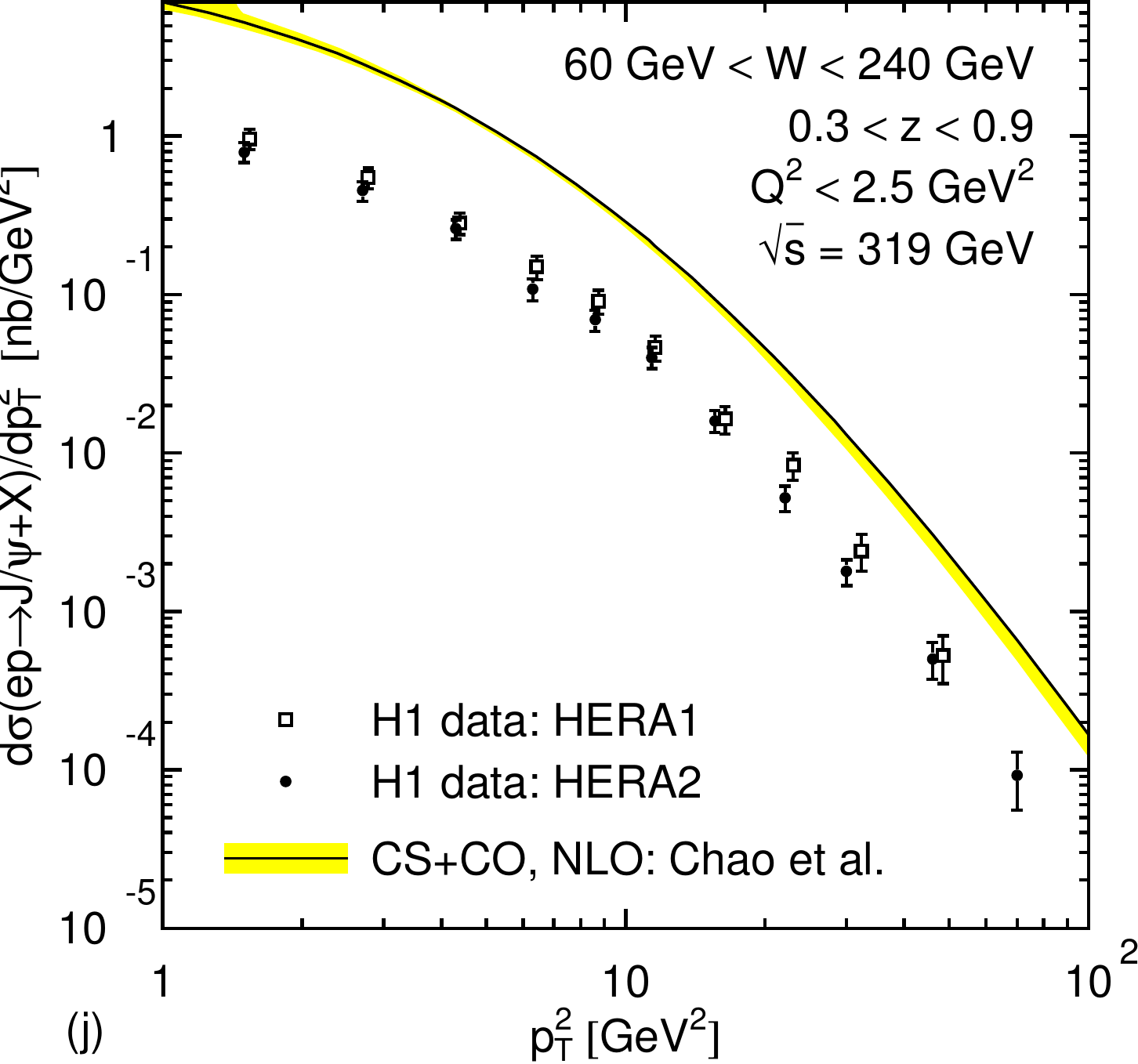}\hspace{2pt}
\includegraphics[width=0.22\linewidth]{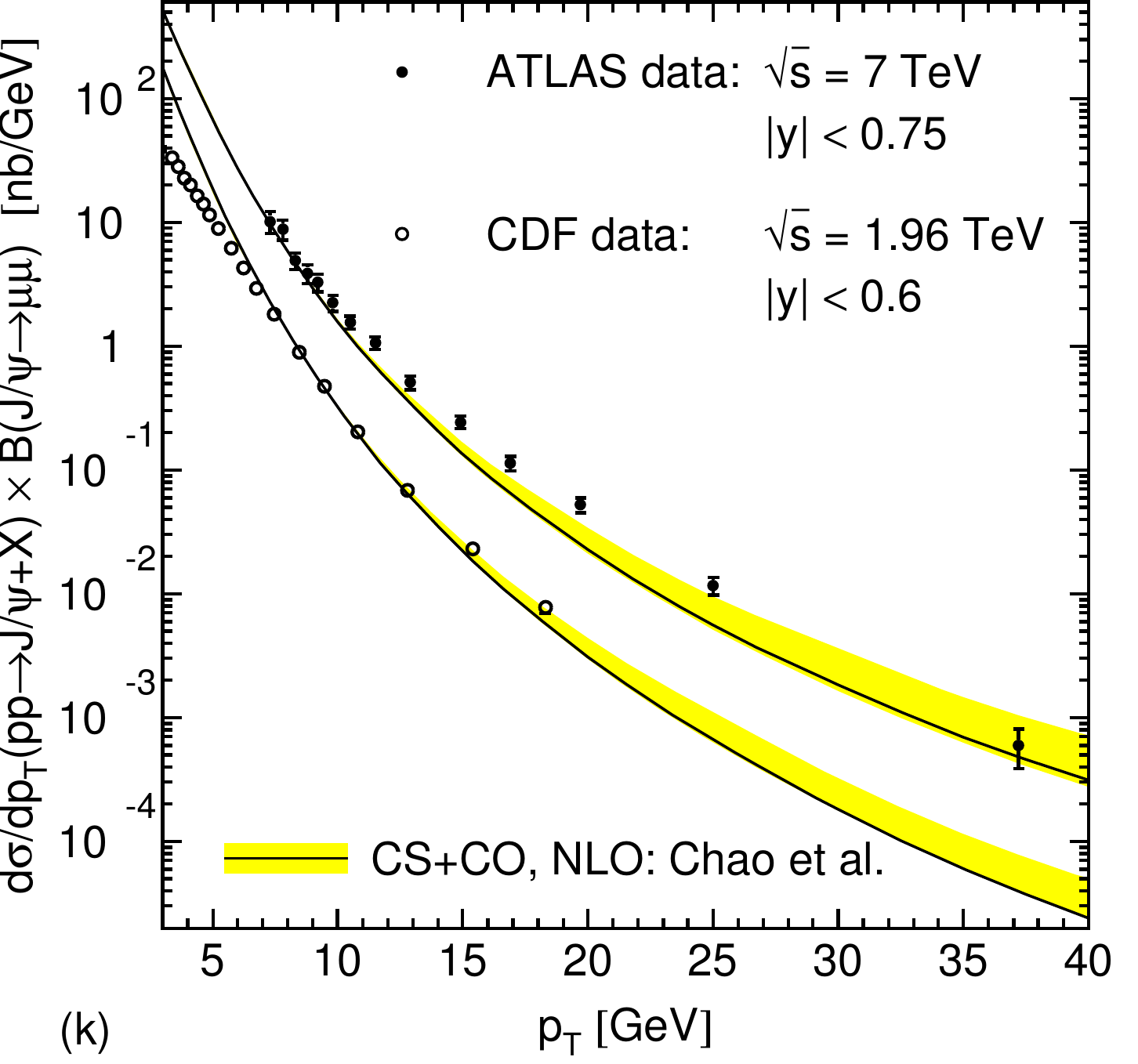}\hspace{2pt}
\includegraphics[width=0.22\linewidth]{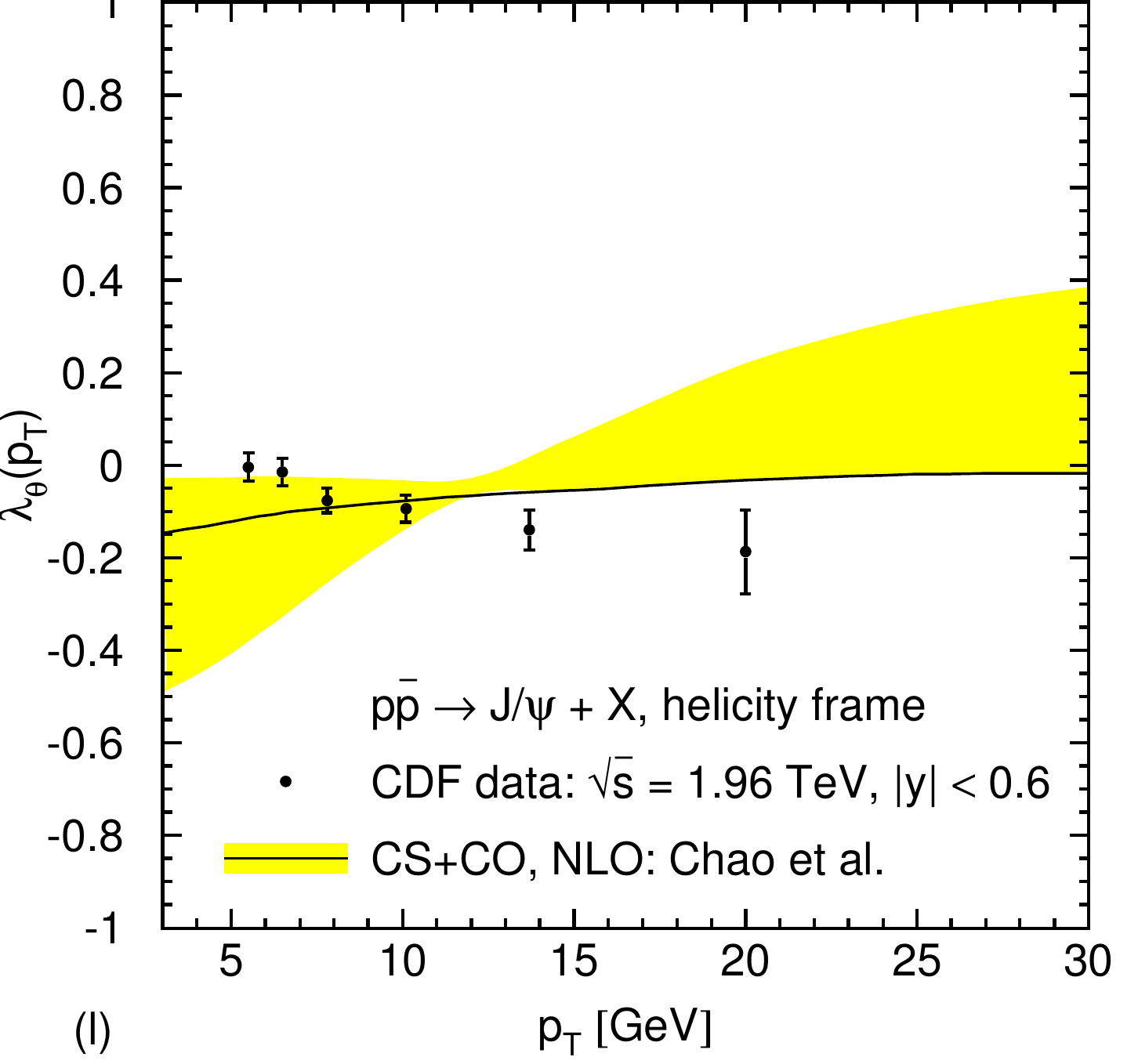}

\caption{The predictions of the $\jpsi$ total $e^+e^-$ cross section measured by Belle \cite{Pakhlov:2009nj}, the transverse momentum distributions in photoproduction measured by H1 at HERA \cite{Adloff:2002ex,Aaron:2010gz}, and in hadroproduction measured by CDF \cite{Acosta:2004yw} and ATLAS \cite{Aad:2011sp}, and the polarization parameter $\lambda_\theta$ measured by CDF in Tevatron run II \cite{Abulencia:2007us}. The predictions are plotted using the values of the CO LDMEs given in \cite{Butenschoen:2011yh}, \cite{Gong:2012ug} and \cite{Chao:2012iv} and listed in Table~\ref{tab:ccompareldmes}. The error bars of graphs a--g refer to scale variations, of graph d also fit errors, errors of graph h according to \cite{Gong:2012ug}. As for graphs i--l, the central lines are evaluated with the default set, and the error bars evaluated with the alternative sets of the CO LDMEs used in \cite{Chao:2012iv} and listed in Table~\ref{tab:ccompareldmes}.
\figPermX{Butenschoen:2012qr}{2013}{World Scientific.}
}
\label{fig:ccomparegraphs}
\end{figure*}

The phenomenological relevance of the NRQCD factorization conjecture is closely tied to the question of whether or not the LDMEs can be shown to be universal. In this section recent works will be reviewed which aim at examining this universality at Next-to-Leading Order (NLO) in $\alpha_s$. In the case of $\chi_{cJ}$, these tests include just the leading order of the NRQCD $v$ expansion, formed by the $n={^3}P_J^{[1]}$ and $n={^3S}_1^{[8]}$ states. In the case of ${^3}S_1$ quarkonia, these tests include the terms up to relative order $O(v^4)$  in the $v$ expansion, namely the $n={^3}S_1^{[1]}$ color singlet state, as well as the $n={^1}S_0^{[8]}$, ${^3}S_1^{[8]}$, and ${^3}P_J^{[8]}$ Color Octet (CO) states, see Table~\ref{tab:cveloscal}. The relativistic corrections involving the $\langle {\cal P}^{H}({^3}S_1^{[1]}) \rangle$ and $\langle {\cal Q}^{H}({^3}S_1^{[1]}) \rangle$ LDMEs are, 
however, not part of these analyses, although they are of order $O(v^2)$ and $O(v^4)$ in the $v$ expansion. There 
are two reasons for that: First, the corresponding NLO 
calculations are far beyond the reach of current techniques, and secondly, they are expected to give significant contributions to hadroproduction only at $p_T\ll m_c$ and for photoproduction only at $z\approx 1$. This behavior is inferred from the behavior at LO in $\alpha_s$ \cite{Paranavitane:2000if,Fan:2009zq} and can be understood by noting that new topologies of Feynman diagrams open up when doing the transition from the ${^3}S_1^{[1]}$ state to the CO states, but not when calculating relativistic corrections: For example, at leading order in $\alpha_s$ the slope of the transverse momentum distribution in hadroproduction is $d\sigma/dp_T \approx p_T^{-8}$ for the ${^3}S_1^{[1]}$ state, compared to $d\sigma/dp_T \approx p_T^{-6}$ for the ${^1}S_0^{[8]}$ and ${^3}P_J^{[8]}$ states and $d\sigma/dp_T \approx p_T^{-4}$ for the ${^3}S_1^{[8]}$ state.

The $O(\alpha_s)$ corrections to the necessary unpolarized short-distance cross sections  of the $n={^1}S_0^{[8]}$, ${^3}S_1^{[1/8]}$, and ${^3}P_J^{[1/8]}$ intermediate states have been calculated for most of the phenomenologically relevant inclusive quarkonium production processes: For two-photon scattering \cite{Klasen:2004az,Butenschoen:2011yh}, $e^+e^-$ scattering \cite{Zhang:2009ym}, photoproduction \cite{Butenschoen:2009zy,Butenschoen:2011yh} and hadroproduction \cite{Gong:2008ft,Ma:2010vd,Ma:2010yw,Butenschoen:2010rq,Wang:2012is}. The polarized cross sections have been calculated for photoproduction \cite{Butenschoen:2011ks} and hadroproduction \cite{Butenschoen:2012px,Chao:2012iv,Gong:2012ug,Gong:2013qka}.

In \cite{Butenschoen:2011yh}, a global fit of the $\jpsi$ CO LDMEs to 26 sets of inclusive $\jpsi$ production yield data from 10 different $pp$, $\gamma$p, $\gamma\gamma$, and $e^+e^-$ experiments was done; see the second column of Table~\ref{tab:ccompareldmes} for the fit results. This fit describes all data, except perhaps the two-photon scattering at LEP \cite{Abdallah:2003du}, reasonably well. This fit is overconstrained, and practically independent of possible low-$p_T$ cuts (unless such high $p_T$ cuts are chosen that all data except hadroproduction drop out of the fit \cite{Butenschoen:2012qh}). Furthermore, the resulting LDMEs are in accordance with the velocity scaling rules predicted by NRQCD; see Table~\ref{tab:cveloscal}. Thus the fit is in itself already a nontrivial test of the NRQCD factorization conjecture, especially since the high-$z$ photoproduction region can now also be well described, which had been plagued by divergent behavior in the earlier Born analyses
\cite{Cacciari:1996dg,Ko:1996xw}.
 However, in \cite{Butenschoen:2012px} it was shown that these CO LDME values lead to 
predictions 
of a strong transverse $\jpsi$ polarization in the hadroproduction helicity frame, which is in contrast to the precise CDF Tevatron run II measurement \cite{Abulencia:2007us}, see Fig.~\ref{fig:ccomparegraphs}d. On the other hand, in \cite{Chao:2012iv} it was shown that both the measured $\jpsi$ hadroproduction cross sections and the CDF run II polarization measurement \cite{Abulencia:2007us} can, even at the highest measured $p_T$ values, be well described when choosing one of the three CO LDME sets listed in columns four through six of Table~\ref{tab:ccompareldmes}. These LDMEs, however, result in predictions for $e^+e^-$ annihilation and photoproduction which are factors four to six above the data, see Fig.~\ref{tab:ccompareldmes}e--f. Third, the calculation \cite{Gong:2012ug} is the first NLO polarization analysis to include feed-down
contributions. To this end, the CO LDMEs of $\jpsi$, $\psi(2S)$ and $\chi_{cJ}$ were fitted to CDF \cite{Acosta:2004yw,Aaltonen:2009dm} and LHCb \cite{Aaij:2011jh,Aaij:2012ag,LHCb:2012af} unpolarized production data with $p_T>7$~GeV; see column three of Table~\ref{tab:ccompareldmes}. These fit results were then used for the predictions of Fig.~\ref{fig:ccomparegraphs}e--h, taking the $\psi(2S)$ and $\chi_{cJ}$ feed-down contributions consistently into account. A similar analysis has recently also been performed for $\Upsilon(1S,2S,3S)$ production \cite{Gong:2013qka}.

The shape of high-$p_T$ $J/\psi$ hadroproduction yield can be nicely described by the ${^1}S_0^{[8]}$ component alone, which automatically yields unpolarized hadroproduction. Since at $p_T>10$~GeV this is already all data available, there is no tension between NRQCD predictions and current data if the validity of the NRQCD factorization conjecture is restricted to high enough $p_T$ values and the ${^3}S_1^{[1/8]}$ and ${^3}P_J^{[1/8]}$ LDMEs are very small or even put to zero, as for example in the sixth column of Fig.~\ref{fig:ccomparegraphs} (set 3). This is also the spirit of \cite{Faccioli:2014cqa}, and of the analysis \cite{Bodwin:2014gia}, in which the NLO short distance cross sections used in \cite{Chao:2012iv} are combined with $c\overline{c}$ production via single parton fragmentation using fragmentation functions at order $\alpha_s^2$ including a leading log resummation.

To summarize, none of the proposed CO LDME sets is able to describe all of the studied $J/\psi$ production data sets, which poses a challenge to the LDME universality. Possible resolutions include the following:
\begin{enumerate}
 \item The perturbative $v$ expansion might converge too slowly.
 \item NRQCD factorization might hold for exclusive, but not inclusive, production.
 \item NRQCD factorization might hold only in the region $p_T\gg M_{\textrm{onium}}$. Currently, photoproduction cross sections are measured only up to $p_T=10$~GeV.
 \item NRQCD factorization might not hold for polarized production.
\end{enumerate}

\subsubsection{Recent calculations of relativistic corrections}

\begin{table}
\caption{Color singlet model predictions for $\sigma(e^+e^-\to \jpsi+\eta_c)$ compared to $B$-factory data \cite{Abe:2002rb,Abe:2004ww,Aubert:2005tj}. As for the theoretical predictions for the leading-order cross section as well as the corrections of order $O(\alpha_s)$, $O(v^2)$, and $O(\alpha_s v^2)$, we compare the results obtained in \cite{He:2007te,Bodwin:2007ga,Li:2013qp}. These calculations mainly differ by different methods of color singlet LDME determinations. As for the values of \cite{Bodwin:2007ga}, the leading-order results include pure QED contributions, the $O(\alpha_s)$ results include interference terms with the QED contributions, and the $O(v^2)$ results include in part a resummation of relativistic corrections, the $O(\alpha_s v^2)$ results do, however, include the interference terms of the $O(\alpha_s)$ and $O(v^2)$ amplitudes only. The short-distance coefficients of the $O(\alpha_s)$ contribution used in \cite{
He:2007te} and \cite{Bodwin:2007ga} were taken over from \cite{Zhang:2005cha}.
 The experimental cross sections refer to data samples in which at least 2, respectively 4, charged tracks were identified.}
\label{tab:cJpsiEtacProduction}
 \begin{tabular}{ccccc}
     \hline\hline
  & He, Fan,& Bodwin, \\
  & Chao \cite{He:2007te} & Lee,Yu \cite{Bodwin:2007ga}& \multicolumn{2}{c}{Li, Wang \cite{Li:2013qp}} \\
  & $\alpha_s(2m_c)$ & $\alpha_s(\sqrt{s}/2)$ & $\alpha_s(\sqrt{s}/2)$ & $\alpha_s(2m_c)$ \\ \hline
  $\sigma_{LO}$ & 9.0~fb &6.4~fb & 4.381~fb & 7.0145~fb\\
  $\sigma(\alpha_s)$ & 8.8~fb & 6.9~fb & 5.196~fb & 7.367~fb \\
  $\sigma(v^2)$ & 2.2~fb &2.9~fb& 1.714~fb & 2.745~fb \\
  $\sigma(\alpha_s v^2)$ &  & 1.4~fb & 0.731~fb & 0.245~fb \\
  sum & 20.0~fb & $17.6^{+8.1}_{-6.7}$~fb & 12.022~fb & 17.372~fb \\ \hline 
  \multicolumn{2}{c}{Belle \cite{Abe:2002rb}} & \multicolumn{3}{c}{$33^{+7}_{-6}\pm9$~fb ($\geq4$ charged tracks)} \\
  \multicolumn{2}{c}{Belle \cite{Abe:2004ww}} & \multicolumn{3}{c}{$25.6\pm 2.8 \pm 3.4$~fb ($\geq2$ charged tracks)}\\
  \multicolumn{2}{c}{BaBar \cite{Aubert:2005tj}} & \multicolumn{3}{c}{$17.6\pm 2.8^{+1.5}_{-2.1}$~fb ($\geq2$ charged tracks)}
\\
  \hline\hline
 \end{tabular}
\end{table}

As explained in the last section, the relativistic corrections of order $O(v^2)$ in the NRQCD $v$ expansion have at leading order in $\alpha_s$ in inclusive hadro- \cite{Fan:2009zq} and photoproduction \cite{Paranavitane:2000if} been shown to be less significant than the CO contributions of order $O(v^4)$ in the NRQCD $v$ expansion. Similarly, the $O(v^2)$ \cite{Bodwin:2003wh} and the technically challenging $O(v^4)$ \cite{Bodwin:2012xc} relativistic corrections to gluon fragmentation into ${^3}S_1$ quarkonia have turned out to be small. The relativistic  $O(v^2)$ corrections to the process $e^+e^-\to \jpsi+gg$ have, however, turned out to be between 20\% and 30\% \cite{He:2009uf,Jia:2009np} relative to the leading order CS cross section, an enhancement comparable in size to the $O(\alpha_s)$ CS correction \cite{Ma:2008gq,Gong:2009kp}. These corrections helped bring the color singlet model prediction for inclusive $\jpsi$ 
production in $e^+e^-$ collisions in rough agreement with experimental data \cite{Pakhlov:2009nj}.

Similarly, in the exclusive process $e^+e^-\to \jpsi+\eta_c$, $O(\alpha_s)$ corrections as well as relativistic corrections of $O(v^2)$ were necessary to bring the color singlet model prediction in agreement with data,
see Table~\ref{tab:cJpsiEtacProduction}. Recently, even $O(\alpha_s v^2)$ corrections to this process have been calculated \cite{Dong:2012xx,Li:2013qp}. For a review of the history of the measurements and calculations of this process, as well as for a description of different methods to determine the LDMEs of relative order $O(v^2)$, we refer to section 4.5.1 of \cite{Brambilla:2010cs}.

As a final point of this section, we mention the interesting work \cite{Xu:2012am} in which relativistic corrections to the process $gg\to \jpsi+g$ via color octet states formally of order $O(v^6)$ were estimated. According to this analysis, at leading order in $\alpha_s$, they might reduce the $O(v^4)$ CO contributions by up to 20 -- 40\% in size.

\subsubsection{Calculations using $k_T$ factorization}

Color singlet model predictions for $J/\psi$ production face many phenomenological problems: Except for $e^+e^-$ annihilation, NLO color singlet model predictions are shown to lie significantly below inclusive $\jpsi$ production data, 1--2 orders of magnitude for hadroproduction and $\gamma\gamma$ scattering, and a factor 3--5 for photoproduction at HERA, see, for example, \cite{Butenschoen:2011yh}. As in photoproduction \cite{Artoisenet:2009xh,Chang:2009uj}, $\jpsi$ polarization in hadroproduction \cite{Gong:2008sn} is at NLO predicted to be highly longitudinal in the helicity frame, in contrast to the CDF measurement at Tevatron run II \cite{Abulencia:2007us}.

According to \cite{Baranov:2010mr,Baranov:2011ib}, these shortcomings can be overcome when the transverse momenta $k_T$ of the initial gluons are retained. The off-shell matrix elements are then folded with unintegrated, $k_T$ dependent, Parton Distribution Functions (uPDFs). The weakest point of this approach is certainly the derivation of the uPDFs from the usual gluon PDFs using varying prescriptions. The latest analyses \cite{Baranov:2010mr,Baranov:2011ib} show very good agreement with $\jpsi$ photoproduction data at HERA \cite{Chekanov:2009ad,Adloff:2002ex,Aaron:2010gz,Chekanov:2002at} and hadroproduction  at the LHC \cite{Khachatryan:2010yr,Aad:2011sp,Aaij:2011jh}. On top of that, the $\jpsi$ is predicted to be largely unpolarized, in 
line with all recent polarization measurements, see paragraph $d$ in Sec.~\ref{subsec:summaryrep}.
As for hadroproduction, the conclusions are however contrary to the author's earlier findings \cite{Baranov:2002cf}, which show longitudinal $\jpsi$ polarization and cross sections an order of magnitude below the CDF production data. They also disagree with the recent work \cite{Saleev:2012hi}, where $\jpsi$ hadroproduction at the LHC was studied in the same way, comparing to the same data \cite{Khachatryan:2010yr,Aad:2011sp,Aaij:2011jh}, even when the same uPDFs \cite{Kimber:2001sc,Watt:2003mx} were used. Here, the color singlet predictions lie again clearly below the data, and the difference was even used to fit the CO LDMEs of NRQCD in a $k_T$ factorization approach.

We note that calculations in the $k_T$ factorization scheme can be performed for any intermediate Fock state of the NRQCD $v$ expansion. On the other hand, even a fully worked out framework of $k_T$ factorization at NLO in $\alpha_s$ could not cure the problem of uncanceled infrared singularities in color singlet model calculations for $P$ wave quarkonia.

\subsubsection{Current trends in theory}

The most prominent candidate theory for heavy quarkonium production is NRQCD, and lots of effort is going on to prove its factorization theorem on the one hand, and to show the universality of the LDMEs by comparison to data on the other.  Since at the moment there are hints that at least to the orders currently considered in perturbation theory, not all data might be simultaneously described by single LDME sets, more effort will be going on to refine NRQCD calculations for specific observables or specific kinematic regimes, such as the low and high $p_T$ limits of the hadroproduction cross sections. For low $p_T$ resummation of large logarithms, the recent work \cite{Sun:2012vc} followed the idea of \cite{Berger:2004cc} to apply the Collins-Soper-Sterman impact parameter resummation formalism \cite{Collins:1984kg}. For high $p_T$ resummation, the factorization theorem of \cite{Kang:2011mg,KMQS-hq1} in terms of single and double parton fragmentation functions, and the soft-collinear effective theory 
approach \cite{
Fleming:2012wy,Fleming:2013qu} can be applied. Other paths may be to apply transverse momentum dependent PDFs in quarkonium production calculations, but the uncertainties inherent to these calculations will still need to be thoroughly investigated, as can be seen from contradicting $k_T$ factorization results. But also in more phenomenologically based models, like the color evaporation model, new predictions are still calculated \cite{Nelson:2012bc}.

\subsection{Future directions}

\label{sec:secC7}

Our understanding of heavy quark hadronic systems improves with the
progress made on experimental measurements of masses, production and
decay rates, the development of suitable effective field theories,
perturbative calculations within these frameworks, and the progress on
lattice gauge theory calculations.

Lattice simulations are obtaining a more and more prominent role in
heavy quark physics. They may compute low-energy matrix elements, 
factorized by effective field theories, appearing in the study of quarkonia below threshold,
improving our understanding of the dynamics of these systems and providing, 
among others, precision determinations of the strong coupling constant at low energies 
and the heavy quark masses. For states at and above threshold, they may eventually 
be able to determine the nature of the $XYZ$ exotic states, including in particular the role
that mixing between tetraquark and multihadron states plays. 
A possible way to address these problems that relies on lattice simulations 
has been very recently proposed in \cite{Braaten:2014ita,Braaten:2014qka}.
Lattice simulations are also required for determining nonperturbative form factors needed in 
extracting the CKM matrix elements $|V_{cb}|$, $|V_{ub}|$, $|V_{cs}|$ and
$|V_{cd}|$ from $B \to D^\star/\pi l \nu$ and $D \to K/\pi l \nu$ decays, respectively. 
Current gaps between lattice determinations and experimental fits of these form factors 
are expected to be removed by further progress in lattice simulations. 
The emergence of ensembles incorporating the effects of dynamical charm quarks 
in lattice calculations will help to establish whether charm sea contributions to
charmonium spectra and to flavor observables are relevant. At the same time, 
the trend to finer lattice spacings (even if currently somewhat displaced 
by a trend to perform simulations at the physical pion mass) is likely to continue 
in the long run and will eventually enable the use of fully relativistic b-quarks, 
which will provide an important cross-check on effective field theories, 
and eventually for some observables replace them.

Rapid progress on the side of effective field theories is currently
happening for any system involving heavy quarks.
Many quantities, like spectra, decays, transitions and production cross sections,  
are computed in this framework with unprecedented 
precision in the velocity and $\alpha_s$ expansions.
Noteworthy progress is happening, in particular, in the field of quarkonium production. 
Here, the recent Snowmass White Paper on ``Quarkonium at the Frontiers of
High Energy Physics''~\cite{Bodwin:2013nua} provides an excellent
summary. Future work will be likely centered around the effort to
search for a rigorous theoretical framework (factorization with a
rigorous proof) for inclusive as well as exclusive production of
quarkonia at various momentum scales.  While a proof of NRQCD
factorization is still lacking, performing global analyses of all
available data in terms of the NRQCD factorization formalism is
equally important, so that the universality of the NRQCD LDMEs can be
systematically tested, which is a necessary condition for the
factorization conjecture. To better test the conjecture, a resummation
of various large logarithms in perturbative calculations in different
production environments are critically needed.

The currently running experiments, in particular BESIII and the LHC
experiments, will at this stage primarily help refine previous
measurements. The LHC will in particular continue to provide
measurements on heavy quarkonium production rates at unprecedented
values of transverse momentum, provide better measurements on
quarkonium polarization, but might also provide more diverse
observables, such as associated production of a heavy quarkonium with
gauge bosons, jets or other particles. The LHC will also continue to
contribute to the studies of $XYZ$ states, and determine the $XYZ$
quantum numbers from amplitude analyses. Studies of $Z_c$ states at
BESIII will continue and provide precise measurements of spin-parities
and resonance parameters from multiple decay channels and amplitude analyses.

\newpage

In the farther future, however, Belle~II is expected to produce more
and better data that will be particularly useful to reduce the
uncertainties on the CKM matrix elements $|V_{cb}|$ and $|V_{ub}|$.
Data from a larger phase space can provide more precise information to
solve the long-standing discrepancy between the inclusive and
exclusive measurements of $|V_{ub}|$. Having about 100~fb$^{-1}$
integrated luminosity from the first Belle~II run at the
$\Upsilon(6S)$ resonance or at a nearby energy will be very exciting
for bottomonium studies. $\Upsilon(6S)$ deserves further studies, in
particular, to clarify if $Z_b$ states are also produced in its
decays, to search for $\Upsilon(6S) \to h_b \pi^+\pi^-$ transitions,
and to measure the $e^+e^- \to h_b \pi^+\pi^-$ cross section as a
function of energy, which should provide important information that is
needed to answer whether $\Upsilon(6S)$ is more similar to
$\Upsilon(5S)$ or to $Y(4260)$ in its properties. With a possible
upgrade of the injection system to increase its energy from current
$11.2$~GeV, Belle~II could access also more molecular states close to
$B^{(*)} \overline{B}^{(*)}$, predicted from heavy quark spin symmetry. 
Belle~II and the LHC upgrade, as well as future higher energy/luminosity $ep$ (electron-ion) 
and $e^+e^-$ (Higgs factory) colliders, will provide precision measurements 
of heavy quarkonium production with more diverse observables in various environments, 
and might thereby challenge our understanding of how heavy quarkonia emerge from high energy collisions.

\clearpage
\section[Chape]{Searching for new physics with precision measurements and computations 
\protect\footnotemark}
\footnotetext{Contributing authors:
S.~Gardner$^{\dagger}$, H.-W.~Lin$^{\dagger}$, Felipe J.~Llanes-Estrada$^{\dagger}$, W.M.~Snow$^{\dagger}$, 
X.~Garcia~i~Tormo, A.S. Kronfeld}

\label{sec:chape}
\subsection{Introduction} \label{sec:secE1}

The scope of the current chapter extends beyond that of QCD. Therefore, 
we begin with a brief overview of the standard model (SM) in order
to provide a context for the new physics searches we describe
throughout. 

The current SM of particle physics is a renormalizable quantum field theory 
based on an exact SU(3)$_{c}\times$SU(2)$_{L} \times$U(1)
gauge symmetry. 
As a result of these features 
and its specific particle content, it contains additional, accidental 
global symmetries,
of which the combination  B$-$L is anomaly free. 
It also preserves the discrete spacetime symmetry CPT, 
but C and P and T are not separately
guaranteed --- and indeed P and C are violated by its explicit construction. 
It describes all the observed interactions of known matter, save for those 
involving gravity, 
with a minimum of 25 parameters. 
These parameters can be taken as the six quark masses, the six lepton masses, the 
four parameters each (three mixing angles and a CP-violating phase) 
in the CKM and Pontecorvo-Maki-Nakagawa-Sakata (PMNS) matrices which describe the 
mixing of quarks and neutrinos,\footnote{The three-flavor PMNS matrix carries
two additional phases if the neutrinos are Majorana.} respectively, under 
the weak interactions, and the five parameters 
describing the gauge and Higgs sectors. 
The SM encodes CP violation in the quark sector not only 
through a phase in the CKM matrix but also through a ``would-be'' 
parameter $\bar\theta$,
which the nonobservation 
of a permanent electric dipole moment of the neutron~\cite{Baker:2006ts} 
limits to 
$\bar\theta < 10^{-10}$ if no other sources of CP violation operate. 

The SM, successful though it is, is incomplete in that it leaves 
many questions unanswered. Setting aside the question of gravity, which is excluded 
from the onset, 
the SM cannot explain, e.g.,  why 
the $W$ and $Z$ bosons have the masses that they do, 
the observed pattern of masses and mixings of
the fermions, nor why there are three generations. It
cannot explain why $\bar\theta$ is so small, 
nor why the baryon asymmetry of the Universe
has its observed value. 
It does not address the nature or even the existence of dark matter and dark energy. 
It has long been thought that the answers to some of these questions could
be linked and, moreover, would find their resolution in new physics 
at the weak scale. The LHC 
is engaged in just such a search for those distinct and new phenomena that
cannot be described within the SM framework. 
In Sec.~\ref{sec:secE2} we review 
current collider efforts and how QCD studies advance and support them. 

Direct searches for new particles and interactions at colliders  
certainly involve precision measurements 
and computations, but discoveries of new physics can also 
be made at low energies through such
efforts. There are two paths: one can discover new physics through 
(i)~the observed failure 
of the symmetries of the SM, or (ii)~the failure of a precision computation to confront
a precision measurement. 
Examples from the first path include searches for permanent 
electric dipole moments (EDMs) and for charged-lepton flavor violation, at 
current levels of sensitivity, as well as searches 
for neutrinoless double beta decay and $n$-$\bar{n}$ oscillations. Prominent 
examples from the second path 
are the determination of the lepton anomalous magnetic moments, 
the $g-2$ of the muon and of the electron. Taken more broadly, the second path 
is also realized 
by overconstraining 
the SM parameters with multiple experiments and trying to find an inconsistency 
among them.
Updated elsewhere 
in this review are determinations of the weak mixing angle $\theta_W$ 
(Sec.~\ref{sec:secB4}) 
and the strong coupling constant $\alpha_s$ 
(Sec.~\ref{sec:secB5}), 
which are under intense scrutiny by the QCD community. We refer to 
 Sec.~\ref{sec:secB4} 
for a discussion of the muon $g-2$. 
In this chapter we review such 
results from quark flavor physics. 

QCD plays various roles in these efforts. In the first case, 
the discovery of whether a SM symmetry 
is actually broken is essentially an experimental question, 
though QCD effects play a key role not only 
in assessing the relative sensitivity of different experiments but also 
in the interpretation of an experimental result in terms of the parameters 
of an underlying
new physics model. In the second case, 
the importance of QCD and confinement physics is clear. 
QCD effects are naturally dominant in all experiments 
searching for new physics that involve hadrons. We emphasize that experiments in 
the lepton sector are not immune to such issues, because hadronic effects are simply 
suppressed by power(s) of the fine-structure constant $\alpha$ --- they 
enter virtually through loop corrections. Their ultimate importance is predicated by 
the precision required to discover new physics in a particular process. 
Generally, for fixed experimental precision, 
a lack of commensurate control over QCD corrections, 
be it in experiments at high-energy colliders or at low energies, can  
jeopardize our search for physics beyond the~SM.

In this document, we consider the broad ramifications of the 
physics of confinement, 
with a particular focus on our ability to assess its impact in the context of QCD. 
This interest drives the selection of the topics which follow. 
We begin with a brief overview of 
the role of QCD in collider physics. This part particularly concerns factorization
theorems and resummation, which is illustrated with a few, select 
examples. Our discussion, however, is not comprehensive, so that we
do not review here the recent and impressive progress on next-to-leading-order (NLO) 
predictions for multi-parton production processes, 
see Ref.~\cite{Bern:2013gka} for a recent example, or the associated development of
on-shell methods, which are reviewed in \cite{Ellis:2011cr,Bern:2007dw}. 
We refer to 
Sec.~\ref{sec:G_ConfBSM} of this document for a terse summary of these
developments. 
Next, we move to the primary focus of 
the chapter, which is the role of 
QCD in the search for new physics in low-energy processes. 
There is a large 
array of possible observables to consider; we refer the reader to a brief, recent 
overview~\cite{Gardner:2012aqa}, as well as to a dedicated suite
of reviews~\cite{Cirigliano:2013lpa,Engel:2013lsa,deGouvea:2013zba,Cirigliano:2013xha,Balantekin:2013tqa,Balantekin:2013gqa,Gardner:2013ama,Haxton:2013aca}.
In this chapter we describe 
the theoretical framework in which such experiments
can be analyzed before delving
more deeply into examples which illustrate the themes we have described. 
We consider, particularly, searches for 
permanent electric dipole moments of the neutron and proton
and precision determinations of $\beta$-decay correlation coefficients. 
We refer the reader to 
 Sec.~\ref{sec:secB4} for a detailed discussion of 
the magnetic moment of the muon. 
We proceed to consider the 
need for and the computation of 
particular nucleon matrix elements 
rather broadly before 
turning to a summary of recent results in flavor-changing processes and
an assessment of future directions.

\subsection{QCD for collider-based BSM searches}\label{sec:secE2}

\subsubsection{Theoretical overview: factorization}

A general cross section for a collider process involving hadrons is
not directly calculable in perturbative QCD. Any such process will
involve, at least, the energy scale of the collision and scales
associated with masses of the hadrons, apart from other possible
scales related to the definition of the jets involved in the process or to
necessary experimental cuts. 
There is therefore an unavoidable dependence on long-distance, 
nonperturbative scales, and one cannot invoke asymptotic freedom to
cope with it. Factorization theorems in QCD allow us to separate, 
in a systematic way,
short-distance and, thus, perturbatively calculable effects from
long-distance nonperturbative physics, which are encoded in
process-independent objects, 
such as the parton distribution functions (PDFs). 
We refer to  Sec.~\ref{sec:lq.struct.PDF-TMD-theory} for 
the theoretical definition of a 
PDF in the Wilson line formalism and a discussion of its 
empirical extraction. 
(A summary of pertinent lattice-QCD results, notably of 
the lowest moment of the isovector PDF, can 
be found in Sec.~\ref{sec:lq.struct.form-factors.lqcd}.) 
Factorization theorems are, therefore, 
essential to QCD calculations of hadronic hard-scattering
processes. The simpler structure of emissions in the soft and
collinear limits, which can generate low-virtuality states, are at the
basis of factorization proofs. Factorized forms for the cross sections
(see the next section and Sec.~\ref{sec:lq.struct.PDF-TMD-theory}
for some examples)
can be obtained via diagrammatic methods in perturbative 
QCD~\cite{Collins:1989gx} or, 
alternatively, 
by employing effective field theories (EFTs) 
to deal with the different scales present in the
process. Soft collinear 
effective theory  (SCET)~\cite{Bauer:2000yr,Bauer:2001yt,Bauer:2002nz,Beneke:2002ph} 
is the effective theory that implements the structure of soft and collinear
interactions at the Lagrangian level, and it has been extensively used in
the last years for many different processes, along with 
traditional diagrammatic approaches. Establishing a factorized form
for the cross section is also the first necessary step to perform
resummations of logarithmically enhanced terms, 
which are key for 
numerical accuracy in certain portions of phase space. 
In the following, we discuss a few 
illustrative examples, which allow us not only to glimpse state-of-the-art
techniques but also to gain an impression of the current challenges.

\subsubsection{Outcomes for a few sample processes}
We begin with single vector-boson ($W/Z/\gamma$) production in hadron-hadron  
collisions in order to illustrate an 
application of the procedure known as threshold
resummation. The transverse momentum, $p_{\rm T}$, 
spectrum for these
processes is known at 
NLO~\cite{Ellis:1981hk,Arnold:1988dp,Gonsalves:1989ar}, 
and there is ongoing work to
obtain the NNLO corrections. This is an extremely challenging
calculation, but even without it 
one can
improve the fixed-order results by including the resummation of
higher-order terms that are enhanced in certain limits. In some cases,
such resummations of the fixed-order results are necessary in order to obtain a
reasonable cross section. In particular, we focus now on the
large-$p_{\rm T}$ region of the spectrum, where enhancements related to 
partonic thresholds can appear. 
By a partonic threshold we mean configurations in which
the colliding partons have just enough energy to
produce the desired final state. In these cases, the invariant mass of
the jet recoiling against the vector boson is small, and the
perturbative corrections are enhanced by logarithms of the jet mass
over $p_{\rm T}$. The idea is that one can expand around the threshold 
limit and resum such
terms. For single-particle production this was first achieved at 
next-to-leading-logarithmic (NLL)
accuracy in \cite{Laenen:1998qw}. In general the cross section
also receives contributions from regions away from the partonic
threshold, but due to the rapid fall-off of the PDFs at large $x$ 
the
threshold region often gives the bulk of the perturbative
correction. 
SCET offers a convenient, well-developed framework in which to perform
such resummations, and allows one to push them to higher
orders. A typical factorized form for the partonic cross section
$d\hat{\sigma}$, for example in the $q\bar{q}\to gZ$ channel, looks schematically as follows
\begin{equation}\label{eq:dsigmapartfact}
d\hat{\sigma} \propto
	H \,
	\int \! \mathrm{d} k \, J_g(m_X^2-(2E_J)k)
	S_{q\bar{q}}(k)\,,
\end{equation}
with $m_X$ and $E_J$ the invariant mass and energy of the radiation
recoiling against the vector boson,
respectively. The jet function $J_g$ describes collinear radiation initiated (in
this case) by the gluon $g$ present at Born level, the soft function $S_{q\bar q}$
encodes soft radiation, and $H$ is the hard function which contains
short-distance virtual corrections. The argument of $J_g$ in
Eq.~(\ref{eq:dsigmapartfact}) can be understood by recalling that the
recoiling radiation $p_X^{\mu}$ is almost massless, i.e., it consists of
collinear radiation $p_J^{\mu}$ and additional soft radiation $p_S^{\mu}$. We can
then write $m^2_X=p_X^2=(p_J^{\mu}+p_S^{\mu})^2=p_J^2+2p_J\cdot p_S$, up to
terms of order $p_S^2\ll p_J^2$; the collinear radiation can be
approximated at leading order as $p_J^{\mu}\sim E_Jn_J^{\mu}$, with $n_J^{\mu}$ a
light-like vector, and we obtain $p_J^2=m_X^2-(2E_J)k$, where
$k\equiv n_J\cdot p_S$ is the only component of the soft radiation that is
relevant in the threshold limit. The hadronic cross section $d\sigma$ is
given by a further convolution with the PDFs $f_a$ as 
\begin{equation}
d\sigma\propto\sum_{ab=q,\bar{q},g}\int
dx_1dx_2f_a(x_1)f_b(x_2)\left[d\hat{\sigma}_{ab}\right] \,,
\end{equation}
where we include a sum over all allowed partonic channels $ab$. 
Resummation has now been achieved at NNLL accuracy in
Refs.~\cite{Becher:2009th,Becher:2011fc,Becher:2012xr} using SCET
techniques, which are based on the renormalization group (RG) 
evolution of the hard, jet, and soft
functions. Some NNLL results obtained 
using diagrammatic methods have also been presented in
\cite{Kidonakis:2012sy}. All ingredients required to achieve
N$^3$LL accuracy within the SCET framework are essentially
known~\cite{Garland:2001tf,Garland:2002ak,Gehrmann:2002zr,Gehrmann:2011ab,Becher:2006qw,Becher:2010pd,Becher:2012za,Becher:2013vva}. A
phenomenological analysis at that unprecedented 
level of accuracy, combined with the 
inclusion of electroweak corrections which are
enhanced by logarithms of the $Z$ or $W$ mass over $p_{\rm T}$~\cite{Becher:2013zua}, can be
expected to appear in the near future. These predictions can then be
used, for instance, to constrain the $u/d$ ratio of PDFs at large $x$ 
(to which we return again 
in Sec.~\ref{par:secEpdf} from a lower-energy point of view), 
and to help estimate the $Z(\to\nu\bar{\nu})+$jets background to 
new heavy-particle searches~\cite{Malik:2013kba} at the LHC. 

The same vector-boson production process but in the opposite limit, i.e., at
low $p_{\rm T}$, is a classic example in which resummation is essential to obtain
reasonable predictions, since the perturbative fixed-order calculation
diverges. An all-orders resummation formula for this cross section at
small $p_{\rm T}$ was first obtained in \cite{Collins:1984kg}. All 
ingredients necessary for NNLL accuracy have been 
computed. Predictions for the cross section at this level of accuracy
are discussed in 
Refs.~\cite{Bozzi:2010xn,Becher:2010tm,Becher:2011xn}. The factorized
formulas for this process are more involved than the corresponding
ones for threshold resummation in the previous paragraph. In the SCET
language, they involve what is sometimes called a ``collinear factorization anomaly''. This means that  
the treatment of singularities present in SCET diagrams
requires the introduction of additional regulators, in addition to the
usual dimensional regularization, such as an analytic phase-space
regularization~\cite{Becher:2011dz}, or, alternatively, one can also use
the so-called ``rapidity renormalization group'' 
formalism~\cite{Chiu:2012ir}, which is based on the regularization of 
Wilson lines. In any case, this generates some additional
dependence (the aforementioned collinear anomaly) 
on the large scale of the process, $Q$, with respect to
what one might otherwise expect.
There are, by
now, well-understood consistency
conditions~\cite{Chiu:2007dg,Becher:2010tm} that restrict the form of
this $Q$ dependence to all orders, and the factorization formula
remains predictive and useful. This nuance is directly related to the definition and
regularization of the TMD PDFs, which appear in the factorization
formula; see Sec.~\ref{sec:lq.struct.PDF-TMD-theory}
for further
discussion of TMD PDFs. Similar issues 
also appear when studying the evolution of double
parton distribution functions in double-parton scattering
(DPS) processes~\cite{Manohar:2012jr}; further discussion on DPS 
is given in Sec.~\ref{par:secEpdf}.

As we discuss in the next section, 
much of the current effort is,
of course, devoted to the study of the Higgs and its properties. Let
us just highlight here one example where good control over QCD effects
is necessary, and for which recent
progress has been significant.

To optimize the sensitivity of the analyses, Higgs-search data is
often separated into bins with a specific number of jets in the final
state. In particular, for the Higgs coupling measurements and spin
studies, the $H\to W^+W^-$ decay channel is quite relevant; but in
this channel there is a large background coming from $t\bar{t}$
production, which after the tops decay can produce a $W^+W^-$ pair
together with two $b$-quark jets. To reduce this background, events
containing jets with transverse momentum above a certain threshold are
rejected, i.e., one focuses on the 0-jet bin, which is also known as 
the jet-veto cross section. This
restriction on the cross section enhances the higher-order QCD
corrections to the process, by terms that contain logarithms of the
transverse-momentum veto scale (typically around 25--30~GeV) over the
Higgs mass. One should be careful when estimating the perturbative
uncertainty of fixed-order predictions for the jet-veto cross section,
since the cancellation of different effects can lead to artificially small
estimations. A reliable procedure to estimate it was presented in
\cite{Stewart:2011cf}, and the outcome is that the perturbative
uncertainty for the jet-veto cross section is around $20\%$, which is
comparable to the current statistical experimental uncertainty and
larger than the systematic one. It is therefore desirable to improve
these theoretical predictions. There has been a lot of progress, 
starting with
\cite{Banfi:2012yh}, which showed that the resummation could be
performed at NLL accuracy, and its authors 
also computed the NNLL terms associated with the
jet radius dependence. Subsequently, resummation of these 
logarithms 
was performed at NNLL 
precision~\cite{Stewart:2013faa,Becher:2013xia,Banfi:2012jm,Tackmann:2012bt,Becher:2012qa}. An
all-orders factorization formula was also put forward in
Refs.~\cite{Becher:2013xia,Becher:2012qa} within the SCET framework;
its adequacy, though, has been questioned in
Refs.~\cite{Stewart:2013faa,Tackmann:2012bt}. In any case, the
accuracy for this jet-veto cross section has significantly improved,
and there is room to continue improving the understanding of jet-veto
cross sections and their uncertainty.

Related to the discussion of the previous paragraph, one would also
like to have resummed predictions for $N$-jet processes, by which we mean any 
process with $N$ hard jets.  Although
there has been important recent
progress~\cite{Becher:2009cu,Becher:2009qa} regarding the structure of
infrared singularities in gauge theories, connecting them to $N$-jet
operators and its evolution in SCET, many multi-jet processes involve
so-called nonglobal logarithms~\cite{Dasgupta:2001sh}. These are
logarithms that arise in observables that are sensitive to radiation
in only a part of the phase space. In general they appear at the NLL
level, and although several explicit computations of these kinds of
terms have been performed, it is not known how to resum them in general. Their
presence, therefore, hinders the way to resummation for general
$N$-jet cross sections. One might be forced to switch to simpler
observables; see, e.g., Ref.~\cite{Stewart:2010tn}, 
to be able to produce predictions at
higher-logarithmic accuracy. 

Giving their present significance, jet studies command a great part 
of the current focus of attention. In particular, driven in part by 
the new possibilities that the LHC offers, the study of jet substructure, 
and jet properties in general, is a growing field. Jet substructure analysis can allow
one, for instance, to distinguish QCD jets from 
jets coming from hadronic decays of
boosted heavy objects, see,
e.g., Refs.~\cite{Kim:2010uj,Thaler:2010tr,Feige:2012vc}. Many other new results
have appeared recently, and one can certainly expect more progress
regarding jet studies in the near future. This will hopefully allow
for improved identification techniques in searches for new heavy particles.

\subsubsection{LHC results: Higgs and top physics}
\label{subsubsec:secE2Higgstop}
The announcement in mid-2012 of the discovery 
of a boson of mass near 125~GeV while searching for
the SM Higgs electrified the world and represents a landmark achievement 
in experimental particle physics. It is decidedly a new physics result
and one which we hope will open a new world to us. 
The discovery raises several key questions: 
What is its spin? Its parity? 
Is it pointlike or composite? One particle or the beginning of a 
multiplet? Does it couple like the SM Higgs to quarks, leptons, 
and gauge bosons? No other significant deviations from SM expectations
have as yet been observed, falsifying many new-physics models. 
Nevertheless, plenty of space remains for new possibilities, both
within and beyond the Higgs sector, 
and we anticipate that resolving whether the new particle is
``just'' the SM Higgs 
will require years 
of effort, possibly extending beyond the LHC. 
The ability to control QCD uncertainties will be essential 
to the success of the effort, as we can already illustrate. 

\paragraph{Higgs production and decay \label{par:secEHiggs}}

\begin{figure}[b]
\includegraphics*[width=\linewidth]{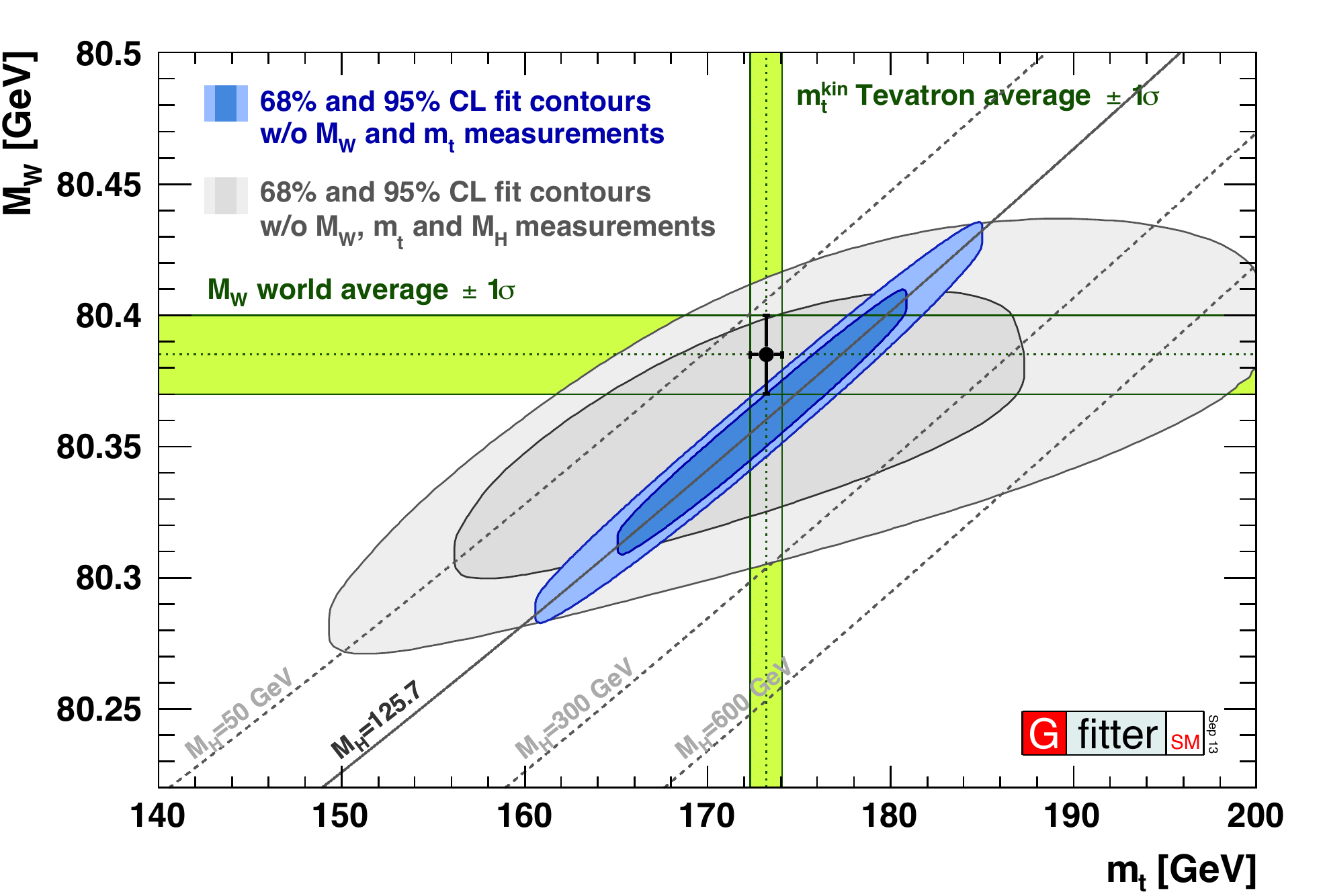}
\caption{Direct and indirect determinations of the 
$W$-boson and $t$-quark masses within the SM 
from measurements at LEP~\cite{ALEPH:2005ab} and the Tevatron~\cite{Aaltonen:2012ra},
and from Higgs mass $M_H$ measurements 
at the LHC~\cite{Aad:2012tfa,Chatrchyan:2012ufa}. 
The nearly elliptical contours indicate 
constraints from global fits to electroweak data, note
\protect{\url{http://cern.ch/gfitter}}~\cite{Flacher:2008zq}, exclusive of 
direct measurements of $M_W$ and $m_t$ from LEP and 
the Tevatron~\cite{Baak:2011ze,Baak:2012kk}. The smaller (larger) contours
include (exclude) the Higgs mass determinations from the LHC. 
We show a September, 2013 update from a similar figure in 
\cite{Baak:2012kk} and refer to it for all details. 
}
\label{fig:Higgsmass}
\end{figure}

The observation of the Higgs candidate 
by the 
\href{http://atlas.ch}{ATLAS} \cite{Aad:2012tfa} and 
\href{http://cern.ch/cms}{CMS} \cite{Chatrchyan:2012ufa}
collaborations was based on the study of the $H\to \gamma\gamma$
and $H\to ZZ \to 4{\ell}$, with ${\ell} \in e,\mu$, channels, due
to the excellent mass resolution possible in these final 
states~\cite{Dissertori:2013,CMS:yva}. The finding was 
supported by reasonably good statistics, 
exceeding 4$\sigma$ significance, in the four-lepton channel, for
which the background is small, whereas the background in $H\to \gamma \gamma$ 
is rather larger. Further work has led to an observed 
significance of $6.7\sigma$ (CMS) in the $H\to ZZ$ channel alone,
and to studies of the $H\to WW$, $H\to bb$, $H\to \tau\tau$ decay
modes as well~\cite{CMS:yva}. It is worth noting that the 
Bose symmetry of the observed two-photon final state
precludes a $J=1$ spin assignment to the new particle; 
this conclusion is also supported by further study of 
$H\to ZZ\to 4 {\ell}$~\cite{Aad:2013xqa}. 
Moreover, the finding is compatible with 
indirect evidence for the existence of a light Higgs boson~\cite{Aaltonen:2012ra}. 
Figure~\ref{fig:Higgsmass} shows a comparison of recent direct
and indirect determinations of the $t$ quark and $W$ gauge boson 
masses; this tests the
consistency of the SM. 
The horizontal and vertical bands result 
from using the observed $W$ (LEP+Tevatron)
and $t$ (Tevatron) masses at 68\% C.L., and 
global fits to 
precision electroweak data, once the $t$ and $W$ direct measurements are excluded,
are shown as well~\cite{Baak:2012kk}. 
The smaller set of ellipses include determinations
of the Higgs mass determinations from the LHC. 

\begin{figure}[b]
\includegraphics*[width=\linewidth]{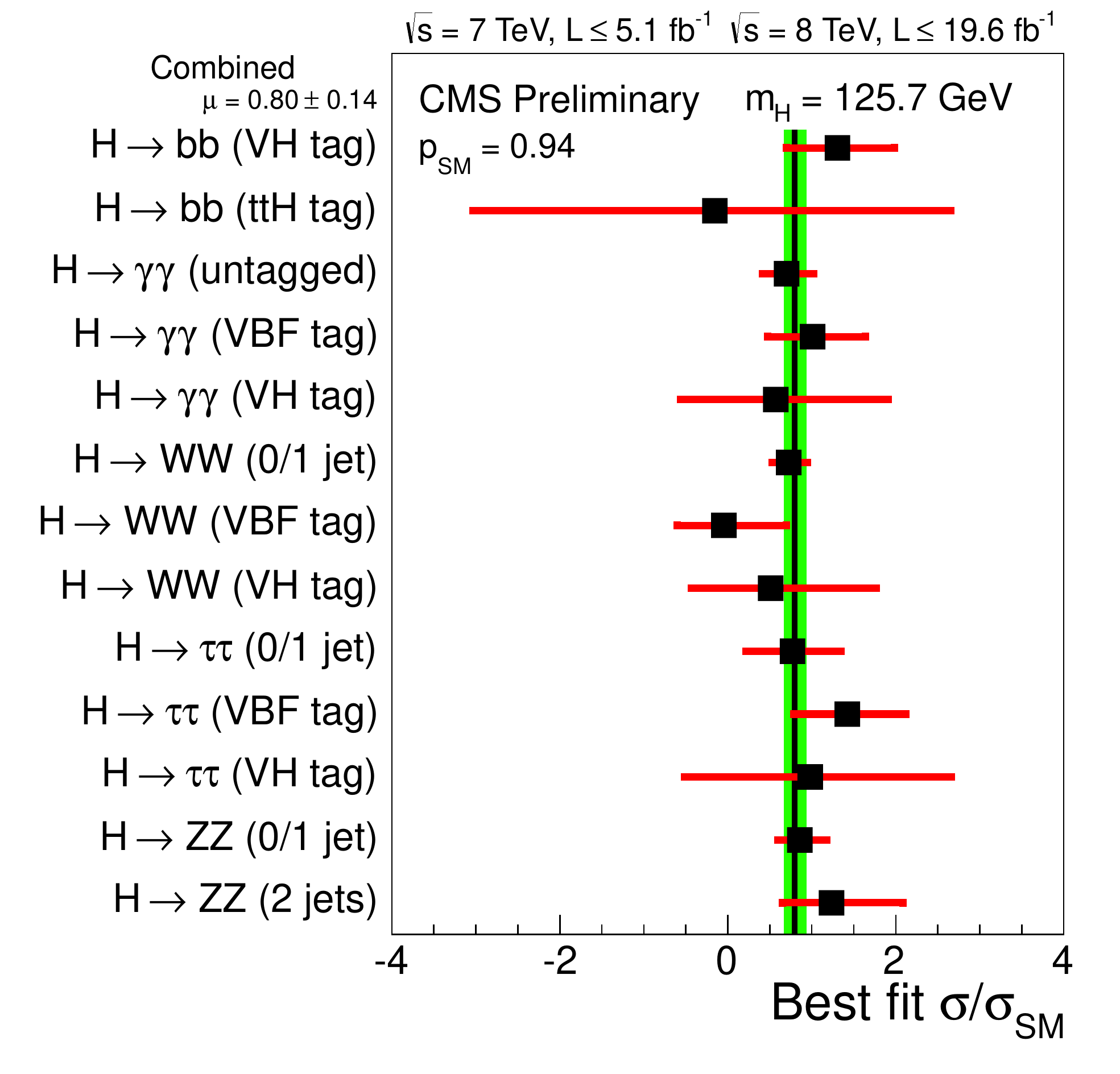}
\caption{Values of $\sigma/\sigma_{\rm SM}$ for particular decay
modes, or of subcombinations therein which target particular 
production mechanisms. The horizontal bars indicate $\pm 1\,\sigma$ errors
including both statistical and systematic uncertainties; the vertical
band shows the overall uncertainty. 
The quantity $\sigma/\sigma_{\rm SM}$ (denoted $\mu(x,y)$ in text) is
the production cross section times the branching fraction, relative
to the SM expectation~\cite{CMS:yva}. 
(Figure reproduced from \cite{CMS:yva}, courtesy of the
CMS collaboration.)
}
\label{fig:Higgsbr}
\end{figure}

We now summarize ongoing studies of the Higgs couplings, as well
as of its spin and parity, 
highlighting the essential role of QCD in these efforts. It is evident
that the Higgs discovery opens a new experimental approach to the
search for new physics, through the determination 
of its properties and couplings that are poorly constrained beyond
the SM~\cite{Brock:2014tja}. 
The theoretical control over the requisite SM cross sections
and backgrounds needed to expose new physics 
becomes more stringent as the constraints sharpen 
without observation of departures from the SM. 
Figure~\ref{fig:Higgsbr} shows the value of 
$\sigma/\sigma_{\rm SM}$, namely, of 
the production cross section times the branching fraction, relative
to the SM expectation~\cite{CMS:yva}, with decay mode and targeted
production mechanism, where the latter includes $gg$, VBF, VH (WH and ZH), and
$t{\bar t}H$ processes. This quantity is usually called $\mu$, and we can 
define, for production mode $X$ and decay channel~$Y$, 
\begin{equation}
\mu(X,Y) \equiv \frac{\sigma(X){\cal B}(H\to Y)}
{\sigma_{\rm SM}(X){\cal B}_{\rm SM}(H\to Y)} \,,
\label{defmu}
\end{equation} 
noting a global average of 
$\mu=0.80\pm 0.14$ for a Higgs boson mass of 125.7~GeV~\cite{CMS:yva}. 
See Ref.~\cite{Heinemeyer:2013tqa} for further
results and discussion and
Ref.~\cite{Boughezal:2013} for a succinct review. 
We note that $pp\to H$ via gluon--gluon fusion is computed
to NNLO $+$ NNLL precision in QCD, with an estimated uncertainty of
about $\pm 10\%$ by varying the renormalization and factorization 
scales~\cite{deFlorian:2012mx,Boughezal:2013}. In contrast, the 
error in the computed partial width of $H\to b\bar b$ is about 6\%~\cite{Denner:2011mq}.
The Higgs partial widths are typically accessed through channels
in which the Higgs appears in an intermediate state, as in 
Eq.~(\ref{defmu}). Consequently, the ratio of the Higgs coupling to a final state $Y$ 
with respect to its SM value, defined as 
$\kappa_Y^2 = \Gamma(H\to Y{\bar Y})/\Gamma_{\rm SM}(H\to Y{\bar Y})$, 
is determined through a multi-channel fit. 
The ability of the LHC to probe $\kappa_Y$
has been forecast to be some 
10--30\%~\cite{Duhrssen:2004cv,Lafaye:2009vr,Boughezal:2013}. 
Estimates instigated by the U.S.-based Community Planning Study (Snowmass 2013) 
support these assessments \cite{Brock:2014tja}, 
comparing the sensitivity of the current stage of the LHC (data samples 
at 7--8 TeV with an integrated luminosity of $20\,{\rm fb}^{-1}$) to staged
improvements at the LHC and to possible new accelerators, such as
differing realizations of a linear $e^+e^-$ collider. 
New backgrounds can appear at the LHC which were not known at LEP; e.g., 
a previously unappreciated background to the Higgs signal in $H\to ZZ$ and 
$H\to WW$, arising from asymmetric internal Dalitz conversion to a 
lepton pair, has 
been discovered~\cite{Gray:2011us}. Nevertheless, even with 
conservative 
assessments of the eventual (albeit known) systematic errors, tests of 
the Higgs coupling to $W$'s or $b$-quarks of sub-10\% precision 
are within reach of the LHC's high luminosity upgrade, with 
tests of sub-1\% precision possible at a $e^+e^-$ collider~\cite{Brock:2014tja}.  
These prospects demand further refinements of the existing SM predictions, with 
concomitant improvements in the theoretical inputs such as 
$\alpha_s$, $m_b$, and $m_c$~\cite{Brock:2014tja}. 

Current constraints on the quantum numbers of the new boson support
a $0^+$ assignment but 
operate under the assumption that it is {\it exclusively} of a particular
spin and parity. Of course admixtures are possible, and they 
can reflect the existence of CP-violating couplings; such possibilities
are more challenging to constrain. Near-degenerate states are
also possible and are potentially discoverable~\cite{Gunion:2012he}. 
ATLAS has studied various, possible spin and parity assignments,
namely of $J^P = 0^-,1^+, 1^-, 2^+$, as alternative hypotheses
to the $0^+$ assignment associated with a SM Higgs, 
and excludes these at a C.L. in excess of 97.8\%~\cite{Aad:2013xqa}. 
In the case of the $2^+$, however, a specific graviton-inspired model is
chosen to reduce the possible couplings to SM particles. 
It is worth noting that QCD effects play a role in these studies 
as well. In the particular example of the $H\to \gamma \gamma$ mode, 
the $J^P$ assignments of $0^+$ and $2^+$ are compared vis-a-vis
the angular distribution of the photons with respect to the $z$-axis
in the Collins-Soper frame~\cite{Aad:2013xqa}. 
The expected angular distribution of the 
signal yields in the $0^+$ case is corrected for interference
effects with the nonresonant diphoton background $gg\to\gamma\gamma$ 
mediated through quark loops~\cite{Dixon:2003yb}.

EFT methods familiar from the study of processes at lower energies 
also play an important role, and can work to disparate ends. 
They can be used, e.g., to describe 
a generalized Higgs sector~\cite{Nyffeler:1995mb}, providing not only a 
theoretical framework for the simultaneous possibility of various SM extensions 
therein~\cite{Buchalla:2013rka,Willenbrock:2014bja} but also a description of its 
CP-violating aspects~\cite{Altmannshofer:2011rm}. 
In addition, such methods can be used to 
capture the effect of higher-loop computations within the Standard Model.
For example, the effective 
vertex ($v$ is the Higgs vacuum expectation value)~\cite{Gounaris:1998ni}
\begin{equation}
\mathcal{L}_{\rm eff}= \alpha_s \frac{C_1}{4v} H F^a_{\mu\nu} F^{a\ \mu\nu}
\end{equation}
couples the Higgs to the two gluons in a SU(3)$_c$-gauge-invariant manner. 
It can capture this coupling in a very efficient way, yielding 
a difference of less than 1\% between the exact and approximate NLO cross
sections for a Higgs mass of less than 200~GeV~\cite{Boughezal:2013}. 
This speeds up Monte Carlo programs, for example.
All short-distance information (at the scales of $M_H$, $m_t$, or 
new physics) is encoded 
in the Wilson coefficient $C_1$, which 
is separately computed in perturbation theory.

\begin{figure}[b]
\includegraphics*[width=\linewidth]{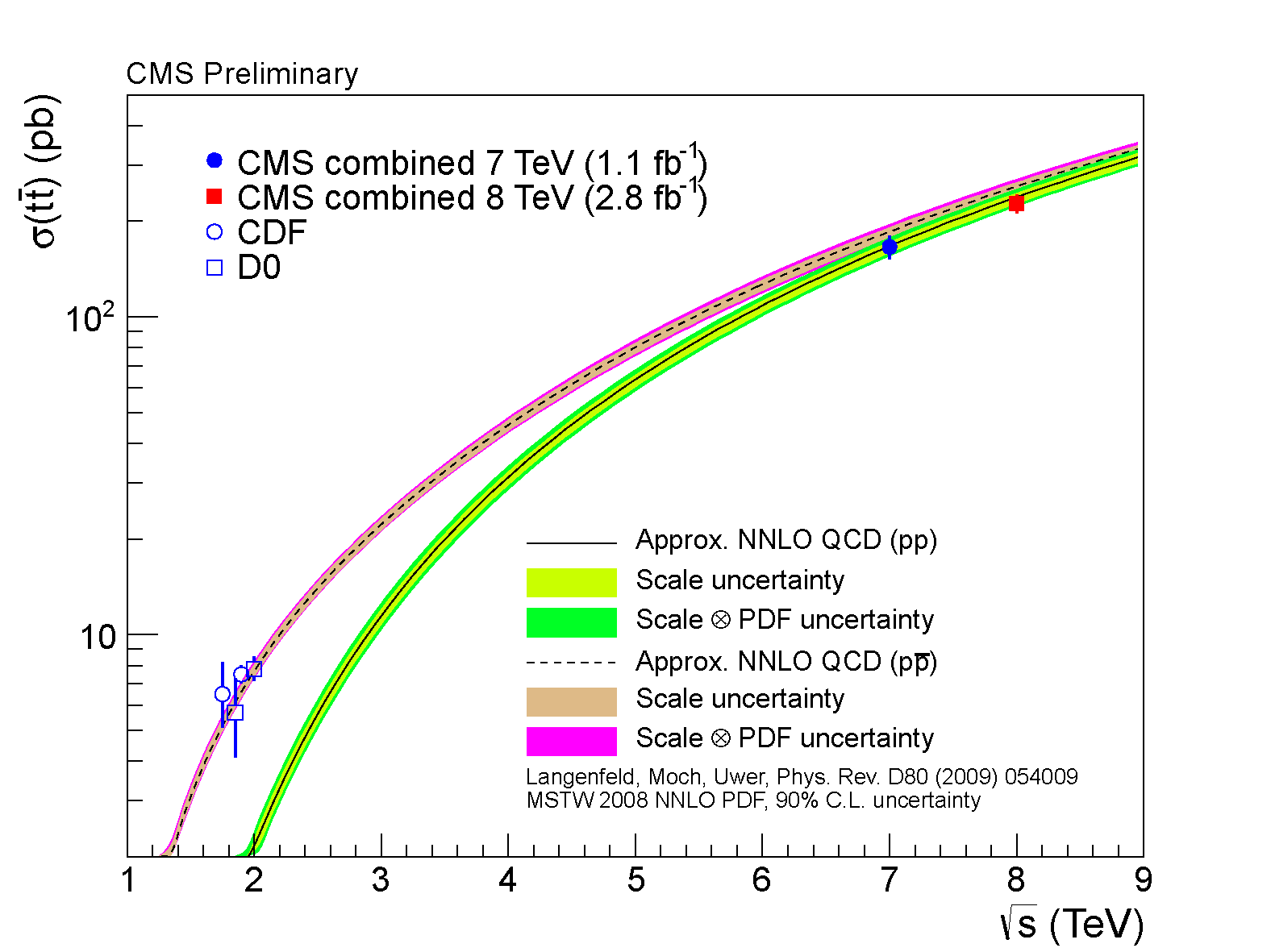}
\caption{Inclusive cross section for top pair production with center-of-mass energy
in $pp$ and $p\bar p$ collisions~\cite{Langenfeld:2009wd}, 
compared with experimental cross
sections from CDF, D0, ATLAS, and CMS~\cite{Silva:2013ira}. 
(Figure reproduced from \cite{Silva:2013ira}, 
courtesy of the CMS collaboration.)
}
\label{fig:new_topproduction}
\end{figure}

\paragraph{Top quark studies\label{par:secEtop}}
From the Tevatron to the LHC, the cross section for top-quark pair production 
$\sigma(t\bar{t})$, in Fig.~\ref{fig:new_topproduction}, grows 
by a factor of roughly $30$ due to the larger phase space; from 
7~pb at the 1.96~TeV center-of-mass (CM) energy of the Tevatron 
to some 160~pb at 
7~TeV and to some 220~pb at 8~TeV. We refer to 
\cite{Langenfeld:2009wd} cross-section predictions at 14~TeV and to 
\cite{Calderon:2013ar} for recent cross-section results
from CMS and ATLAS. 

A good part of $t\bar t$  production is near threshold, 
with a small relative velocity between the two heavy quarks. 
A nonrelativistic, fixed-order organization of 
the perturbative series is appropriate. Supplementing such a NNLO calculation 
with a resummation of soft and Coulomb corrections 
at NNLL accuracy, a computation of $\sigma(t\bar{t})$ at the LHC (7~TeV) 
of $10$~pb precision has been reported~\cite{Beneke:2011mq,Beneke:2012wb,Beneke:2013}. 
More generally, the predictions show a residual theoretical uncertainty
of some $3$--$4\%$, with an additional $4$--$4.5\%$ uncertainty
from the PDFs and the determination of $\alpha_s$~\cite{Beneke:2012wb,Beneke:2013}.
Measurements of the $t\bar t$ inclusive cross section can thus
be used to extract the top-quark mass, yielding a result of 
$m_t=171.4 \stackrel{+5.4}{{}_{-5.7}}\,{\rm GeV}$~\cite{Beneke:2012wb}, 
in good agreement with the direct mass determination from the Tevatron,
$m_t=173.18 \pm 0.56\,({\rm stat.})\,\pm 0.75\,({\rm syst.})\,
{\rm GeV}$~\cite{Aaltonen:2012ra},  
but less precise. The measurement of near-threshold $t\bar t$ production
at an $e^+e^-$ collider, in contrast, can reduce the precision with 
which $m_t$ is known 
by a factor of a few, spurring further theoretical 
refinements~\cite{Beneke:2013jia,Hoang:2013uda}. Moreover, in this case, the connection 
to a particular top mass definition is also crisp.

\paragraph{Collider searches for new particles\label{par:secEnew}}

ATLAS and CMS continue to search for the new
physics effects expected in various extensions of the SM. 
All searches, thus far, yield results compatible with the SM. 
Certain efforts concern searches 
for high mass $t\bar t$ resonances, such as could
be generated through a high mass (leptophobic) $Z^\prime$ or Kaluza-Klein gluon, 
or searches for top $+$ jet resonances, such as could
be generated through a high mass 
$W^\prime$~\cite{Erdmann:2013,Gauthier:2013,Aad:2012em,ATLAS:2012qe,Marionneau:2013fna}.  
Experimental collaborations face a new problem in collecting 
large top samples at the higher LHC energies: 
often the $t$ and $\bar{t}$ fly away together in a boosted frame, so that the SM 
decay with visible particles 
\begin{equation}
t\bar{t}\to Wb W\bar{b}\ \  (\to 6\ {\rm jets\ or \ \to 2 \ jets} + 2 {\rm  leptons} )
\end{equation}
contains several jets that may overlap
yielding ``fat jets,'' for which new algorithms are being 
developed~\cite{Almeida:2011ud}.

The constraints are sharpest for $t\bar t$ resonances, which decay into 
lepton pairs, with exclusion limits of 2.79~TeV at 95\% C.L. for a
$Z^\prime$ (with SM-like couplings) decaying into $e^+e^-$. 
In contrast, the 95\% C.L. exclusion limit on a 
leptophobic $Z^\prime$ decaying into $t\bar t$ is greater than
1.5~TeV~\cite{Marionneau:2013fna}.
The parity programs at JLab (note, e.g., 
\href{http://hallaweb.jlab.org/experiment/HAPPEX}{HAPPEX} and 
\href{http://www.jlab.org/qweak}{Q-weak})
and 
\href{http://www.prisma.uni-mainz.de/mesa.php}{MESA} at Mainz are geared towards 
searches for similar objects, in complementary regions of parameter space, 
 through the precision measurement
of parity-violating asymmetries 
at low momentum transfers~\cite{Kumar:2013yoa,Androic:2013rhu}. 
Moreover, a unique window on the possibility of a leptophobic $Z^\prime$ 
can come from the study of 
parity-violating deep inelastic scattering of polarized electrons from
deuterium~\cite{Buckley:2012tc}. 

Significant indirect constraints exist on the possibility of
an extra chiral generation of quarks 
from the observation of 
$H\to \gamma\gamma$~\cite{Kuflik:2012ai}, as well as through the
apparent production of the Higgs through 
$gg$ fusion. 
Direct searches are mounted, however, for certain ``exotic'' variants
of the extra generation hypothesis, 
be they vector-like quarks, or quarks with unusual electric
charge assignments~\cite{Gauthier:2013,ATLAS:2012qe}. All 
searches thus far are null, and $(5/3)e$-charged up quarks, e.g., 
are excluded for masses below 700 GeV at 95\% C.L.~\cite{Marionneau:2013fna}. 

Because no new particle (beyond the Higgs-like particle) has yet appeared 
in the mass region below 1~TeV, direct searches for 
a new resonance $R$ will likely extend to higher mass scales. 
This will push the QCD inputs needed for PDF fits to the
limits of currently available phase space, and it is worth
exploring the prospects for better control of such quantities. 
Precision determinations of the particle properties and couplings
of the particles we know also drive a desire to understand the PDFs
as accurately as possible. 
We also refer to Sec.~\ref{sec:lq.struct.PDF-TMD-theory} for a discussion of PDFs
and their uncertainties. 

\subsubsection{Uncertainties from nucleon structure and PDFs}
\label{par:secEpdf}

In order to produce a previously unknown particle $R$, 
the colliding partons in the initial state, as in for instance 
$g(x_1) g(x_2)\to R +X$,  
must each carry a significant fraction of the proton's momentum. 
This makes constraining parton distribution functions at large Bjorken $x$, 
particularly for $x>0.5$, 
ever more important as the mass of $R$ increases. 
As we have seen, the PDF and scale uncertainties 
are the largest uncertainties in the predicted inclusive 
$t\bar t$ cross section. Such uncertainties are also important to the 
interpretation of ultra-high--energy neutrino events observed 
at Ice Cube~\cite{Chen:2013dza}, whose rate may exceed that of the SM. 
There is currently an effort~\cite{Accardi:2009br,Accardi:2011fa,Owens:2012bv} 
to investigate this issue by combining the traditional 
CTEQ fits in the large-$x$ ($x\to 1$) region with JLab data 
at lower energies. These efforts will likely wax with importance
in time because, 
the 12~GeV upgrade at JLab will allow greatly expanded
access to the large-$x$ region~\cite{Dudek:2012vr}. Various complications 
emerge as $x\to 1$, and
it is challenging to separate the additional contributions that arise. 
In particular,
large logarithms, the so-called Sudakov double logarithms, 
appear in the $x\to 1$ region, and they need to be resummed in order to 
get an accurate assessment of the cross section. 
To this end the $x\to 1$ region has been subject to 
extensive theoretical investigation, both in traditional approaches based
on factorization theorems~\cite{Sterman:1995fz,Berger:2003zh} 
and in effective 
field theory~\cite{Manohar:2003vb,Becher:2006nr,Idilbi:2006dg,Chen:2006vd,Becher:2006mr}. 
Moreover, studies of 
deep inelastic scattering in nuclei require the assessment of 
Fermi-motion effects as well. The former issue is skirted in traditional global 
fits, based on structure functions in leading-twist, collinear factorization, by 
making the cut on the hadronic invariant mass $W$ large, such as 
in~\cite{Martin:2009iq} for which $W^2 \ge 15\,{\rm GeV}^2$.
Here, $W^2 =M^2 + Q^2 (1-x)/x$. 
The global-fit approach in~\cite{Accardi:2009br,Accardi:2011fa,Owens:2012bv} 
includes both large-$x$ and nuclear corrections and allows the $W$ cut to be relaxed
to $W\sim 1.7\,{\rm GeV}$~\cite{Accardi:2013pra}. 

To obtain the $d$ quark distribution, for example, 
one uses the data on the unpolarized structure function 
$F_2$, e.g., from deep inelastic scattering on the 
proton and neutron, to find  
\begin{equation}
\frac{d(x)}{u(x)} = \frac{4 F_{2n}(x) - F_{2p}(x)}{4F_{2p}(x) -F_{2n}(x)} \,, 
\end{equation}
where, for brevity, 
we suppress the $Q^2$ dependence.
Since there are no free neutron targets, the experiments
are performed with few-body nuclei, either the deuteron or $^3$He. 
For $x$ above $x\simeq 0.5$, 
the nuclear corrections become large. 
The CTEQ-JLab fits employ a collinear factorization formula
\begin{eqnarray}
F_{2d}(x,Q^2) &=& 
\sum_{N=p,n} \int dy S_{N/A}(y,\gamma) F_2(x/y,Q^2) \nonumber \\ 
&&+ \Delta^{\rm off}(x,Q^2) \,,
\end{eqnarray}
where the deuteron structure function is computed 
from the parametrized nucleon $F_2$, 
a modeled off-shell correction $\Delta^{\rm off}$, and 
a nuclear smearing function $S_{N/A}$, computed from 
traditional nuclear potential theory based on the 
Paris, Argonne, or CD-Bonn interactions. 
There is clearly room for QCD-based progress in these computations.
The notion of \cite{Accardi:2012pba,Owens:2012bv} is that
data on the $W^\pm$ 
charge asymmetry from the Tevatron~\cite{Abazov:2008qv,Aaltonen:2009ta} 
can be used 
to fix the $d(x)/u(x)$ ratio at large $x$, and then precision nuclear
experiments can be used to fix the nuclear corrections. 
Future JLab experiments, which are less sensitive to nuclear effects, 
can then be used to test the procedure~\cite{Accardi:2012pba}. 

Of course, the higher energy run of the LHC at 14~TeV, scheduled for 2015,
should also lower the $x$ needed for a given energy reach. 
Taking 2~TeV as the reference CM energy for a gluon-gluon collision, 
doubling the LHC energy from 7 to 14~TeV increases  
the parton luminosity by a factor of 50~\cite{Quigg:2011zu}, making the 
new physics reach at ${\cal O}(1\,\text{TeV})$ 
less sensitive to the large $x$ behavior of the PDFs. At 14~TeV the parton
luminosity (taking this as a crude proxy for $x$) of the 2~TeV gluon-gluon
subprocess in the 7~TeV collision is found at a CM energy of 
3.3~TeV~\cite{Quigg:2011zu}. Sorting out the PDFs in the large-$x$ region
may prove essential to establishing new physics. 

Another issue 
for new physics searches and Higgs physics is 
double-parton scattering~\cite{Manohar:2012pe,Manohar:2012jr}.
Two hard partons collide if they coincide within a transverse area of size $1/Q^2$ 
out of the total $1/\Lambda_{\rm QCD}^2$. 
The flux factor being $1/\Lambda_{\rm QCD}^2$, 
the probability of one hard collision scales as $\hat{\sigma}_1 \propto 
({1}/{\Lambda_{\rm QCD}^2}) ( {\Lambda_{\rm QCD}^2}/{Q^2})$. 
The probability of a double collision 
in the same $pp$ event (this is not the same as pile up, which is the aftermath of 
multiple, nearly simultaneous $pp$ events) 
is thus power-suppressed,  
$\hat{\sigma}_2 \propto ({1}/{\Lambda_{\rm QCD}^2}) ( {\Lambda_{\rm QCD}^2}/{Q^2})^2$. 
The rate is small
but still leads to a background about three times the signal 
in Higgs processes such as 
$pp\to WH\to l\bar{\nu}b\bar{b}$~\cite{Bandurin:2010gn}. 
It also entails power corrections to double Drell-Yan processes, 
an important 
background to four-lepton Higgs decays. 
Like-sign $W^+W^+$ production has long been recognized as a viable 
way to identify double-parton 
scattering~\cite{Kulesza:1999zh,Gaunt:2010pi} because this final state 
is not possible in single-parton scattering unless 
two additional jets are emitted (due to charge and quark-number conservation). 
It comes to be dominated by 
double scattering when the particle pairs come out almost 
back-to-back (typically $|{\bf p}_{1{\rm T}} + {\bf p}_{2{\rm T}}| \sim \Lambda_{\rm QCD}$).

One might suppose the differential cross-section for double-parton scattering could
be described as~\cite{Paver:1982yp}  
\begin{multline} 
\frac{d\sigma^{DPS}}{dx_1dx_2dx_3dx_4} 
\propto \\
\!\! \int \!\!d^2z_\perp F_{ij}(x_1,x_2,z_\perp)
F_{kl}(x_3,x_4,z_\perp) \hat{\sigma}_{ik} \hat{\sigma}_{jl} \,,
\label{facDPS}
\end{multline}
employing a distribution-like function $F$ to describe 
the probability of finding the two partons in the proton at $z_\perp$ 
from each other in the plane perpendicular to the momentum, 
with given momentum fractions $x_i$. Quantum interference is  
intrinsic to this process, however, 
so that 
some knowledge of the proton at the wavefunction or amplitude level is 
needed, as a purely probabilistic description is insufficient. 
We refer to \cite{Manohar:2012jr} for a detailed analysis. 

\subsubsection{Complementarity with low-energy probes}

Searches for unambiguous signs of 
new physics at high-energy colliders have so far proved null; it may
be that new physics appears at yet higher energy scales or that it is more weakly coupled
than has been usually assumed. 
In the former case, a common theoretical framework, 
which is model-independent and contains few assumptions, 
can be used to connect the constraints from collider observables to those
from low-energy precision measurements; we 
provide an overview thereof in the next section. 
In the latter case, an explicit BSM model is required to connect
experimental studies at high and low energy scales, and the 
minimal supersymmetric standard model (MSSM) is a particularly popular example. 
The impact of permanent electric dipole moment (EDM) searches at low energies, for example, 
on the appearance of CP-violating terms in the softly broken supersymmetric sector of the MSSM 
and its broader implications 
have been studied for 
decades~\cite{Ellis:1982tk,Buchmuller:1982ye,Polchinski:1983zd,Pospelov:2005pr,Ellis:2008zy,Li:2010ax}. Computations of the various QCD matrix elements which appear 
are important to the assessing the 
loci of points in parameter
space which survive these constraints; we discuss the state of the art, albeit
in simpler cases, in Sec.~\ref{par:secEedmlat}. 

In the event that new physics 
is beyond the reach of current colliders, 
the connection between experimental probes at the highest and lowest energies
mentioned is particularly transparent and certainly 
two-way. Although collider experiments limit new-physics 
possibilities at low energies, it is also the case that low-energy experiments
limit the scope of new-physics at colliders. 
Before closing this section, we consider an example of how 
a model-independent approach employing effective Lagrangian 
techniques can be used in the 
top-quark sector as well~\cite{Gounaris:1996vn}. Usually
such techniques are employed assuming the accessible energy to be 
no larger than the $W$ mass~\cite{Buchmuller:1985jz,Grzadkowski:2010es}. 
In particular, 
we consider the possibility that the top quark itself could have a permanent 
(chromo)electric or (chromo)magnetic dipole moment. 
This is particularly natural 
if the top quark is a composite 
particle~\cite{Kaplan:1991dc}, and the large top-quark mass suggests that the 
effects could well be large~\cite{Agashe:2006wa}. 
Although such effects could potentially be probed directly through spin 
observables~\cite{Fajfer:2012si}, 
severe constraints from the neutron EDM also operate~\cite{Kamenik:2011dk,Ayazi:2013cba}, 
to yield a severe constraint on the chromoelectric 
top-quark operator through its effect 
on the coefficient 
$w$  of  the Weinberg three-gluon operator
\begin{equation} \label{Weinbergthreegluon}
{\mathcal L}_{W3g}= -\frac{w}{6} f^{abc} \varepsilon^{\mu\nu\lambda\rho} (F^a)_{\mu \sigma}
(F^b)^\sigma_\nu (F^c)_{\lambda \rho}
\end{equation}
at low energies~\cite{Kamenik:2011dk}, where $f^{abc}$ are SU(3) structure 
constants. 
Turning to the specific numerical details, 
the QCD matrix element of the Weinberg operator in the neutron is needed, 
and the QCD sum rule calculation of \cite{Demir:2002gg} has been employed 
to obtain the limits noted~\cite{Kamenik:2011dk}. (See Secs.~\ref{par:secEedmeft}
and \ref{par:secEedmlat} for further discussion of matrix elements for EDMs.) 
Stronger limits on the color-blind dipole moments, however, come from 
$b\to s \gamma$ and $b\to s {\ell}^+{\ell}^-$ decays~\cite{Hewett:1993em,Kamenik:2011dk}.
In the face of such constraints, the space of new-physics models to be 
explored at the LHC is significantly reduced~\cite{Kamenik:2011dk,Ayazi:2013cba}, 
and presumably can be sharpened further, even in the absence of additional 
experimental data, if the nonperturbative matrix element can be more accurately
calculated. In the sections to follow we will find further examples of low-to-high
energy complementarity. 

\subsection{Low-energy framework for the analysis of BSM effects}\label{sec:secE3}

The SM leaves many questions unanswered, and the best-motivated models of
new physics are those which are able to address them. Commonly, this is realized
so that the more fundamental theory has the SM as its low-energy limit. 
It is thus natural to analyze the possibility of physics beyond the SM within an 
effective field theory framework. To do this 
we need only assume that we work at some energy $E$ below the scale $\Lambda$ at which
new particles appear. Consequently for $E< \Lambda$ any new degrees of freedom
are ``integrated out,'' and the SM is amended by higher-dimension operators
written in terms of fields associated with SM particles~\cite{Appelquist:1974tg}. 
Specifically, 
\begin{equation}
{\cal L}_{\rm SM} \to {\cal L}_{\rm SM} + 
\sum_i \frac{c_i}{\Lambda^{D-4}} {{\cal O}^D_i} \,,
\label{eq:frame}
\end{equation}
where the new operators ${\cal O}_i^D$ have dimension $D>4$. 
We emphasize that ${\cal L}_{\rm SM}$ contains a dimension-four operator,
controlled by $\bar \theta$, that can also engender CP-violating effects,
though they have not yet been observed. 
The experimental limit on the neutron EDM 
implies $\bar\theta<10^{-10}$~\cite{Baker:2006ts}, though 
the underlying reason for its small value is unclear. 
This limitation is known as the ``strong CP problem.'' If its resolution is
in a new continuous symmetry~\cite{Peccei:1977hh} that is 
spontaneously and mechanically broken at low energy, then there is a new
particle, the axion~\cite{Weinberg:1977ma,Wilczek:1977pj}, which we may yet
discover~\cite{Asztalos:2009yp,Graham:2013gfa}. 
The higher-dimension operators include
terms which manifestly break SM symmetries and others which do not. 

Since 
flavor-physics observables constrain the appearance of 
operators that are not SM invariant to energies
far beyond the weak scale~\cite{Isidori:2010kg,Bona:2007vi,Charles:2004jd}, 
it is more efficient to organize the higher-dimension terms so that only those 
invariant under SM electroweak gauge symmetry are included. 
Under these conditions, and setting aside B- and L-violating operators, 
the leading-order 
(dimension-six) terms in our SM extension 
can be found in~\cite{Buchmuller:1985jz,Grzadkowski:2010es}. 
Nevertheless, this description does not capture all the
possibilities usually considered in dimension six because of 
the existence of neutrino mass. The latter has been established
beyond all doubt~\cite{Beringer:2012zz}, though the need for the inclusion of 
dynamics beyond that in the SM to explain it has, as yet, not been
established. To be specific, we can use the Higgs mechanism to generate their 
mass.\footnote{Alternatively, a dimension-five operator, which is SM electroweak 
gauge invariant but L-violating, can be used~\cite{Weinberg:1979sa}. 
After electroweak symmetry breaking, this term yields a Majorana neutrino mass term 
$(Y^2 v^2/\Lambda) \nu_L^T C \nu_L$, in which $Y$ is the hypercharge and $C$ is 
the charge-conjugation operator, 
and makes neutrinoless double-$\beta$ decay possible.}
Since the neutrinos are all light in mass, to explore the consequences
of this possibility,  
we must include three right-handed neutrinos explicitly in our description 
at low energies~\cite{Cirigliano:2012ab}. 
Finally, if we evolve our
description (valid for $E<\Lambda$)  to the low energies ($E \ll M_W, \Lambda$) 
appropriate to the study of the weak decays of
neutrons and nuclei, we recover precisely ten independent terms, just as
argued long ago by Lee and Yang starting from the assumption of Lorentz
invariance and the possibility of parity nonconservation~\cite{Lee:1956qn}. 
The latter continues to be the framework in which new physics searches 
in $\beta$-decay are analyzed, as discussed, e.g., 
in~\cite{Severijns2006dr,Severijns:2011zz,Cirigliano:2013xha,Gonzalez-Alonso:2013uqa}.

In order to employ the low-energy quark and gluon operator framework we have discussed 
in a chiral effective theory in nucleon degrees of freedom, 
nucleon, rather than meson, matrix elements need to be computed. 
Nucleon matrix elements are generally more computationally demanding 
than meson matrix elements in lattice QCD, 
since the statistical noise 
grows with Euclidean time $t$ as $\exp[(M_N-3 M_\pi /2)t]$ 
for each nucleon in the system. 
Thus, results with high precision in the nucleon sector 
lag those in the meson sector. Furthermore, extrapolating 
to the physical light-quark masses is more challenging for baryons, 
since chiral perturbation theory converges more slowly. 
The latter issue is likely to be brought under control in the near 
future, as ensembles of lattices begin to be generated 
with physical $u$ and $d$ (and $s$ and $c$) quark masses. 
This should greatly reduce the systematic uncertainties. 
Other systematics, such as finite-volume effects, 
renormalization, and excited-state contamination 
can be systematically reduced by improved algorithms 
and by increasing the computational resources devoted to the calculations.
We refer to Sec.~\ref{sec:lq.struct.form-factors.lqcd} for additional 
discussion. 

One interesting idea from experimental physics
is to perform ``blind'' analyses, so that the true result is hidden 
while the analysis is performed. Concretely what this means is 
that the result should only be revealed after all 
the systematics have been estimated. This technique has begun to be employed
in lattice-QCD calculations, notably in the computation 
 of the exclusive semileptonic decay matrix elements
needed to determine the CKM matrix elements 
$|V_{cb}|$ and $|V_{ub}|$~\cite{Bernard:2008dn,Bailey:2008wp}.
It would be advantageous to 
implement this approach in 
lattice-QCD calculations of nucleon matrix elements as well, 
so that an analysis of systematic effects
could be concluded on grounds independent of the specific result found. 
Blind analysis would help in ensuring an extremely careful 
analysis of systematics, and we hope the lattice 
community will choose to follow this approach 
in the next few years. 

We now turn to the analysis of particular low-energy experiments
to the end of discovering physics BSM and the manner in which 
theoretical control over confinement physics can support or limit them.

\subsection{Permanent EDMs} \label{sec:secE4}

\subsubsection{Overview \label{par:secEedmintro}}

The (permanent) EDM of the neutron 
is a measure of the distribution of positive and negative 
charge inside the neutron; it is nonzero if a slight offset in the arrangement of
the positive and negative charges exists. 
This is possible if interactions are present
which break the discrete symmetries of parity P and 
time reversal T. In the context of the CPT theorem, 
it also reflects the existence of CP violation, 
i.e., of the product of charge conjugation C and parity P, as well.  
Consequently, permanent 
EDM searches probe the possibility of new sources of CP violation
at the Lagrangian level. 
The EDM $\mathbf{d}$ of a nondegenerate system 
is proportional to its spin $\mathbf{S}$, and it is nonzero if the energy of the
system shifts in an external electric 
field $\mathbf{E}$, 
with an interaction energy proportional to $\mathbf{S}\cdot \mathbf{E}$. 

As we have already noted, 
the SM nominally possesses two sources of CP violation, 
the single phase $\delta$ in the Cabibbo-Kobayashi-Maskawa (CKM) matrix and
the coefficient $\bar\theta$ which controls 
the T-odd, P-odd product of the gluon field-strength tensor and its dual,
namely $\bar{\theta} ({\alpha_s}/{8\pi}) F^a \tilde{F}^a$. 
Experimental studies of CP violation in the B system have shown that 
$\delta\sim{\cal O}(1)$~\cite{Charles:2004jd,Bona:2007vi}, 
whereas neutron EDM limits have shown that
the second source of CP violation does not appear to operate. 
Even if a physical mechanism exists to remove the appearance of 
$\bar\theta$, higher-dimension operators from physics BSM may still 
induce it, so that we 
use experiment to constrain this second source, as well as CP-violating
effects arising from other BSM operators. 

The CKM mechanism of CP violation does give rise to nonzero 
permanent EDMs; however, the first nontrivial contributions
to the quark and charged lepton EDMs come in three- and four-loop order (for massless
neutrinos), 
respectively, so that for the down quark
$|d_d| 
\sim 10^{-34}\, e\hbox{-}{\mathrm{cm}}$~\cite{Khriplovich:1985jr,HEP-PH/9704355}. 
Nevertheless, there exists a well-known, long-distance  chiral enhancement of
the neutron EDM (arising from a pion loop and controlled by $\log(m_\pi/M_N)$), 
and estimates yield 
$|d_n| \sim 10^{-31}$--$10^{-33}\, e\hbox{-}{\mathrm{cm}}$~\cite{Gavela:1981sk,Khriplovich:1981ca,He:1989xj}, 
making it relatively larger but still
 several orders of magnitude below the 
current experimental sensitivity. It is worth noting that  
the nucleon's intrinsic flavor content can also modify an EDM 
estimate~\cite{Ellis:1996dg,ARXIV:1202.6270,Fuyuto:2012yf}. 
Finally, if neutrinos are massive Majorana particles, then 
the electron EDM induced by the CKM matrix can be greatly enhanced, 
though not sufficiently to make it experimentally observable~\cite{Ng:1995cs}. 
(Neutrino mixing and Majorana-mass dynamics can also augment the 
muon EDM in the MSSM in a manner which evades $e$-$\mu$ universality, motivating
a dedicated search for $d_\mu$~\cite{Babu:2000dq}.) 
A compilation of the results
from various systems is shown in Table~\ref{edm:tab:limits}. 

\begin{table}
\caption{Upper limits on EDMs ($|d|$) from different experiments. For the
``Nucleus'' category, the EDM values are of the $^{199}$Hg atom that contains 
the nucleus.
No {\it direct} limit yet exists
on the proton EDM, though such could be realized through a storage-ring experiment.
Here we report the best inferred limit in brackets, which is
determined by asserting that the $^{199}$Hg 
limit is saturated by $d_p$ exclusively. 
Table adapted from \cite{Kronfeld:2013uoa}.}
\label{edm:tab:limits}
\begin{tabular}{llc}
\hline\hline\noalign{\smallskip}
 Category & EDM Limit ($e\hbox{-}{\mathrm{cm}}$)  &  SM Value ($e\hbox{-}{\mathrm{cm}}$)   \\
\noalign{\smallskip}\hline\noalign{\smallskip}
 Electron             & $1.0\times10^{-27}\,(90\%\, {\rm C.L.})$~\cite{Hudson:2011zz}  &   10$^{-38}$       \\
 Muon             &  $1.9\times10^{-19}\,(95\%\, {\rm C.L.})$~\cite{Bennett:2008dy} &   10$^{-35}$       \\
 Neutron             & $2.9\times10^{-26}\,(90\%\, {\rm C.L.})$~\cite{Baker:2006ts}   &   10$^{-31}$       \\
 Proton             &  $[\,7.9\times10^{-25}\,]\quad\quad\quad\quad$~\cite{Griffith:2009zz} &  10$^{-31}$       \\
 Nucleus     & $3.1\times10^{-29}\,(95\%\, {\rm C.L.})$~\cite{Griffith:2009zz}   &   10$^{-33}$      \\
\noalign{\smallskip}\hline\hline
\end{tabular}
\end{table}

\subsubsection{Experiments, and their interpretation and implications}\label{par:secEedmexp}

The last few years have seen an explosion of interest in experimental approaches 
to searches for electric dipole moments of particles composed of light quarks and leptons. 
This increased scientific interest has developed for many reasons. 
First, the power of the existing and achievable constraints from EDM searches on 
sources of CP violation BSM 
has become more and more widely recognized. Moreover, 
other sensitive experimental tests of ``T'' invariance 
come from particle decays and reactions in 
which the observables are only motion-reversal odd 
and thus do not reflect true tests of time-reversal 
invariance~\cite{Sachs:1987gp}. 
Such can be mimicked  
by various forms of final-state effects which eventually limit their sensitivity. In contrast, 
the matrix element associated with an intrinsic particle EDM has definite transformation 
properties under time reversal because the initial state and the final state are the same.
The consequence is that an EDM search is one of the 
few true null tests for time-reversal invariance. 
Consequently an upper bound on an EDM 
constitutes a crisp, non-negotiable limit, and  
a positive observation of an EDM at foreseeable 
levels of sensitivity would constitute 
incontrovertible evidence for T violation. Moreover, since
the SM prediction is 
inaccessibly small, as shown in Table~\ref{edm:tab:limits}, it
would also speak directly to the existence of new physics. 
Popular models of new physics 
at the weak scale generate EDMs greatly in excess of SM 
expectations, and the parameter space of these models is already strongly constrained 
by current limits. Consequently, 
even null results from the next generation of EDM experiments would be interesting, for
these would give hints as to the energy scale at which new physics could be. 

Such null results could also damage beyond repair certain theoretical explanations
for generating the baryon asymmetry of the universe 
through the physics of the electroweak phase transition. 
Two of the famous Sakharov conditions for the generation 
of the baryon asymmetry (namely, B violation and a departure from thermal equilibrium) 
are already present in the SM, in principle. 
For a Higgs mass of some 125~GeV, however, SU(2)
lattice gauge-Higgs theory simulations, as in~\cite{Aoki:1999fi}, e.g., reveal that 
the electroweak phase transition is not 
of first order. The lack of a sufficiently 
robust first-order phase transition can also be problematic in BSM models. 
Nevertheless, new mechanisms, or sources,  of CP (or T) and 
C violation in the quark sector 
could make baryon production much more effective. Existing EDM constraints 
curtail possible electroweak baryogenesis scenarios in the MSSM 
severely~\cite{Cirigliano:2006dg,Carena:2008vj,Carena:2012np}, 
and an improvement in the experimental bound on $d_n$ 
by a factor of $\sim$100 could rule out the
MSSM as a model of electroweak 
baryogenesis~\cite{Morrissey:2012db,Cirigliano:2009yd,Li:2010ax}. 
This outcome would thereby favor supersymmetric models beyond the MSSM, 
such as
in~\cite{Pietroni:1992in,Davies:1996qn,Huber:2000mg,Kang:2004pp,Menon:2004wv,Huber:2006wf,Huber:2006ma,Profumo:2007wc,Blum:2010by,Carena:2011jy,Kumar:2011np}, or possibly
mechanisms based on the two-Higgs doublet model (2HDM)~\cite{Shu:2013uua},
or mechanisms which are not tied to the weak scale, 
such as leptogenesis, or dark-matter mediated scenarios. 
Consequently, people have come to recognize that a measurement 
of an EDM in any system, regardless of its complexity, is of fundamental interest. 
Since there are many different possibilities for generating an EDM 
at a microscopic level,
many experiments are likely to be needed to localize 
the fundamental source of any EDM once observed. 
New ideas for EDM measurements abound and have come from scientific 
communities in atomic, 
molecular, nuclear, particle, and condensed-matter physics. 

Compact overviews of this field
 can be found in~\cite{Chupp:2013,Kronfeld:2013uoa}, 
whereas a recent theoretical 
review can be found in~\cite{Engel:2013lsa}. 
The most stringent limits on particle EDMs 
come from atomic physics measurements in $^{199}$Hg~\cite{Griffith:2009zz}. 
However, it is known that, in the pointlike, nonrelativistic limit, 
the electron cloud of an atom shields 
any EDM which might be present in the nucleus --- making the atomic EDM 
zero even if the nuclear EDM were not. This ``no-go'' result is known as 
Schiff's theorem~\cite{Schiff:1963zz}. 
As a consequence, the fantastic upper bound on the EDM in this atom
places a much weaker constraint on the EDM of its nucleons.

Atomic and molecular 
physicists have long sought systems in which the EDM 
could be amplified rather than shielded by electron effects; such an
 amplification can indeed occur 
in certain polar molecules~\cite{Regan:2002,Hudson:2011zz}. 
Gross enhancements also exist in certain heavy atoms whose relativistic motion evades
Schiff's theorem, yielding an EDM which
scales as $Z^3\alpha^2$~\cite{Sandars:1965,Sandars:1966}. 
More recently it has been recognized that atoms whose nuclei possess octupole 
deformations~\cite{Spevak:1996tu} can have particularly enhanced atomic EDMs, 
by orders of magnitude over 
$^{199}$Hg~\cite{Dobaczewski:2005hz}, 
in part through the resulting mixing of certain nearly-degenerate atomic energy levels. 
Even such enhancements do not defeat Schiff's theorem completely, though they can come
close. 
To be suitable for an EDM experiment, it is also necessary to be able 
to polarize sufficiently large ensembles of nuclei in order 
to perform the delicate NMR frequency-difference measurements 
typically needed to detect EDMs. Such needs, in concert with the 
desired enhancements, 
lead one to consider 
certain heavy radioactive nuclei such as radon and radium.
Recently, the first direct evidence of octupole deformation 
in $^{224}$Ra has been established 
through measurements of Coulomb excitation of 2.85~MeV/amu 
rare-isotope beams at REX-ISOLDE (CERN)~\cite{rf:Gaffney2013}, 
strengthening the confidence in the size of the Schiff moment
in like systems, whose computation 
is dominated by 
many-body calculations in nuclear and atomic physics. 
Generally, in the presence of rigid octupole deformation, as observed in $^{224}$Ra, 
 the computation of
the Schiff moment is expected to be more robust~\cite{Engel:2013lsa}. 
This underscores the discovery potential of an EDM measurement in such systems. 
Progress towards an EDM measurement in $^{225}$Ra, e.g., is ongoing~\cite{Holt:2010upa}, and 
the sensitivity of an eventual EDM limit 
could be greatly increased through the enhanced isotope production 
capability of a megawatt-class 1 GeV proton linac~\cite{Kronfeld:2013uoa}.

EDM searches on simpler objects such as the neutron, proton, or deuteron, e.g., are 
of course more directly interpretable in terms of the 
fundamental sources of CP violation 
at the quark level. The theoretical interpretation of these systems in chiral 
effective theory has been under intense development~\cite{Mereghetti:2010tp,Maekawa:2011vs,deVries:2011re,deVries:2011an,deVries:2012ab}. 
Many experiments to search 
for a neutron EDM are in 
progress~\cite{Golub:1994,Baker:2010zza,Burghoff:2011xk,Altarev:2012uy}, 
of which the nEDM-SNS experiment under development at ORNL
is the most ambitious~\cite{Golub:1994}. 
Its ultimate goal is to improve the 
sensitivity by more than two orders of magnitude beyond the present 
90\% CL bound of some $3\times 10^{-26}\, e\hbox{-}{\mathrm{cm}}$~\cite{Baker:2006ts}. 
This limit already constrains, e.g., the CP-violating phases 
in minimal supersymmetric models to assume unnaturally 
small values, or to make the masses of the supersymmetric partner particles
larger than previously anticipated, or to make the spectrum of partner particles 
possess unexpected degeneracies. 
 These experiments are broadly similar in experimental 
strategy to atomic physics approaches. 

Over the last few years 
a qualitatively new approach to the measurement of particle EDMs 
using charged particles in storage rings~\cite{Farley:2003wt}, exploiting the 
large electric fields present in such environments, 
has come under active development. 
Such an experiment would have the advantage of enlarging the 
spectrum of available species to include charged particles, and the 
ability to allow a coherent effect to accumulate over many revolutions around a ring. 
A variety of operators can generate an EDM, so that 
stringent EDM measurements on the proton, neutron, and 
other light nuclei are complementary and can help unravel the underlying
CP-violating mechanism if a signal is seen. The theoretical insights to be
gained have been studied 
carefully~\cite{deVries:2011re,deVries:2011an,Bsaisou:2012rg}.
An experimental difficulty of this approach is that one loses 
the clean electric-field flip used in previous experiments on 
electrically neutral objects to reduce systematic errors. 
Instead one must typically measure a rotation of the plane of 
polarization of a transversely polarized particle in the ring and to 
develop other methods to deal with systematic errors, as discussed in~\cite{Kronfeld:2013uoa}. 
Measurements in existing storage rings to quantify these instrumental issues are in progress. 

Finally one can consider constraints on the EDMs of leptons. 
The muon EDM can be limited in part as a byproduct of the muon $g-2$ 
measurements~\cite{Bennett:2008dy}, 
and the heavier mass of the muon amplifies its sensitivity to 
certain new-physics possibilities. 
Nevertheless, at anticipated levels of sensitivity, such 
experiments constrain CP-violating sources which do not simply scale with the mass of the muon. 
Such flavor-blind CP-violating contributions 
to the muon EDM are already severely constrained by 
electron EDM limits; rather, direct 
limits probe the possibility of lepton-flavor 
violation here as 
well~\cite{Babu:2000dq,Ibrahim:2001jz,Hiller:2010ib}. 
The electron EDM possesses 
stringent limits from atomic and molecular physics measurements, 
and in addition there are many promising approaches under development, 
which could achieve even higher levels of sensitivity. 
These range from 
solid-state systems at low temperature~\cite{rf:Shapiro1968,Eckel:2012aw}
to new experiments with cold molecules~\cite{ACME:2013}. Indeed, the ACME
collaboration, using ThO, has just announced a limit on $d_e$ an order of magnitude
smaller than any ever achieved before~\cite{ACME:2013de}. 
These constraints are important in themselves and 
are also needed to interpret the source of an EDM if observed 
in an atomic physics experiment.  

\subsubsection{EFTs for EDMs: the neutron case}\label{par:secEedmeft}

We now consider how 
sources of CP violation beyond the SM can generate a
permanent EDM at low energies. 
Noting~\cite{Pospelov:2005pr},  
we organize the expected contributions
in terms of the mass dimension of the possible CP-violating operators, 
in quark and gluon degrees of freedom, appearing in an effective field
theory with a cutoff of $\sim 1$~GeV:
\begin{eqnarray}  
{\cal L} &=&
\frac{\alpha_s \bar\theta}{8\pi} 
\epsilon^{\alpha\beta\mu\nu}F_{\alpha\beta}^a F_{\mu\nu}^a \nonumber \\ 
&& - \frac{i}{2} \sum_{i\in u,d,s}\left( d_i \bar \psi_i F_{\mu\nu}\sigma^{\mu\nu} \gamma_5 \psi_i 
+
 {\tilde d}_i \bar \psi_i F_{\mu\nu}^a T^a\sigma^{\mu\nu}\gamma_5 \psi_i \right) \nonumber\\
&& + \frac{w}{3} f^{abc} \epsilon^{\nu\beta\rho\delta} (F^a)_{\mu\nu} (F^b)_{\rho\delta}
(F^c)_\beta^{\mu} \nonumber \\
&& + \sum_{i,j} C_{ij} (\bar \psi_i \psi_i)(\bar \psi_j i\gamma_5\psi_j) 
+ \dots 
\label{edm:eqn:Leff}
\end{eqnarray}
with $i,j\in u,d,s$ unless otherwise noted --- all heavier degrees of freedom 
have been integrated out.
The leading term is the dimension-four strong~CP term already discussed, 
proportional to the parameter $\bar\theta$, though it can also be 
induced by higher-dimension operators 
even in the presence of axion dynamics~\cite{Pospelov:1997uv,Pospelov:2005pr}
so that we retain it explicitly. 
The terms in the second line of Eq.~(\ref{edm:eqn:Leff}) appear to be 
of dimension five, but their chirality-changing nature implies that a 
Higgs insertion, of form $H/v$, say, is needed to make
the operator invariant under SU(2)$_{L}\times$U(1) symmetry. 
(See Eq.~(3.1) in~\cite{Bhattacharya:2012bf} for an explicit expression.)
Therefore 
these operators, which determine 
the fermion EDMs $d_i$ and quark chromo-EDMs (CEDM)
$\tilde d_i$, 
are suppressed by an additional factor containing a large mass scale
and should be regarded as dimension-six operators in numerical effect. 
The remaining terms in Eq.~(\ref{edm:eqn:Leff}) are 
the dimension-six Weinberg three-gluon 
operator from Eq.~(\ref{Weinbergthreegluon}) with coefficient
$w$, and CP-violating four-fermion operators, characterized by coefficients $C_{ij}$. 
Turning to \cite{Grzadkowski:2010es}, we note that 
after electroweak symmetry breaking there are also chirality-changing 
four-fermion operators which, analogously, are of dimension eight numerically once
SU(2)$_{L}\times$U(1) symmetry is imposed. 
Various extensions of the SM can generate 
the low-energy constants which appear, 
so that, in turn, EDM limits thereby constrain the new sources of CP violation which appear
in such models. In connecting the Wilson coefficients of 
these operators and hence models of new physics 
to the low-energy constants of a 
chiral effective theory in meson and nucleon degrees of freedom 
requires the evaluation of nonperturbative 
hadron matrix elements. Parametrically, we have~\cite{Pospelov:2005pr}
\begin{eqnarray}
d_n &=& d_n({\bar \theta},d_i, {\tilde d}_i,w,C_{ij}) \nonumber \\
{\bar g}_{\pi NN}^{(i)} \,,
&=& 
{\bar g}_{\pi NN}^{(i)}({\bar \theta},d_i, {\tilde d}_i,w,C_{ij}). 
\label{edm:eqn:match}
\end{eqnarray} 
Several computational aspects 
must be considered in connecting a model of new physics at the TeV scale to the
low-energy constants of Eq.~(\ref{edm:eqn:Leff}). 
After matching to an effective theory in SM degrees of freedom,
there are QCD evolution and operator-mixing effects, as well as flavor thresholds, 
involved in realizing the Wilson coefficients at a scale of $\sim$1~GeV. 
Beyond this, the hadronic matrix elements must be computed. 
A detailed review of all these issues can be found in \cite{Engel:2013lsa}. 
Typically QCD sum rule methods, or a SU(6) quark model, 
have been employed in the computation of the matrix elements~\cite{Pospelov:2005pr};
for the neutron, 
we note \cite{ARXIV:0806.2618} for a comparative review of different methods.
Lattice gauge theory can also be used to compute the needed proton 
and neutron matrix elements, and the
current status and prospects for lattice-QCD calculations are 
presented in the next section. 
We note in passing that $d_n$ and $d_p$ have also been analyzed 
in chiral perturbation 
theory~\cite{Pich:1991fq,Ottnad:2009jw,Guo:2012vf}, 
as well as in light-cone QCD~\cite{Brodsky:2006ez}. We refer
to Sec.~\ref{sec:subsecB36} for a general discussion of chiral
perturbation theory in the baryon sector. 

We turn now to the evaluation of the requisite hadron matrix
elements with lattice QCD. 

\subsubsection{Lattice-QCD matrix elements}\label{par:secEedmlat}

\begin{figure}[b]
\includegraphics*[width=\linewidth]{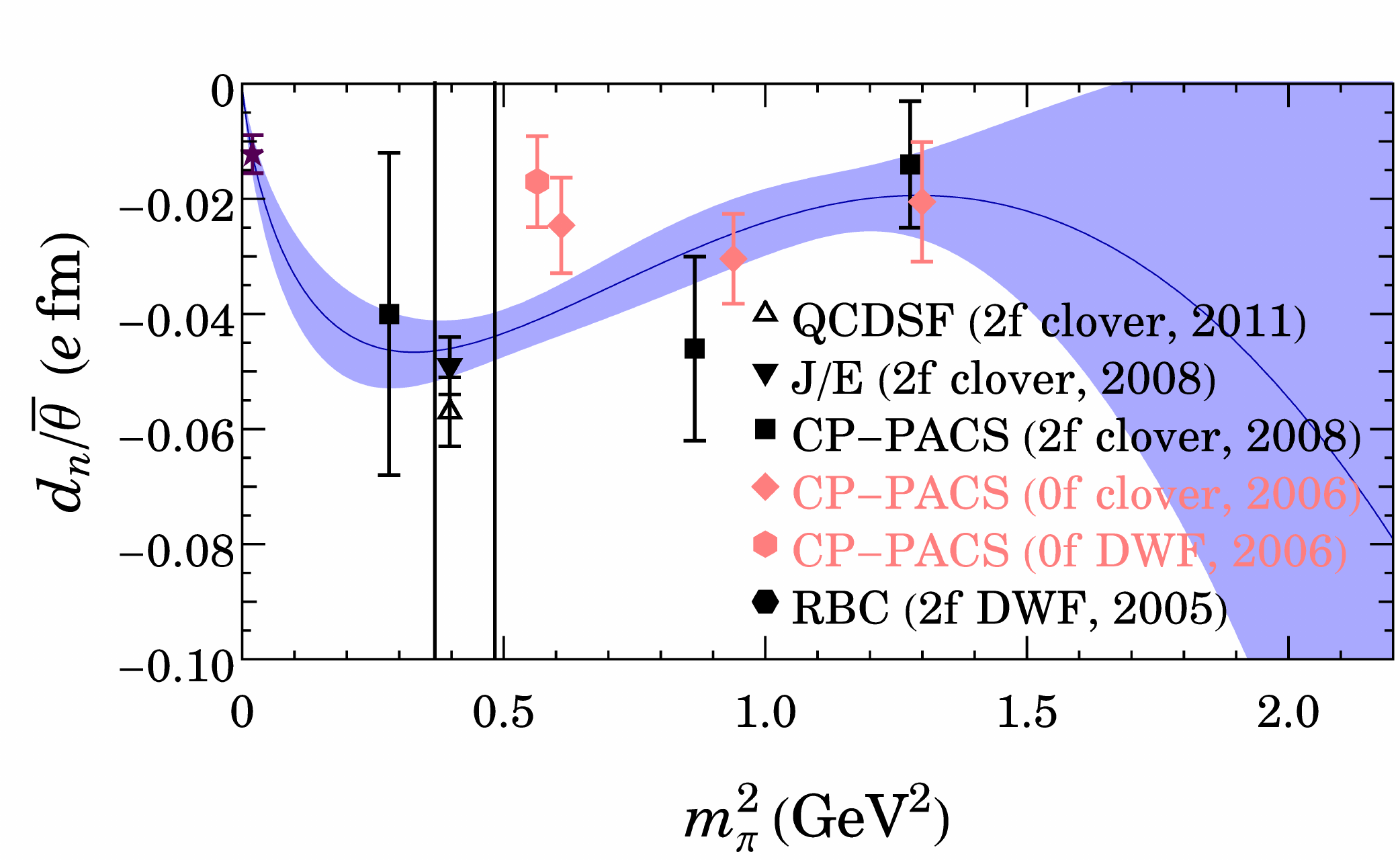}
\caption{Summary of the latest dynamical calculations of 
the neutron EDM~\cite{Berruto:2005hg,Shintani:2005xg,Shintani:2006xr,Aoki:2008gv,Shintani:2008nt,Schierholz:2011}
$d_n$ as a function of $m_\pi^2$ from a nonzero $\bar\theta$ term in QCD.
The band is a global extrapolation at 68\% CL combining all the lattice points
(except for \cite{Schierholz:2011}) each weighted
by its error bar. The leftmost star indicates the value at the physical pion mass.
Figure taken from \cite{Lin:2011cr}.
}
\label{fig:nedmsum}
\end{figure}

To generate a nonzero neutron EDM, one needs interactions that violate CP 
symmetry, and the 
CP-odd $\bar\theta$-term in the SM is one possible example. 
The most common type of lattice-QCD EDM calculation 
is that of the neutron matrix element of the operator associated
with the leading $\bar\theta$ term.
A recent combined analysis gives $O(30\%)$ 
in the statistical error alone, noting Fig.~\ref{fig:nedmsum} 
for a summary, 
so that 
the precision of lattice-QCD calculations needs to be greatly improved. 
All-mode averaging (AMA) has been proposed to improve the current 
statistics even at near-physical pion mass~\cite{Blum:2012uh}. 

There are currently three main approaches to computing these matrix elements 
using lattice QCD. One is a direct computation, 
studying the electromagnetic form factor $F_3$ under the
 QCD Lagrangian including the CP-odd $\theta$ term (as adopted by RBC, J/E, and
CP-PACS (2005)~\cite{Schierholz:2011,Shintani:2008nt,Aoki:2008gv,Shintani:2006xr,Berruto:2005hg})
\begin{eqnarray}
\left.\langle n \mid {\mathcal J}^{\rm EM}_\mu \mid n 
\rangle\right|_{\cancel{\hbox{CP}}}^{\theta}
     &=& \frac{F_3(q^2)}{2 M_n} \bar u q_\nu \sigma^{\mu\nu} \gamma_5 u \nonumber \\
d_n &=& \lim_{q^2\to0} \frac {F_3(q^2)}{2 M_n}\,,\label{eq:F3}
\end{eqnarray} 
where $J^{\rm EM}_\mu$ is the electromagnetic current, 
$\bar u$ and $u$ are appropriate spinors for the neutron, and $q$ is  the transferred momentum. 
This requires an extrapolation of the form factors to $q^2=0$, which can 
introduce systematic error and exacerbate the statistical error. 
Another method is 
introducing an external static and uniform electric field and 
looking at the energy difference 
induced between the two spin states of the nucleon 
at zero momentum (by CP-PACS~\cite{Shintani:2006xr}), one can infer $d_n$.  
Or, finally, one can compute the product of the anomalous magnetic 
moment of neutron $\kappa_N$ and $\tan(2\alpha)$ (by QCDSF~\cite{Schierholz:2011}), 
where $\alpha$ 
is the $\gamma_5$ rotation of the nucleon spin induced by the CP-odd source. 
A summary of $d_n$ 
calculations from dynamical lattice QCD is shown in 
Fig.~\ref{fig:nedmsum}, 
where the results are given as a function of the pion mass used in the calculation. 
Combining all data and extrapolating to the physical pion mass yields 
$d_n^{\rm lat}=(0.015 \pm 0.005) \bar\theta$~
$e\hbox{-}{\mathrm{fm}}$~\cite{Lin:2011cr}, 
which 
is the starred point in the figure.  
Further and more precise calculations from various groups are currently 
in progress, using improved techniques to reduce the statistical error, 
such as the aforementioned AMA~\cite{Blum:2012uh}. 

It should also be possible to compute the nucleon matrix elements of 
higher-dimension operators, 
such as the quark electric dipole moment (qEDM) and 
the chromoelectric dipole moment (CEDM). 
This will require us to extend lattice-QCD 
calculations to such cases~\cite{Bhattacharya:2012bf}, and we now discuss the 
prospects. 

\paragraph{Quark Electric Dipole Moment \label{par:secEedm}}
In this case, the neutron EDM is induced by nonzero quark electric dipole moments, 
which are related to the following matrix elements of the hadronic part of the
first of the effectively 
dimension-six operators in 
Eq.~(\ref{edm:eqn:Leff}): 
\begin{multline}
\left.\langle n \mid {\mathcal J} ^{\rm EM}_\mu \mid n \rangle\right|_{\cancel{\hbox{CP}}}^{\rm qEDM} =
\\
 q^\nu \left( \frac{d_u+d_d}2 \right) \left\langle n \left| \bar \psi \sigma_{\mu\nu} \psi \right| n  \right\rangle \\ 
 + q^\nu \left( \frac{d_u-d_d}2\right) \left\langle n
  \left| \bar \psi \sigma_{\mu\nu} \tau_3 \psi
      \right| n \right\rangle\,.\label{eq:qEDM}
\end{multline}
The nucleon matrix elements can be accessed through direct lattice-QCD 
calculations with isoscalar and isovector tensor charges. 
There are several existing lattice-QCD 
calculations of the isovector tensor charge; see Fig.~\ref{fig:gTS-eSeT}. 

\paragraph{Chromoelectric Dipole Moment \label{par:secEcedm}}
In this case a direct calculation of the chromoelectric dipole moments would be more 
challenging on the lattice, since it requires the calculation of a four-point 
Green function. 
Only a few such calculations have previously been attempted. 
One way to avoid this problem would be to apply the Feynman-Hellmann 
theorem by introducing an external electric field $E$ to extract the matrix elements: 
\begin{multline}
\!\!\!\!\!
\left.\langle n \mid {\mathcal J}^{\rm EM}_\mu \mid n \rangle\right|_{\cancel{\hbox{CP}}} 
=  \\
\!\!\!\!\!\!\!\!\!\! \left\langle n
    \left| J^{\rm EM}_\mu \left( {\tilde d}_u \bar u \sigma_{\nu\kappa} u + {\tilde d}_d \bar
       d\sigma_{\nu\kappa} d\right) G^{\nu\kappa} \right| n \right\rangle =  \\
\!\!\!\!\!\!\!\!\!\!   \left.\frac\partial{\partial A_\mu(E)}\right|_{E=0}\! \! \! \!
         \left.\left\langle n \left| \left( {\tilde d}_u \bar u \sigma_{\nu\kappa} u + {\tilde d}_d \bar
          d\sigma_{\nu\kappa} d\right) G^{\nu\kappa} \right| n \right\rangle\right|_E\, ,\\
\end{multline}
where $A_\mu(E)$ refers to the corresponding vector potential and 
$G^{\nu \kappa}$ is shorthand for $(F^a)^{\nu\kappa} T^a$. 
Similar techniques have been widely implemented in lattice QCD 
to determine the strangeness contribution to the nucleon mass; one only needs to combine the idea with 
a nucleon matrix element calculation. Although as of the time of this review, no lattice calculation of the 
chromoelectric dipole moment has been attempted, we are optimistic that it will be explored within the next 
few years.

Currently, lattice-QCD calculations on $d_n$ due to the leading $\theta$ term have 
statistical errors at the level of 30\% after a chiral extrapolation combining 
all existing dynamical data. More updates and precise calculations from various 
groups are currently in progress, including improved numerical techniques 
that will significantly reduce the errors. Within the next 5 years, 
lattice QCD should be able to make predictions of better than 10\% precision, 
and one can hope that percent-level computations will be available on a ten-year timescale.

Outside the leading-order $\theta$ term, there are plans for calculating the 
dimension-six operator matrix elements by the 
\href{http://www.phys.washington.edu/users/hwlin/pndme/index.xhtml}{PNDME}
collaboration. 
The matrix elements relevant to the quark electric dipole moments are rather 
straightforward, involving isovector and isoscalar nucleon tensor matrix elements. 
The latter one requires disconnected diagrams with extra explicit quark loops. They 
will require techniques 
similar to those already used to determine the strangeness contribution 
to the nucleon mass and the 
strange spin contribution to nucleon. However, the chromoelectric dipole 
moment is more difficult still, since it requires a four-point Green function. 
One alternative method we have considered 
would be to take a numerical derivative with the magnitude of 
the external electric field~\cite{Bhattacharya:2012bf}. 
We should see some preliminary results soon.

\subsection{Probing non-$(V-A)$ interactions in beta decay}
\label{sec:e:ProNonInt}
The measurement of non-SM 
contributions to precision neutron (nuclear) beta-decay measurements 
would hint to the existence of 
BSM particles at the TeV scale; if new particles exist, 
their fundamental high-scale interactions would appear at low energy in the 
neutron beta-decay Hamiltonian as new terms, where we recall Eq.~(\ref{eq:frame})
and the opening discussion of Sec.~\ref{sec:secE3}. 
In this case the new terms are most readily revealed by their symmetry; they can 
violate the so-called $V-A$ law of the weak interactions. 
Specifically, in dimension six, the effective Hamiltonian takes the form 
\begin{eqnarray}
H_{\rm eff} &=& G_F \Bigg( J_{V-A}^{\rm lept} \times J_{V-A}^{\rm quark} \nonumber \\
&+& \sum_i \epsilon_i \hat{O}_i^{\rm lept} \times \hat{O}_i^{\rm quark} \Bigg),
\label{eq:nonVmA}
\end{eqnarray}
where $G_F$ is the Fermi constant, $J_{V-A}$ is the 
left-handed current 
 of the indicated particle, 
and the sum includes operators of non-$(V-A)$ form which represent 
physics BSM.
As we have noted, 
the new operators will enter with coefficients 
controlled by the mass scale of new physics; this is similar 
to how the dimensionful Fermi constant 
gave hints to the masses of the $W$ and $Z$ bosons 
of the electroweak theory prior to their discovery. 
Matching this to an effective theory at the nucleon 
level, the ten terms of the effective Hamiltonian are independent, 
linear combinations of the
coefficients of the Lee-Yang Hamiltonian~\cite{Lee:1956qn,Cirigliano:2012ab}. 
Since scalar and tensor structures (controlled by 
$\epsilon_S$ and $\epsilon_T$ in $\beta$ decay)
do not appear in the SM Lagrangian, signals in these channels at current levels of
sensitivity 
would be clear signs of BSM physics. 
In neutron decay, 
the new operators of Eq.~(\ref{eq:nonVmA}) yield, in particular, the following low-energy coupling constants $g_T$ and $g_S$ 
(here, multiplied by proton and neutron spin wavefunctions):
\begin{eqnarray}
g_T \bar{u}_n \sigma_{\mu\nu} u_p &=& \langle n | \overline{u}\sigma_{\mu\nu} d | p \rangle \\
g_S \bar{u}_n u_p &=& \langle n | \overline{u} d | p \rangle \,.
\end{eqnarray}
Lattice QCD is a perfect theoretical tool to determine these constants precisely.

The search for BSM physics proceeds experimentally by either measuring the Fierz 
interference term $b$ (i.e., $b m_e/E_e$) 
or the neutrino asymmetry parameter $B$ 
(i.e., $B(E_e){\bf S}_n \cdot {\bf p}_\nu$)
of the neutron differential decay rate~\cite{Cirigliano:2013xha}. 
The Fierz term can either be measured 
directly or indirectly, the latter 
through 
either its impact on 
the electron-neutrino correlation $a {\bf p}_e\cdot {\bf p}_\nu$ 
or on 
the electron-momentum correlation with neutron-spin, $A{\bf S}_n\cdot {\bf p}_e$. 
Here, ${\bf S}_n$ and ${\bf p}_{\ell}$ denote the neutron spin and 
a lepton momentum (${\ell} \in (e,\nu)$), respectively. We note, neglecting
Coulomb corrections, $b=(2/(1+3\lambda^2))(g_S {\rm Re(\epsilon_S)} - 12
\lambda g_T {\rm Re}(\epsilon_T))$~\cite{Jackson1957zz}, 
where $\lambda=g_A/g_V\approx 1.27$. 
Assessing $b$ through $a$ or $A$
employs an asymmetry measurement, reducing the impact of possible 
systematic errors. The Fierz term is nonzero only if scalar or tensor currents appear, 
whereas the latter contribute to the magnitude of~$B$.

There are several upcoming and planned experiments worldwide to measure the 
correlation coefficients in neutron 
decay, 
with plans to probe 
$b$ and $B_{\rm BSM}$ 
up to the $\sim 10^{-3}$ level or better, 
and they include PERC~\cite{Dubbers:2007st} at the FRM-II, PERKEOIII 
at the 
ILL~\cite{PERKEOIII:2009}, UCNB~\cite{Plaster:2008si} and UCNb~\cite{UCNb} at LANL, 
Nab at ORNL~\cite{Pocanic:2008pu,abBA}, and ACORN~\cite{Wietfeldt:2005wz} at NIST --- 
indeed the PERC 
experiment~\cite{Dubbers:2007st}
has proposed attaining $10^{-4}$ precision. 
Models of QCD give rather loose bounds on $g_S$ and $g_T$; 
for example, $g_S$ is estimated to range between 0.25 and 1~\cite{Bhattacharya:2011qm}. 
Consequently, 
lattice-QCD calculations of these quantities need not be terribly precise 
to have a dramatic positive impact. 
Indeed, determining $g_{S,T}$ 
to 10--20\% (after summing all systematic 
uncertainties)~\cite{Bhattacharya:2011qm} is 
a useful and feasible goal. 
The obvious improvement in the ability to 
limit the coefficients of the underlying non-$(V-A)$ interactions 
speaks to its importance. 

\begin{figure}
\begin{center}
\includegraphics*[width=.48\textwidth]{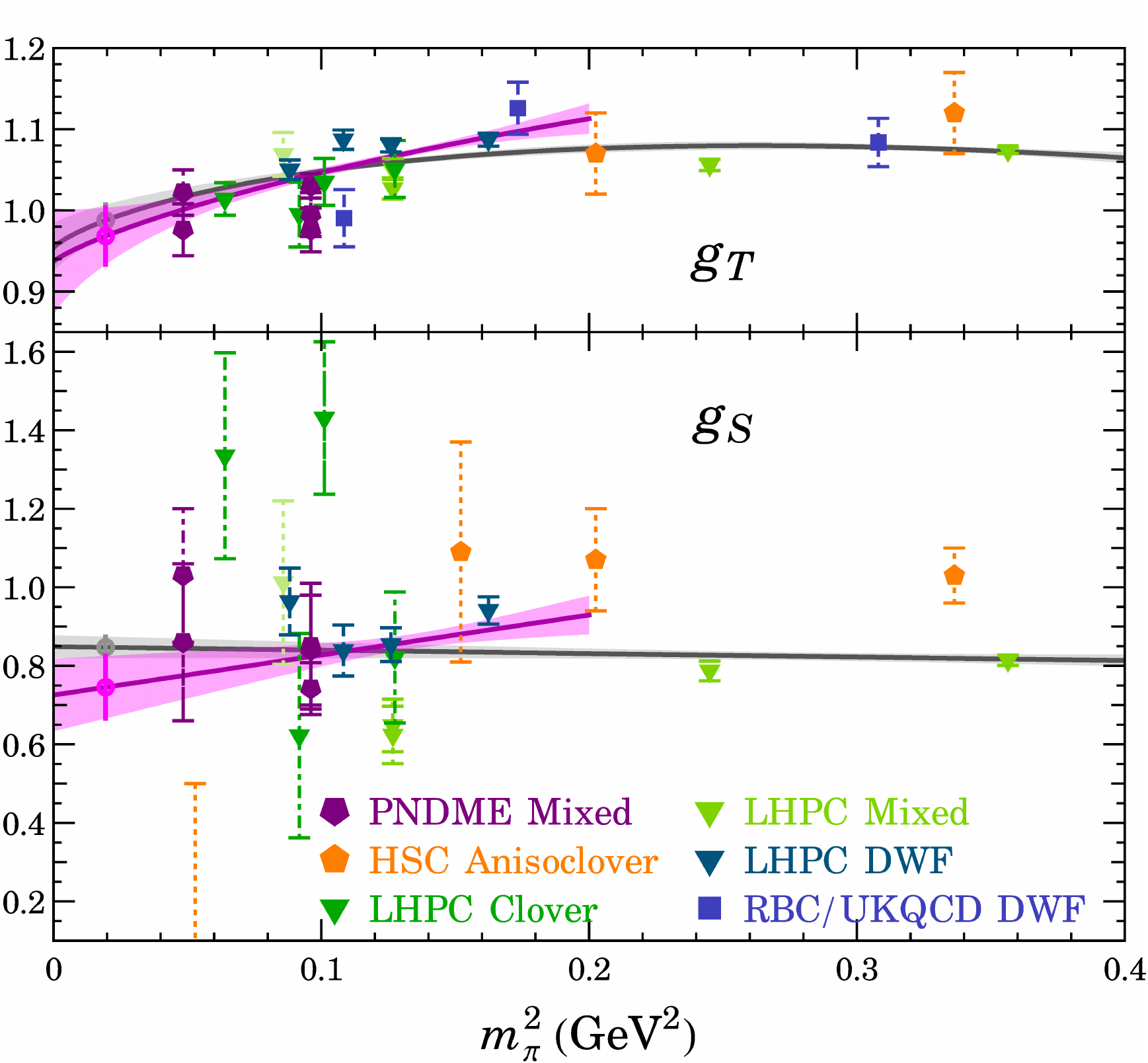}
\includegraphics*[width=.48\textwidth]{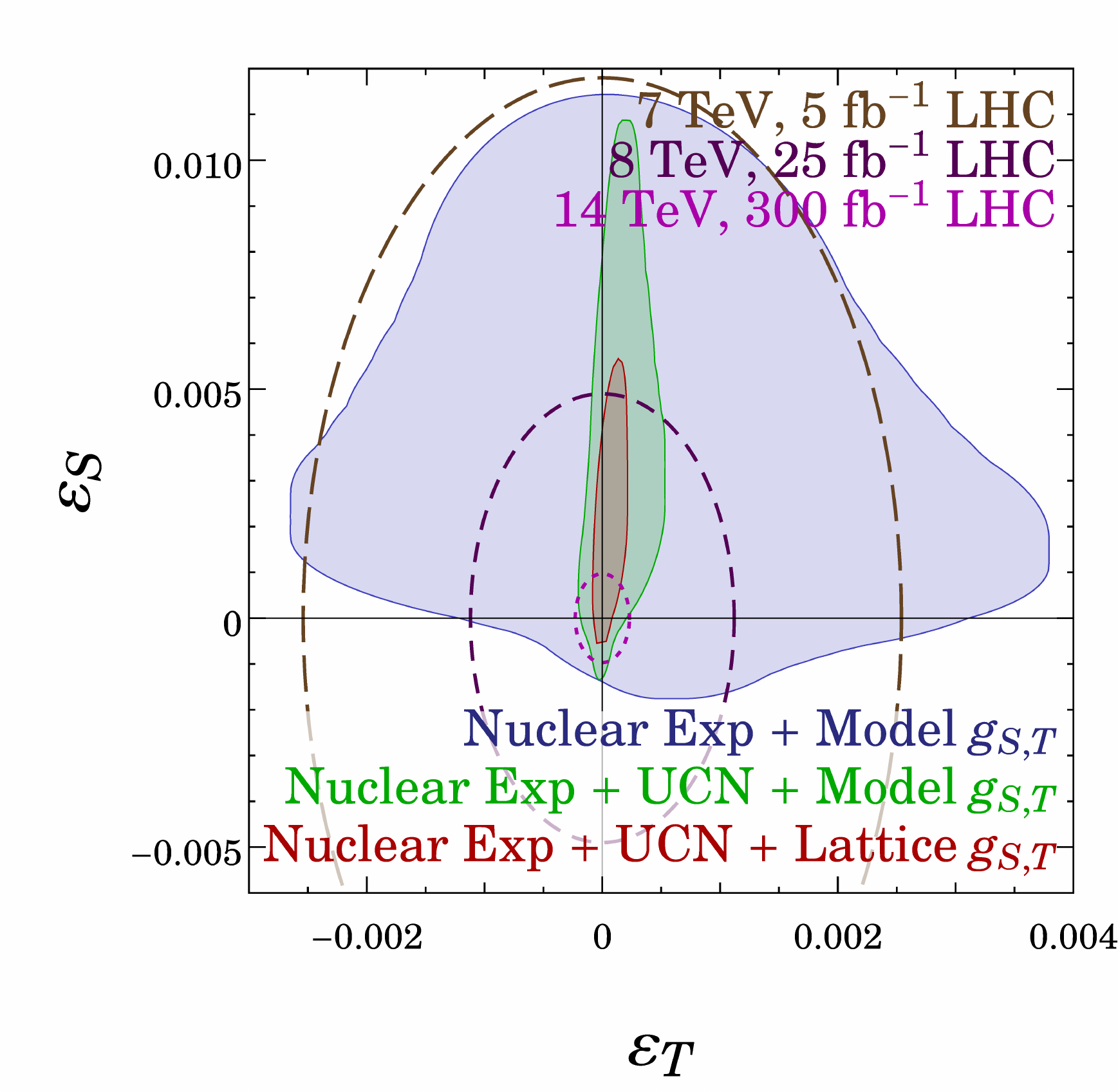}
\end{center}
\vspace{-0.5cm}
\caption{
Figures adapted from \cite{Lin:2012ev}. 
(Upper) Global analysis of all $N_f=2+1(+1)$ lattice calculations of $g_T$ (upper) and 
$g_S$ 
(lower)~\cite{Yamazaki:2008py,Aoki:2010xg,Bratt:2010jn,Bhattacharya:2011qm,Green:2012ud}
with $m_\pi L >4$ and  $m_\pi T > 7$ cuts to avoid 
systematics due to small spatial or temporal extent. 
The leftmost points are the extrapolated values at the physical pion mass. 
The two bands show extrapolations with different upper 
pion-mass cuts: $m_\pi^2 <0.4$ and  $m_\pi^2 <0.2$. The $m_\pi L<4$ data points 
are marked faded within each calculation; the lattice spacings for each point 
are denoted by a solid line for $a\leq 0.06$~fm, dashed: $0.06 < a \leq 0.09$~fm, 
dot-dashed: $0.09 < a \leq 0.12$~fm, and dotted: $a > 0.12$~fm.
 (Lower) The allowed $\epsilon_{S}$-$\epsilon_{T}$ parameter region using 
different experimental and theoretical inputs. The outermost (green),  
middle (purple), and innermost (magenta) dashed lines 
are the constraints from the first LHC run~\cite{CMS:2012fba}, 
along with near-term expectations, 
running to a scale of 2~GeV to 
compare with the low-energy experiments. The inputs for the low-energy experiments 
assume that limits (at 68\% CL) of $|b|< 10^{-3}$ and $|B_{\rm BSM} -b| < 10^{-3}$ 
from neutron
$\beta$ decay and a limit of $g_T \epsilon_T < 2\times 10^{-4}$ 
from $^6$He $\beta$ decay~\cite{Knecht:2012jb}, which is
a purely Gamow-Teller transition. These low-energy experiments probe $S,T$ 
interactions through possible interference terms and yield constraints on 
${\rm Re}(\epsilon_T)$ and ${\rm Re}(\epsilon_S)$ only. 
 }\label{fig:gTS-eSeT}
\end{figure}

The PNDME collaboration 
reported the first lattice-QCD results for 
both $g_S$ and $g_T$~\cite{Bhattacharya:2011qm} and gave the first estimate 
of the allowed region of $\epsilon_{S}$-$\epsilon_{T}$ 
parameter-space when combined with the 
expected experimental precision; we will return to this point in a moment. 
The latest review, from~\cite{Lin:2012ev}, contains a summary of these charges; 
to avoid the unknown systematics coming from finite-size artifacts, 
data with $M_\pi  L \leq 4$ and $M_\pi  T \leq 8$ are omitted, 
as shown on the lower part of Fig.~\ref{fig:gTS-eSeT}.
This figure includes updated calculations of $g_{S,T}$ 
after~\cite{Bhattacharya:2011qm} from the PNDME and LHP 
collaborations. Reference~\cite{Lin:2012ev} uses the chiral formulation 
given in~\cite{Detmold:2002nf} and~\cite{Green:2012ej} for the tensor and 
scalar charges, respectively, to extrapolate to the physical pion mass.
We see that the PNDME points greatly constrain the uncertainty 
due to chiral extrapolation in both cases and obtain 
$g_T^{\rm lat}= 0.978 \pm 0.035$ and $g_S^{\rm lat}= 0.796 \pm 0.079$, 
where only statistical errors have been reported. 

As previously mentioned, the tensor and scalar charges can be combined 
with experimental data to determine the allowed region of parameter space 
for scalar and tensor BSM couplings. 
Using the $g_{S,T}$ from the model estimations and combining with the existing nuclear 
experimental data,\footnote{For example, nuclear beta 
decay $0^+ \rightarrow 0^+$ transitions 
and other processes, such as the $\beta$ asymmetry in 
oriented $^{60}\mbox{Co}$ decay, the longitudinal 
polarization ratio between the 
Fermi and Gamow-Teller transitions in $^{114}\mbox{In}$ decay, the positron 
polarization in polarized $^{107}\mbox{In}$ decay, 
and the beta-neutrino correlation parameters in 
nuclear transitions.}
we get the constraints shown as the outermost band of the lower part 
of Fig.~\ref{fig:gTS-eSeT}.
Combining anticipated (in the shorter term) results from $\beta$-decay
and existing measurements, and again, using the model inputs of $g_{S,T}$, we see the 
uncertainties in $\epsilon_{S,T}$ are significantly improved. 
(A limit on $g_T \epsilon_T$ also comes from radiative pion decay, but it can be evaded
by cancellation and has been omitted~\cite{Bhattacharya:2011ah,Cirigliano:2013xha}.) 
Finally, using our present lattice-QCD values of the scalar and tensor charges, 
combined with the anticipated precision of the experimental bounds on the deviation of 
low-energy decay parameters from their SM values, we find the constraints 
on $\epsilon_{S,T}$ are further 
improved, shown as the innermost region. These upper bounds on the effective 
couplings $\epsilon_{S,T}$ would 
correspond to lower bounds for the scales $\Lambda_{S,T}$ 
at 5.6 and 10~TeV, respectively, 
determined using naive dimensional analysis ($\epsilon_i \sim (v/\Lambda_i)^2$ 
with $v \sim 174~{\rm GeV}$), 
for new physics in these channels. 

\begin{figure}[b]
\includegraphics*[width=0.7\linewidth]{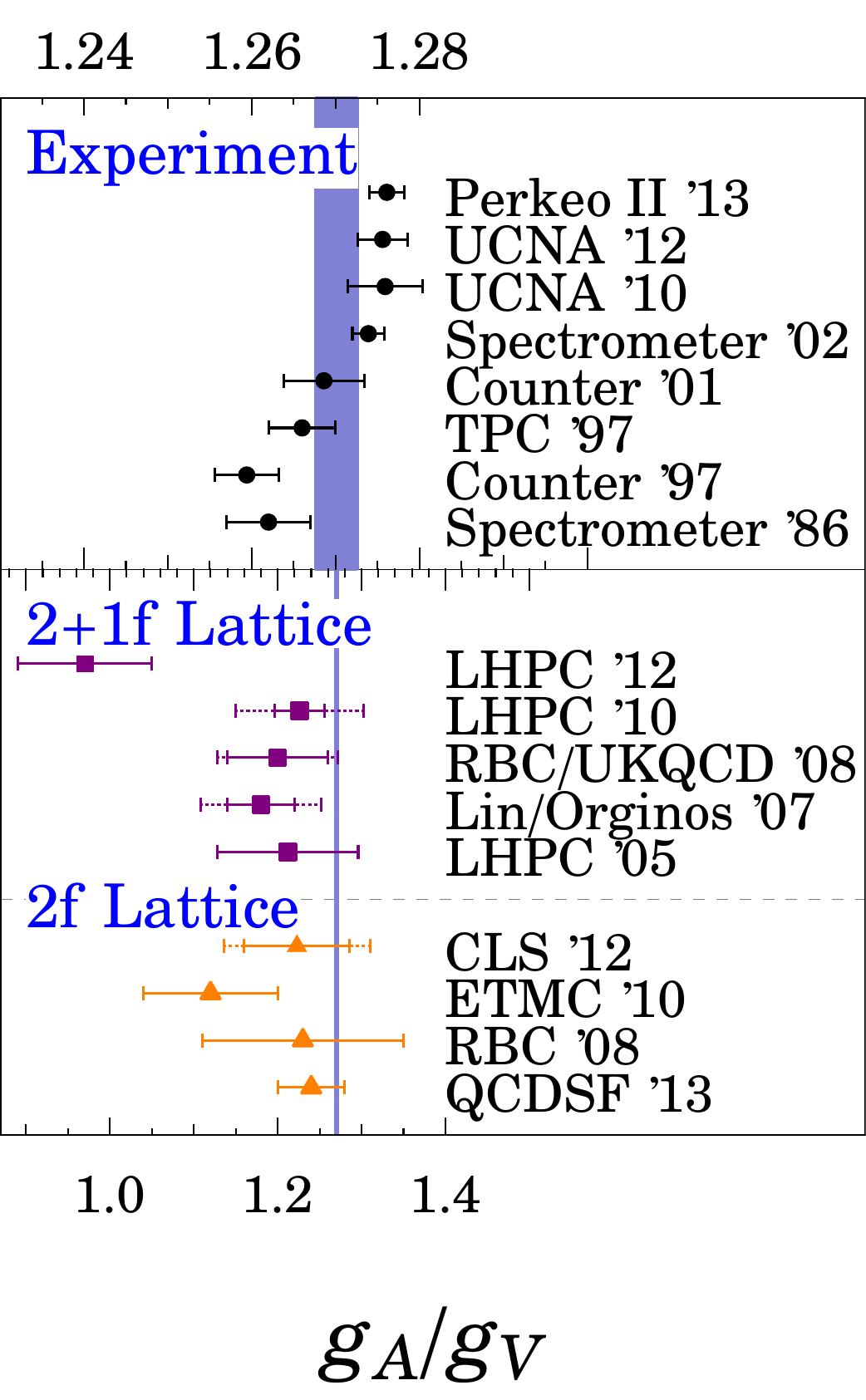}
\caption{Compilation of $g_A$ determined from experiment (top) and lattice QCD (bottom) 
adapted from Ref.~\cite{Bhattacharya:2012bf}. 
The lower panel shows $g_A$ values after extrapolating  to the physical pion mass collected from dynamical 2+1-flavor and 2-flavor lattice
calculations using $O(a)$-improved fermions~\cite{Khan:2006de,Alexandrou:2010hf,Pleiter:2011gw,Capitani:2012ef,Capitani:2012gj,Brandt:2011sj,Horsley:2013ayv,Lin:2008uz,Yamazaki:2008py,Aoki:2010xg,Edwards:2005ym,Edwards:2006qx,Gockeler:2011ze,Lin:2011sa,Owen:2012ts,Green:2012ej,Dinter:2011sg,Alexandrou:2013joa}
A small discrepancy persists: while calculations continue to tend towards 
values around 1.22 with a sizeable error, 
the experimental values are converging towards $1.275 \pm 0.005$. 
A significant lattice effort will be necessary to reduce the
systematics and achieve total error at the percent level.
}
\label{fig:gacomp}
\end{figure}

There is a complication, 
however, that should be noted. The analysis of neutron $\beta$ decay requires 
a value of the neutron axial vector coupling 
$g_A$ as well (similar considerations operate for Gamow-Teller
nuclear transitions); presently, this important quantity 
is taken from experiment because theory cannot 
determine it well enough, as illustrated in Fig.~\ref{fig:gacomp}. 
This topic is also addressed 
in Sec.~\ref{sec:lq.struct.form-factors.lqcd}; here we revisit possible
resolutions. 
A crucial direction for lattice QCD is to reexamine the systematics 
in the nucleon matrix elements, a task that was somewhat neglected in the past 
when we struggled to get enough computing power to 
address merely statistical errors. 
Resources have improved, and many groups have investigated 
the excited-state contamination, and this seems to be under control in recent years. 
However, the results remain inconsistent with experiment, and more extensive studies 
of finite-volume corrections with high statistics will be carried out 
in the future. In addition, the uncertainty associated with extrapolating to a 
physical pion mass should be greatly improved within the next year or two. 
Overall, we believe 
$g_A$ will be calculated to the percent-level or 
better (systematically and statistically) in the next few years. 
It is worth noting that a blind analysis 
should be easy to carry out for $g_A$ since it is an overall constant 
in the lattice three-point correlators. Nevertheless, it is currently 
the case that poorly understood systematics can affect the lattice-QCD 
computations 
of the nucleon matrix elements, and those of $g_A$ serve as an explicit
example. However, those uncertainties are not so large that
they undermine the usefulness of the $g_S$ and $g_T$ results. 
As we have shown, 
the lattice computations of these quantities need not be very precise to be useful. 

There are also other $\beta$-decay nucleon matrix elements induced 
by strong-interaction effects which 
enter as recoil corrections at ${\cal O}(E/M)$, where $E$ is the electron
energy scale and $M$ is the nucleon mass. The weak magnetic coupling $f_2$ 
can be determined using the conserved-vector-current (CVC) hypothesis 
(though a fact in the SM) 
and the isovector nucleon magnetic moment, though this prediction, 
as well as that of the other matrix elements
to this order, is modified by isospin-breaking effects. 
This makes it useful to include errors in the 
assessment of such matrix elements, 
when optimizing the parameters to be determined from experiment. Such a scheme 
has been developed, after that
in~\cite{Charles:2004jd}, in~\cite{Gardner:2013aya}, and the 
impact of such theory errors on the ability
to resolve non-$(V-A)$ interactions has been studied, 
suggesting that it is important to study the induced tensor term $g_2$ and other
recoil-order matrix elements using lattice QCD as well. The study of~\cite{Gardner:2013aya}
shows that it is also crucial to measure the neutron lifetime extremely well, 
ideally to ${\cal O}(0.1\,{\rm sec})$ precision, in order to falsify the 
$V-A$ law and establish the existence of physics BSM in these processes. 
We refer the reader to Sec.~\ref{sec:secEtaun} for a perspective on the
neutron lifetime and its measurement. 

Second-class currents 
gain in importance in neutron decay precisely because it is a
 mixed transition --- and because BSM effects are already known to be so small. 
 No direct lattice-QCD study of these isospin-breaking couplings has yet been done, 
but a few previous works have tried to estimate their size 
in hyperon decay~\cite{Guadagnoli:2006gj,Sasaki:2008ha,Lin:2008rb}. 
Perhaps particularly interesting is the analysis of the 
process $\Xi^0 \to \Sigma^+ {\ell}\bar \nu$, 
 in which the second-class current terms emerge
as SU(3)$_f$ breaking effects. 
In this case,~\cite{Sasaki:2008ha} 
 $|f_3(0)/f_1(0)| = 0.14 \pm 0.09$
 and
 $|g_2(0)/f_1(0)| = 0.68 \pm 0.18$; this exploratory calculation is made
in the quenched approximation with a relatively heavy pion mass of 539--656~MeV. 
Nevertheless, this decay is a strict 
analog of the neutron decay process, with the $d$ valence quark replaced
by $s$, so that one can {\it estimate}
the size of $g_2/g_A$ in neutron decay by scaling the earlier results by a factor of 
$m_d/m_s \sim 1/20$~\cite{Gardner:2013aya}. 
Ultimately, one can foresee results with reduced
uncertainties from direct calculations 
on physical pion mass ensembles, using the variation of the up and down 
quark masses to resolve the second-class contributions in neutron decay. 

High-energy colliders can constrain $\epsilon_S$ and $\epsilon_T$ in 
the manner shown in Fig.~\ref{fig:gTS-eSeT}. Unfortunately, as shown 
in~\cite{Bhattacharya:2011qm,Bhattacharya:2013ehc}, 
the CDF and D0 results do not provide
useful constraints in this context.
The limits shown follow from estimating 
the $\epsilon_{S,T}$ 
constraints from LHC current bounds and near-term expectations through an effective Lagrangian 
\begin{equation}
{\cal L} = -\frac{\eta_S}{\Lambda_S^2}V_{ud}(\overline{u}d)(\overline{e}P_L\nu_e)
    -\frac{\eta_T}{\Lambda_T^2}V_{ud}(\overline{u}\sigma^{\mu\nu}d)
     (\overline{e}\sigma_{\mu\nu}P_L\nu_e)\,,
\end{equation} 
where $\eta_{S,T}=\pm 1$ to account for the possible sign of the couplings at 
low-energy. The high-energy bounds are scaled down to 
a scale of 2~GeV to compare with low-energy predictions. 
By looking at events with high transverse mass from the LHC 
in the $e\nu+X$ 
channel and comparing with the SM $W$ background, 
the authors of~\cite{Bhattacharya:2013ehc}
estimated 90\%-C.L. constraints on $\epsilon_{S,T}$ 
based on existing data~\cite{CMS:2012fba}, 
$\sqrt{s}=7$~TeV $L=10\mbox{ fb}^{-1}$ (the outermost (green) line) and
the anticipated (null result) data sets at  $\sqrt{s}=8$~TeV $L=25\mbox{ fb}^{-1}$ 
(the middle (purple) line) and $\sqrt{s}=14$~TeV $L=300\mbox{ fb}^{-1}$ 
(the innermost (magenta) line) 
of the lower panel in Fig.~\ref{fig:gTS-eSeT}. 
The low-energy experiments
can  potentially yield much sharper constraints. 

There is plenty of room for further improvements of lattice-QCD calculations 
of $g_{S,T}$. 
Currently, 
there are fewer lattice calculations of $g_T$ and $g_S$, and the 
errors are 
10\% and 30\%, respectively. Ongoing calculations are improving control over the 
systematics due to chiral extrapolation and finite-volume effects. 
In addition, we expect more collaborations will compute these quantities, 
and near-future work will substantially reduce the errors. 
In particular, there is presently no chiral perturbation theory formula for the
extrapolation to a physical pion mass, 
and operator matching is done either at tree- or one-loop level. 
Work is underway to reduce these errors, and we expect 
results with 5\% errors (including all systematics) on a five-year timescale.

\subsubsection{The role of the neutron lifetime}\label{sec:secEtaun}

The neutron lifetime value provides important input to test weak-interaction theory in the charged-current
sector~\cite{Dubbers:2011ns}.
It is also important for Big-Bang nucleosynthesis (BBN), which is becoming more and more important for
constraining many BSM physics scenarios which produce new contributions to the relativistic particle energy
density~\cite{Gardner:2013ama}.
BBN predicts the primordial abundances of the light elements (H, He, D, Li) in terms of the baryon-to-photon
ratio $\eta$, together with nuclear physics input that includes eleven key nuclear reaction cross-sections,
along with the neutron lifetime~\cite{Burles:2000zk}.
As primordial neutrons are protected against $\beta$-decay by fusing with protons into deuterons and then
into $^4$He, a shorter neutron lifetime results in a smaller $^4$He abundance ($Y_{p}$).
The dependence of the helium abundance on changes in the neutron lifetime, the ``effective'' number
of light neutrinos $N_{\rm eff}$, and the baryon-to-photon ratio are: $\delta Y_{p}/Y_{p}=+0.72 \delta
\tau_{n}/\tau_{n}$, $\delta Y_{p}/Y_{p}=+0.17 \delta N_{\rm eff}/N_{\rm eff}$, and $\delta Y_{p}/Y_{p}=+0.039
\delta \eta/ \eta$ ~\cite{Mathews:2004kc, Iocco:2008va}.
With the precise determination of $\eta$ from WMAP~\cite{Hinshaw:2012aka} and now PLANCK~\cite{Ade:2013qta},
the 0.2--0.3\% error on the BBN prediction for Y$_p$ is now dominated by the uncertainty in the neutron
lifetime.
At the same time astrophysical measurements of the helium abundance ($Y_{p}=0.252 \pm
0.003$~\cite{Cyburt:2008kw, Iocco:2008va}) from direct observations of the H and He emission lines from
low-metallicity regions are poised for significant improvement.
Astronomers are now in a position to re-observe many of the lowest abundance objects used for nebular
$^{4}$He abundance determinations over the next 3-5 years and will continue to find additional ultralow
abundance objects~\cite{Skillman:2013oaa}.
Sharpening this test of BBN will constrain many aspects of nonstandard physics scenarios.

Measurements of the neutron lifetime had been thought to be approaching 
the 0.1\% level of precision (corresponding to a $\sim$ 1~s uncertainty) 
by 2005, with the Particle Data Group~\cite{Yao:2006px}
reporting $885.70\pm 0.85$~s. However, several neutron lifetime results 
since 2005 using ultracold neutron measurements in 
traps~\cite{Serebrov:2004zf, Serebrov:2007ve, Ezhov:2009, Pichlmaier:2010zz} 
reported significantly different results from the earlier PDG average: the latest 
PDG value ($880.1 \pm 1.1$~s)~\cite{Beringer:2012zz} includes all these measurements, 
with the uncertainty scaled up by a factor of $2.7$. The cause of this many-sigma 
shift has not yet been resolved. The large discrepancies between the latest 
lifetime measurements using ultracold neutrons in material bottles make it 
clear that systematic errors in at least some previous measurements have been 
seriously underestimated, and precision measurements using alternative techniques 
are badly needed~\cite{RevModPhys.83.1173}. The latest update~\cite{Yue:2013qrc} 
from a Penning trap neutron lifetime experiment in a cold neutron 
beam~\cite{Dewey:2003hc, Nico:2004ie} gives 
$\tau_{n}= 887.7 \pm 1.2\,({\rm stat.})\,\pm 1.9\,({\rm sys})\,{\rm s}$. 
In addition to continued measurements using the Penning trap technique, 
neutron lifetime measurements with ultracold neutrons now concentrate on 
trapping the neutrons using magnetic field gradients in an attempt to avoid 
what people suspect to be uncontrolled systematic errors from surface effects 
in material traps. A recent experiment at Los Alamos using a magneto-gravitational 
trap that employs an asymmetric Halbach permanent magnet array~\cite{Walstrom:2009} 
has observed encouraging results~\cite{Salvat:2013gpa}.

\subsection{Broader applications of QCD} \label{sec:secE6}

Nucleon matrix elements, and lattice-QCD methods,  are key to a broad sweep of 
low-energy observables which probe how precisely we understand the nature of things. 
We now consider a range of examples, to illustrate the breadth of the possibilities. 

\subsubsection{Determination of the proton radius}

The charge radius of the proton $r_p$ 
has not yet been precisely calculated in lattice QCD because the computation of
disconnected diagrams with explicit quark loops is required. (In the case of
the isovector charge radius ($r_p - r_n$) the disconnected diagrams cancel, so 
that this quantity could be more precisely calculated than $r_p$.) 
Rather, 
it is currently determined from the theoretical analysis of experimental results. 
There has been great interest 
in $r_p$ because the determination of this
quantity from the 
study of the Lamb shift 
in muonic hydrogen~\cite{Pohl:2010zza,Antognini:1900ns}, 
yielding~\cite{Antognini:1900ns} 
\begin{equation}
r_p^{(\mu H)} = 0.84087 \pm 0.00039 \,{\rm fm} \,,
\end{equation}
is some $7\sigma$ 
different from the value
in the CODATA-2010 compilation~\cite{Mohr:2012tt}, determined from
measurements of hydrogen spectroscopy ($r_p^{(e H)}$)
and electron-proton ($r_p^{(e p)}$)
scattering. 
The incompatibility of the various extractions 
offers a 
challenge to both theory and 
experiment. 

We note that $r_p^{(ep)}$ is by no means a 
directly determined quantity, because 
two-photon exchange effects do play a numerical role as well. 
Such corrections also appear in the context of the muonic-hydrogen analysis, 
though the effects 
turn out to be too small to explain the discrepancies. 
For example, 
a dispersive re-evaluation~\cite{Gorchtein:2013yga} of such 
hadronic effects 
based on experimental input (photo- and electroproduction of 
resonances off the nucleon and high-energy pomeron-dominated cross-section) 
yields a contribution of $40\pm 5$~$\mu$eV  to the muonic hydrogen Lamb shift.
Even if the error were underestimated 
for some unknown reason, its order of magnitude
is insufficient to resolve the 300~$\mu$eV discrepancy between 
direct measurement 
of the muonic Lamb shift~\cite{Pohl:2010zza,Antognini:1900ns} 
and its expectation 
determined from QED theory and conventional spectroscopy. 
Such difficulties have prompted much
discussion~\cite{Jentschura:2010ej}, and 
we refer to 
Sec.~\ref{sec:lq.struct.proton-radius} for further details. 
It is still too speculative to state that we are confronting a violation 
of universality in the couplings of the electron and the muon, but 
hope that hadron contributions to the two-photon exchange between 
the muon and the proton would resolve the issue seems misplaced. 
Nevertheless, a viable BSM model does exist which would permit
the discrepancy to stand~\cite{Batell:2011qq}. It 
predicts the existence of new parity-violating
muonic forces which potentially can be probed through experiments using
low-energy muon beams, notably through 
the measurement of a parity-violating asymmetry in elastic scattering
from a nuclear target. Unfortunately, this picture cannot easily 
explain the
existing muon $g-2$ discrepancy~\cite{Batell:2011qq}. 
Disagreement between theory and experiment lurks there also,
but the precision of the discrepancy is 
two orders 
of magnitude smaller than in the muonic Lamb shift case. 
Indeed the muon $g-2$ result constrains new, muon-specific 
forces~\cite{Karshenboim:2014tka}. Planned studies of $\mu p$ 
and $e p$ scattering at PSI
should offer a useful direct test on the universality of 
lepton-proton interactions~\cite{Gilman:2013eiv}.

\subsubsection{Dark-matter searches}

Various threads of astronomical evidence reveal that we live
in a Universe dominated by dark matter and 
dark energy~\cite{Beringer:2012zz}. It 
is commonly thought that dark matter could be comprised of an 
as yet unidentified weakly-interacting massive particle (WIMP). 
Such particles in the local solar neighborhood of our own Milky Way galaxy 
can be constrained or discovered through low-background experiments which 
search for anomalous recoil events involving the 
scattering of WIMPs from nuclei~\cite{Drukier:1983gj,Goodman:1984dc}. 
Supersymmetric models offer
a suitable candidate particle, the neutralino, which can be made 
compatible with all known astrophysical constraints~\cite{Ellis:1983ew,Jungman:1995df}. 
The neutralino is made stable by introducing a discrete symmetry, $R$ parity, 
that forbids its decay. An analogous discrete symmetry can be introduced in 
other, nonsupersymmetric new-physics 
contexts, such as in ``little Higgs'' models~\cite{Freitas:2009jq}, 
to yield an identical effect
--- generally, one can introduce 
a dark-matter parity that renders the dark-matter candidate  
stable. 

WIMP-nuclear interactions
mediated by $Z^0$ exchange were long-ago ruled out~\cite{Falk:1994es,Jungman:1995df}, 
so that the WIMP of supersymmetric models is commonly regarded 
as a neutralino. 
Current experiments probe the possibility 
of mediation by Higgs exchange. 
Consequently, the spin-independent neutralino-nucleon 
cross section is particularly sensitive to the strange scalar 
density, namely, the value of 
$y=2 \langle N | \bar s s | N \rangle/
\langle N | \bar u u + \bar d d | N \rangle$,
noting \cite{Ellis:2008hf} and references therein, 
because the 
Higgs coupling increases with quark mass. The value of this
quantity impacts the mapping of the loci of supersymmetric parameter space
to the exclusion plot of WIMP mass versus the WIMP-nucleon cross section. 
Earlier studies relate $y$ to the $\pi N$ sigma term $\Sigma_{\pi N}$ via 
$y=1- \sigma_0/\Sigma_{\pi N}$ for fixed 
$\sigma_0 \equiv m_l 
\langle N | \bar u u 
+ \bar d d - 2 \bar s s | N \rangle$~\cite{Ellis:2008hf}, where 
$m_l\equiv(m_u+m_d)/2$, suggesting that 
the predicted neutralino-nucleon cross section depends strongly 
on the value of this phenomenological quantity~\cite{Giedt:2009mr}. 
Its impact can be remediated, however, without recourse to assumptions
in regards to SU(3)-flavor breaking; e.g., 
as shown in~\cite{Crivellin:2013ipa}, 
the couplings to the $u$- and $d$-quarks 
can be analyzed directly within the framework of $SU(2)$ 
chiral perturbation theory (ChPT), permitting, in addition,  
control over isospin-breaking effects. 

The matrix elements 
$m_s \langle N | \bar s s | N \rangle$ and 
$\Sigma_{\pi N} \equiv m_l\langle N | \bar u u + \bar d d | N \rangle$
can also be computed directly in lattice-QCD, via different techniques, and the sensitivity 
to $\Sigma_{\pi N}$ is lessened~\cite{Giedt:2009mr}. 
Several lattice-QCD groups have addressed this problem, and new results
continue to emerge~\cite{Junnarkar:2013ac,Young:2013nn,Alexandrou:2013uwa}. 
The spin-independent
WIMP-nucleon cross section can be predicted to much better precision than 
previously thought, though the cross section tends to be smaller 
than that previously assumed~\cite{Giedt:2009mr}, 
diminishing the new physics reach of 
a particular WIMP direct detection experiment. 
Heavier quark flavors can also play a significant role in 
 mediating the gluon coupling to the Higgs, 
and hence to the neutralino, and the 
leading contribution in the heavy-quark limit is 
well-known~\cite{Shifman:1978zn,Jungman:1995df} --- and this treatment 
should describe elastic scattering sufficiently well. Nevertheless, 
the nonperturbative scalar charm matrix element should also be considered, and 
it has also been recently evaluated~\cite{Freeman:2012ry}. 
We note, moreover, in the case of heavy WIMP-nucleon scattering, that 
the renormalization-group evolution from the weak to typical hadronic scales 
also plays a numerically important role~\cite{Hill:2011be}. 

The effects of the nuclear medium 
in mediating effects beyond the impulse approximation (for scalar-mediated interactions)
have also been argued to be important~\cite{Prezeau:2003sv}. This possibility has been
recently investigated on the lattice, and the effects actually appear
rather modest~\cite{Beane:2013kca}. Nevertheless, two-body exchange currents,
which appear in chiral effective theory, can be important in regions of WIMP
parameter space for which the usual WIMP-nucleon interaction is 
suppressed~\cite{Cirigliano:2012pq}. For a study in spin-dependent 
WIMP-nuclear scattering see~\cite{Menendez:2012tm}. 

\subsubsection{Neutrino physics}

The physics of QCD also plays a crucial role in the analysis of neutrino experiments, 
particularly through the axial-vector form factor of the nucleon (and of nuclei). 
The value of the axial coupling of the nucleon $g_A$, which is precisely measured
in neutron $\beta$-decay, is key to the crisp interpretation 
of low-energy neutrino experiments such as SNO~\cite{Ahmad:2002jz}. 
In higher energy experiments, however, 
the $q^2$ dependence of the axial form factor becomes important. 
In particular, elucidating the axial mass $M_A$, which 
reflects the rate at which the form factor changes with $q^2$, is crucial to the 
interpretation of neutrino oscillation 
experiments at ${\cal O}(1~{\rm GeV})$, an energy
scale typical of accelerator-based studies. 
Commonly the value of $M_A$ is assessed experimentally 
by assuming the form factor can be described by a dipole approximation, 
\begin{equation}
G_A^{\rm dipole}(q^2) = \frac{g_A}{\left[1- q^2/M_A^2\right]^2} \,, 
\end{equation}
and the nuclear effects, at least for 
neutrino quasi-elastic scattering, are assessed within
a relativistic Fermi gas model, though final-state interactions of
the produced hadrons in the nucleus can also be included. A consistent
description of the neutrino--nuclear cross sections with beam energy and 
nuclear target is essential for future investigations of the neutrino
mass hierarchy and CP violation in long-baseline experiments
(LBNE, T2K, NO$\nu$A, CNGS). 
Within this framework, tension exists in the empirically determined 
values of $M_A$~\cite{Bhattacharya:2011ah}. Moreover, 
recent studies at 
\href{http://www-boone.fnal.gov}{MiniBoone} 
have illustrated that the framework itself appears to be
wanting~\cite{AguilarArevalo:2007ab,AguilarArevalo:2013hm}. 
Current and future studies at \href{http://minerva.fnal.gov}{MINER$\nu$A} can 
address these
deficiencies by measuring the neutrino (and antineutrino) 
reaction cross sections with various nuclei~\cite{Fiorentini:2013ezn,Fields:2013zhk}. 
Model-independent analyses of experimental data have also been 
developed~\cite{Bhattacharya:2011ah} and have explored ways in which to relax
the usual dipole parametrization of the axial form factor of the 
nucleon, as it is only 
an approximation. 
Nevertheless, a computation of the $q^2$ dependence
of the nucleon axial form factor within QCD is greatly desired. 

The value of 
$M_A$ can be estimated from the nucleon isovector axial form factor 
by a fit of its $q^2$ dependence to a dipole form. Alternatively, 
$r_A$, the axial radius, is calculated by taking the derivative of the form 
factors near $Q^2=0$, and they are linked through $r_A^2={12}/{M_A^2}$ (in the
dipole approximation). The quantity $r_A$ is ultimately of greater interest as
it is not tied to a dipole form. Lattice-QCD calculations of axial form factors,
as well as of vector form factors, tend to yield smaller slopes
and, thus, prefer a larger value of 
$M_A$~\cite{Alexandrou:2007xj,Yamazaki:2007mk,Lin:2008uz,Bratt:2010jn}.
This tendency may stem from 
a heavy pion mass or finite volume
effects.

\subsubsection{Cold nuclear medium effects}

Many precision searches for new physics are undertaken within nuclear environments, be they dark-matter
searches or studies of neutrino properties, and so far there is no universal understanding nor theoretical
control over nuclear corrections.
A common assumption is that the WIMP, or neutrino, interactions in the nucleus are determined by the sum of
the individual interactions with the nucleons in the nucleus, as, e.g.,
in~\cite{Fitzpatrick:2012ix,Anand:2013yka}.
This impulse-approximation picture treats nuclear-structure effects independently of
the particle-physics interaction with a single nucleon.
Nevertheless, single-particle properties can be modified in the nuclear medium, and evidence for such effects
range from low-to-high energy scales.
For example, at the lowest energy scales, the possibility of quenching of the Gamow-Teller strength in nuclei
(with respect to its free-nucleon value) has been discussed for some
time~\cite{Brown:1985zz,Chou:1993zz,Osterfeld:1991ii,Rapaport:1994nr}, though its source is unclear.
It may be an artifact of the limitations of nuclear shell-model calculations\footnote{A recent determination
of the Gamow-Teller matrix element in $^6$He~\cite{Knecht:2011ir} decay is consistent with {\it ab initio}
calculations --- no quenching of the Gamow-Teller strength has been observed.}~\cite{Haxton:1999vg} or a
genuine effect, possibly arising from meson-exchange currents in nuclei~\cite{Menendez:2011qq}.
At larger energies, in deep-inelastic lepton scattering from nuclei, medium effects are long established,
most famously through the so-called EMC effect noted in $F_2$ structure function data~\cite{Aubert:1983xm}.
At ${\cal O}(1\,{\rm GeV})$ energy scales important for accelerator-based, long-baseline neutrino
experiments, medium effects have also been observed in the studies of $3.5~{\rm GeV}$ neutrino-nuclear
interactions in the MINER$\nu$A experiment~\cite{Fiorentini:2013ezn,Fields:2013zhk}.
The inclusion of two-nucleon knock-out in addition to quasielastic scattering appears to be needed to explain
the observed neutrino-nuclear cross sections at these energies~\cite{Formaggio:2013kya}.
This is a challenging energy regime from a QCD viewpoint; the interactions of ${\cal O}(1\,{\rm GeV})$
nucleons are not suitable for treatment in chiral effective theory or perturbative QCD.

In-medium effects may also help explain older puzzles. 
For example, the 
\href{http://www-e815.fnal.gov}{NuTeV} 
experiment~\cite{Zeller:2001hh} yields a value of 
$\sin^2\theta_W$ in neutrino-nucleus scattering 
$\sim 3\sigma$ 
away from the SM expectation. This anomalous result can be explained,
at least in part, 
by QCD effects, through corrections arising from modifications of 
the nuclear environment~\cite{Cloet:2009qs}.

Theoretical insight into these problems may prove essential to the
discovery of new physics. 
Unfortunately, multibaryon systems are complicated to calculate in lattice QCD 
due to a rapid increase in statistical noise. 
An analogous, albeit simpler, system using many pions has been the subject of 
an exploratory study. This first lattice-QCD attempt 
to measure many-hadron modifications of the 
hadronic structure in a pion ($\pi^+$) medium 
uses pion masses ranging 290--490~MeV at 2 lattice spacings~\cite{Detmold:2011np}. 
The preliminary result indicates strong medium corrections to the first 
moment of the pion quark-momentum fraction. With recent improvements to the 
efficiency of making quark contractions,  
which was one of the bottlenecks preventing lattice QCD 
from accessing even just twelve-quark systems, 
we expect to see development toward structure calculations for light nuclei albeit at 
heavier pion masses within the next few years. 

\subsubsection{Gluonic structure}

In the current global fit of the unpolarized parton distribution functions (PDFs) 
the gluonic contribution plays an enormously important 
role --- roughly half of the nucleon's
momentum is carried by glue. 
However, 
gluonic structure has been notoriously difficult to calculate with 
reasonable statistical signals in lattice QCD, even for just the first moment. 
Despite these difficulties, 
gluonic structure has been re-examined recently, with new work providing 
approaches and successful demonstrations that 
give some hope that the problem can be addressed. 
Both $\chi$QCD~\cite{Liu:2012nz,Deka:2013zha} and QCDSF~\cite{Horsley:2012pz} 
(note also \cite{Alexandrou:2013tfa}) have 
made breakthroughs with updated studies of gluonic moments 
in quenched 
ensembles with lightest pion masses of 480~MeV. 
The two groups attack the problem using different techniques and show 
promising results, with around 15\% uncertainty when extrapolated to the physical 
pion mass. These methods are now being applied to 
gauge ensembles with dynamical sea quarks,  
and we expect to see updated results within a few years. Similar
methods are also now used to probe the role of glue in the angular momentum of the 
proton~\cite{Liu:2012nz,Deka:2013zha}.


Let us conclude this section more broadly and note that, 
in addition to these known effects, 
lattice-QCD matrix elements are also important to 
experiments which have not yet observed any events, 
such as $n$-$\bar{n}$ oscillations~\cite{Buchoff:2012bm} or proton 
decay~\cite{Aoki:2013yxa}. 
Lattice-QCD calculations can provide low-energy constants 
to constrain the experimental search ranges. The potential to search for 
new physics using these precision nucleon matrix elements during the LHC era and 
in anticipation of future experiments at Fermilab make lattice-QCD calculations 
of nucleon structure particularly timely and important.

\subsection{Quark flavor physics} \label{sec:secE8}

The majority of the SM parameters have their origin in the 
flavor sector. The quark and lepton masses vary widely, which is an enduring puzzle. 
In this section we review 
studies of flavor and CP violation in the quark sector, usually probed through the  
weak decays of hadrons. 
In the SM the pattern of observed quark flavor and CP violation is captured by 
the CKM matrix, and the pattern is sufficiently distinctive that by 
overconstraining its parameters with multiple experiments 
and by employing accurate calculations, there is hope that an inconsistency 
between them (and therefore new physics) will ultimately emerge. 
At a minimum, this effort would allow the extraction of the CKM parameters 
with ever increasing precision.
Extensive reviews of this issue already exist; we note the massive efforts of the 
Heavy Flavor Averaging Group~\cite{Amhis:2012bh} and the PDG~\cite{Beringer:2012zz} 
for experimental matters, as well as 
similar reviews of 
lattice-QCD results~\cite{Laiho:2009eu,Colangelo:2010et,Aoki:2013ldr}. 
Thus, we concentrate 
here on a few highlights suggested by very recent progress or promise of principle.
We turn first, however, to two topics in non-CKM flavor physics which link to 
searches for BSM physics at low energies. 

\subsubsection{Quark masses and charges}

\paragraph{ Light quark masses}
The pattern of fermion masses has no explanation in the SM, but if the lightest quark
mass were to vanish, the strong CP problem would disappear. Thus in light of our
discussion of permanent EDMs and the new sources of CP violation 
that those experimental 
studies may reveal, it is pertinent to summarize the latest lattice-QCD results for
the light quark masses. Current computations work in the isospin limit ($m_u=m_d$),
treating electromagnetism perturbatively.
Turning to the compilation of the second phase of the 
Flavour Lattice Averaging Group (FLAG2)~\cite{Aoki:2013ldr},  we note with 
$N_f=2+1$ flavors (implying that a strange sea quark has been included) 
in the ${\overline{\rm MS}}$ scheme at a renormalization scale of $\mu=2$~GeV
that
\begin{eqnarray} 
{m}_s &=& (93.8\pm 1.5 \pm 1.9)~{\rm MeV} \,,\, \nonumber \\
{m}_{ud} &=& (3.42\pm 0.06\pm 0.07)~{\rm MeV} \,, 
\label{lqmass}
\end{eqnarray}
with $m_{ud} \equiv (m_{u} + m_{d})/2$, 
where the first error comes from averaging the lattice results and
the second comes from the neglect of charm (and more massive) sea quarks. 
The $m_s$ average value 
employs the results of \cite{Bazavov:2009fk,Durr:2010vn,Durr:2010aw,Arthur:2012opa}
whereas the $m_{ud}$ average value employs the results of 
\cite{Bazavov:2010yq,Durr:2010vn,Durr:2010aw,Arthur:2012opa}. 
To determine 
$m_u$ and $m_d$ individually additional input is needed. A study of 
isospin-breaking effects in chiral perturbation theory yields 
an estimate of $m_u/m_d$; this with the lattice results of Eq.~(\ref{lqmass}) 
yields~\cite{Aoki:2013ldr} 
\begin{eqnarray} 
{m}_u &=& (2.16\pm 0.09 \pm 0.07)~{\rm MeV} \,,\,  \nonumber \\
{m}_{d} &=& (4.68\pm 0.14 \pm 0.07)~{\rm MeV} \,,
\end{eqnarray}
where the first error represents the lattice statistical and systematic
errors, taken in quadrature, and the second comes from uncertainties in 
the electromagnetic corrections.\footnote{We note, too, that the neutron-proton 
mass difference has now been
computed within a self-consistent lattice-QCD calculation~\cite{deDivitiis:2013xla}:
$[M_n - M_p]_{\rm QCD} ({\overline{\rm MS}},\,2\,{\rm GeV}) = 2.9 \pm 0.6 \pm 0.2\,{\rm MeV}$.}
The electromagnetic effects could well deserve closer scrutiny. 

Nevertheless, it is apparent 
the determined up quark mass is definitely nonzero. This conclusion is not a 
new one, even if the computations themselves reflect the latest technical 
advances, and it is worthwhile to remark on the ($m_u=0$) proposal's enduring appeal. 
Ambiguities in the determination of $m_u$ have 
long been noted~\cite{Kaplan:1986ru,Banks:1994yg}; particularly, 
Banks et al.~\cite{Banks:1994yg} 
have argued that both the real and imaginary
parts of $m_u$ could be set to zero if there were an accidental U(1) symmetry
predicated by some new, spontaneously broken, horizontal symmetry. This
would allow $\delta$ of the CKM matrix to remain large at the TeV scale, 
while making $\bar\theta$ small. In this picture, a nonzero $m_u$ still exists
at low scales, but it 
is driven 
by nonperturbative QCD effects (and the strong CP problem can still be solved
if its impact on the EDM is sufficiently small). 
That is, in this picture $m_u$ is 
zero at high scales but is made nonzero at low scales 
through additive renormalization~\cite{Banks:1994yg}. Namely, 
its evolution from its low-scale value $\mu_u$ to a high-scale value $m_u$ ($m_u=0$)
would be determined by 
\begin{equation}
\mu_u = \beta_1 m_u + \beta_2 \frac{m_d m_s}{\Lambda_{\rm QCD}} +\dots  \,,
\label{E:eq:upMass}
\end{equation}
where $\beta_1$ and $\beta_2$ are dimensionless, scheme-dependent constants. 
This solution has been argued to be untenable because 
$m_d$ and $m_s$ are guaranteed not to vanish 
(independently of detailed dynamical calculations) by simple 
spectroscopy and no symmetry distinguishes the 
$m_u=0$ point~\cite{Creutz:2003xc,Creutz:2013xfa}. 
This makes the notion of a zero up-quark mass ill-posed~\cite{Creutz:2013xfa} 
within strict QCD, though this does not contradict the proposal in 
\cite{Banks:1994yg}, precisely because their analysis takes the second term of 
Eq.~(\ref{E:eq:upMass}) into account.

In~\cite{Banks:1994yg}, $\mu_u$ on the left-hand side of Eq.~(\ref{E:eq:upMass}) 
was argued to
hold for the mass parameter of the chiral Lagrangian.
The pertinent question is whether 
it applies to the masses of the QCD Lagrangian, obtained from lattice 
gauge theory.
Because $\beta_2$ is scheme dependent, the answer depends 
on details of the lattice determination.
Still, there is no evidence that the additive renormalization term is large
enough to make the $m_u=0$ proposal phenomenologically viable~\cite{Sharpe:2013}. 
The proposal of~\cite{Banks:1994yg} could be independently falsified 
if the residual ${\rm Im}(m_u)$ effects at low energies could be shown 
at odds with the existing neutron EDM bounds. 

\paragraph{ Quark charges}
In the SM with a single generation, electric-charge quantization 
(i.e., unique $U(1)_Y$ quantum number assignments) is predicated by 
the requirement that the gauge anomalies cancel, and ensures that both the neutron
and atomic hydrogen carry identically zero electric charge. There has been 
much discussion of the fate of electric charge quantization upon the inclusion
of new physics degrees of freedom, prompted by the experimental 
discovery that neutrinos have mass --- we refer to \cite{Foot:1992ui} for a review. 
For example, enlarging the SM with a gauge-singlet fermion, or right-handed neutrino, 
breaks the uniqueness of the hypercharge assignments, so that electric charge is
no longer quantized unless the new neutrino is a Majorana 
particle~\cite{Babu:1989tq,Babu:1989ex}. 
This outcome can be understood 
in terms a hidden $B-L$ symmetry which is broken 
if the added particle is 
Majorana~\cite{Babu:1989tq,Babu:1989ex,Foot:1990uf,Foot:1990mn}. 
More generally, electric charge quantization is not guaranteed in theories for which 
the Lagrangian contains anomaly-free
global symmetries which are independent 
of the SM hypercharge Y~\cite{Foot:1992ui}.\footnote{Consequently charge quantization
is not guaranteed in a three-generation SM because the difference in family 
lepton numbers is anomaly free~\protect{\cite{Foot:1990uf,Foot:1992ui}}. 
Nevertheless, the 
neutron and atomic hydrogen remain electrically neutral.} 
If neutrinoless double $\beta$ decay or neutron-antineutron oscillations are
observed to occur, then the puzzle of electric-charge quantization is solved. 
Alternatively, if the charge neutrality of the neutron or atomic hydrogen
were found to be experimentally violated, then it would suggest 
neither neutrinos nor neutrons
are Majorana particles. 
Another pathway to charge quantization could lead from 
making the SM the low-energy 
limit of a grand unified theory, though this is not guaranteed even if
such a larger theory occurs in nature. 
Ultimately, then, 
searches for the violation of charge neutrality, both of the neutron
and of atoms with equal numbers of protons and electrons, probe for the
presence of new physics at very high scales~\cite{Arvanitaki:2007gj}. 
Such a violation could also connect to the existence of new sources of 
CP violation~\cite{Witten:1979ey}. 
Novel, highly sensitive, 
experiments~\cite{Arvanitaki:2007gj,DurstbergerRennhofer:2011ap} are under
development, and 
plan to better existing limits by orders of
magnitude. 

\begin{figure}
\begin{center}
\includegraphics*[width=\linewidth]{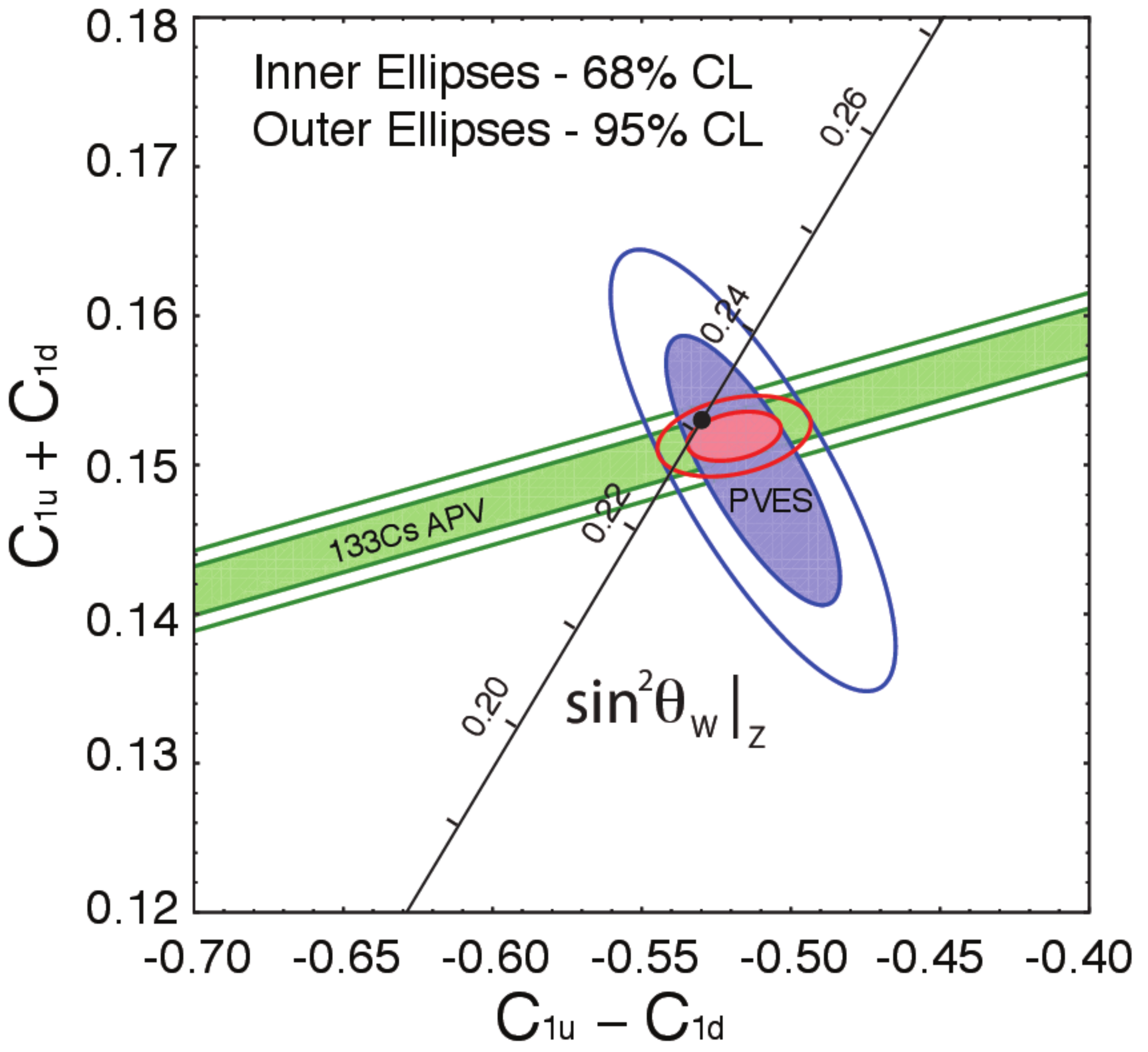}
\end{center}
\vspace{-0.5cm}
\caption{
Constraints on the (neutral-current) weak couplings of the $u$ and $d$ quarks
plotted in the $C_{1u}$--$C_{1d}$ plane. The band 
refers to the limits from APV, whereas the more vertical 
ellipse represents a global fit to the existing PVES data 
with $Q^2< 0.63~({\rm GeV})^2$. 
The small, more horizontal ellipse refers to the constraint determined from 
combining the APV and PVES data. The SM prediction as a function of 
$\sin^2\theta_W$ in the ${\overline{\rm MS}}$ scheme appears as a diagonal line; 
the SM best fit value is $\sin^2\theta_W=0.23116$~\cite{Beringer:2012zz}. 
Figure taken from~\cite{Androic:2013rhu}, and we refer to it for
all details.}
\label{fig:qweak}
\end{figure}

Experimental measurements of parity-violating electron scattering (PVES) 
observables yield
significant constraints on the weak charges of the quarks and leptons, 
probed through the neutral current. 
Recently, the weak charge of the proton $Q_W^p$ has been measured in 
polarized $\vec{e}$--$p$ elastic scattering at 
$Q^2=0.025~({\rm GeV})^2$~\cite{Androic:2013rhu}. This result, 
when combined with measurements of atomic parity violation (APV), 
yields the weak charge of the neutron $Q_W^n$ as well. 
The associated limits on the weak couplings of the quarks 
are shown in Fig.~\ref{fig:qweak}. For reference, we note that 
$Q_W^p=-2(2C_{1u} + C_{1d})$, where $C_{1i} = 2 g_A^e g_V^i$ and 
$g_A^e$ and $g_V^i$ denote the axial electron and vector quark couplings, 
respectively. The plot depicts an 
alternate way of illustrating the 
constraints on the $Q^2$ evolution of $\sin^2\theta_W$ 
in the ${\overline{\rm MS}}$ scheme 
discussed in Sec.~\ref{sec:secB4}. 

Nonperturbative QCD effects enter in this context as well, and we pause
briefly to consider the extent to which they could limit the sensitivity of
BSM tests in PVES. One notable effect is 
the energy-dependent radiative correction which arises from the 
$\gamma$-$Z$ box diagram. Dispersion techniques can be used to evaluate
it~\cite{Gorchtein:2008px,Tjon:2009hf,Sibirtsev:2010zg,Rislow:2010vi,Gorchtein:2011mz,Blunden:2011rd,Hall:2013hta}, and the correction 
is demonstrably large, contributing to some 8\% of $Q_W^p$ in 
the SM~\cite{Beringer:2012zz}. Currently the dispersion in its assessed
error is greater than that in the predicted central value, though the expected 
error can be refined through the use of additional PDF data~\cite{Hall:2013hta}.
Charge-symmetry-breaking effects in the nucleon 
form factors may eventually prove significant as well but are presently
negligible as they should represent a $\lesssim 1\%$ 
correction~\cite{Dmitrasinovic:1995jt,Miller:1997ya,Miller:2006tv,Kubis:2006cy,Miller:2013nea}. 
The implications of such theoretical errors, which appear manageable at current
levels of sensitivity, 
could eventually be assessed in a framework analogous to that recently
developed for neutron decay observables~\cite{Gardner:2013aya}.

\subsubsection{Testing the CKM paradigm}

We begin by presenting the moduli 
of the elements $V_{ij}$ of the CKM matrix 
determined in particular charged-current processes, using
the compilation of~\cite{Beringer:2012zz}:
\begin{eqnarray} \label{CKMmatrix}
\begin{array}{|ccc|}
|V_{ud}| & |V_{us}| &  |V_{ub}|    \\
|V_{cd}| & |V_{cs}| &  |V_{cb}|  \\
|V_{td}| & |V_{ts}| &  |V_{tb}| 
\end{array}
\propto \quad\quad\quad \quad\quad\quad 
\\ \nonumber 
\begin{array}{|ccc|} 
0.97425\pm 0.00022 & 0.2252 \pm 0.0009 & 0.00415 \pm 0.00049 \\
0.230 \pm 0.011   & 1.006 \pm 0.023  & 0.0409 \pm 0.0011 \\
0.0084 \pm 0.0006 & 0.0429 \pm 0.0026 &  0.89 \pm 0.07 
\end{array} \,.
\end{eqnarray}
Quark intergenerational mixing is 
characterized by the parameter $\lambda \approx |V_{us}|$, 
with mixing of the first (second) and third generations scaling as 
${\cal O}(\lambda^3)$ (${\cal O}(\lambda^2)$)~\cite{Wolfenstein:1983yz}. 
The CKM matrix $V_{\rm CKM}$ 
can be written in terms of 
the parameters $\lambda, A, {\bar \rho}, 
{\bar \eta}$; thus parametrized it is unitary to all 
orders in $\lambda$~\cite{Buras:1994ec,Charles:2004jd}. 
We test the SM of CP violation by determining whether all CP-violating
phenomena are compatible with a universal value of $({\bar\rho}, {\bar\eta})$
(note~\cite{Beringer:2012zz} for the explicit connection to $\delta$). 

\begin{figure}[b]
\includegraphics*[width=\linewidth]{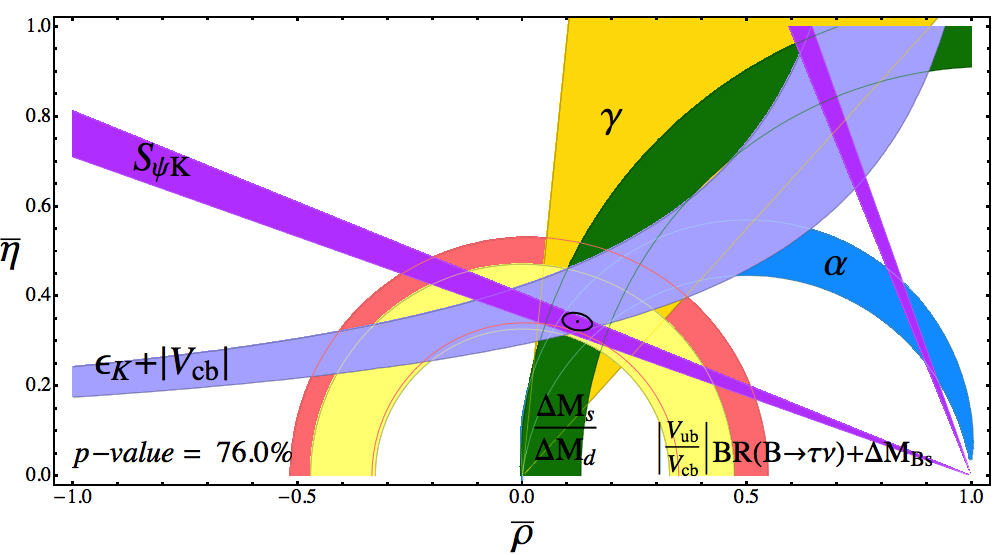}
\caption{Precision test of the SM mechanism of CP violation in charged-current processes 
realized through the comparison of the parameters $\bar\rho$ and $\bar\eta$ determined
through various experimental observables and theory inputs from lattice QCD. The experimental
inputs are as of September 2013, and the lattice inputs are derived from published results 
through April 30, 2013; the figure is an update of those in Ref.~\cite{Laiho:2009eu}. 
}
\label{fig:rhoeta}
\end{figure}

Current constraints are illustrated in Fig.~\ref{fig:rhoeta}. The so-called unitarity 
triangle in the ${\bar\rho}$--${\bar\eta}$ plane has vertices located at  
$(0,0)$, $(1,0)$, and $(\bar\rho_{\rm SM},\bar\eta_{\rm SM})$. 
The associated 
interior angles at each vertex are 
$\gamma (\phi_3)$, $\beta (\phi_1)$, and $\alpha (\phi_2)$, respectively.
The CP asymmetry $S_{\psi K}$ is 
realized through the interference of $B^0$-$\bar{B}^0$ 
mixing and direct decay into $\psi K$ and related modes. It 
is $\sin 2\beta$ in the SM up to hadronic uncertainties which 
appear in ${\cal O}(\lambda^2)$. 
The other observables require 
hadronic input of some kind to determine the parameters of interest; 
lattice-QCD calculations are essential to realize the precision of the
tests shown in Fig.~\ref{fig:rhoeta}. 
The constraints thus far are consistent with the SM of CP violation; 
the upper $S_{\psi K}$ band in the $\bar{\rho}-\bar{\eta}$ plane arising from a
discrete ambiguity has been ruled out by the determination that 
$\cos 2\beta > 0$ at 95\% C.L.~\cite{Aubert:2004cp}. 
Experimental studies of CP violation in the $B$ system continue, and we note
an improved constraint on $\gamma$ of $\gamma=67^\circ\pm12^\circ$ 
from LHCb~\cite{LHCb:2013gka}, which is consistent with the SM and with earlier
B-factory determinations~\cite{Aaij:2013zfa}.
Certain, early anomalies in B-physics observables
can be explained by a possible fourth SM-like generation~\cite{Hou:2006mx,Hou:2010mm}, 
and it remains an intriguing idea. 
Its existence, however, is becoming less and less consistent with experimental data.
Direct searches have yielded nothing so far~\cite{Chatrchyan:2012fp}, and
a fourth SM-like generation is disfavored by the observation of the Higgs, 
and most notably of $H\to\gamma\gamma$, as well~\cite{Lenz:2013iha,Kuflik:2012ai}. 
Flavor and CP violation are well-described by the CKM matrix~\cite{Isidori:2010kg}, 
so that it has become popular to build BSM models of the electroweak scale 
which embed this feature. That is, flavor symmetry is broken only by the 
standard Yukawa couplings of the SM; this paradigm is 
called Minimal Flavor Violation 
(MFV)~\cite{D'Ambrosio:2002ex,Buras:2003jf,Altmannshofer:2008hc}.

The structure of the CKM matrix can also be tested by determining whether the
empirically determined elements are compatible with unitarity. 
Figure~\ref{fig:rhoeta} illustrates that 
unitarity is maintained if probed through the angles determined from CP-violating
observables, that is, e.g., 
$\alpha+\beta+\gamma=(178^{+11}_{-12})^\circ$~\cite{Beringer:2012zz}. 
The most precise unitarity test comes 
from the first row~\cite{Beringer:2012zz}, namely, of whether 
$\Delta_u\equiv |V_{ud}|^2 + |V_{us}|^2 + |V_{ub}|^2 -1$ is nonzero. The
contribution of $|V_{ub}|^2$ is negligibly small at current 
levels of sensitivity, and for the last, several years the
uncertainty has been dominated by that in $|V_{us}|$~\cite{Beringer:2012zz}.
This situation changed, however, in 2013 with 
new, precise calculations of kaon decay parameters in lattice QCD becoming
available~\cite{Bazavov:2013vwa,Dowdall:2013rya,Bazavov:2013maa}. 
The quantity $|V_{ij}|^2 (\delta |V_{ij}|)^2$ 
determines the impact of a CKM matrix element on the unitarity test, and by this 
measure that of $|V_{us}|$ and $|V_{ud}|$~\cite{Hardy:2008gy} 
are now comparable~\cite{Dowdall:2013rya,Bazavov:2013maa}. Consequently,
the earlier result~\cite{Beringer:2012zz} 
\begin{equation}
\Delta_u = -0.0001\pm 0.0006 \,, 
\end{equation}
becomes, using the average value of $f_{K^\pm}/f_{\pi^\pi}$ in 
 QCD with broken isospin from~\cite{Aoki:2013ldr}, 
$\Delta_u =0 \pm 0.0006$. By averaging over computations in 
$N_f=2+1+1$, $N_f=2+1$, and $N_f=2$ ensembles the improvements associated with 
the included (published at that time)  
$N_f=2+1+1$ computation~\cite{Bazavov:2013vwa} is
muted, begging the question of whether it is 
appropriate to average calculations which differ in their 
quenching of heavier sea quarks.\footnote{Indeed Ref.~\cite{Aoki:2013ldr} typically 
declines to employ such averages.} The use of the most precise kaon
results yields a tension with CKM unitarity~\cite{Dowdall:2013rya,Bazavov:2013maa}. 
The value of $|V_{us}|$ can also
be determined from $\tau$ decay, and the situation there is quite different. 
The inclusive $\tau$ decay data yields a value of $|V_{us}|$ which 
is less precisely determined but still different from the one assuming 
three-flavor CKM unitarity by three sigma~\cite{Amhis:2012bh,Aoki:2013ldr}; 
more theoretical~\cite{Beneke:2008ad,Caprini:2009vf,Menke:2009vg,Boyle:2013xw} 
and experimental work 
will likely be needed to determine the origin of the discrepancy. We refer
to Sec.~\ref{sec:secB5} for a discussion of the determination $\alpha_s$ in 
hadronic $\tau$ decays, needed for a determination of $|V_{us}|$. 

The most precise determination 
of $|V_{ud}|$, $|V_{ud}|=0.97425 \pm 0.00022$~\cite{Hardy:2008gy}, , 
comes from the study of 
superallowed ($0^+\to 0^+$) 
transitions in nuclei. Its error is dominated by theoretical 
uncertainties, particularly from Coulomb corrections in the nuclear matrix
elements and other nuclear-structure-dependent effects~\cite{Hardy:2008gy} 
and from the evaluation of the $\gamma$-$W$ 
box diagram~\cite{Marciano:1985pd,Czarnecki:2004cw,Marciano:2005ec}. The 
assessment of nuclear Coulomb corrections~\cite{Hardy:2008gy} 
has been criticized as incomplete~\cite{Miller:2008my,Miller:2009cg}, though 
it has been experimentally validated in a superallowed decay for which the corrections
are particularly large~\cite{Melconian:2011kk}. 
Another unitarity test comes
from the second row; this can either be accessed directly through 
determinations of the $V_{ij}$ or indirectly though the leptonic width
of the $W$, for which the hadronic uncertainties are trivially small. 
The former procedure yields 
$\Delta_c \equiv 
|V_{cd}|^2 + |V_{cs}|^2 + |V_{cb}|^2 -1 = 0.04 \pm 0.06$~\cite{Aoki:2013ldr}, 
whereas the latter 
yields $\Delta_c = 0.002 \pm 0.027$~\cite{Beringer:2012zz}, making the
indirect method more precise. 

Theory plays a key and indeed expanding role in making 
all these tests more precise, so that increasingly 
the comparison between theory and experiment becomes a test field for QCD. 
We now consider some of the theory inputs in greater detail. 

\paragraph{ Theory inputs for $V_{us}$ \label{par:secEvus}}
Until very recently, 
the error in $V_{us}$ dominated that of 
the first-row CKM unitarity test.
Here, we consider different
pathways to $V_{us}$ through meson decays; as we have noted, such efforts
parallel the extraction of $V_{us}$ from $\tau$ 
decays~\cite{Gamiz:2002nu,Gamiz:2013wn,Pich:2013lsa,Boyle:2013xw}. 

Typically, $V_{us}$ has been determined through $K\to \pi {\ell} \nu$ ($K_{{\ell}3}$) 
decays and for which 
the following formula for the decay width applies~\cite{Cirigliano:2011ny}: 
\begin{eqnarray}
\Gamma(K_{{\ell} 3}) = \frac{G_F^2 m_K^5}{128\pi^3} C_K^2 S_{EW} 
(1+\delta_{\rm SU(2)}^{K\pi} + \delta_{\rm EM}^{K\ell}) \\ \nonumber
\times |V_{us}|^2 |f_+^{K^0\pi^-}(0)|^2 I_{K\ell} \,,
\label{genK}
\end{eqnarray} 
which includes various electroweak, electromagnetic, and
isospin-breaking corrections, 
in addition to the phase space integral $I_{K\ell}$ and other known factors. 
We have separated in the second line two of the most interesting ones. 
The first is the wanted CKM matrix element, and the second is a hadronic 
form factor to be 
evaluated at zero-momentum transfer. The form factors $f_{\pm}^{K\pi}(t)$ 
are determined by the QCD matrix elements 
\begin{eqnarray} \label{eq:secE:ff} \nonumber
&& \langle\pi(p_\pi)|{\bar s}\gamma_\mu u| K(p_K)\rangle = \\
&& (p_\pi + p_K)_\mu f_+^{K\pi}(t) + (p_\pi - p_K)_\mu f_-^{K\pi}(t) \,,
\end{eqnarray}
where $t=(p_K- p_\pi)^2$, and we note 
\begin{equation}
\delta_{\rm SU(2)}^{K \pi} = (f_+^{K\pi}(0)/f_+^{K^0\pi^-}(0))^2 -1 \,.
\end{equation}
There are, in principle, five different widths to be determined, in 
$K^\pm_{e3}, K^\pm_{\mu 3}, K_{L\,\!e3}, K_{L\,\!\mu3}$, and $K_{S\,\!\mu3}$ decay, 
 and the 
corrections in each can differ. Moreover, 
real-photon radiation 
also distinguishes the various processes, and it must be
treated carefully to determine the experimental decay widths~\cite{Cirigliano:2011ny}. 
Great strides have been made in the analysis of the various
corrections~\cite{Antonelli:2010yf,Cirigliano:2011ny,Flavia}, which are 
effected in the context of chiral perturbation theory,
and it is reasonable to make a global average of 
the determinations of $V_{us} f_+(0)$ in the various modes~\cite{Beringer:2012zz}. 
Progress also continues to be made on the experimental front,  
there being new measurements of $K^\pm\to \pi^0 l^\pm \nu$ 
by the NA48/2 experiment at CERN~\cite{Lamanna:2013}. 
The updated five-channel average is 
$f_+(0)|V_{us}|=0.2163 \pm 0.0005$~\cite{Moulson:2013wi}. 
The $t$ dependence of the form factor is 
embedded in the evaluation of $I_K^{\ell}$ in Eq.~(\ref{genK}). 
NA48/2 has selected events with one charged lepton and two photons that reconstruct 
the $\pi^0$ meson and extract form factors that they fit 
with either a quadratic polynomial in $t$, or 
a simple pole ansatz (be it scalar or vector),
\begin{equation}
f_{+,0}(t) = \frac{m_{v,s}^2}{m_{v,s}^2-t}\,,
\end{equation}
where $f_0(t)=f_+(t) + (t/(m_K^2 - m_\pi^2))f_-(t)$. 
A good fit is obtained with $m_v= 877 \pm 6$ MeV and $m_s=1176 \pm 31$ MeV; 
these quantities do not precisely correspond to known particles 
but are of a reasonable magnitude. 
We detour, briefly, to note that 
the systematic error in the precise choice of fitting form can be mitigated 
through considerations of analyticity and crossing 
symmetry~\cite{Boyd:1994tt,Hill:2006bq}; the 
latter permits the use of experimental 
data in $\tau \to K \pi \nu$ decays~\cite{Hill:2006bq} to constrain the fitting
function. 
Finally we turn to the determination of $f_+(0)$, 
for which increasingly
sophisticated lattice-QCD calculations have become available. 
Noting~\cite{Aoki:2013ldr}, we report 
$[N_f =2]$~\cite{Lubicz:2009ht} and 
$[N_f =2 +1]$~\cite{Bazavov:2012cd} results: 
\begin{multline}
f_+(0) = 0.9560 \pm 0.0057 \pm 0.0062 \quad [N_f =2] 
\\
f_+(0) = 0.9667 \pm 0.0023 \pm 0.0033 \quad [N_f =2 +1]
\end{multline}
as well as~\cite{Bazavov:2013maa}
\begin{equation}
f_+(0) = 0.9704 \pm 0.0024 \pm 0.0022 \quad [N_f =2 +1 +1]\,.
\end{equation}
Using the last value for $f_+(0)$, which attains a physical value 
of the pion mass, 
and those of 
$|V_{us}| f_+(0)$ and $V_{ud}$ we have reported, yields 
$\Delta_u=-0.00115\pm 0.00040 \pm 0.0043$, where the first (second) error
is associated with $V_{us}$ ($V_{ud}$), 
and roughly a $2\sigma$ tension with
unitarity~\cite{Bazavov:2013maa}.

As a final topic we consider the possibility of determining 
$V_{us}/V_{ud}$ from the ratio of 
$K_{\ell 2 (\gamma)}$ and $\pi_{\ell 2 (\gamma)}$ decay widths with the use of 
the decay constant ratio $f_K/f_\pi$ computed in lattice QCD~\cite{Marciano:2004uf}. 
This method competes with the $K_{\ell 3}$ decays in precision. 
In a recent development, 
the isospin-breaking effects which enter can now be computed
using lattice-QCD methods as well; the  method is based on the expansion of
the Euclidean functional integral in the terms of the up-down mass 
difference~\cite{deDivitiis:2011eh,deDivitiis:2013xla}. Generally, the
separation of isospin-breaking effects in terms of up-down quark mass
and electromagnetic contributions is one of convention, because the quark 
masses themselves accrue electromagnetic corrections which diverge in the 
ultraviolet~\cite{deDivitiis:2011eh,Lubicz:2013xja}. 
Technically, however, the pseudoscalar 
meson decay constants are only defined within pure QCD, so 
that 
\begin{equation}
\frac{f_{K^+}}{f_{\pi^+}} = \frac{f_{K}}{f_{\pi}} \left( 1 + \delta_{\rm SU(2)}\right) \,,
\end{equation}
where $f_K/f_\pi$ are evaluated in the isospin-symmetric ($m_u=m_d$) limit. 
Thus we can crisply compare the ChPT 
determination of $\delta_{\rm SU(2)}$~\cite{Cirigliano:2011tm} 
with a completely different nonperturbative 
method. Namely, noting~\cite{Lubicz:2013xja}, 
we have 
$\delta_{\rm SU(2)}^{\rm ChPT}=-0.0021\pm 0.0006$~\cite{Cirigliano:2011tm}, whereas 
$\delta_{\rm SU(2)}^{\rm lattice}=-0.0040\pm 0.0003 \pm 0.0002$~\cite{deDivitiis:2013xla} and
$\delta_{\rm SU(2)}^{\rm lattice}=-0.0027\pm 0.006$~\cite{Dowdall:2013rya}. 
Thus tension exists in the various assessments of SU(2)-breaking effects, and 
it will be interesting to follow future developments. 

\paragraph{$B$ and $D$ form factors \label{par:secEBDff}} 
Lattice-QCD methods can also be used for the 
computation of the $B$ and $D$ meson form factors in exclusive semileptonic decays, yielding 
ultimately additional CKM matrix elements once the appropriate partial widths are 
experimentally determined. Generally, CKM information can be gleaned from both
exclusive and inclusive (to a final state characterized by a quark flavor $q$, 
as in $B\to X_q {\ell} \nu$ decay) 
$B$ meson decays, and different theoretical methods figure in each. In the inclusive
case, the factorization of soft and hard degrees of freedom is realized using 
heavy quark effective theory, and the needed nonperturbative ingredients are determined
through fits to data. As we have noted, 
lattice-QCD methods can be employed in the exclusive channels, 
and the leptonic process $B\to \tau \nu$, along
with a lattice-QCD computation of the decay constant $f_B$, also yields $|V_{ub}|$, 
though this pathway is not yet competitive with other methods. 
We refer to Sec.~\ref{sec:chapc} 
for a detailed discussion, though
we note that tension continues to exist between the various determinations of
$|V_{ub}|$ and $|V_{cb}|$.
In particular, an exclusive extraction from $B\to \pi l \nu$ decay has been made 
using form factors computed with 
$[N_f = 2+1]$ dynamical quark flavors by 
HPQCD~\cite{Dalgic:2006dt} and FNAL/MILC~\cite{Bailey:2008wp}. A
simultaneous fit of the lattice and experimental form factors to determine
their relative normalization $|V_{ub}|$ yields~\cite{Aoki:2013ldr} 
$|V_{ub}|=0.00337 \pm 0.00021$ (BaBar~\cite{Lees:2012vv}) 
and $|V_{ub}|=0.00347 \pm 0.00022$ (Belle~\cite{Ha:2010rf}). 
These values remains below the inclusive determination 
of $0.00440\pm 0.00025$~\cite{Amhis:2012bh}. 
The two determinations remain to be reconciled, perhaps by
better measurements separating the background charm decays
of the $B$, since the theoretical determination in terms of
\begin{equation}
\frac{d\Gamma(\bar{B}^0\to \pi^+l\bar{\nu})}{dq^2} = 
\frac{G_F^2 | {\bf p}_\pi|^3}{24\pi^3} | V_{ub}|^2 | f_+(q^2)|^2
\end{equation}
seems crisp, though it could be aided perhaps by better resolution of 
the $q^2$ dependence of the form factor(s).

A similar situation is found in comparing the exclusive
and inclusive determinations of $|V_{cb}|$ where there remains
roughly a $2\sigma$ tension between the results. This parameter is 
important in many instances, for example in tightening 
constrained-MFV models~\cite{Buras:2013raa}.
The exclusive extraction 
requires determining the form factors of
${d\Gamma(B\to (D/D^*)+ l\nu)}/{d({\bf v}_b \cdot {\bf v}_c)}$ at the
zero recoil point. Only a single calculation, of the $B\to D^*$ form factor, 
currently satisfies the FLAG criteria~\cite{Aoki:2013ldr}. This result,  
the 2010 FNAL/MILC calculation~\cite{Bailey:2010gb}, employing $N_f=2+1$ dynamical
quark flavors, yields
$|V_{cb}|_{\rm exc}=0.003955 \pm 0.000072 \pm 0.000050$~\cite{Aoki:2013ldr}, 
where the errors denote lattice and non-lattice (experiment and non-lattice
theory) uncertainties, respectively. 
This is to be compared with
$|V_{cb}|_{\rm inc}=0.004242 \pm 0.000086$~\cite{Gambino:2013rza}; the two
results are discrepant at about $2.3\sigma$. New lattice calculations
of the $B\to D^{(*)}$ form factors are in progress; presumably this will 
improve the situation considerably. Alternatively, 
the possibility of higher-order effects in the heavy-quark
expansion for inclusive B decays, particularly those
due to  ``intrinsic charm,'' 
have been discussed~\cite{Brodsky:2001yt,Bigi:2005bh,Bigi:2009ym}, though 
their magnitude has not yet been established. 

In charm decays, the determinations of $|V_{cd}|$ and $|V_{cs}|$ via leptonic
and semileptonic modes are in reasonably good agreement. 
The $|V_{cd}|$ determinations are all consistent
within errors, whereas the $|V_{cs}|$ in leptonic and semileptonic modes 
disagree at $1.2 \sigma$~\cite{Aoki:2013ldr}. Using the 
results, e.g., for 
$f_+^{D\pi}(0)|V_{cd}|$ and $f_+^{DK}(0)|V_{cs}|$ from~\cite{Amhis:2012bh}, 
and the form factor calculation of the only $N_f=2+1$ lattice calculation
to satisfy FLAG criteria in each case (\cite{Na:2011mc} and 
\cite{Na:2010uf}, respectively), yields 
$|V_{cd}|=0.2192 \pm 0.0095 \pm 0.0045$ and
$|V_{cs}|=0.9746 \pm 0.0248 \pm 0.0067$~\cite{Aoki:2013ldr}.  
For reference, from
neutrino scattering one has  $|V_{cd}|=0.230\pm 0.011$~\cite{Beringer:2012zz}. 

\paragraph{Tests of lepton-flavor universality in heavy-light decays \label{par:secEBuniv}} 
Heavy-light semileptonic processes 
can also be used to challenge the SM with minimum theory 
input, through tests of lepton-flavor 
universality~\cite{Fajfer:2012vx,Becirevic:2012jf,Becirevic:2013zw}. In particular, we 
recall from Sec.~\ref{sec:chapc} 
that BaBar has measured the ratio 
\begin{equation}
R(D) \equiv \frac{{\cal B}(B\to D \tau\nu)}{{\cal B}(B\to D{\ell}\nu)}
= 0.440\pm 0.058 \pm 0.042 \,,
\end{equation}
with ${\ell}\in e, \mu$, 
and substituting the $D$ for a $D^*$ yields 
$R(D^*)=0.332\pm 0.024 \pm 0.018$~\cite{Lees:2012xj}. These ratios
are in excess of SM predictions, at $2.0\sigma$ and $2.7\sigma$, 
respectively, and the  apparent, observed violation of lepton-flavor
universality can be mediated by a new charged Higgs boson~\cite{Lees:2012xj}. 
The measured ratio of ratios, however, appears to be odds with the Type II 
two-Higgs-doublet model~\cite{Lees:2012xj}, though there are many other BSM 
possibilities which can generate an effect~\cite{Fajfer:2012jt,Celis:2013bya,Sakaki:2013bfa}. 
The ratio $R(D)$ has been revisited by the FNAL/MILC collaboration
to find $R(D)=0.316\pm 0.014$~\cite{Bailey:2012jg}), 
a value of some $1.7\sigma$ smaller than the 
BaBar result if the errors are combined in quadrature. Nevertheless, 
their study illustrates the importance of the computation of the scalar form
factor to the prediction of $R(D)$, and we look forward to future results 
in regard to $R(D^*)$. We note that combining the $R(D)$ and $R(D^*)$ 
experimental results currently 
yields a disagreement of $3.4\sigma$ with the SM~\cite{Lees:2012xj}. 
Lattice-QCD methods will also no doubt be important to evaluating the success
of a particular BSM model in confronting the experimental values of $R(D)$ and $R(D^*)$. 

\begin{figure}[b]
\includegraphics*[width=\linewidth]{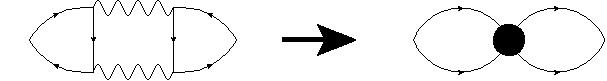}
\caption{The box diagram for neutral meson mixing 
with double $W$ exchange (as in the SM) 
reduces at low energy to the matrix element 
of a contact four-quark operator that yields the ``bag parameter.''}
\label{fig:bagparameter}
\end{figure}
\paragraph{ Neutral meson mixing and bag parameters \label{par:secEbag}} 
The nonperturbative matrix element associated with neutral-meson mixing, 
depicted schematically in Fig.~\ref{fig:bagparameter}, is 
termed the bag parameter $B_{\rm mes}$; it captures the deviation of
the operator matrix element from its vacuum insertion value for which $B_{\rm mes}=1$. 
In the kaon system, it is essential to an understanding of 
$|\epsilon_K|$~\cite{Beringer:2012zz}, 
the parameter which characterizes CP violation in $K^0$--$\bar{K}^0$ mixing, 
and whose interpretation in terms of CKM parameters has languished for decades. 
In the $K^0$--$\bar{K}^0$ system $B_K$ is given by 
\begin{equation}
B_K = \frac{\langle \bar{K}^0\arrowvert {\mathcal{O}}_{LL}^{\Delta S=2} \arrowvert 
K^0\rangle}{\frac{8}{3}f_K^2 m_K^2}\, ,
\label{eq:bag}
\end{equation} 
at some scale $\mu$, where 
${\mathcal{O}}_{LL}^{\Delta S=2} = (\bar s \gamma^\mu (1-\gamma_5) d)(\bar s 
\gamma_\mu (1-\gamma_5) d)$, and from which the renormalization-group-invariant (RGI)
quantity $\hat{B}_K$ can be determined~\cite{Antonio:2007pb}. Several 
$N_f=2+1$ calculations now exist, and their average (specifically 
of~\cite{Durr:2011ap,Kim:2011qg,Bae:2011ff,Laiho:2011np,Arthur:2012opa})
yields
$\hat{B}_K=0.766 \pm 0.0010$~\cite{Aoki:2013ldr}. 
Since $\hat{B}_K$ is now known to some 1.3\%, the ability to interpret $\epsilon$
has changed dramatically for the better. This improvement is captured
in the width of the $|\epsilon_K|$ band in Fig.~\ref{fig:rhoeta}, 
and the dominant residual uncertainties in its interpretation come from that
in $|V_{cb}|$, which enters as $|V_{cb}|^4$, 
 and in the perturbative contribution from $c\bar c$ quarks~\cite{Brod:2011ty}. 
Concerning new physics searches, 
the computation of a complete set of $|\Delta| S=2$ hadron 
operators for $K^0$--$\bar{K}^0$ mixing is underway 
by several collaborations, including the ETMC~\cite{Bertone:2012cu} and 
RBC/UKQCD~\cite{Boyle:2012qb}. 
This should help constrain 
the couplings to and masses of 
additional particles beyond the SM.

The parameter which characterizes 
direct CP violation in the 
kaon system, ${\rm Re}(\epsilon'/\epsilon)$, is definitely nonzero, 
${\rm Re}(\epsilon'/\epsilon)=(1.67\pm0.23)\times 10^{-3}$~\cite{Beringer:2012zz}, 
and probes ${\rm Im}(V_{td}V^*_{ts})$, 
though large theoretical hadronic uncertainties beset its interpretation. 
This arises due to the approximate cancellation of the gluonic 
and electroweak penguin contributions, exacerbating the role of 
isospin-violating effects~\cite{Donoghue:1986nm,Buras:1987wc}, 
so that effects beyond $\pi^0-\eta,\eta^\prime$ mixing 
can also play an important role~\cite{Gardner:1999wb,Wolfe:2000rf,Cirigliano:2003nn}. 
The recent strides in lattice-QCD computations has spurned progress in 
this system as well~\cite{Blum:2011ng,Blum:2012uk}, 
though an analysis of its isospin-breaking effects
would seem beyond the scope of current ambitions.

Mixing in the $B^0_{s}\bar{B}^0_{s}$ system has now also been 
established~\cite{Aaij:2011qx,Aaij:2013mpa}, and the comparison of its
measured mixing parameters with those of the 
$B^0_{d}\bar{B}^0_{d}$ system 
yields a precision test of the SM. 
The mass difference between the weak-interaction eigenstates in the SM
is given by
\begin{equation}
\Delta M_{q}|_{\rm SM} \propto |V^*_{t(q)}V_{tb}|^2 M_{B_{q}} f_{B_{q}}^2
\hat{B}_{B_{q}} \,,
\end{equation}
with $q\in (d,s)$ and where we define $B_{B_q}$ after Eq.~(\ref{eq:bag}), noting 
$\hat{B}_{B_{q}}$ is the RGI quantity. The constant of proportionality is common
to the two systems, so that the ratio $(\Delta M_s/\Delta M_d)_{\rm SM}$ is determined
by $\xi^2 M_{B_s}/M_{B_d}$, where the nonperturbative parameter
\begin{equation}
\xi=\frac{f_{B_{s}} \sqrt{B_{B_{s}}}}{f_{B_{d}} \sqrt{B_{B_{d}}}}\,
\end{equation}
can be computed in lattice QCD. 
Its deviation from unity measures the size of SU(3)$_f$ breaking. 
There have been several $N_f=2+1$ lattice-QCD calculations of this quantity, 
but only one passes all the FLAG criteria~\cite{Aoki:2013ldr}, so that
we report $\xi=1.268 \pm 0.063$~\cite{Bazavov:2012zs}. 
Confronting $\Delta M_q$ directly is a much more challenging task, though 
this, as well as the matrix elements of all five (leading-dimension) operators that can
generate $B^0_{q}$--$\bar{B}^0_{q}$ mixing 
system are under analysis~\cite{Bazavov:2012zs}. 
Such efforts are crucial to determining to what extent BSM efforts 
operate in the $B^0_{q}\bar{B}^0_{q}$ system (and in the $K^0\bar{K}^0$ system by comparison) 
and to constrain the models which could generate them~\cite{Buras:2012jb}. 

\subsubsection{New windows on CP and T violation \label{par:secEcpt}}

We now review some recent results in the study of CP and T violation. 
Several new results concern searches for direct CP violation in systems for which
such observables are parametrically small in the SM. 
The key issue is how large the latter can possibly be; 
can the observation of a larger-than-expected
CP asymmetry be an imprimatur of new physics? 
Direct CP violation, such 
as a rate asymmetry in $B\to f$ and $\bar B \to \bar f$ decays, follows
from the quantum interference of two amplitudes, typically 
termed ``tree'' and ``penguin'' 
as per their quark-flow diagrams, that differ in both 
their strong and weak phase. 
Unfortunately 
the tree-penguin interference effects which give rise to the CP
asymmetries are notoriously challenging to calculate and can be subject to 
nonperturbative enhancement. It is probably better to be cautious in 
considering 
a larger-than-expected CP asymmetry as evidence of new physics. To give a context
to this assessment we offer a terse overview of the theory of nonleptonic
$B$-meson decays, though the essential ideas apply to the analysis of
nonleptonic $D$ and $K$ decays as well. 

Early exploratory studies employed ``vacuum saturation''~\cite{Bauer:1986bm}
or ``generalized factorization''~\cite{Ali:1998eb,Ali:1998gb} to evaluate 
the matrix elements of the dimension-six operators which appear in $B$ decays, though
such work had conceptual and computational limitations~\cite{Buras:1998us}. 
Alternatively, the large energy release of $B$ decays to light hadrons motivates 
the use of flavor-symmetry-based (of the $u$, $d$, $s$ valence quarks) 
strategies, relating experimental data in various final states 
to determine the ill-known amplitudes, as in, e.g., \cite{Silva:1993sv}.
Such strategies are approximate and may fail as constraints on new physics become
more severe. 

The effective Hamiltonian for $|\Delta B|=1$ processes at the $b$-quark
mass scale has been known to NLO precision for 
some time~\cite{Buras:1991jm,Buchalla:1995vs}. 
The construction of a ``QCD factorization,'' based on the combined use
of the heavy quark and strong-coupling-constant expansions, was a major
step forward --- the 
scale dependence of the decay amplitude (combining the pieces from the
Wilson coefficients with those from the evaluation of the matrix elements of
the associated local operators) was shown, for the 
first time, to vanish in NLO precision
and at leading power 
in the heavy-quark expansion~\cite{Beneke:1999br,Beneke:2000ry}. 
The approach has been applied to a sweep of two-body $B$-meson
decays and works fairly well~\cite{Beneke:2001ev,Beneke:2003zv}, 
though there are some systemic problems. 
The theory has difficulties confronting empirical branching ratios 
in modes for which the tree amplitude is suppressed, and it has
trouble confronting CP-asymmetries. Systematic study suggests that
the power corrections (in the heavy-quark mass) are phenomenologically
important~\cite{Cheng:2009eg,Cheng:2009cn}, 
though explicit studies of NNLO corrections in 
$\alpha_s$~\cite{Bell:2009nk,Bell:2009fm,Beneke:2009ek}
have also eased tension with the data in the explicit modes 
studied~\cite{Jager:2013wba}. Other approaches 
include the use of SCET~\cite{Bauer:2004tj,Arnesen:2006vb,Chay:2007ep} and of $k_T$ 
factorization~\cite{Keum:2000wi,Keum:2000ph,Li:2005kt,Li:2009wba}; 
the latter does not take the heavy-quark limit. 
Possible troubles with power corrections in the heavy-quark mass do not 
bode well for the analysis of hadronic $D$ decays, though the $k_T$ factorization
approach can be and has been used~\cite{Li:2012cfa}.

\paragraph{ Three-body decays and Dalitz plot analyses \label{par:secEdalitz}} 
Two-body decays have been much studied in the analysis of CP-violating observables
in the $B$ system, but the large phase space available in the decay of $B$ mesons 
to light hadrons makes three-body final states far more copious. 
Such final states are theoretically more difficult to handle because 
factorization theorems in QCD of exclusive 
heavy-light decays exist for two-body final 
states~\cite{Beneke:1999br,Beneke:2000ry,Bauer:2004tj}. 
However, flavor-based analyses, such as an isospin analysis of 
$B\to \rho\pi$~\cite{Snyder:1993mx},  
can nevertheless be successful in extracting CKM phase information, 
even in the presence of SM isospin-breaking effects such as
$\pi^0-\eta,\eta^\prime$ mixing~\cite{Gardner:2001gc}. 
A great deal of information is encoded in the Dalitz plot of the final state
and can yield new pathways to the identification of 
direct CP 
violation~\cite{Gardner:2002bb,Gardner:2003su,Petrov:2004gs,Bediaga:2009tr,Bediaga:2012tm,Sahoo:2013mqa}, 
both in the 
$B$ and $D$ system. 
Interpreting the Dalitz plot requires good control of the manner in which
the various resonances can appear, making the use of simple 
Breit-Wigner forms insufficient. Rather, constraints from low-energy
chiral dynamics, including those of unitarity and analyticity,
such as realized in Omn\`es-based approaches, 
 should be brought to bear~\cite{Gardner:2001gc}. The latter are
gaining ground~\cite{Schneider:2012ez,Daub:2012mu,Niecknig:2012sj}.

Recently LHCb has reported an enhanced signal of CP violation in 
$B^\pm\to K^\pm\pi^+\pi^-$ and $B^\pm \to K^\pm K^+ K^-$ 
final states~\cite{Aaij:2013sfa,Aaij:2013bla}, and theoretical 
work has concerned, e.g., 
whether the effects are consistent with $U$-spin symmetry~\cite{Bhattacharya:2013cvn}, 
as well as the particular Dalitz-plot interference mechanisms 
needed to explain them~\cite{Bediaga:2013ela}. 

\paragraph{ CP violation in the $B_s$ system \label{par:secEBs}} 
Recently, the LHCb collaboration has made a series of measurements probing the decays of 
$B_s^0 (\bar B_s^0)$ mesons to different CP-eigenstates, specifically 
$J/\psi \phi$~\cite{LHCb:2011aa}, 
$J/\psi f_0$~\cite{LHCb:2011ab}, 
$J/\psi \pi^+\pi^-$~\cite{LHCb:2012ad,Aaij:2013oba}, and $J/\psi K^+K^-$~\cite{Aaij:2013oba}. 
We note that ATLAS has studied $B_s\to J/\psi \phi$ as well~\cite{Aad:2012kba}. The CP-violating
phase $\phi_s$, which is determined by the argument of the ratio of the off-diagonal
real and dispersive pieces in $B_q^0$--$\bar B_q^0$ mixing, and other mixing parameters
are also determined, which include the average $B_s^0$ decay width $\Gamma_s$ and the
width difference $\Delta \Gamma_s$. 
The measurement of $B_s \to J/\psi \phi$ alone leaves a two-fold 
ambiguity in  $(\phi_s,\Delta \Gamma_s)$ but this can be resolved by the study
of the $J/\psi K^+K^-$ final state with the invariant mass of the $K^+K^-$ pair. 
The latest LHCb results are~\cite{Aaij:2013oba}: 
\begin{eqnarray}
\phi_s &=& 0.07 \pm 0.09 \pm 0.01 \,{\rm rad} \quad [J/\psi K^+K^-]\,, \\
\phi_s &=& 0.01 \pm 0.07 \pm 0.01 \,{\rm rad} \quad [{\rm combined}]\,,  
\end{eqnarray}
where ``combined'' refers to 
a combined fit of $J/\psi K^+K^-$ and $J/\psi \pi^+\pi^-$ events. 
The enumerated errors are statistical and systematic, respectively. 
In the SM, ignoring subleading penguin contributions, 
$\phi_s = -2\beta_s$, where 
$\beta_s= {\rm arg}\left( - {V_{ts} V_{tb}^\ast}/{V_{cs} V_{cb}^\ast}\right)$, 
and indirect global fits assuming the SM yield 
$2\beta_s=0.0364\pm 0.0016\,{\rm rad}$~\cite{Charles:2011va}. The $\phi_s$  result, as well
as those for $\Gamma_s$ and $\Delta\Gamma_s$ are compatible with 
SM expectations~\cite{Charles:2011va,Lenz:2011ti}. 
CP violation has been observed in the $B_s^0$ system, however, in 
$B_s^0\to K^-\pi^+$ decay~\cite{Aaij:2013iua}. 

\paragraph{ CP violation in the $D$ system \label{par:secED}} 
There has been much interest in probing CP violation in the $D$ system since
the common lore is that a CP asymmetry in excess of $10^{-3}$ in 
magnitude 
would be a signal of new physics~\cite{Hiller:2012wf}. 
The $D$ meson is produced copiously by $e^+e^-$ machines at the $\psi(3770)$
resonance, as well as  at higher-mass 
resonances such as the $\psi(4040)$ or $\psi(4160)$ that can be used at 
BES-III.  
It is also a common end-product of the fragmentation of a c-quark at the LHC. 

Much discussion has been sparked by a claim of evidence 
for direct CP violation in $D$ decays by the LHCb 
collaboration~\cite{Aaij:2011in}, 
and there has been ongoing discussion as to how large SM CP-violating effects
can really be, given theoretical uncertainties 
in the long-distance physics
which can enter~\cite{Franco:2012ck,Brod:2012ud}. 
We can construct 
a CP asymmetry in the usual way: 
\begin{equation}
A_{\rm CP}= \frac{\Gamma(D^0\to h^+h^-)-\Gamma(\bar{D}^0\to 
h^+h^-)}{\Gamma(D^0\to h^+h^-)+\Gamma(\bar{D}^0\to  h^+h^-)} \,,
\end{equation}
from which the direct and indirect (via $D^0\leftrightarrow\bar{D}^0$ mixing) 
contributions can be separated, since both $\pi\pi$ and $KK$ channels are available. 
Thus we form the direct CP asymmetry 
$\Delta A_{\rm CP} = A_{\rm CP}(K^+ K^-) - A_{\rm CP}(\pi^+ \pi^-)$, for 
which LHCb~\cite{Aaij:2011in} reports  
$(-0.82\pm 0.21 \pm 0.11)\%$, 
and the CDF result using the full Run II data set
is comparable in size: $(-0.62\pm 0.21 \pm 0.10)\%$~\cite{Collaboration:2012qw}. 
An update of the earlier LHCb analysis using a much larger data set yields
$\Delta A_{\rm CP} = (-0.34 \pm 0.15 \pm 0.10)\%$~\cite{LHCb:2013dka}, 
which is much smaller. Moreover, an independent LHCb measurement based on 
$D^0$'s from semileptonic $b$-hadron decays yields
$(0.49 \pm 0.30 \pm 0.14)\%$~\cite{Aaij:2013bra}. Thus the early evidence
remains unconfirmed. 
The possibility of 
direct CP violation in the charm sector is of enduring 
interest~\cite{Bianco:2003vb,Grossman:2006jg}, 
however, and the search goes on. 

\begin{figure}[b]
\includegraphics*[width=0.8\linewidth]{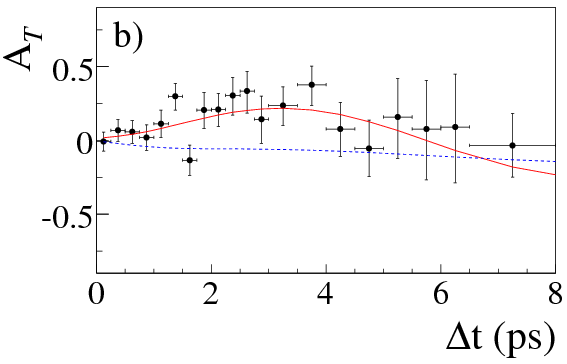}
\caption{One of four T-violating asymmetries reported by the BaBar 
collaboration~\cite{:2012kn}.}
\label{fig:Tviolation}
\end{figure}

\paragraph{ Observation of T violation in the $B$ system \label{par:secETviol}} 
In a separate development, an observation of direct 
T-violation has been claimed~\cite{:2012kn}.
Its presence is expected because the CPT theorem of local, Lorentz-invariant 
quantum field theory implies the existence of T violation in the presence of CP violation. 
Direct measurement of a fundamental T-violating effect in hadronic processes 
 is a bit tricky, however, 
because it 
requires being able to compare an S-matrix element 
$S_{f,i}$ to its reciprocal $S_{i_T,f_T}$ in which $i_T$ and $f_T$ are the time-reversed
states of $i$ and $f$. 
It is challenging to prepare the requisite states, so that 
robust ``detailed balance'' tests of T are rare. 
A nonzero permanent EDM, of course, would display 
a fundamental violation of T-invariance in a stationary state, 
and experimental limits are becoming more
stringent. The CPLEAR collaboration~\cite{Angelopoulos:1998dv} observed
a difference in the rate of $K^0 \to \bar K^0$ and $\bar K^0 \to K^0$, 
where the initial $K^0$ ($\bar K^0$) 
is identified by its associated production with a 
$K^+$ ($K^-$) in $p \bar p$ collisions and the final-state 
$\bar K^0$ ($K^0$) is identified through the sign of the lepton charge in semileptonic decay. 
This has been questioned as a direct test of time-reversal 
violation~\cite{Wolfenstein:1999TRV,Wolfenstein:2000ju}
because (i) the constructed asymmetry is independent of time  and 
(ii) unitarity considerations reveal that if more $K^0$ goes to $\bar K^0$ than 
$\bar K^0$ goes to $K^0$ 
this can only occur if more $\bar K^0$ decays to $\pi\pi$ than 
$K^0$, making the appearance of particle decay (which is irreversible) 
essential to the effect. 
In regards to a detailed balance study of T in the B system, a theory 
proposal~\cite{Bernabeu:2012ab} has been recently implemented by 
the BaBar collaboration~\cite{:2012kn}; this is a much richer system
than that studied by CPLEAR. The initial and final 
states are pairs of neutral $B$ mesons, be they in the 
flavor-eigenstate basis $B^0$, $\bar{B}^0$, or in the CP eigenstate basis 
$B^{{\rm CP}+}$, $B^{{\rm CP}-}$. The two reactions whose rates 
are compared are the neutral meson oscillations between states in the 
 two different bases, e.g., 
\begin{equation}\label{E6:transition}
\bar{B}^0\to B^{{\rm CP}-}\ \ \ ; \ \   B^{{\rm CP}-}\to \bar{B}^0 \ .
\end{equation}
To prepare the initial state, BaBar makes use of 
quantum entanglement in the reaction $e^-e^+\to \Upsilon(4S) \to B\bar{B}$. 
Because the intermediate vector $\Upsilon(4S)$ state has definite b-flavor 
(0) and 
 CP ($+$), one chooses to make a measurement of 
either the CP or flavor of one of the two $B$-mesons, and this leaves its entangled $B$ 
partner in a CP or a flavor eigenstate. The partner of the tagged B is left to propagate 
and then the opposite measurement, of either flavor or CP, is made 
on the second $B$ meson. This second $B$ must have undergone the transition in 
Eq.~(\ref{E6:transition}) since it was produced as an eigenstate of either CP or flavor, 
but it is detected as an eigenstate of the other variable. 
The two reactions can at last be compared.
It remains to be said what measurements reveal the CP content or the flavor content of the 
neutral $B$. The flavor of a $B$ meson can be tagged by the sign of the lepton charge in 
semileptonic decay, whereas its CP  
can be tagged by using 
$B\to J/\psi K_S$ (CP$=-$) or $B\to J/\psi K_L$ (CP$=+$) decays, 
noting that direct CP violation in these decays is both ${\cal O}(\lambda^2)$
and $\alpha_s$ suppressed, note \cite{Faller:2008zc} for an explicit estimate.
An example  outcome of the experiment
is reproduced in Fig.~\ref{fig:Tviolation}. 
Splendidly, the use of entanglement allows both of the 
reservations~\cite{Wolfenstein:1999TRV,Wolfenstein:2000ju} levied against
the CPLEAR experiment to be set to rest: $A_T$ changes sign with that of 
$\Delta t$, and unitarity does not require particle decay to make $A_T$ nonzero. 
Moreover, in these observables the T and CP transformations are distinct~\cite{:2012kn}. 
Consequently, we 
conclude that BaBar has indeed observed direct T-violation 
 in these reactions. The CP-violating asymmetry is of the same magnitude 
as the T-violating one, so that the outcome is compatible with CPT symmetry. 
Since direct CP violation is possible in the CP tag, this cross-check gives
direct support to our theoretical assessment that penguin pollution in the 
$b\to s c{\bar c}$ modes is small.

\subsubsection{Rare decays \label{par:secErare}}

Rare decays of heavy mesons offer another class of useful ``null'' tests for 
BSM searches. 
The SM predictions tend to be exceeding small, and 
improving the experimental limits on their decay rates sharpens
the constraints on models of physics BSM. We 
refer to Sec.~\ref{sec:chapc} for a discussion
of rare charm decays. 
In this section 
we focus on $B_q\to \mu^+\mu^-$ 
decay, with $q\in (d,s)$, because 
these decays are expected to occur at enhanced rates in the MSSM at large 
$\tan \beta$~\cite{Babu:1999hn}. The decay  
$B_s \to \mu^+ \mu^-$ has recently been observed for the 
first time~\cite{Aaij:2012ac,:2012ct}, at a 
rate compatible with SM expectations. 
As reviewed in~\cite{Gamiz:2013waa}, there
are different ways to compute the $B_q\to \mu^+\mu^-$ decay rates
within the SM, using distinct nonperturbative parameters computed in lattice QCD. 
For example, modern computations of the bag parameters
$\hat{B}_s=1.33 \pm 0.06$ and $\hat{B}_d=1.26 \pm 0.11$~\cite{Gamiz:2009ku}  serve to update 
the SM prediction for the $B_q \to \mu^+\mu^-$ branching ratios; 
specifically, 
${\cal B}(B_q\to \mu^+\mu^-) = [{\rm known}\, {\rm factors}]/\hat{B}_q$. 
Using empirical values of the lifetimes and 
$\Delta \Gamma_s=0.116\pm 0.019\,{\rm ps}^{-1}$, one gets~\cite{Gamiz:2013waa} 
\begin{eqnarray}
{\cal B}(B_s\to\mu^+\mu^-) &=& (3.65\pm 0.20)\times 10^{-9} \,; \\
{\cal B}(B_d\to\mu^+\mu^-) &=& (1.04\pm 0.09)\times 10^{-10} \,.
\end{eqnarray}
We note that the alternate pathway uses the
lattice-QCD meson decay constants $f_{B_q}$ 
and gives branching ratios which are in excellent agreement~\cite{Gamiz:2013waa}. 
The experimental values are~\cite{Aaij:2012ac,:2012ct}  
\begin{eqnarray}
{\cal B}(B_d\to\mu^+\mu^-)                & < &  0.81 \times  10^{-9}\,[90\%\,{\rm C.L.}] \, ; \\ 
{\cal B}(B_s\to\mu^+\mu^-)                & = &  3.2^{+1.5}_{-1.2} \times 10^{-9} \,,
\end{eqnarray}
and the comparison with the SM expectations seems to leave little room for new physics. 
In particular, the MSSM at large $\tan\beta$ is quite constrained~\cite{Haisch:2012re}.
Belle-II will hopefully be able to improve their experimental sensitivity to the extent
that they can probe down to the SM level in both channels. 
Interestingly, the ratio
\begin{equation}
\frac{{\cal B}(B_s\to\mu^+\mu^-)\Delta m_d \tau_d \hat{B}_s}{{\cal B}(B_d\to\mu^+\mu^-)\Delta m_s \tau_s \hat{B}_d}
\end{equation}
still leaves room for significant new physics effects. 
The ATLAS collaboration~\cite{Aad:2012pn} is addressing this, 
employing as a benchmark the well-known $B\to J/\psi K$ decay 
as a reference in order to compute the branching fractions.

To conclude our discussion of flavor physics, we observe that 
all the quark flavor and CP violation currently observed in nature appears to be 
controlled by the CKM matrix~\cite{Isidori:2010kg}. We have considered a broad sweep
of low-energy observables, 
many of which are only statistics limited in their sensitivity, 
and for which theoretical uncertainties are under sufficient control to permit
the discovery of departures from the CKM paradigm and indeed of 
physics BSM. 

\subsection{Future Directions} 

\label{sec:secE9}

Popular models of physics BSM are becoming 
increasingly constrained through null results from direct
searches at collider energies as well as from indirect searches
realized from precision measurements at lower energies. 
This sweep of negative results nevertheless allows us to come to
at least one positive conclusion, for we have established 
beyond doubt that the dominant mechanism of flavor and CP
violation within the quark sector is due to the CKM matrix. 
CP violation may well exist in the neutrino sector as well, and 
with effort we should have the knowledge we need in regards
to the interactions of neutrinos with matter in order to discover
whether it does. We may also have discovered the mechanism by
which elementary particles accrue mass, though it may take
decades to establish whether the couplings of the Higgs are 
as predicted in the SM or not. 
Irrespective of this, and in contrast, continuing 
null (or contradictory) results in regards to particle dark matter
yields no positive conclusion, for dark matter, and dark energy
for that matter, have no explanation within the SM. 
Nevertheless, the astrophysical observations which 
led to their articulation are both robust and concrete. 
There is undoubtedly new physics to explain, and possibly an expansion 
of the SM that we can empirically establish to explain it. 

It is entirely possible that the physics BSM for which we search 
will fit within the context of a model that we know. This means
that the sweep of experiments we have considered are the right ones
and that we need only be able to interpret experiments of 
enhanced sensitivity. We have offered a suite of experimental
observables for which that is the
case. In that class, there are, most transparently, 
 various null tests, such as searches for permanent EDMs, or 
for neutrinoless double-beta decay. In the case of EDMs
we have considered how robust nonperturbative methods in QCD, 
be their origin in lattice QCD or in effective field theory, 
can be used to interpret the experimental results in various
systems if discoveries are ultimately made. In the case of
neutron EDM matrix elements in lattice-QCD 
of nonleading dimension operators, 
the detailed methods  are still under development. 
This theoretical
control also extends to measurements of nonzero quantities
to higher precision, such as that of the anomalous
magnetic moment of the muon, or of the parameters of meson
mixing, or of the neutral weak couplings of the quarks in PVES. 

It is also possible that the explanations we seek will surprise
us, that the BSM 
models of ultimate use have as yet to be invented. 
Since little in regards to dark matter is established, 
this is quite possible and supports broader thinking 
in regards to possible 
experiments~\cite{Graham:2011qk,Graham:2013gfa,Gardner:2008yn,Gardner:2013aiw}. 
Nevertheless, the
nonperturbative tools we have discussed for the control of
QCD will undoubtedly continue to play an important role. 

Although we have illustrated through many examples in a sweep of contexts 
that lattice-QCD can play and has played a key role in the 
search for physics BSM, its utility has nevertheless been limited
to particular classes of problems. That is, it has been restricted to 
systems for which the nonperturbative dynamics can be
captured by the matrix elements of local operators (and typically of low operator
dimension) and for which disconnected insertions, or quark loops, play a minimal
role in the dynamics. Concretely, then, we have used lattice-QCD methods to greatest
effect in the analysis of flavor-changing weak decays to leptonic and 
semileptonic final states. Let us then conclude with a perspective
on the possibility of extending lattice-QCD methods to particle decays with
nonleptonic final states~\cite{Kronfeld:2013lsa}. 
It is worth emphasizing that such a generalization
would be key to the study of systems with enhanced, long-distance effects,
such as $D{\bar D}$ mixing~\cite{Golowich:1998pz}, or the 
study of rescattering effects in hadronic 
$B$ (or $D$ or $K$) decays~\cite{Donoghue:1996hz,Suzuki:1999uc}. Ultimately
the limitations of lattice-QCD in this regard stem not from the use of
discrete spacetime per se, but rather from a famous ``no-go'' 
theorem~\cite{Maiani:1990ca}: it is generally not possible to analytically continue
a 3-point Green function computed in Euclidean space back to Minkowski space. 
A possible resolution to this puzzle relies on the structure of the S-matrix; 
e.g., 
the S-matrix and the energy-levels of two-particle systems at finite volume 
are closely tied~\cite{Luscher:1986pf}. 
An early application of these ideas was to systems
with nearly elastic interactions in the final state~\cite{Lellouch:2000pv}.
Systems with inelastic interactions are more interesting, however, and recently
progress has been made to understand inelastic 
scattering in a finite volume~\cite{Hansen:2012tf,Briceno:2012yi,Doring:2012eu}. 
Such are the first
steps towards the 
complete analysis of nonleptonic decays (or of $D{\bar D}$ mixing) in QCD, 
and we relish such prospects. 

\clearpage
\section[Chapd]{Deconfinement \protect\footnotemark}
\footnotetext{Contributing authors: P.~Foka$^{\dagger}$,  H.~Meyer$^{\dagger}$, R.~Vogt$^{\dagger}$, A.~Vuorinen$^{\dagger}$, P.~Arnold, N.~Brambilla, P.~Christakoglou, P.~Di~Nezza, J.~Erdmenger, Z.~Fodor, M.A.~Janik, A.~Kalweit, D.~Keane, E.~Kiritsis, A.~Mischke, G.~Odyniec}
\label{sec:chapd}

A robust prediction of Quantum Chromodynamics (QCD) is that
at a certain value of temperature (or energy density),
hadronic matter undergoes a transition to
a deconfined state of quarks and gluons,
known as the Quark-Gluon Plasma (QGP).
By now, numerical simulations of lattice QCD have convincingly
shown that this transition is in fact not a true phase transition but
instead a rapid crossover that takes place at temperatures around 160 MeV.
In the same temperature region, chiral symmetry is additionally restored
up to a small explicit breaking due to nonzero quark masses. The physics of these
two conceptually distinct but almost concurrent transitions has been
the subject of intense activity in the theory community. The study of
the transition region has subsequently been extended to nonzero
baryon chemical potential $\mu_B$, corresponding to a nonzero average
value of the net baryon density in the system. Increasing the chemical
potential from zero, the transition may strengthen and eventually
become a first order phase transition, signaling the presence of a
so-called critical point on the QCD phase diagram. An alternative
scenario, potentially without a critical point, is that the
crossover from hadronic to QGP matter becomes broader with
$\mu_B$. The existence of a critical point would
establish a remarkable universality link between QCD matter 
and condensed matter physics. Indeed, a prediction of 
universality is that many properties of
quark matter near the critical point would be the same same as
in a large class of condensed matter systems near their respective
critical points.

Experimentally, heavy-ion collisions make it possible to study
strongly interacting matter under extreme conditions in the laboratory.
Several facilities
contribute to understanding the details of the QCD phase transition,
mapping out different regions of temperature and baryon chemical 
potential in the QCD phase diagram. At the top RHIC and LHC collider 
energies, the produced matter is characterized by very small net baryon 
densities and high temperatures, while future facilities at FAIR and 
NICA are planned to explore the phase diagram at high baryon 
chemical potential and lower temperature.

After the first experimental efforts in the 1970s at LBNL and JINR and 
intense theoretical and experimental research at different facilities and energies
from GSI SIS to BNL AGS and CERN SPS,
an assessment of the SPS program was presented in 2000 
\cite{CernPress,Heinz:2000bk}.
The essence of the assessment, based on the results of half a dozen experiments 
at the SPS \cite{Schukraft:1997mv, Antinori:2003hw, Satz:2004zd}, 
was that a new state of matter was produced in the SPS energy regime, featuring 
some of the most important predicted characteristics of a QGP (thermalization, 
chiral symmetry restoration, deconfinement).
The continuation of the heavy-ion program 
at RHIC at BNL \cite{Arsene:2004fa, Back:2004je, Adams:2005dq, Adcox:2004mh}
and at the CERN SPS  \cite{Specht:2010xu} 
confirmed and further refined the first SPS results. 
A comprehensive analysis of the first years' data from all RHIC experiments 
(BRAHMS \cite{Adamczyk:2003sq}, PHENIX \cite{Adcox:2003zm}, PHOBOS \cite{Back:2003sr}
and STAR \cite{Ackermann:2002ad}) led to an assessment in 2005 \cite{Arsene:2004fa, Back:2004je, Adams:2005dq, Adcox:2004mh, BnlPress}
establishing the existence of the sQGP (where s stands for ``strongly interacting''). 
The produced matter was found to behave like an extremely strongly-interacting, 
almost perfect liquid with minimal shear viscosity, absorbing much of the 
energy of fast partons traversing it \cite{Jacobs:2004qv, Muller:2006ee}.
After the discovery phase for the QGP and its qualitative characterization 
was well under way, the LHC \cite{Evans:2008zzb} took over with a primary 
objective of continuing and expanding the quantitative precision measurements 
begun at RHIC, taking advantage of the much increased energy and luminosity. 
First results \cite{Muller:2012zq} came quickly, confirming the RHIC 
observations and exploring the properties 
of this new state of matter in the higher energy regime.
While ALICE \cite{Aamodt:2008zz} was designed as a dedicated experiment 
to study typical heavy-ion observables 
\cite{Carminati:2004fp,Alessandro:2006yt}, all other LHC experiments, 
ATLAS \cite{Aad:2008zzm}, CMS \cite{Chatrchyan:2008aa} and 
LHCb \cite{Alves:2008zz} also participate in the heavy-ion program, 
contributing to the detailed characterization of the produced matter
(with LHCb taking part in the $p$-nucleus part of the program).

Detailed studies of the QGP produced in nuclear collisions at LHC and RHIC 
have already shown that this new state of matter has unique properties and 
presents challenging questions to theory \cite{Muller:2013dea}.  While theory 
has no complete answers yet, great advances have been made toward 
developing frameworks in which such questions can be addressed. Thus, 
experimental data can be used to clarify those properties of hot QCD matter 
that  cannot yet be reliably predicted by QCD.
A particular problem hindering the theoretical interpretation of the 
experimental results is the extremely rapid and complex dynamical 
evolution of the produced system.
Typically, instead of a microscopic theory,
effective descriptions are employed,
ranging from relativistic hydrodynamics to Monte-Carlo transport simulations and
simplified models.

Despite these challenges, the field is currently
advancing towards a ``standard model of heavy-ion collisions''. The
initial collision of the two nuclei is thought to result in the
formation of a dense, nonequilibrium QCD plasma
which rapidly thermalizes. The expansion and cooling of the
near-thermal QGP is described by hydrodynamics until thermal freezeout
produces a hadronic resonance gas. At this point, although the
chemical composition of the produced particles is approximately fixed
(chemical freezeout), the spectral distributions still evolve until
kinetic freezeout.
As a way to test the emerging qualitative picture, a number of
experimental observables have been employed to probe the properties of
the produced medium as well as the space-time evolution of the
system.
A non-exhaustive list of experimental observables, related to the properties 
of the QGP that we expect to determine from these studies, can be summarized 
as follows \cite{Muller:2013dea}: (i) The equation of state of the produced 
matter is reflected in the spectra of the emitted particles and lattice QCD 
can reliably compute these quantities. (ii) Microscopic properties, such as the 
QGP transport coefficients, are related to the final-state flow pattern and 
the energy loss of high-$p_T$ partons. Those include the shear viscosity,
the coefficient ($\hat{q}$) governing the transverse momentum diffusion of a fast parton,
the coefficient of linear energy loss, and
the diffusion coefficient of a heavy quark in matter. 
Currently, lattice gauge theory cannot reliably calculate these dynamical 
quantities.
(iii) The dissolution of bound states of heavy quarks in the QGP is governed
by static color screening, which can be reliably calculated on the lattice.
(iv) The electromagnetic response function of the QGP is reflected in the
emission of thermal photons and lepton pairs. While it is difficult to 
calculate this dynamical quantity on the lattice, some progress has been made recently.

This interplay of theory and experiment, as well as the complementarity 
between different approaches, particularly essential for advances 
in the heavy-ion field, is reflected in this chapter.  
We review recent progress in the study of the deconfined phase of QCD, on 
both the theoretical \cite{Muller:2013dea} and experimental sides 
\cite{Schukraft:2013wba}, pointing out current challenges and open questions. 
Thus we mostly present recent advances from the LHC era, not attempting 
a review of the field. The review of first results at LHC  \cite{Muller:2012zq}, 
followed as a basis, is also a source of primary literature. The material was 
updated following the fast progress reported at major conferences, from 
QM2012 \cite{QM2012web} to more recent ones \cite{hp2013}. In anticipation 
of new interesting results presented at QM2014 \cite{QM2014web}, the reader 
is referred to the upcoming presentations and publications.

In this chapter, we concentrate on finite temperatures, leaving the case of 
cold and dense (nuclear) matter to Chap.~\ref{sec:chapf}. 
We begin by reviewing what is known about the equilibrium properties of the
theory, in particular the part of the phase diagram explored by lattice QCD
calculations in Sec.~\ref{sec:d:MapQcdPha}. In connection with the phase
diagram, we describe the status of the Beam Energy Scan (BES) at RHIC
and briefly touch upon ``event-by-event'' studies which employ
fluctuations and correlations to search for critical behavior.
From low transverse momentum particles, we can
infer the bulk properties of the created matter and
the dynamical evolution of the system. The main
aspects of the hydrodynamic description of a near-thermal QGP are
reviewed in Sec.~\ref{sec:d:NeaEquPro} together with experimental
results on the bulk properties and collective behavior of the system.
Our current theoretical
understanding of the different stages of the collision prior to the
formation of the QGP is discussed in Sec.~\ref{sec:d:TheWeaStr}. 
Experimental results and our current theoretical understanding of the particle
multiplicity and entropy production are discussed in Sec.~\ref{sec:5.new}.
The high energies and luminosities of modern colliders, in particular
the LHC, allow detailed studies of ``hard probes''. These are produced
by hard scatterings at early times during the initial stage of the
collision ($t\sim 1/Q$) and can therefore be regarded as external
probes of the nature and properties of the QGP. The current status of
the theoretical and experimental efforts concerning these probes is
reviewed in Sec.~\ref{sec:HardAndHeavy}. We begin with an
introduction to the theory of hard probes, starting with nuclear matter
effects which provide the baseline for understanding the modification
of these probes in hot matter. We provide brief theoretical overviews
of energy loss in hot matter and of quarkonium suppression. We then turn
to recent experimental results on high-$p_{T}$ particle and jet
production as well as heavy-flavor production. Recent results on \pA\ 
collisions, studied in order to disentangle initial- from final-state effects, 
are discussed in Sec.~\ref{chapd:pPbEXP}.
In addition to lattice QCD, which is best suited for the
regime of low baryon density and static observables, theoretical
frameworks have been developed to address the dynamical properties of
the QGP. 
As an alternative to weak-coupling methods, strong-coupling calculations
involving gauge/gravity duality have provided a different paradigm
for the QGP studies at the temperatures explored in heavy-ion collisions.
More generally, a number of effective field theories (EFTs) have been
developed in the last decades to address different physical regimes 
and observables: Hard Thermal Loop (HTL) EFT, Electrostatic QCD (EQCD), 
Magnetostatic QCD (MQCD), Hard Thermal Loop NRQCD and Hard Thermal Loop 
(pNRQCD). They establish the link between perturbative calculations and 
strong-coupling calculations and allow precise definition and systematic 
calculation of quantities of great physical impact (such as the heavy 
quark-antiquark potential at finite temperature).
In Sec.~\ref{sec:d:LatQcdAds}, we compare and contrast several results 
for bulk thermodynamics and transport quantities computed within these
frameworks. In Sec.~\ref{sec:d:ImpTheFie}, we discuss recent progress in
thermal field theory calculations in the context of hot matter in the early 
universe -- a closely-related area where progress is often
directly tied to advances in heavy-ion physics. Finally, in 
Sec.~\ref{sec:d:ChiMagEff} we present experimental results on the 
chiral magnetic effect, 
while a theoretical review of this phenomenon is given in 
Sec.~\ref{sec:d:ChiMagEff}. 
We end with a discussion of open questions and future directions 
for the field in Sec.~\ref{sec:5.9}.


\subsection{Mapping the QCD phase diagram} \label{sec:d:MapQcdPha}

The QCD phase diagram as a function of temperature $T$ and baryon
chemical potential $\mu_{B}$ is expected to have a rich structure. In
this section, we discuss the bulk properties of quark matter 
in the region of small to moderate baryon chemical potential, 
$0\leq \mu_B\lesssim 1\,{\rm GeV}$, which can be explored
experimentally in heavy-ion collisions. In particular, lattice QCD allows for first
principles calculations of equilibrium quantities at $\mu_B=0$.  
To extend these studies
to moderate values of the baryon chemical potential, 
$\mu_B \lesssim 3T$,
various methods have been recently used ~\cite{Philipsen:2011zx}.
For the phase structure at higher baryon densities
see Sec.~\ref{critpt_th} and Chapter~\ref{sec:chapf}. 

The equation of state is an important input in 
the hydrodynamic calculations that have been successful
in describing the evolution of the expanding matter 
created in relativistic heavy-ion collisions (see Sec.~\ref{sec:d:NeaEquPro}). 
Quantifying the equation of state and the associated quark number
susceptibilities~\cite{Borsanyi:2011sw,Bazavov:2012jq} below the
transition temperature is important for testing the freezeout
mechanism and the Hadron Resonance Gas
\cite{BraunMunzinger:2011ta} description of hadronic matter up to the
crossover temperature.  The cumulants of the quark number
distributions also provide information about the presence of a
critical point in the QCD phase diagram if a sufficient number of them
are known \cite{Borsanyi:2011sw,Bazavov:2012jq}; see Sec.
\ref{critpt_th}.

While educated guesses as to the qualitative behavior of the equation
of state have been around for a long time, it has been determined with
precision on the lattice~\cite{Borsanyi:2010cj} only in the last five
years. 
At low temperatures, the matter can be
described in terms of a dilute hadron gas.  
The passage from a bulk hadronic state at low temperatures to
a quark-gluon plasma phase at high temperatures was found to be an
analytic crossover in lattice QCD calculations~\cite{Aoki:2006we}.
A rapid rise of the
entropy density, $s$, occurs around a temperature of $\sim
160$~MeV~\cite{Borsanyi:2010cj}.  This can be
interpreted as a transition to partonic degrees of freedom.  Above
400~MeV, $s/T^3$ has weak temperature dependence and is expected
to reach the Stefan-Boltzmann limit at asymptotically high temperatures.
The fact that lattice data are still below the ideal gas limit is an indication that
interactions are still important at high $p_T$.  
Agreement with the perturbative equation of state has been
established at high temperatures in the (numerically less demanding) 
pure gluon
plasma~\cite{Borsanyi:2012ve}.  Some recent results on the equation of state
 and the quark number susceptibilities are discussed
below in Sec.~\ref{latticemuB0}.  We refer the reader
to~\cite{DeTar:2009ef} for a more complete introduction to
finite-temperature lattice calculations.

At vanishing baryon chemical potential, the integrand in the standard
path integral expression for the QCD partition function is real and positive 
once the quark fields are integrated out analytically. 
This integrand can therefore be interpreted as a probability distribution for 
the gluon fields. The high-dimensional integral can then be estimated by 
importance-sampling Monte-Carlo methods: the gluon fields are sampled in 
such a way that  the probability of occurrence of a field configuration 
is proportional to the value 
of the integrand evaluated on that configuration.
When nonzero baryon chemical potential is introduced on the lattice,
the integrand becomes complex.  In this case, Monte-Carlo methods based
on the importance sampling of field configurations no longer apply. 
The phase of the integrand can be absorbed into the observables, but 
its fluctuations from configuration to configuration lead to uncontrollably 
large cancellations. This numerical challenge is known as the
``sign'' problem.   It is only recently that ways of overcoming
this difficulty have been developed, including overlap-improving
multi-parameter reweighting \cite{Fodor:2001pe,
  Fodor:2001au,Fodor:2002sd}, Taylor expansion \cite{Allton:2002zi}
and analytic continuation from imaginary to real chemical potential
\cite{deForcrand:2002ci}.  While the transition initially exhibits
little sensitivity to the baryon chemical potential, some of these
calculations suggest that the phase transition is no longer a
crossover beyond a certain critical value of $\mu_B$, but instead
becomes a first order transition (Sec.\ \ref{critpt_th}).  There is
strong experimental interest in discovering this critical point.
Recent studies are described in Sec.~\ref{expBES}.

\subsubsection{Precision lattice QCD calculations at finite-temperature}  
\label{latticemuB0}

In precision lattice QCD calculations, two aspects are particularly
important. First of all, physical quark masses should be used.  While
it is relatively easy to reach the physical value of the strange quark
mass, $m_s$, in present day lattice simulations, it is much more
difficult to work with physical up and down quark masses $m_{u,d}$,
because they are much smaller: $m_s/m_{u,d} \approx 28$ (Chapter~\ref{sec:chapb}). In
calculations with $m_s/m_{u,d} < 28$, the strange quark mass is
usually tuned to its approximate physical value while the average up
and down quark masses are larger than their physical values. Second,
the characteristics of the thermal transition are known to suffer from
discretization errors~\cite{Borsanyi:2010cj,Bazavov:2009zn}. The only
way to eliminate these errors is to take smaller and smaller lattice
spacings and systematically extrapolate to vanishing lattice spacing
(and thus to the continuum limit).
It is computationally very demanding to fulfill both conditions. There are 
only a few cases for which this has been achieved. Within the staggered 
formalism of lattice QCD (see for instance \cite{DeTar:2009ef} for a description
of different lattice fermion actions), there are full results 
on quantities such as 
the nature of the transition \cite{Aoki:2006we}, the transition 
temperature for vanishing and small chemical potential 
\cite{Aoki:2006br,Borsanyi:2010bp,Bazavov:2011nk,Endrodi:2011gv}, 
the equation of state \cite{Borsanyi:2010cj} and 
fluctuations \cite{Borsanyi:2011sw,Bazavov:2012jq}.

\begin{figure}\begin{center}
\includegraphics[width=0.45\textwidth]{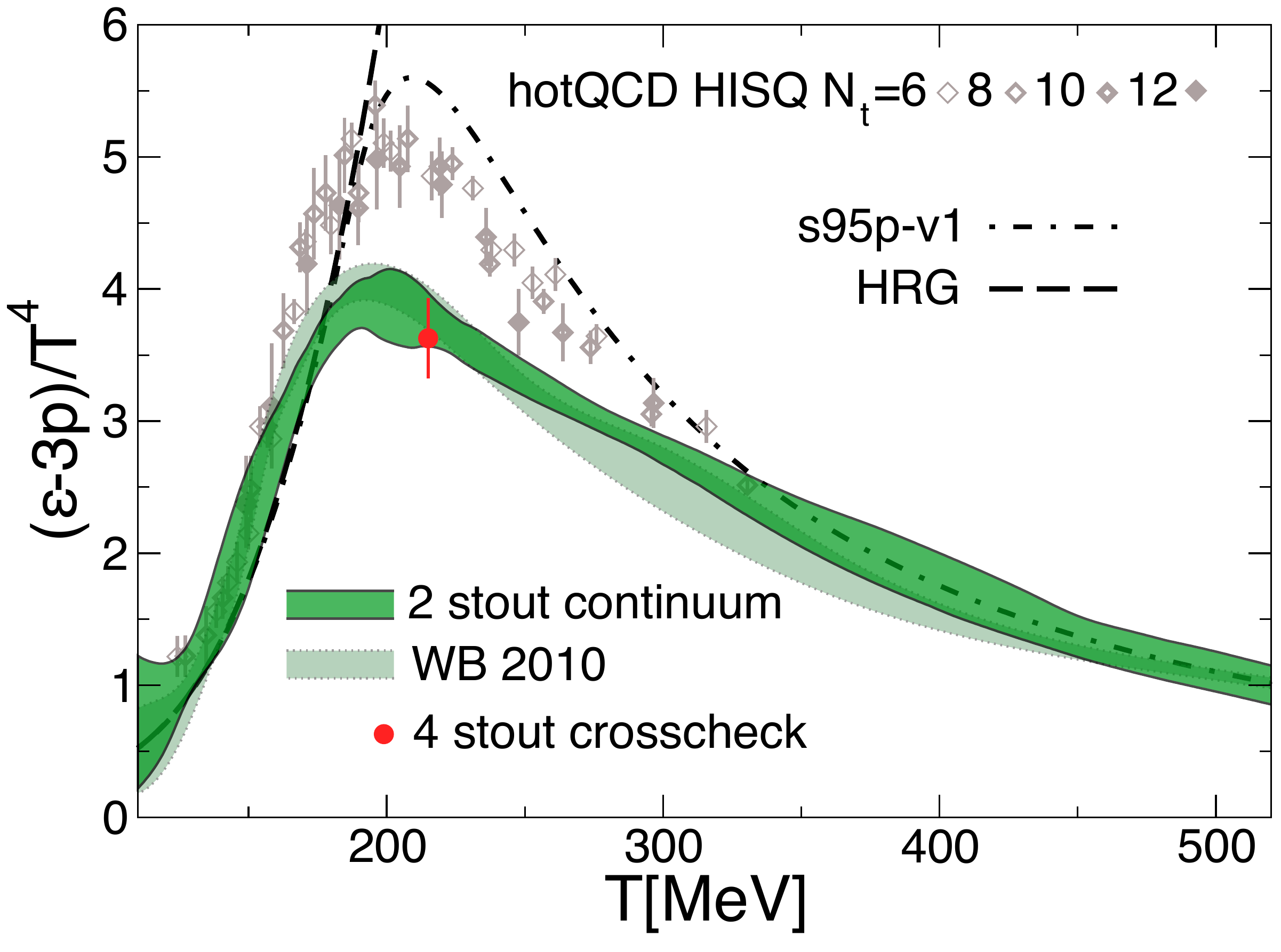}
\hspace*{2mm}
\includegraphics[width=0.45\textwidth,viewport=70 29 400 303,clip]{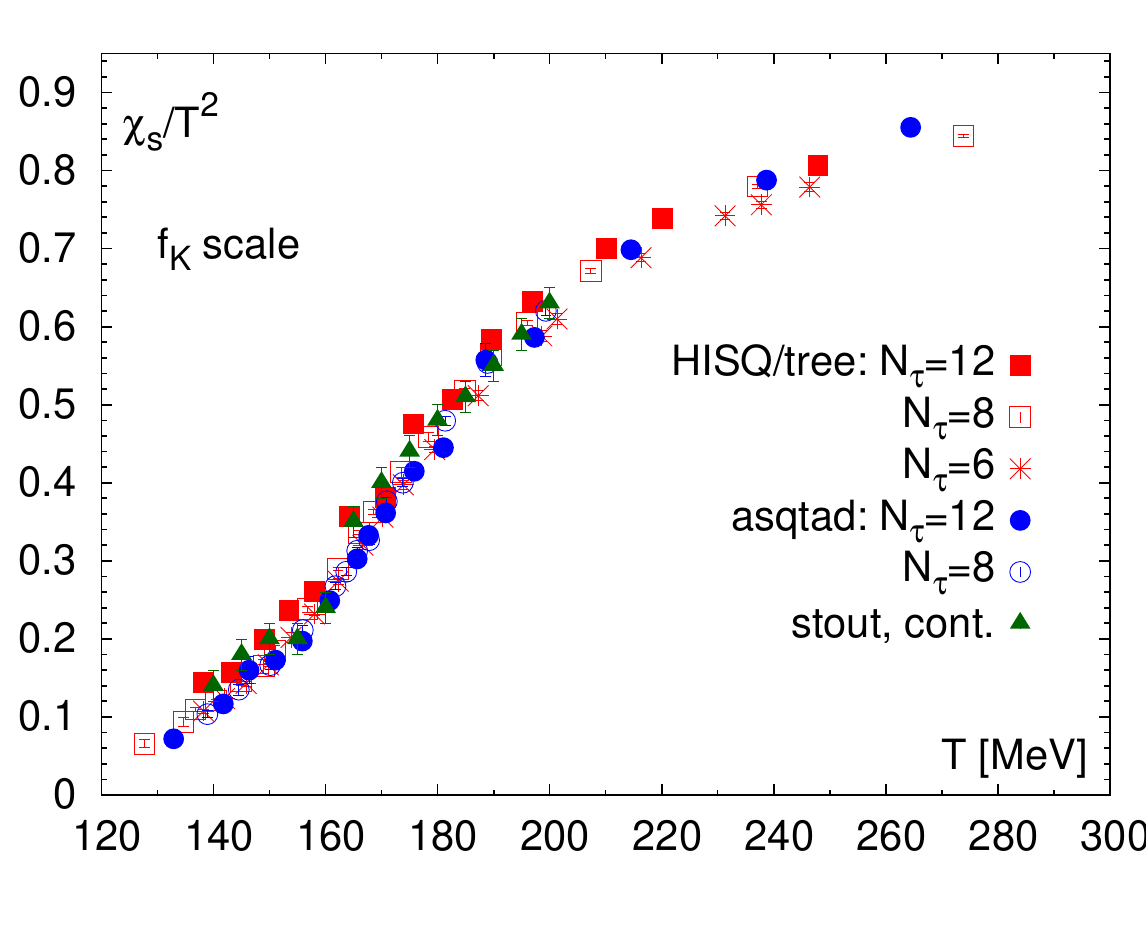}
\caption{\label{fodor:fig}
(Top) Comparison of the equations of state obtained by the Wuppertal-Budapest group (shaded region) and the HotQCD Collaboration (points). 
There is still a sizable discrepancy between the results. From \cite{Borsanyi:2013bia}.
(Bottom) The strange quark susceptibilities calculated by the two groups. In the continuum limit, the results agree. From \cite{Bazavov:2011nk}.}
\end{center}\end{figure}

\vskip 3 mm
\noindent
\paragraph{Status of the equation of state}~ The first step in obtaining any 
trustworthy result in QCD thermodynamics is to determine the temperature 
of the QCD transition. Its value was disputed for some years, but 
it is a great success for the field of lattice QCD that the results 
from two independent groups using different lattice discretizations 
now completely agree \cite{Aoki:2006br,Borsanyi:2010bp,Bazavov:2011nk}. 
Since the transition is a crossover, 
the precise value of the transition temperature depends on the chosen definition, but a typical
value based on the chiral condensate and the associated susceptibility
is 155 MeV with a (combined statistical and systematic) uncertainty of $\sim 3 \%$.
The next important 
step is the determination of the equation of state. There are various 
calculations with different fermion formulations, see Ref.~\cite{Umeda:2012er} for 
a calculation using Wilson fermions.  The current most precise 
results have been obtained with staggered quarks.
In these calculations, the light and strange quark masses take their 
(approximate) physical values. There is still a discrepancy in the 
equation of state in the literature. The Wuppertal-Budapest group 
obtained \cite{Aoki:2005vt} a peak value of the trace anomaly
at $(\epsilon-3p)/T^4 \sim 4$, confirmed later in ~\cite{Borsanyi:2010cj}. 
The HotQCD Collaboration typically finds higher values for the peak value of 
the trace anomaly, see Ref.~\cite{Petreczky:2012gi}. The top panel of 
Fig.~\ref{fodor:fig} compares the results from the two groups. Still more work is 
needed to clarify the source of the difference. 

\vskip 3 mm
\noindent
\paragraph{Susceptibilities from lattice QCD}~ Fluctuations and correlations of
conserved charges are important probes of various aspects of
deconfinement. This is because fluctuations of conserved charges are
sensitive to the underlying degrees of freedom which could be hadronic
(in the low-temperature phase) or partonic (in the high-temperature
phase). Fluctuations of conserved charges have primarily been studied using
different staggered actions. The two most complete calculations have been
carried out by the Wuppertal-Budapest group and by the HotQCD
Collaboration \cite{Borsanyi:2011sw,Borsanyi:2013hza,Bazavov:2012jq,Bazavov:2012vg}. 
The bottom panel of Fig.~\ref{fodor:fig} compares results on the strange quark
number susceptibility.
The fluctuations are small at low temperatures because strangeness is
carried by massive strange hadrons (primarily by kaons). This region is 
described by the Hadron Resonance Gas model \cite{BraunMunzinger:2011ta}.  
Strangeness fluctuations rise sharply through the transition region, as the
strange quarks are no longer bound. At the highest temperatures shown,
the susceptibility approaches unity.

The strange quark susceptibility has been determined to high precision. 
Other quantities and, in particular, higher cumulants are under investigation 
by many lattice groups.  High quality results are expected in the near future.

\begin{figure}\begin{center}
\includegraphics[width=6.5cm]{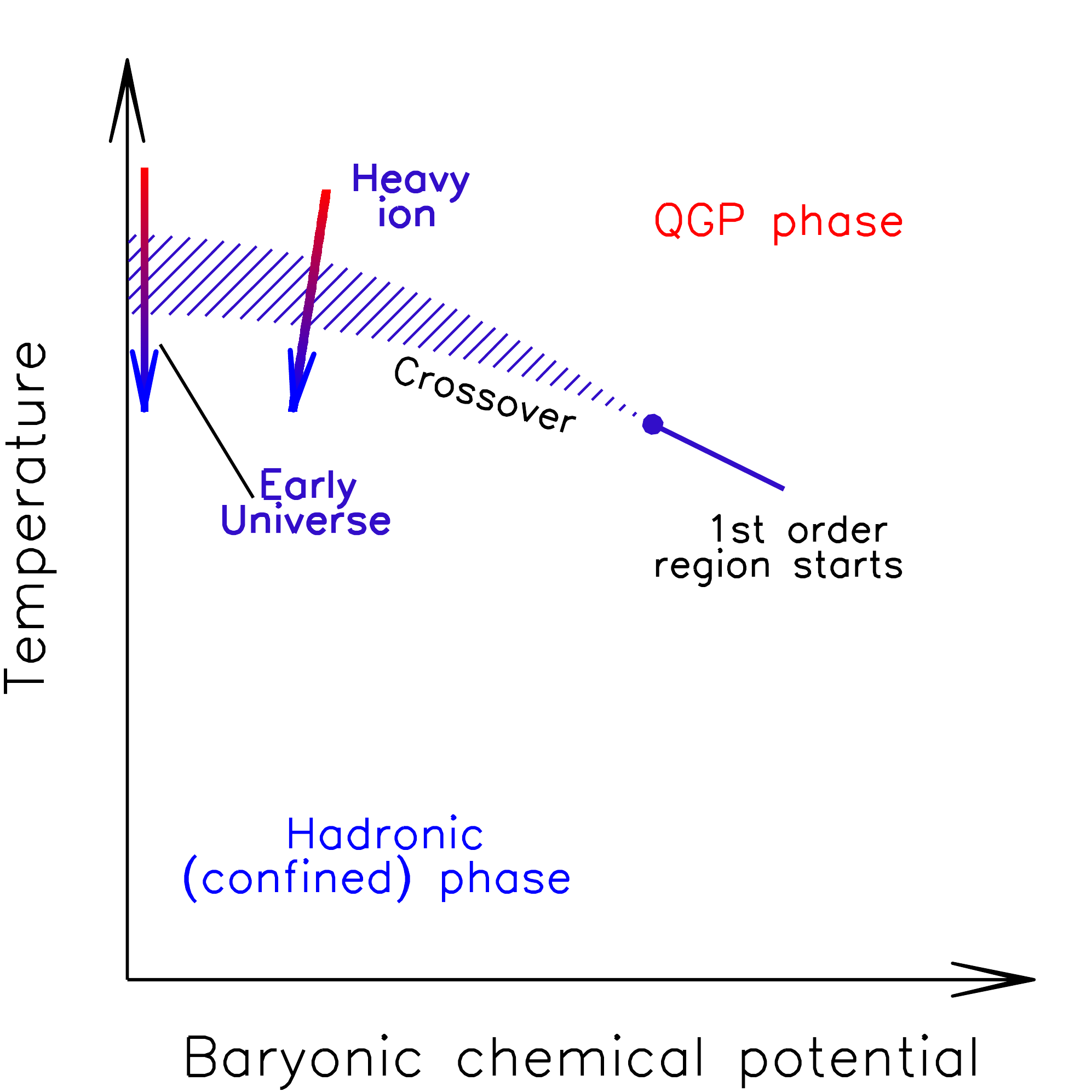}
\hspace*{5mm}
\includegraphics[width=6.5cm]{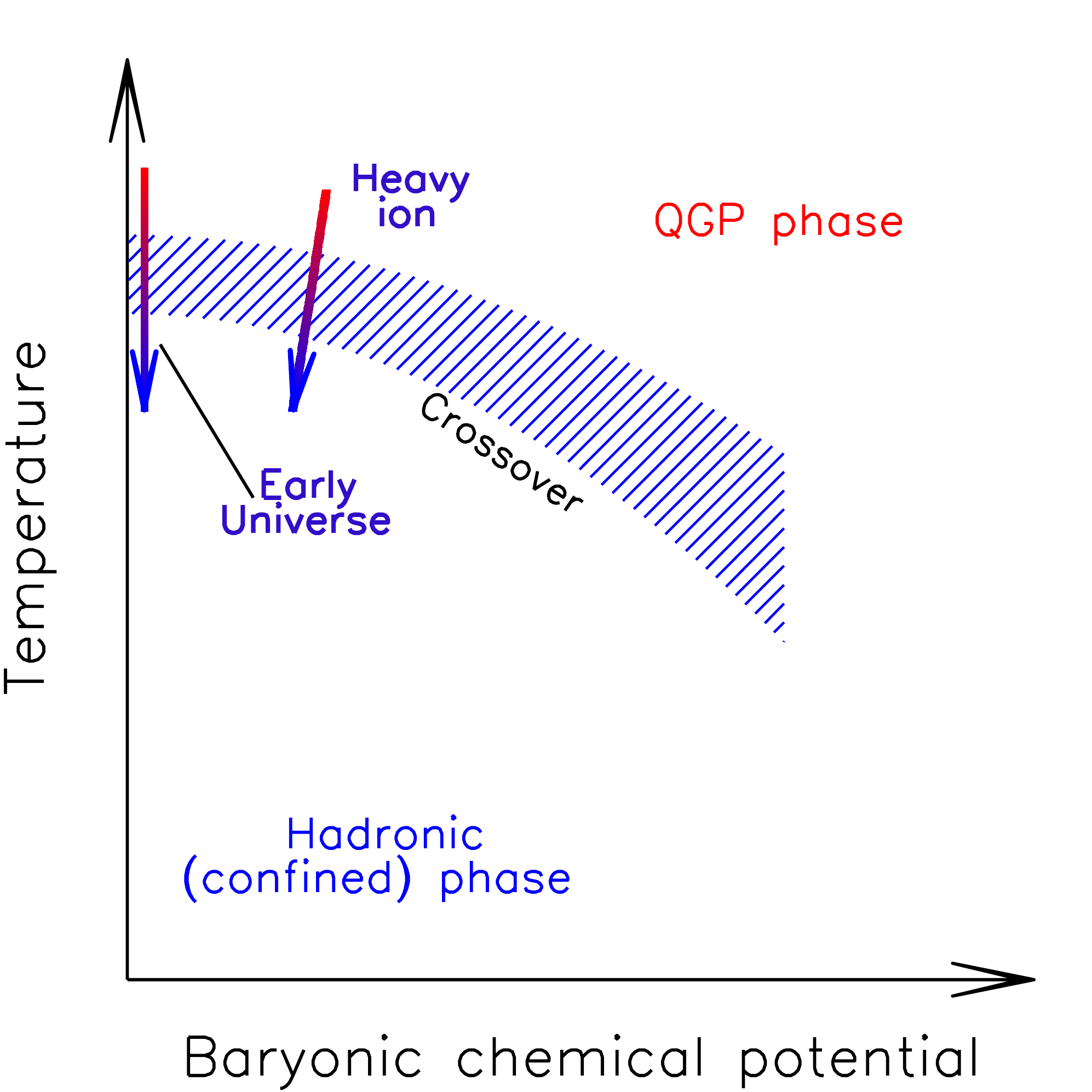}
\caption{\label{phasediag}
Two scenarios for the phase diagram of QCD for small to moderate baryon chemical potential $\mu_B$. In the upper panel, the phase diagram contains a critical point in this region, while in the bottom it does not. From \cite{Endrodi:2011gv}.}
\end{center}\end{figure}

\subsubsection{A critical point in the QCD phase diagram? \label{critpt_th}}

A number of interesting properties of QCD matter follow from the
assumption that a critical point exists in the phase diagram.  
Two scenarios for the phase diagram of QCD for small to moderate baryon chemical potential $\mu_B$
are presented in Fig.~\ref{phasediag}.
In the first, the phase diagram contains 
a critical point in this region, while in the second it does not.
The critical point is the end of a line of first order transition and, as such, is similar to
the critical point in the water-vapor system.  The universality class
is the three-dimensional Ising model class, Z(2).  The dynamic
universality class is that of model `H' of the classification 
\cite{Hohenberg:1977ym},
corresponding to the liquid-gas phase transition~\cite{Son:2004iv}.


Does the critical point exist? There is no firm answer yet from
the theory side. Chiral models remain inconclusive,
see~\cite{Fukushima:2010pp} for a recent discussion and references
therein.  Two kinds of lattice results speak in favor of it. One is
the reweighting result \cite{Fodor:2004nz}, obtained on a coarse
lattice.  The other is the Taylor expansion of the pressure
\cite{Gavai:2008zr,Datta:2012pj}. When all Taylor coefficients have
the same sign, a radius of convergence can be estimated which gives
the location of the critical point.  However  
a large number of terms are needed to convincingly
establish the existence of the critical
point~\cite{York:2011km,Karsch:2010hm}.

What speaks against a critical point relatively close to the $\mu_B=0$
axis is the study of the width of the transition region as a function
of $\mu_B$ using a Taylor series around $\mu_B=0$
\cite{Endrodi:2011gv}. It shows that the width is initially
practically independent of $\mu_B$. This result goes in the same
direction as the study of de Forcrand and
Philipsen~\cite{deForcrand:2006pv,deForcrand:2008vr}, who tracked the
chiral critical surface in the parameter space of light and strange
quark masses and the chemical potential, $(m_l,m_s,\mu_B)$.  A point
on the surface corresponds to a set of parameters for which the
thermal phase transition is second order. In the plane $\mu_B=0$, a
critical line separates the origin (where the transition is first
order) from the point of physical quark masses (where the transition
is a crossover).  At small $\mu_B$ they showed that the critical
surface recedes away from the point $(m_l^{\rm phys},m_s^{\rm
  phys},\mu_B)$ indicating that at physical quark masses the
transition becomes weaker upon switching on a small chemical
potential.  It is not excluded however that the chiral critical
surface $(m_l^{\rm crit}(\mu_B),m_s^{\rm crit}(\mu_B),\mu_B)$ bends
over again.  The critical point would be given by the conditions
$m_{l,s}^{\rm crit}(\mu_B^{\rm crit}) = m_{l,s}^{\rm phys}$.  These
results both suggest that, if a critical point exists, it lies beyond
about $\mu_B\simeq 500\,{\rm MeV}$~\cite{Endrodi:2011gv,Philipsen:2011zx}.

\subsubsection{Experimental exploration of the QCD phase diagram}
\label{expBES}


By varying $\sqrt{s_{NN}}$ in heavy-ion
reactions, experiments can scan a large region of the phase diagram.
The systems created at different values of $\sqrt{s_{NN}}$ have
different trajectories in the $T-\mu_{B}$ plane and may pass through
the critical point.
There have been two experimental programs so far to search for the
critical point and signatures of a phase transition.  Both programs
employ an energy scan over a region of relatively low center of mass
energies.  

The first such systematic study was
performed within the CERN SPS beam energy scan program between
1998 and 2002.  This scan, covering five values of $E_{\rm beam}$,
was primarily undertaken by the NA49 \cite{Afanasev:1999iu} experiment with participation
from NA45 and NA57 \cite{SPS}.  This program is currently being extended by NA49's 
successor, NA61.  After finishing the \ppcoll\ and Be+Be measurements, 
data will be taken with the larger systems Ar+Ca 
and Xe+La.
The second program, currently active, is the beam energy scan program at RHIC.  
The STAR collaboration \cite{Ackermann:2002ad} is, as described below, taking data over a similar 
$\sqrt{s_{NN}}$ range as that of the SPS.  As a collider experiment, STAR 
has the advantage that its acceptance around midrapidity 
does not depend on $\sqrt{s_{NN}}$. 
The PHENIX collaboration has placed its emphasis on higher energies,  
$\sqrt{s_{NN}} \geq 39$ GeV \cite{Adcox:2003zm}.  

The NA49 experiment at the SPS carried out, in fixed-target mode, the
first beam energy scan at energies ranging from $\sqrt{s_{NN}} = 17.2$
GeV down to $6.2$ GeV
\cite{Alt:2007hk,Alt:2007jq,Anticic:2008aa,Alt:2008ab,Anticic:2011am,Anticic:2012cf}.
The NA49 collaboration has published various inclusive measurements which 
they have interpreted as hinting at the onset of deconfinement near 
$\sqrt{s_{NN}} = 7.7$ GeV. These measurements include, among others, the ``horn" 
effect, which is a local peak in the $K/\pi$ ratio as a function of $E_{\rm beam}$, and the ``dale" 
phenomenon, which is a minimum in the width of the 
pion rapidity density, compared to a reference expectation, as a function of $E_{\rm beam}$ 
\cite{Alt:2007hk,Alt:2007jq,Anticic:2008aa,Alt:2008ab,Anticic:2011am,Anticic:2012cf}.
This SPS-based program will be taken over by NA61 experiment.  

Establishing whether or not a critical point exists is a top priority.
The divergence of susceptibilities of conserved quantities such as 
baryon number, charge, and strangeness at the critical point 
translate into critical fluctuations in the multiplicity distributions 
and can be studied experimentally \cite{Stephanov:2008qz,Stephanov:2011pb}. 
Generally speaking, one is looking for a qualitative change in these
observables as a function of baryon chemical potential $\mu_B$. 
Therefore, experimental studies focus on the behavior of multiplicity 
fluctuation-related observables in small steps of beam energy. At first, experimental investigations were limited to the 
second moments of multiplicity distributions, which are proportional to the square of the correlation 
length $\xi$ \cite{Stephanov:2008qz}. In heavy-ion collisions, 
the latter is estimated to be small, $\sim$ 2--3 fm \cite{Berdnikov:1999ph}, 
in the vicinity of a critical point. Therefore, the higher moments of 
event-by-event multiplicity distributions are preferred; the higher 
the order of the moment, the more sensitive it is to the correlation length 
of the system, e.g., the third moment (skewness) $S\sim\xi{^{4.5}}$ and the 
fourth moment (kurtosis) $\kappa{^2}\sim\xi{^7}$ \cite{Stephanov:2008qz}.  
Measurements of higher moments of event-by-event identified-particle 
multiplicity distributions, and their variation with centrality and beam 
energy, provide a direct link between experimental observables 
and lattice QCD calculations.

The exploratory phase, Phase I, of the Beam Energy Scan (BES) program at RHIC was completed in 2011, with data taken at $\sqrt{s_{NN}} = 39$, 27, 19.6, 11.5 and 7.7 GeV. All data taken by the STAR detector below the RHIC injection energy $\sim 20$ GeV are affected by large statistical errors, increasing steeply with decreasing energy. Together with larger data sets at 62, 130 and 200 GeV, these measurements provided an initial look into the uncharted territory of the QCD phase diagram.  

The BES program goals \cite{Aggarwal:2010cw} are focused on three
areas.  The first, and least complicated, is a scan of
the phase diagram at different $\sqrt{s_{NN}}$ to vary the values of 
$\mu_B$ and $T$ to determine at which energy (if any) the
key QGP signatures reported at the highest RHIC energies 
\cite{Adams:2005dq, Adcox:2004mh} are no longer observed.  
The disappearance of a single QGP signature as the energy is decreased
would not be convincing evidence
that the border between confinement and deconfinement has been reached at that energy
since other phenomena, unrelated to
deconfinement, could result in similar effects, or else the sensitivity to
the particular signature could be reduced at lower energies.
However, the modification or disappearance of several signatures
simultaneously would constitute a more compelling case. 
  
A second goal is the search for critical fluctuations, e.g., measured in 
net-proton multiplicity distributions, associated with
a strong increase in various susceptibilities, expected in the
vicinity of a critical endpoint. However, finite size effects tend to
wash out this critical behavior, making it difficult to predict 
the signatures of the critical fluctuations quantitatively.

A third proposed goal is to find evidence of the softening of the equation 
of state as the system enters a
mixed phase (such as a speed of sound in medium well below the ideal
$1/\sqrt{3}$).  Promising observables in this search
include the directed flow $v_1$ and elliptic flow $v_2$ (i.e., the first and second Fourier 
coefficients for the azimuthal anisotropy relative to the reaction plane; 
see Sec. \ref{chapd:azimuthal_ani} for a more complete discussion),  
and these flow measurements are for charged particles as well as identified
protons, net protons, and pions.  Other relevant measurements are 
azimuthally-sensitive 
particle correlations. 

The STAR BES Phase I results discussed below
\cite{Kumar:2012fb,Pandit:2012mq,McDonald:2012ts,Das:2012yq} 
have made it possible to close in on some of the goals outlined
above. It is very encouraging that the performance of both the collider 
and the experiments was excellent throughout the entire energy range explored 
to date. Phase I Energy Scan data allowed STAR to extend the $\mu_B$ range of 
RHIC from a few tens of MeV up to $\sim 400$~MeV. The critical region in 
$\mu_B$ has been predicted to span an interval of 50 to 100 MeV 
\cite{Gavai:2008zr,Gavai:2004sd,Gavai:2010zn,Gupta:2009mu,Asakawa:2008ti,Costa:2008gr,Costa:2007ie}. 

As to the first goal, the violation of constituent quark
number scaling and the disappearance of high $p_{\rm T}$ hadron suppression
\cite{Kumar:2012fb,Pandit:2012mq,McDonald:2012ts,Das:2012yq} suggest
that hadronic interactions dominate over partonic interactions
when the collision energy is decreased below the measured energy point 
at $\sqrt{s_{NN}} = 11.5$ GeV.

\begin{figure} 
\begin{center}
\includegraphics[width=0.50\textwidth]{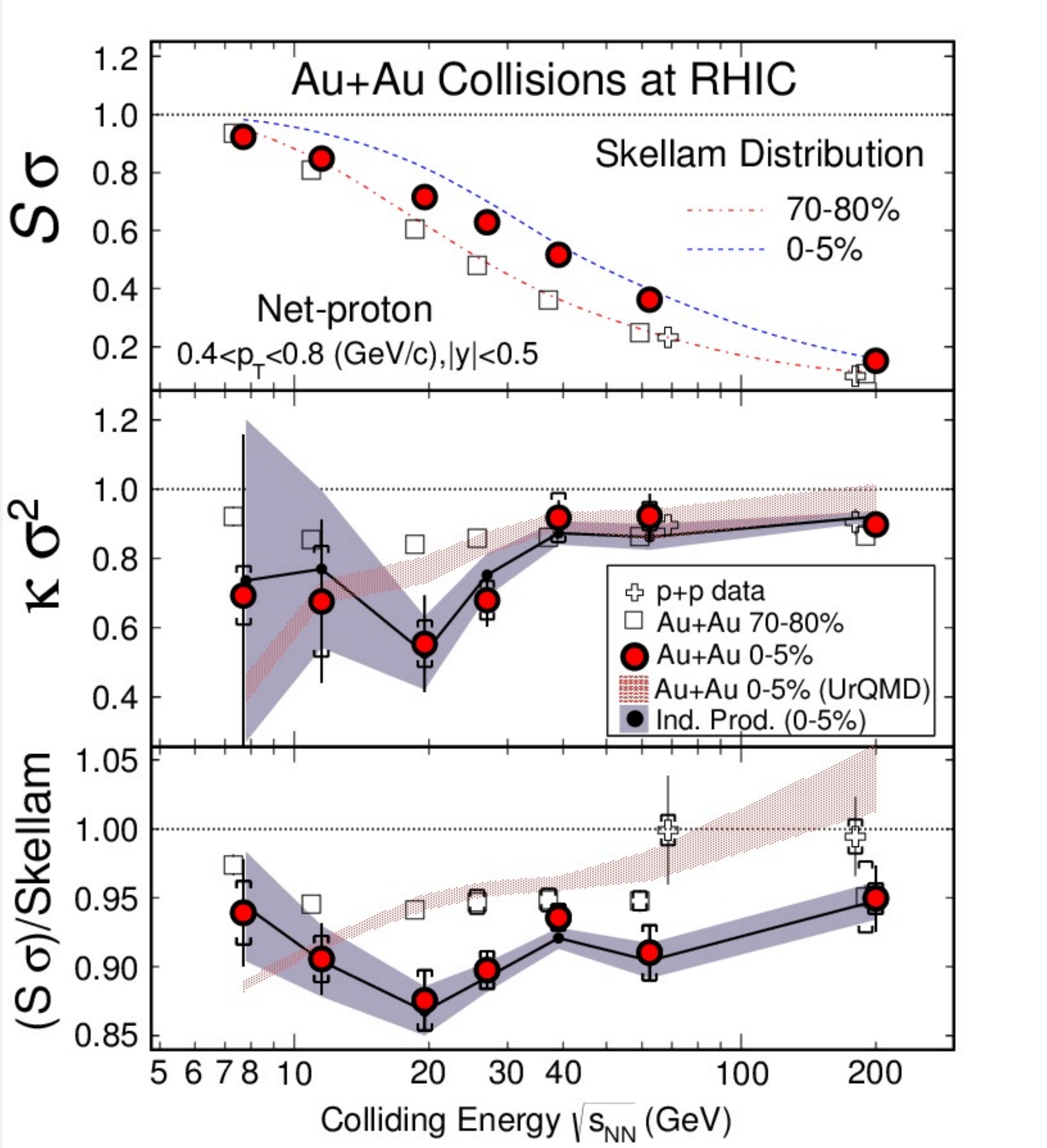}
 \caption{STAR's measurements of $\kappa\sigma^2$, $S\sigma$ and $S\sigma/$Skellam as a function of beam energy, at two different centralities. A Skellam distribution is the difference between two independent Poisson distributions \cite{Adamczyk:2013dal}. Results from \pp\ collisions are also shown. One shaded band is an expectation based on assuming independent proton and antiproton production, and the other shaded band is based on the UrQMD model. From \cite{Adamczyk:2013dal}.}
\label{HighMoments:fig}
\end{center}
\end{figure}

In order to address the second goal, higher 
moments of the net-proton distribution (a proxy for net-baryon number) 
are considered to be the best suited observables in the search 
for a critical point. Figure~\ref{HighMoments:fig} demonstrates that the 
measurements from BES Phase I do not deviate from expectations 
based on assuming independent production of protons and antiprotons 
\cite{Adamczyk:2013dal}. However, there is a considerable gap in $\mu_B$, of 
the order of 110 MeV, between the beam energy points at 11.5 GeV and 19.6 GeV. 
Based on common estimates of the extent of the critical region in $\mu_B$, which 
could well be of the same order, it is a valid 
concern that BES Phase I could have missed it.  Therefore, at the beginning of 2014, the STAR collaboration 
started to run Au+Au collisions at 14.6 GeV. 

In terms of the third goal, 
the first signals of possible softening of the equation of state
were also observed.  In particular the directed flow of protons 
and net protons within $7.7 < \sqrt{s_{NN}} < 200$~GeV 
\cite{Kumar:2012fb,Pandit:2012mq,McDonald:2012ts,Das:2012yq} bears a striking
similarity to hydrodynamic simulations with a first-order phase
transition \cite{Stoecker:2004qu}. The implications of these
measurements for understanding the QCD phase structure are however not yet 
resolved.

The statistics collected during Phase I of BES are insufficient for 
final conclusions on the program goals.  
Therefore, STAR proposed precision measurements in Phase II to map out the 
QCD phase diagram with an order of magnitude increase in statistics,
planned around 2018 and 2019. 

There is also a plan to run STAR in fixed-target mode
concurrently with collider mode during BES Phase II. 
With a fixed-target program in STAR, the range of accessible values of baryon 
chemical potential could be extended from $\mu_{B} \sim 400$~MeV up
to $\sim 800$~MeV at $\sqrt{s_{NN}} \sim 2.5$~GeV.

This wide-ranging experimental effort must be accompanied by advances
in theory. 
The detailed evolution of the matter produced at RHIC, and its transformation 
from hadronic to partonic degrees of freedom and back again, are not
understood.  Simulations employing models with and without a phase transition
as well as with and without a critical point over the BES range are
important to guide the experimental program and interpret the results. For
example, it is necessary to know whether or not STAR net-proton directed flow
measurements at BES energies can be explained by hadron physics only.
While there is no qualitatively viable hadronic explanation based on current
models, tighter scrutiny is needed to convincingly exclude such a description.
Therefore, more predictions of measurable observables related to the location 
of the critical point and/or
phase boundaries should be made. In particular, the behavior of observables
in simulations that incorporate a first-order phase transition needs further
study.  For example, a mean-field potential
can be constructed to implement a first-order phase transition in
transport models. Overall, significant progress has been made up to this 
point, but the additional detailed data expected from BES Phase II will be essential 
for completing the program goals, while parallel theoretical progress will be 
equally vital.



\subsection{Near-equilibrium properties of the QGP}\label{sec:d:NeaEquPro}

\subsubsection{Global event characterization}

In ultrarelativistic heavy-ion collisions, the majority of the produced 
particles are emitted with transverse momenta below a few GeV/$c$. Precision 
studies of their production characterizes the dynamic evolution of the bulk 
matter created in the collision. 
Measurements of the multiplicity distribution are related to the initial 
energy density. Identified particle yields and spectra reflect
the conditions at and shortly after hadronization. The space-time evolution 
of the particle-emitting source and its transport properties are accessible 
experimentally through particle correlations.
In this section, we briefly describe some of the relevant observables and
recent results. The experimental overview presented in this section is largely based
on the review by M\"uller, Schukraft and Wyslouch \cite{Muller:2012zq}, which summarized the
first results of \PbPb\ data taking at the LHC and extends it with the latest findings based
on increased statistics and more refined analyses.

\begin{figure}
\begin{center}
\includegraphics[width=0.45\textwidth]{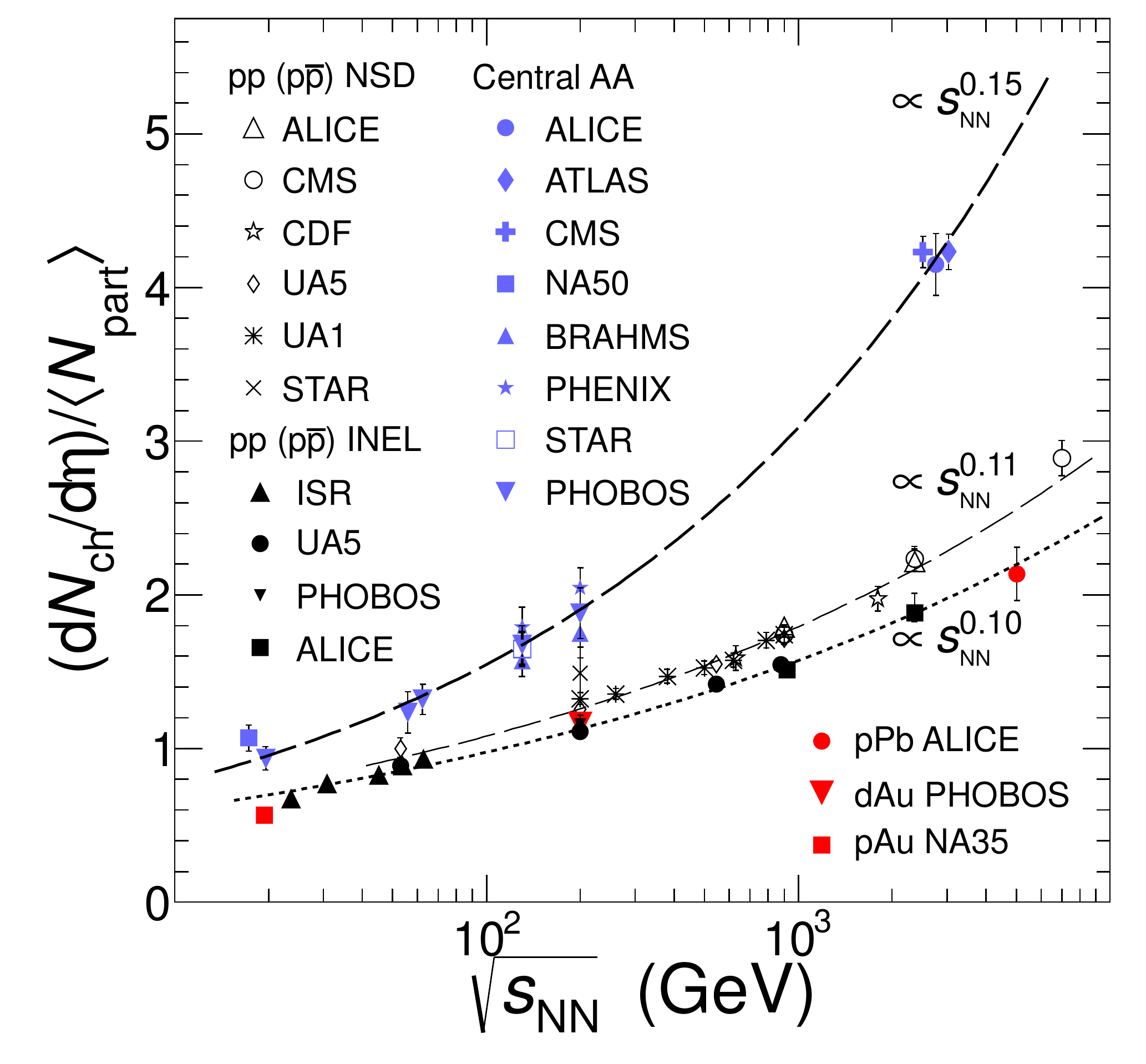}
\caption{Charged particle pseudorapidity density at midrapidity, \dndh, per participant as a function of $\sqrt{s_{NN}} $ for \ppcoll, 
$pA$ and $AA$. From \cite{PhysRevLett.110.032301}.}
\label{fig:dNdeta}       
\end{center}
\end{figure}

\vskip 3 mm
\paragraph{Multiplicity}~ 
Particle production at low $p_T$ cannot be calculated from first principles with currently available theoretical tools.
Despite the availability of RHIC data before the LHC startup, the predictions 
for particle multiplicities varied widely. Figure~\ref{fig:dNdeta} presents
a summary of the charged particle pseudorapidity density per
participant measured in \ppcoll, $pA$ and $AA$ collisions \cite{PhysRevLett.110.032301, PhysRevLett.105.252301}. While the energy dependence 
of $dN_{ch}/d\eta$ in non-single diffractive (NSD) \ppcoll\ and $pA$ collisions follows a power law, $s_{NN}^\alpha$ with $\alpha = 0.1$, the $AA$ data
shows a much steeper dependence that can be best described with $\alpha = 0.15$. 

This behavior underlines
the fundamental differences of bulk particle production in $AA$ with respect to $pp$ and $pA$ collisions and
provides an essential constraint for models, see Sec.~\ref{chapd:pPbEXP}. A comparison between data and theoretical models can be found in \cite{ATLAS:2011ag,Muller:2012zq}.
In addition, the multiplicity distribution in $AA$ collisions has also been studied 
employing holographic approaches, as discussed in Sec.~\ref{sec:5.new}.

\vskip 3 mm
\paragraph{Energy density}~ The measured  \dndh~can be related to the initial
energy density of the system using the Bjorken hydrodynamic model 
\cite{Bjorken:1982qr}, based on a longitudinal, isentropic expansion.  The energy density reached in the initial stage ($\tau_{0}=1$ fm/$c$) of a central
\PbPb\ collision at the LHC of about $\epsilon = 15$~GeV/fm$^3$
\cite{Krajczar:2011zza} is almost three times higher than the one
reported at RHIC 
\cite{Arsene:2004fa,Adcox:2004mh,Back:2004je,Adams:2005dq,Adler:2004zn} and well 
above the critical energy density required for the predicted phase transition 
to a deconfined state of quarks and gluons of about 0.7 GeV/fm$^3$ \cite{Karsch:2000kv}. 

\begin{figure}
\begin{center}
\includegraphics[width=0.45\textwidth]{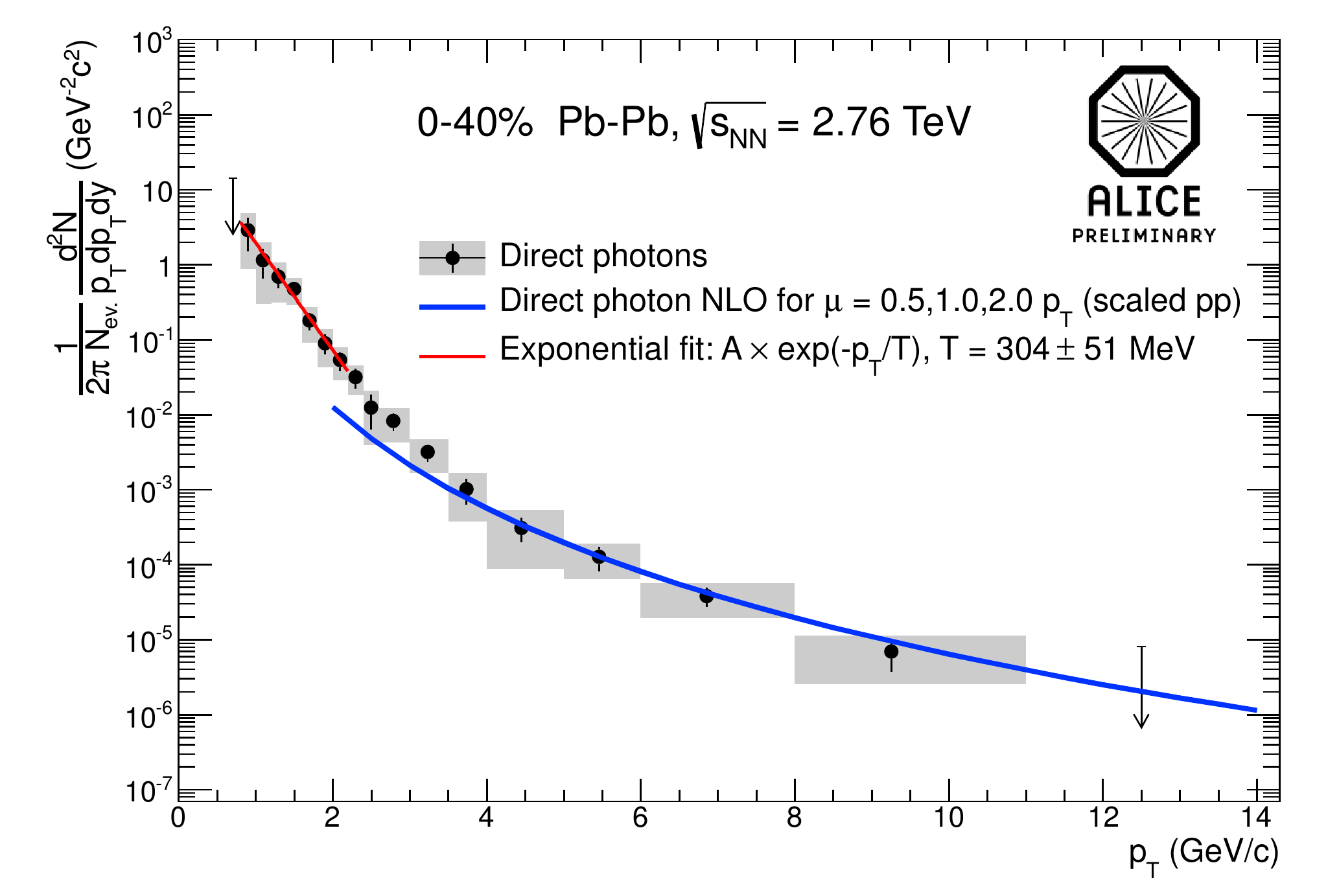}%
\end{center}
\caption{
The direct photon \Pt\ spectrum with the NLO prediction at high \Pt\ and an exponential fit at low \Pt. From \cite{Safarik:2013zza}.}
\label{fig:DirectPhoton}
\end{figure}

\vskip 3 mm
\paragraph{Initial temperature}~
This relative increase of energy density from RHIC to the LHC implies a corresponding
initial temperature at the LHC of $\approx 300$ MeV for central \PbPb\
collisions.  An experimental access to this temperature is given 
by the measurement of thermal photons, emitted in the initial stage of the collision.
The \Pt\ spectrum of direct photons, measured
by the ALICE Collaboration using $\gamma$ conversions in the 40$\%$
most central \PbPb\ collisions \cite{Safarik:2013zza}, is shown in 
Fig.~\ref{fig:DirectPhoton}.
The spectrum is reproduced by the NLO pQCD prediction
for \ppcoll\ collisions, scaled by the number of binary collisions
at \Pt~$>4$~GeV/$c$.
Below 2~GeV/$c$ there is an excess attributed to thermal photons.  An
exponential fit in the range $0.8<p_T<2.2$~GeV/$c$ yields an inverse
slope parameter $T=(304 \pm 51$) MeV. The quoted uncertainties include
both statistical and systematic uncertainties. The LHC value of this effective
temperature is about 40$\%$ higher than that measured in a similar analysis by PHENIX
\cite{Adare:2008ab} and is clearly above the expected phase transition
temperature of about 160 MeV. Before firm conclusions can be drawn from these measurements, two important
considerations have to be taken into account. First, the measurement of the thermal photon spectrum is 
experimentally very demanding \cite{Wilde:2012wc}. Despite the impressive precision already achieved \cite{Abelev:2012cn},
further refined analyses are expected in the future. In particular, a more precise estimation of the 
detector material budget, needed for the determination of the photon conversion probability, 
are expected to further reduce the experimental uncertainties. Furthermore, the thermal photon spectrum is
obtained by subtracting the decay photon spectrum which is obtained by a complicated cocktail calculation. 
While the contribution from the $\pi^{0} \rightarrow \gamma\gamma$ decay can be based on the measured
spectrum, the contribution from unmeasured meson yields in \PbPb\ collisions (such as $\eta$, 
$\eta'$, $~\omega$, $~\rho^{0}$) have to be interpolated from $m_{T}$-scaling. 
Second, more rigorous theoretical analyses of 
the ALICE data \cite{Klasen:2013mga} are ongoing, which also include Doppler blue-shift corrections 
of the temperature due to the radially expanding medium \cite{Shen:2013vja}. 

\vskip 3 mm
\paragraph{System size and lifetime}~ The space-time evolution of the
expanding system is studied using identical pion interferometry
techniques known as Hanbury-Brown Twiss (HBT) correlations
\cite{Lisa:2005dd}. At LHC energies, the measurement in the 5$\%$ most 
central \PbPb\ collisions 
shows that the homogeneity volume at freezeout (when strong
interactions cease) is 5000~fm$^3$, twice as large as the volume
measured at RHIC. The total lifetime of the system (the time between the initial nucleon-nucleon collisions
and freezeout) is approximately 10 fm/$c$, 30$\%$ larger than at RHIC \cite{Aamodt:2011mr}. 
The extracted volume increases linearly as a function of charged particle multiplicity. Extrapolation to \dndh~shows that, in this limit, the system size coincides with the initial volume of
a Pb nucleus and its lifetime vanishes \cite{Muller:2012zq}. 
Hydrodynamic models correctly describe the evolution with center-of-mass
energy from RHIC to LHC as well as the dependence of the individual radius
parameters on the pair momentum, which is sensitive to radial flow \cite{Frodermann:2007ab,Bozek:2011ph,Karpenko:2012yf}.
Measurements with kaons and protons are being carried out to test whether the 
collective motion includes heavier mesons and baryons. In addition, baryon correlations are
sensitive to mutual strong interactions, which are poorly
known, especially for baryon-antibaryon pairs. The parameters of this
interaction can be deduced from two-particle baryon correlations \cite{Szymanski:2012qu,Kisiel:2014mma}, 
which constitute a powerful way to obtain such information.

As a reference the same measurement was carried out in smaller systems ($pp$, $pA$). Particular attention was given to high multiplicity events where collectivity was predicted to arise in some models \cite{Werner:2011fd,Bozek:2013df}. The extraction of femtoscopic radii in such systems is complicated due to the presence of other correlation sources, i.e.,  mini-jets and energy and momentum conservation. Monte-Carlo models have to be used to account for these effects \cite{Aamodt:2011kd}. Other methods, such as three-pion correlations, are by construction less sensitive to such background, due to the usage of higher order cumulants \cite{PhysRevC.89.024911}. The analysis shows that the radii in small systems depend on multiplicity and pair momentum but not on collision energy. The overall magnitude is smaller than in collisions of heavy ions at comparable multiplicity. Although decrease of radii with pair momentum is observed it is of different nature as compared to heavy-ion collisions. Therefore qualitatively new features are observed in HBT of small systems, that still require theoretical investigation.

\vskip 3 mm
\paragraph{Particle spectra in different \Pt~ranges}
Transverse momentum spectra are sensitive to different underlying physics
processes in different \Pt\ domains. In a crude classification, three separate
regions can be identified: low, intermediate and high $p_{T}$. 
At $p_{T} < 2$~GeV/$c$, the bulk matter dynamics can
be described by relativistic hydrodynamic models. Even at LHC energies, 
more than 95$\%$ of all particles are produced within this \Pt\
range. While the
spectral shape reflects the conditions at kinetic freezeout (where particle
momenta are fixed), the integrated particle yields reflect the conditions at
chemical freezeout (where particle abundances are fixed).
At $p_{T} > 8$~GeV/$c$, partons from hard scatterings interacting with the medium 
dominate the spectrum. At intermediate
\Pt, the data reflect an interplay of soft and hard processes.  The
energies available at the LHC open up the possibility for detailed
measurements over an extended \Pt\ range, up to hundreds of GeV$/c$ in some cases.  
Understanding the interplay of soft and hard processes and the onset of hard processes remains a
theoretical challenge. We discuss some low and intermediate \Pt\ results in the remainder of this section.
Hard processes are discussed in Sec.~\ref{sec:HardAndHeavy}.

\vskip 1 mm
\paragraph{Low \Pt\ }~ The spectra of identified charged hadrons ($\pi$, $K$ 
and $p$), measured
in the 5$\%$ most central \PbPb\ collisions at the LHC \cite{Abelev:2012wca} for $0.1 <$~\Pt~$< 4.5$~GeV/$c$ and at midrapidity, $|y| < 0.5$, are harder than the ones
measured in central \AuAu\ collisions at $\sqrt{s_{NN}} = 200$~GeV
at RHIC \cite{Abelev:2008ab,Adler:2003cb}, reflecting the
stronger radial flow at the LHC.  A blast-wave fit of the spectra
\cite{Schnedermann:1993ws} yields a kinetic freezeout
temperature $T_{kin} = 96 \pm 10$~MeV, similar to the one at RHIC,
and a collective radial flow velocity, $\langle \beta_{T}
\rangle = 0.65 \pm 0.02$, 10$\%$ higher than the one at RHIC.  When compared 
to hydrodynamic calculations
\cite{Shen:2011eg,Karpenko:2011qn,Karpenko:2012yf,Bozek:2011ua,Werner:2012xh}, 
the data are in better agreement with calculations including
rescattering during the hadronic phase. 
Similar behavior is observed in other centrality classes \cite{Abelev:2013vea}.

The conditions at chemical freezeout, where particle abundances are
fixed, are characterized by the chemical freezeout temperature
($T_{ch}$) and baryochemical potential ($\mu_{B}$) and are
determined from measured particle yields in thermal model
calculations.  
Recent comparisons of the ALICE measurements in central \PbPb\ collisions with
thermal models \cite{Andronic:2008gu,Cleymans:2006xj} show the best agreement
between data and theory calculations at vanishing baryochemical potential, 
$\mu_{B}\approx 1$~MeV, and at a chemical freezeout temperature of 
$T_{ch} \approx 156$~MeV, lower
than the value $T_{ch} \approx 164$~MeV predicted before the LHC startup. 
This difference was caused by an overestimate of the proton yield in the model 
for higher chemical freezeout temperatures. 
The remaining tension between the fit and the proton yield at 
$T_{ch} = 156$~MeV is 23\% (2.9$\sigma$).  This might be further reduced 
by construction of a more complete hadron spectrum within the thermal model 
\cite{Andronic:2008gu}.  Additional data analyses will clarify the experimental 
significance of the observed effect. Several possible explanations for these 
deviations have been suggested.  In particular large baryon-antibaryon 
annihilation rates in the late hadronic phase could be the source of some
lower baryon yields \cite{Steinheimer:2012rd}.  
Such annihilation processes are reflected also in ALICE femtoscopic measurements of $p \bar{p}$ and $\Lambda\bar{\Lambda}$ correlations \cite{Adare:2013oaa,Szymanski:2012qu}, however the yield modification cannot be directly obtained from such considerations.
The influence of these effects on the thermal parameters extracted from the
data has been quantified based on UrQMD \cite{Becattini:2012xb}. At lower 
center of mass energies, this approach improves the agreement between the
experimentally reconstructed hadrochemical equilibrium points in the (T, $\mu_{B}$) plane  
and the parton-hadron phase boundary recently predicted by lattice QCD \cite{Endrodi:2011gv, Kaczmarek:2011zz}.
Other possible explanations are based on nonequilibrium thermal 
models \cite{Petran:2013qla} or a flavor 
dependent freezeout temperature, as indicated by recent 
lattice QCD calculations \cite{Bellwied:2013cta}.


\vskip 1 mm
\paragraph{Particle composition at intermediate \Pt}~ To probe how the
interplay of soft and hard processes affects the particle composition
at intermediate \Pt, baryon-to-meson ratios such as
$\Lambda/K_{s}^{0}$ and $p/\pi$ are studied \cite{Abelev:2013xaa,Adams:2006wk,Zhang:2007st}.  
An enhancement of $\Lambda/K_{s}^{0}$, relative to the measured ratio in \ppcoll\
collisions, was first observed at RHIC (the so-called
baryon anomaly) \cite{Adcox:2001mf,Vitev:2001zn}. The ALICE $\Lambda/K_{s}^{0}$ data, measured up
to $p_{T} \approx 6$~GeV/$c$, confirm that the effect persists at the LHC, 
is slightly stronger than at RHIC and extends to higher \Pt.  Comparisons of the
$\Lambda/K_{s}^{0}$ ratio with models shows that the strong rise of the ratio 
at low \Pt\ can be described by relativistic hydrodynamic models.
The EPOS model describes the effect over the entire \Pt\ range and for all 
studied centrality classes \cite{Werner:2012sv}.
In contrast to other models, it connects soft and hard processes by a 
mechanism in which jet-hadrons, produced inside the fluid, pick up quarks 
and antiquarks from the thermal matter rather than creating $q \bar q$ pairs by 
the Schwinger mechanism.

\begin{figure}
\begin{center}
\includegraphics[width=0.48\textwidth]{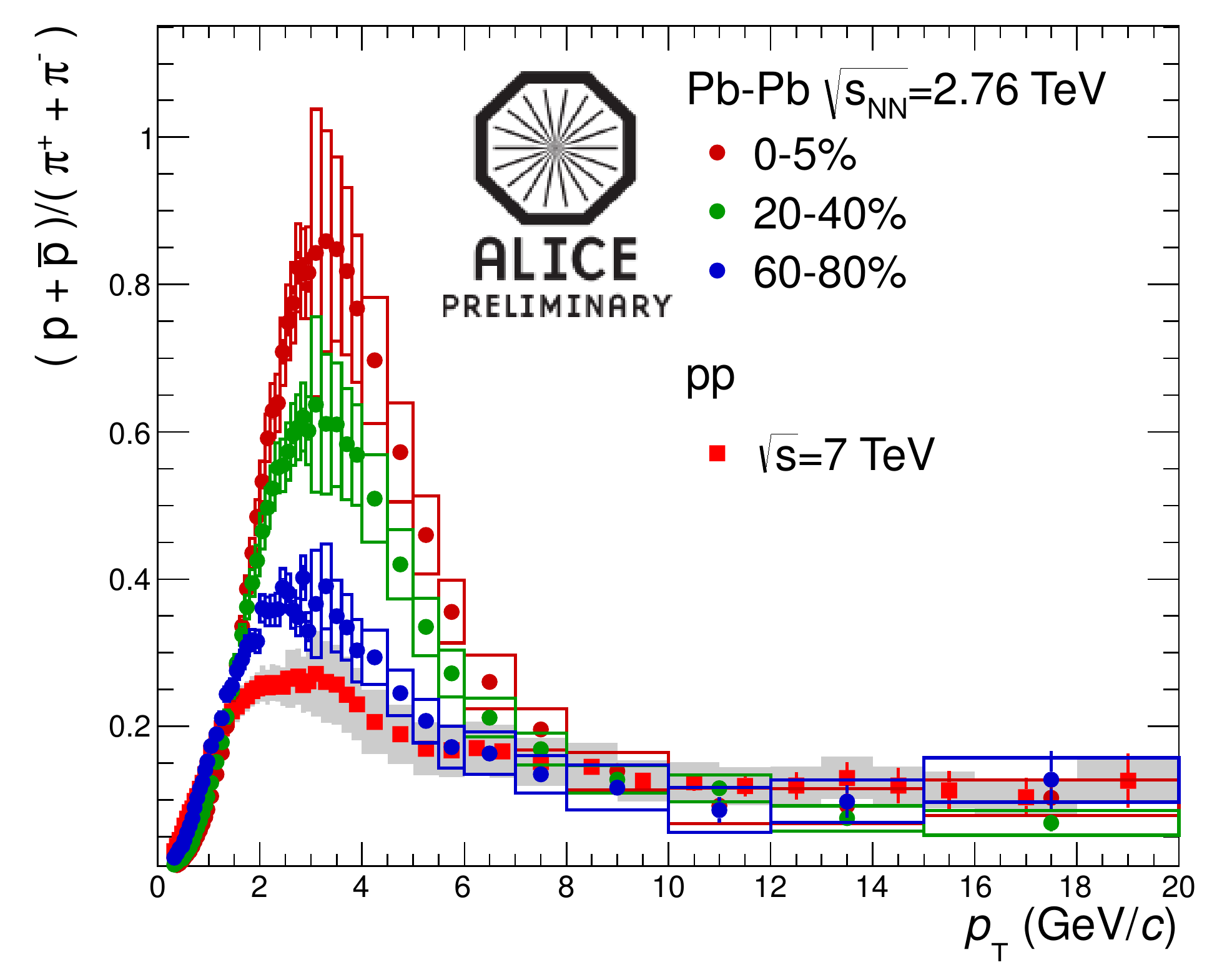}
\caption{The $p/\pi$ ratio measured for several centrality classes in \PbPb\ collisions relative to \ppcoll\ results at $\sqrt{s_{NN}} = 2.76$~TeV. From \cite{OrtizVelasquez:2012te}.}
\label{fig:pTopiRatio}
\end{center}
\end{figure}

Figure~\ref{fig:pTopiRatio} shows the $p/\pi$ ratio as a function of
\Pt\ measured up to $p_{T} \approx 20$~GeV/$c$ in several centrality bins 
in \PbPb\ collisions compared to \ppcoll\ results.  At $p_{T} \approx 3$~GeV/$c$, 
the $p/\pi$ ratio in the 5$\%$ most central \PbPb\ collisions is a factor
3 larger than the \ppcoll\ ratio.  At higher \Pt, the enhancement is 
reduced and, above 10~GeV/$c$, the \PbPb\ ratio becomes compatible with the
\ppcoll\ value.  In the most peripheral bin, 60--80\%, the $p/\pi$ ratios in
\PbPb\ and \ppcoll\ collisions are comparable over most of the measured
\Pt\ range.

As is the case for the $\Lambda/K_{s}^{0}$ ratio, the observed anomalous 
baryon to meson enhancement can be attributed to the effect of radial flow 
that pushes heavier particles to higher \Pt.  However, it seems to extend 
beyond the region where radial flow is applicable.
This enhancement was also interpreted as possibly caused
by the recombination of quarks into hadrons \cite{Fries:2003kq}. 
Further studies involving different other observables 
are expected to disentangle the different effects.

\subsubsection{Azimuthal anisotropies}
\label{chapd:azimuthal_ani}
Measurements of azimuthal particle anisotropies probe collective
phenomena that are characteristic of a bulk system such as the one
expected to be created in heavy-ion collisions \cite{Heinz:2013th}.
In non-central collisions, anisotropic pressure gradients, developed
in the overlap region of the two colliding nuclei, transform the 
initial spatial anisotropy into an observed momentum anisotropy, through
interactions between the produced particles, leading to an
anisotropic particle distribution $dN$/$d\varphi$.  This anisotropy is
usually quantified via a Fourier expansion of the azimuthal
distribution \cite{Voloshin:1994mz}.  The
Fourier (or flow) coefficients, $v_{n}$, dependent on \Pt\ and
pseudorapidity, are given by
\begin{equation}
v_n = \langle \cos \big [n(\varphi - \Psi_{n}) \big] \rangle \,\, ,
\label{Eq:Fourier}
\end{equation}
where $n$ is the order of the flow harmonic, $\varphi$ the azimuthal
angle of the particle and $\Psi_n$ the azimuthal angle of the initial
state spatial plane of symmetry for harmonic $n$.  The isotropic (or
angle averaged) component is known as radial flow ($v_0$) 
while the $v_{1}$ coefficient is referred to as directed flow.
The second Fourier coefficient, $v_2$, is the elliptic flow.  In this case 
$\Psi_{2} \approx \Psi_{RP}$ where $\Psi_{RP}$ is the
angle of the reaction plane, defined by the beam direction and the impact
parameter plane.  Elliptic flow has been extensively studied 
as a measure of collective
phenomena in bulk matter in contrast to a superposition of
independent $NN$ collisions, where particle momenta would be uncorrelated
relative to the reaction plane.

Higher-order odd harmonics, $n \geq 3$, had previously been neglected because they
were expected to be zero due to symmetry.  However, the
statistical nature of individual nucleon-nucleon collisions can lead
to highly irregular shapes of the reaction region and thus the corresponding
initial energy and pressure distributions 
\cite{Sorensen:2010zq,Alver:2010gr}, resulting in event-by-event fluctuations 
in the elliptic flow direction and magnitude, as well as in all other harmonics.
Different experimental methods are used to measure the symmetry plane
angles and the $v_{n}$ coefficients, via two- and
higher particle correlations 
\cite{Poskanzer:1998yz,Voloshin:2008dg,Snellings:2011sz}.
Each coefficient is sensitive to different
effects, allowing a comprehensive study of fluctuations and
non-flow contributions.

The first measurements at the LHC \cite{Aamodt:2010pa} confirmed
hydrodynamic predictions and indicated that the system created in
\PbPb\ collisions at $\sqrt{s_{NN}}=2.76$ TeV still behaves like a 
strongly-interacting, almost
perfect, liquid with minimal shear viscosity to entropy ratio,
$\eta/s$, similar to the one at RHIC 
\cite{Back:2004je,Arsene:2004fa,Adcox:2004mh,Adams:2005dq,Muller:2006ee}. 

Further differential studies of the anisotropic flow coefficients involve 
the quantitative extraction of the transport coefficients of the medium. 
A precise determination is currently hampered by poor knowledge of the
initial state of the collision, along with a significant number of
other, smaller, theoretical uncertainties \cite{Luzum:2012wu}. 
One of the key uncertainties is the description of the initial-state geometry. 
The studies of higher-order flow components, in particular the triangular flow $v_3$
\cite{Alver:2010gr}, have provided new input to reduce these uncertainties.
A complementary approach was provided by CMS studies of ultra-central 
collisions, 0--0.2\% \cite{Roland:2013aaa}, where the initial-state 
eccentricities are defined by fluctuations of the participant geometry.  
Additional constraints
on $\eta/s$ were obtained by studying $v_{2}$ as a function of centrality
and \Pt\ for different particle species.
Comparison with models typically yields $\eta/s \approx (1-2.5)/4\pi$
\cite{Heinz:2013th}, close to the lower bound conjectured by
AdS/CFT for a good relativistic quantum fluid \cite{Kovtun:2004de}.  
Recent results \cite{Heinz:2013th} show that ``IP-Glasma" initial conditions
\cite{Schenke:2012hg} and average values of $\eta/s \approx  0.2$ for \PbPb\
collisions at the LHC and 0.12 for \AuAu\ collisions at RHIC, 
provide a good description of the majority of the data \cite{Gale:2012rq}.

\vskip 3 mm
\paragraph{$v_{n}$ measurements from RHIC to LHC}~ Compared to RHIC, the
LHC has significantly extended the azimuthal anisotropy measurements both in
pseudorapidity and \Pt. The ALICE and ATLAS results up to 
$p_{T} \approx 20$~GeV/$c$ show the same trends as the CMS data, which extends 
the $v_2$ measurement up to $p_{T} \approx 60$~GeV/$c$
\cite{Safarik:2013zza,Voloshin:2012fv,Roland:2013aaa}.

In general, the
integrated $v_{2}$ increases by 20--30$\%$ at midrapidity and 
$\approx30$\% at forward rapidity relative to RHIC, in agreement with
hydrodynamic calculations \cite{Luzum:2010ag}. The $n=3$ coefficient, $v_{3}$, shows a weak centrality dependence with a
similar magnitude in central and peripheral collisions.  In central collisions,
the magnitude of $v_{2}$ is similar to that of $v_{3}$.  These measurements 
confirm that $v_{2}$ is driven by geometry while $v_{3}$ is dominated by 
initial-state fluctuations. The latter also generates the finite $v_{2}$ in the
most central collisions which approximates the ideal case of zero impact 
parameter.  

The fourth-order harmonic was measured with respect to the second- and fourth-order event planes, $v_{4}(\Psi_2)$ and $v_{4}(\Psi_4)$ 
\cite{Safarik:2013zza,Voloshin:2012fv}. The difference between the
results for the two event planes is entirely due to fluctuations in the 
fourth-order harmonic flow and, as such, provides important constraints on the 
physics and origin of the flow fluctuations. 

At LHC energies, the large particle multiplicities produced in each event 
also allow a determination of the flow coefficients in individual events. 
The ATLAS collaboration has measured $v_n$ for $2<n<4$ event-by-event \cite{Aad:2013xma}.
Comparisons with a Glauber-based geometric model \cite{Miller:2007ri} and a model that includes corrections to the 
initial geometry due to gluon saturation effects \cite{Drescher:2006pi} fail to describe the
experimental data consistently over most of the measured centrality range.

In addition to the integrated value of $v_{2}$, valuable information can also be
determined from the dependence of $v_2$ on transverse momentum
and particle mass. The shape of the \Pt-differential 
anisotropic flow is determined by different underlying physics processes in 
the various \Pt\ regions. The behavior of the bulk matter for 
$p_{T}  < 1$--$2~\text{GeV}/c$
is mostly determined by hydrodynamic flow which 
exhibits a typical ``mass splitting'' \cite{Huovinen:2001cy} induced by the 
collective radial expansion of the system.  While the effect is cumulative 
over the lifetime
of the system, it has a significant contribution from the partonic phase.
However, hadronic rescattering in the late stages might mask the
information from the early stage.  

The measured value of $v_{2}$ reaches 
a maximum around $p_{T} \approx 2$~GeV/$c$ and 
slowly decreases until it approaches zero for $p_{T} \approx$
40--60 ~GeV/$c$ as measured by CMS.
At $p_{T} > 10$~GeV/$c$ the elliptic flow results are well
described by extrapolation of the WHDG model \cite{Wicks:2005gt} to LHC 
energies \cite{Horowitz:2011gd}, which takes into account 
collisional and radiative energy loss in an expanding medium. In this model, 
the anisotropy is controlled by the energy loss mechanism. Similar to \RAA, 
only a minor dependence on particle type is expected in this region. 

\begin{figure}
\begin{center}
\includegraphics[width=0.48\textwidth]{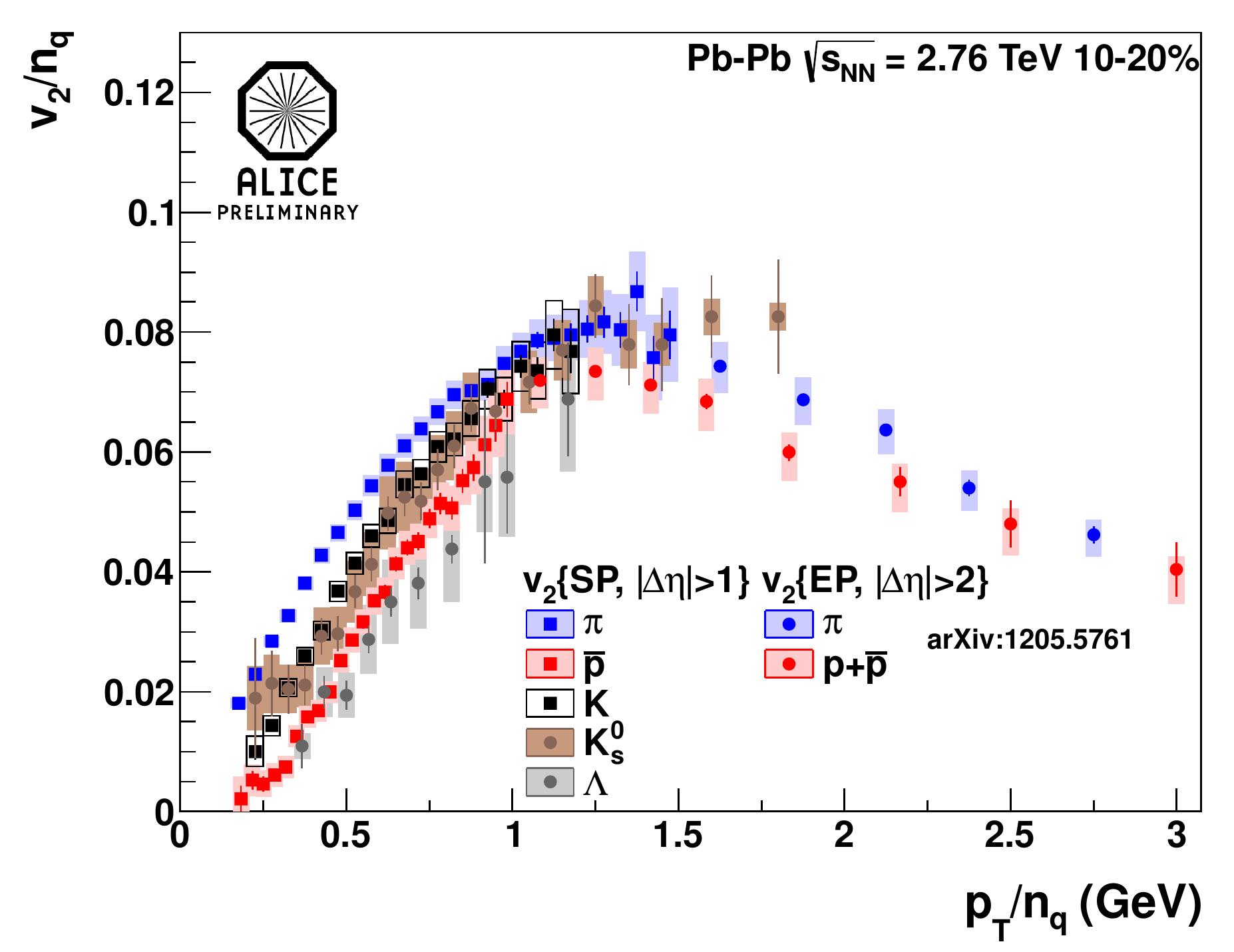}
\includegraphics[width=0.48\textwidth]{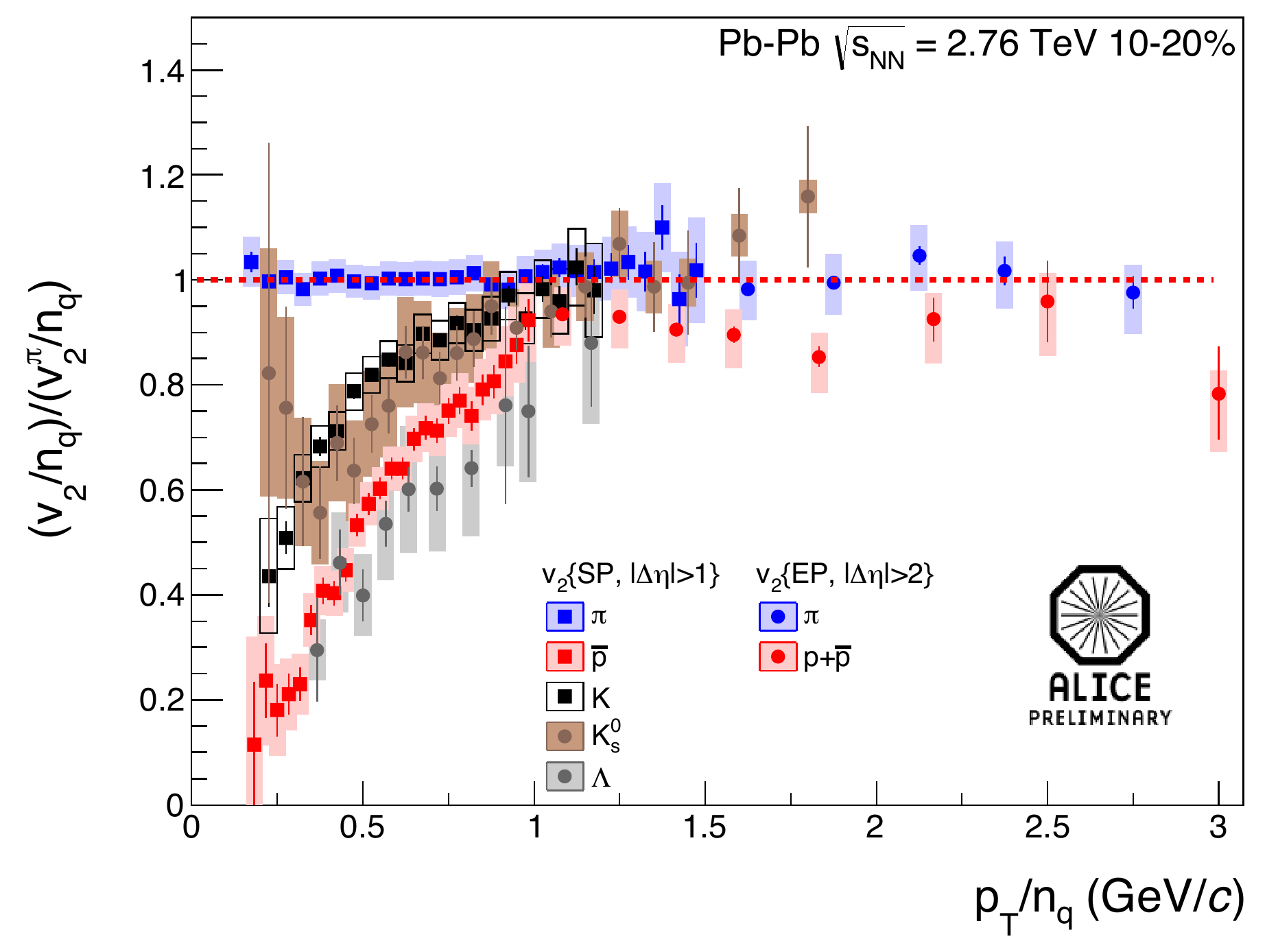}
\caption{(Top) 
Elliptic flow coefficient $v_{2}$ as a function of the transverse momenta scaled by the number of constituent quarks in \PbPb\ collisions 
at $\sqrt{s_{NN}}=2.76$~TeV. (Bottom) The same data are shown normalized to the polynomial fit to the pion elliptic flow. From \cite{Noferini:2012ps}. }
\label{fig:QuarkNumberScalingv2}
\end{center}
\end{figure}


\vskip 3 mm
\noindent
\paragraph{Quark number scaling}~At RHIC, it was observed that all baryons
exhibited a similar anisotropic flow pattern; with the ratio of baryon to meson $v_2$ being 3:2,  
see Ref.\cite{Abelev:2007rw} and references therein. These findings suggested that hadron formation at 
intermediate \Pt\ 
is dominated by quark coalescence at the end of the partonic evolution
\cite{Voloshin:2002wa}.  PHENIX has observed
that the scaling is broken when plotted as a function of the transverse
kinetic energy at $KE_T/n > 1$ GeV in all but the most central collisions \cite{PhysRevC.85.064914}.
ALICE has subsequently
studied quark number scaling for a number of identified particles in different
centrality ranges.
Also at the LHC, quark scaling appears to be broken for transverse momenta
per number of constituent quarks, \Pt$/n_q$, 
below 1 GeV/$c^2$.
At higher \Pt, the scaling appears to hold at the 20\% level as shown in Fig.~\ref{fig:QuarkNumberScalingv2}. 
The significance of this scaling and
the size of the violations needs further study.
These theoretical and experimental investigations are of particular importance 
as a picture of anisotropic quark flow and subsequent hadronization via coalescence 
has been related to deconfinement by some authors \cite{Wu:2010zzj, Molnar:2003ff, Lin:2002rw, Fries:2003vb,PhysRevLett.91.172301}.
 
\subsubsection{Transport coefficients \& spectral functions: theory}\label{sec:d:TraCoeSpe}

A comparison of RHIC and LHC heavy-ion data with the results of viscous 
hydrodynamic simulations for quantities such as the elliptic flow seems to 
imply a remarkably small value of the shear viscosity of the QGP 
(see Sec.~\ref{chapd:azimuthal_ani}). While the 
quantitative value depends somewhat on the details of the simulation, in 
particular on the initial conditions, it is widely accepted that 
the shear viscosity to entropy ratio $\eta/s$ is rather close to the 
value $1/(4\pi)$~\cite{Kovtun:2004de} found
in strongly-coupled gauge theories with gravity duals.
The existing (full leading order) 
weak-coupling prediction for this ratio is considerably larger for 
reasonable values of $\alpha_s$ \cite{Arnold:2003zc}. 

Thus, it is very important to develop nonperturbative first-principles tools 
to compute the shear viscosity in QCD. 
More generally, transport coefficients such as the shear viscosity can be regarded as the low-energy constants of hydrodynamics,
which describes slow, long-wavelength departures from equilibrium of a thermal system.
The values of the transport coefficients, however, must be computed in the underlying microscopic theory
---  QCD in the case of the quark-gluon plasma.
In strongly coupled gauge theories, important progress has been made employing the gauge/gravity 
correspondence; see Sec.~\ref{sec:d:LatQcdAds} as well as 
Refs.~\cite{Son:2007vk,Gubser:2009md,CasalderreySolana:2011us}. 
This correspondence provides a paradigm diametrically opposite
to the quasiparticle picture that underlies weak-coupling
calculations.  The relative ease with which real-time physics can be extracted 
from the gauge/gravity correspondence at strong coupling, such as with
the methods of Ref.~\cite{Son:2007vk}, is particularly impressive. 
Although non-supersymmetric and conformally
broken quantum field theories have been investigated using the gauge/gravity 
correspondence (see for instance \cite{Gubser:2008yx,Gursoy:2010fj}), 
no exact QCD dual has been constructed to date, hence the phenomenological 
predictions obtained from gauge/gravity techniques must be regarded as
semi-quantitative at best.

On the other hand, the lattice QCD framework is ideally suited to
reliably determine the equilibrium characteristics of the QGP, such as
the equation of state, see also Sec.~\ref{sec:d:MapQcdPha}.  Because 
numerical lattice gauge theory employs
the Euclidean formalism of thermal field theory,
dynamical properties are normally only
accessible through analytic continuation, posing a considerable
numerical challenge; see Ref.~\cite{Meyer:2011gj} for a recent review.

Spectral functions encode important dynamical properties of the medium. For
instance, the photon and dilepton production rates in the QGP are
proportional to the spectral function of the conserved vector
current. Hydrodynamic modes and quarkonium states show up as peaks
whose widths are proportional to the rate at which these excitations
dissipate. In lattice QCD, the spectral function is obtained by solving the integral
equation
\begin{equation}
\label{eq:GErho}
G(\tau,\boldsymbol{k},T)  = \int_0^\infty d\omega \; 
\rho(\omega,\boldsymbol{k},T) \;
\frac{\cosh[\omega(\frac{1}{2T}-\tau)]}{\sinh[\frac{\omega}{2T}]} \, \, ,
\end{equation}
given the Euclidean correlator $G$ at a discrete set of points $\tau$ with a 
finite statistical accuracy. When Euclidean correlation functions are known numerically instead of 
analytically, the determination of the spectral function involves the solution 
of a numerically ill-posed inverse problem.
Compared to nonrelativistic systems such as cold Fermi gases, such as in 
Ref.~\cite{Wlazlowski:2012jb}, QCD has the added difficulty that 
correlation functions are strongly divergent at short distances.
In spite of these difficulties, with good numerical data and the help of prior 
analytic information, including effective field theory, sum rules and the 
operator product expansion, the gross features of the spectral function
$\rho$ can be determined.  
In practice, however, the temperature scale imposes a limit on the
frequency resolution.
Then the identification of bound states or transport peaks, substantially
narrower than the temperature, cannot be formulated in a model independent way.
An accurate and reliable calculation of the Euclidean correlators nevertheless 
remains an important goal for lattice QCD, not least because they can be used 
to test various analytic methods; see Sec.~\ref{sec:d:LatQcdAds}.

At zero temperature, an one-to-one 
correspondence exists between the spectral function below
inelastic thresholds and stationary observables, thus making the spectral function 
directly accessible to lattice QCD~\cite{Meyer:2011um}.  
A typical example is the possibility to calculate the  $\rho$-channel 
spectral function~\cite{Meyer:2011um,Luscher:1991cf}
in the elastic regime without an explicit analytic continuation. 
Whether a similar correspondence can be constructed for the nonequilibrium properties
of the QGP along these lines has yet to be determined. 
In particular, the finite volume used in the lattice simulations plays a 
crucial role in relating stationary observables to dynamical quantities at 
$T=0$ and the volume effects on the thermal spectral function should be investigated.


In the following, we briefly discuss several channels of interest.
Consider first the spectral function of the conserved vector current.
For a generic frequency, NLO perturbative calculations are available, 
including quark mass corrections~\cite{Burnier:2012ze}.
For the light quark flavors, the vector channel is related to 
the production of real photons and lepton pairs in the thermal medium. 
A recent NLO calculation of the thermal photon production 
rate showed that the  convergence rate is reasonably good~\cite{Ghiglieri:2013gia}.
The dilepton rate for an invariant mass on the order of the typical thermal momentum 
has been computed at NLO even for a nonvanishing spatial momentum \cite{Laine:2013vma}.
Extensive phenomenological studies have been carried out in order to compare 
different spectral function calculations to heavy-ion data~\cite{Rapp:2011is}. 
The low energy part of the experimental dilepton spectrum was found to be 
dominated by the contribution from the confined phase. Lattice 
results  have been reported in the continuum limit of the quenched 
approximation \cite{Ding:2010ga}, as well as with dynamical quarks 
at a single lattice spacing \cite{Brandt:2012jc,Amato:2013naa}. In the thermodynamic limit, 
the thermal part of the spectral function is constrained by a sum 
rule~\cite{Bernecker:2011gh}.  In the chirally-restored phase of QCD with two 
massless quarks, the isovector-vector and axial-vector correlators are exactly 
degenerate so that the thermal generalization of the Weinberg sum 
rule \cite{Kapusta:1993hq} is trivially satisfied.

In the shear \cite{Meyer:2007ic,Meyer:2008sn} and bulk 
\cite{Meyer:2010ii,Meyer:2007dy} channels, lattice QCD data is so far only 
available for pure Yang-Mills theory. This is due to the need for very high 
statistics in the flavor singlet channels which can only be reached in the 
computationally faster Yang-Mills case. In the bulk channel, the operator 
product expansion and a sum rule have also been used to further constrain 
the spectral function \cite{Meyer:2010ii}.  In the shear channel, the 
corresponding sum rule remains incompletely known due to the complicated 
structure of contact terms (the correlator has a stronger short distance 
singularity here than in the bulk or vector channels). A more systematic 
derivation of sum rules and the operator product expansion predictions of 
the asymptotic behavior of the spectral functions is thus required
\cite{CaronHuot:2009ns,Romatschke:2009ng}.  There has recently been substantial 
progress in perturbative calculations of the shear \cite{Zhu:2012be} and 
bulk \cite{Laine:2011xm} channel spectral functions. The convergence of the 
perturbative results for the Euclidean correlators is good, particularly in
the shear channel. These calculations provide very useful information that 
can eventually be combined with numerical lattice data. 

We briefly consider the idealized problem of heavy-quark diffusion in the QGP
in the static limit, $m_{q}\to\infty$. An NLO 
perturbative calculation \cite{CaronHuot:2008uh} is available; unfortunately
the convergence rate turns out to be poor. The 
main quantity of interest, the momentum diffusion coefficient, $\kappa$, can 
be extracted with Heavy Quark Effective Theory \cite{CaronHuot:2009uh}
as well as with lattice QCD \cite{Meyer:2010tt,Francis:2011gc,Banerjee:2011ra}. 
Since the physical observable is essentially reduced to a pure gluonic
one, it is expected to be accurately computed in pure Yang-Mills theory. An important 
advantage of this channel over those discussed above is that no sharp 
features are expected in the spectral function~\cite{CaronHuot:2009uh}, even at
weak coupling, which makes the inverse problem better defined. 
The most important next steps will be to determine the normalization of the
chromoelectric field operator nonperturbatively and to take the
continuum limit of the Euclidean correlator before attacking the
inverse problem. Whether the operator product expansion and a possible
sum rule can also be useful here is not yet clear and
deserves further investigation.


\subsection{Approach to equilibrium}

A major challenge for the theoretical description of heavy-ion collisions is to follow the evolution of the system from its initial state to a near-equilibrium plasma, the behavior of which can be approximated by hydrodynamics. To describe this equilibration process, it is necessary to solve a strongly time-dependent system away from both asymptotically weak and strong coupling. In this section, we describe recent developments in this direction, covering early perturbative work as well as holographic results.
	
\subsubsection{Thermalization at weak and strong coupling}
\label{sec:d:TheWeaStr}

Conceptually, relativistic heavy-ion collisions evolve in steps. The initial nuclear collision liberates partons, which become a nonequilibrium quark-gluon plasma (or liquid), which in turn equilibrates to form a quark-gluon plasma in approximate local equilibrium. Near-equilibrium hydrodynamics then describes the evolution of the plasma from deconfinement until the time that the system begins to hadronize \cite{Song:2010aq}. The stage of the system from the initial collision through the nonequilibrium plasma has been called the ``glasma''. This term arises from the description of the initial nuclei in terms of the color glass condensate, a state characterized by the presence of strong color fields and the over-occupation of soft gluon modes. The transition between the nonequilibrium and equilibrium plasma may in turn be investigated using methods generalized from traditional plasma physics to non-Abelian gauge theories.

Important recent progress has been made in understanding the processes by which a pre-equilibrium QGP approaches equilibrium at high energies or weak coupling.  The situation is complicated by the fact that even in the limit where the gauge coupling is treated as arbitrarily small, the initial color fields are strong enough to make the dynamics of the system nonperturbative. The same is true even at later times for filamentary instabilities which result in the growth of chromomagnetic fields large enough to compensate for the small coupling. Thus, quantifying how the equilibration time of the system depends on the coupling appears to require a combination of analytic weak-coupling techniques and classical real-time lattice simulations. The latter, at weak coupling, correctly treat the nonlinear dynamics of the classical fields representing large soft gluon occupation numbers. 

Competing analytic scenarios for the equilibration process in non-Abelian gauge theories include the bottom-up picture of Ref.~\cite{Baier:2000sb}, as well as the newer proposals of Refs.~\cite{Kurkela:2011ub,Kurkela:2011ti}, emphasizing the role of plasma instabilities, and Ref.~\cite{Blaizot:2011xf} involving formation of a gluonic Bose-Einstein condensate. Very recently, classical simulations of SU(2) lattice gauge theory were carried out in a longitudinally-expanding system \cite{Berges:2013eia}. It was found that, independent of the initial conditions, the system always appears to approach an attractor solution with scaling exponents consistent with the bottom-up solution \cite{Baier:2000sb}. However, closely related work \cite{Gelis:2013rba} challenges this outcome and instead suggests fast isotropization of the system.

Once the weak coupling thermalization mechanism has been qualitatively understood, this insight needs to be translated into quantitative predictions. A calculation of the transition of the system from pre-equilibrium to equilibrium could then be coupled to weak-coupling glasma calculations of the creation of the initial pre-equilibrium plasma to provide a complete picture of the dynamics. Recent progress in calculating the seeding and development of instabilities in the glasma \cite{Dusling:2011rz} is an encouraging development in this direction.

Equilibration of the QGP can also be studied in an altogether different and highly complementary limit, i.e.,~in strongly coupled QCD-like plasmas that have a dual gravity description. A clear distinction between the equilibration, isotropization, and hydrodynamization processes of the plasma has been achieved in this limit 
\cite{Arnold:2004ti,Chesler:2009cy,Heller:2011ju,Chesler:2011ds}. Formally, the success of hydrodynamics only depends on the isotropization of the stress tensor (i.e.,~the pressure) in the local fluid frame and not necessarily on thermal equilibration, while viscous hydrodynamics accounts for small deviations from an isotropic pressure.  The observation that hydrodynamics may be a very good approximation even in situations where the anisotropy is not small \cite{Chesler:2009cy,Heller:2011ju} was a surprise, and is not yet completely understood. This may be of quite some phenomenological relevance, as viscous hydrodynamic simulations of heavy-ion collisions reveal significant pressure anisotropy at some stages of the collision.

In the future, it is necessary to understand why hydrodynamics seems to provide an accurate description at earlier times and in a wider range of systems than naively expected.  In the case of strong coupling, some of the approximations inherent in the holographic calculations listed above, such as the conformal invariance of the field theory and the limits of infinite 't Hooft coupling and $N_c$, should be relaxed.  To this end, the equilibration of an ${\mathcal N}=4$ SYM plasma was studied at large but finite coupling  \cite{Steineder:2012si,Steineder:2013ana,Stricker:2013lma}, showing a clear weakening of the usual top-down pattern of holographic thermalization.

\subsubsection{Multiplicities and entropy production}\label{sec:5.new}

The particle multiplicities in heavy-ion collisions can be estimated in several ways.  Event generators determine multiplicities from their models of soft particle production followed by fragmentation and hadronization
\cite{Wang:1991hta,Gyulassy:1994ew,Deng:2010mv,Deng:2010xg,ToporPop:2011wk,ToporPop:2010qz,Barnafoldi:2011px,Pop:2012ug,homepage,Lin:2004en,Werner:2012sv}. 
A more first-principles QCD approach comes from color glass condensate (CGC),
a saturation-based description of the initial state in which nuclei in a high-energy
nuclear collision appear to be sheets of high-density gluon matter. In this 
approach, gluon production can be described by $k_{T}$-factorization which
assumes an ordering in intrinsic transverse momentum rather than momentum
fraction $x$, as in collinear factorization. The unintegrated gluon density 
associated with $k_T$ factorization is related to the color dipole forward scattering amplitude which satisfies the JIMWLK evolution equations
\cite{JalilianMarian:1997jx,JalilianMarian:1997gr,Iancu:2000hn}. 
In the large $N_c$ limit, the coupled JIMWLK equations simplify to the Balitsky-Kovchegov (BK) equation \cite{Balitsky:1995ub,Kovchegov:1999yj,Kovchegov:1999ua,Balitsky:2006wa}, a closed-form result for the rapidity
evolution of the dipole amplitude.  The running coupling corrections to the leading log BK equation, rcBK, have been phenomenologically successful in describing the rapidity/energy evolution of the dipole \cite{Albacete:2010sy}.  The initial condition still needs to be modeled, typically by a form motivated by the McLerran-Venugopalan model \cite{McLerran:1993ni,McLerran:1993ka,McLerran:1994vd} with parameters constrained by data ~\cite{Albacete:2012xq}.  The impact parameter dependent dipole saturation
model (IP-Sat) \cite{Kowalski:2003hm,Tribedy:2010ab,Tribedy:2011aa} is a refinement of the dipole
saturation model that reproduces the correct limit when the dipole radius
$r_T \rightarrow 0$.  It includes power corrections to the collinear DGLAP evolution and
should be valid where logs in $Q^2$ dominate logs of $x$.
It should be noted that all of the above approaches involve some parameter tuning at some energy to predict results for other energies; for details and further model references see Ref.~\cite{Albacete:2013ei}.

\begin{figure} 
\begin{center}
\includegraphics[width=0.48\textwidth]{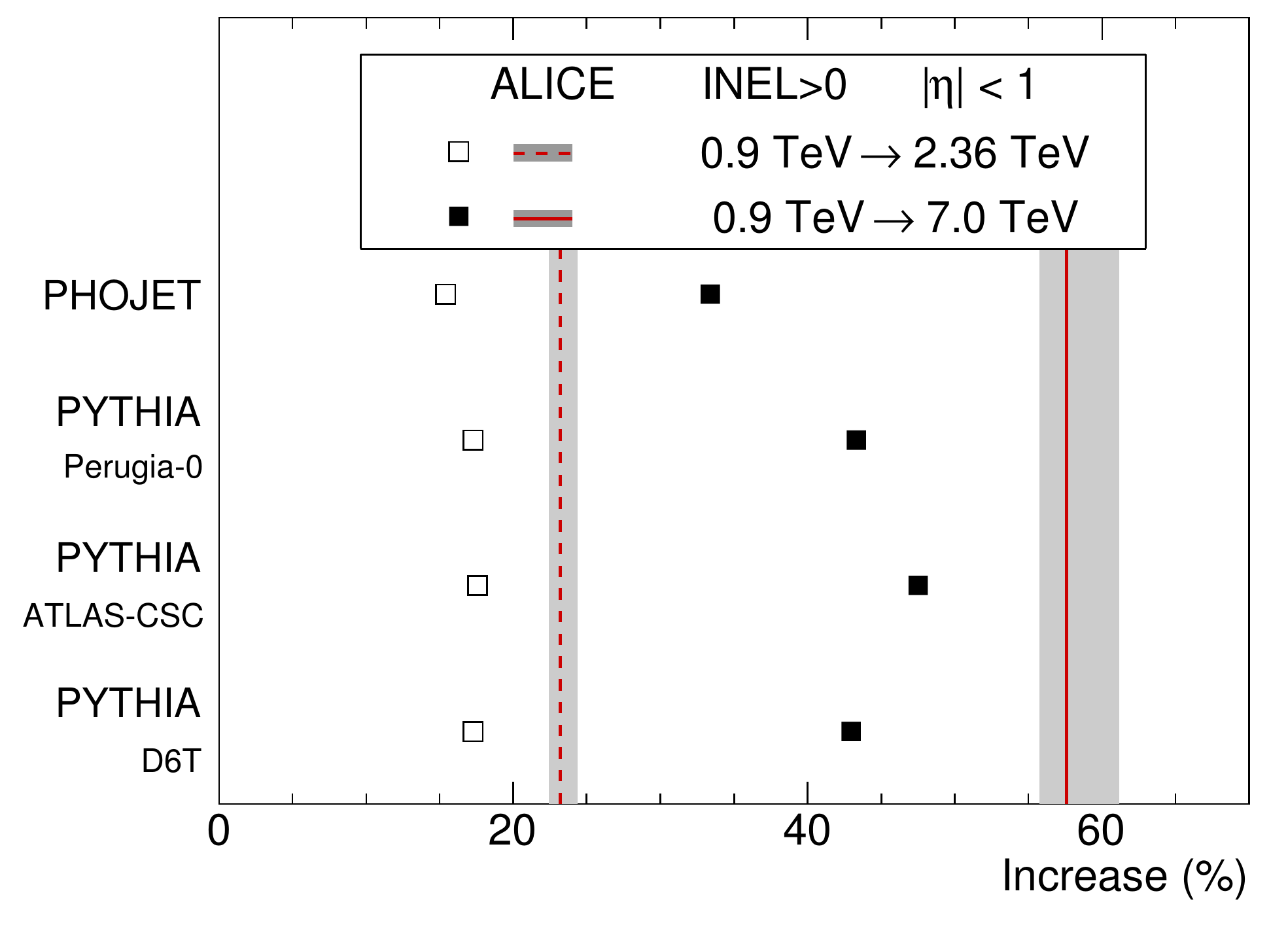}
\caption[]{The relative increase of the charged particle pseudorapidity density for inelastic collisions having at least one charged
particle in $|\eta| < 1$, between $\sqrt{s}=0.9$ and 2.36 TeV (open
squares) and between $\sqrt{s}= 0.9$ and 7 TeV (full squares),
is shown for various models. The corresponding ALICE measurements are
shown by the vertical dashed and solid lines.  The width of the shaded
bands correspond to the statistical and systematic uncertainties added in quadrature 
\protect\cite{Aamodt:2010pp}.
}
\label{fig:dndeta_pp}
\end{center}
\end{figure}

\begin{figure} 
\begin{center}
\includegraphics[width=0.48\textwidth]{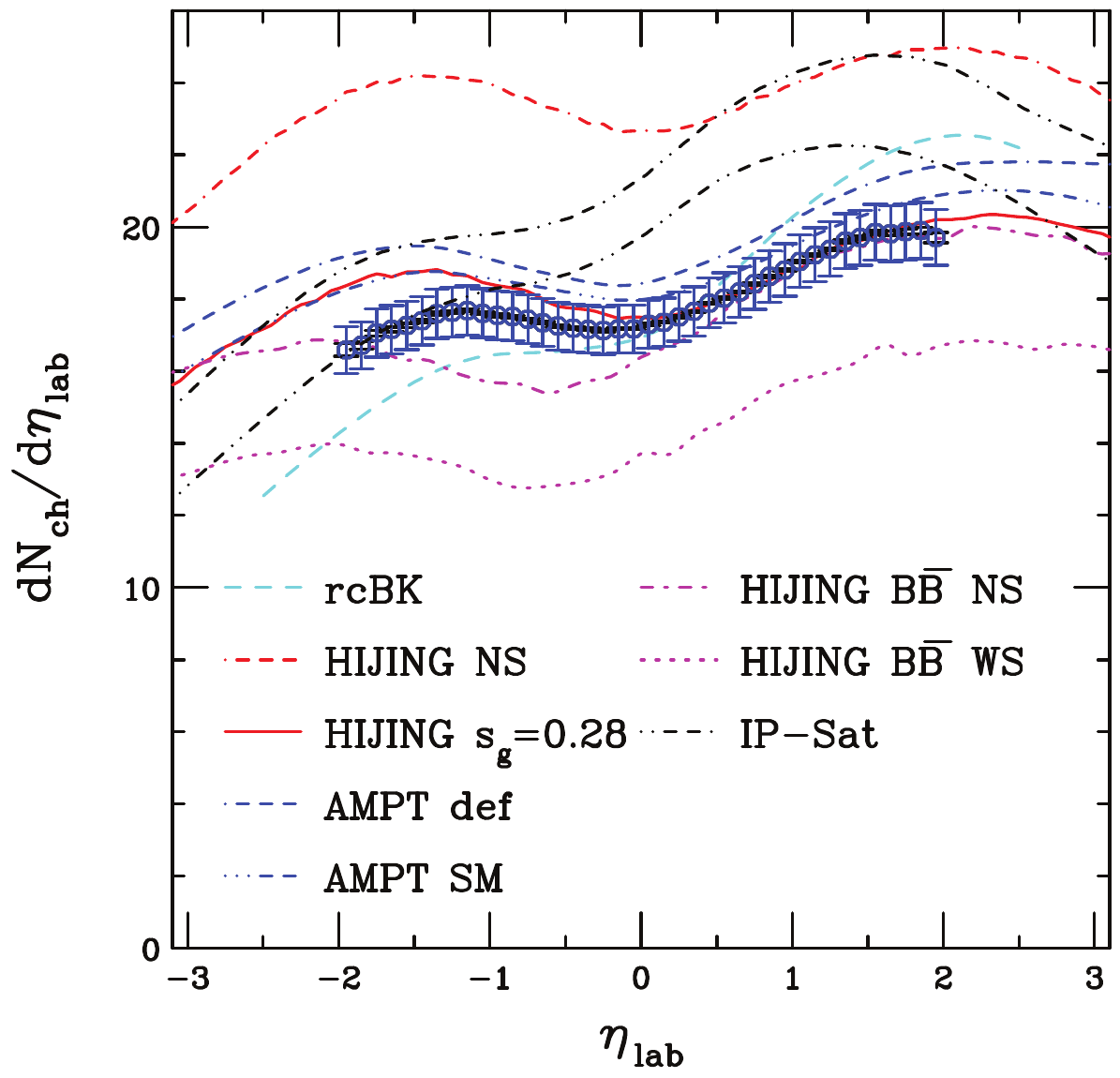}
\caption[]{Charged particle pseudorapidity distributions for \pPb\
collisions at $\sqrt{s_{NN}}= 5.02$ TeV in the laboratory frame. 
A forward-backward asymmetry between the proton and lead hemispheres is clearly visible
with the $Pb$ remnant going into the direction of positive pseudorapidity. 
The rcBK (dashed cyan)
result is from Ref.~\protect\cite{Albacete:2012xq}.  The IP-Sat result is
shown as the dot-dot-dash-dashed black band.
The HIJING2.1 result without (NS, 
dot-dash-dash-dashed red) and with shadowing ($s_g = 0.28$, solid red)
and the HIJINGB$\overline{\rm B}$ result without (dot-dashed magenta) and
with shadowing (dotted magenta) are also shown.  Finally, the 
AMPT-def (dot-dash-dash-dashed blue) and AMPT-SM
(dot-dot-dot-dash-dash-dashed blue) are given.  The ALICE results from
Ref.~\protect\cite{ALICE:2012xs} are given on the right-hand side.  The
systematic uncertainties are shown in blue, the statistical uncertainties are
too small to be visible on the scale of the plot.  From~\protect\cite{Albacete:2013ei}.
}
\label{fig:dndeta_cm_lab}
\end{center}
\end{figure}

\begin{figure} 
\begin{center}
\includegraphics[width=0.48\textwidth]{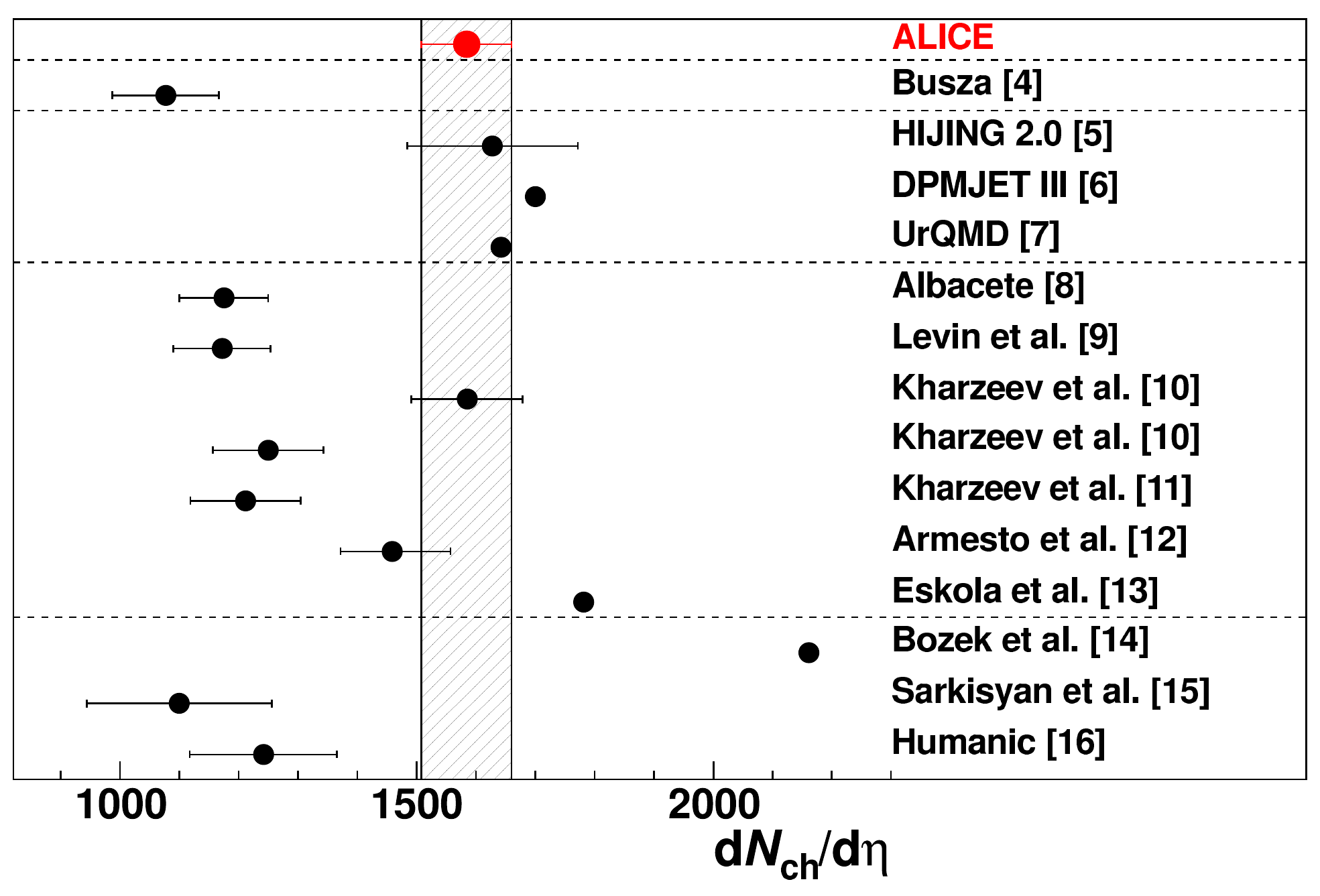}
\caption[]{The charged particle pseudorapidity distributions in \PbPb\ 
collisions at $\sqrt{s_{NN}} = 2.76$ TeV \protect\cite{Aamodt:2010pb} are compared to
model predictions. The horizontal dashed
lines group similar theoretical approaches. For the model references see 
\protect\cite{Aamodt:2010pb}.}
\label{fig:dndeta_PbPb}
\end{center}
\end{figure}

Figures~\ref{fig:dndeta_pp}-\ref{fig:dndeta_PbPb} show model predictions of
the charged particle multiplicity densities in $pp$, $p+$Pb, and Pb+Pb 
collisions compared to data.  

Figure~\ref{fig:dndeta_pp} shows a comparison of charged particle pseudorapidity density
in $pp$ collisions at $\sqrt{s} = 0.9$, 2.36  and 7 TeV, measured by the ALICE 
Collaboration \cite{Aamodt:2010pp}. The results for the relative increase of the 
$dN_{ch}/d\eta$ in $|\eta| < 1$ between 0.9 and 2.36 TeV and between
0.9 and 7 TeV were compared to models.  Three different PYTHIA tunes were
compared, along with PHOJET results.  The Perugia-0 tune and PHOJET were
chosen because they exhibited the largest difference in multiplicity 
distributions at very low multiplicities.  All the models underpredicted the
observed relative increase.

The shapes and magnitudes of the pseudorapidity distributions predicted by 
models are compared to the $p+$Pb test beam data in 
Fig.~\ref{fig:dndeta_cm_lab}.  While several of the calculations are in 
relatively good agreement with the value of $dN_{ch}/d\eta$ at
$\eta_{lab} = 0$, the shapes are generally not compatible with that of the
data.  The rcBK result was calculated assuming the same rapidity to 
pseudorapidity transformation in $pp$ as in $p+$Pb collisions.
Another choice, based on the number of participants in the Pb nucleus would
lead to a flatter distribution, more compatible with the data \cite{Adrianpriv}.
Most of the event generator results disagree with both the shape and magnitude
of the data except for AMPT and HIJING2.0 with shadowing ($s_g = 0.28$ in Fig.~\ref{fig:dndeta_cm_lab}).

Finally, Fig.~\ref{fig:dndeta_PbPb} compares several classes of model 
predictions to the Pb+Pb data at midrapidity, $|\eta|< 0.5$ 
\cite{Aamodt:2010pb}.  The result, $dN_{ch}/d\eta = 1584 \pm 4 ({\rm stat.}) \pm 76 ({\rm syst.})$, 
is a factor of 2.2 larger than the 200 GeV Au+Au result
at RHIC.  All model calculations shown in Fig.~\ref{fig:dndeta_PbPb}
describe the RHIC results.  However, most of them underpredict the
Pb+Pb data by $\sim 25$\%, 
including empirical extrapolations from lower energy data (labeled
Busza); many saturation-based models (only one of the estimates from 
Kharzeev et al. agrees with the data); 
an extrapolation based on Landau hydrodynamics (maximum compression) (labeled
Sarkisyan et al.); and hadronic rescattering (labeled Humanic).
The event generator results in the upper part of Fig.~\ref{fig:dndeta_PbPb}
are generally in relatively good agreement with the data.  Calculations based
on hydrodynamics generally overpredict the data: a
hybrid hydrodynamics and phase-space saturation calculation
(labeled Eskola et al.) overpredicts the multiplicity by 7\% while
a hydrodynamic model with a multiplicity scaled from $pp$ collisions
(labeled Bozek et al.) overestimates the result by 40\%.  This comparison
illustrates that, even if model calculations are tuned to results at one
energy, agreement with higher energy data is not guaranteed.  As 
Fig.~\ref{fig:dndeta_cm_lab} showed, predicting the average multiplicity at
one rapidity also does not guarantee that the full pseudorapidity dependence
can be reproduced.

Finally, multiplicities can be estimated with holographic methods.  In the dual gravity description, local thermalization involves the formation of a horizon.  The area of this
horizon controls the final-state multiplicities \cite{kt1}. 
Gravitational techniques, pioneered by Penrose \cite{Penrose:1969pc}, have provided useful tools for estimating the formation of horizons and given bounds for the related multiplicities using the concept of
``trapped surfaces''.
Several calculations in the dual of ${\mathcal N}=4$ SYM theory have analyzed the formation of trapped surfaces in collisions of planar shock waves \cite{sw1,sw2,sw3,sw4,sw5,sw6,sw7,sw8}. In  \cite{wr}, it was shown that the entropy released during the collision is 60\% larger that the bound obtained from trapped surface calculations, a result that is independent of the collision energy due to the conformality of the system.

\begin{figure}[!th]
\centering
\includegraphics*[width=0.48\textwidth,height=5.0cm]{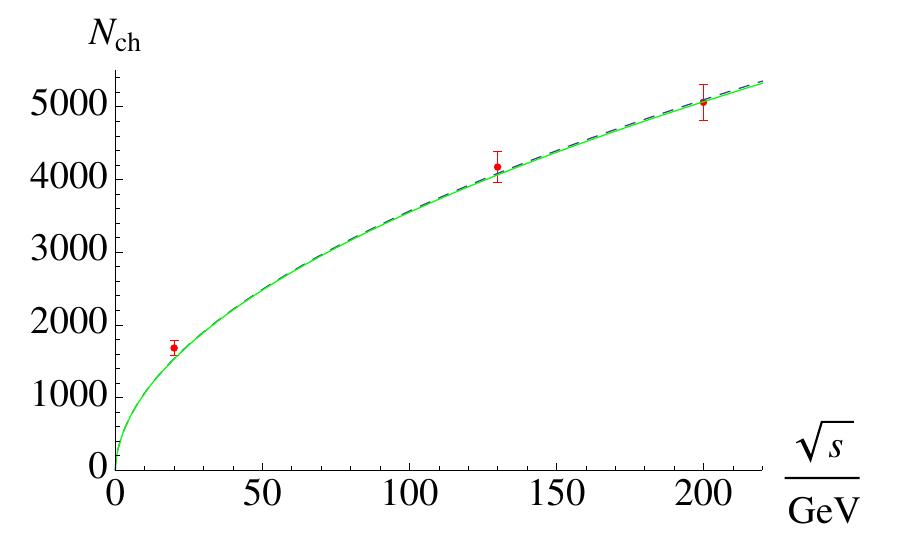} \\
\vspace{0.5cm}
\includegraphics*[width=0.48\textwidth,height=5.0cm]{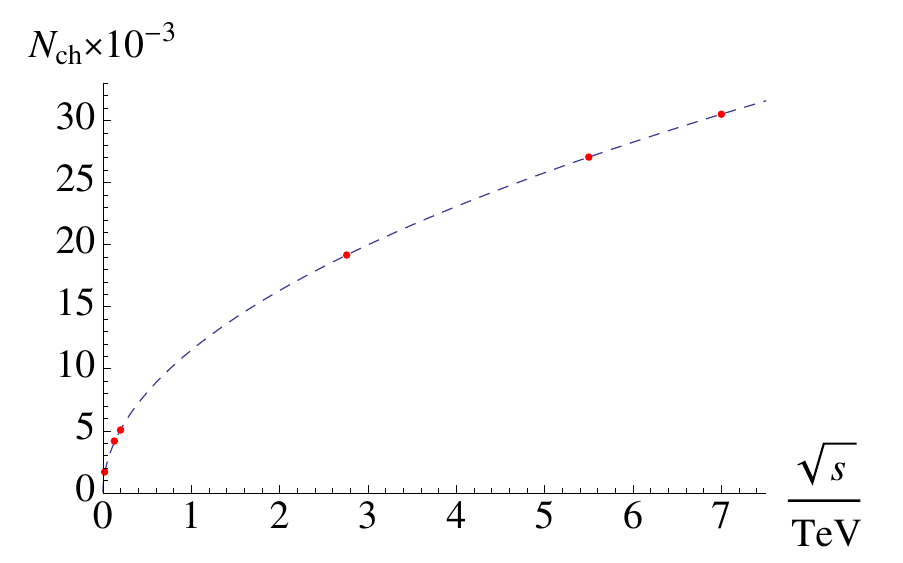}
\caption{The total multiplicity as a function of center-of-mass energy measured in \AuAu\ collisions at RHIC (top) and \PbPb\ collisions at the LHC (bottom). The points on the top figure correspond to RHIC data while the dashed curve shows a prediction from the IHQCD scenario \cite{sw8}. In the lower figure, the dashed line is a prediction of the same holographic calculation, extended to higher energies.  The red point at $2.76$ TeV corresponds to the ALICE 
measurement  \cite{Collaboration:2011rta}, while the other 
red points are predictions for future LHC runs.}
\label{P1}
\end{figure}

In the simplest models involving planar shock waves in an AdS$_5$ space-time, the entropy and the total multiplicities scale as $N_{tot} \sim s^{1/3}$. If the running coupling is simulated with an ultraviolet (UV) cutoff for the trapped surface, the energy dependence of $N_{tot}$ changes to $N_{tot}\sim s^{1/6}$, indicating that violation of conformal invariance may affect the dynamics of heavy-ion collisions.  This issue was studied in two different approaches \cite{sw8}: ``AdS-$Q_s$'', where an explicit UV cutoff is introduced at $r=1/Q_s$ (see \cite{sw8} for details), and Improved Holographic QCD (IHQCD), where conformal invariance is broken due to a dynamical dilaton field \cite{ihqcd1,ihqcd2,ihqcd3}.  The RHIC data point was fit to determine the constant parameter that scales the calculated multiplicity.  Thus, the energy dependence of the
multiplicity is fixed. The upper part of Fig.~\ref{P1} indicates that other RHIC multiplicities are successfully reproduced.
 A subsequent extension to LHC energies is shown in the lower part of Fig.~\ref{P1}. The
 red points are predictions for 2.76, 5.5 and 7 TeV \PbPb\ collisions. The 2.76 TeV 
 result is in agreement with the ALICE result  \cite{Collaboration:2011rta}. As seen in Fig.~\ref{P1}, the agreement of the IHQCD calculation with data is good.

\subsection{Hard processes and medium induced effects} \label{sec:HardAndHeavy}

\subsubsection{Introduction}

The high energies reached in heavy-ion collision experiments at RHIC and the 
LHC allow precision studies of hard processes involving high momentum or mass 
scales. Such probes originate from partonic scatterings in the very initial 
stage of the collision and thus are sensitive to the state of the system at
early times.

A crucial issue in the study of heavy-ion collisions is employing an
appropriate reference system which would disentangle medium effects from 
vacuum expectations.  Proton-proton collisions provide the vacuum reference,
as it was verified at lower energies and then at the LHC.
However, these hard probes are also subject to the state of the 
nuclear matter systems, when no hot matter is produced.
To this end, hard probes have been studied in $d+$Au collisions
at RHIC and, most recently, in \pPb\ collisions at the LHC to separate
initial-state from final-state matter effects.  
A discussion on the theory of the initial state effects in nuclear collisions can be found in
Sec.~\ref{RVpPb}, while experimental results on \pPb\ collisions at the LHC
are presented in Sec.~\ref{chapd:pPbEXP}.

A detailed analysis of phenomena such as parton energy loss 
via collisions and medium-induced gluon radiation offers new insight into 
the most fundamental properties of hot QCD matter and constitutes an important 
subfield of heavy-ion physics.  Perturbative calculations of radiative energy 
loss \cite{Majumder:2010qh} generally predict that the energy loss of a parton 
should be proportional to the Casimir eigenvalue of its color charge \cite{Dokshitzer:2001zm}. This 
implies that gluons should lose approximately twice as much energy as quarks.
In addition, the energy lost by heavy quarks should be reduced by the so-called 
dead-cone effect, i.e., the suppression of gluon radiation at small angles 
\cite{Dokshitzer:2001zm}.  However, strongly-coupled gauge theories applying 
the AdS/CFT conjecture often predict that energy loss has a stronger 
dependence on the path length of the probe through the medium. These 
different scenarios, as well as other issues related to the theoretical description of  
parton energy loss, are discussed in Sec.~\ref{Sec:ELOSSTHEORY}.

Some of the best known hard probes are the quarkonium states.
Bound states of charm and bottom quarks are predicted to be suppressed in 
heavy-ion collisions as a consequence of ``melting'' due to color screening in 
a QGP. Suppression of the $J/\psi$ was first predicted by Matsui and Satz
\cite{Matsui:1986dk} in 1986.  This idea was later developed into a sequential 
pattern of suppression for all heavy quarkonium states since the magnitude of 
suppression should depend on their binding energy with the most strongly bound 
$\Upsilon$(1S) state showing only little modification. However, other cold
and hot matter effects may also contribute, see Secs.~\ref{RVpPb} for further discussion.

Experimentally, the focus of hard probes has mainly been on high \Pt\ hadrons,
heavy flavors and quarkonium states.  Manifestations of parton energy loss
were first observed as strong suppression of back-to-back-emission of high
\Pt\ hadrons  at RHIC. The higher energies of the LHC allow these studies to
be expanded to much higher \Pt\ as well as fully reconstructed jets, as 
discussed in Sec.~\ref{Sec:ELOSSEXP}.  The mass dependence of parton energy
loss, as well as other open heavy flavor observables are also described in
Sec.~\ref{Sec:HEAVY}, along with results on quarkonium production and 
suppression.  Early intriguing
results emerging from the LHC \pPb\ program are also 
presented.



\subsubsection{Theory of hard probes}
\subsubsection*{Nuclear matter effects in \pA\ collisions}
\label{RVpPb}

As discussed in the introduction to this section, a reliable reference for
heavy-ion results is critical for understanding the strength of plasma effects
relative to non-plasma effects, referred to here as cold nuclear matter effects.
These effects are in addition to the vacuum reference obtained in proton-proton
collisions.  They have been studied already in fixed-target interactions in
addition to higher energy measurements in $d+$Au and \pPb\ collisions at the 
RHIC and LHC colliders.  In this section, effects important for the cold nuclear
matter baseline are introduced and discussed.  We do not discuss 
results of the highest multiplicity \ppcoll\ and \pPb\ collisions,
for those, see Sec.~\ref{chapd:pPbEXP}.

There are several important cold nuclear matter effects that need to be taken
into account when determining the strength of deconfinement effects on a 
particular final state.  The most general, affecting all production processes, 
is the modification of the parton distributions in nuclei, often
referred to as shadowing.  This effect is well known, starting from the 
EMC effect at relatively large Bjorken $x$ \cite{Ashman:1992kv} 
and studied further at lower $x$ in nuclear deep-inelastic scattering (nDIS) 
experiments at SLAC \cite{Gomez:1993ri}, CERN 
\cite{Amaudruz:1995tq,Arneodo:1995cs,Arneodo:1996rv}, HERA \cite{Airapetian:2007vu,Airapetian:2011jp}, 
and 
Fermilab \cite{Adams:1995is}.
Given the fixed-target nature of these experiments, only moderately low values
of $x$ ($x \geq 0.01$) are reached at perturbative momentum transfers 
($Q^2 > 1$~GeV$^2$).  These data are augmented by Drell-Yan hadroproduction 
data at higher $Q^2$ and moderate $x$ \cite{Alde:1990im,Vasilev:1999fa}.

Since the nDIS experiments probe only charged parton densities, the nature and
magnitude of the effect on the nuclear gluon density was known only from the 
$Q^2$ evolution of the structure function \cite{Arneodo:1996ru}
and the momentum sum rule, see e.g. Ref.~\cite{Eskola:2009uj}.  While data from the RHIC collider have extended the
range in $x$ and $Q^2$, in particular through $\pi^0$ production \cite{Adler:2006wg},
they have not directly probed the gluon density.  One possible
experimental means of probing the nuclear gluon density is through
ultraperipheral collisions at the LHC \cite{Baltz:2007kq}.  In these collisions, 
the nuclei do not touch and only interact electromagnetically so
that $J/\psi$ photoproduction involves the low $x$ gluon density in a single
nucleus.  The ALICE collaboration has already published such data and shows 
that this method can eliminate certain shadowing parameterizations 
\cite{Abelev:2012ba,Abbas:2013oua}.

The effects of shadowing in nuclei are parameterized by various groups using
global fitting methods similar to those used to evaluate the parton densities 
in the proton, see Chapter~\ref{sec:chapb}.  The first such parameterizations
were rather crude, involving only a single leading order (LO) modification for
quarks, antiquarks and gluons as a function of $A$ and $x$ but independent of 
$Q^2$ \cite{Mueller:1985wy}.  Greater levels of sophistication
have been introduced until, currently, LO and NLO sets are available with up
to 31 error sets, evolving quarks, antiquarks and gluons separately with
$Q^2$.  Some recent sets are EPS09 \cite{Eskola:2009uj}, DSSZ \cite{deFlorian:2011fp}, 
HKN07 \cite{Hirai:2007sx}  and FGS10 \cite{Frankfurt:2011cs}.  Regardless of the level of 
sophistication and general agreement between different sets on the valence and 
sea quark densities in nuclei, the uncertainty on the
gluon density in the nucleus remains large without general agreement on the
best fit shape.

Quark-dominated production processes in nuclear collisions also exhibit a
dependence on the relative neutron-proton content of the nucleus (isospin).  
For some final states, the change in production rates with nuclei related to 
isospin is as strong or stronger than that due to shadowing \cite{Vogt:2000hp,Accardi:2004be}.
The high energies of the LHC allow studies of these effects at higher $Q^2$
than ever before with low to moderate values of $x$, such as for vector boson production 
\cite{Vogt:2000hp,Accardi:2004be,Guzey:2012jp}.  
Such data is available already for $W$ and $Z^0$
production in Pb+Pb collisions at $\sqrt{s_{NN}} = 2.76$ TeV from the 
ATLAS \cite{Milov:2012pd} and CMS \cite{Chatrchyan:2011ua,Chatrchyan:2012nt} collaborations.

Another significant unknown relating to nuclear shadowing is its dependence on
impact parameter or collision centrality.   Fixed-target data were presented 
as a function of $A$ and did not try to distinguish between nuclear 
interaction points.  One exception was an experiment studying gray
tracks in emulsion which did see hints of an impact parameter dependence
\cite{Kitagaki:1988wc}.  The impact parameter dependence was neglected in most previous
parameterizations, the exception being the
FGS parameterizations based on diffractive data \cite{Frankfurt:2011cs}.  Instead, 
assumptions based on either a linear
dependence on path length through the nucleus or the nuclear density were
introduced \cite{Emel'yanov:1998df}.  Only recently has data emerged to challenge the 
assumption of a linear dependence.  The PHENIX
$d+$Au $J/\psi$ data suggested a stronger than linear dependence \cite{Adare:2010fn}.
These results prove challenging for the recent EPS09s spatially-dependent 
modifications which retain up to quartic powers in
the expansion of the centrality dependence as a function of path length for
$A$-independent coefficients \cite{Helenius:2012wd}.  Instead these data suggest that 
shadowing is concentrated in the core of the nucleus
with radius of $R\sim 2.4$~fm with a relatively sharp surface, a width of
$d\sim 0.12$~fm \cite{McGlinchey:2012bp}.  These
studies need to be backed up with more data over more final states.

A second cold matter effect is energy loss in medium.  This has been treated
as both an initial-state effect, related to soft scatterings of the projectile 
parton in the nucleus before the hard scattering
to produce the final-state particle, and a final-state effect where the
produced parton scatters in the medium.  Initial-state energy loss has been
studied in Drell-Yan production \cite{Gavin:1991qk}.
The effect has generally been found to be small, too small to be effectively
applied to $J/\psi$ production at large Feynman $x$ ($x_F$) \cite{Wohri:2011zz}.  
In addition, there is an inherent ambiguity when applying
initial-state energy loss to Drell-Yan production since most groups
parameterizing the nuclear parton densities include these same Drell-Yan data 
to extract the strength of shadowing on the
antiquark densities \cite{Eskola:2009uj}.  Also, by forcing the loss to be large 
enough to explain the high $x_F$
behavior of $J/\psi$ production in fixed-target interactions \cite{Leitch:1999ea}
violates the upper bound on energy loss
established by small angle forward gluon emission \cite{Brodsky:1992nq}.  
More recently, it has been proposed that
rather than an initial-state effect, cold matter energy loss should be treated
as a final-state effect, with scattering of the produced final-state with 
gluons in the medium \cite{Arleo:2012rs}.  This would eliminate
the ambiguity of shadowing relative to initial-state energy loss in Drell-Yan
production and, indeed, eliminate the need to introduce energy loss effects 
on Drell-Yan production completely.  The final-state energy loss in
$pA$ collisions is currently implemented for quarkonium production as a
probability distribution dependent on the energy loss
parameter. The effect modifies the $x_F$ and $p_T$ distributions in a rather
crude fashion since the quarkonium distribution in $pp$ collisions is 
parameterized as a convolution of factorized power laws, $ \propto (1-x)^n
(p_0^2/(p_0^2 + p_T^2))^m$, rather than using a quarkonium production model 
\cite{Arleo:2012rs,Arleo:2013zua}.  
It has yet to be implemented for other processes.

As previously mentioned, initial-state energy loss in the medium can be
connected to transverse momentum kicks that broaden the $p_T$ distributions 
in nuclei relative to those in $pp$ collisions.
This can be related to the Cronin effect \cite{Kluberg:1977bm}
and was first seen for hard processes in fixed-target
jet production \cite{Corcoran:1990vq}.

Nuclear absorption, which affects only quarkonium states, involves breakup
of the nascent quarkonium state in cold nuclear matter
\cite{Gerschel:1988wn}.  Thus it is a final-state
effect.  The matter that causes the state to break up is typically assumed to
be nucleons only.  However, $J/\psi$ suppression in nuclear collisions was
also attributed to breakup with produced particles called comovers.  

Absorption is the only effect we have discussed that is related to the size
and production mechanism of the interacting state and can be described by a
survival probability, $S_A^{abs} = \exp\{ -\int_z^\infty dz' \rho_A(b,z')
\sigma^C_{abs}(z-z')\}$  where $z'$ is the production point and $z$ is the
dissociation point; $\rho_A(b,z')$ is the nuclear matter density; and 
$\sigma^C_{abs}$ is the effective absorption cross section for quarkonium
state $C$ \cite{Vogt:2001ky}.  Because the
quarkonium states have different radii, $\sigma^C_{abs}$ is e.g.
dependent upon the final-state size so that $\sigma^{\psi'}_{abs} 
\approx 4\sigma^{J/\psi}_{abs}$ \cite{Povh:1987ju}. 
Color singlet quarkonium states
are assumed to grow from their production point until they reach their 
asymptotic size, typically outside the nucleus \cite{Blaizot:1989de,Gavin:1990gm}.  
In this case, the survival
probability is less than unity for rapidities where the state can hadronize 
in the interior of the nucleus but equal to unity for all rapidities where
the state only reaches its final-state size outside the target.  Color octet
quarkonium states can interact strongly inside the target but, if they
convert to the final color-singlet quarkonium state inside the target before
interacting and dissociating, they will interact as singlets, giving a 
different suppression pattern.  The color octet to singlet conversion depends
on the proper time after production \cite{Kharzeev:1995br,Arleo:1999af}
and is most important for rapidities which
the quarkonium state can interact in the interior of the nucleus and, again,
is inactive when the state hadronizes outside the nucleus.

Previous studies have shown the absorption cross section to depend on
rapidity (or $x_F$) as well as the nucleon-nucleon 
center of mass energy, $\sqrt{s_{NN}}$ with stronger absorption at lower
energies \cite{Lourenco:2008sk}.  
Increased effective absorption at backward rapidity may be due
to interaction or conversion inside the target while increased effective
absorption at forward rapidity may be due to energy loss.  However, some finite
value of $\sigma^C_{abs}$ is assumed for all rapidities.  The $J/\psi$
has been most studied.  Larger effects, at least at midrapidity, have been
seen for the $\psi'$ \cite{Leitch:1999ea}.  
Such effects on $\Upsilon$ production may also be
expected with stronger nuclear effects on the $\Upsilon(2S)$ and $\Upsilon(3S)$
relative to the $\Upsilon(1S)$ \cite{Alde:1991sw}.

Interactions with comovers, while first thought to be an important effect in $AA$ collisions
\cite{Gavin:1988hs,Vogt:1988fj,Horvath:1988qd},
were later assumed to be small and, indeed, negligible \cite{Kharzeev:1994pz}.  
More recent data on
$\psi'$ production as a function of the number of binary nucleon-nucleon
collisions, $N_{coll}$, in $d+$Au collisions at $\sqrt{s_{NN}} = 200$ GeV
shows a very strong dependence on $N_{coll}$ for the $\psi'$ compared to
almost no effect on the $J/\psi$ \cite{Adare:2013ezl}.  
Since the $\psi'$ mass 
is only $\sim 50$ MeV/$c^2$ below the $D \overline D$ threshold, 
interactions with comoving hadrons and/or partons could
easily break up the $\psi'$ but not the $J/\psi$.  Unfortunately the charmonium
production rates have not been measured in conjunction with charged hadron
multiplicity at RHIC.  However, such data exist for $pp$ collisions 
at the LHC and show that the
$J/\psi$ multiplicity increases with the charged particle multiplicity at
both mid- and forward rapidity \cite{Abelev:2012rz}.  
If the $\psi'$ exhibits similar behavior, then
one might further expect stronger $\psi'$ suppression in higher multiplicity
(larger $N_{coll}$) collisions.

\subsubsection*{Energy loss theory}
\label{Sec:ELOSSTHEORY}

The theory of parton energy loss in hot matter has come a long way from the
``jet quenching'' predictions by Bjorken and others 
\cite{Bjorken:1982tu} describing radiative energy loss by a
fast parton.  As discussed later in Sec.~\ref{Sec:ELOSSEXP}, experimentally
the field has gone from studies of leading particle suppression at RHIC to
true jet suppression at the LHC.  Ongoing experimental studies address the
influence of color charge and quark mass on the magnitude of the effect; the
relative contributions of radiation and collisional (elastic) loss; the 
dependence on the thickness of the medium; and, in the case of
jets, where the lost energy goes (related to the dependence on the jet cone
radius).  Here we describe some of the pQCD approaches to parton energy loss,
some remaining open questions, and new approaches in the context of gravity 
dual theories.

The pQCD approaches have been summarized in detail in 
Ref.~\cite{Armesto:2011ht}.  They are known by a number of acronyms including
AMY \cite{Arnold:2001ms,Arnold:2002ja}, ASW \cite{Wiedemann:2000za,Salgado:2003gb,Armesto:2003jh}, BDMPS \cite{Baier:1996kr,Baier:1996sk,Baier:1998kq,Zakharov:1996fv,Zakharov:1997uu}, DGLV \cite{Gyulassy:1999zd,Gyulassy:2000er,Djordjevic:2003zk}, HT \cite{Wang:2001ifa,Majumder:2009zu} and WHDG \cite{Wicks:2005gt}.
They differ with respect to modeling the medium,
the kinetic approximations taken into account, and the treatment of multiple
gluon emission.  We will briefly mention the differences, for full details,
see Ref.~\cite{Armesto:2011ht}.

There are several ways of modeling the medium that the fast parton passes
through.  The simplest is to treat the medium as a collection of scattering
centers with the parton undergoing multiple soft scatterings.  A particular
approach in this treatment is the opacity expansion which depends on the
density of scattering centers (or, equivalently, the parton mean-free path)
and the Debye screening mass.  This expansion includes the power-law tail of the
QCD scattering cross section, resulting in shorter formation times for the
radiation compared to multiple soft scatterings alone.  The medium has also 
been characterized by matrix elements of gauge field operators, in particular
in the higher-twist approach.  These higher-twist matrix elements are 
factorized into the nuclear parton densities and matrix elements describing
the interaction of the partons with the medium in terms of expectation values
of field correlation functions.  Finally, the medium has been formulated
as a weakly-coupled system in thermal equilibrium.  In this case, all the
properties are specified by the temperature and baryon chemical potential.
This approach is really valid only in the high temperature regime, $T \gg T_c$.

All the approaches, however, make similar assumptions about the kinematics
of the medium.  They assume that the initial parton and the radiated gluon
follow eikonal trajectories with both the parton energy, $E$, and the emitted
gluon energy, $\omega$, much greater than the transverse momentum exchanged
with the medium, $q_T$: $E \gg q_T$, $\omega \gg q_T$.  They also
assume that the gluon energy is much larger than its transverse momentum, 
$k_{T}$: $\omega \gg k_{T}$.  In the case of massive quarks, this 
constraint leads to the ``dead-cone'' effect where gluon radiation is suppressed
for angles where $k_T/\omega < M/E$ \cite{Dokshitzer:2001zm}.  
Finally, they all assume some sort of localized
momentum transfer with a mean-free path much larger than the screening length:
$\lambda \gg 1/\mu_D$.

Multiple gluon emission is treated differently in the models.  Some assume
a Poisson probability distribution for the number of emitted gluons with an
energy distribution following a single gluon emission kernel.  This procedure
can lead to a distribution of energy loss that does not conserve energy if
the degradation of the parent parton momentum is not dynamically updated. 
Interference between medium-induced and vacuum radiation is included but 
the parton fragments in vacuum.  Other approaches take a coupled evolution
procedure with rate equations or medium-modified DGLAP evolution.  The emission
probability changes as the jet energy degrades, decreasing the path length
through the medium.  

In most approaches, the energy loss is characterized by the transport
coefficient $\hat{q}$, the mean of the squared transverse momentum
exchanged with the medium per unit path length.  The pQCD approaches 
described above were compared and contrasted for the simplified
``brick'' problem in Ref.~\cite{Armesto:2011ht}.
This problem involves a uniform, finite 
block of quark-gluon plasma surrounded by vacuum. The goal was to study the 
energy lost by a high-energy parton produced inside the brick which travels
a distance $L \sim 2$ fm through it before exiting into the vacuum. This setup 
provides a useful test bed for model comparison because it separates the 
conceptual differences from other complications inherent in heavy-ion 
collisions, such as modeling the hydrodynamic flow. The aim was to develop a 
``master'' formalism which could reproduce all other representations in 
limiting cases.  Thus each group's results could be reproduced by turning 
approximations on and off, making it possible to examine the physical processes
occurring as well as quantitatively assess which approximations are the most 
robust. This goal has not quite been achieved, though some progress has 
been made. 

There are several technical issues not mentioned previously 
that need to be taken into 
account when comparing models.  The first is the approximation that 
bremsstrahlung radiation (and/or pair production) is nearly collinear to the 
initial high-energy parton.  This may, however, not always be the case in
relevant situations \cite{Armesto:2011ht} which may be sensitive to soft and 
non-collinear gluon bremsstrahlung.  Some more recent formulations have
incorporated non-collinear radiation and have at least roughly accounted for 
the accompanying kinematic constraints.  However a universal treatment is 
still lacking.

Next, systematic organization of corrections to energy loss calculations has not
yet been achieved.  An illustration of this is the Landau-Pomeranchuk-Migdal 
(LPM) effect, which accounts for the difference between the gluon formation time
and the time between scatterings in the medium.  In a dense medium, a 
high-energy parton undergoes multiple scatterings before a bremsstrahlung gluon
forms.  While the treatment of the LPM effect in the collinear approximation
is understood (it diagrammatically corresponds to the resummation of an 
infinite class of diagrams), the systematization of corrections to these
calculations order-by-order in perturbation theory is unknown. There have been 
some recent attempts to organize these corrections employing Soft Collinear 
Effective Theory \cite{D'Eramo:2010ak,Ovanesyan:2011kn,D'Eramo:2012jh}, but it has not
yet been accomplished. In the same framework in \cite{Benzke:2012sz} a gauge 
invariant definition of the jet quenching parameter has been obtained enabling to relate 
it  to the quark-antiquark static potential \cite{Laine:2012ht,Benzke:2012sz}.
 Recently a first step towards calculating jet quenching 
via lattice simulation has been 
undertaken
\cite{Laine:2013apa}.

We now turn to a somewhat more fundamental issue concerning these calculations. 
Most derivations of jet energy loss assume that the coupling between the 
initial high-energy parton and the two subsequent daughter partons is weak 
during high-energy bremsstrahlung or pair production: $\alpha_{s}(Q_T) 
\ll 1$. The relevant scale in the coupling, $Q_T$, is the transverse 
momentum between the two daughter partons.  In thick media $Q_T$ scales 
only weakly with the initial parton energy, $Q_T \sim (\hat q E)^{1/4}$. 
The squared transverse momentum gained per unit length as the parton traverses
the medium, $\hat q$, is, however, a characteristic of the medium. For 
realistic jet energies, $\alpha_{s}$ might indeed be relatively small but not
very small.  It is thus important to understand the size and nature of the 
corrections to the weak-coupling limit.

A QCD-like toy model in which the question of scales can be (and, indeed, has 
been) investigated is again the large-$N_{c}$ ${\cal N} = 4$ SYM theory. As a 
warm-up for more complicated problems in jet energy loss, the stopping distance 
of a high-momentum excitation in the plasma can be calculated. For  
${\cal N} = 4$ SYM and QCD, the answer is (up to logs) that the maximum 
stopping distance scales with energy as $E^{1/2}$, see Ref.~\cite{Arnold:2009ik} 
for explicit QCD results. For ${\cal N}{=}4$ SYM, the calculation may, however, 
also be carried out at strong coupling.  In this case, application of the 
AdS/CFT duality leads to an energy dependence of $E^{1/3}$. It is unknown how 
the $E^{1/2}$ dependence for weak coupling transforms to $E^{1/3}$ at strong
coupling.  Understanding this transition may also help to understand how
to treat the problem of small but not very small $\alpha_{s}(Q_T)$ of real QCD. It may well be that a key element in the resolution of this open puzzle will be an efficient use of effective field theory techniques.

In holographic investigations of energy loss, another particularly
straightforward problem is the determination of the drag force felt by a heavy 
quark traversing a strongly-coupled ${\mathcal N}=4$ SYM plasma  
\cite{her,lrw,gub1}. In the simplest formulation of the problem \cite{her}, 
the quark is represented by an open string hanging from the boundary, where
the string endpoint, attached to a D-brane, is being pulled along a given
spatial direction with constant velocity $v$. The equations of motion of the 
string are solved and the radial profile of the trailing string found as it 
moves through a black hole background representing the deconfined heat bath. 
The energy absorbed by the string is calculated and the drag force is
found to scale with the square root of the 't Hooft coupling.

Since the appearance of the original works on heavy quark energy loss at strong coupling \cite{Herzog:2006gh,CasalderreySolana:2006rq,Liu:2006ug},
the picture has been improved and expanded. An important development has been the study of the stochastic nature of the system analogous to the dynamics of heavy particles in a heat bath, giving rise to Brownian motion. This diffusive process was first considered in a holographic setting \cite{tea}, employing the Schwinger-Keldysh formalism. Subsequently, a study of the (quantum) fluctuations of the trailing string has provided information about heavy quark momentum broadening as it moves through the plasma \cite{gubser,Casal}. The 
stochastic motion has also been formulated as a Langevin process associated with the correlators of string fluctuations \cite{deboer,sonteaney}. These developments are closely related to the determination of transport coefficients in the holographic picture, see Sec.~\ref{sec:d:TraCoeSpe}.

In most experiments, heavy quarks move at relativistic velocities. Therefore, it is necessary to also study the relativistic Langevin evolution of a trailing string in the ${\cal N}=4$ case \cite{iancu}. A similar study in non-conformal theories, in particular IHQCD, was performed in 
\cite{l1,l2}. 

Finally, a salient feature of the above picture involves the presence of a string world-sheet horizon with a Hawking temperature $T_s$, distinct from that of the strongly-coupled plasma. In the conformal case, $T_s = T(1-v^2)^{1/4}\leq T$ where $v$ is the velocity of the heavy quark. This temperature controls the world-sheet ensemble of the trailing string, which is not in thermal equilibrium with the surrounding plasma.  


\subsubsection*{Quarkonium interaction at finite temperature and quarkonium suppression}
\label{Sec:JPSITHEORY}

Since the pioneering paper of  Matsui and Satz \cite{Matsui:1986dk}, 
the suppression of quarkonium  in a hot medium has been  considered 
one of the cleanest  probes of deconfined matter, detected as a suppressed 
yield in the easily accessible dilepton decay channel (see e.g. 
Refs.~\cite{Brambilla:2010cs,Brambilla:2004wf}).
However, quarkonium  suppression as a diagnostic tool of hot media has 
turned out to be quite challenging for several reasons.
On one hand, the effect has to be carefully disentangled from  
nuclear matter effects (discussed in the first subsection of Sec. 
\ref{RVpPb} and from recombination effects (relevant at least for 
charmonium suppression in colliders, particularly at LHC, see the discussion 
in  Sec. \ref{Sec:ELOSSEXP} in the quarkonium subsection). On the other hand,   
the level of quarkonium suppression measured in heavy-ion collisions has 
to be defined with respect to a clean baseline (at colliders, suppression 
has been investigated employing the nuclear modification factor $R_{AA}$, 
defined as the quarkonium yield in nucleus-nucleus 
collisions divided by the corresponding yield in $pp$, scaled by the number 
of binary collisions, see Sec. \ref{Sec:ELOSSEXP}  
for a discussion) and the contribution of decays 
from the excited states to lower-lying states has to be disentangled from 
the measured yield to extract the direct yield.
Additionally, it is critical  to understand the way that heavy quarks 
interact in the hot medium and what this 
brings to quarkonium suppression.
 
Originally, Matsui and Satz argued that, in a deconfined medium, the 
interaction between the heavy quark and the heavy antiquark would be 
screened, leading to the dissolution of the quarkonium state at a 
sufficiently  high temperature.  The naive expectation was that the static 
$Q\overline{Q}$ potential would be screened by 
$\exp \{- m_D (T) r\}$ where $m_D$ is the Debye mass, the temperature-dependent
 inverse of the screening length of the  chromoelectric interaction and 
$r$ is the distance between the quark and antiquark.
Thus quarkonia states would function as an effective thermometer for the 
medium, dissociating at different temperatures, depending on their radii.
In particular, for temperatures above the transition temperature, $T_c$, 
the range of the heavy quark 
interaction would become comparable to the bound state radius. 
Based on this general observation, 
one would expect that the charmonium states, as well as the excited 
bottomonium states, do not remain bound at
temperatures above the deconfinement transition.  This effect is referred 
to as quarkonium {\it dissociation} or quarkonium {\it melting}.

However up to recently no proper tool
for defining and calculating the quarkonium 
potential at finite $T$ was developed. 
Most prior investigations were 
performed with phenomenological potentials
inspired by lattice calculations of the $Q \overline Q$ free energy.
The free energy was chosen because,
in the zero-temperature limit, it coincides (up to small corrections) 
with the zero-temperature potential,
while it flattens at finite $T$ and long distance, consistently with  
 screening \cite{Kaczmarek:2005ui,Kaczmarek:2005gi,Bazavov:2012fk}.

On the lattice, the free energy is extracted from the calculation of 
quark-antiquark Polyakov loop correlators. There are singlet and octet 
channels that are gauge dependent.  An average gauge-independent free 
energy can also be defined.
The three above-mentioned lattice free energies do not exhibit the same 
dependence on the $Q\overline{Q}$ separation distance
and thus lead to different binding energies when used as 
phenomenological potentials in the Schr\"odinger equation 
\cite{Philipsen:2008qx}. There are many papers in the literature either
employing the singlet free energy or the corresponding internal energy 
as phenomenological
potentials to calculate quarkonium binding energies at finite $T$ 
(see e.g. Refs.~\cite{Rapp:2008tf,Kluberg:2009wc})
or reconstructing the lattice meson correlation functions from the 
Schr\"odinger wavefunctions  \cite{Mocsy:2007jz}
to understand which approach is better.

Lacking a comprehensive theoretical framework, other effects have often 
been included in addition to screening of the potential, such as the 
break up of the bound state by inelastic gluon collisions (gluodissociation)
\cite{Kharzeev:1994pz,Xu:1995eb,Bhanot:1979vb,Peskin:1979va}      
or by light partons in the medium (quasi-free dissociation) 
\cite{Grandchamp:2001pf,Grandchamp:2002wp,Rapp:2008tf}.

Information about the behavior of the quarkonium bound state at finite $T$ 
can also  be obtained directly from the spectral function. On the lattice 
this quantity is accessible via   calculations 
of the corresponding Green's function employing the maximum entropy method 
(MEM) \cite{Ding:2012sp,Karsch:2012na}
The challenges of this  approach  have been discussed in Sec.~\ref{sec:d:TraCoeSpe}.

It is therefore very important to find a QCD-based theoretical 
framework that can provide a precise definition of the finite temperature
$Q\overline{Q}$ potential and thus an unambiguous calculational tool.
Such a definition has been obtained recently for weak-coupling through 
construction of appropriate effective field theories (EFT).

First \cite{Laine:2006ns,Laine:2007qy}, the static potential was calculated 
in the regime $T \gg 1/r \simg m_D$  by performing an analytical continuation 
of the Euclidean Wilson loop to real time. The calculation was done in 
weak-coupling resummed perturbation theory. The imaginary part of the gluon 
self energy gives an imaginary part to the static $Q \overline Q$
potential and hence a thermal width to the quark-antiquark bound state 
(see also \cite{Beraudo:2007ky}).
Subsequently, an EFT framework for finite-temperature quarkonium in real time 
was developed \cite{Brambilla:2008cx}  
(see \cite{Escobedo:2008sy, Escobedo:2010tu} for results in QED)  
working in real time and at small coupling $g$, $g T \ll T$, and for the  
velocity $v$ of the quark in the bound state of order  $v \sim \als$  
(expected to be valid for tightly bound states: $\Upsilon(1S)$, $J/\psi$, ...~).

The EFT description starts from the observation that 
quarkonium in a medium is characterized by different energy and momentum scales.
As previously explained in Sec. \ref{sec:subsecC11} beyond
the scales typical of nonrelativistic bound states
($m_Q$, the heavy quark mass; $m_Qv$, the scale of the typical inverse 
distance between the heavy quark and antiquark; $m_Qv^2$, the scale of the 
typical binding energy or potential energy and $\Lambda_{\rm QCD}$)  there are 
thermodynamical scales ($T$, the temperature; $m_D$, the Debye mass, 
$\sim gT$ in the perturbative regime)
and lower scales such as the magnetic scale that we neglect in the following.

\begin{figure} 
\begin{center}
\includegraphics[width=0.45\textwidth]{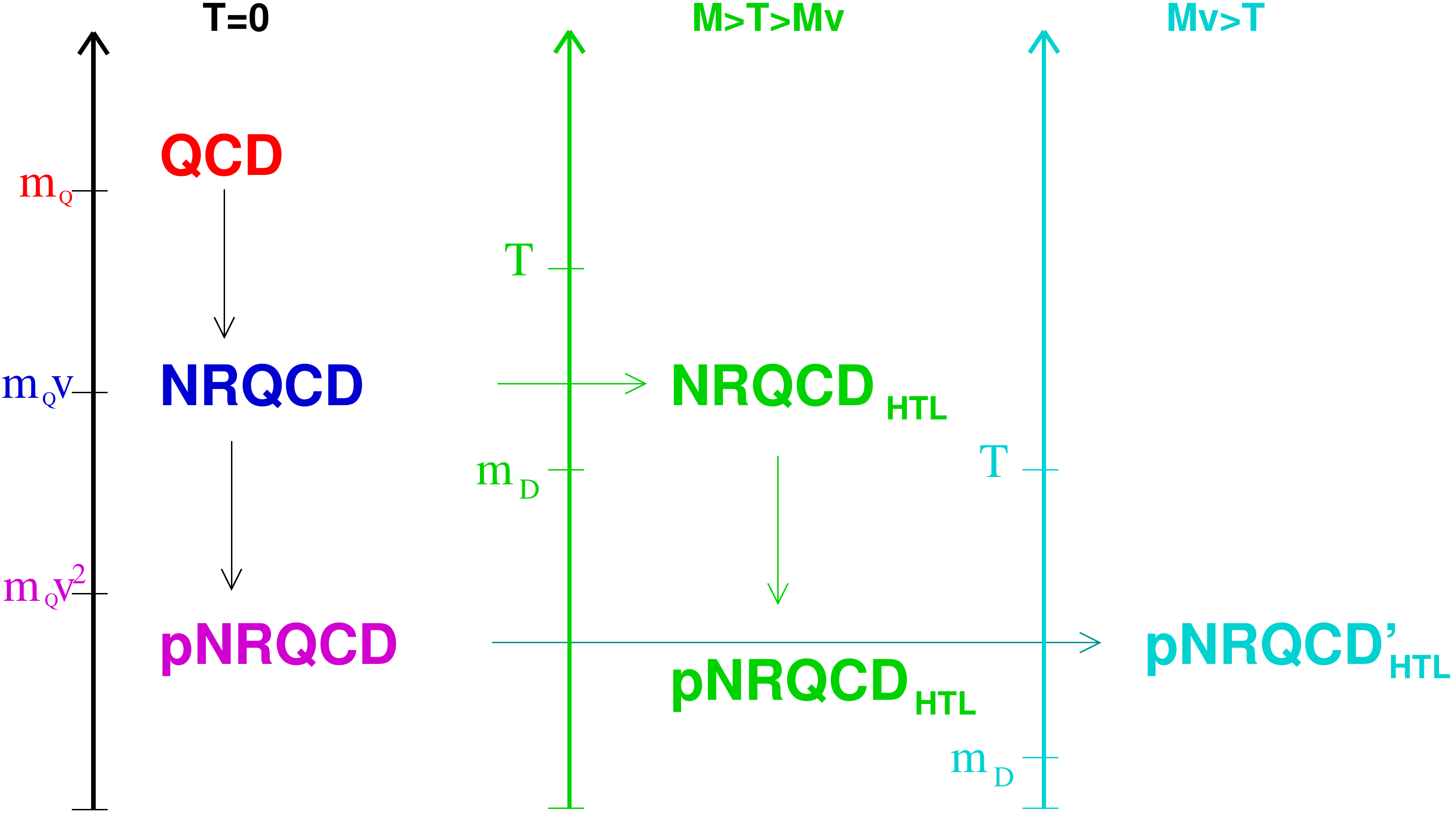}
\caption{Hierarchies of EFTs for quarkonium at zero temperature 
(see Sec. \ref{sec:subsecC11} and Ref.~\cite{Brambilla:2004jw})
and at finite temperature 
\cite{Brambilla:2008cx,Escobedo:2008sy,Vairo:2009ih,Brambilla:2010vq,Escobedo:2010tu}.
If $T$ is the next relevant scale after 
$m_Q$, then integrating out $T$ from NRQCD leads to an EFT called 
NRQCD$_{\rm HTL}$, because it contains the hard thermal loop (HTL) Lagrangian.
Subsequently, integrating out  the scale $m_Qv$ from NRQCD$_{\rm HTL}$ leads to 
a thermal version of pNRQCD called pNRQCD$_{\rm HTL}$. If the next relevant 
scale after $m_Q$ is $m_Qv$, then integrating $m_Qv$ out from NRQCD leads to 
pNRQCD. If the temperature is larger than $m_Qv^2$, 
then $T$ may be integrated out from pNRQCD, 
leading to a new version of pNRQCD$_{\rm HTL}$  \cite{Vairo:2009ih}. From \cite{AVairocourtesy}.}
\label{figeft}
\end{center}
\end{figure}

If these scales are hierarchically ordered, physical observables can be 
systematically expanded in the ratio of such scales. At the level 
of the Lagrangian, this amounts to substituting QCD with a hierarchy of EFTs
which are equivalent to QCD order by order in the expansion parameters.
At zero temperature in Sec. \ref{sec:subsecC11}, the two nonrelativistic 
EFTS that follow from QCD by integrating out the scales 
$m_Q$ (NRQCD) and $m_Qv$ (pNRQCD) have been discussed. 
At finite $T$ 
different possibilities for the scale hierarchies arise.  The corresponding EFTs are shown in
Fig.~\ref{figeft}. 

In the EFT, the interaction potential $V$ is clearly defined and a  
structured power counting  to calculate the quarkonium energy and width 
is provided.
The potential follows from integrating out all contributions from modes
with energy and momentum larger than the binding energy.
For temperatures smaller than the binding energy the potential is simply the 
Coulomb potential. Thermal corrections 
affect the energy and induce a thermal width to the quarkonium state which 
may be relevant for describing the quarkonium in-medium modifications at 
relatively low temperatures.
For temperatures larger than the binding energy, the potential acquires 
both real and imaginary thermal contributions.

This QCD-based description has resulted in a {\it paradigm shift} 
in our understanding of quarkonium properties in a weakly-coupled plasma. 
The following pattern is observed \cite{Brambilla:2008cx,Escobedo:2008sy,Vairo:2009ih,Brambilla:2010vq,Escobedo:2010tu,Laine:2006ns,Laine:2007qy}:
\begin{itemize}
\item{}  The thermal part of the potential has a real and  an imaginary part. 
The imaginary part of the potential smears out the bound state 
peaks of the quarkonium spectral function, leading to their dissolution at 
lower temperatures than those required for the onset of Debye screening in 
the real part of the potential (see, e.g. \cite{Laine:2008cf}).
Thus quarkonium dissociation appears to  be a consequence of the 
appearance of a thermal decay width rather than being due to the  color screening of
the real part of the potential: the thermal decay width  may become 
as large as the binding energy at a lower temperature than that 
at which color screening sets in.
\item{} Two mechanisms contribute to the thermal decay width: the imaginary 
part of the gluon self energy, induced by the Landau-damping phenomenon
(also present in QED)  
\cite{Laine:2006ns} and the quark-antiquark color singlet 
to color octet thermal break up (a new effect, specific to QCD)  \cite{Brambilla:2008cx}.
These two mechanisms are related to the previously described gluodissociation
\cite{Kharzeev:1994pz,Xu:1995eb,Bhanot:1979vb,Peskin:1979va} 
and quasi-free dissociation   
\cite{Grandchamp:2001pf,Grandchamp:2002wp,Rapp:2008tf}. 
The EFT power counting establishes which dissociation mechanism dominates 
parametrically in which temperature regime.  Landau damping
dominates for temperatures where the Debye mass $m_D$ is larger than the 
binding energy $E_B$ while the singlet to octet break up dominates for $m_D < E_B$.  
The distinction between the two dissociation mechanisms holds at leading order.
Both can be calculated by cutting appropriate diagrams in the relevant EFTs. 
See \cite{Brambilla:2011sg,Brambilla:2013dpa}
for results relating  the  quarkonium widths to the in-medium or vacuum 
cross sections that correct or complement the previously-used approximations 
and phenomenological formulas.
\item{} The resulting color singlet thermal potential, $V$, is neither the 
color-singlet quark-antiquark free energy  
\cite{Brambilla:2010xn}
nor the internal energy. It has an imaginary part and may contain divergences
that eventually cancel in physical observables \cite{Brambilla:2008cx}.
\item{}  Temperature effects can be other than screening,
typically they may appear as power law or logarithmic
corrections \cite{Brambilla:2008cx,Escobedo:2008sy}.
\item{} The dissociation temperature behaves
parametrically as  $\pi T_{\rm melting} \sim m_Q g^{4\over 3}$ 
\cite{Escobedo:2008sy,Laine:2008cf,Escobedo:2010tu}.  
\end{itemize}

In particular, in  Ref.~\cite{Brambilla:2010vq} 
heavy quarkonium energy levels and decay widths in a quark-gluon plasma, 
at a temperature below the quarkonium melting temperature satisfying the hierarchy  
$m_Q \gg m_Q\alpha_s \gg \pi T \gg m_Q\alpha_s^2 \gg m_D$
have been calculated to order $m_Q \alpha_s^5$.
This hierarchy may be relevant for the lowest-lying bottomonium states 
($\Upsilon(1S)$, $\eta_b$) at the LHC, for which it may hold
$m_b \approx 5 \; \hbox{GeV} \; > m_b\alpha_s \approx 1.5 \; \hbox{GeV} \; >  
\pi T \approx 1 \; \hbox{GeV} \; >  m_b\alpha_s^2 \approx 0.5  \; \hbox{GeV} 
\; \simg  m_D$.
In this situation, the dissociation width grows linearly with temperature.
Then the mechanism underlying the decay width is the color-singlet to 
color-octet thermal break-up, implying the tendency of quarkonium to decay 
into a continuum of color-octet states.
This behavior \cite{Brambilla:2010vq,Escobedo:2010tu}  
is compatible with the data  (($\Upsilon(1S)$ seems not to be yet 
dissociated at LHC) and with finite $T$ NRQCD lattice calculations 
\cite{Aarts:2010ek,Aarts:2013kaa,Kim:2013seh}.

Even if the above-described theory holds only for weak coupling, it has had 
a more general impact  on our understanding of the physics since, for the first
time, it provides a coherent, systematic theoretical framework.
A key feature of the potential obtained in this picture is that it contains 
a sizable imaginary part encoding the decoherence effects caused by 
interactions with the medium. The impact of such an imaginary part has been 
studied \cite{Margotta:2011ta,Mocsy:2013syh} but a fully consistent 
phenomenological description of quarkonium suppression is yet to appear.
Additional effects that are just beginning to be considered are the 
effect of an anisotropic medium 
\cite{Strickland:2014eua,Burnier:2009yu,Philipsen:2009wg}    
and the relative velocity between the quarkonium state and the medium
\cite{Escobedo:2011ie,Escobedo:2013tca,Aarts:2012ka,Brambilla:2011mk}.

The next step would be to generalize these results to strong coupling.
Initial investigations have been made recently on the lattice \cite{Rothkopf:2011db,Burnier:2013nla} but a
complete EFT description is still lacking.
Preliminary work includes study of the Polyakov loop and Wilson loop correlators and their relation to
singlet and octet correlators in perturbation theory \cite{Brambilla:2010xn} and in general.
The nontrivial renormalization properties of the cyclic Wilson loop have been investigated
\cite{Berwein:2012mw,Berwein:2013xza}, making it possible to determine which combinations of correlators are
suitable for lattice calculations.

It may be possible to calculate the behavior of the potential at 
strong-coupling using holographic correspondence.  However, the imaginary part of the potential,
responsible for the thermal decay width, was not predicted in
AdS/CFT-inspired
calculations.  After this effect was identified in perturbative
calculations
\cite{Albacete:2008dz,Hayata:2012rw,Fadafan:2013bva}, 
it was also obtained using holographic methods.

Some of the outstanding questions in quarkonium theory include whether 
quarkonium and heavy quarks are indeed external probes of the medium,
the connection of the magnitude of their flow and the diffusion coefficients 
in EFTs, and quantification of the importance of recombination effects.
The experimental state of the art regarding these questions is discussed in
the quarkonium subsection of  Sec.~\ref{Sec:ELOSSEXP}.


\subsubsection{Experimental results on hard probes}
\label{Sec:ELOSSEXP}

The details of the production and propagation of high \Pt\ and high mass probes
can explore the mechanisms of parton energy loss and deconfinement in the medium
and shed light on the relevant physical mechanisms and the microscopic properties
of the medium. In addition, the underlying event, even if considered
as a background contribution to the hard probes, is an important element of
the hadronic environment consisting of complex contributions,
spanning over nonperturbative and perturbative QCD and
including sensitivities to multiscale and low $x$ physics.

Experimentally, several methods are used to address such questions, generally
through comparison of the relative production of single particles or fully 
reconstructed jets in nuclear collisions to expectations from a superposition 
of independent nucleon-nucleon collisions.

In particular, jet production is decoupled from the 
formation of the medium and can be considered an external probe traversing the 
hot medium. Due to their early production, jets are well calibrated probes:
the production rates can be calculated using pQCD in the vacuum 
because their large energy scale minimizes cold nuclear matter effects.

At the LHC, high-\Pt\ hadron production is dominated by gluon fragmentation. 
The gluons have a larger color-coupling than light quarks, thus gluon energy 
loss is expected to be larger. Moreover, heavy quarks with \Pt\ lower than
or equivalent to the quark mass should have
less gluon radiation and thus a smaller suppression than light quarks.  This is
discussed further in the subsection dedicated to heavy flavors.

To quantify suppression effects, the nuclear modification factor, \RAA, is 
widely used. It is defined as the ratio of yields in \AAcoll\ collisions to 
those in \ppcoll, scaled by the number of binary 
collisions,
\begin{equation}
R_{AA} = \frac{(1/N_{evt.}^{AA})d^2N_{ch}^{AA}/dp_{{T}}d\eta}{\langle 
N_{coll} \rangle (1/N_{evt.}^{pp}) d^2N_{ch}^{pp}/dp_{T}d\eta} \, \, ,
\label{Eq:raa}
\end{equation}
where the average number of binary nucleon-nucleon collisions, $\langle 
N_{coll} \rangle$, is given by the 
product of the nuclear overlap function, $T_{AA}$, calculated in the Glauber 
model \cite{Miller:2007ri}, and the inelastic $NN$ cross section, 
$\sigma_{in}^{NN}$.  The collision centrality is often quantified in
terms of the number of nucleon participants, $N_{part}$, also calculated in 
the same Glauber framework.
In the absence of nuclear effects, \RAA\ is unity by construction.
In addition to \RAA, the quenching effects can be quantified using the 
central-to-peripheral ratio, \RCP, defined as the ratio of the 
per-event jet yield in a given centrality bin normalized by the number of $NN$
collisions in the same centrality bin to the
same quantity in a more peripheral bin, typically 60--80\%.

Differential measurements include: $\gamma$+jet, hadron+jet, and dijet
spectra; angular correlations; azimuthal anisotropies; jet shapes; and 
fragmentation functions. Measurements of the azimuthal 
anisotropy, $v_2$, can probe thermalization at low \Pt , while at high \Pt\ 
the path length dependence of energy loss can be studied. 
The measurement of the 
reaction plane allows more differential measurements such as the study of
\RAA\ ``in-'' and ``out-of-plane'' (i.e., along the short and long axes of the 
almond-shaped overlap region of the two nuclei in semi-central collisions).
Azimuthal spectra of dijet events in different centrality bins as well as
separation of leading and sub-leading jets all
allow further insight into the path length dependence of energy loss and the 
redistribution of the quenched jet energy.

\subsubsection*{High \Pt observables}

\paragraph{Charged hadrons and bosons}

\begin{figure}
\begin{center}
\includegraphics[width=0.48\textwidth]{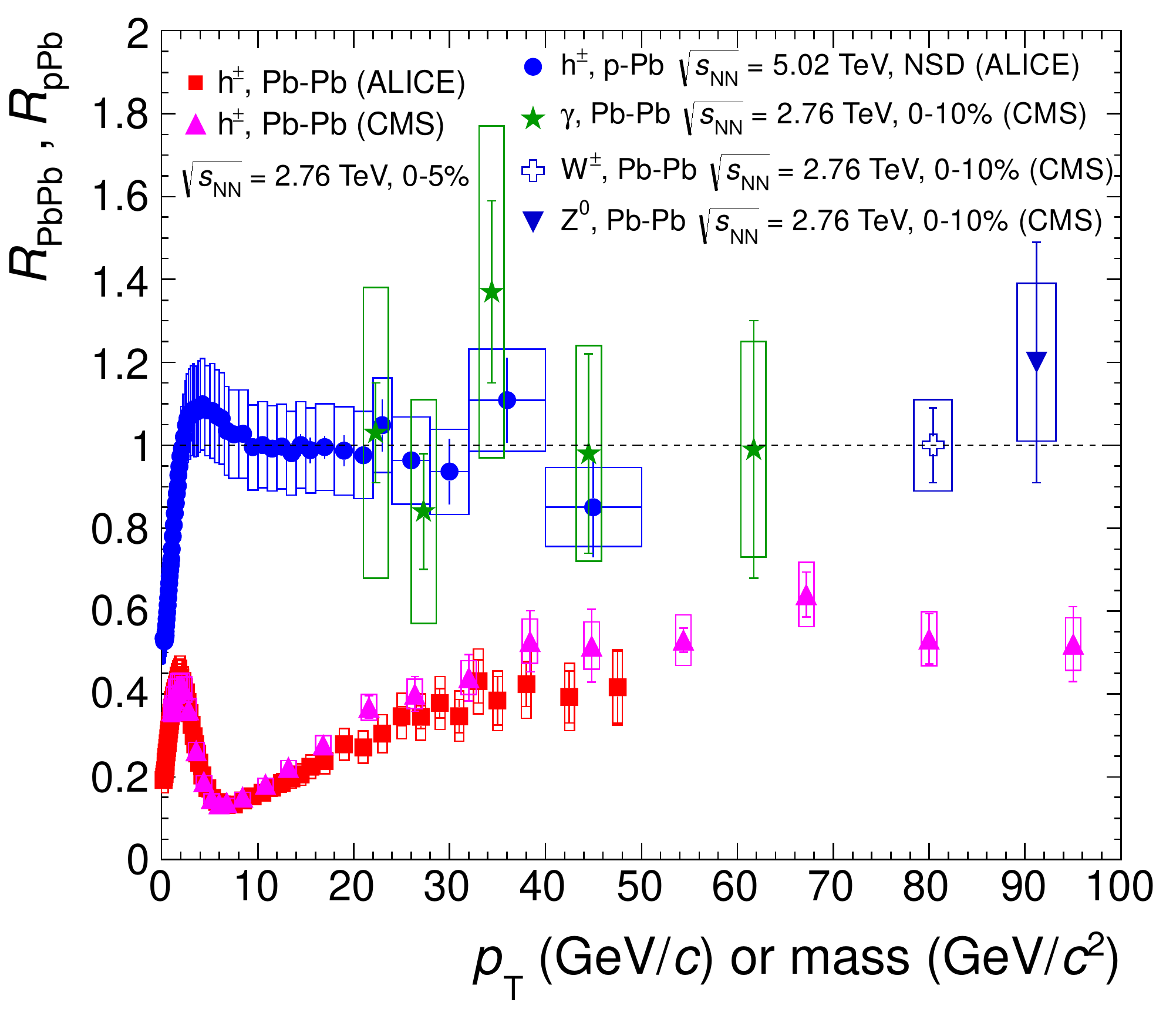}
\caption{ The \RAA\ for charged particles in the 5\% most central \PbPb\ collisions at $\sqrt{s_{NN}} =
2.76$~TeV is compared for ALICE and CMS \cite{CMS:2012aa}.
The results are also compared to those for \W and \Z ~bosons as well as isolated photons measured by CMS
\cite{Chatrchyan:2011ua,Chatrchyan:2012nt,Chatrchyan:2012vq}.
The \RpPb\ for \pPb\ collisions at $\sqrt{s_{NN}} = 5.02$ TeV measured by ALICE is also shown
\cite{ALICE:2012mj,Abelev:2014dsa}.}
\label{fig:RAA}
\end{center}
\end{figure}

Inclusive measurements can give the first indication of the existence 
of a hot and dense medium. One of the most complete pictures of interactions 
of hadrons and electroweak bosons with the medium is shown in
Fig.~\ref{fig:RAA} for the charged particle \RAA\
in central \PbPb\ collisions at $\sqrt{s_{NN}} = 2.76$~TeV \cite{ALICE:2012mj,Abelev:2014dsa,CMS:2012aa}
compared to the \RAA\ of \W, \Z\ \cite{Chatrchyan:2012nt,Chatrchyan:2011ua,Chatrchyan:2012xq,ATLAS-CONF-2013-106}
and isolated photons \cite{Chatrchyan:2012vq} at the same
energy.  The charged particle
\RpPb\ from \pPb\ collisions at $\sqrt{s_{NN}} = 5.02$~TeV
is also shown \cite{Aad:2012ew}. 
Understanding the detailed structure of these ratios is the subject of intense discussions among theorists 
and experimentalists \cite{CasalderreySolana:2012bp}.

The peak in $R_{\rm PbPb}^{ch}$ at $p_{{T}} \approx 2$~GeV/$c$
can be interpreted as a manifestation of radial collective
flow~\cite{Muller:2012zq}.  Energy loss causes a pile up
at low \Pt\ which is enhanced by flow. This is also supported by the results
obtained looking at identified hadrons and their mass ordering effects, as discussed later.
At $p_{T} \approx$ 5--7 GeV/$c$, $R_{\rm PbPb}^{ch}$  
falls to a minimum of $\approx 0.13$, lower than the minimum at RHIC \cite{PhysRevC.88.024906}
of $R_{AA}^{ch} \approx 0.2$, indicating a slightly larger suppression at the LHC.
Above 7~GeV/$c$, \RAA\ increases to $\approx 0.4$ at $p_{T} > 30$~GeV/$c$
and remains relatively constant thereafter,
showing that the medium can quench even very high \Pt\ particles.  
One possible explanation is that a constant energy loss shifts the entire \Pt\ spectrum
to lower \Pt.
In general, the low \Pt\ region reflects an interplay of soft physics effects (shadowing, saturation,
Cronin, flow, etc.) which are still under investigation \cite{Albacete:2013ei}.

The LHC measurements confirm and extend the experimental signatures of partonic 
energy loss first observed in the 5\% most central \AuAu\ collisions at 
$\sqrt{s_{NN}} = 200$~GeV at 
RHIC~\cite{Arsene:2004fa,PhysRevLett.88.022301} where the measured signals include suppression of 
single hadrons and modification of dihadron angular correlations \cite{Agakishiev:2010ur}. 
At the LHC, hadron production cross sections are several
orders of magnitude higher that those at RHIC, allowing measurements over a 
wider \Pt\ range and giving access to multidimensional studies of 
cross-correlated observables.

The \RAA\ distributions have also been compared to model calculations employing the 
RHIC data to calibrate the medium density.  They implement several different 
energy loss mechanisms~\cite{Armesto:2011ht,Horowitz:2011gd,Salgado:2003gb,Chen:2011vt,Chatterjee:2011dw,Renk:2011gj,Majumder:2010ik}.
Some of them can qualitatively reproduce the increase of \RAA\ with \Pt.
This rise can be understood as a decrease of the fractional energy loss 
of the parton with increasing \Pt, reflecting the weak dependence of pQCD radiative 
energy loss on parton energy.
The observed trend is semi-quantitatively described by several models of QCD
energy loss.
The differences between the results presented in Refs.~\cite{Armesto:2011ht,Horowitz:2011gd,Salgado:2003gb,Chen:2011vt,Chatterjee:2011dw,Renk:2011gj,Majumder:2010ik} and elsewhere are under systematic investigation.  They may arise from
poorly-controlled aspects of leading order collinear gluon radiation.
A complete picture of energy loss at next-to-leading order is under study but
difficult to achieve.
Further details and open questions related to the theory of energy loss
are discussed in Sec.~\ref{Sec:ELOSSTHEORY}.

The measurements (Fig.~\ref{fig:RAA}) also show that isolated photons and \W\ and \Z\ bosons, 
which do not carry color charge, are not suppressed. This is consistent with 
the hypothesis
that the observed charged hadron suppression is due to final-state 
interactions with the hot and dense medium.
Further input comes from the \pPb\ data which were expected to clarify initial- from final-state
effects, as discussed in Sec. \ref{RVpPb}. 
First results of \RpPb\ data from the \pPb\ pilot run at
$\sqrt{s_{NN}} = 5.02$~TeV \cite{ALICE:2012mj,Abelev:2014dsa} are
compared to the \PbPb\ results in Fig.~\ref{fig:RAA}.
The \pPb\ measurement was performed for non single diffractive collisions
in the pseudorapidity range $|\eta_{cms}|<0.3$. 
In this minimum-bias sample, with no further constraints on multiplicity, the data show 
no strong deviation from scaling with the number of binary 
nucleon-nucleon collisions. This is in agreement with the hypothesis 
that the strong suppression of hadron production at
high \Pt\ observed in central \PbPb\ collisions is not due to initial-state
effects, supporting the production of hot quark-gluon matter in \PbPb\ collisions
\cite{ALICE:2012mj,Abelev:2014dsa}.

The observed trends qualitatively resemble those of \RdAu\ at RHIC.
At low \Pt, suppression may be related to parton shadowing or saturation
while the rise at $p_{T} \approx 4$~GeV/$c$ may be a manifestation of 
the Cronin effect which originates from multiple scattering during the initial 
phase of the collision. 

However, further extensive analysis of the LHC data reveal different aspects.
In ATLAS, per-event inclusive hadron yields were 
measured in different centrality and rapidity regions, demonstrating strong dependence of the Cronin 
peak not only on centrality, but also on rapidity~\cite{ATLAS-CONF-2013-107}. Measurements with 
fully reconstructed jets reveals a strong reduction of the jet yield in the proton-going direction
in more central collisions relative to peripheral collisions~\cite{ATLAS-CONF-2013-105}. The reduction becomes
more pronounced with increasing jet \Pt\ and at more forward proton-going rapidities. When the jet $R_{CP}$ 
is measured as function of the full jet momentum, $p_{T} \cosh y$, the rapidity variation factors out, 
reducing the \RAA\ measured in all rapidity intervals to the a single curve. 

Results from CMS \cite{CMS-PAS-HIN-12-017,hp2013} extend the charged particle \RpPb\
up to $p_{T} \approx 130$ GeV/$c$. The value of \RpPb\ rises
above unity for $p_{T} > 30$ GeV/$c$, near the onset of the gluon
antishadowing region but significantly larger than predicted.

Furthermore, the data reveal different trends when the measurements are performed in the 
low- or high-multiplicity samples. Additional intriguing features, observed for high-multiplicity
events, are discussed in Sec. \ref{chapd:pPbEXP}.
In particular, indications of collective behavior are seen 
for several distributions in the high-multiplicity sample.  Currently this is 
a puzzle that is actively being pursued both by experimentalists and theorists.

\paragraph{Identified hadrons}

In order to set additional constraints on energy loss,
the nuclear suppression factor has been studied for identified particles.
At the LHC, measurements of identified particle \RAA\ include
light and strange hadrons, isolated photons, \Z, \W, \D, \jpsi\ and $\Upsilon$.
The suppression of individually reconstructed prompt and nonprompt \jpsi\ (from $B$ decays, 
identified by displaced vertex techniques) is discussed further in 
the subsection dedicated to heavy flavor.

\begin{figure}
\begin{center}
\includegraphics[width=0.48\textwidth]{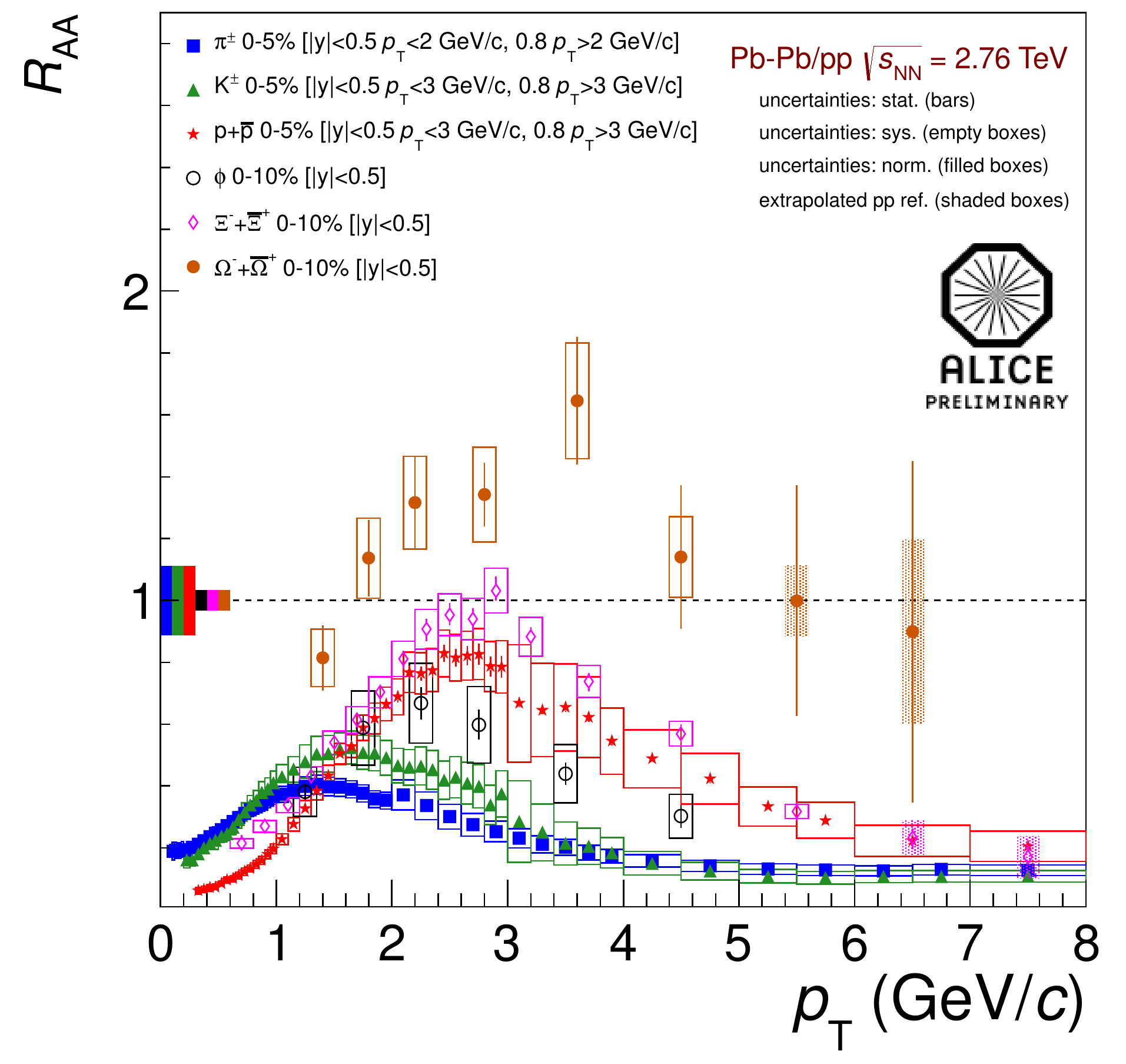}
\caption{Nuclear modification factors \RAA\, at midrapidity versus \Pt\ and centrality, for light and strange particles: pion, kaon, proton (covering low and high \Pt\ up to 8 GeV/$c$), $\phi$, $\Xi$ and $\Omega$. From \cite{Knospe:2013tda}.}
\label{fig:alimass}
\end{center}
\end{figure}

The ALICE \RAA\ for identified
pions, kaons and protons up to $p_{T} \approx 20$~GeV/$c$, 
confirms the observations at RHIC
and shows that a hierarchy of suppression is observed at low \Pt.  
In order to better understand the influence of rescattering effects, 
\RAA\ of resonances and stable hadrons 
have been also measured, Fig.~\ref{fig:alimass}. 
Of particular interest is the measurements of $\phi$ which is a meson with the mass of the proton
and can discriminate mass and quark content effects.  
The nuclear modification factor the \RAA\ of $\phi$ appears to follow \RAA\ of $\Xi$ and $\Omega$ for $p_{T} \leq 2.5$ GeV/$c$, and lies between the \RAA\ of mesons 
($\pi$ and $K$) and \RAA\ of baryons ($p$ and $\Xi$) at high \Pt\ \cite{Knospe:2013tda}.
More generally, the meson results cluster around a lower value of \RAA\ than the protons,
reflecting strong radial flow \cite{Muller:2012zq}. For $p_{T} \geq$ 8--10 GeV/$c$,
the suppression seems to be the same for different particle species,
indicating that the medium effects are similar for all light hadrons.

A detailed systematic study of charged hadron spectra and \RAA\ as a function of centrality was also carried out for \AuAu\ and \dAu\ collisions at $\sqrt{s_{NN}} = 200$~ GeV \cite{PhysRevC.88.024906}.  Baryon enhancement is present in both systems. In \dAu\ collisions, the Cronin enhancement has long been known to be stronger for baryons than for mesons, however, for the first time the results present clear evidence for a strong centrality dependence of this effect. 
In \AuAu\ collisions, the baryon enhancement has been attributed to parton recombination 
at hadronization.  When combined with the mass dependence of $v_2$ measured at the LHC, there is a strong indication that the mass effect observed in \pPb\ collisions has a collective final-state origin.  A similar but weaker effect was also observed by PHENIX \cite{Muller:2012zq,hp2013}.
In general, the measurements of identified hadrons over a wide \Pt\ range, have also
the potential to address modifications of the jet fragmentation functions. 

\paragraph{Reconstructed jets}

\begin{figure}
\begin{center}
\includegraphics[width=0.48\textwidth]{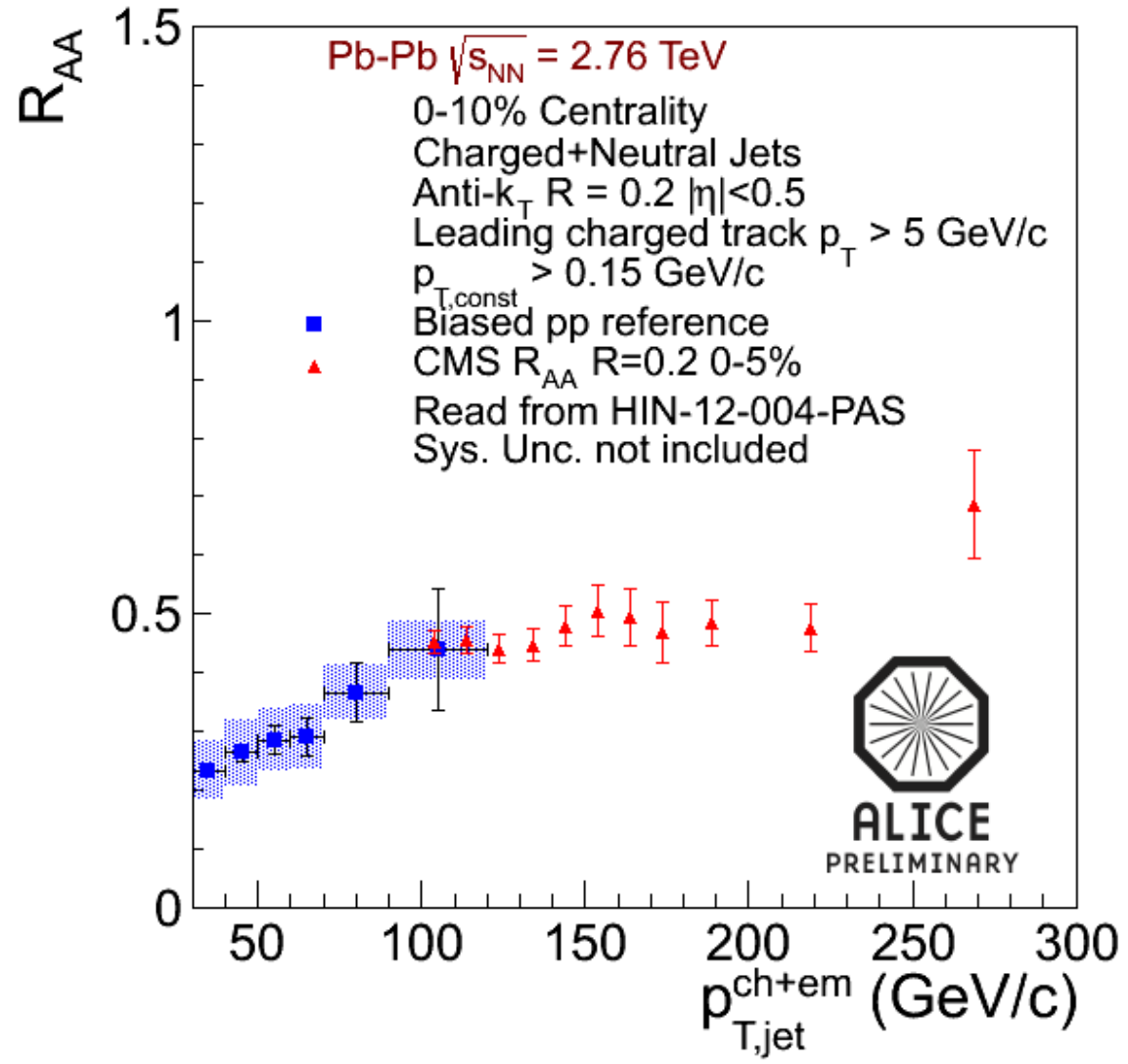}
\caption{
The jet \RAA\ over a wide \Pt\ range measured by ALICE and CMS in central \PbPb\ collisions at $\sqrt{s_{NN}} = 2.76$~TeV. From ALICE \cite{Reed:2013rpa} and CMS \cite{CMS:2012rba}.}
\label{fig:alicms}
\end{center}
\end{figure}

Fully reconstructed jets available over a wide \Pt\ range
at $\sqrt{s_{{NN}}}=2.76$ TeV at the LHC confirm and extend the 
suppression pattern observed for charged particles. Figure~\ref{fig:alicms} presents 
the ALICE \RAA\ results covering low $p_{T}$ down to $\approx$ 30--40 GeV/$c$ 
\cite{Verweij2013421}
and the CMS measurements~\cite{CMS:2012rba} up to
$p_{T} \approx 270$~GeV/$c$. 
Good agreement is observed in the overlapping \Pt\ region.
Similar results have been
obtained by the ATLAS collaboration \cite{Aad:2012vca} .

The complementarity of these results, together with 
combined systematic studies over the widest available \Pt\ range and 
employment of particle identification 
at low \Pt\ explore different aspects of energy loss. 
Note that, although the original 
parton energy is better reconstructed in a jet than 
by tagging only a fast hadron, 
the single inclusive jet suppression is similar
to that of single hadrons. This can be understood if parton energy loss is
predominantly through radiation outside the jet cone radius used in the jet 
reconstruction algorithm.

Jet reconstruction in heavy-ion collisions is challenging due to the high-multiplicity environment.
However, dedicated algorithms and background subtraction techniques have been optimized to reconstruct all the particles resulting from the hadronization of the parton along its trajectory within a fixed jet-cone radius \cite{Cacciari:2008gp,Abelev:2012ej}.

In \PbPb\ collisions, the strongest jet suppression is observed for the most 
central events.  A clear centrality dependence is observed in successively
peripheral events with decreased suppression (larger \RAA) in peripheral
collisions.  In particular, imposing a minimum fragmentation bias on single 
tracks of 0.150 GeV/$c$,
ALICE explored the low \Pt\ region (30--110 GeV/$c$) \cite{Verweij2013421}
finding $R_{AA} \sim 0.4$ for a jet cone radius $R = 0.3$.
At higher \Pt, $\sim 200$ GeV/$c$
for ATLAS \cite{Aad:2012vca} and $\sim 300$ GeV/$c$ for CMS, $R_{{AA}} =0.5$, almost independent of
jet \Pt. These results imply that the full jet energy cannot be captured for
$R<0.3$ in heavy-ion collisions. 

\begin{figure}
\begin{center}
\includegraphics[width=0.48\textwidth]{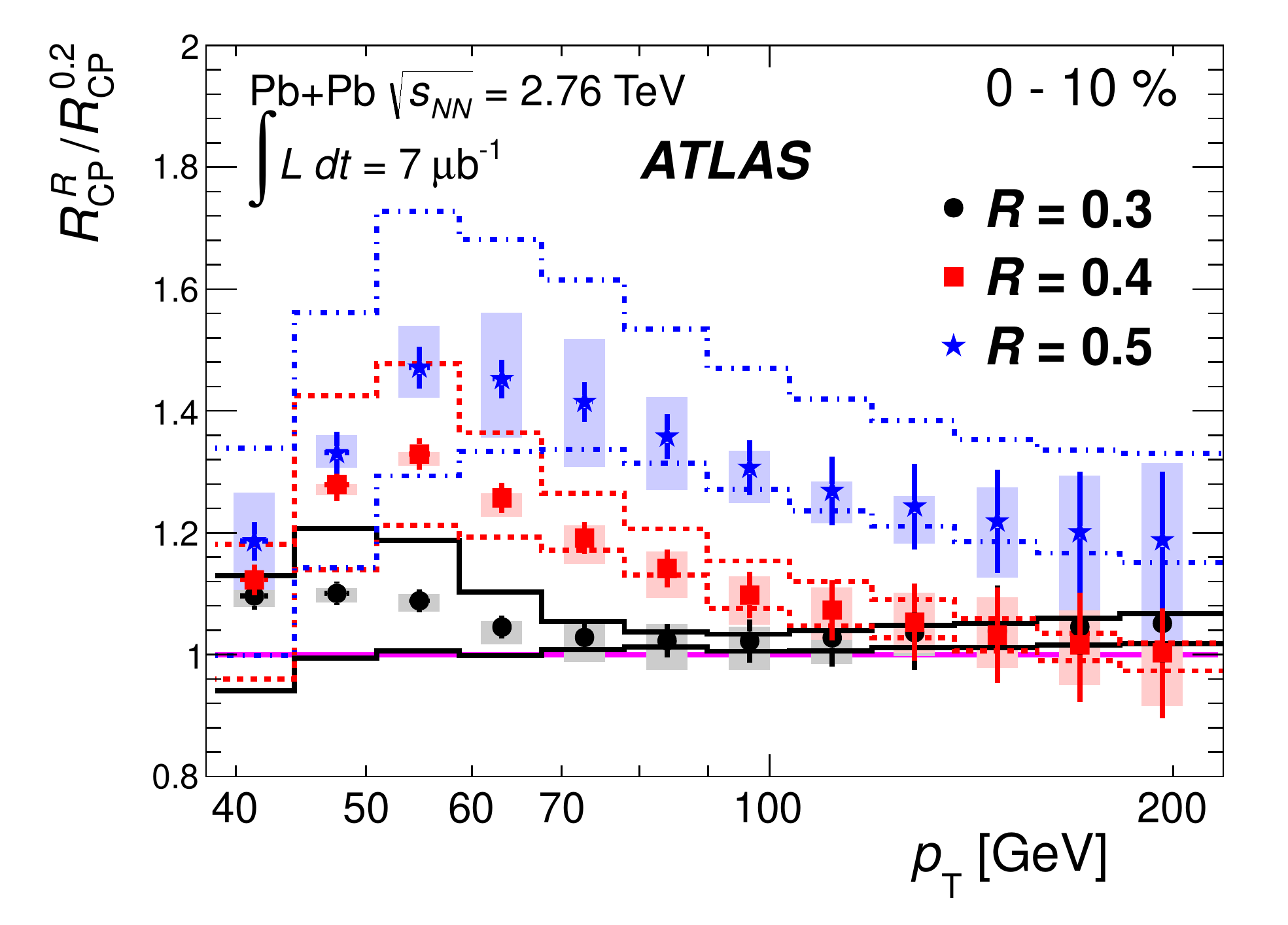}
\caption{The ratios $R_{CP}^R/R_{CP}^{R=0.2}$ for $R=0.3$, $0.4$ and $0.5$ as a function of jet \Pt\ in the 0--10\% centrality bin. The bars show the statistical uncertainties, the lines indicate fully correlated uncertainties and the shaded boxes represent partially correlated uncertainties between different \Pt\ values. From \cite{Aad:2012vca}. }
\label{fig:atlascp}
\end{center}
\end{figure}

The same conclusion can be reached by studying the jet \RCP, as shown
in Fig.~\ref{fig:atlascp}. 
For $p_{T} < 100$ GeV/$c$, the ratio
$R_{{CP}}^R/R_{{CP}}^{{R=0.2}}$, for $R = 0.4$ and 0.5, differs from unity beyond 
the statistical and systematic uncertainties, indicating a clear jet broadening.
However, for $R \leq 0.4$ at $p_{T} > 100$~GeV/$c$,
the ratio is consistent with jet production in vacuum over all centralities.
This may be interpreted as an indication that the jet core remains intact
with no significant jet broadening observed within the jet cone resolution
\cite{Aad:2012vca}.

\paragraph{Path length dependence of the energy loss}

\begin{figure}
\begin{center}
\includegraphics[width=0.48\textwidth]{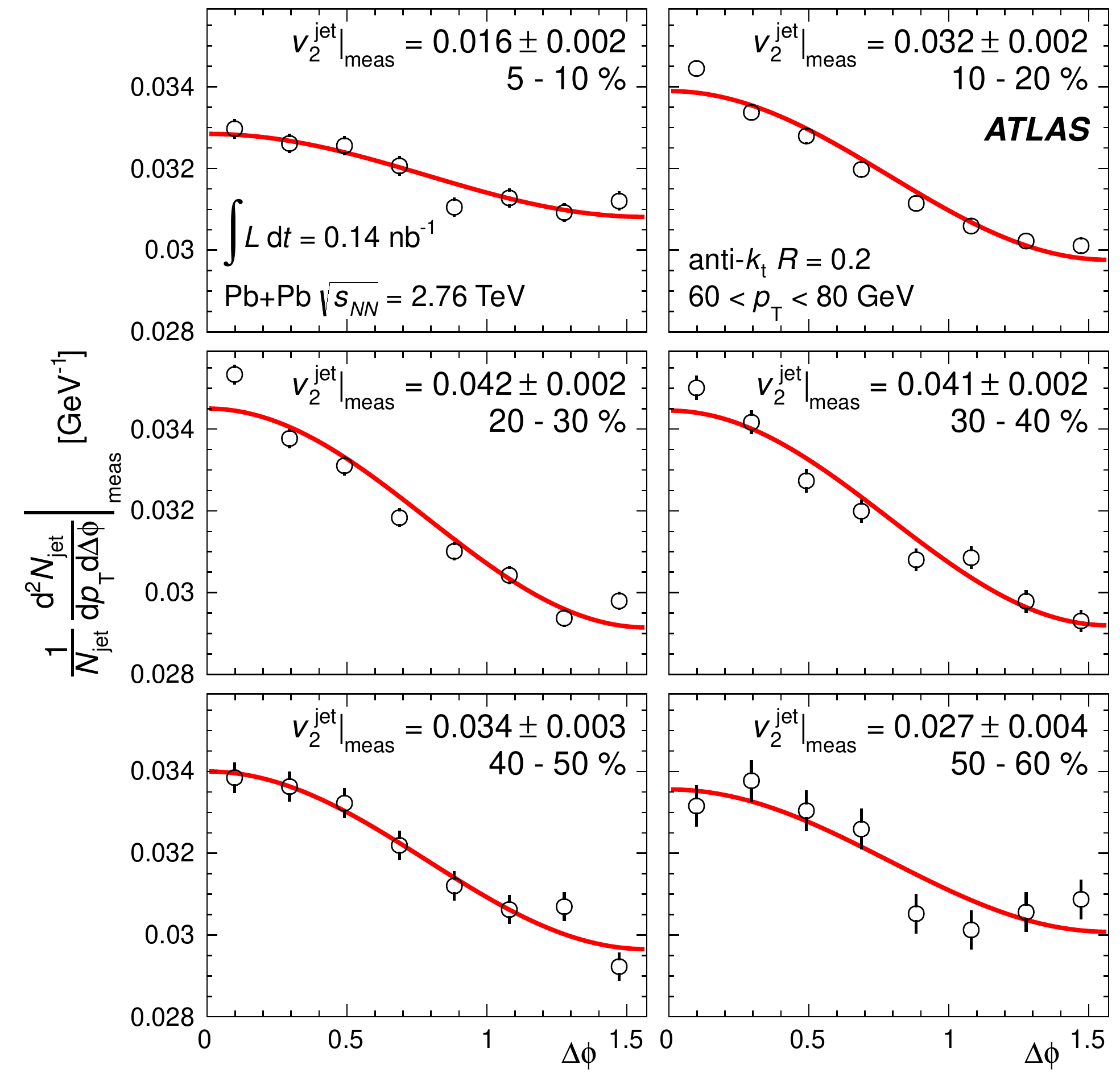}
\caption{The $\Delta \phi$ dependence of measured jet yield in the 60 $<$ \Pt\ $<$ 80 GeV/c interval for six ranges of collision centrality. The yields are normalized by the total number of jets in the \Pt\ interval. The solid curves are a fit to the data \cite{PhysRevLett.111.152301}. }
\label{fig:Atlas-Dphi}
\end{center}
\end{figure}

Measurements of inclusive jet suppression as a function of azimuthal separation
with respect to the event plane, $\Delta \phi$, makes possible an estimate of the path-length
dependence of energy loss for the first time. A measurement of the variation of the 
jet yield as a function of the distance traversed through the matter
can provide a direct constraint on the relative
theoretical models. 
Figure~\ref{fig:Atlas-Dphi} shows the variations in the jet yield as a function of $\Delta \phi$ at different centralities for $60 < p_{T} < 80$ GeV/$c$ for fully reconstructed jets measured 
by ATLAS ~\cite{PhysRevLett.111.152301}.
The observed azimuthal variation amounts to a reduction of 10\%--20\% in the jet yields between in-plane and out-of-plane directions establishing a clear relationship between jet suppression and the initial nuclear geometry and confirming that 
jet suppression is stronger in the direction where the parton traverses the
greatest amount of hot medium. 

The azimuthal anisotropy of charged particles with respect to the event plane 
has been studied by CMS over the widest \Pt\ range, up to $\sim 60$ GeV/$c$.
The results \cite{Chatrchyan:2012xq} show a rapid rise of the anisotropy
to a maximum at $p_{T} \sim 3$ GeV/$c$ with a subsequent decrease in all
centrality and $\eta$ ranges. A common trend in the centrality
dependence is observed over a wide \Pt\ range, suggesting a potential connection to the
initial-state geometry.

\paragraph{Correlations}

\begin{figure*}
\begin{center}
\includegraphics[width=\textwidth]{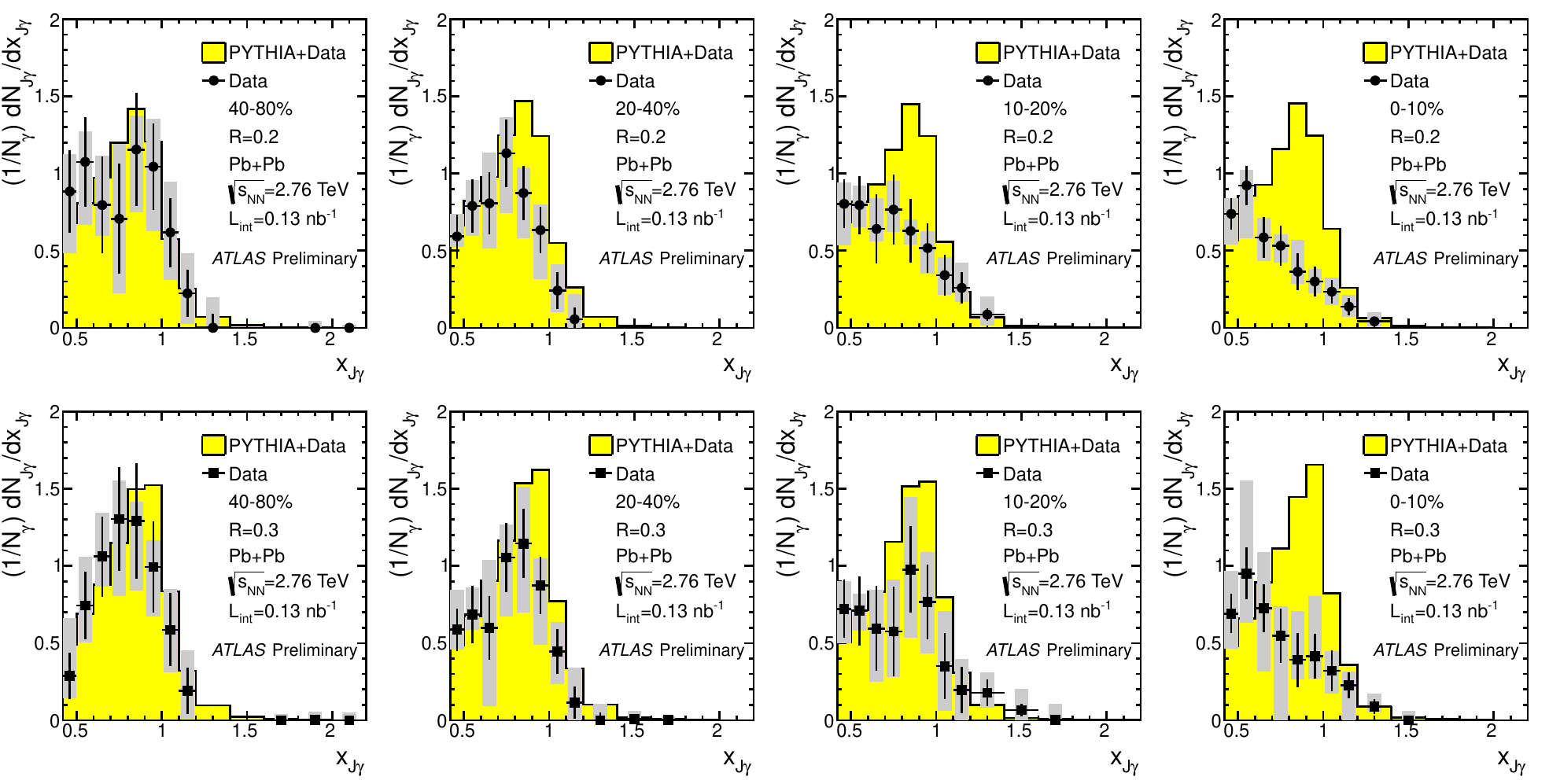}
\caption{The distribution of the mean fractional energy carried by a jet 
opposite an isolated photon, $x_{\rm{J\gamma}}$, in \PbPb\ collisions 
(closed symbols) compared with PYTHIA ``true jet''/``true photon'' distributions 
(yellow histogram) embedded into simulated background heavy-ion events. The rows represent jet cone radii $R=0.2$ (top) and 
$R=0.3$ (bottom). The columns represent different centralities with increasing
centrality from left to right. The error bars represent statistical errors 
while the gray bands indicate the systematic 
uncertainties~\cite{ATLAS:2012cna}. }
\label{fig:ATLAS_Fig_3}
\end{center}
\end{figure*}

Inclusive jet measurements provide only limited information because
the initial jet energy is unknown.
The magnitude of the energy lost by jets can be measured by studying 
boson-jet correlations, assuming that the boson
momentum represents the initial jet momentum. As already shown in 
Fig.~\ref{fig:RAA}, the electroweak gauge bosons, which do not
carry color charge, are unaffected by the medium and therefore
retain the kinematics of the initial hard scattering 
\cite{Chatrchyan:2011ua,Chatrchyan:2012nt,Chatrchyan:2012vq}.
This suggests that identifying the correlations between isolated photons and 
jets is one of the key methods of determining 
the energy of the parton which generated the jet 
\cite{Chatrchyan:2012gt,ATLAS:2012cna}. 
Measurements of the photon \cite{Chatrchyan:2012vq,Chatrchyan:2012nt}, 
\Z\ \cite{Aad:2012ew} and \W\ \cite{ATLAS:2012zla} production rates 
are shown to scale with 
the nuclear overlap function, $T_{AA}$, in Fig.~\ref{fig:RAA}.  In addition the shapes of the
\Pt\ and rapidity distributions are unmodified in
\PbPb\ collisions.  Updated \RAA\ measurements for bosons at higher statistics and
in various decay channels were presented in \cite{hp2013}.
As an example, Fig.~\ref{fig:ATLAS_Fig_3} shows the mean fractional 
energy distribution carried by the jet opposite a photon, $x_{\rm{J\gamma}}$, 
in \PbPb\ collisions \cite{ATLAS:2012cna} compared to PYTHIA simulations
(yellow histogram) embedded into simulated background heavy-ion events. As the centrality 
increases, the distribution shifts toward smaller $x_{\rm{J\gamma}}$, suggesting 
that more and more of the jet momentum distribution falls below a minimum 
$x_{\rm{J\gamma}}$. In contrast, the PYTHIA ratio of the ``true jet'' to ``true photon''
distribution exhibits no centrality dependence.
Similar results are obtained from CMS with photon+jet 
events~\cite{Chatrchyan:2012gt} and from ATLAS for \Z+jet \cite{ATLAS:2012ena} 
and confirmed in \cite{hp2013} with higher statistics.

\paragraph{Jet fragmentation}

\begin{figure}
\begin{center}
\includegraphics[width=0.48\textwidth]{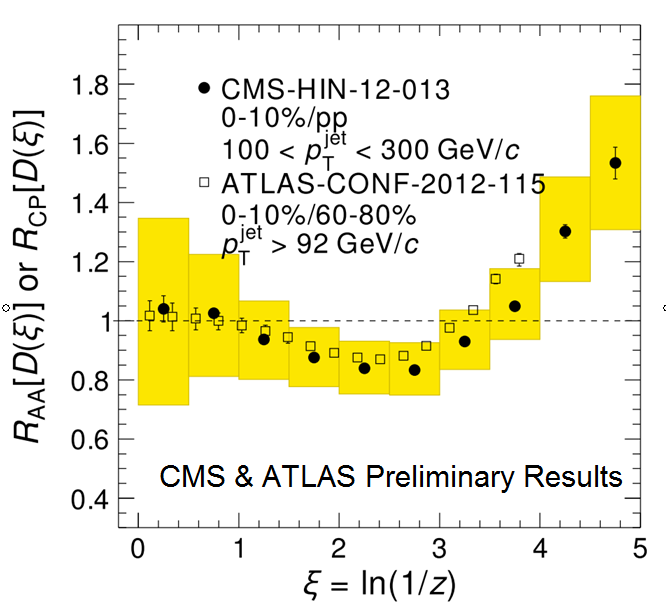}
\caption{The ratio of jet fragmentation functions measured in \PbPb\ and 
\ppcoll\ collisions in two centrality bins as a function of the scaling variable $\xi =\ln (1/z)$, with $z = (p^{\rm track}_{\parallel}/p^{\rm jet})$
where $p^{track}_{\parallel}$ is the momentum component of the track along the jet axis, and $p^{jet}$ is the magnitude of the jet momentum. From CMS \cite{CMS:2012wxa} and ATLAS \cite{ATLAS:2012ina}.}
\label{fig:FFcms}
\end{center}
\end{figure}

Jet structure in the medium has been studied through the fragmentation 
functions and dijet transverse momentum imbalance by means of hard momentum 
cuts on charged particles at \Pt\ $>$ 4~GeV/$c$ and jet cone radii $R<0.3$. 
Figure~\ref{fig:FFcms} shows the ratios of the
fragmentation functions measured in \PbPb\ and \ppcoll\ collisions at
$\sqrt{s_{NN}} = 2.76$ TeV. The fragmentation is measured with respect to the 
final-state jet momentum (after energy loss). 
The results show that the longitudinal
structure of the jet does not change in the high \Pt\ $>$ 100~GeV/$c$ region
where the measurement has been performed
\cite{Chatrchyan:2012gw,CMS:2012wxa,ATLAS:2012ina}.
However, the trend suggests 
a softening of the fragmentation function in the
most central collisions if softer particles (\Pt\ $>$ 1~GeV/$c$) 
are included \cite{CMS:2012wxa}.

In addition to the longitudinal structure of the jet, its transverse structure 
can also be studied. In central \PbPb\ events a significant
shift of the transverse momentum imbalance of the leading jet and its recoil 
partner is observed for
$\Delta\phi_{1,2}> 2\pi/3$ with respect to \ppcoll\ collisions. 
The shift, which changes monotonically with centrality, does not show a 
significant dependence on the leading jet \Pt\ \cite{Chatrchyan:2012nia}. 
The implication for the absolute magnitude of energy loss should be extracted 
employing realistic models \cite{Chatrchyan:2012nia}. 

\paragraph{Jet structure}

\begin{figure*}
\begin{center}
\includegraphics[width=\textwidth]{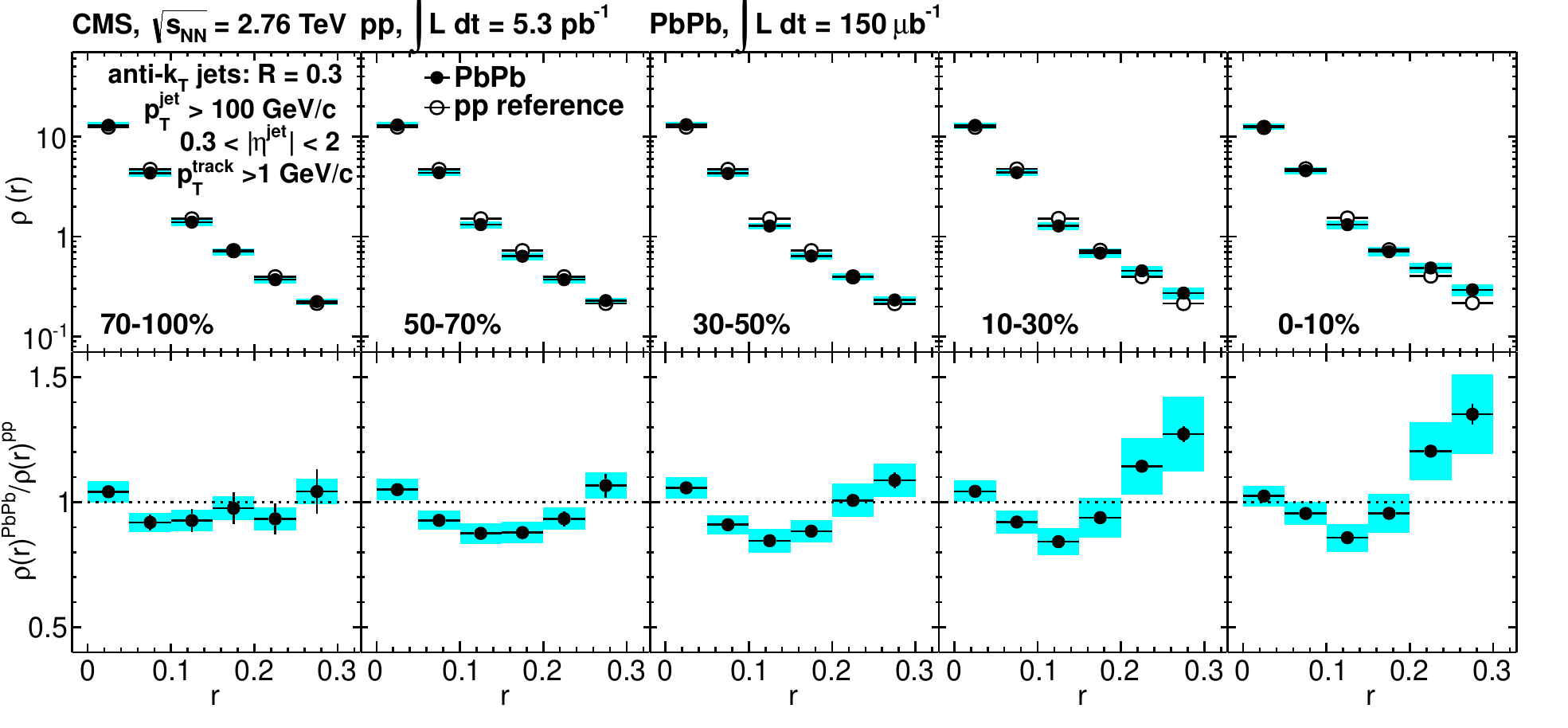}
\caption{The differential jet shapes, $\rho(r)$, in \PbPb\ and \ppcoll\ 
collisions determined by CMS shown as a function of 
the annular regions in the jet cone $r$, in steps of $\Delta R$ = 0.05.
The \PbPb\ data are shown as the filled points 
while the open circles show the \ppcoll\ reference. In the bottom
panel, the ratio of the \PbPb\ to \ppcoll\ jet shapes is shown for annular
regions in the jet cone, from the center to the edge of the jet cone radius 
$R$. The band represents the total
systematic uncertainty \cite{Chatrchyan:2013kwa}.}
\label{fig:jshape}
\end{center}
\end{figure*}

The QGP is expected to modify the jet shape both because of
parton interactions with the medium and because soft particle production in
the underlying event adds more particles to the jet.
Thus the energy flow inside a jet, sensitive to the characteristics of the
medium traversed by the jet, can be studied through jet shape analysis which
should then widen due to quenching effects. CMS has measured the 
average fraction of the jet transverse momentum within annular regions
of $\Delta R$=0.05 from the inner part of the jet to the edge of the jet cone.  
Correcting for the underlying
event and all instrumental effects in central collisions, moderate jet 
broadening in the medium is observed for $R = 0.3$
\cite{Chatrchyan:2013kwa}. The effect increases for more central
collisions.  This is consistent with the concept that energy lost by jets is
redistributed at large distances from the jet axis, outside the jet cone, 
see Fig.~\ref{fig:jshape}.
For an update on theoretical
developments at high \Pt, see Ref.~\cite{CasalderreySolana:2012bp}.

As discussed previously, the structure of high energy jets at the LHC is 
unmodified: the radiated energy is carried by low \Pt\ particles at large
distance away from the jet axis \cite{Chatrchyan:2011sx}.
Models suggest different behaviors in the jet core and outer regions of
the jet cone due to the different couplings
to the longitudinally-flowing medium or to turbulent color field leading to 
eccentric jet structure.
ALICE extended the study of the centrality dependence of shape evolution in 
the near-side correlation peak to the low and intermediate \Pt\ regions
by measuring the width of the peak in $\Delta \eta$ (longitudinal) and 
$\Delta \phi$ (azimuthal) directions \cite{Morsch2013281}.
The width in $\Delta \eta$ shows a strong centrality dependence,
increasing by a factor $\approx 1.6$ from peripheral to central \PbPb\ events,
while the width in $\Delta \phi$ is almost independent of centrality.
The AMPT model calculations \cite{Lin:2004en,Xu:2011fi}, which take into account
collective phenomena,
exhibit similar behavior, indicating that the observed effects reflect 
collectivity. Such behavior is expected in models taking into account  
interaction of the fragmenting jet with the longitudinally-flowing medium 
which distorts a jet produced with an initial conical profile \cite{Abreu:2007kv}.

To further study the interplay of soft (flow) and hard processes (jets) and 
how they affect hadrochemistry, the particle composition in jet-like structures
was investigated by ALICE.  They studied the $p/\pi$ ratio
in $\Delta \eta$ - $\Delta \phi$ space
relative to a trigger particle.
It is found that in the ``near-side'' peak region, the ratio is consistent 
with expectations from \ppcoll\ collisions, estimated using PYTHIA. In the 
``bulk'' region the ratio is compatible with that obtained for 
non-triggered events, a factor of 3--4 increase compared to \ppcoll.
The hadrochemistry result suggests that there is no significant medium-induced 
modification of particle ratios within jets and the enhancement of the inclusive $p/\pi$ 
ratio observed in minimum bias \PbPb\ collisions is a result of bulk processes 
and not jet fragmentation \cite{Veldhoen2013306}.


\subsubsection*{Heavy flavors}
\label{Sec:HEAVY}

Because heavy quarks are produced in the very early stage of the collision, 
they probe the properties of the QCD medium while traversing it.  Open heavy-flavor measurements
are used to investigate details of the energy loss,  
thermalization, and the hadronization mechanism.   Quarkonium, hidden heavy flavor bound
states, are sensitive to the temperature of the system and the deconfinement mechanism.

In this section we first focus on \D\ and $B$ meson production 
($B$ mesons are identified through their decay to \jpsi\ after they have passed through the medium). 
The role of these mesons as probes, 
rather than relying on their semileptonic decays, has come into maturity at
the LHC.  Single lepton measurements, 
while useful, do not generally allow a clean flavor separation.  Thus
we concentrate on \D\ and $B$ measurements at the LHC, with reference to RHIC
results where appropriate. 
We first discuss the measurement of the nuclear modification factor \RAA\ and 
azimuthal anisotropy $v_2$ of heavy flavor in the bulk medium.
We then discuss correlation studies as well as some of the first \pPb\ results on 
open heavy flavor.  The rest of the section is devoted to a discussion of quarkonium results.

\paragraph{Mass hierarchy of \RAA}

The nuclear modification factor \RAA\ of heavy-flavored particles has been 
measured up to rather high \Pt\ 
and can thus provide information about parton energy loss in the medium.  
Based on QCD predictions of parton energy loss, see Sec.~\ref{Sec:ELOSSTHEORY} and
Sec.~\ref{Sec:ELOSSEXP}, 
pions, primarily originating from gluons and light quarks, should be more 
suppressed than charm particles which are, in turn, expected to be more 
suppressed than particles containing bottom quarks.  Thus a hierarchy of 
suppression is expected with $R_{AA}^B > R_{AA}^\D> R_{AA}^{ch}$.

\begin{figure} 
\begin{center}
\includegraphics[width=0.48\textwidth]{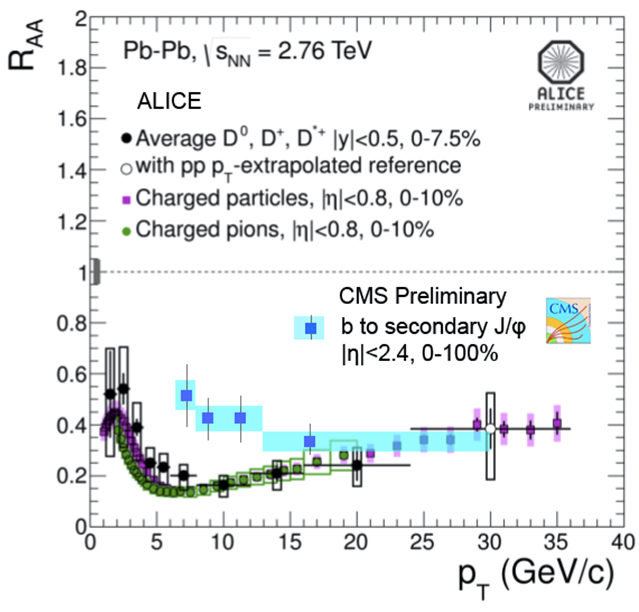}
\end{center}
\caption{
Transverse momentum dependence of the nuclear modification factor \RAA\ for prompt \D\ mesons measured by ALICE as the average of the relevant factors for \Dze, \Dpl, and \Dstp\ at midrapidity in central \PbPb\ collisions at $\sqrt{s_{NN}} = 2.76$~TeV, compared to the \RAA\ of charged hadrons and pions \cite{Grelli:2012yv}. The b-quark energy loss, via nonprompt \jpsi\ from $B$-hadron decays by CMS is also shown \cite{Chatrchyan:2013exa}. From ALICE \cite{Grelli:2012yv} and CMS \cite{Chatrchyan:2013exa}.}
\label{fig:RaaHeavyFlavorD}
\end{figure}

Figure~\ref{fig:RaaHeavyFlavorD} presents the \RAA\ of charged hadrons, charged pions,
prompt \D\ and  $B$-hadron decays via nonprompt \jpsi .  
The \RAA\ for prompt \D\ mesons is
calculated as the average of the relevant contributions from 
\Dze, \Dpl, and \Dstp\ \cite{ALICE:2012ab} for the 
7.5$\%$ most central \PbPb\ collisions at the LHC~\cite{delValle:2012qw,Grelli:2012yv}.
A suppression factor of 4--5 is observed, corresponding to a minimum \RAA\ of
$\approx 0.2$ at \Pt\ = 10~GeV/$c$.
An increase of the \RAA\ with \Pt\ may be expected for a power-law spectrum with
an energy loss equivalent to a constant fraction of the parton momentum if the
exponent in the power law increases with \Pt. 
To test the predicted hierarchy of suppression, the results are compared to the 
charged hadron \RAA\ and found to be very similar.
At \Pt~$<8$~GeV/$c$ the average \RAA\ for prompt \D\ mesons is slightly higher 
than the charged particle \RAA\ (although still within the systematic 
uncertainties), showing a weak indication that $R_{AA}^\D > R_{AA}^{ch}$.
At higher \Pt, the \D\ meson \RAA\ is similar to that of charged hadrons 
\cite{delValle:2012qw,Grelli:2012yv}.
The b-quark energy loss, via nonprompt \jpsi\ from $B$-hadron decays, has been measured
by CMS \cite{Chatrchyan:2013exa} showing
a steady and smooth increase of the suppression as \Pt\ increases and remaining always above the
\D\ mesons.
Similar measurements have been published also by ATLAS in particular open heavy-flavor production
via semileptonic decays to muons as a function of centrality \cite{ATLAS-CONF-2012-050}.
However, more data are still needed to draw final conclusions about the light hadron and
charm meson hierarchy of energy loss.

\begin{figure} 
\begin{center}
\includegraphics[width=0.48\textwidth]{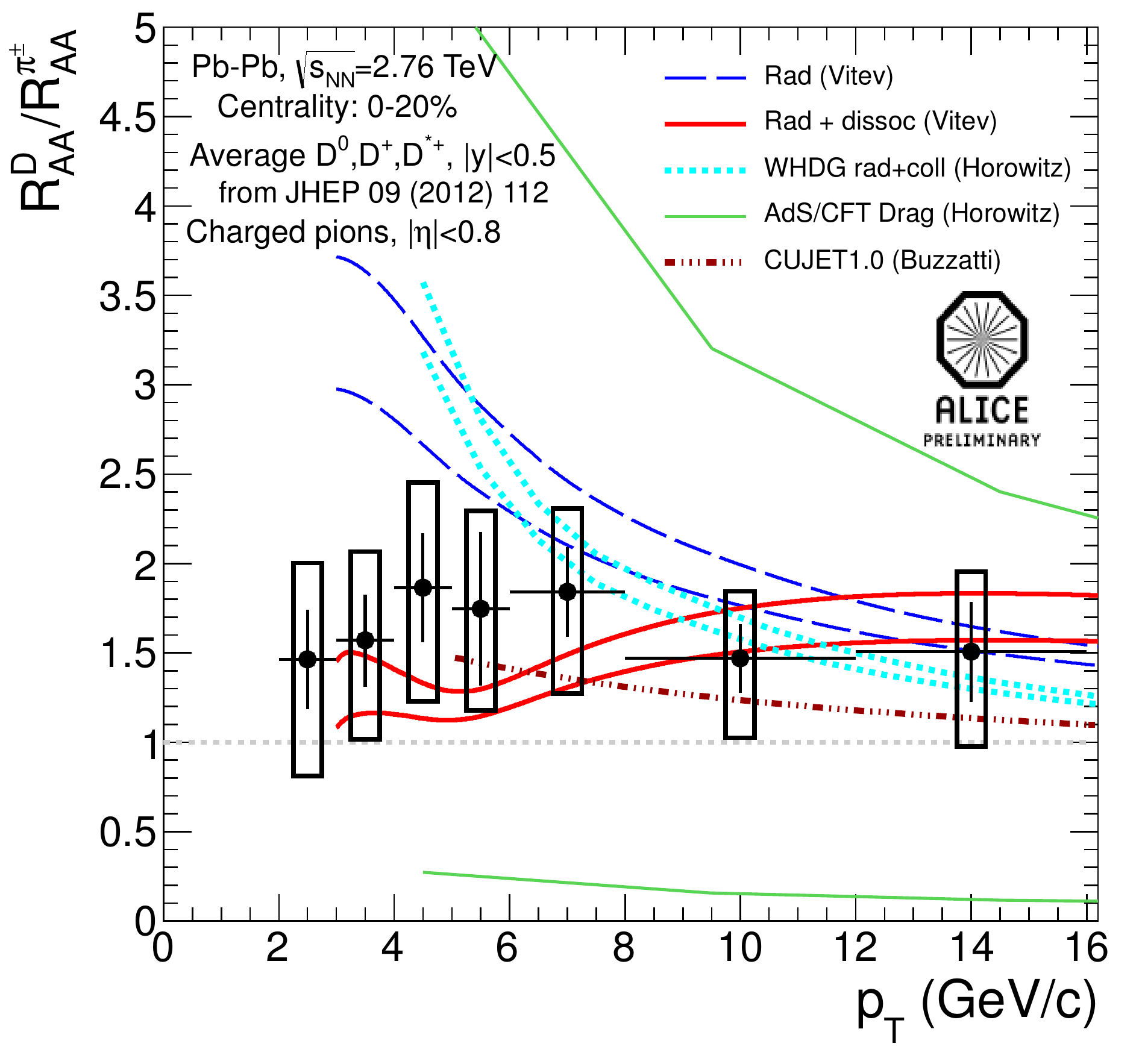}
\end{center}
\caption{ Transverse momentum dependence of the ratio of \RAA\ for \D\ mesons to pions \cite{Grelli:2012yv}. The data are compared to the following model predictions: Rad (Vitev) and Rad+dissoc (Vitev) \cite{Sharma:2009hn,He:2011pd}, WHDG \cite{Horowitz:2011cv}, AdS/CFT Drag \cite{Horowitz:2011wm}, CUJET \cite{Buzzatti:2011vt}. From \cite{Andronic:2013awa}.}
\label{fig:Anton1}
\end{figure}

To better quantify the difference between the \RAA\ of \D\ mesons and 
charged pions, Fig.~\ref{fig:Anton1} shows the ratio $R_{AA}^\D/R_{AA}^{\pi^\pm}$.
The ratio is larger than unity so that $R_{AA}^\D > R_{AA}^{\pi^\pm}$. 
The model comparisons, also presented, show that a consistent description of 
energy loss for light and heavy quarks is a challenge to theory. 
As seen in Fig.~\ref{fig:Anton1}, partonic energy loss models achieve a good 
description at high \Pt\ while the low \Pt\ region is generally not well 
described.  The similarity between light and charm hadron \RAA\ at high \Pt\
is perhaps not so surprising because, in the region where $p_{T} \gg m_Q$,
the heavy quark is effectively light as well.  However, at low to intermediate
values of \Pt, mass effects become important and it becomes more challenging 
to model these results.
In general, more data and quantitative comparison with models is required to 
understand how the relative small difference between the \RAA\ for light hadrons
and heavy flavor can be accommodated by theory in the region where 
$p_{T} \gg m$ does not hold.  This behavior could be due to large elastic 
energy loss in the strongly-coupled quark-gluon plasma 
\cite{Horowitz:2012cf,Renk:2011aa} or the persistence of
heavy resonances within the medium \cite{He:2011qa}.  Recent studies have
shown that calculations involving strong coupling 
\cite{Horowitz:2012cf,Renk:2011aa}, fixed
at RHIC energies, do not extrapolate well to LHC, neither for light nor heavy
flavors.  Enhanced elastic scattering with resonances in a partly confined medium
\cite{Nahrgang:2013xaa} seems a promising scenario.

\begin{figure} 
\begin{center}
\includegraphics[width=0.48\textwidth]{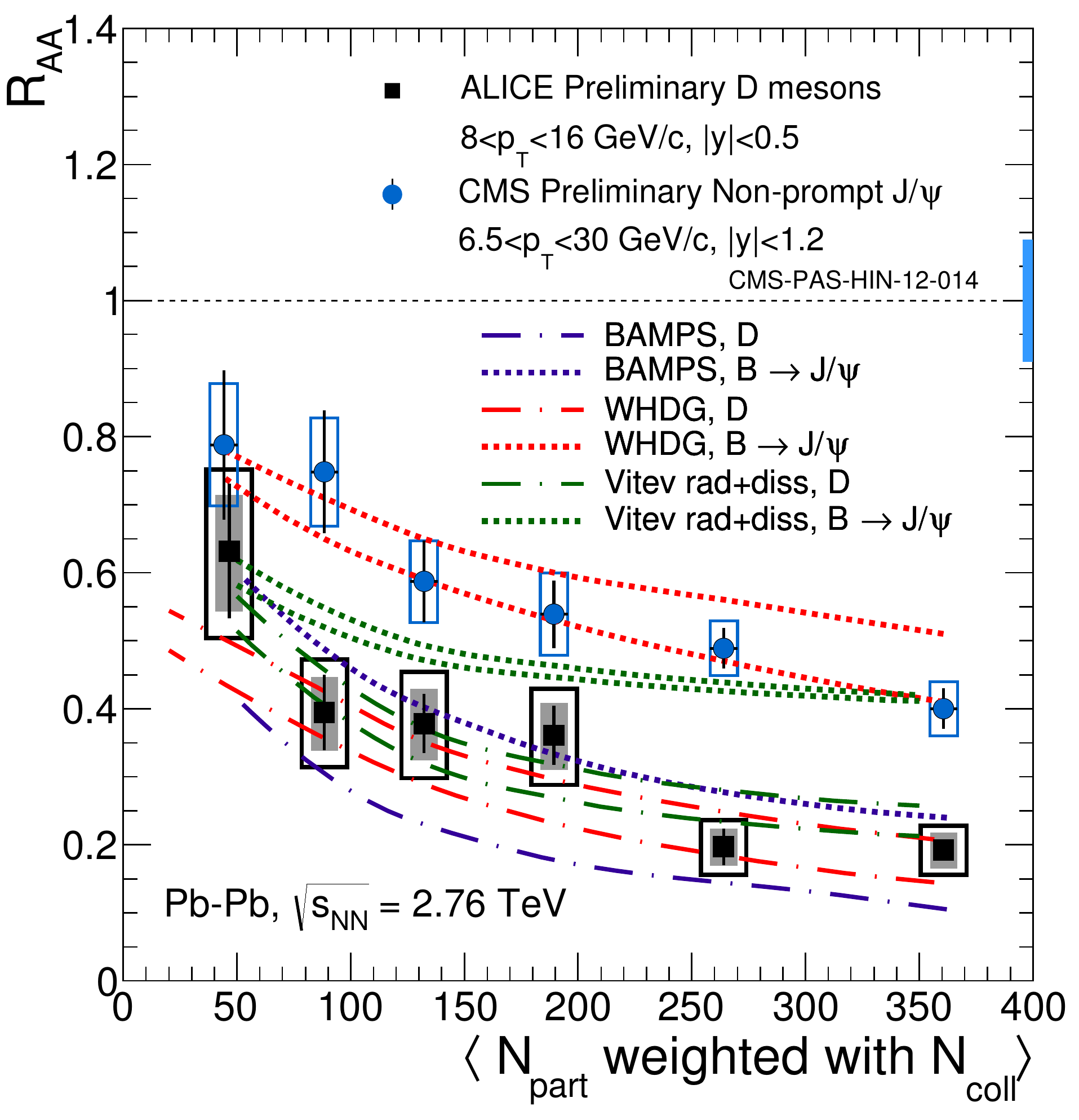}
\end{center}
\caption{Centrality dependence of the charm and bottom hadron \RAA \cite{Stocco.SQM2013,CMS-PAS-HIN-12-014}. The data are compared to BAMPS \cite{Fochler:2011en}, WHDG \cite{Horowitz:2011cv} and Vitev et al. \cite{Sharma:2009hn} model calculations. From~\cite{Grelli:2012yv}.}
\label{fig:Anton2}
\end{figure}

Further indications of the flavor dependence of \RAA\ are shown in 
Fig.~\ref{fig:Anton2}, which presents the centrality dependence of the charm 
and bottom hadron \RAA\ at intermediate \Pt, where \RAA\ exhibits a shallow minimum.
The ALICE data on prompt charmed hadrons are compared to CMS measurements of 
\jpsi\ from $B$-hadron decays to \jpsi. These nonprompt \jpsi\ results were
the first to directly show $B$-meson energy loss.  A compatible \Pt\ range for \D\ 
($\langle p_{{T}}^{\D}\rangle \approx 10$ GeV/$c$) and $B$-mesons 
($\langle p_{{T}}^{B}\rangle \approx 11$ GeV/$c$) has been 
chosen for more direct comparison. These results provide the first clear 
indication of the mass dependence of \RAA.

Similar to inclusive hadrons, jet modification in high-energy heavy-ion collisions is expected to depend on the flavor of the fragmenting parton. To disentangle this flavor dependence, heavy-quark jets have been studied. CMS measured $b$-quark jet production relative to inclusive jets in $pp$ and Pb+Pb collisions at $\sqrt{s_{NN}} = 2.76$ TeV \cite{Chatrchyan:2013exa}. The measurement is in the range 
$80 < p_{T} < 200$ GeV/$c$.  The measured values are comparable to those predicted by PYTHIA vacuum simulations.  The Pb+Pb $b$-jet fraction is also compatible with the $pp$ $b$-jet fraction, within sizeable uncertainties. The measurement is sufficiently precise to demonstrate that $b$-jets are subject to jet quenching, although a precise comparison of light- and heavy-quark jet quenching would require a reduction of the statistical and systematic uncertainties.

\begin{figure}
\begin{center}
\includegraphics[width=0.48\textwidth]{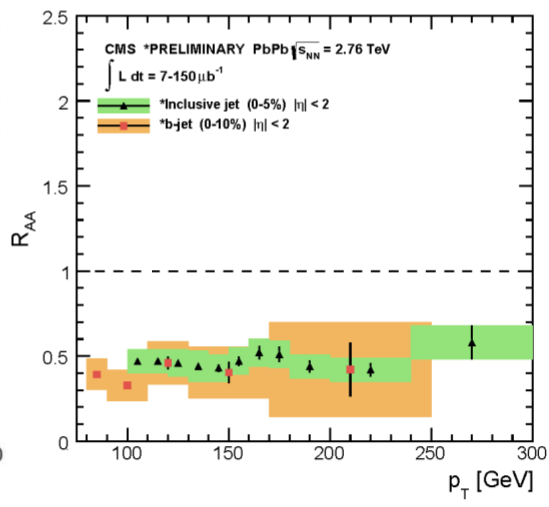}
\end{center}
\caption{
The inclusive \cite{CMS:2012rba} and $b$-jet \cite{Chatrchyan:2013exa} \RAA\  is compared as a function of \Pt\ in the most central Pb+Pb collisions.}
\label{fig:cmsbjet}
\end{figure}

These results from $b$-jet studies, together with those from $B$ decays cover a
wide \Pt\ range and provide a consistent picture.  In general the mass effect seems to be as predicted: 
at intermediate \Pt, bottom is less suppressed than charm, whereas at very high \Pt\ ($E/m\gg1$)
$b$-jets and inclusive jets are similarly modified \cite{Schukraft:2013wba}.

The data are compared to several model calculations. 
Out of the calculations shown, only the WHDG result \cite{Horowitz:2011cv} is 
compatible with \RAA for both \D\ and $B$ hadrons.  WHDG
also achieve results in agreement with the \D\ to $\pi^\pm$ \RAA\ shown in
Fig.~\ref{fig:Anton1} for $p_{T} > 6$~GeV/$c$.  The calculations by
Vitev {\it et al.} \cite{Sharma:2009hn}
Rad (Vitev) and Rad+dissoc (Vitev) \cite{Sharma:2009hn,He:2011pd}, 
especially those labeled ``Rad+dissoc'' in 
Fig.~\ref{fig:Anton1}, agree well with $R_{AA}^\D/R_{AA}^{\pi^\pm}$.  However,
they overpredict the $B$ meson suppression as seen in Fig.~\ref{fig:Anton2}.
The limitation of some calculations to describe the ratio of heavy-to-light \RAA\, shown in
Fig.~\ref{fig:Anton1} for $p_{T} < 8$~GeV/$c$, may be expected, because,
in this range, charm mass effects may still play a role.

The measurements of \D\ mesons with $u$ and $d$ quarks  have recently been 
complemented with the first  measurement of charm-strange, \D$_{s}$, mesons in 
\PbPb\ collisions by the ALICE collaboration \cite{Innocenti:2012ds}. 
Since the \D$_s$ contains both
charm and strange quarks, neither of which exist in the initial state, these
mesons can probe the details of the hadronization mechanism \cite{Kuznetsova:2006bh,He:2012df}.
For example, if in-medium hadronization is predominantly responsible for 
hadronization at low momentum, the relative production of strange to 
nonstrange charm hadrons should be enhanced.  The measurement shows that
\RAA\ for \D$_{s}$  at $8 < p_{{T}} < 12$ GeV/$c$ 
is compatible with that for \D\ mesons, 
with a suppression factor of 4--5 for $p_{{T}} > 8$ GeV/$c$.
In the lower \Pt\ bin, the \D$_{s}$ \RAA\ seems to show an intriguing increase relative to that
of \Dze\ but the current experimental uncertainties need to be improved 
before any conclusive comparison can be made.

\paragraph{Heavy-flavor azimuthal anisotropy}

Further insight into the properties of the medium can be obtained by investigating 
the azimuthal anisotropies of heavy-flavor hadrons. If heavy quarks 
re-interact strongly with the medium, heavy-flavor hadrons should inherit 
the azimuthal anisotropy of the medium, similarly to light hadrons. Measurements 
of the second Fourier coefficient $v_2$ at low \Pt\ can provide information on the degree of thermalization, while,
at high \Pt\ can give insight to the energy loss mechanism. 

\begin{figure} 
\centerline{\includegraphics[width=.49\textwidth]{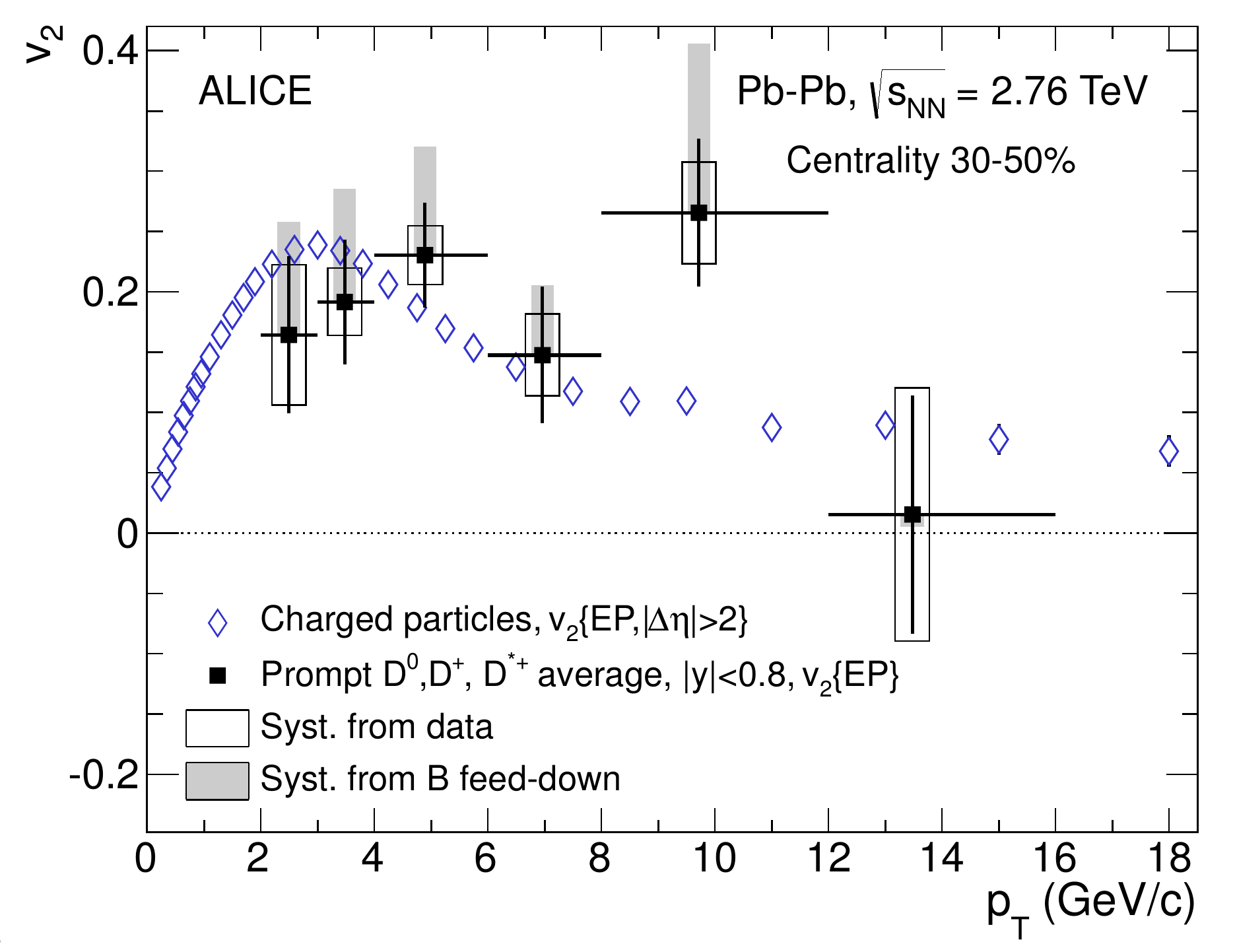}} 
\caption{
The transverse momentum dependence of $v_2$ for \D\ mesons in the 30--50\% 
centrality bin relative to that of inclusive charged hadrons. From \cite{PhysRevLett.111.102301}.}
\label{fig:Anton3}
\end{figure}

Recent measurements of the prompt \D\ meson $v_2$ as a function of \Pt\ in the
30--50\% centrality bin are shown in Fig.~\ref{fig:Anton3}.
A finite $v_2$ value with a significance of 3$\sigma$ is observed for
$2 < p_{T} < 8$~GeV/$c$,
compatible with that of light hadrons within the uncertainties, showing that 
\D\ mesons interact strongly with the medium.
However, higher statistics measurements covering lower \Pt\ are needed to draw 
firm conclusions about charm quark thermalization in the hot medium created at 
the LHC. 

Further differential measurements include the study of the \Dze\ \RAA\
in the in- and out-of-plane azimuthal regions \cite{Caffarri:2012wz}.
The results indicate larger suppression in the out-of-plane azimuthal region, 
as expected, due to the longer path length traversed through the medium in this case.

Current model comparisons which include both radiative and elastic (collisional)
energy loss can explain the high-\Pt\ \RAA\ data in the region where 
$p_{T} \gg m_Q$.  However, energy loss alone is insufficient for describing
the low-\Pt\ \RAA\ and $v_2$ results.  Models which incorporate recombination
or in-medium resonance formation can better describe this region
where mass effects could be important.  

\begin{figure} 
\begin{center}
\includegraphics[width=0.48\textwidth]{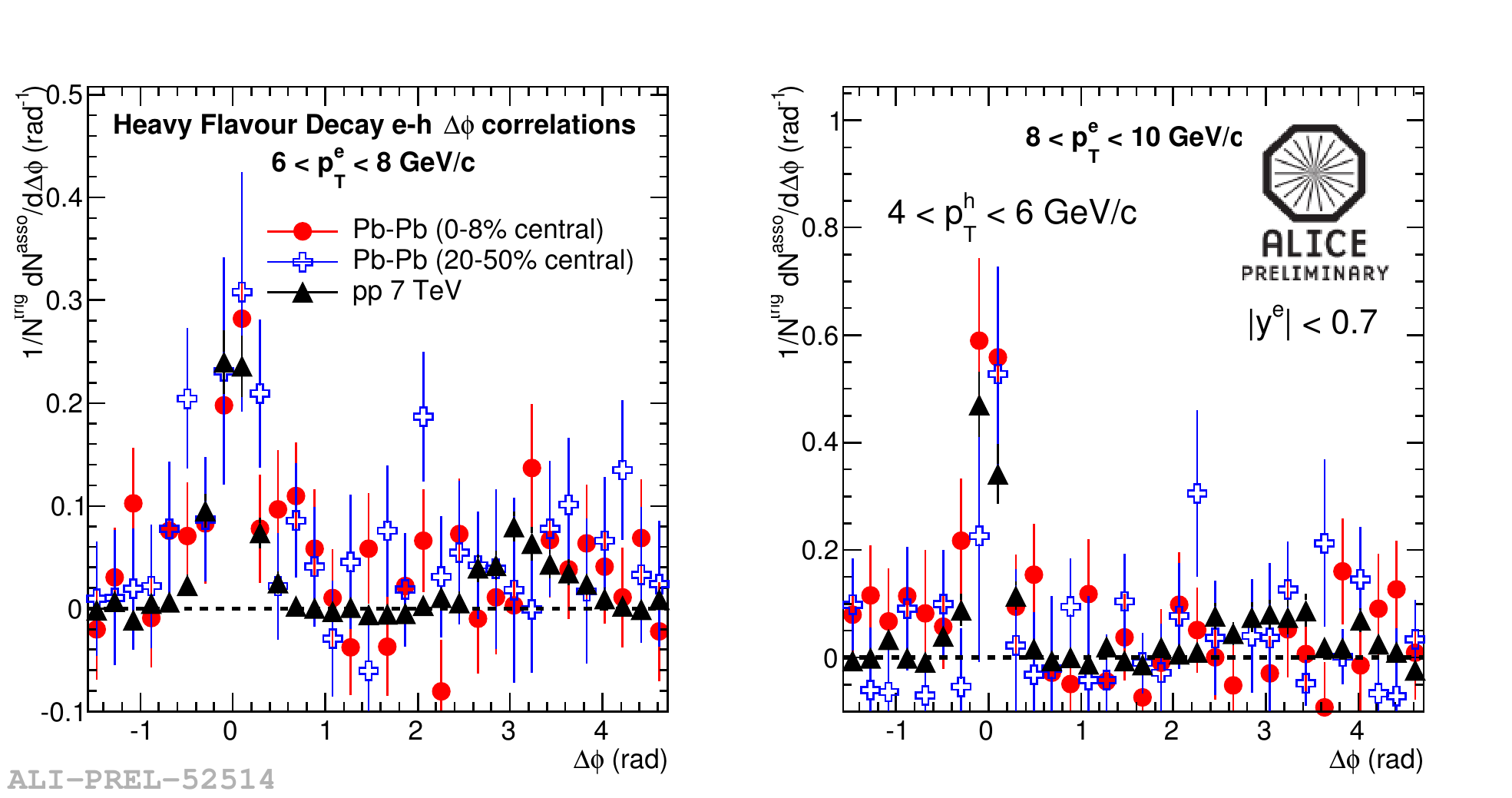}
\end{center}
\caption{Azimuthal angular
correlations $\Delta\phi(\mathit{HFE},h)$ for $4<p_{T}^{h}<6$ GeV$/c$ in $0-8\%$ (red) and $20-50\%$ (blue) central Pb--Pb collisions at $\sqrt{s_{NN}} = 2.76$ TeV compared to $pp$ collisions at $\sqrt{s} = 7$ TeV (black).
From \cite{DeepaThomasfortheALICE:2013rfa}.}
\label{fig:AndreNearSidePeak}
\end{figure}
%
\begin{figure} 
\begin{center}
\includegraphics[width=0.48\textwidth]{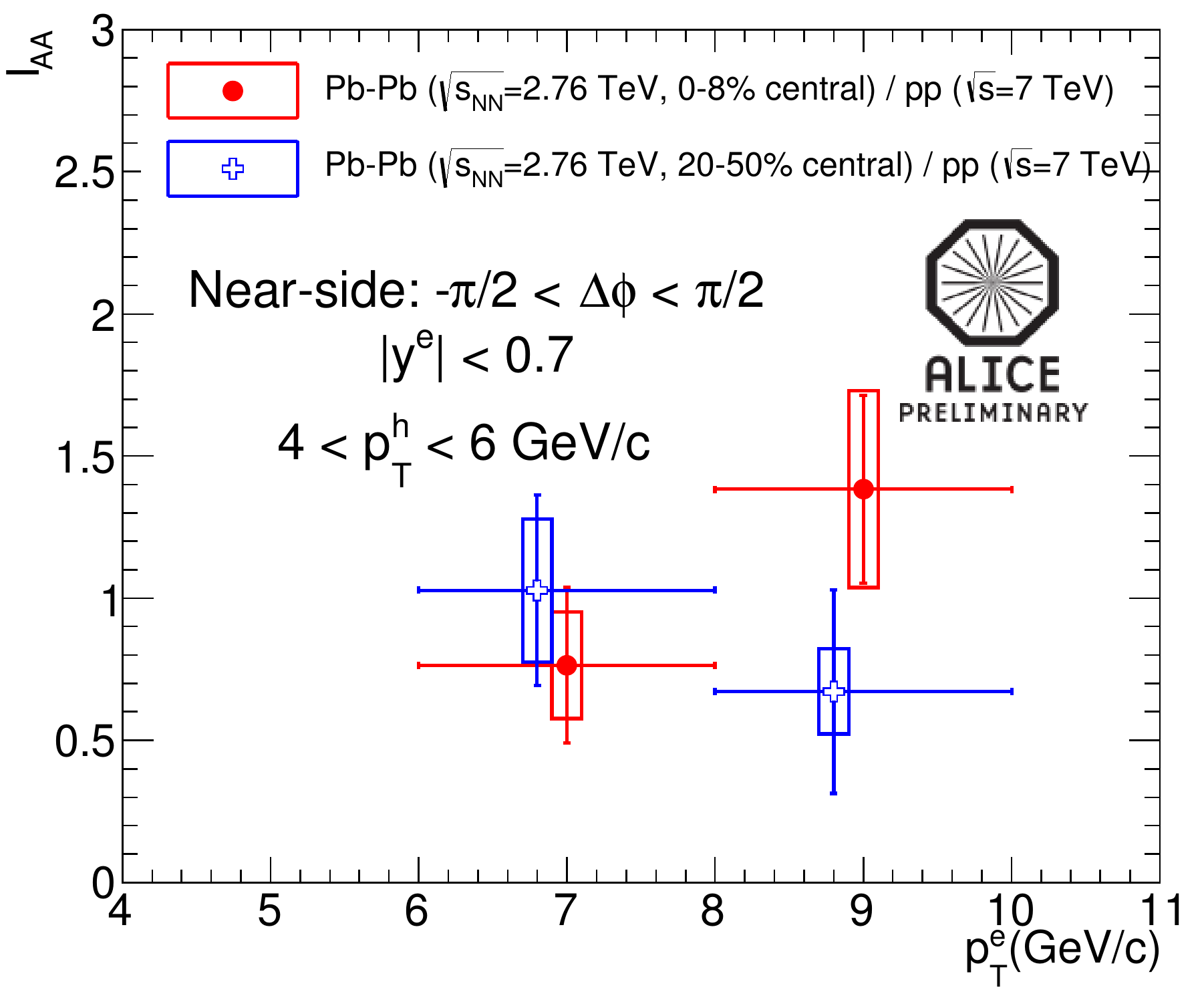}
\end{center}
\caption{The 
$I_{AA}$ of the near-side $\Delta\phi({\rm HFE, hadron})$ correlation 
in the 0--8\% and 20--50\% most central \PbPb\ collisions \cite{Thomas.SQM2013}. }
\label{fig:IAA}
\end{figure}

\paragraph{Flavor correlations}

To a great extent, correlations between heavy quarks survive the fragmentation 
process in proton-proton interactions.  On the other hand, in heavy-ion 
collisions, the medium alters the fragmentation process, so that observables 
are sensitive to the properties of the medium. It has been shown that 
the fragmentation function, which describes how the parton momentum is 
distributed among the final-state hadrons, is most suited for these detailed 
studies.  Flavor conservation implies that heavy quarks are always produced 
in pairs.  Momentum conservation requires that these pairs are correlated in 
relative azimuth ($\Delta\phi$) in the plane perpendicular to the colliding 
beams.  Since heavy flavors are produced in $2 \rightarrow 2$ ($gg \rightarrow
Q \overline Q$) and $2 \rightarrow 3$ (e.g., $gg \rightarrow Q \overline
Q g$) processes, the azimuthal correlation is not strictly back-to-back.

One method of exploiting this pair production characteristic is to measure the 
correlation  of electrons from semileptonic decays of heavy-flavor hadrons 
(HFE) with charged hadrons.
Figure \ref{fig:AndreNearSidePeak} shows the $\Delta\phi({\rm HFE, hadron})$ 
distribution measured by the ALICE Collaboration  \cite{Thomas.SQM2013}.
A distinct near-side correlation is observed. 

The ratio of the measured \PbPb\ correlation relative to the \ppcoll\ 
correlation, $I_{AA}$, 
\begin{eqnarray}
I_{AA} = \frac{\int_{\phi_1}^{\phi_2} d \Delta \phi (dN_{AA}/d \Delta \phi)}
{\int_{\phi_1}^{\phi_2} d \Delta \phi (dN_{pp}/d \Delta \phi)} \, \, ,
\end{eqnarray}
on the near side ($-\pi/2 < \Delta \phi  < \pi/2$)
as a function of the electron trigger \Pt\
is shown in Fig.~\ref{fig:IAA}.  An excess, $I_{AA} > 1$, may be expected at 
high electron \Pt\ in central collisions due to the depletion and
broadening of the correlation signal in the medium.
These results agree with previous measurements at RHIC \cite{Adare:2010ud}. 
However, so far they are statistics limited and more precision data is 
needed, both at RHIC and the LHC, 
to draw final conclusions. Simulation studies suggest that the 5.5 TeV \PbPb\ data, expected after 2015, 
should be sufficient for these studies.

\paragraph{Heavy flavor in \pPb\ collisions}

To quantitatively understand \AAcoll\ results in terms of energy loss, it is 
important to disentangle hot nuclear matter effects from initial-state effects 
due to cold nuclear matter, such as the modification of the parton distribution 
functions in the nucleus \cite{Eskola:2009uj}, discussed in Sec.~\ref{RVpPb}, 
and saturation effects in the heavy-flavor sector \cite{Fujii:2006ab}.
Initial-state effects can be investigated by measuring \D\ production in \pPb\ 
collisions.

\begin{figure} 
\begin{center}
\includegraphics[width=0.48\textwidth]{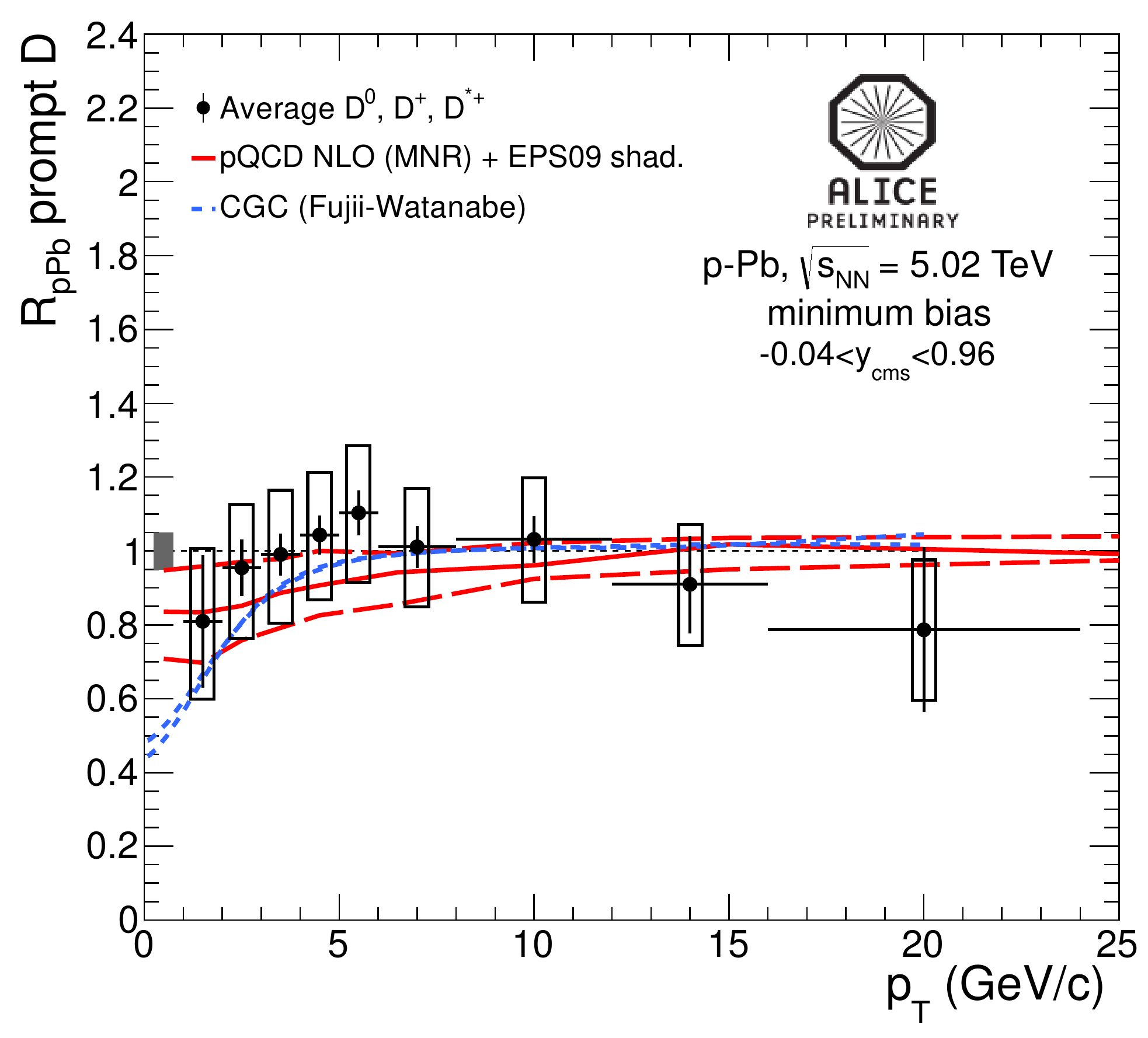}
\end{center}
\caption{Average \Dze, \Dpl, and \Dstar\ \RpPb\ \cite{Luparello.SQM2013}
compared with NLO pQCD \cite{Mangano:1991jk,Eskola:2009uj}
and CGC calculations \cite{FujiWatanabe.private}. From \cite{Luparello.SQM2013}.}
\label{fig:Daverage}
\end{figure}

The nuclear modification factor of the averaged prompt \Dze, \Dpl\ and \Dstar\ 
mesons in minimum bias \pPb\ collisions at $\sqrt{s_{NN}} = 5.02$~TeV is compatible with 
unity within systematic uncertainties over the full \Pt\ range, see 
Fig.~\ref{fig:Daverage}. 
The data are compared with pQCD calculations based on the exclusive NLO
heavy-flavor calculation \cite{Mangano:1991jk}
employing the EPS09 modifications of the parton distribution 
functions \cite{Eskola:2009uj} and also with a color glass condensate-based 
calculation \cite{Fujii:2013yja}. 
Both models describe the data within the uncertainties
indicating that the strong suppression observed in central \PbPb\ interactions 
is a final-state effect.

\paragraph{Quarkonium results}

We now turn to recent results on quarkonium, bound states of ``hidden'' charm
(\jpsi\ and $\psi'$) and bottom ($\Upsilon(1S)$, 
$\Upsilon(2S)$ and $\Upsilon(3S)$).  
As discussed in Sec. \ref{Sec:JPSITHEORY}
the dissociation of the quarkonium states due to color 
screening in the QGP is one of the classic signatures of 
deconfinement~\cite{Matsui:1986dk}.
The sequential suppression of the quarkonium states results from 
their different typical radius
providing a so-called ``QCD thermometer'' \cite{Digal:2001ue}.  In this 
scenario excited states such as the $\Upsilon(2S)$, are more
suppressed than the \jpsi\ while the $\Upsilon(1S)$, the most tightly bound
quarkonium $S$ state, is the least suppressed, as shown by the CMS Collaboration
\cite{PhysRevLett.107.052302}. 

The nuclear modification factor \RAA\ has been measured at mid- and forward
rapidity in \PbPb\ collisions at $\sqrt{s_{NN}} = 2.76$~TeV.  The \RAA\ can be
quantified either in terms of collision centrality, generally presented as a function 
of the number of nucleon participants, $N_{part}$, or as a function of the
quarkonium \Pt\ in a given centrality bin.  While we present the quarkonium \RAA\
here, we note that a comparison to \ppcoll\ may not be the most relevant baseline
for quarkonium.  Instead, quarkonium suppression should be normalized relative
to open heavy flavor results in the same acceptance \cite{Satz:2013ama} because
a similar suppression pattern for \jpsi\  and \D\ mesons may not be indicative of
Debye screening but of another mechanism, such as parton energy loss, particularly
at high \Pt.

\begin{figure} 
\begin{center}
\includegraphics[width=0.48\textwidth]{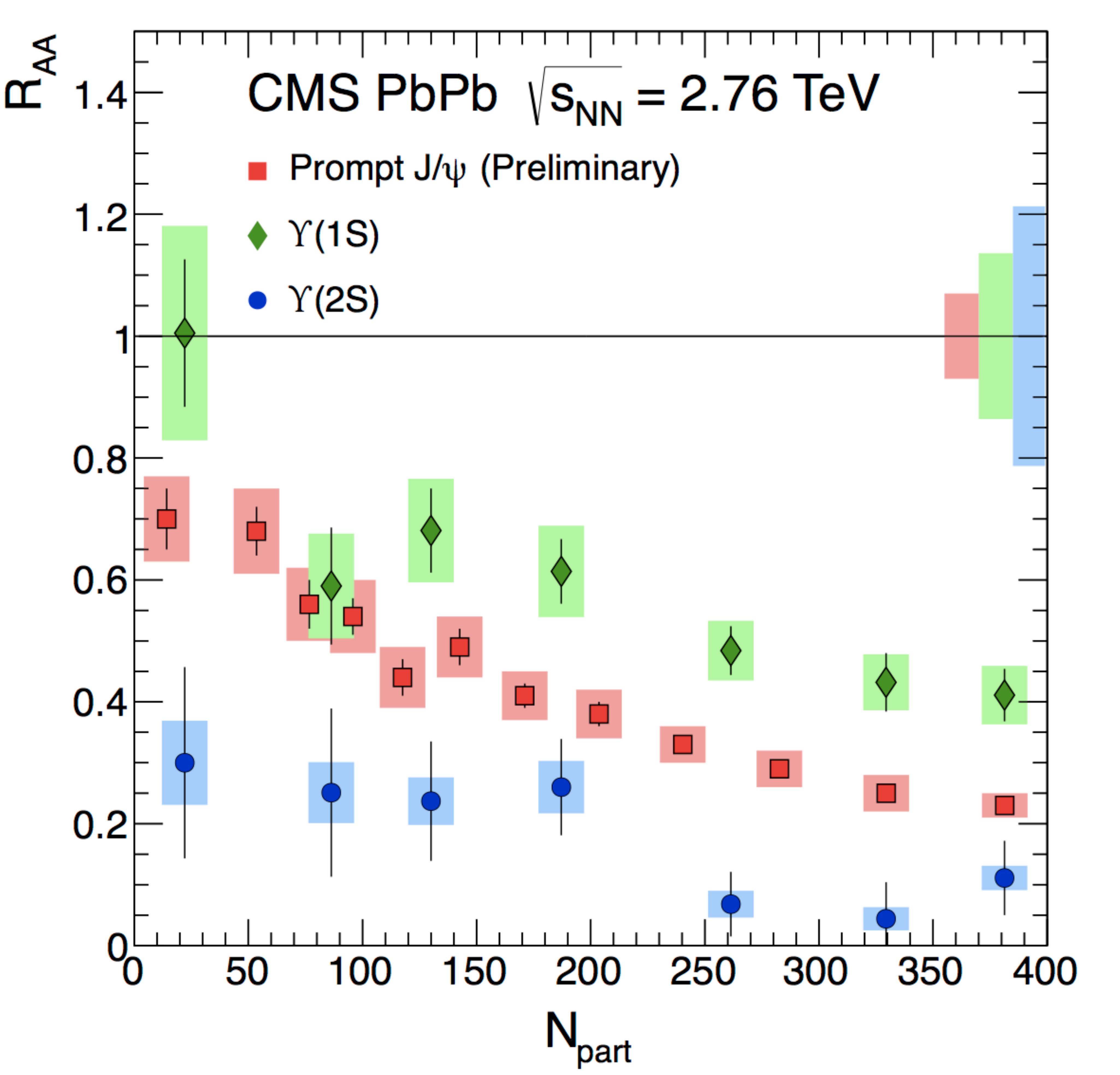}
\end{center}
\caption{The 
nuclear modification factor \RAA\ for prompt \jpsi, $\Upsilon(1S)$ and
$\Upsilon(2S)$ at midrapidity as a function of the number of participants in 
\PbPb\ collisions at $\sqrt{s_{NN}} = 2.76$~TeV measured in the dimuon
channel by the CMS Collaboration. From \cite{Arnaldi:2012bg}.}
\label{fig:mischke1}
\end{figure}

The CMS $\Upsilon$ and \jpsi\ results,
shown as a function of $N_{part}$ in Fig.~\ref{fig:mischke1}, 
indicate that the sequential melting 
scenario appears to hold.  The $\Upsilon(1S)$ is least suppressed while the $\Upsilon(2S)$
is almost completely suppressed in the most central collisions.  The prompt \jpsi\ result,
with the $J/\psi$s from $B$ decay removed, is intermediate to the two.
Note, however, that the prompt \jpsi\ measurement
is at higher \Pt, $p_{T} > 6$ GeV/$c$, than those of the $\Upsilon$ states, 
available for $p_{T} > 0$.  

\begin{figure} 
\begin{center}
\includegraphics[width=0.48\textwidth]{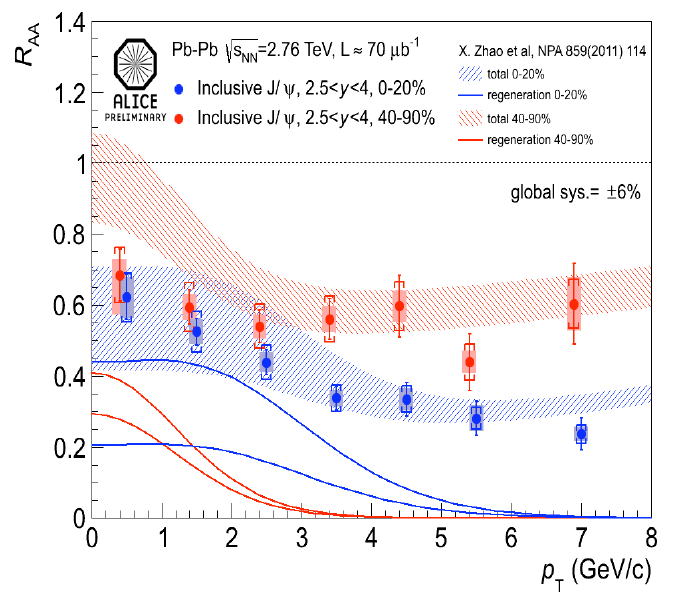}
\end{center}
\caption{Inclusive \jpsi\ $R_{AA}$ in the dimuon channel at forward rapidity in two different centrality bins measured by ALICE \cite{Scomparin:2012vq}. The curves show transport model calculations \cite{Zhao:2011cv}.}
\label{fig:mischke2}
\end{figure}

The sequential suppression pattern described above may be affected by
regeneration of the quarkonium states due to the large $Q \overline Q$
multiplicity at LHC energies, either in the QGP or at chemical freezeout 
\cite{Andronic:2003zv,BraunMunzinger:2000px,Thews:2005vj,Liu:2009nb,Zhao:2011cv}.
Such regeneration might lead to enhancement of the quarkonium
yields in some regions of phase space, as we now discuss.

The ALICE Collaboration has measured \jpsi\ suppression at midrapidity with
electrons and at forward rapidity in the dimuon channel.  The suppression has
been studied as a function of centrality and \Pt. The results indicate that  
inclusive \jpsi\ production is less suppressed at low \Pt, even at forward
rapidity, see Fig.~\ref{fig:mischke2}, which was not observed at the lower RHIC energy. 
In general, the ALICE measurements show that for collisions with $N_{part} >
100$ (the 40--90\% centrality bin), \RAA\ is almost constant as a
function of \Pt\ while the overall suppression is less than that observed in
the most central RHIC collisions.  In the 20\% most central collisions, 
\RAA\ decreases with \Pt, as also shown in Fig.~\ref{fig:mischke2}, similar to 
the RHIC measurements \cite{PhysRevC.84.054912,PhysRevC.86.064901,PhysRevLett.107.142301,PhysRevLett.111.202301}.
A smaller suppression is observed at $p_{T} < 2$~GeV/$c$ than at higher
\Pt\ ($5 < p_{{T}} < 8$~GeV/$c$), especially in more central collisions, 
as also seen in Fig.~\ref{fig:mischke2}~\cite{Ref:AliceRaaJPsi}.
The ALICE results also suggest that the
midrapidity measurements (not shown) exhibit less suppression in 
central collisions than those at forward rapidity \cite{Ref:AliceRaaJPsi}.

The results shown in Fig.~\ref{fig:mischke2} are qualitatively in agreement 
with quarkonium regeneration, where the effects are expected to be important 
in central collisions, particularly at low \Pt\ and midrapidity.  Remarkably,
these results suggest that regeneration may still be important at forward
rapidity.  While this needs to be thoroughly checked before firm 
conclusions are drawn, comparison with transport and statistical model 
calculations suggest that a sizable regeneration component is needed to describe
the low-\Pt\ data.  Further details on the measurements and model comparisons 
can be found in Ref.~\cite{Andronic:2013awa}.

\paragraph{\jpsi\ azimuthal anisotropy}

\begin{figure} 
\begin{center}
\includegraphics[width=0.48\textwidth]{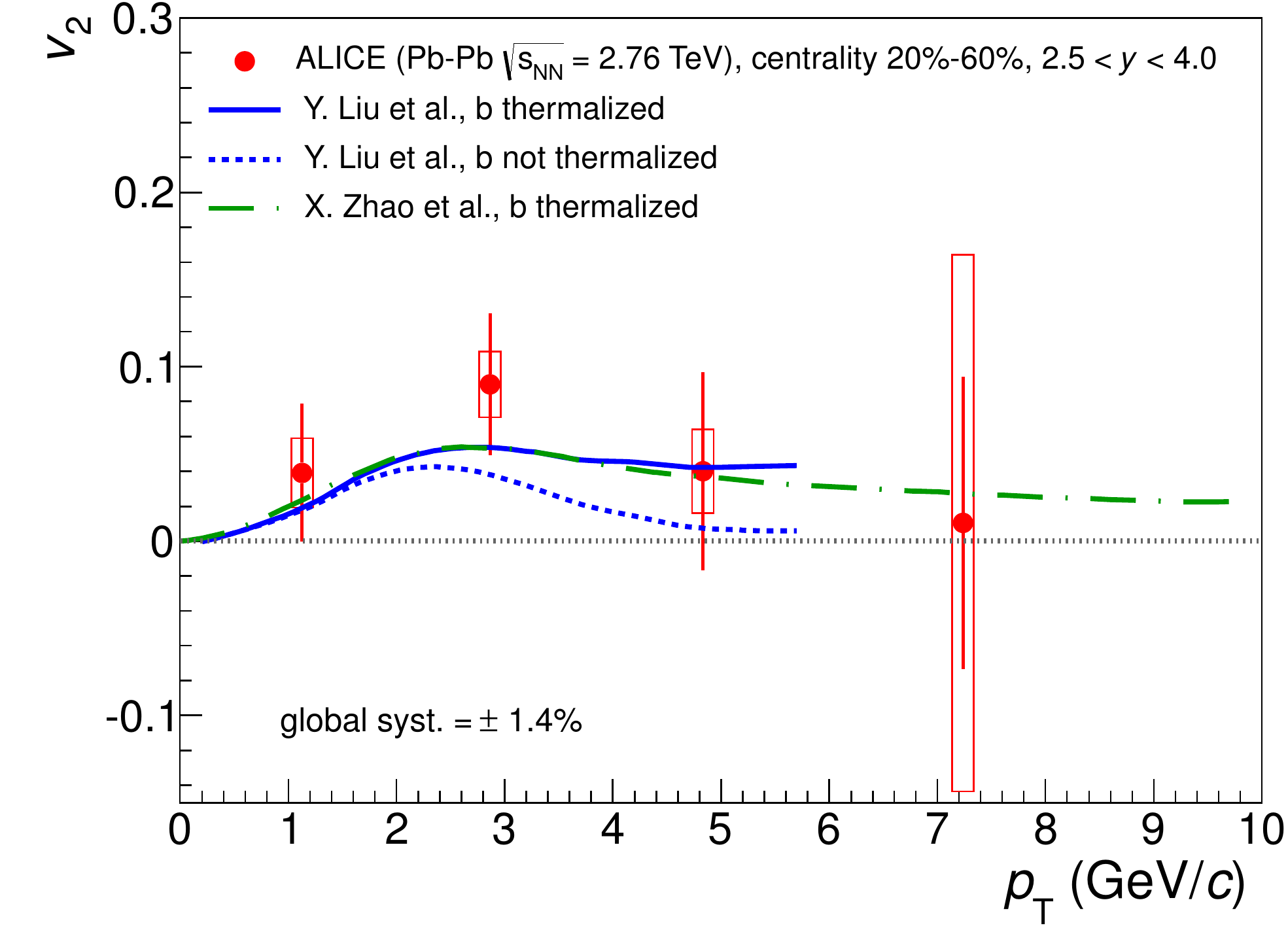}
\caption{\jpsi\ second Fourier coefficient $v_2$ for the 20--60\% centrality range as a function of \Pt. The ALICE data \cite{ALICE:2013xna} are compared with transport model predictions \cite{Liu:2009gx,Zhao:2012gc}. From \cite{ALICE:2013xna}.}
\label{fig:v2JPsi}
\end{center}
\end{figure}

The ALICE Collaboration has studied the \jpsi\ azimuthal anisotropy at forward 
rapidity. The results are shown in Fig.~\ref{fig:v2JPsi} for the 20--60$\%$ 
centrality bin.  This first measurement of inclusive \jpsi\ $v_2$ 
at the LHC shows a hint of a nonzero value
in a somewhat narrower \Pt\ range than that of the \D\ mesons.
This measurement suggests that the \jpsi\ may also follow the collective behavior of the
bulk QGP at low \Pt.  
These results are in agreement with expectations from kinetic and 
statistical hadronization models which require thermalization of the charm 
quarks in the QGP. The calculations differ as to whether or not the $b$ quarks
responsible for nonprompt \jpsi\ production thermalize in the medium.
For more details, see Ref.~\cite{ALICE:2013xna}.

\paragraph{\jpsi\ \RAA\ in \pPb\ collisions}

\begin{figure} 
\begin{center}
\includegraphics[width=0.48\textwidth]{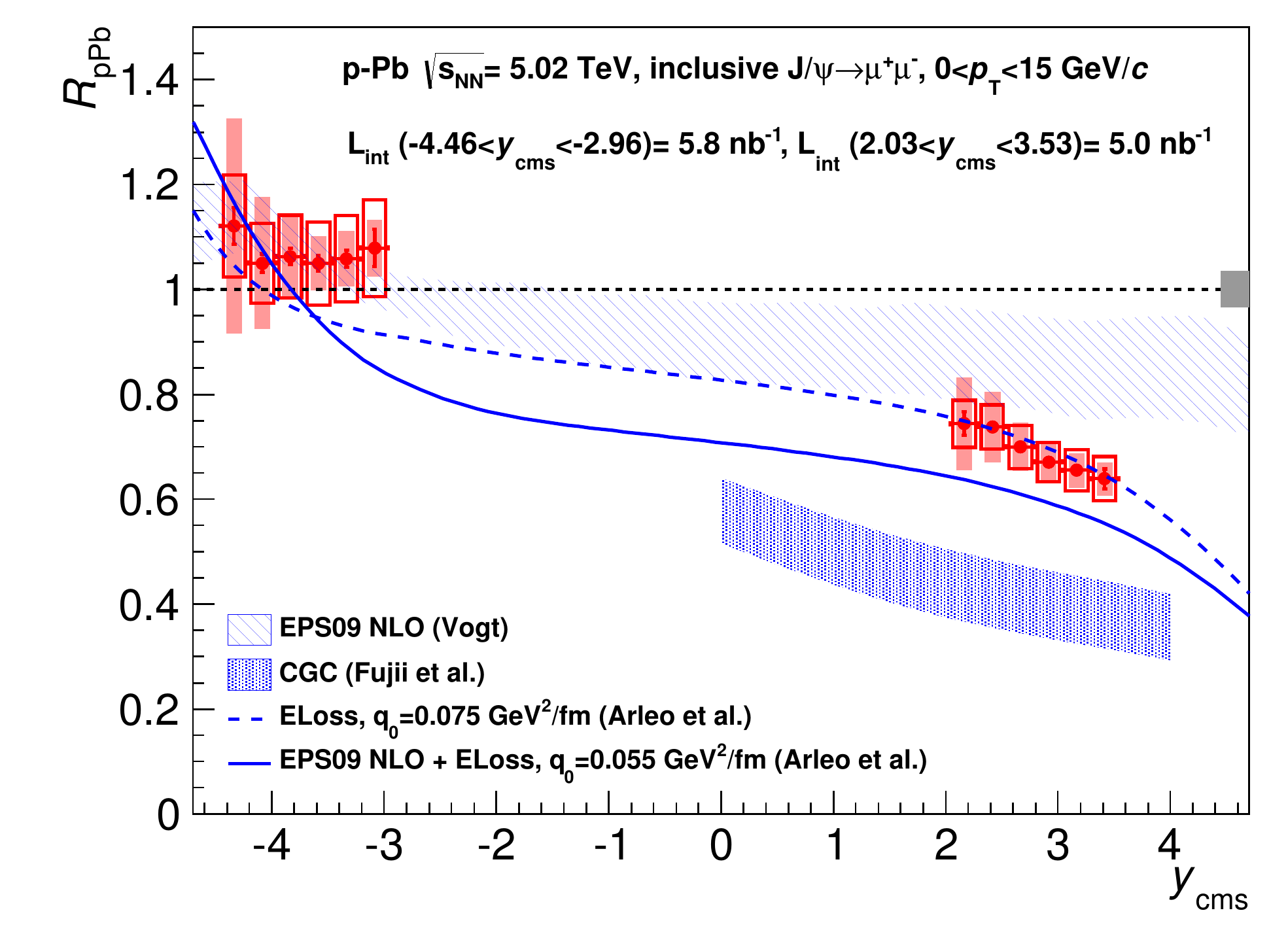}   
\caption{The nuclear modification factors for inclusive \jpsi\ production at $\sqrt{s_{NN}} = 5.02$ TeV measured by the ALICE Collaboration \cite{Bruno.SQM2013}. Calculations from several models \cite{Albacete:2013ei,Fujii:2013gxa,Arleo:2012rs} are also shown. From \cite{GiuseppeEugenioBrunofortheALICE:2013pba}.}
\label{fig:RpA_RAp}
\end{center}
\end{figure}

We now discuss the \jpsi\ results in \pPb\ collisions at the LHC.
The rapidity dependence of the nuclear modification factor
\RpPb\ measured by ALICE
is shown in Fig.~\ref{fig:RpA_RAp}  \cite{Abelev:2013yxa}. 
The LHCb result \cite{Aaij:2013}, in a narrower rapidity window, agrees well with 
the ALICE measurement.
While there is a suppression relative
to \ppcoll\ at forward rapidity, no suppression is observed in the backward
region.  There is good agreement with predictions based on nuclear
shadowing with the EPS09 parameterization alone 
\cite{Albacete:2013ei,Ferreiro:2013pua},  as
well as with models including a contribution from coherent partonic energy loss 
\cite{Arleo:2012hn}.  Whether shadowing only or shadowing with energy loss 
is the correct description requires more data and smaller uncertainties.  The
largest experimental uncertainty is due to the \ppcoll\ interpolation.
The CGC prediction clearly overestimates the suppression.
These results suggest that no significant final-state absorption effects on the 
\jpsi\ are required to explain the data, providing an important baseline
for the interpretation of heavy-ion collision results.

\subsection{Reference for heavy-ion collisions}
\label{chapd:pPbEXP}


One of the most powerful tools in heavy-ion physics is the comparison of $AA$ data
with \ppcoll\ or $pA$ reference data in order to disentangle initial- from final-state effects. 
The \RpPb\ measurements shown in Sec.~\ref{Sec:ELOSSEXP} for charged hadrons and 
heavy flavor are typical examples of this approach. It is, however, based on the assumption that
final-state effects are absent in the elementary collision systems. In the LHC and RHIC energy regime,
this assumption is non-trivial due to the relatively large number of produced
particles and is currently under experimental investigation.

Similar to the measurement in Pb+Pb collisions, the \Pt-integrated charged particle
density distribution measured as a function of $\eta$ in \pPb\ provides essential constraints \cite{PhysRevLett.110.032301}: 
models that include shadowing \cite{Xu:2012au} or saturation \cite{Dumitru:2011wq,Tribedy:2011aa}
predict the total measured multiplicity to within 20\% (see also Fig.~\ref{fig:dNdeta} and Fig.~\ref{fig:dndeta_cm_lab}). 
A closer look at the $\eta$-dependence reveals that saturation models tend to overpredict the difference in multiplicity in the
Pb direction relative to the multiplicity in the proton direction, see Sec.~\ref{sec:5.new}.
Other models like \cite{Barnafoldi:2011px} which consider the effects of strong longitudinal 
color fields, predict too much suppression when shadowing is included and 
too little when it is not.
By tuning the gluon shadowing in \dAu\ collisions at RHIC, DPMJET 
\cite{Roesler:2000he} and HIJING 2.1 \cite{Xu:2012au}, obtain 
multiplicities that are close to the data. 
Recent ATLAS preliminary results~\cite{ATLAS-CONF-2013-096} on the centrality dependence of the 
charged particle multiplicity production in \pPb\ can provide further constraints on model predictions. 
In particular, the data seems to be correctly described by the prediction  
of Ref.~\cite{PhysRevLett.111.182301}.

\begin{figure}
\begin{center}
\includegraphics[height=5.4cm,width=7.5cm]{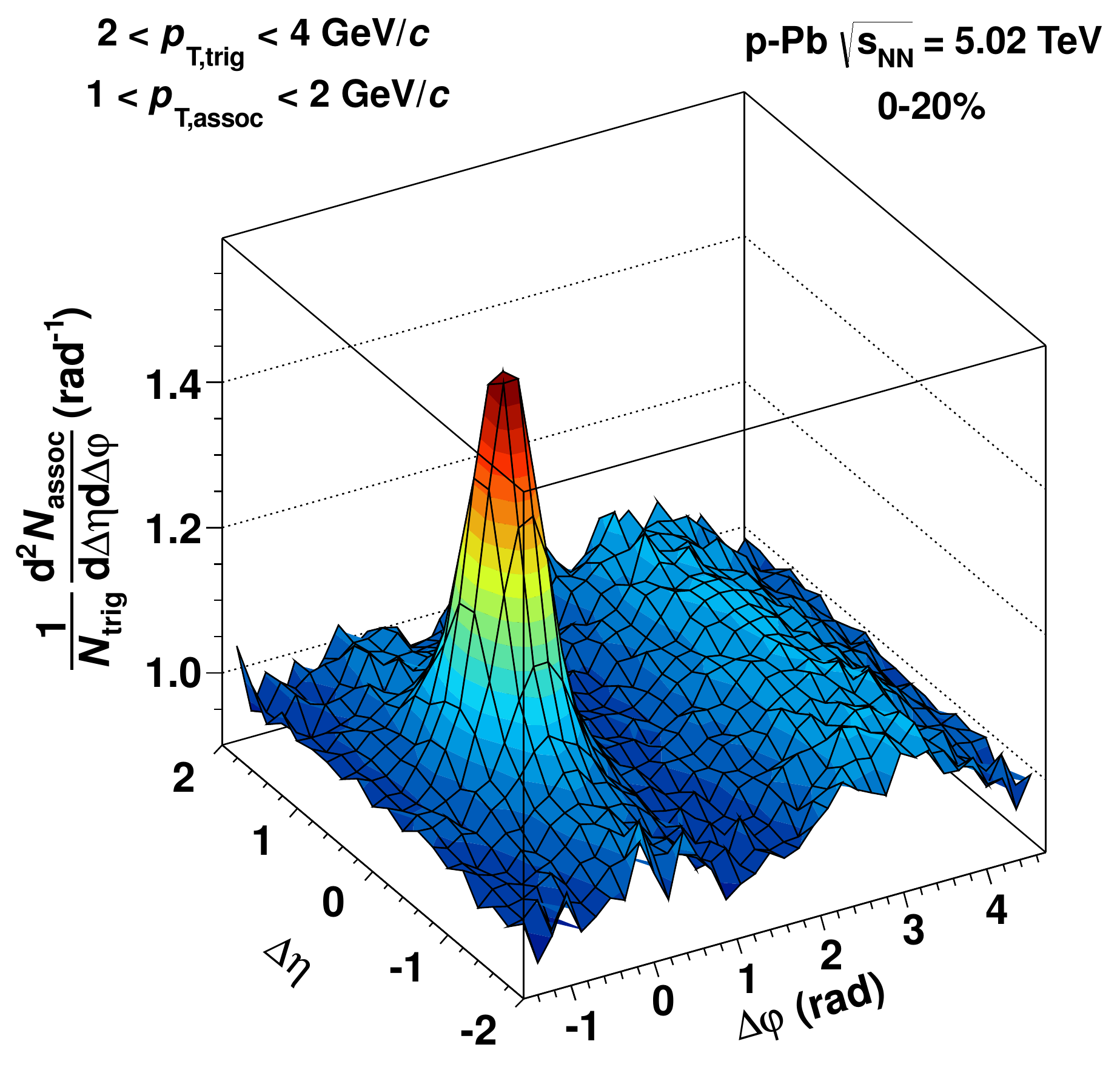}
\includegraphics[height=5.4cm,width=7.5cm]{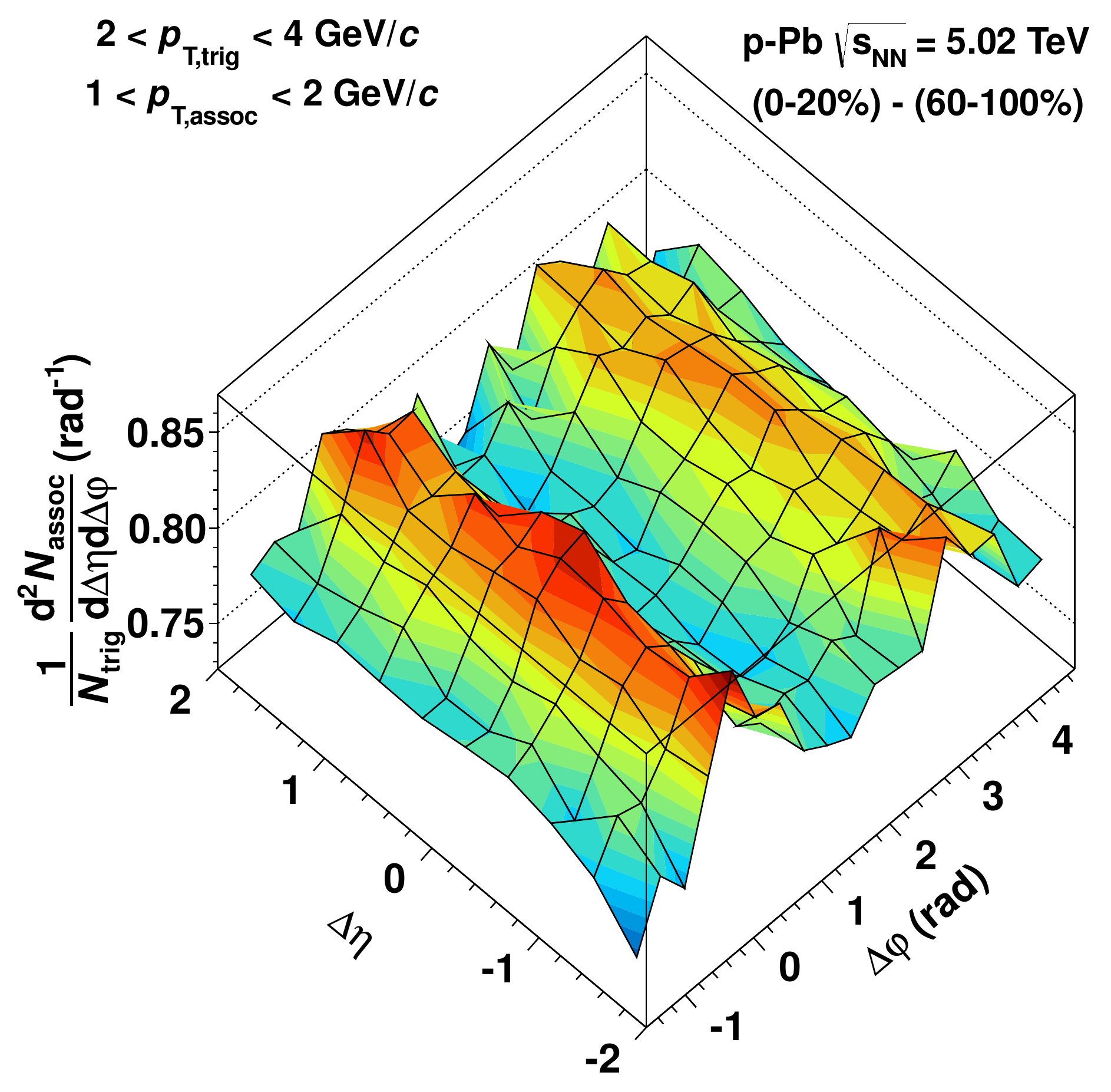}
\caption{(Top) The associated yield per trigger particle in $\Delta\varphi$ and $\Delta\eta$ for pairs of charged particles with $2<$\Pt$<$~4~GeV/$c$ for the trigger particle and $1<$\Pt$<$~2~GeV/$c$ for the associated particle in \pPb\ collisions at 5.02 TeV for the 0--20\% event multiplicity class. (Bottom) The same quantity after subtraction of the associated yield obtained in the 60--100\% event class. From \cite{Abelev:2012ola}.}
\label{fig:DoubleRidge}       
\end{center}
\end{figure}

In addition to the studies of the minimum bias data samples, typical observables used to characterize heavy-ion collisions can be studied as a function of multiplicity in \pp\ and \pPb\ collisions.
 In particular,
at the high LHC energies the particle multiplicity in the high-multiplicity classes of elementary collisions are comparable to e.g., \CuCu\ collisions at RHIC energies. 

Some of the most surprising results in elementary collision systems at the LHC have been obtained by measuring two-particle correlations in 
high-multiplicity events. In particular, $\Delta\eta$-$\Delta\phi$ distributions exhibit several structures that arise from different physics mechanisms;
here, $\eta$ and $\phi$ denote pseudorapidity and azimuthal angle,
while $\Delta$ forms the difference between the trigger particle and the associated particle.
In \ppcoll\ \cite{Khachatryan:2010gv} as well as \pPb\ collisions \cite{Chatrchyan:2013nka,ABELEV:2013wsa,Aad:2013fja,Aad:2012gla}, a novel ridge-like correlation structure elongated in rapidity
has been observed for particles emitted within an azimuthal angle close to that of the trigger particle. This region in phase space with $\Delta\phi \approx 0$ is often referred to as the ``near side''. In \pPb\ collisions, effects originating from the interplay of multiple $NN$ collisions are separated
from those arising from a superposition of individual $NN$ collisions by subtracting the distributions
of low-multiplicity events, with $N_{\rm part} \sim 2$, from the ones of high-multiplicity.
As shown in Fig.~\ref{fig:DoubleRidge}, this procedure removes the jet peak close to $\Delta\eta \approx 0$ on the near-side and reveals the presence of the same ridge structure on the ``away side'' ($\Delta\phi \approx \pi$) with similar magnitude \cite{ABELEV:2013wsa}. In heavy-ion reactions, the double-ridge structure has been interpreted as originating from collective phenomena such as elliptic flow. Several theoretical explanations of these observations have been put forward, including those based on saturation models \cite{hp2013} and hydrodynamic
flow \cite{Li:2012hc}. However, the application of hydrodynamic models to small systems such as \pPb\ is complicated because uncertainties related to initial-state geometrical fluctuations and viscous corrections may be too large for hydrodynamics to be a reliable framework \cite{Bzdak:2013zma}.

To clarify the situation, the mass dependence of the ridge effect has been investigated \cite{ABELEV:2013wsa}. Indeed, an ordering of pions, kaons, and protons was found, which is reminiscent of similar observables in \PbPb\ collisions (see also Fig.~\ref{fig:QuarkNumberScalingv2}). This behavior was successfully predicted by the EPOS event generator \cite{Werner:2013ipa}. The model is founded on the parton-based Gribov Regge theory, in which the initial hard and soft scatterings create flux tubes that either escape the medium and hadronize as jets or contribute to the bulk matter, described in terms of hydrodynamics.

Significant insights into the origin of the azimuthal correlations in small collision systems
has been provided by CMS by studying two- and four-particle azimuthal correlations,
particularly in the context of hydrodynamic and color glass condensate models. 
A direct comparison of the correlation data between \pPb\ and \PbPb\ collisions has been measured
as a function of particle multiplicity and transverse momentum. The observed correlations were quantified in terms of the integrated near-side associated yields and azimuthal anisotropy Fourier harmonics ($v_n$).  Multiparticle correlations were also directly investigated over a wide range of pseudorapidity as well as in full azimuth \cite{Chatrchyan:2013nka}.

Exploiting the excellent particle identification capabilities at low momentum ALICE is measuring untriggered $\Delta\eta$-$\Delta\phi$ correlations of pions, kaons and protons. Qualitatively new features, as compared to non-identified correlations, are observed for kaons and protons. In particular the influence of the local conservation of strange and baryon quantum numbers was found to be large \cite{Graczykowski2014,Janik:2014cua}. The effects are not well reproduced by Monte-Carlo models \cite{Graczykowski2014,Janik:2014cua,Bencedi:2014xka}. This measurement can shed new light on the process of fragmentation and hadronization in elementary collisions. Such studies were initiated in the 70's by several groups, including Richard Feynman \cite{Field:1977fa}.

A careful analysis of the mean transverse momenta of charged particles \cite{Abelev:2013bla} and of the spectral shapes of $\pi$, $K$, $p$ and $\Lambda$ production \cite{Abelev:2013haa,Chatrchyan:2013eya} as a function of event multiplicity have been pursued in order to investigate the presence of radial flow. In both cases, hydrodynamics-based models, like EPOS, yield a reasonable description of the data. The same holds true for the blast-wave picture \cite{Schnedermann:1993ws}, in which the simultaneous description of the identified particle spectra shows similar trends as in Pb+Pb collisions. The observed baryon-to-meson ratios show an enhancement at intermediate transverse momenta which is even more pronounced in high multiplicity collisions. This behavior is phenomenologically reminiscent of the evolution of the same observable with centrality in \PbPb\ collisions (see also Fig.~\ref{fig:pTopiRatio}).

At the same time, detailed comparisons with PYTHIA8 \cite{Corke:2010yf} show that other final state mechanisms, such as color reconnection \cite{Ortiz:2013yxa}, can mimic the effect of collectivity. In particular, the evolution of \Pt\ distributions in PYTHIA8 from those generated in \ppcoll\ collisions follows a trend similar to the blast-wave picture for \pPb\ or \PbPb\ collisions, even though no hydrodynamic component is present in the model. It will therefore be challenging 
to differentiate between these two scenarios. Systematic comparisons of identical observable in $pp$, \pPb\, and \PbPb\ collisions will help clarify the situation. 

However, in order to be able to perform quantitative comparisons between the different collision systems,
the centrality determination in \pPb\ collisions needs to be carefully addressed. In general, centrality classes are defined as percentiles of the multiplicity distributions observed in different sub-detectors covering disjunct pseudorapidity ranges. In contrast to $AA$ collisions, the correlation between the centrality estimator and the number of binary collisions $N_{\rm coll}$ is not very pronounced: the same value of $N_{\rm coll}$ contributes to several adjacent centrality classes. In particular at the LHC, several technical, conceptually different methods are being developed and investigated in order to reduce the influence of these fluctuations and to provide a reliable estimate of 
$N_{\rm coll}$ \cite{ATLAS-CONF-2013-096,ALICE:2012xs}. For the time being, systematic comparisons between experiments and collision systems can rely on multiplicity classes similar to \ppcoll\ collisions.

Based on latest results, as for example the ones presented at \cite{hp2013}, 
it is clear that \pPb\ collisions can serve not only as a reference to the more 
complex $AA$ systems, but also provide new insights on these
issues and, more generally, on the QCD structure.


\subsection{Lattice QCD, AdS/CFT and perturbative QCD}\label{sec:d:LatQcdAds}


One of the major questions in quark-gluon plasma physics is whether a 
weak-coupling based description works at temperatures of a few hundred MeV, 
relevant for heavy-ion collisions, or whether the system should be described 
using strong-coupling techniques. In the strong-coupling limit, the 
gauge/gravity correspondence provides a computational scheme radically 
different from traditional field theory tools, applicable to large-$N_c$ 
${\cal N}=4$ SYM theory and various deformations thereof.  Numerous 
calculations have demonstrated that non-Abelian plasmas behave very differently 
at weak and strong coupling. In particular, while the weakly-coupled system 
can be described by a quasiparticle picture, at strong coupling the poles of 
retarded Green's functions give rise to quasinormal-mode spectra where the 
widths of the excitations grow linearly with energy. The different couplings 
lead to strikingly different predictions for many collective properties of the 
plasma.  Perhaps the best known example is the extremely low shear viscosity 
of the system \cite{Kovtun:2004de}.

While the weakly- and strongly-coupled pictures of a non-Abelian plasma are 
clearly contradictory, it is a priori uncertain which observables and physical 
processes in a real QGP fall into which realm. One clean way to address this 
question for equilibrium quantities is to compare advanced perturbative and 
gauge/gravity predictions to nonperturbative lattice QCD simulations of the
same quantities.  In this section, we compare several quantities (often from 
pure Yang-Mills theory), for which lattice, perturbative and holographic 
predictions exist. On the holographic side, we begin our discussion from the ${\mathcal N}=4$ SYM theory, moving later to bottom-up gravity duals for non-supersymmetric 
large-$N_c$ Yang-Mills theory.

\subsubsection{Weakly and strongly coupled (Super) Yang-Mills theories}

We begin with the equation of state.  The current best-controlled lattice-QCD 
calculation in the high-temperature regime is found in 
Ref.~\cite{Borsanyi:2012ve}. By $T\approx 2-4 T_c$, the pressure ($p$), entropy 
density ($s$) and energy density ($e$) are all in agreement with perturbative 
predictions. The trace anomaly $e-3p$ has on the other hand traditionally been a more problematic quantity, with perturbative calculations missing the famous peak structure at low temperatures, but a recent calculation employing Hard-Thermal Loop perturbation theory (HTLpt) finds agreement for it already at around $T\approx 2T_c$ \cite{Haque:2014rua}.  On the holographic side, the pressure of an ${\cal N}=4$ SYM 
plasma at infinitely strong coupling is known to be equal to 3/4 the value in
the noninteracting limit, similar to that of the equation of state
at $T\approx 2 T_c$ found in lattice QCD. It is therefore not surprising that in bottom-up 
holographic models of Yang-Mills theory, good quantitative agreement is found 
for nearly all thermodynamic observables close to the transition temperature,
see Sec.~\ref{holoter}.

\begin{figure} 
\begin{center}
\includegraphics[width=0.48\textwidth]{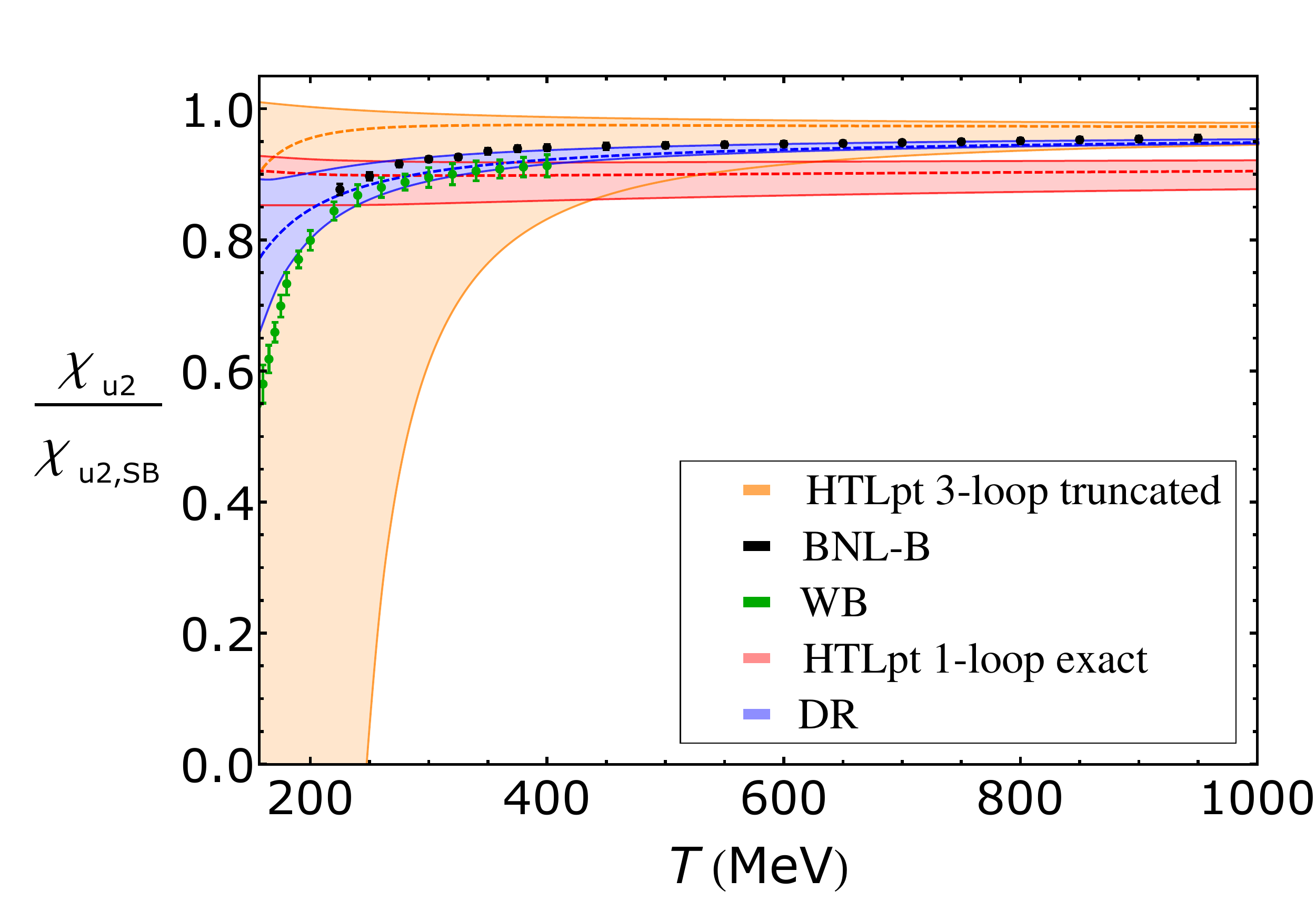}
\end{center}
\caption{The second order light quark number susceptibility evaluated in two different schemes of resummed 
perturbation theory (``dimensional reduction-inspired resummation'' and HTLpt compared with recent lattice 
results from the BNL-Bielefeld (BNL-B) and Wuppertal-Budapest (WB) collaborations. 
From ~\cite{Mogliacci:2013mca}.} 
\label{fig:susc}
\end{figure}

To probe the region of nonzero quark density, technically very demanding for 
lattice QCD, one typically studies quark number susceptibilities, i.e.,
derivatives of the pressure with respect to the quark chemical potentials, 
evaluated at $\mu_q=0$. Continuum-extrapolated lattice data is currently 
available only up to $T\approx 400$ MeV \cite{Borsanyi:2011sw,Bazavov:2012jq}, 
but even below this temperature impressive agreement with resummed perturbation 
theory has been observed, see Refs.~\cite{Andersen:2012wr,Mogliacci:2013mca,Haque:2013sja} as well as Fig.~\ref{fig:susc}.  This can be understood from the 
fermionic nature of the observable, and similar conclusions have indeed been 
drawn for the full density-dependent part of the pressure 
\cite{Vuorinen:2003fs,Haque:2012my}. Very few gauge/gravity results exist for 
these quantities due to the supersymmetry of the SYM theory; one exception is, 
however, the study of off-diagonal susceptibilities found in 
Ref.~\cite{CasalderreySolana:2012np}.

Spatial correlation functions, which reflect the finite correlation lengths of 
the non-Abelian plasma, are another set of interesting observables.  Although 
both perturbative \cite{Laine:2009dh} and holographic \cite{Bak:2007fk} 
predictions for these quantities exist, a systematic precision lattice-QCD 
study of this screening spectrum is still missing, even in pure Yang-Mills 
theory.  At distances much shorter than $1/T$, the correlations of local 
operators effectively reduce to the corresponding vacuum correlators. This 
contribution can, however, be subtracted nonperturbatively \cite{Meyer:2008dt}, 
allowing a prediction of the short distance behavior of the correlation
function in the operator product expansion.  Such a lattice calculation was 
carried out \cite{Iqbal:2009xz} for the components of the energy-momentum 
tensor.  Strikingly, the morphology of the vacuum-subtracted correlator of the 
scalar operator, $G^a_{\mu\nu}G^{\mu\nu a}$, was found 
to be closer to strongly-coupled ${\cal N}=4$ SYM theory than to weakly-coupled 
Yang-Mills theory. This prompted a higher-order calculation of the relevant 
Wilson coefficients in Yang-Mills theory \cite{Laine:2010tc} that, while 
displaying less than optimal convergence properties, drove the analytic 
prediction towards the lattice data. It was also pointed out that considering 
static ($\omega=0$) rather than equal-time correlators is technically more 
favorable for perturbative computations~\cite{Laine:2010fe}, suggesting that 
new lattice calculations should be performed to aid the comparison.

A closely related quantity, also directly accessible by lattice methods, is 
the Euclidean imaginary-time correlation function. These correlators play an 
important role in constraining the corresponding spectral functions, needed to 
calculate transport coefficients, but can also be subjected to a much more 
straightforward (and less model-dependent) test: direct comparison with 
the corresponding perturbative predictions
\cite{Zhu:2012be,Laine:2011xm,Burnier:2012ze}. Extensive continuum-extrapolated 
calculations are needed to make precise comparisons, achieved in only a few 
cases so far.  For example, when the continuum limit of the 
isovector-vector channel is taken in the quenched approximation 
\cite{Ding:2010ga}, an 8--9\% deviation from the massless tree-level prediction 
was found at $\tau=1/(2T)$. These calculations are, however, quite time 
intensive; some estimated computational times as a function of lattice spacing can be found in Ref.~\cite{Meyer:2011gj}.

In weak coupling, there is typically no major difference in the complexity of 
determining finite-temperature Green's functions in the Minkowski-space and 
Euclidean formulations. Indeed, the thermal spectral function is a particularly 
versatile quantity since it allows direct determination of a number of other 
correlators. At the moment, results have been determined up to NLO in 
several channels.  Some relevant operators include the electromagnetic current 
generated by massless \cite{Baier:1988xv,Gabellini:1989yk,Altherr:1989jc} and 
massive quarks \cite{Burnier:2008ia}; the color electric field 
\cite{Burnier:2010rp}; the scalar and pseudoscalar densities 
\cite{Laine:2011xm}; and the shear component of the energy-momentum tensor 
\cite{Zhu:2012be}. The last two results have so far been obtained only for pure 
Yang-Mills theory. In Ref.~\cite{Laine:2011xm} the calculation of NLO spectral 
functions in the bulk channel was significantly refined and systematized. In 
particular, it was shown how these quantities can be reduced to sums of 
analytically-calculable vacuum components and rapidly-converging 
finite $T$ pieces. Very recently, this work was further generalized to account 
for nonzero external three-momenta \cite{Laine:2013vpa}, extending the 
applicability of the results to particle 
production rates in various cosmological scenarios, see Sec.~\ref{sec:d:ImpTheFie}. 

In the absence of lattice data on the Minkowski-space spectral functions, the 
perturbative results can be tested in three different ways:  deriving 
imaginary time Green's functions and comparing them to lattice results, as 
discussed above; verifying and refining nonperturbative sum rules 
\cite{Romatschke:2009ng,Meyer:2010gu}; and direct comparison to gauge/gravity 
calculations. The latter path was taken in Ref.~\cite{k2}, where the bulk and 
shear spectral functions of bottom-up Improved Holographic QCD (IHQCD), 
described later, were seen to accurately reproduce the short-distance (UV) 
behavior of the NLO perturbative Yang-Mills results 
\cite{Laine:2011xm,Zhu:2012be}.  The imaginary time correlators obtained from 
the holographic spectral functions were also seen to be in rather good accord 
with current lattice data.

Finally, we note that meson spectral functions can be calculated rather 
straightforwardly holographically, even at finite density. Gauge/gravity 
duality predicts that meson bound states survive above the deconfinement 
temperature and that their decay is related to a first-order transition within 
the deconfined phase \cite{Babington:2003vm,Mateos:2006nu,Erdmenger:2007cm}. 
For ground state mesons, the new transition temperature is proportional to the 
meson mass so that heavy quarkonia survives at higher temperatures. These 
results are interesting to compare to those of other 
approaches, see Ref.~\cite{Satz:2013ama} and references therein. 

\subsubsection{Holographic breaking of scale invariance and IHQCD \label{holoter}}
\def\pa{\partial}
\def\la{\lambda}

While ${\mathcal N}=4$ SYM theory provides an interesting toy model for strong 
interactions, to approach QCD, breaking of scale invariance must be 
incorporated into the dual-gravity description. There are two classes of 
successful string-inspired models that, beyond modifying
the metric, also introduce a dynamical dilaton field $\phi$, dual to the 
Yang-Mills scalar operator Tr$[F^2]$ \cite{ihqcd1,ihqcd2,ihqcd3,gn}. They both 
can be expressed as a five-dimensional action,
\be
S=M_p^2 N_c^2\int d^5 x\sqrt{g}\left[R-{4\over 3}(\pa\phi)^2+V(\phi)\right],
\ee
where the potential $V(\phi)$ is responsible for the running of the 't Hooft 
coupling, dual to $\la= e^{\phi}$.  IHQCD \cite{ihqcd1,ihqcd2,ihqcd3} is 
constructed so that the theory is dual to pure Yang-Mills theory at both zero 
and finite temperature while the formulation of Ref.~\cite{gn} 
only reproduces the gluon dynamics at finite temperature while it is gapless 
at $T=0$. In the remainder of this section, we focus on IHQCD and its salient 
features.

\begin{figure}
\centering
\includegraphics[width=0.48\textwidth]{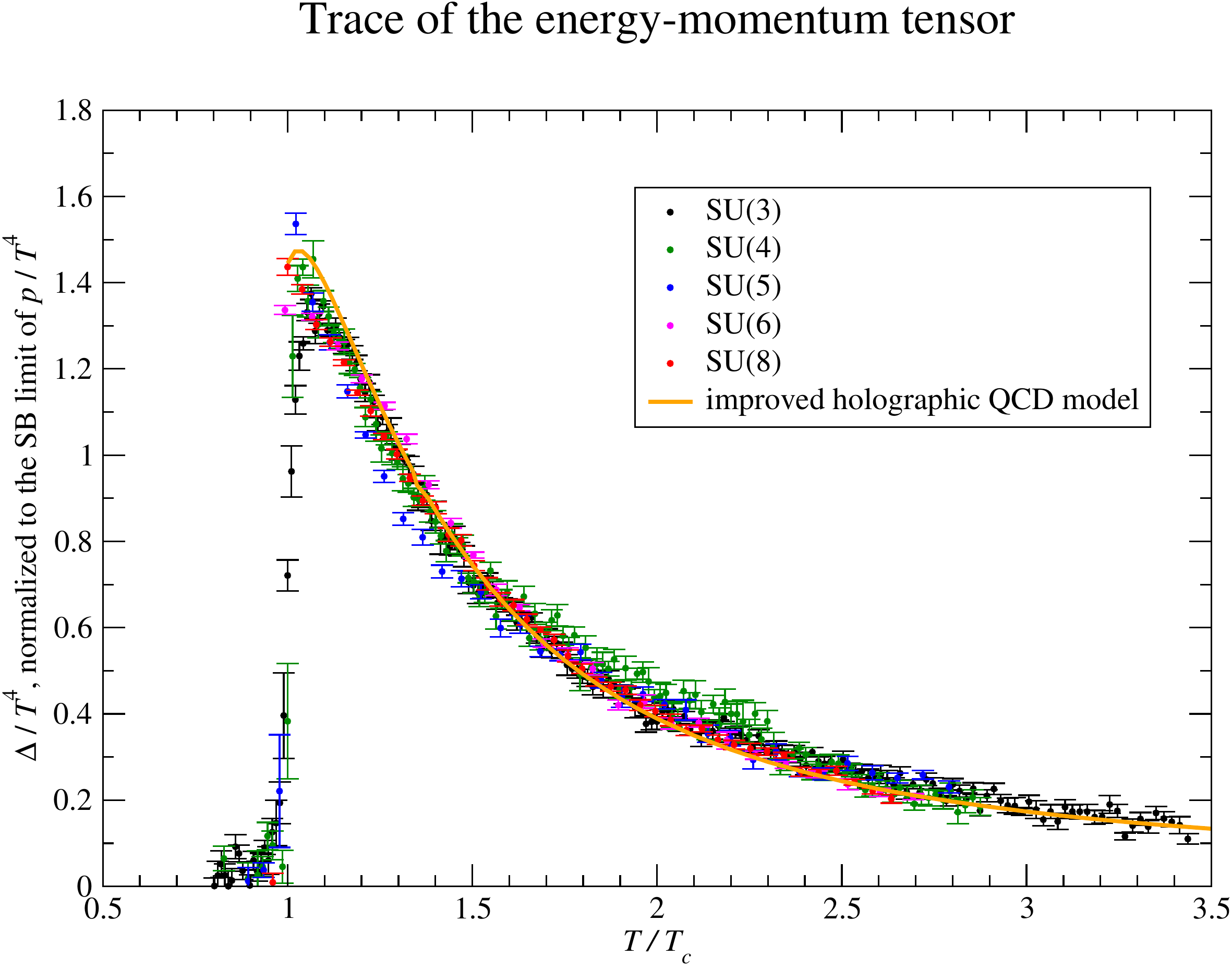}%
\caption{The conformal trace anomaly, $e-3p$, of SU($N_c$) Yang-Mills theory, normalized by $N_c^2T^4$. The points with uncertainties are from lattice calculations \protect\cite{panero}, while the yellow line corresponds to the IHQCD prediction \protect\cite{data}. From \protect\cite{panero}.}
\label{P4}
\end{figure}

For a holographic model to properly account for the UV asymptotic behavior of SU($N_c$) 
Yang-Mills theory, the potential must have a regular expansion in the limit 
$\la\to 0$,
\be 
V(\la)\simeq {12\over \ell^2}\left[1+V_1\la+V_2\la^2+{\cal O}(\la^3)\right],
\ee
where the coefficients $V_i$ are in one-to-one correspondence with the 
perturbative $\beta$-function of the theory, $\beta(\la)$. The long distance 
(IR) asymptotic behavior, $V(\la)\sim \la^{4\over 3}\sqrt{\log \la}$, is 
responsible for the presence of a mass gap and a linear glueball spectrum as 
$\la\to\infty$ \cite{ihqcd2}. Fitting $V_1$ and $V_2$, it 
is possible to accurately reproduce both the $T=0$ glueball spectrum 
and the thermodynamic behavior \cite{data}. This is demonstrated in 
Fig.~\ref{P4}, where the trace anomaly in IHQCD is compared to high precision 
lattice results, evaluated for several values of $N_c$ \cite{panero,lucini}.

In addition to bulk thermodynamic quantities, several transport coefficients 
have been determined in the deconfined phase of IHQCD. While the shear 
viscosity to entropy ratio is found to be the same as in ${\mathcal N}=4$ SYM 
theory, the bulk viscosity, $\zeta$, is also finite in IHQCD 
\cite{bulk}. Recently, these calculations have been extended to cover the full 
frequency dependence of the corresponding spectral densities \cite{k1,k2,k3}, 
revealing good agreement with lattice data. The Chern-Simons diffusion rate 
has also been determined within IHQCD and shown to be about 30 times larger 
than previous estimates based on ${\mathcal N}=4$ SYM \cite{cs}.

Finally, (unquenched) flavor dynamics have recently been added to IHQCD in the Veneziano limit \cite{jk} and the conformal phase transition identified  
as a Berezinsky-Kosterlitz-Thouless type topological transition. Preliminary investigations of the 
corresponding spectra have indicated the presence of Miransky 
scaling\footnote{exponential scaling at the quantum critical point}, the 
absence of a dilaton in the walking regime, and the presence of a substantial 
$S$ parameter\footnote{scale dependence of the difference between the vector 
and axial-vector vacuum polarization amplitudes in technicolor} \cite{jk1}. 
The finite-temperature phase diagrams have the expected forms with an 
additional surprise in the walking regime \cite{jkt}.

Beyond IHQCD, the finite density landscape of QCD has been studied by extending 
the model of Ref.~\cite{gn} by the addition of an extra U(1) gauge field 
\cite{dgr1,dgr2}. In particular, it was found that the phase diagram exhibits 
a line of first-order phase transitions which terminates at a 
second-order critical endpoint, much as expected in $N_c = 3$ QCD.


\subsection{Impact of thermal field theory calculations on cosmology}\label{sec:d:ImpTheFie}

Systematic techniques developed for problems in hot QCD 
may find direct or indirect use in cosmology. Some particular cases are 
detailed in the following.

In cosmology, one compares the rate of expansion of the
universe with the equilibration rate.
The expansion (or Hubble) rate is determined
from the equation of state of the matter that fills 
the universe via the Einstein
equations.  The equilibration rate depends on 
the microphysical processes experienced by a particular excitation. 
Cosmological ``relics'', such as dark matter or baryon asymmetry, 
form if a particular equilibration rate 
falls below the expansion rate.  For example, the cosmic 
microwave background radiation arises when photons effectively 
stop interacting with the rest of the matter. A cartoon equation
for these dynamics is
\begin{equation}
  \dot{n} + 3 H n  
   = 
   - \Gamma^{ }_{ } 
 (n - n^{ }_{\mbox{\scriptsize eq}}) + 
 O(n - n^{ }_{\mbox{\scriptsize eq}})^2 ,
\end{equation} 
where $n$ is the relevant number density, 
$n^{ }_{\mbox{\scriptsize eq}}$ is its equilibrium value, 
$H$ is the Hubble constant and $\Gamma^{ }_{ }$ the
microscopic interaction rate. 

Similar rates also play a role in heavy-ion experiments: the QCD 
equation of state determines the expansion
rate of the system while microscopic rates determine how fast probes
interact with the expanding plasma. 

An apparent difference between cosmology and heavy-ion collisions 
is that, in the former, weak and electromagnetic interactions
play a prominent role whereas, in heavy-ion collisions, strong interactions
dominate. However, in a relativistic plasma even weak 
interactions become strong:  obtaining formally consistent results 
requires delicate resummations and, even then, the results may 
suffer from slow convergence. 

The development and application of resummation techniques 
in hot QCD or cosmology can benefit both fields.
For example, techniques~\cite{Kajantie:1995dw} originally
applied to computing the QCD equation of 
state~\cite{Braaten:1995jr} 
have been employed to compute the equation of state of full Standard
Model matter at very high 
temperatures~\cite{Gynther:2005dj,Laine:2006cp}. 
Techniques
for computing transport coefficients~\cite{Jeon:1995zm,Arnold:2006fz} have
led to the determination of some friction coefficients in 
cosmology~\cite{Bodeker:2006ij,Laine:2010cq}. 
Techniques developed for computing 
the photon/dilepton production rate from a hot QCD plasma  
near~\cite{Arnold:2001ms} or  
far from~\cite{CaronHuot:2009ns}
the light cone can be applied to 
computation of the right-handed neutrino production rate in 
cosmology~\cite{Anisimov:2010gy,Besak:2012qm,Salvio:2011sf,Laine:2011pq}. 
(In some cases, such as the rate of anomalous chirality 
changing processes or chemical equilibration of heavy 
particles, methods originating in cosmology 
\cite{Bodeker:1999gx,Hisano:2006nn}  
were later applied to heavy-ion collisions \cite{Moore:2010jd,Bodeker:2012zm}.)
Also in these cases it may help to combine different QCD effective field theories (EFTs). In 
\cite{Biondini:2013xua}
an EFT for nonrelativistic Majorana particles, which combines
heavy-quark EFT and Hard Thermal Loop EFT, has been developed  and applied to the case of
heavy Majorana neutrino decaying in a hot and dense plasma of Standard Model particles, whose temperature
is much smaller than the mass of the Majorana neutrino but still much larger than the
electroweak scale. It may have applications to a variety of different models involving
nonrelativistic Majorana fermions.

Apart from these methodological connections, there are also
direct physics links between hot QCD and cosmology.
In these cases, QCD particles do not themselves
decouple from equilibrium: their collective dynamics provides
a background for the evolution of other perturbations present in the medium. 
For instance, the QCD epoch 
could leave an imprint on the gravitational wave 
background~\cite{Schwarz:1997gv}, 
or on the abundance of 
cold~\cite{Hindmarsh:2005ix} or 
warm~\cite{Laine:2008pg} 
dark matter. In the case of dark matter, not only the equation of state 
but also various spectral functions, 
estimated from lattice simulations~\cite{Brandt:2012jc}, 
could play a role~\cite{Asaka:2006rw}.

An example of an outstanding issue in cosmology is 
a first-principles ``leptogenesis'' computation of right-handed neutrinos
in different mass and coupling regimes. It would 
be interesting to find hot QCD analogues for such CP-violating phenomena. 
 

\subsection{The chiral magnetic effect} \label{sec:d:ChiMagEff}

Parity (P) as well as its  combination with charge conjugation (C) are 
symmetries known to be broken in the weak interaction. 
In the strong 
interactions, however, both P and CP are conserved except by the $\theta$ term, making the strong CP 
problem one of the remaining puzzles of the Standard Model. The possibility 
to observe parity violation in the hot and dense hadronic matter produced in
relativistic heavy-ion collisions has been discussed for 
many years \cite{Kharzeev:2004ey}. 
In the vicinity of the 
deconfinement phase transition, the QCD vacuum could create 
domains that could introduce \cp-violating effects \cite{Kharzeev:2004ey}. For a critique regarding the observability of those effects in heavy-ion collisions see \cite{Khriplovich:2010wq}. These 
effects could manifest themselves as charge separation along the direction of 
the angular momentum of the system or, equivalently, along the direction of 
the strong magnetic field, $\approx 10^{18}$~G, created in semi-central 
and peripheral heavy-ion collisions perpendicular to the reaction plane.  
This phenomenon is known as the \cme~(CME). Due to 
fluctuations in the sign of the topological charge of these domains, the 
resulting charge separation, averaged over many events, is zero. This makes 
the observation of the CME possible only in P-even observables, 
expressed by correlations between two or more particles.

\begin{figure} 
\includegraphics[width=\linewidth]{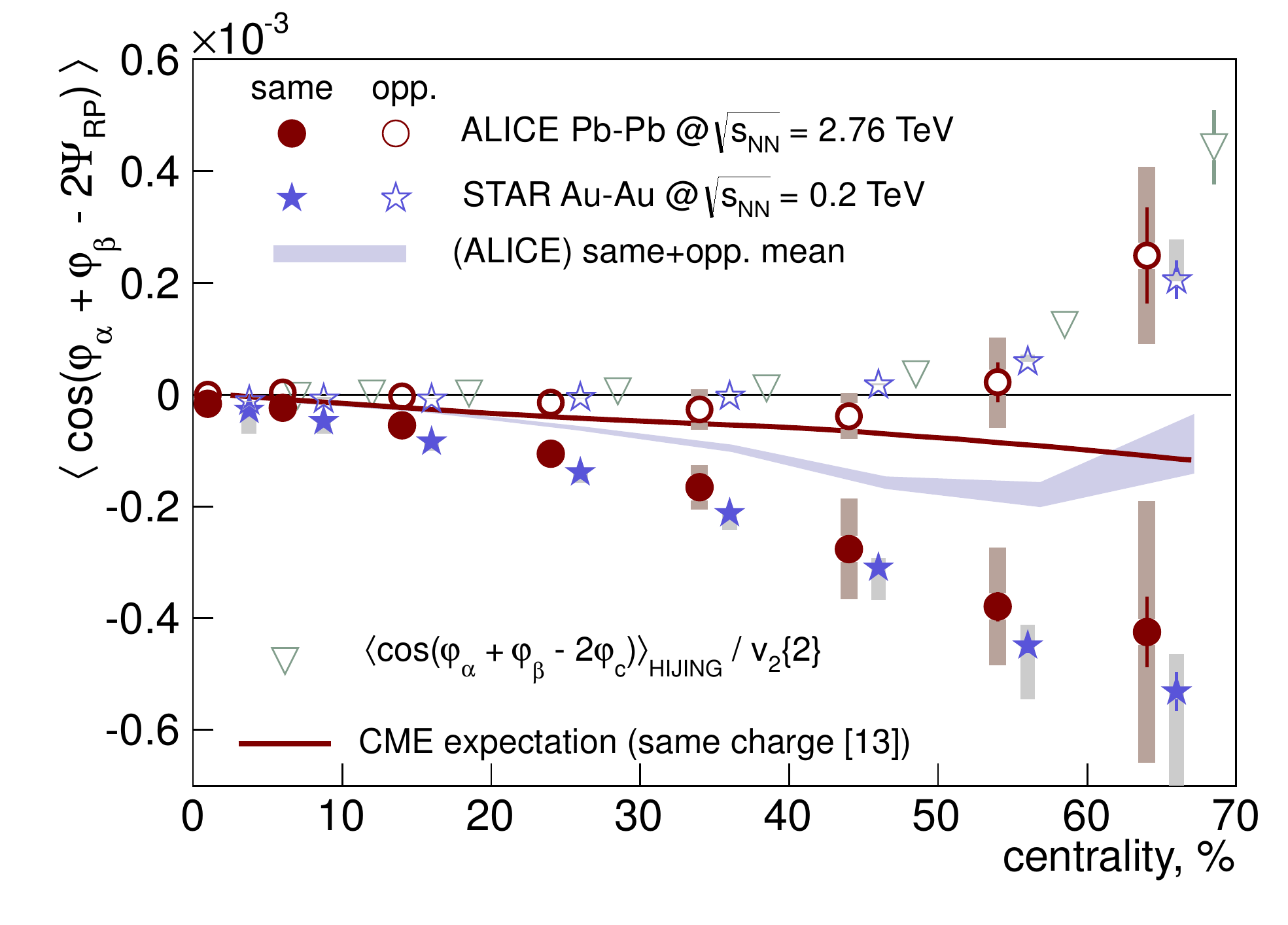}
\caption{The centrality dependence of the correlator $\langle \cos(\mathsf{\varphi}_{\alpha} + \mathsf{\varphi}_{\beta} -2\Psi_{\rm RP}) 
\rangle$. From \cite{Abelev:2012pa}.}
\label{fig:comparisonRP}
\end{figure}

The CME has been studied both at RHIC and LHC employing the three-particle
correlator $\langle \cos (\varphi_\alpha + \varphi_\beta - 2\Psi_{\rm RP}) 
\rangle$.  Here $\varphi_i$ is the azimuthal emission angle of particles with 
charge or type $i$ and $\Psi_{\rm RP}$ is the orientation of the reaction plane.  
The correlator probes the magnitude of the expected signal while concurrently
suppressing background correlations unrelated to the reaction plane.

The STAR Collaboration published the first results from \AuAu\
collisions at $\sqrt{s_{NN}} = 0.2$ TeV, consistent with
CME predictions \cite{Abelev:2009ac}. ALICE has studied these same 
correlations at midrapidity in \PbPb\ collisions at $\sqrt{s_{NN}} = 2.76$ TeV 
\cite{Abelev:2012pa}.  The ALICE analysis was performed over the full minimum 
bias event sample recorded in 2010 ($\sim 13$M events).  

Figure~\ref{fig:comparisonRP} presents the correlator $\langle 
\cos(\varphi_{\alpha} + \varphi_{\beta} - 2\Psi_{\rm RP}) \rangle$ measured by STAR 
and ALICE as a function of the collision centrality compared to model 
calculations. The ALICE points, filled and open circles for pairs with the 
same and opposite charges, respectively, indicate a significant difference 
not only in the magnitude but also in the sign of the correlations 
for different charge combinations, consistent with the qualitative expectations 
for the \cme. The effect becomes more pronounced in peripheral collisions. The 
earlier STAR measurement in \AuAu\ collisions at $\sqrt{s_{NN}} = 200$~GeV, 
represented by stars, is in good agreement with the LHC measurement. 

The solid line in Fig.~\ref{fig:comparisonRP} shows a prediction of same-sign 
correlations due to the CME.  The model does not predict the absolute magnitude 
of the effect and describes the energy dependence from the duration and time
evolution of the field.  It significantly underestimates the magnitude of the
same-sign correlations at the LHC \cite{Toneev:2010xt}.  Other recent models 
suggest that the magnitude of the CME might be independent of energy 
\cite{Kharzeev:2007jp,Zhitnitsky:2010zx}.

Other effects, unrelated to the CME, may also exhibit a correlation signal.
Results from the HIJING event generator, which does not include P violation, 
are also shown (inverted triangles), 
normalized by the measured value of $v_2$.  Because no 
significant difference between same and opposite-sign pair correlations is 
present in the model, they are averaged in Fig.~\ref{fig:comparisonRP}.  The 
finite effect in HIJING can be attributed to jet correlations, unrelated to 
the reaction plane.  Another possible explanation for the behavior of the 
correlator comes from hydrodynamics.  If the correlator has an out-of-plane, 
charge independent, component arising from directed flow fluctuations, the 
baseline could be shifted \cite{Teaney:2010vd}.
The sign and magnitude of these correlations is given by the shaded
band in Fig.~\ref{fig:comparisonRP}.

The measurements, including a differential analysis, will be extended to higher 
harmonics and identified particle correlations. These studies 
are expected to shed light on one of the remaining fundamental questions 
of the Standard Model.

The CME also occurs within AdS/CFT approaches. In
AdS/CFT the CME is closely related to an anomaly \cite{Son:2009tf}. A
related observable is the chiral vortical effect where the angular momentum of
non-central collisions generates helicity separation between particles.
This appears naturally from supergravity within AdS/CFT
\cite{Erdmenger:2008rm,Banerjee:2008th}.
The chiral vortical effect can also arise
from a current generated in the presence of a gravitational
vortex in a charged relativistic fluid and was found to
be present even in an uncharged fluid
\cite{Landsteiner:2011cp,Landsteiner:2011iq}.
In the case of two U(1) charges, one axial and one vector, the
CME appears formally as a first-order transport coeffcient in the vector
current. In this case, there is evidence that the CME
depends on $v_2$ \cite{Gahramanov:2012wz}.
Finally, topological charge fluctuations can generate an
axial chemical potential that splits the effective masses
of vector mesons with different circular polarizations in
central heavy ion collisions, complementary to the CME in
noncentral heavy-ion collisions
\cite{Andrianov:2012hq,Andrianov:2012dj}.
 

\subsection{Future directions} \label{sec:5.9}

Based on the current results and open questions, as detailed in this chapter, several of the main
experimental issues can be addressed in the short to medium term using the facilities currently in operation.
An extended list of experimental measurements has to await long-term upgrades and planned new facilities.
The driving force behind the forthcoming developments in heavy-ion physics are the existing and planned
experimental heavy-ion programs.
These include the collider experiments at the LHC and RHIC, as well as future programs at fixed-target
facilities either under construction or in the planning stage.
The physics opportunities and goals are somewhat different at each facility, offering a complementarity that
can be exploited.

Indeed, after the first two years of ion runs at the LHC and further results 
from RHIC, the field has made significant progress.  Detailed, 
multi-differential measurements have shown that the produced system can still 
be described by hydrodynamics in the new energy domain of the LHC.  Thus its 
bulk macroscopic properties can be characterized.
Moreover, significant progress has been made in determining the microscopic 
properties of the QGP (shear viscosity and plasma opacity) with increased 
precision.  Detailed studies of identified particles and extended 
measurements of heavy flavors have introduced new input to the topics of
thermalization and recombination.


In addition, the study of proton-proton and proton-nucleus collisions,
used as reference baselines for comparison to heavy-ion results,
also revealed some unexpected findings.
The very first discovery at the LHC was related to the appearance of the 
``ridge'' structure in high multiplicity $pp$ events, associated with 
long-range correlations. A similar but much stronger feature appeared in 
the proton-nucleus data, attracting great interest.  It is clearly very 
important to identify whether the ridge phenomena is of 
hydrodynamic or saturation origin and how it relates to similar phenomena 
occurring in nuclear collisions.  While saturation physics can explain the 
qualitative appearance of the
ridge phenomena in proton-proton and proton-nucleus collisions, hydrodynamic
flow could further collimate the signal, particularly in nucleus-nucleus 
collisions \cite{Dumitru:2010iy}.


In particular, novel, high-resolution methods to probe the early times of 
the evolution of $AA$ collisions, developed at RHIC and the LHC, need to be 
pursued with higher statistics and greater precision.
The corresponding observables are parton attenuation in the early partonic 
medium, and higher flow moments. The latter are being extended by the analysis 
of single events which reflect the primordial evolution without the ensemble 
averages that can blur the resulting picture.
Such measurements, as well as sophisticated developments from both the 
theoretical and experimental sides have brought some fundamental aspects of 
QCD within reach.
First, it has become possible to probe a primordial phase
founded on gluon saturation physics --- the so-called Color Glass 
Condensate (CGC) that arises at asymptotically high gluon densities.
Next, it was recognized that the QGP is not a weakly-coupled
parton gas but a strongly-coupled, near-ideal liquid with a
very low ratio of (shear) viscosity to entropy density ($\eta/s$),
close to the theoretical lower limit derived from
quantum gauge field theory. 
The attenuation of leading partons in the medium
is characterized by the parton transport coefficient $\hat{q}$ which
quantifies the medium-induced energy loss.
The fundamental quantities of $\eta/s$ and $\hat{q}$ are related, as detailed 
in Ref.~\cite{PhysRevLett.99.192301}:
a large value of $\hat{q}$ implies a small $\eta/s$, indicative of 
strong coupling.  Moreover both quantities can be addressed within the 
so-called AdS/CFT conjecture
\cite{HLiu,Maldacena:1999}, by a dual weakly-coupled string theory.


The experimental methods and avenues of theoretical research, pioneered 
at RHIC, could reach their full promise at the LHC
with further increases in luminosity and runs at the top design energy,
leading to greater precision for drawing crucial theoretical conclusions.
Indeed, within the currently-approved LHC schedule, an order of magnitude higher
statistics is expected to be collected, necessary for the description
of statistics-limited phenomena such as the differential study of higher
harmonic particle 
flow and high \Pt\ ``jet quenching''.


The measurements and the conclusions reached employing jets arising
from high energy partons revealed the richness of these high \Pt\ probes which
access not only the properties of the medium, but also properties of the 
strong interaction. Jets put constraints on the amount of energy loss
in the medium and the dependence on the parton type which can disfavor some 
current models.  At the LHC, jets are more clearly defined 
and better separated from the background, both in single and in dijet 
production than at RHIC, due to the larger cross section for hard processes.
However, the correlation between hard jets and soft particles in the underlying 
event remains difficult to describe by any currently-known mechanism, even if 
it turns out to be factorizable in QCD.  In-depth studies of the energy
redistribution within a jet or of the low \Pt\ particles emitted far away 
from the jet axis, together with the precise description and modeling of
the modification of the jet fragmentation functions and jet shape,
could unveil the properties of the QGP. These studies could also clarify why, 
in the case of high \Pt\ jet suppression, jet cone radii of up to $R=0.8$ are 
unable to capture all the radiated energy.  A better understanding of the 
large average energy imbalance, also seen in the golden $\gamma +$jet channel,
and of the angular correlations between jets that are, surprisingly, not 
strongly modified in the range $40 < p_T < 300$ GeV/$c$, could be obtained by 
extensive studies of dijet events.


In addition, the higher luminosity will enable precision studies of
quarkonium suppression, dramatically increasing our
understanding of the interactions of hard particles with
the thermal medium. Finally, the study of the thermalization
and chiral symmetry restoration will be considerably enhanced by measurements
of thermal dilepton and photon radiation, as well as the determination
of vector meson spectral functions. 

Furthermore, recent developments have advanced our knowledge of $AA$ collisions
at comparatively modest center of mass energies, where lattice QCD predictions 
at finite baryon chemical potential $\mu_B$ locate the parton-hadron phase
boundary. First indications of a critical point need to be clarified by
further systematic studies. Such investigations are also
fundamental for the characterization of the QCD phase diagram.
Complementary research is planned and is being conducted at lower center of mass
energies and temperatures. Data from the beam energy scan at the RHIC collider 
at larger baryon densities will contribute to the search for a critical point 
on the QCD phase diagram.  New fixed-target experiments
will increase the range of energies available for the studies of hot,
baryon-dense matter.
The CERN SPS will remain the only fixed-target facility capable of delivering
heavy-ion beams with energies greater than 30~GeV/nucleon, making studies of
rare probes at these energies feasible.
At the FAIR accelerator complex under construction at GSI, Darmstadt, 
heavy-ion experiments are being prepared to explore the QCD phase diagram 
at high baryon chemical potential with unprecedented sensitivity and precision.
Finally, the NICA project at JINR, Dubna, will complement these programs.
In particular, these new low-energy facilities are being built to study 
compressed baryonic matter at high baryon density and (comparatively) low 
temperatures where the matter may undergo a first-order phase transition. 
In these systems, the produced matter is more closely related to neutron stars.


On the theory side, important progress is expected in both phenomenology and
pure theory. A well-coordinated phenomenological effort is clearly needed to
fully exploit the current and
future precision data from the facilities mentioned above.

Indeed, the new reference data from \pPb\ collisions at the LHC have presented
some unique challenges for phenomenology. Potential new QCD phenomena could be 
unveiled by solving the ridge puzzle in \pPb\ collisions. One promising
proposed method is to employ multiparticle methods in order to access and 
measure collective phenomena which can discriminate
between initial-state (CGC) and final-state (hydro) mechanisms.

The behavior of the low-$x$ gluon
density in nuclei needs to be better understood, both in the shadowing and
saturation pictures. In addition, the question of whether the high-multiplicity
events in \ppcoll\ and $pA$ collisions can be described in terms of cold
nuclear matter or whether they should be thought of as having created a hot
medium is one that will come to the fore.

In addition to phenomenology, establishing the quantitative properties of 
a deconfined quark-gluon plasma from first principles, both in and out of
thermal equilibrium, continues to be a fundamental theory goal, requiring a
combination of lattice, perturbative and effective field theory methods. In
this context, the major challenges will be to extend equilibrium thermodynamic
calculations on the lattice to larger quark densities; to obtain
accurate nonperturbative predictions of the QGP transport properties; and to
further increase understanding of the dynamics that lead to the apparent
early thermalization in heavy-ion collisions.



\newpage

Putting the heavy-ion program in a broader context,
the LHC is the high-energy frontier facility not only of particle physics but 
also of nuclear physics, with an extensive, well-defined program.
The active RHIC program, complementary and competitive,
continues to map the phase diagram of nuclear matter at lower temperatures
pursuing the search for a tricritical point.
Continuation and strengthening of the SPS fixed-target program is under 
discussion. Furthermore, the two new low-energy facilities (FAIR at GSI and 
NICA at JINR) are being built to explore 
the part of the phase diagram at the other extreme from the colliders. Thus, 
the global heavy-ion physics program can fully map the QCD phase diagram, 
spanning these two limits.  Thus strongly-interacting matter under extreme 
conditions, such as those prevailing in the early universe (LHC, RHIC) as well
those similar to the conditions in the interior of neutron stars (FAIR, NICA)
can be studied in the laboratory.

In summary,
the exploration of the phases of strongly-interacting matter is one of the 
most important topics of contemporary nuclear physics.
The study of strong-interaction physics, firmly rooted in the Standard Model, 
has already brought surprises and discoveries as well as showcased the
potential of heavy-ion research
which is expected to keep on providing new and interesting results.

\clearpage
\section[Chapf]{Nuclear physics and dense QCD in colliders and compact stars \protect\footnotemark}
\footnotetext{Contributing authors: M.~Alford$^{\dagger}$, T.~Cohen$^{\dagger}$, L.~Fabbietti$^{\dagger}$, A.~Schmitt$^{\dagger}$, K.~Schwenzer}
\label{sec:chapf}

In this chapter we discuss open problems and future directions in nuclear physics (addressing issues concerning dense nuclear matter as well as  
low-density and vacuum nuclear interactions) and high-density quark matter, both of which are relevant for the physics of compact stars. 
The composition of the inner core of a compact star is not known. Constraints on the Equation Of State (EOS) of the star
core are imposed by the measured radii and masses, but several scenarios are possible. These scenarios vary from considering only neutrons and
protons as constituting the inner core, assuming the presence of hyperons, or a kaon condensate, or the
existence of a dense quark matter core. These different hypotheses are discussed from an experimental and theoretical point of view in the following 
subsections.
The chapter is divided into three subsections. In Sec.\ \ref{sec:secF1} we focus on accelerator experiments that can shed light onto 
kaon--nucleon and hyperon--nucleon interactions in a dense medium and the implications for neutron stars, such as the thickness of the neutron star crust
via measurements of neutron-rich nuclei. In the second part, 
Sec.\ \ref{sec:secF2}, we discuss theoretical attempts to understand the nucleon--nucleon interaction from QCD. In particular, we address 
promising directions in lattice QCD, effective field theory methods, and the large-$N_c$ approach. Finally, in Sec.\ \ref{sec:secF3}, we mostly discuss dense 
quark matter, starting from asymptotic densities. We discuss several theoretical approaches and come back to compact stars to address various astrophysical 
observables and their relation to the microscopic physics of dense matter.     

\subsection{Experimental constraints on high--density objects}
\label{sec:secF1}
The study of high density objects can be pursued among other methods by investigating hadron-hadron collisions at accelerators.
On the one hand, heavy--ion collisions at moderate kinetic energies ($E_\text{KIN}=\,1\--8$ AGeV \footnote{AGeV= GeV per nucleon.}) lead to the formation
of a rather dense environment with $\rho=\,\text{2--7}\,\rho_0$ (with $\rho_0=\,0.172 \,\mathrm{fm^{-3}}$ being the normal nuclear density) which can be characterized in terms of its global properties and the
interactions among the emitted particles. Normally, the density reached in the collisions as a function of the
incoming energy is extracted from transport calculations. In these kinds of experiments, one of the goals is to determine the EOS for nuclear matter
and extract constraints for the models of neutron stars.
On the other hand, the understanding of the baryon--baryon and meson--baryon interaction as a function of the system density 
should be complemented by the study of elementary reactions that give access to the interaction in the vacuum and serve as a
fundamental reference.
Important references are delivered by the measurement of kaonic-atoms and hypernuclei, as described in the following paragraphs.
Aside from the measurement of strange hadrons, novel measurements of the properties of neutron-rich nuclei can constrain the thickness
of the external crust of neutron stars.
\subsubsection{Results from heavy--ion collisions}
The EOS for nuclear matter relates the pressure of the system to its internal energy, 
 density, and temperature and is fundamental for the modelling of different astrophysical objects.
Indeed, by knowing the EOS of a certain state of matter, hypotheses about the content of dense astrophysical objects can be put forward and the mass to radius
 relationship can be extracted starting from the EOS and exploiting the Tolman-Oppenheimer-Volkoff equations \cite{PhysRev.55.374}.
 A more detailed description of the extraction of the mass and radius of neutron stars is given in Sec.\ \ref{sec:secF3}.
 From the experimental point of view, one of the tools used to study dense systems are heavy--ion collisions at accelerator facilities.
Transport calculations \cite{Hartnack:2011cn} indicate that in the low and intermediate energy range ($E_\mathrm{lab} =\, \text{0.1--2}\,\mathrm{AGeV}$) nuclear densities between $\text{2--3}\,\rho_0$ are accessible while the highest
baryon densities (up to $8\rho_0$) can be reached increasing the beam kinetic energy up to $10\,\mathrm{GeV}$.
The EOS of nuclear matter is normally characterized by the incompressibility parameter which is expressed as:
\begin{equation}
K=\left.\, 9\rho_0^2 \frac{d^2E}{d\rho^2}\right|_{\rho=\,\rho_0}.
\end{equation}
Hence, if the system energy is parametrized as a function of the system density, the parameter $K$ represents the curvature of this function at normal nuclear density and is a measure of the evolution of the system energy as a function of the density. 
The boundary between a soft and a stiff EOS is set around a value of $K=\,200\,\mathrm{MeV}$, with values below and above $200\,\mathrm{MeV}$ corresponding to a soft and stiff EOS respectively, with predictions for a rather stiff EOS corresponding to $K=\,380\,\mathrm{MeV}$ \cite{Danielewicz:2002pu}.
Increasing the stiffness of the EOS translates into an increased pressure of the system. The experimental variables used to characterize the EOS are linked to the system pressure.

The collective properties of the fireball formed in heavy--ion collisions for different kinetic energies are linked to the system pressure and they have been studied to derive the compressibility of
 nuclear matter at the achieved aforementioned densities \cite{Danielewicz:2002pu}. 
The compressibility of the matter formed right after the ion's collision can be related
 to the variation of the mean value of the $x$-component (assuming that the $z$-component is parallel to the beam direction) of the particles momenta. A larger resulting pressure 
 on the emitted particles correspond to a stiffer EOS and also to larger values of the sideward forward-backward deflection parameter $F$ that measures the variation of the average $p_X$ component\footnote{$F=\frac{d \langle p_X \rangle /A}
 {dp_T}
 \mid_{Y=Y_{CM}}$, where $Y_{CM}$ is the center of mass rapidity of the nucleus-nucleus system.}.  
 
\begin{figure}[t]
\vspace{0cm}
\includegraphics*[width=9cm]{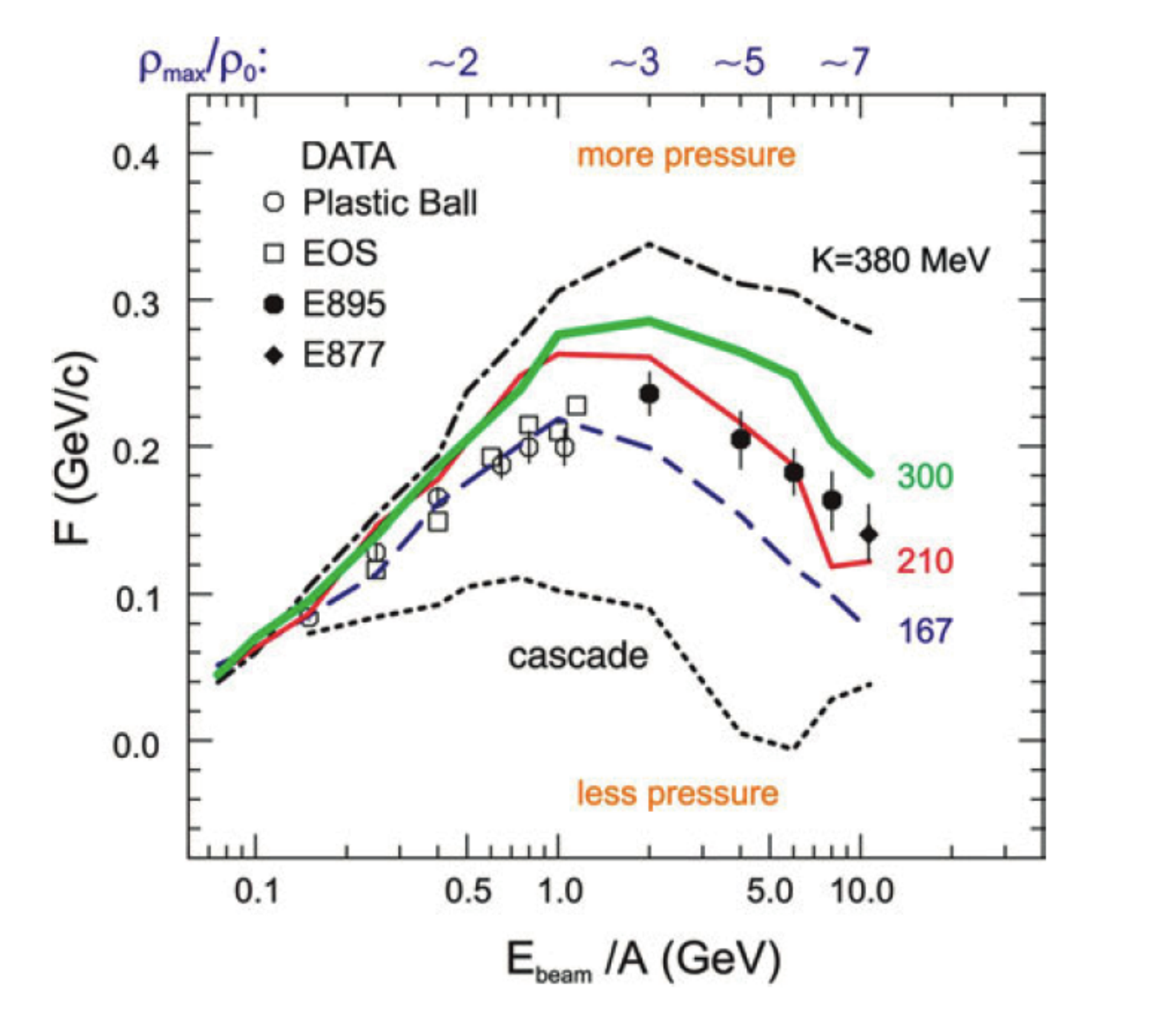}
   \caption{Sideward flow excitation function for Au+Au collisions. Data and transport calculations are represented by symbols and lines, 
    respectively \cite{Danielewicz:2002pu}.}
\label{fig:FDaniel}
\end{figure}
Figure~\ref{fig:FDaniel} shows the measure of the sideward forward-backward deflection $F$ for charged particles produced in Au+Au
 collisions for several beam energies. The maximal density reached for each setting is indicated in the upper horizontal axis. 
Lines represent simulations assuming different EOS and the comparison to the data points favors a compressibility parameter of 
$K\approx\,\text{170--210}\, \mathrm{MeV}$ which translates into a rather soft EOS for normal nuclear matter.
One can see that a single EOS is not sufficient to reproduce all the experimental data and that the stiffness of the system increases 
as a function of the density.
This observation suggests that a the EOS of nuclear matter could depend on the system density and a transition from a softer
to a stiffer EOS might occur. \\
The extraction of the EOS from the measurement of the flow of charged particles produced in heavy--ion collisions 
is limited by the following factors.
The different transport models, that are used to compare the experimental observables measured in
heavy--ion collisions at intermediate energy ($E_\text{KIN}<\,10$ AGeV), solve the Boltzmann transport 
equation with the inclusion of a collisional term modeling the heavy--ion collisions as the superposition of binary nucleon--nucleon interactions.
Normally a solid knowledge of the elementary nucleon--nucleon, $\Delta\--$nucleon and $\pi\--$nucleon cross-sections is needed as an input for 
transport models. Unfortunately, the processes involving a $\Delta$ or a neutron are either not measurable exclusively or have not been measured yet.
Moreover, one has to consider that the particle equations of motion in transport models generally do not contain any dependence on the system temperature,
 which is certainly not negligible in heavy--ion collisions. The maximal temperature of the colliding systems can be estimated via
statistical models of particle hadronization, and it already reaches $100\,\mathrm{MeV}$ for $E_\text{KIN}=\,1$--2 GeV \cite{Agakishiev:2010rs}.

The properties of kaons ($K$) and anti-kaons ($\bar{K}$) in the nuclear medium have also been the object of numerous investigations, since the possible 
existence of a $\bar{K}$ condensed phase in dense nuclear matter and thus in the interior of compact neutron stars was pointed out by Kaplan and 
Nelson \cite{Nelson:1987dg}.
This idea originates from the fact that various theoretical approaches, based on effective chiral models
where $K$/$\bar{K}$  and nucleons are used as degrees of freedom, agree qualitatively in predicting density-dependent modifications in mass and coupling 
constants for $K$ and $\bar{K}$. This results in the growth (drop) of the effective mass of $K$ ($\bar{K}$) with
increasing nuclear matter density \cite{Fuchs:2005zg,Schaffner:1995th,Ramos:2000dq}. Indeed, the scalar part of the mean potential is attractive for both $K$-types ,
while the vector part is slightly repulsive for $K$ ($V=\,20\--30\,\mathrm{MeV}$ at $\rho=\,\rho_0$ and $\vec{p}=\,0$) and yet
attractive for $\bar{K}$ ($V=\,-\text{50--150}$ MeV at $\rho=\,\rho_0$ and $\vec{p}=\,0$).
The effective $\bar{K}$ mass is then expected to undergo a substantial reduction in the presence of dense nuclear matter,
up to the point where strangeness-violating decays of protons into neutrons, $\bar{K}$ and neutrinos occur.  This 
process might set in starting at rather high baryonic densities ($\rho=\,3\--4\,\rho_0$) and could lead to the creation of an 
equilibrated condensate of $\bar{K}$ and neutrons in the interior of neutron stars.

The investigation of the kaon properties produced in heavy--ion collisions at intermediate
energy was successfully carried out in the 1990s by the KaoS \cite{Forster:2003vc} and FOPI \cite{Crochet:2000fz} collaborations
 and more recently by the HADES collaboration \cite{Agakishiev:2010zw} by measuring heavy--ion collisions with beam kinetic 
 energies up to $\mathrm{1.9\,GeV}$.
In this energy regime, strange hadrons are produced to a large extent below the nucleon--nucleon energy threshold and hence
mainly by secondary collisions, which are used as reservoirs to gather the necessary energy to produce strange hadrons.
These secondary collisions are more probable during the initial, denser phase of the collision, where the baryonic
matter undergoes the highest compression and hence strange hadrons are highly sensitive to 
possible repulsion/attraction in dense baryonic matter \cite{Pal:2001nk}.
The nuclear matter EOS determined by comparing  the $K$ multiplicities produced in heavy (Au+Au) and light (C+C) colliding
 systems as a function of the kinetic energy is found to be rather soft \cite{Forster:2007qk}.
 Indeed, the $K^+$s are produced in the initial dense phase of the collisions and since they do not undergo absorption, because the
 imaginary part of their spectral function is rather small,
 the scaling of the production rate with increasing incoming kinetic energy, which corresponds to an increasing compression of
 the system, can be used to tag the compressibility parameter and hence the EOS.
 Additionally to the effects linked to the compressibility of pure nuclear matter, by looking at the $K$ observable one should
 consider the effect of the repulsive potential between nucleons and $K$.
 It is very difficult to disentangle the properties of nuclear matter and its interaction with $K$ and $\bar{K}$ with the same observable.
 This is the reason why in addition to the $K$ yields other observables have also been taken into account.

 The standard method to study the potential effects on $K$ and $\bar{K}$ production consists in analyzing either their collective flow or in the study of the 
 $p_T$ spectra\footnote{$p_T$ is the momentum component perpendicular to the beam axis.} in different rapidity intervals \cite{Crochet:2000fz,Barth:1997mk}.
First we discuss the flow observable.
 Taking as a reference the reaction plane formed by the distance of closest
 approach of the two colliding nuclei and the beam direction, the particle emission angle with respect
 to this plane is considered. The azimuthal anisotropies in the collective expansion, also called
 anisotropic flow, are usually characterized by a Fourier expansion of the azimuthal distribution 
 of the produced particles: 
 \begin{equation}
 v_n=\,\langle \langle \cos n(\phi-\Psi_R) \rangle \rangle,
 \end{equation}
where $\Psi_R$ and $\phi$ represent the orientation of the reaction plane and the azimuthal angle of the particle with respect to the
reaction plane. The two averages of the $\cos$ function run over the number of particle per event and over the total number of events.
The resulting parameters $v_1$ and $v_2$ are known as direct and elliptic flow, respectively.
Direct and elliptic flow for $K$ and $\bar{K}$ have been recently
 measured by the FOPI \cite{Ritman:1995td} collaboration in Ni+Ni collisions at $\mathrm{1.9\,GeV}$ kinetic energy and compared to the different
  transport models. Preliminary results show that the expected sensitivity of the direct flow $v_1$ in the target region is weaker than 
  predicted by transport calculations including a strongly attractive $\bar{K}N$ potential. 
  For $K$ the qualitative behavior of $v_1$ is described by the transport models including a slightly average repulsive
  potential of $20\,\mathrm{MeV}$.
  One has to mention here that the major limitation of the transport models in the description of $K$ and $\bar{K}$ flow is
the fact that the momentum dependence of $v_1$ and $v_2$ is far from being well modelled for this energy regime \cite{Hartnack:2011cn}.
One has to consider that the approximation made so far by the transport models used for these comparisons,
   in which the interacting $K$-nucleus potential depends linearly upon the system density, is certainly much too simplistic.
   In this respect recent developments of the GiBUU \cite{Larionov:2007rv} model includes a more realistic chiral potential for the $K$, but first
   tests are only now being carried out with data extracted from proton--induced collisions. The next step would be to extend this model
   to heavy--ion collisions. 
    To summarize the $K$ and $\bar{K}$ flow results a slight repulsive potential is confirmed for the $K$
    produced in heavy--ion collisions at intermediate energies but unfortunately no evidence of a strongly attractive potential for 
    $\bar{K}$ could be observed  within the statistical sensitivity of the data. 

The doubly differential analysis of the $p_T$ spectra for $K$ shows a better consistency. There, by looking at the experimental  
$p_T$ distributions
for different rapidity intervals, a systematic shift towards higher momenta is observed \cite{Benabderrahmane:2008qs,Agakishiev:2010zw}. 
This shift is thought to be due to the repulsive potential felt by $K$ in the nuclear medium.
So far, the experimental findings about the $p_T$ distributions indicate a repulsive average potential for $\mathrm{K^0}$ and $\text{K}^+$, estimated to be
 between $\mathrm{20\,MeV}$ \cite{Benabderrahmane:2008qs} and $\mathrm{40\, MeV}$ \cite{Agakishiev:2010zw} at $\rho=\,\rho_0$ but stays
 rather controversial for $K^-$ \cite{Fuchs:2005zg}. 
 Further measurements with $\pi$-beams planned at GSI already in 2014 and at
  JPARC  starting from 2013 should allow a more quantitative determination of this potential. In particular, elementary reactions are needed to 
  provide the transport models with solid inputs for all reaction channels.

It has been mentioned that the understanding of the $\bar{K}$ properties in dense nuclear matter is still far from being properly tagged down. 
So far the scarce amount of data for $\bar{K}$ produced in heavy--ion collisions at intermediate energies
has hampered this study. The existing data by the  FOPI collaboration about the $\bar{K}$ flow are unfortunately not
definitive, as mentioned above.
More experiments are needed in this direction.

Meanwhile, new theoretical developments have been carried out towards a more realistic treatment of the $\bar{K}$ in-medium spectral function \cite{Tolos:2013qv}. 
There, unitarized theories in coupled channels based on chiral dynamics \cite{Lutz:1997wt,Korpa:2004ae} and meson-exchange models \cite{Ramos:1999ku} are discussed, with particular emphasis on the novel inclusions of higher-partial waves beyond the s-wave in the meson--baryon coupling \cite{Tolos:2006ny}.
In such calculations all possible meson--baryon coupled channels are considered to compute the final $\bar{K}$ spectral function in the medium, including effects such as the Pauli blocking in medium, and the self-consistent consideration of the $\bar{K}$ self-energy, the self-energies of the mesons and baryons in the intermediate states.
Within this approach, an attraction of the order of $-50\,\mathrm{ MeV}$ at normal nuclear matter density is obtained.
This kind of calculations should be implemented in transport models to extract predictions for the $\bar{K}$ properties as a function of the system density.

Unfortunately in other approaches the low density approximation is employed to describe the broadening of the
imaginary  part of the $\bar{K}$ spectral function. This approach does not suit the complex behavior expected
for $\bar{K}$ in the medium \cite{Korpa:2004ae}. 

\subsubsection{The $K$--nucleon interaction in vacuum}

The issue of the $K$--nucleon interaction has also been addressed in an alternative way in recent years.
Indeed, so-called kaonic bound states, formed from a $\bar{K}$ sticking to two or more nucleons, have been predicted by theory
\cite{PhysRevC.65.044005} and shortly afterwards observed in experiments \cite{Agnello:2005qj,Kienle:2011mi}.
This idea originates from the first studies that Dalitz and colleagues did on the intrinsic nature of the $\Lambda(1405)$ resonance
 in the 1960's \cite{PhysRevLett.2.425}, when they proposed a description of this particle as a molecular state of a $\bar{K}$--p and $\mathrm{\pi\text{--}\Sigma}$ poles 
 interfering with each other. Since the $\Lambda(1405)$ is, at least partially, a $\bar{K}$--p bound state, it was
  natural to investigate the possibility of adding one or more nucleons and still finding a bound state.
 The binding energy and the width of this so-called kaonic cluster would reveal the strength of the $\bar{K}$--nucleon interaction in 
 vacuum.
As far as the $\Lambda(1405)$ is concerned, several approaches based on chiral effective field theory do describe this resonance 
as a molecular system emerging naturally from coupled channels calculations of meson--baryon pairs with $\mathrm{S=\,-1}$
 \cite{Ikeda:2012au,Mai:2012dt}.  These models are constrained above the $\bar{K}$N threshold by scattering data and 
 by the very precise measurement of the $\mathrm{\bar{K}p}$ scattering length at the threshold extracted from the
kaonic-hydrogen data measured by the SIDDHARTA collaboration \cite{Bazzi:2011zj}. The underlying concept for such an
 experiment is to determine the shift and width of the ground levels in kaonic hydrogen and deuterium caused by the strong 
 interaction between the $\bar{K}$ and the nuclei. Therefore, the X-rays emitted in transitions of the $\bar{K}$ to the ground level are
 measured. By comparing the measured X-ray energies with the values expected from QED only the strong
interaction induced shift and width are obtained. The measurement of the kaonic-deuterium is planned by the SIDDHARTA 
 collaboration at DA$\Phi$NE after an upgrade of the experimental apparatus and will also be pursued at JPARC. These
  experiments will allow the determination of the isospin-dependent scattering lengths, currently strongly under debate from the side of theory.

To this end it is clear that within this approach the whole $\bar{K}$ dynamics in nuclear matter is strongly influenced by the 
presence of the $\Lambda(1405)$ resonance.
Several experiments have been carried out, employing either stopped kaons and antikaons, or beams of these particles, real photons, and protons, 
 to study the properties of the $\Lambda(1405)$ and to search for kaonic bound states.
The molecular nature of the $\Lambda(1405)$ is supported by the observation of different 
spectral function distributions measured with different initial states \cite{PhysRevLett.95.052301}. Different production mechanisms
 correspond indeed to different coupling strength of the poles leading to molecule formation.

The spectral shape of the $\Lambda(1405)$ resonance measured from its decay into $(\Sigma\pi)^0$ pairs has been reconstructed by the CLAS \cite{Brooks:2000ws}
collaboration \cite{Moriya:2013hwg} for the reaction $\gamma+\text{p}\rightarrow \Lambda(1405) + \text{K}^+$ and for  9 different settings of 
the photon energy varying from $\mathrm{2}$ to $\mathrm{2.8\,GeV}$. The experimental data have been discussed in terms of 
phenomenological fits to test the possible forms and magnitude of the contributing amplitudes. Two $I=0$ amplitudes with an 
additional single $I=1$ amplitude parametrized with Breit-Wigner functions work very well to model all line shapes simultaneously.
This comes as a surprise since the $I=0$ poles are very different than those obtained by coupled channel calculations 
\cite{Mai:2012dt,Ikeda:2012au}, and the existence of a bound state in the $I=1$ channel is very controversial. An alternative 
strategy has been proposed in \cite{Roca:2013av} to describe the CLAS data in the $\Sigma^0\pi^0$ decay channel by varying 
the chiral coefficients for the meson--baryon coupling amplitudes. This variation is motivated by the fact that higher order calculations
of the chiral amplitude (most of the models include only the Weinberg-Tomosawa term for the interaction) might lead to significant corrections.
 This empirical approach delivers a reasonable description of the data but does not explain the presence of
 the $I=1$ bound state in the charged decays.
 
The $\Lambda(1405)$ signal reconstructed from the reaction $\mathrm{p+p\rightarrow \Lambda(1405)+p+K^+}$ and the 
successive decay into the two charged channels $\Sigma^{\pm}\pi^{\mp}$ has been recently analyzed by the HADES collaboration 
 \cite{Agakishiev:2011qw}. There, a shift of the spectral function associated to the $\Lambda(1405)$  has been observed and the 
 maximum of the distribution is found $\mathrm{20\,MeV}$ lower than the nominal value of $1405\,\text{MeV}$. This effect, which strongly differs 
  from the CLAS results, shows clearly the molecular character of the $\Lambda(1405)$ resonance and is yet to be fully understood 
from a theoretical point of view. It should be mentioned that the shifted pole towards lower masses might indicate that in the p+p entrance channel the $\pi\Sigma$ pole couples stronger to the $\Lambda(1405)$ formation \cite{Siebenson:2013rpa}. This is so far only a speculation that should be verified within a solid theoretical calculation, but could nevertheless strongly modify the $\bar{K}$ dynamic in the nuclear medium.

Following this line of thought and assuming that the $\bar{K}$--p pole dominates in the formation of the $\Lambda(1405)$, the smallest of
 the kaonic clusters ($\mathrm{ppK^-}$) could be obtained by adding an additional proton.   
The experimental evidence for kaonic bound states is partly strongly criticized but the signal measured with stopped $K$ by the
 FINUDA collaboration in the $\mathrm{\Lambda}$--p final state seems rather robust \cite{Agnello:2005qj}. 
One of the critical reviews of this work \cite{Magas:2006ee} emphasizes the role played by the one nucleon and two nucleons 
absorption reactions (K$^- +\mathrm{N}\rightarrow \Lambda \text{--N--} \pi$ 
or K$^- +\mathrm{N}\rightarrow \Lambda \text{--N--N--}\pi$) and
its contribution to the measured $\mathrm{\Lambda-p}$ final state.
Recent measurement by the KLOE collaboration \cite{VazquezDoce:2011zz} shows the feasibility of the exclusive measurement of the one--nucleon absorption. These results should be quantitatively compared to theoretical prediction and be included in the further analysis of the KEK \cite{Suzuki:2007pd} and future JPARC experiments on this subject.

The findings by the FINUDA \cite{Agnello:2005qj} and DISTO \cite{Kienle:2011mi} collaborations about the signature of the smallest 
of the kaonic cluster $\mathrm{ppK^-}$ do agree on the reported value for the binding energy, which is found to be about 
$\mathrm{100\,MeV}$, but differ strongly on the state width ($60$ to $\mathrm{100\,MeV}$). The great majority of the
 theoretical models predict the existence of such cold and dense objects as well, but the landscape of the binding energies and
 widths is rather broad with intervals of $\mathrm{9-90\,MeV}$ and $\mathrm{35-110\,MeV}$ respectively 
 \cite{Gal:2011zz,Ikeda:2008ub,Shevchenko:2006xy,PhysRevC.79.014003}.
 
 Concerning the findings in p+p reactions, the contribution of the $\mathrm{N^*}$ resonance to the analyzed final state and the 
 interference
 effects among different resonances has not yet been taken into account. A global study of the available $\mathrm{p+p
 \rightarrow p+\Lambda + K^+}$ data sets within a Partial Wave Analysis (PWA) should clarify the situation and quantify the 
 contributions of the non-exotic and exotic sources to the final state.
 As far as the upcoming experiments at JPARC are concerned, two main issues should be mentioned. First of all, the one$\--$ and
  two$\--$ nucleon absorption should be measured exclusively and, second, spin observables would be necessary to separate 
  different contributions of the measured spectra.
 
 Summarizing the situation for the $\bar{K}$ and its link to dense objects, the measured data show some evidence for a strong
 $\mathrm{\bar{K}N}$ binding but more quantitative information is still needed.

\subsubsection{Hyperon--nucleon interaction}

The hypothesis of a $\bar{K}$ condensate in neutron stars is complemented by a scenario that foresees hyperon production
and coexistence with the neutron matter inside neutron stars.
At present, the experimental data set on the $\Lambda$N and $\Sigma$N interactions consists of not more than 850
spin-averaged scattering events, in the momentum region
from $200$ to $\mathrm{1500\, MeV}$, while no data are available for
hyperon--hyperon scattering.
This case can be approached by the measurement of hypernuclei.
The $\Lambda$N effective interaction has been determined from reaction spectroscopy where 
hadronic final states are used to determine the hyper nuclei binding energies and 
$\gamma$-ray spectroscopy on the hyper nuclei decay.
The reaction spectroscopy provides access to the central part of the $\Lambda$N potential at zero momentum while 
the fit of $\gamma$-ray data on p-shell hypernuclei allows the determination the contribution by the spin--spin term in the $\Lambda$N interaction
 \cite{Botta:2012xi}. There, hypernuclei are produced employing secondary meson beams and primary electron beams, and  the 
reaction spectroscopy results for several nuclei are consistent with calculations including an average
attractive $\Lambda$-nucleus potential of $\mathrm{ReV_{\Lambda}\approx\,-30\,MeV}$, with $\mathrm{ReV_{\Lambda}}$ representing the real part
of the optical potential. The spin-spin corrections depend
on the nuclear species and amount to about $\mathrm{1\,MeV}$.
Of particular interest in this context is the recent observation of the neutron-rich hypernucleus $^{6}_{\Lambda}H$ 
\cite{Agnello:2011xr}. 
Despite the fact that the measured yield amounts only to three events, the extracted binding energy allowed testing some 
models of the $\mathrm{\Lambda NN}$ interaction and excluding a strongly attractive one. Future measurements in this direction 
are planned at the JPARC
facility where an unprecedented intensity of  kaon-beams will be achieved in the next years. 
A different approach to study hypernuclei is to use projectile fragmentation reactions of heavy--ion beams. In such reaction, 
a projectile fragment can capture a hyperon produced in the hot participant region to produce a hypernucleus.
Since a hypernucleus is produced from a projectile fragment, isospin and mass values of the produced hypernuclei can
be widely distributed.  The life-time and binding energies of the so-produced hypernuclei can be studied by the techniques
developed in heavy--ion collisions experiments with fixed target set-ups. A pilot experiment reported in \cite{Rappold:2013fic} shows the 
feasibility of this technique. An extensive program based on this method is planned at FAIR (Facility for Antiproton and Ion Research).
Sigma hypernuclei do not exist for times longer than $10^{-23}$ s, due to the strong $\mathrm{\Sigma-N\rightarrow\Lambda-N}$ conversion
but the analysis of $\Sigma$ formation spectra \cite{Bart:1999uh} shows that the average $\Sigma$--Nucleon potential
is repulsive $\mathrm{ReV_{\Sigma}(\rho_0)\approx +(10-50) MeV}$.
Moreover, that $\Lambda$ hyperons can be also analyzed in heavy--ion collisions too.
The extracted kinematic variables can be then compared to transport models. So far, the results obtained at intermediate energies (Ni+Ni, $E_{\text{KIN}}=\,1.93$ AGeV) 
\cite{Merschmeyer:2007zz}
show that $\Lambda$-hyperons exhibit a different behavior compared to protons, if one looks at the flow pattern of the two particle species. Systematic studies as a function
of the particle momentum and system density  that could lead to the extraction of an interacting potential  between the $\Lambda$-hyperon and the nucleons participating 
in the reaction have still to be carried out.

\subsubsection{Implications for neutron stars}
\begin{figure}[t]
\vspace{0cm}
\includegraphics*[width=9cm]{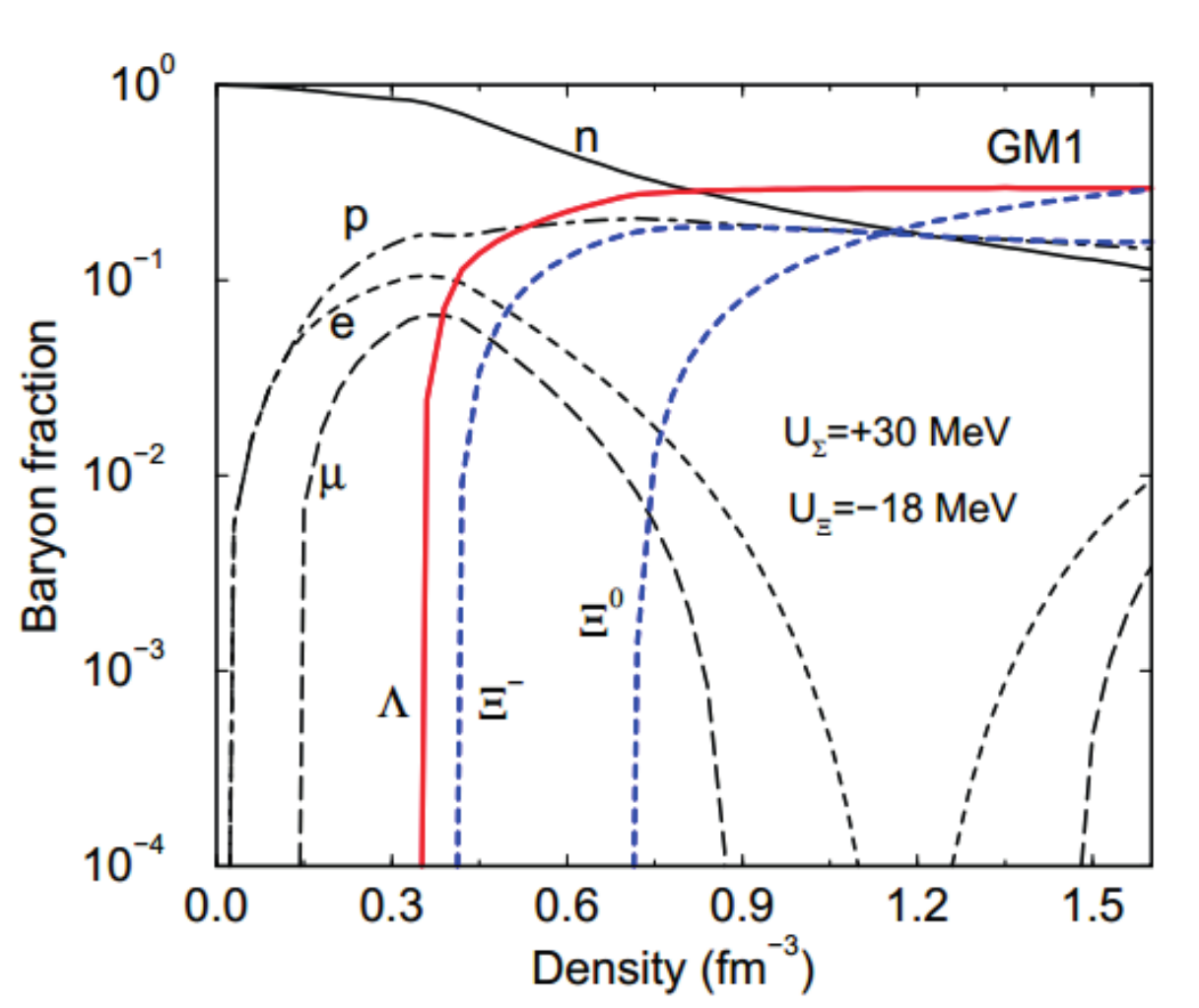}
   \caption{The fraction of baryons and leptons in neutron star matter for a RMF \cite{SchaffnerBielich:2008kb} calculation with weak hyperon--hyperon interactions .}
\label{fig:FhypNS}
\end{figure}
Returning to neutron stars, the presence of $\bar{K}$s in their core would soften the EOS, leading to upper limit for the maximal mass
of the stars that are lower than the one observed. 
This way, we might also
doubt the results about the compressibility of nuclear matter extracted from heavy--ion collisions. Indeed, most of the models used to describe these data do not contain an explicit 
dependence on the temperature of the system and in heavy--ion collisions the so called thermal contribution can influence the results. 
Two years ago the discovery of a neutron star of about two solar masses \cite{Demorest:2010bx} turned the situation upside down.
Indeed, such massive neutron stars are neither compatible with a soft equation of state nor with the presence
of a $\bar{K}$ condensate inside the star.
In this context, a theoretical work \cite{Fiorilla:2011sr} suggests that the
inner part of neutron stars  is composed by normal nuclear matter and that the maximal densities reached for these objects do not exceed $4\,\rho_0$.
Following this line of thought, a prediction of a rather stiff equation of state for normal nuclear matter has been put forward. The EOS for a finite system made only from neutrons and a small fraction of protons, including also three-body forces,  has been calculated and the two solar-masses objects have been assigned a radius varying from $11$ to $\mathrm{14\, km}$ depending on the EOS constraints. The maximal density within neutron stars associated with this calculation does not exceed $\mathrm{3\--4\,\rho_0}$.
Others suggest a transition from a soft to a stiffer EOS happening at densities between $\mathrm{3\--5\,\rho_0}$ \cite{Sagert:2011kf}. 
There the authors have stressed the importance of measuring the radius of small neutron stars, with mass near $\approx\,$1.4 solar masses, to verify the EOS 
for these systems, which density is supposed to be more compatible with conditions produced in heavy--ion collisions at intermediate energies.
This scenario is very model dependent and mainly based on the compressibility extracted from kaon data.
On the other hand, the hypothesis that only plain nuclear matter might constitute the core of the stars is in conflict with the very
 likely case that $\Lambda$ and $\Sigma$ hyperons might appear starting at densities around $\mathrm{3\--4\,\rho_0}$ and 
 hence influence the EOS \cite{SchaffnerBielich:2008kb}. Figure~\ref{fig:FhypNS} shows the fraction of baryons and leptons as 
 a function of the system density in neutron star matter. One can clearly see the appearance of the $\Lambda$ hyperons already 
 at the density $\rho=\,\,2.3\,\rho_0$. Their presence should enhance the cooling of the neutron stars via direct URCA\footnote{Urca is the name of the casino in Rio de Janeiro
  where G. Gamow and M. Sch\"onberg discussed for the first time neutrino-emitting processes responsible for the cooling of neutron stars.} processes 
 driven by hyperons, but in case of large modifications of the hyperon mass in the dense environment a coexistence of neutrons 
 and hyperons could be favored. The scenario with plain neutron-like matter up to large densities seems in this context 
 rather improbable.
On the other hand most probably, not anti-kaons but hyperons play a leading role in dense and cold systems, making the study of the 
interaction of the hyperons with nucleons as a function of the relative distance, temperature, and density of the surrounding 
system fundamental.
Hyperons created in dense nuclear matter have already been studied extensively, but so far the kinematics of the 
hyperon reconstructed in heavy--ion collisions at intermediate energies \cite{Merschmeyer:2007zz,Agakishiev:2010rs} was only 
compared to either the Boltzmann-like distribution to describe the kinematic freeze-out or to a statistical thermal model to infer upon the 
chemical freeze out.
The future perspectives, aside from the hypernuclei measurements, foresee a detailed analysis of the double differential kinematic observables ${p_T},\,Y_{CM}$ for $\Lambda$ hyperons produced in proton- and pion-induced reactions at kinetic energies around $2$ GeV and Au+Au 
collision at $1.25$ AGeV to extract the effect of the average $\Lambda$--nucleon interaction as a function of the system density.   Moreover, $
\Lambda$--p correlations can be studied in elementary reactions to infer on the distance dependence of the interaction which is so far not known at all.

\subsubsection{Neutron--rich nuclei}
\begin{figure}[t]
\vspace{0cm}
\includegraphics*[width=9cm]{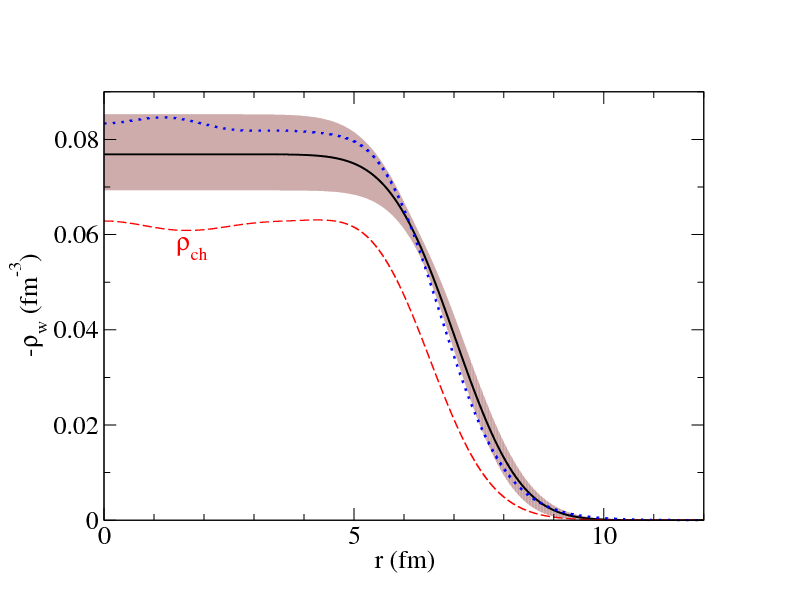}
   \caption{Weak charge density $\mathrm{\rho_W(r)}$ of $\mathrm{^{208}Pb}$ that is consistent with the PREX result (solid black line) 
   \cite{Abrahamyan:2012gp}. The brown error band shows the incoherent sum of experimental and model errors. The red dashed curve is the 
   experimental (electromagnetic) charge density $\rho_{\text{ch}}$.}
\label{fig:Fprex}
\end{figure}

The quest for the properties of neutron-rich matter and associated compact objects has also been addressed recently 
by parity violating scattering experiments with electron beams impinging on neutron-rich nuclei \cite{Horowitz:2013vxa}.
This method has the advantage of being completely free from contributions by the strong interaction and provides a model independent probe of the 
neutron density in nuclei with a large neutron excess. By measuring the asymmetry in the scattering of electrons with different helicity,  one 
can first extract the weak form factor. This is the Fourier transform of the weak charge density. 
Considering that the neutron weak charge is much larger than the proton weak charge, and applying corrections for the Coulomb distortion,
 the spatial distribution of the matter densities can be derived to the weak charge density. The black line in Fig.~\ref{fig:Fprex} 
  shows the extracted weak charge density extracted within the Helm model \cite{Helm:56} on the base of the asymmetry measured  by 
  the PREX experiment at JLab  \cite{Abrahamyan:2012gp}. The brown error band shows the incoherent sum of 
 experimental and model errors and the red dashed line shows the measured charge density \cite{Frois:1977hr}.
 
 The point neutron density can be deduced from the weak charge density and the matter radius $R_{\text{n}}$ can be determined.
The difference between the charge and matter radius of $\mathrm{^{208}Pb}$ and has found $R_{\text{n}}-R_{\text{p}}=\,0.33^{+0.16}_{-0.18}
$  \cite{Abrahamyan:2012gp}, being $R_{\text{n}}$ and $R_{\text{n}}$ the matter and charge radii of the nucleus respectively. Future measurements 
are planned to reduce the error to $\mathrm{0.05\, fm}$. There is a strong correlation between the 
$R_{\text{n}}$ and the pressure in neutron stars at densities of $\frac{2}{3}\rho_0$, hence this measurement can constraint further the EOS of neutron-rich 
matter. Indeed a larger internal pressure in neutron stars would push the neutrons against the surface increasing $R_{\text{n}}$. For this reason an 
accurate measurement of the matter radius can better constraint the nuclear EOS, and in general, a larger value of $R_{\text{n}}$ is linked with a stiffer EOS.
Moreover, the symmetry energy $s$ of nuclear matter, which plays an important role when one departs from a symmetric situation in the 
number of protons and neutrons, is also correlated with $R_{\text{n}}$. In particular, the variation of $s$ 
with the system density is found to be strongly correlated with $R_{\text{n}}$ \cite{Horowitz:2013vxa}. In case of a large value of $s$ 
for large system densities, the non--negligible fraction of protons in the system would stimulate URCA processes and hence cool the neutron star more rapidly. 

An alternative method to determine the matter radius was proposed in \cite{Lenske:2005nt}, in which antiproton collisions with nuclei would be exploited. 
There the extraction of the matter radius is achieved by measuring the antiproton-neutron annihilation cross section, but it is found to be rather model 
dependent. The produced $A-1$ nucleus  after the $\mathrm{\bar{p}}$ annihilation can be detected by exploiting the Schottky technique \cite{Bergstrm:52694} 
in a storage ring.
Such an experimental method has been proposed as a part of the FAIR project \cite{Beller:2006kr}, where antiprotons of about $\mathrm{500\, Mev}$ can be 
stored and then collided with nuclei. The intact $A-1$ nuclei can further circulate in the storage ring and be detected via the Schottky technique.
This method would be an interesting alternative to the parity violating measurements with electron beams.

\subsection{Nucleon-nucleon interactions and finite nuclei from  QCD}
\label{sec:secF2}

As issues of the nuclear equation of state will be dealt with later, this subsection concentrates on finite nuclei and nucleon-nucleon interactions.    Since electromagnetic and weak effects are typically rather small for light nuclei,  the holy grail of theoretical nuclear physics is to understand nuclear phenomena in terms of QCD.  This task is quite daunting because QCD is a difficult theory that must be treated using non-perturbative methods.  However, lattice QCD has begun to emerge as a truly precision tool to deal with many nonperturbative problems in QCD.  Unfortunately, nuclear phenomena are not yet  in this class of problems.  For a variety of  technical reasons, problems in nuclear physics are particularly difficult to pursue on the lattice.   While there has been significant recent effort in attacking nuclear problems using lattice methods, and insights into nuclear problems can be gleaned from the present day calculations, there remains some distance to go.   Given this situation, there is an interest in seeing whether one can learn {\it something} about nuclear physics from QCD without solving the theory.  One method which has been pursued over the decades is to use ``QCD-motivated'' models to attack problems in nuclear physics.  This approach has one highly problematic feature---it is difficult or impossible to tell what parts of a result come from QCD and what parts from an {\it ad hoc} model.  As such this approach will not be discussed here.  An alternative way forward is to consider systematic expansions based on counting rules which encode basic features of QCD.  Two approaches of this sort will be discussed here: an effective field theory approach which can, in principle, encode the underlying approximate chiral symmetry of QCD and the other is the large $N_c$ limit of QCD  and the $1/N_c$ expansion (where $N_c$ is the number of colors).  While both of these approaches are interesting they do have important limitations.
 
 \subsubsection{Lattice QCD and nuclear physics}
 \label{sec:LatQcdNucl}
 
 As we have noted, nuclear physics problems are intrinsically difficult  to compute on the lattice.  There are numerous reasons for this.  The natural energy scales in nuclear physics are much smaller than in hadronic physics, so that calculations need to be done with much higher accuracy than in hadronic physics to determine phenomenologically relevant results.  For example, an energy measurement with an accuracy of 1 MeV is a 0.1\% measurement of the nucleon's mass but a 50\% measurement of the deuteron's binding energy.  This means, for instance, that extrapolating to the physical point for the pion mass can have a particularly large effect.  Moreover, since the systems are bigger than for single hadrons, finite volume effects can be significant unless large lattices are used.  Furthermore, signal-to-noise problems are expected to be more severe for systems which require many propagators.  Finally, these calculations are simply more involved than  calculations for single hadrons since they  involve a large number of contractions.
 
 Despite these challenges, there are major efforts to study the nucleon-nucleon interaction and bound states from the lattice, and significant progress has been made.   One of these is centered in the NPLQCD collaboration.  Reference \cite{NPLQCD:2013} gives a good idea of the state of the art for this work.  The goals of NPLQCD are to compute   observables of relevance to nuclear physics directly from the lattice.  The approach can be used to compute the binding energy of light nuclei directly.  While this is intellectually straightforward, the problem is technically challenging.  Since the most basic interaction in nuclear physics is between two nucleons, it is also of interest to extract information about the nucleon-nucleon interaction.   The most natural observables associated with this interaction describe nucleon-nucleon scattering as well as properties of the deuteron bound state.    Some scattering observables, such as phase shifts, can be obtained by the standard approach of relating scattering observables to the energy levels in a box \cite{Luscher:1986pf,Luscher:1990ux}.   Another class of observables of interest are hypernuclei and  hypernucleon-nucleon scattering.  The method is equally suitable for the study of these.  While the current state of the art does not yet allow for computations in the regime of physical pion masses, serious calculations of real interest are being done---for example, a computation of the nucleon-nucleon scattering lengths in a world of exact $SU(3)$ flavor symmetry and the computation of the binding energies of light nuclei and hypernuclei \cite{NPLQCD:2013,NPLQCD:2012}.    Moreover, there is a clear path forward for this line of research, and one might expect this approach in time to lead to results which are directly applicable to the physical world. The techniques that have been developed are interesting in part because they are applicable to problems where experiments are difficult.  Reliable {\it a priori} calculations are critical for  resolving such issues as  whether a bound H-dibaryon exists \cite{NPLQCD:2013}. The principal open question with this approach is just how far it can be pushed in practice.

 Another approach has been pushed by the HAL QCD collaboration.  It is in many ways far more ambitious than the NPLQCD approach.  However, the scope of its ambition pushes the approach to a more problematic premise.  Unlike the NPLQCD approach, this does not strive to compute nuclear observables directly from the lattice calculations.  Rather, the underlying philosophy is to attempt to extract a nucleon-nucleon interaction in the form of a non-local potential from QCD  which is supposed 
to be usable in few-body and many-body calculations.  A review of this approach can be found in \cite{HALQCD}.  There is an important theoretical issue about the foundations of this approach.  Namely,  the extent to which the interaction so obtained is capable of accurately describing many-nucleon systems.
 
 \subsubsection{Effective field theory approach}
 
 The initial drive underlying this strand of research was to encode the underlying approximate chiral symmetry of QCD into nuclear calculations in a systematic way in much the same way that chiral perturbation theory is used in hadronic physics.  However, in nuclear physics there are low energy scales that are not the direct result of chiral symmetry; for example, the large scattering lengths in the nucleon-nucleon system which are not the direct result of chiral physics.  Thus, underlying the approach is the idea that one can build 
both the result of chiral dynamics {\it and} the other light scales of nuclear physics into an effective field theory (EFT).  
The main challenge with this approach is the calculation of physical observables (such as nuclear binding energies) from the EFT since unlike in chiral perturbation theory, this EFT must be used in a nonperturbative context.  A recent review of this approach can be found in \cite{EFT1}.

The principal  new development in the last several years has been the development of a lattice based approach to calculations within chiral effective field theory 
\cite{EFT1}.   The approach requires novel numerical techniques which are quite different from those in lattice QCD. This has already been applied to nuclear systems as heavy as ${}^{12}C$ and has proved amenable to calculations of excited states, including the Hoyle state in ${}^{12}C$ \cite{Hoyle}.  

There are open issues in this field of both practical and  theoretical  significance.  
On the practical level, the principal issue is the extent to which this method can be pushed to describe heavier nuclei.  The basic theoretical question concerns its status as an EFT reflecting the underlying chiral structure of QCD.  While there is a power counting scheme at the level of the interaction, the nonperturbative nature of the calculations mixes the various powers.  An interesting and important question is the extent to which one can estimate {\it a priori} the scale of the effects of neglected higher order terms on the nuclear observables based on power counting principles.   

 \subsubsection{Large $N_c$ limit and the $1/N_c$ expansion}
 
 The $1/N_c$ expansion around the large $N_c$ limit of QCD has yielded qualitative insights and some semi-quantitative results in hadronic physics.  It is natural to ask whether it will be useful in nuclear physics.  It is probably true that its principal value is of theoretical rather than of phenomenological value for most problems in nuclear physics.  The reason is that the nucleon-nucleon force is much larger in a larger $N_c$ world than in the world of $N_c=3$ and many of the delicate cancellations which occur in nuclear physics are spoiled.  For example, nuclear matter is believed to be a crystal at large $N_c$ while it is  thought to be a Fermi liquid in the physical world.  It is important to recognize these limitations since many models are justified, at least implicitly, only at large $N_c$, where mean-field approaches become exact.
 
 On the theoretical side,  there is interest in understanding nuclear physics in QCD-like theories, even in domains which are rather far from the physical world.  Large $N_c$ gives access to such worlds.  However, it is typically not possible to solve theories directly even given the simplifications due to the large $N_c$ limit.  Recently, however, the total nucleon-nucleon total cross section in QCD at momenta far above the QCD scale was shown to be calculable \cite{NN} and given by 
 \begin{equation}
\sigma^{\rm total} = \frac{2 \pi \, \log^2(N_c)}{m_{\pi}^2} \, .
\end{equation}
This result follows from the fact that the nucleon-nucleon interaction is strong in the large $N_c$ world and thus the cross section is fixed by the mass of the lightest particle in the theory which acts to fix that  range at which this intrinsically large interaction becomes weak.  Unfortunately, it is of little phenomenological relevance as the corrections are of relative order $1/\log(N_c)$; the predicted cross section is a factor of three to four larger than the phenomenological one at energies of a few GeV.   There have also been recent results on nuclear matter which are valid in a world in which both  the number of colors is large and  the  quark masses are well above the QCD scale \cite{NMNc1,NMNc2}.  While the world in which we live is far from  the combined large $N_c$ and heavy quark limits, the study of such a world is interesting since this represents a system for which a QCD-like theory is tractable.

The principal open issue in the field is whether there are phenomenologically relevant predictions in nuclear physics which are obtainable in practice from large $N_c$ analysis.

\subsection{Dense matter: theory and astrophysical constraints}
\label{sec:secF3}

\subsubsection{Ultra-dense QCD and color-flavor locking}

We do not have much knowledge from rigorous first-principles calculations about the QCD phase diagram in the plane of temperature and baryon chemical potential.
The region of cold and dense matter turns out to be especially challenging. 
In the extreme limit of infinite density the system becomes tractable because the average energies and momenta of the particles are large and
asymptotic freedom allows us to use weak-coupling methods. This is analogous to the high-temperature regime discussed in Chapter~\ref{sec:chapd}.
Since the quark-quark interaction 
is attractive in the anti-triplet channel, the standard Bardeen-Cooper-Schrieffer (BCS) argument for superconductivity then tells us that 
the quark Fermi surface is unstable with respect to the formation of a quark Cooper pair condensate (this is ``color superconductivity'' -- for a review 
see \cite{Alford:2007xm}).
If we consider three-flavor quark matter\footnote{If we work at asymptotically large densities, we must in principle also 
consider the heavy $c$, $b$, and $t$ quarks. However, for compact star interiors we are interested in densities where the quark chemical potential is 
much lower than the masses of these quarks.}, the ground state is the Color-Flavor Locked (CFL) state \cite{Alford:1998mk}, where the three flavors pair in a very 
symmetric way. 
The symmetry breaking pattern is
\begin{eqnarray} 
SU(3)_c \times SU(3)_L\times SU(3)_R \times U(1)_B  \nonumber \\[2ex]  
\to SU(3)_{c+L+R} \times \mathbb{Z}_2 \, , \label{FCFLpattern}
\end{eqnarray}
where $SU(3)_c$ is the color gauge group, $SU(3)_L\times SU(3)_R$ the chiral flavor group (which is an exact symmetry at ultra-high densities where the 
quark masses are negligible compared to the quark chemical potential, $m_u,m_d,m_s\ll\mu$), and $U(1)_B$ is the symmetry associated with baryon number conservation.
The unbroken symmetry is a global $SU(3)$ of simultaneous rotations in color and flavor space, hence the name color-flavor locking. In particular, CFL
breaks chiral symmetry by an order parameter in a manner very similar to that seen in the vacuum where the order parameter is a chiral quark-antiquark condensate. 
We conclude that chiral symmetry of QCD is spontaneously broken at low {\em and} high densities. 
The low-energy degrees of freedom in CFL quark matter are Goldstone modes: one exactly massless superfluid phonon and eight light pseudoscalar mesons analogous to the 
pions and kaons. The quarks are gapped by their Cooper pairing; so, the phenomenology of the CFL phase at low energies is dominated by the Goldstone bosons.

Because of the spontaneous breaking of the color gauge symmetry, the gluons in the CFL phase acquire a Meissner mass, just like the photon in an ordinary 
electronic superconductor. More precisely, seven of the gluons plus one combination of the eighth gluon and the photon become massive, while the orthogonal combination of the eighth gluon and the photon remains massless. In the gauge sector, the infrared physics of the CFL phase thus reduces to an Abelian theory. 

The CFL phase has many interesting properties, some of which have been worked out and some of which should be determined in the future. The phenomenology of the CFL phase is relevant for compact stars; see the discussion below and in Sec.~\ref{sec:Fastro}. Of course, the matter in a compact star is in a region of the QCD phase
diagram that is far from being asymptotically dense. In fact, one can estimate that the perturbative weak-coupling calculation of the CFL energy gap is reliable 
only for $\mu\gtrsim 10^8\,{\rm MeV}$ \cite{Rajagopal:2000rs}. This corresponds to densities 15-16 orders of magnitude larger than those in the center of compact stars. It is thus
important to ask what the ground state of dense quark matter at these much lower densities is. Finding the answer to this question is a major challenge, and the problem is currently unsolved. The difficulties of this problem and approaches that 
have been applied and may be applied in the future are explained in the next two subsections.    

\subsubsection{Moderately dense QCD}
\label{sec:Fmoderate}

Phases of QCD at moderate densities can be studied from 
two different perspectives. Either ``from below,'' by investigating dense nuclear matter and extrapolating results to higher densities (see Sec.~\ref{sec:secF1}) 
or ``from above,'' starting from CFL and asking what are the next phases down in density. Here we shall  
take the latter approach. As we reduce the density, we encounter two complications. First, we 
leave the safe grounds of asymptotic freedom and have to deal with a strongly coupled theory. Currently, there are no reliable methods in QCD to apply to
this problem, and we have to rely on the alternative approaches discussed below. Second, the particularly symmetric CFL state will be disrupted because at the densities of interest the strange quark mass can no longer be neglected because its
density-dependent value, which lies between the current mass $\sim 100\,{\rm MeV}$ and the vacuum constituent mass $\sim 500\,{\rm MeV}$, is not
small compared to quark chemical potentials of the order of $(400-500)\,{\rm MeV}$ inside a compact star. Thus the Fermi momenta of up, down, and strange 
quarks are no longer equal: it is energetically more costly now to have strange quarks in the system, and hence the strange quark Fermi momentum becomes smaller. In the standard 
BCS pairing, however, it is crucial that the Fermi momenta of the quarks that form Cooper pairs are identical. Since CFL pairing relies on the attractiveness of the pairing between quarks of different flavors, this Fermi momentum mismatch imposes a kind of stress on the pairing. A simplified version of this problem was already discussed
in the context of electronic superconductivity by Clogston and Chandrasekhar in the 1960's \cite{Clogston:1962,Chandrasekhar:1962}. In this case, the superconducting
state becomes disfavored with respect to the unpaired state when $\delta \mu > \Delta/\sqrt{2}$, where $\delta\mu$ is the difference in chemical potential
of the two fermion species that form Cooper pairs and $\Delta$ the quasiparticle energy gap. In quark matter, $\delta \mu$ is determined by $m_s^2/\mu$. 
However, the situation in a compact star is more complicated than in an electronic superconductor 
because we are dealing not with 2 (spin up and down) but with $2N_f N_c=18$ fermion species (antiparticles can be neglected since they are strongly blocked in the 
presence of the Fermi sea)
and because the conditions of electric and color neutrality 
impose constraints on the system. Nevertheless, the general expectation that it becomes ``harder'' for the quarks to form Cooper pairs in the presence of a 
non-negligible strange quark mass remains true. 

The most radical possibility for the system to respond to the stress would be not to form any Cooper pairs. There are other options, however, 
which constitute viable candidates for 
matter in the core of compact stars. First of all, CFL may survive in a modified version, by producing a $K^0$ condensate (relieving the stress by producing 
negative strangeness), where the $K^0$ is the lightest of the (pseudo-)Goldstone modes of the chiral symmetry breaking in CFL \cite{Son:1999cm,BedaqueSchaefer}. 
The resulting phase, usually called CFL-$K^0$, has interesting phenomenological properties that are being worked out in a series of studies, see for instance 
\cite{Alford:2007qa,Alford:2008pb,Alford:2009jm}. In a way, CFL-$K^0$ is the ``mildest'' modification of the CFL phase. Larger values of the 
strange quark mass (more precisely of $m_s^2/\mu$ compared to the energy gap $\Delta$) lead to more radical modifications and eventually to a
breakdown of CFL. Continuing our journey down in density (at zero temperature)
we next expect the Cooper pairs to break in certain directions in momentum space, 
spontaneously breaking rotational symmetry. 
In general, such a phase can be thought of as a compromise between the fully paired and fully unpaired phases: 
the energy cost of forming Cooper pairs with zero total momentum in all directions becomes too large, but it is still preferable to form Cooper pairs in certain 
directions, if the kinetic energy cost is sufficiently small.
Counter-propagating currents arise, of the kaon condensate on the one hand and the unpaired fermions
on the other hand; hence, this phase is termed curCFL-$K^0$ \cite{Schafer:2005ym,Kryjevski:2008zz,Schmitt:2008gn}. With even larger mismatches, counter-propagating
currents appear in more than one direction; as a result the system spontaneously breaks translational invariance and crystalline structures become possible, where
the gap $\Delta$ varies periodically in space and vanishes along certain surfaces \cite{Alford:2000ze,Rajagopal:2006ig,Mannarelli:2006fy}. Further increasing
$m_s^2/\mu$, the CFL pairing pattern may break down, and pairing only between up and down quarks (``2SC phase'') \cite{Bailin:1983bm} or single-flavor pairing 
in the spin-one channel \cite{Schafer:2000tw,Schmitt:2004et} become candidates for the ground state. 

This journey down in density has been done by varying the ``parameter'' $m_s^2/\mu$ and by relying on effective theories, phenomenological models, etc., but not
on first-principles QCD calculations. In QCD, the only dimensionful parameter is $\mu$, while $m_s$, $\Delta$, and the strong coupling constant are functions of $\mu$ that are unknown 
in the strongly coupled regime. In other words, it is
currently not known how the above sequence of phases translates into the QCD phase diagram. It is conceivable that the CFL phase (or variations of it) persists down to 
densities where the hadronic phase takes over. In this case, the intriguing possibility of a quark-hadron crossover might be realized 
\cite{Schafer:1998ef,Hatsuda:2006ps,Schmitt:2010pf}. Or, there may be one or several of the above more exotic color superconductors between the CFL phase and
the hadronic phase. It is fair to say that a major improvement of current theoretical tools is needed to settle these questions unambiguously. We shall discuss
some of the tools used so far and promising theoretical directions for the future in the next subsection. A complementary line of research is provided by 
astrophysics, where candidate phases can be potentially ruled out from properties of compact stars, see Sec.~\ref{sec:Fastro}.

\subsubsection{Theoretical approaches and challenges}

Let us discuss the theoretical tools available for studying dense QCD matter and their potential and perspective for future research. We start with the ones that 
have already been employed extensively to obtain the above sketched picture of the QCD phase structure and then turn to more novel approaches. 

\paragraph{Perturbative QCD}
We have already mentioned the regime of applicability of perturbative QCD, which is limited to densities many orders
of magnitude larger than the densities in the interior of compact stars.
Although distant from the physically relevant regime, this is a secure base
from which we extrapolate down in density just as we use results from nuclear physics to extrapolate up in density. An extrapolation over many orders 
of magnitude seems bold, but the value it gives for the energy gap of color superconductivity is comparable to the result
obtained from a phenomenological Nambu--Jona-Lasinio (NJL) model whose parameters are fit to {\em low}-density properties.
The extrapolation of perturbation theory down to $\mu\simeq 400\,{\rm MeV}$  yields $\Delta\simeq 20\,{\rm MeV}$ \cite{Son:1998uk,Pisarski:1999tv} 
(using a strong coupling constant $g\simeq 3.5$, suggested by the two-loop QCD beta function), 
while NJL calculations
suggest $\Delta\simeq (\text{20 -- 100})\,{\rm MeV}$. 
Given the completely different theoretical origins of these results, their approximate agreement is remarkable. 

In perturbative calculations, the magnitude of the zero-temperature energy gap $\Delta$ translates into a critical temperature $T_c$ for color-superconductivity 
via a BCS-like relation. In BCS theory, $T_c = (e^\gamma/\pi)\,\Delta \simeq 0.57\,\Delta$, where $\gamma$ is the Euler-Mascheroni constant. In some color superconductors
this relation is modified \cite{Schmitt:2002sc}, in the CFL phase $T_c = 2^{1/3}(e^\gamma/\pi)\,\Delta$, but it is still true that the critical temperature is given by 
a numerical factor of order one times the zero-temperature gap. Therefore, one consequence of the above estimate of $\Delta$ is that the critical temperature of the CFL
phase (and of other spin-zero color superconductors) is larger than the typical temperature of a neutron star.  

\paragraph{Effective theory of CFL}
One can construct an effective Lagrangian for the low-energy degrees of freedom of CFL \cite{BedaqueSchaefer}, like the chiral Lagrangian for low-density mesons. This effective theory does not tell us if and at what density CFL is replaced
by another phase, but it can be used to compute properties of CFL and 
CFL-$K^0$ in terms of a small number of unknown couplings in the Lagrangian.
This has been done for transport properties such as bulk and shear viscosities \cite{Manuel:2004iv,Alford:2007rw,Alford:2008pb,Bierkandt:2011zp}.
Since its form is determined by the symmetries of CFL, 
the effective theory must be valid for all densities where CFL, or any other phase with the same
symmetry breaking pattern, exists (at energies far below the critical temperature of CFL). Hence, if CFL 
persists down to densities of astrophysical interest we 
can determine, at least qualitatively, the properties of matter at these 
densities. Quantitative predictions are still subject to uncertainties since up to now
the only way we can estimate the parameters of the Lagrangian is by matching to perturbative high-density results.

\paragraph{Hydrodynamics}
Efforts to connect neutron star observables with the properties of their interior often involve calculating
transport properties that characterize the hydrodynamics of cold dense QCD matter.
(The hydrodynamics of hot QCD matter is an active research field with relevance for heavy-ion collisions, see
Chapter~\ref{sec:chapd}, and it will be interesting to see whether and how these two research lines can benefit
from each other).
In a neutron star, hydrodynamics becomes important for instance in the discussion of $r$-mode
instability\footnote{Oscillatory modes of a compact star are classified according to the restoring force.
In the case of $r$-modes or rotational modes, this is the Coriolis force.} \cite{Andersson:1997xt},
asteroseismology \cite{2011arXiv1111.0514W}, discussed below, and dynamical effects of the magnetic field
\cite{2012arXiv1203.3590L}.
In particular, it is desirable to understand {\em superfluid} hydrodynamics since superfluidity appears in
nuclear matter as well as in quark matter.
In quark matter, only the CFL phase (and its variants) is superfluid, because of the spontaneous breaking of
baryon number conservation, see Eq.~(\ref{FCFLpattern}).
Superfluidity of the CFL phase manifests itself for instance in the presence of three different bulk
viscosity parameters \cite{Bierkandt:2011zp}.
For applications to neutron stars one must deal with the fact that
in some cases the mean free path of the superfluid phonons can be 
comparable to or even larger than the size of the star, as discussed in \cite{Mannarelli:2012eg,Mannarelli:2013hm} with emphasis on applications in 
cold atomic trapped gases.
It is also valuable to formulate the hydrodynamics of CFL in the hydrodynamical framework that is used by astrophysicists. 
A first step in this direction has been made recently in connecting the relativistic two-fluid formalism of 
superfluidity with microscopic physics \cite{Alford:2012vn,Alford:2013koa}. 
Like proton-neutron matter, CFL may also be a complicated multi-fluid system if kaons
condense, i.e., in the CFL-$K^0$ phase. In this case it is not only $U(1)_B$ that is spontaneously broken (by the Cooper pair condensate), but also strangeness 
conservation (by the kaon condensate). Interesting fundamental questions regarding superfluidity arise because strangeness is not conserved when 
the weak interactions are taken into account, i.e., one has to understand whether some superfluid phenomena can persist even if the underlying $U(1)$ is only an 
approximate symmetry.

\paragraph{Nambu--Jona-Lasinio (NJL) model}
In the NJL model, the gluonic interaction between the quarks is replaced by a simple four-fermion interaction. 
Because of its relative simplicity, and because it is well suited to incorporate Cooper pairing (it was developed originally in this context) as well
as the chiral condensate, it has been frequently used to gain some insight into the phase structure of dense quark matter; see for instance 
\cite{Buballa:2003qv,Ruster:2005jc,Blaschke:2005uj,Abuki:2005ms}, and, for extensions including the Polyakov loop, Refs.~\cite{Fukushima:2003fw,Megias:2004hj,Ratti:2005jh,
Fukushima:2008wg}. These studies are very useful since they point out possible phases and phase transitions. However, they 
are ultimately of limited predictive power because their results depend  strongly (even qualitatively) on the chosen values of the parameters such as the coupling 
constants and because the model is not a controlled limit of QCD. 

\paragraph{Ginzburg-Landau (GL) studies}
A GL theory is an effective theory of the order parameter. It is like
the low-energy effective theory described above, except that
it includes the ``radial'' degree
of freedom, which corresponds to the magnitude of the order parameter, hence
it can describe the transition at which the order parameter becomes non-zero. Whereas GL theory is valid in the vicinity of the second-order thermal 
(``melting") phase transition where the correlation length diverges, the aforementioned effective theory of CFL is valid away from this transition line so that these 
two approaches nicely complement each other.
GL theories are commonly used to describe phase transitions in condensed matter 
physics, and, along with NJL models, have also been 
applied to phase transitions in dense quark matter.
A GL theory has been used to show that gauge field fluctuations yield a correction to the critical temperature of color superconductivity
and render this phase transition first order \cite{Giannakis:2004xt}. Also, the significance of the axial anomaly for a possible crossover between 
nuclear and CFL quark matter has been investigated within GL theory \cite{Hatsuda:2006ps}.  As with NJL models, the GL theory can be very useful as a guideline 
for the phases of dense QCD, especially since ordinary condensed matter physics tells us that the phase diagram 
can be expected to be very rich; however, if we are interested in full QCD, it can at best be a first step towards more elaborate studies. 

We now discuss some theoretical approaches which have only recently been considered for the study of dense matter and which may shed light on the open 
problems from different angles, but which all have to deal with difficult theoretical challenges. 

\paragraph{Lattice QCD}
Lattice gauge theory is currently the most powerful method to determine equilibrium properties of the QCD vacuum and its excitations (see Sec.\ \ref{sec:chapd}). 
However, as explained in that chapter, at finite density the usual probabilistic sampling method fails because of the sign problem. Therefore, 
there is currently no input from lattice QCD to the questions we discuss here. Several groups are trying to find ways around the sign problem \cite{Aarts:2008wh,Fromm:2011qi,Gattringer:2012df}. For instance, it
has been shown that in a combined strong coupling and hopping expansion, an effective theory can be derived for which the sign problem is relatively harmless 
\cite{Fromm:2011qi}. Currently this method is restricted to unphysically large quark masses. In this limit, first indications for a nuclear matter onset have been 
obtained \cite{Fromm:2012eb}. It will be very interesting to see whether this approach can be extended to more realistic quark masses, and 
whether it can eventually tell us something about realistic, dense nuclear and quark matter from first principles. For instance, one might try to study quark and 
nucleon Cooper pairing and its phenomenological consequences.     

\paragraph{Large-$N_c$ QCD}
The number of colors $N_c$ is a useful ``knob'' that, if set to a sufficiently large value, deforms QCD into a simpler (albeit not simple) theory; see for instance 
Sec.~\ref{sec:secF2} and Chap.~\ref{sec:chapg}. 
For the study of moderately dense QCD, the $N_c\to \infty$ limit is, like the
asymptotic $\mu\to\infty$ limit discussed above, a more accessible regime from
which we can extrapolate (admittedly with the chance of missing
important physics) to the regime of interest. The difference is that for $N_c\to\infty$ we leave the theory of interest, while for $\mu\to\infty$ we stay within QCD. The gross features of the large-$N_c$
QCD phase diagram are known, and it has been argued that nuclear matter at large $N_c$ (called ``quarkyonic matter'') behaves quite differently from $N_c=3$ nuclear 
matter \cite{McLerran:2007qj}. It is an interesting, unsolved question whether quarkyonic matter survives for $N_c=3$ QCD, and several studies have 
addressed this question, for instance, within NJL-like models \cite{Fukushima:2008wg}. It is also known that for very large $N_c$ quark-hole pairing is favored over 
quark-quark pairing and thus the CFL phase is replaced by a so-called chiral density wave \cite{Shuster:1999tn}. These results seem to indicate that, at least for dense matter, $N_c=3$ is very different from $N_c=\infty$.

\paragraph{Gauge/gravity correspondence}
The gauge/gravity correspondence has become an extremely popular tool to study strong-coupling physics.
It has relevance to heavy-ion physics (see Sec.~\ref{sec:chapd}) and
to dense matter (see Sec.~\ref{sec:chapg}). One can
introduce a chemical potential in a gauge/gravity calculation, and this provides a tractable system of dense matter with strongly-coupled interactions.
The model that currently comes closest to QCD is the Sakai-Sugimoto model \cite{Sakai:2004cn}, which completely
breaks supersymmetry and contains confinement/deconfinement and chiral phase transitions. In the context of dense matter, it has been used
to compute phase structures in the presence of
finite $\mu$ and $T$ \cite{Horigome:2006xu,Bergman:2007wp} and 
in a background magnetic field \cite{Bergman:2008qv,Preis:2010cq}. 
For nuclear matter, however,
its relevance for QCD is questionable, since it has been shown that holographic nuclear matter behaves quite differently from ordinary nuclear
matter \cite{Kaplunovsky:2010eh,Preis:2011sp}; in particular, the nuclear matter onset in the Sakai-Sugimoto model is second order, indicating the absence of a binding energy. One of the 
reasons for this and other differences to QCD is the large-$N_c/N_f$ limit to which most of these studies are constrained. Their relevance for dense QCD is thus 
debatable for reasons discussed in the previous paragraph. 
It would be very interesting, but also very challenging, to lift the constraint 
of large $N_c/N_f$ in these holographic studies; see \cite{Burrington:2007qd,Nunez:2010sf} for pioneering work in this direction. 
It would also be interesting to study color superconductivity in gauge/gravity duality. First steps in this direction within a ``bottom-up'' approach have been done \cite{Basu:2011yg}, resulting in phase diagrams that resemble qualitatively the expected phase structures of dense QCD.

\subsubsection{Dense matter and observations of compact stars}
\label{sec:Fastro}

Matter at densities of several times nuclear ground state density is very difficult to study experimentally (see Sec.~\ref{sec:secF1}). Dedicated collider experiments 
in the coming years at FAIR (Darmstadt) and NICA\cite{nica} (Dubna) will help to extend our experimental reach further into the region of high densities, although it will remain 
a challenge to produce dense matter that is cold enough to
exist in the deconfined quark phases discussed above. We therefore turn our attention to  compact stars, which are the only place in the universe where we might find 
cold nuclear or even quark matter. After black holes, compact stars are the densest objects in nature. They are the remains of massive ordinary stars after 
the nuclear fusion process runs out of fuel, and the gravitational attraction in the collapsing core can only be compensated by the Pauli pressure of the strongly 
interacting constituents. They have masses of more than a solar mass at radii of the order of 10 km and can thereby reach up to 10 times the density reached in atomic 
nuclei, corresponding to a baryon chemical potential up to 1.5 GeV. These densities are large enough that they could contain phases of dense quark matter in their 
interior, but are far below the asymptotic densities described above.

To learn something about dense phases of QCD from astrophysical observations, 
we need to compute properties of candidate phases and see whether the astronomical
observables are able to discriminate between these candidate phases. This would
allow us to put constraints on the structure of the QCD phase diagram.\footnote{It is usually assumed that the general theory of relativity gives the 
correct description of the intense gravitational field of a neutron
star, and mass-radius measurements are then used to constrain the equation of state.
However, one can alternatively assume an equation of state and obtain constraints on
the gravitational coupling \cite{Dobado:2011gd} or on deviations from general relativity
\cite{Arapoglu:2010rz}.}
 For instance, we would like to understand whether compact stars are made of nuclear matter only
(neutron stars), whether they contain a quark matter core with a nuclear mantle (hybrid star), or whether they are pure quark stars. 
Here we will limit ourselves to discussion of a few very interesting recent measurements which nicely demonstrate how we can obtain constraints on dense matter from compact stars and what is needed in the future to 
make these constraints more stringent. For a broader pedagogical review see for instance \cite{Schmitt:2010pn}.

\paragraph{Mass-radius relation and the $2\,M_\odot$ compact star}
The mass-radius function $M(R)$ of a compact star is determined, via the
Tolman-Oppenheimer-Volkov equation, by the EOS of the
dense matter of which it is made. Therefore measurements of masses and radii
provide information about the EOS of nuclear and perhaps quark matter.
The $M(R)$ curve has a maximum mass which is larger if the EOS is
stiffer, i.e., has stronger repulsive interactions, and smaller if the EOS is soft. In order to use measurements of $M(R)$ to learn about dense
matter, we need to calculate the EOS for the various different forms of matter that we think might be present. Such calculations
are not well controlled, particularly at densities above nuclear density,
and the results have large uncertainties. However, in general, one can say that
matter with larger number  of degrees of freedom tends to be softer and yield
smaller maximal masses, or, more precisely, if one adds new degrees of freedom to the system, the interactions must become stronger in order to 
achieve the same maximal mass. Model calculations confirm that
hyperons and/or meson condensates in nuclear matter decrease the 
maximum mass of neutron stars, see for instance Refs.~\cite{Li:2010yc,Weissenborn:2011ut}. Also quark matter has more degrees of freedom than ordinary nuclear matter, 
suggesting a softer equation of state. However, it is not known whether
this effect can be compensated by the strength of the interactions. 

The heaviest neutron stars observed to date are the pulsars PSR J1614-2230 and PSR J0348+0432, which have been
determined to have masses $M=(1.97\pm 0.04)M_\odot$ \cite{2010Natur.467.1081D} and $M=(2.01\pm 0.04)M_\odot$ \cite{2013Sci...340..448A}, respectively. Both 
results are remarkably precise (achieved by measuring Shapiro delay in a nearly edge-on binary system in the first case, and by a precise determination of the 
white dwarf companion mass in the second case).
The large value of the mass rules out several proposed EOS for dense matter \cite{Weissenborn:2011qu,Massot:2012pf} and strongly
constrains the quark matter EOS \cite{Alford:2013aca}.

\begin{figure}[tb]
 \hspace{-1.5cm} \includegraphics*[width=10cm,angle=180]{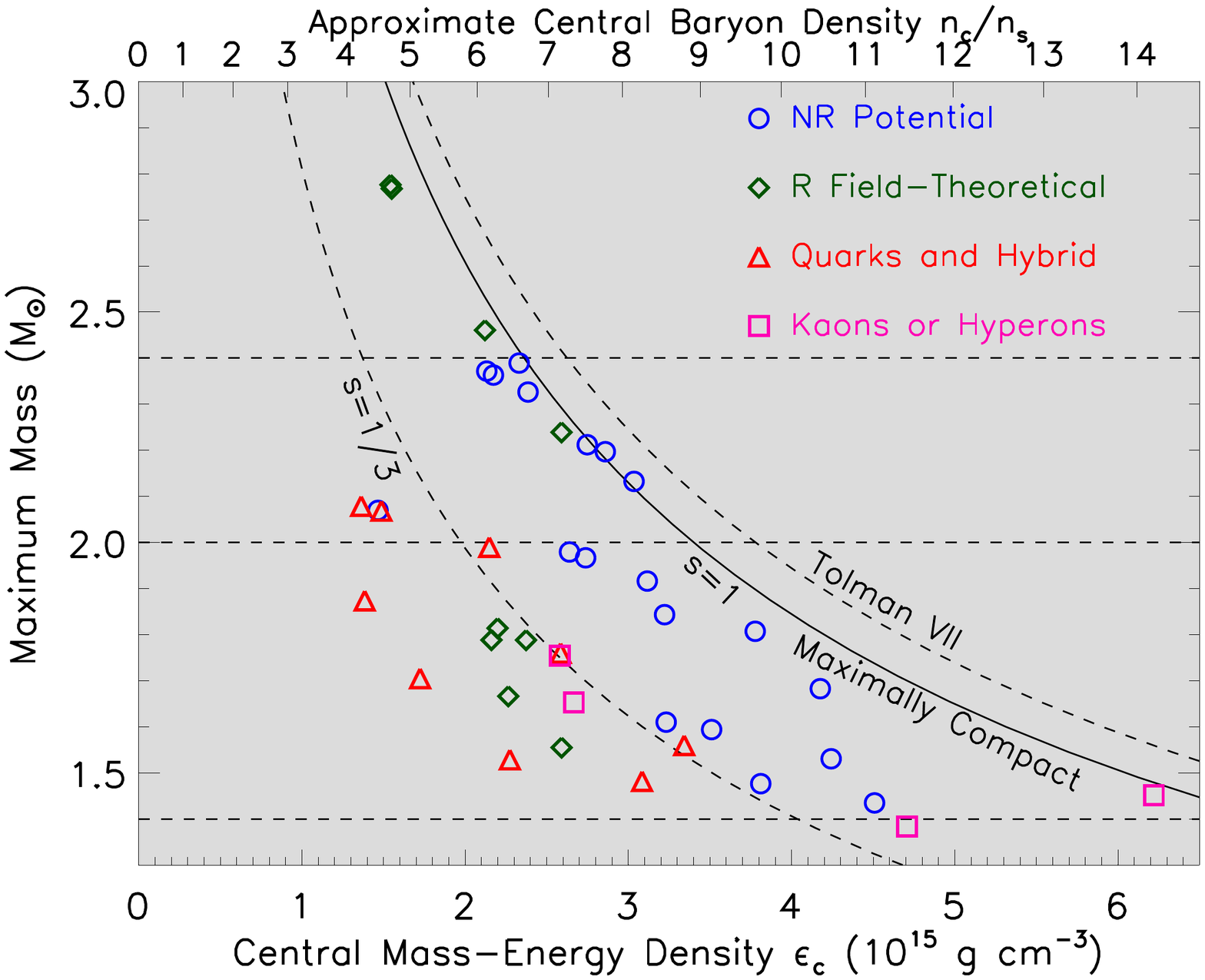}
   \caption{A given compact star mass (vertical axis) implies an upper bound for the energy density (lower horizontal scale)
and the baryon density (upper horizontal scale) in the center of the star. 
For instance, $M\simeq 2\,M_\odot$ (see middle horizontal dashed line) allows a central baryon density of no more than 
about 9 times nuclear ground state density. An even heavier star would {\em decrease} this upper bound. The solid line that gives this bound is obtained by assuming 
a ``maximally compact'' equation of state of the form $P=s(\epsilon - \epsilon_0)$ with $s=1$. Independent of $\epsilon_0$  
one finds $\epsilon_{\rm max}M_{\rm max}^2=1.358\times 10^{16}\,{\rm g}\,{\rm cm}^{-3}M_\odot^2$, which defines the solid line. 
The various points are calculations within different models and matter compositions. They confirm the limit set by the solid curve and show
that equations of state with pure nuclear matter tend to give larger maximal masses than more exotic equations of state. Details can be found in 
\cite{Lattimer:2010uk}, where this figure is taken from.}
\label{fig:Fmaxmass}
\end{figure}

In this area we look forward to both theoretical and observational improvements.
Future observations may yield even heavier stars, and more accurate measurements
of radius along with mass, giving a more accurate idea of the $M(R)$ curve
for compact stars. In Fig.~\ref{fig:Fmaxmass} we show some theoretical results for various
equations of state for neutron stars, hybrid stars, and quark stars together with a general constraint for the maximal density in the center of the star
that can be obtained from a given mass measurement \cite{Lattimer:2010uk}.   
It is important for theorists to improve our understanding of cold, dense, strongly interacting quark matter, for instance with better perturbative calculations, such
as in \cite{Kurkela:2009gj,2014ApJ...781L..25F} where the equation of state up to order $\alpha_s^2$ has been worked out, or non-perturbative studies building on 
a Dyson-Schwinger approach \cite{Chen:2008zr,Klahn:2009mb,Chen:2011my}.

\paragraph{Cooling rate and the fast cooling of Cas A}
While different phases of matter may have very similar equations of state,
which is a bulk property, they may be distinguished by their neutrino
emissivity, which is more sensitive to the low-energy excitations. Since
neutrino emission is the dominant cooling mechanism of a neutron star less
than a million years old, measurements of cooling give information about
neutrino emissivity and hence about the phases present inside the star, in
particular about superfluidity. Unpaired matter can more easily produce
neutrinos and antineutrinos via beta decay: its emissivity varies as some
power of temperature.  In contrast, superfluid matter with an energy gap
$\Delta$ in the quasiparticle spectrum shows an exponential suppression of the
emissivity $\propto e^{-\Delta/T}$ for small temperatures
$T\ll\Delta$ \cite{Yakovlev:2000jp,Jaikumar:2005hy,Schmitt:2005wg}. However, the emissivity of a superfluid can be enhanced -- even 
compared to
unpaired matter -- at temperatures below, but close to, the critical temperature
due to continual pair breaking and formation (PBF) of the Cooper pairs \cite{Flowers:1976ux,Kolomeitsev:2008mc,Steiner:2008qz}.

There has been a noteworthy  recent observation
of the isolated neutron star in the Cassiopeia A (Cas A) supernova remnant \cite{Heinke:2010cr}. It is the youngest known neutron star of the Milky Way with 
an age of 330~yr.
Recent analysis shows that the temperature of this star has decreased from 
$2.12\times 10^6\, {\rm K}$ to $2.04\times 10^6\,{\rm K}$, i.e., by about 4\%, during 2000 to 2009 \cite{Heinke:2010cr}. This
is a surprisingly fast cooling process, which implies a high neutrino
emissivity during that time period.

\begin{figure}[t]
\vspace{-6.5cm}
\includegraphics*[width=9cm]{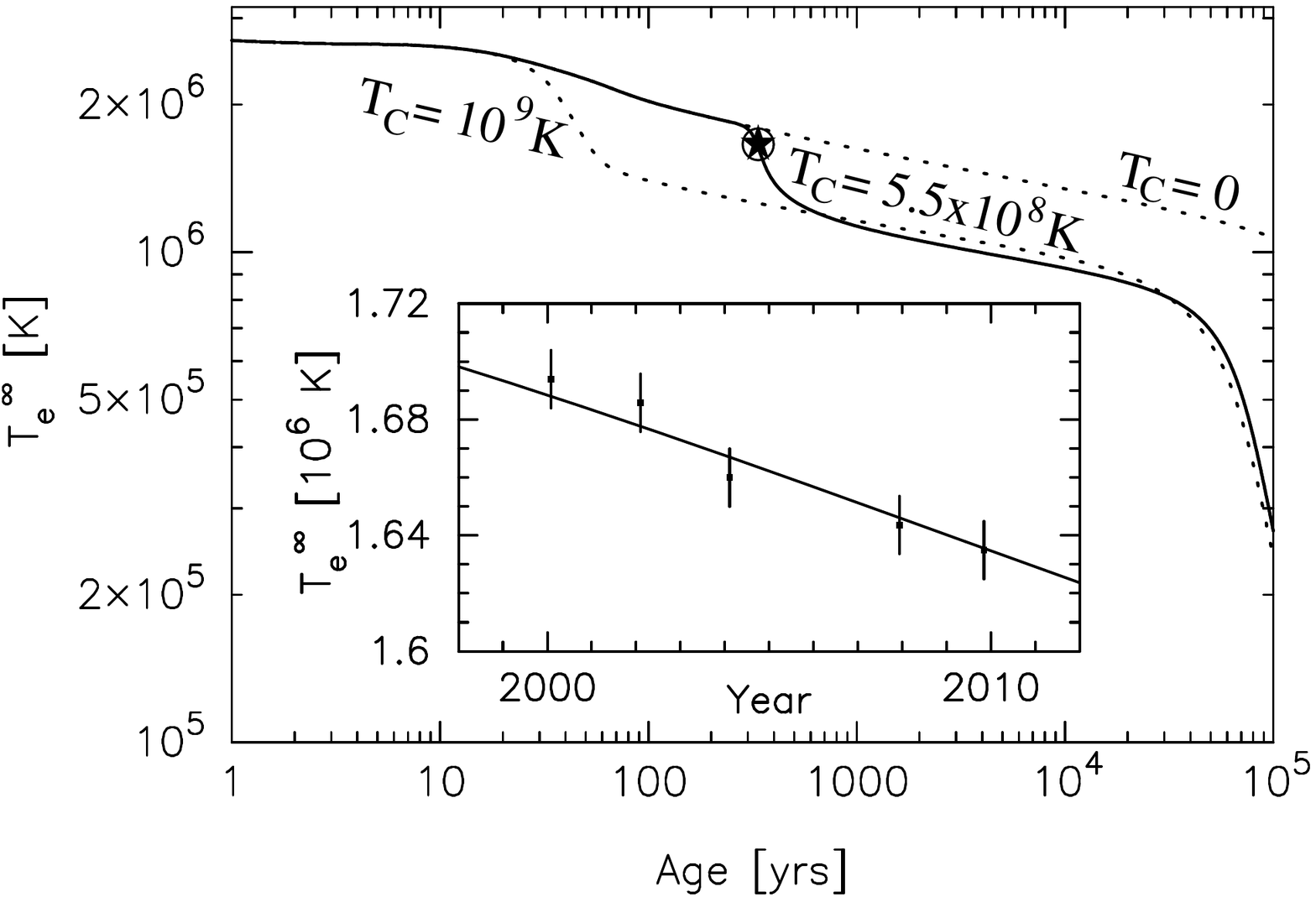}
   \caption{Red-shifted effective temperature versus age of the Cas A neutron star: data (encircled star and points with error bars in the zoom-in) and 
theoretical curves based on the PBF process for various critical temperatures $T_c$ for neutron superfluidity (more precisely the maximal $T_c$, since
$T_c$ depends on density). The solid line, also shown in the zoom-in matches the data points, while larger or smaller values for $T_c$ would lead to 
an earlier or later start of the rapid cooling period. Figure taken from \cite{Page:2010aw}.}
\label{fig:FCasA}
\end{figure}

It has been conjectured that the PBF process mentioned above
might be responsible for the rapid cooling of the Cas A star \cite{Page:2010aw,Shternin:2010qi}, in the following way. Before the rapid cooling began, 
the core of the star contained 
superconducting protons (ensuring that it cooled slowly) and
unpaired neutrons. When the temperature in the core reached the critical temperature $T_c$ for neutron superfluidity in the $^3P_2$ channel, 
the PBF process began to occur in that region, accelerating
the cooling process. It is therefore conjectured that in Cas A
we are observing the superfluid transition of neutrons in real time. 
This explanation assumes, as theorists have predicted
\cite{Dean:2002zx,Muther:2005cj}, that
 the critical temperature is strongly density dependent. 
There is then for an extended time period 
a slowly expanding shell in the core at which the temperature is close to $T_c$ and where the efficient PBF cooling mechanism operates. 
In Fig.~\ref{fig:FCasA} we show how this can explain the data 
\cite{Page:2010aw}\footnote{See also http://chandra.harvard.edu/photo/2011/casa/\\coolCANSv7\_lg\_web.mov for a movie of the cooling process.}.
Although several assumptions go into this interpretation, 
it is a nice example how an astrophysical observation can 
yield constraints
on microscopic parameters such as the critical temperature for neutron superfluidity. This ``measurement of $T_c$" becomes particularly interesting because 
of the enormous uncertainties in the theoretical calculation of $T_c$ from nuclear physics \cite{Baldo:1998ca,Schwenk:2003bc,Khodel:2004nt}.

Alternative scenarios have been discussed in the literature as well, and it is an interesting problem for future studies to either support or rule out the various
possible explanations. For instance, taking into account certain medium corrections to the cooling process appears to 
explain the data without any transition into the superfluid phase, i.e., with a much smaller critical temperature for neutron 
superfluidity \cite{Blaschke:2011gc,Blaschke:2013vma}. It has also been suggested that a color-superconducting quark matter core may explain the cooling data 
\cite{2011arXiv1109.1080N,Sedrakian:2013rr}. In the scenario of \cite{Sedrakian:2013rr}, the rapid cooling is due to a transition from the 
so-called 2SC phase -- plus a phenomenologically imposed gap for the blue quarks that usually remain unpaired in 2SC -- to a crystalline color superconductor 
where there are unpaired fermions that enable efficient neutrino emission. 

\paragraph{Gravitational wave emission and compact star seismology}
Another way to directly probe the interior composition of compact stars is
to study how the dense matter inside damps global oscillation modes, using
seismological methods similar to those employed to learn about the interior
composition of the earth or the sun. Particularly relevant are ``$r$-modes''
\cite{Andersson:1997xt,Andersson:2000mf}, which in the absence of damping are
unstable and grow spontaneously until they 
reach their saturation amplitude. They then cause the star to
spin down by emitting gravitational waves. The observable consequences are
(a) direct detection of the waves \cite{Alford:2012yn,Alford:2014pxa} in next-generation detectors such as 
advanced LIGO \cite{ligo} and VIRGO \cite{virgo} or the planned Einstein telescope;
(b) stars should not be found in the ``instability region'' in
spin-temperature space. Mapping the instability region would tell us about
the interior viscosity, because it is the viscosity that
limits the occurrence of r-modes
to an instability region over a range of temperatures and sufficiently large
frequencies. How quickly stars exit from the instability region
is determined by the saturation amplitude of the r-modes, which is restricted by astrophysical observations \cite{Mahmoodifar:2013quw,Alford:2013pma}. Several possible saturation  
mechanisms have been proposed \cite{Lindblom:2000az,Arras:2002dw,Alford:2010gw,Bondarescu:2013xwa,Haskell:2013hja,Gusakov:2013jwa}.
Although the actual mechanism and amplitude remains uncertain,
no currently proposed mechanism yields amplitudes that are low enough that a fast-spinning star would not spin down too fast to be consistent with the stringent astrophysical bounds. Thus no pulsars are expected to be found
within the $r$-mode instability region in frequency-temperature space.

The current state of observations is that there are many
millisecond pulsars in low mass X-ray binaries (LMXBs) which 
have been spun up by accretion from a companion star and now
lie well within the
instability region that is predicted for a standard neutron star
without enhanced damping mechanisms such as crust-core rubbing \cite{Lindblom:2000gu}.
This conclusion is based on accurate measurements of their frequencies
and reasonable estimates of their temperatures from  a
spectral fit to their quiescent X-ray radiation.
This means there must be additional damping mechanisms that suppress the
$r$-modes, such as structural aspects in the star's crust or the presence
in the star of novel phases such as a superfluid and/or superconductor,
or quark or hyperonic matter \cite{Haskell:2012}. Semi-analytic
expressions have now been derived that
make it possible to estimate the many uncertainties in our predictions
of the instability region, and it has been shown that
many of the relevant macroscopic observables are remarkably
insensitive to quantitative microphysical details, but can nevertheless
distinguish between qualitatively different forms of dense matter
\cite{Alford:2010fd}. A promising recent development is that by a detailed
understanding of the pulsar evolution the r-mode instability could
be connected to timing data of radio pulsars \cite{Alford:2012yn} using novel dynamic instability regions \cite{Alford:2013pma} that confirmed the above picture. The extensive data for these old and very stable sources is among the most precise
data in physics \cite{Manchester:2004bp}, which could allow a clear distinction
between different forms of dense matter. Finally, even if the saturation amplitude is very low so that r-modes would not affect the spindown evolution, they would still strongly heat the star \cite{Alford:2013pma} so that this scenario could be falsified by future x-ray observations.

\paragraph{X-ray bursts and the physics of the neutron star crust}
Nuclear matter at lower densities is found in the crust of neutron stars,
so surface phenomena that are sensitive to the behavior of the crust
give information about this form of matter. A better understanding of the
crust is also valuable for understanding the core, since
measurements of the temperature of the star's core are based on surface
phenomena (quiescent x-ray spectra) combined with models of heat transport
through the crust.

Observations of X-ray bursts are a valuable source of information about the
crust. The bursts arise from light elements, which are accreted onto the
neutron star surface, then gradually sink down
and at a critical pressure and density are
explosively converted into the heavier nuclei that form the star's crust
\cite{Chamel:2008LRR}. Detailed observations of the resultant
X-ray emission and subsequent cooling can then, 
in principle, be used to
constrain the parameters of theoretical models of the crust. This requires
calculations of the expected behavior using those models, including
detailed dynamical understanding of the various
transport properties within the crust.  Recent progress in this
direction includes the development of
effective theories for the crust \cite{Cirigliano:2011tj},
which describe the dynamics of the low energy degrees of freedom given by
electrons, lattice phonons and (in the inner crust)
Goldstone bosons arising from neutron superfluidity.
It is worth noting that understanding the nuclear fusion and capture
reactions underlying X-ray bursts 
is also relevant for understanding
the nucleosynthesis of heavy elements.


\subsection{Future directions} 
\label{sec:secF4}
Throughout the chapter we have pointed out various important future directions for nuclear physics and dense (nuclear and quark) matter
with applications to astrophysics.
 As far as the kaon--nucleon interaction is concerned new results
  of the $K$- and $\bar{K}$-- nucleons interaction at $\rho_0$ are going to be delivered by the GSI and JPARC experiments,
  together with high-statistics and high-precision measurement of all the collective observables from heavy--ion collisions at 10--30~AGeV
with CBM at FAIR which will test higher densities. Further experiments will be carried out mainly at JPARC to search for kaonic bound states
and the planned precision measurement of kaonic--deuterium with SIDDHARTA in FRASCATI and at JPARC will deliver conclusive information on
the isospin dependence of the kaon--nucleon scattering length. All these results together should deliver solid density dependent constraints for models 
that hypothesize the presence of antikaons in the inner-core of neutron stars.
On the other hand, we have pointed out the importance of hyperons in understudying dense and compact objects. New measurements
 of single and double-$\Lambda$ hypernuclei at the upcoming JPARC facility and more data on hyperon production in hadron-hadron collisions
 are expected from HADES and the future FAIR experiments. As a complement to these measurements,
  a more precise determination of the matter radius of neutron-rich nuclei at JLab and MAMI
are planned. These measurements impose important constraints on the thickness of the neutron crust in neutron stars and also 
on the boundaries of possible phase transition when going to the inner part of the star.

Possible ways towards a better theoretical understanding of dense matter include solving or mitigating the sign problem in lattice QCD as well as 
combining established theories (perturbative QCD, effective theory of CFL, hydrodynamics) or models (NJL) with more novel approaches (gauge/gravity correspondence).
A major future direction for the field of theoretical nuclear physics is
the continued push to obtain  {\it ab initio} calculations of nuclear
properties from QCD via lattice methods.  
\pagebreak
In the near future, we can expect more, and more precise, astrophysical data from compact stars (mass, radius, temperature, X-ray bursts, possibly gravitational waves)
which should be compared with predictions from QCD or effective theories/models (equation of state, transport properties).

\clearpage
\section[Chapa]{Vacuum structure and infrared QCD: confinement and 
chiral symmetry breaking \protect\footnotemark}
\footnotetext{Contributing authors: M.~Faber$^{\dagger}$, M.~I.~Polikarpov$^{\dagger}$, R.~Alkofer,
  R.~H\"ollwieser, V.~I.~Zakharov}
\label{sec:chapa}

The Standard Model of particle physics is formulated as a quantum
theory of gauge fields, describing weak and electromagnetic
interactions by electroweak theory and strong interactions by Quantum
Chromodynamics (QCD). Quantum field theories have a very well
developed perturbation theory for weak couplings. Processes of
elementary particles at high energies are characterized by asymptotic
freedom, by decreasing strength of the strong interaction with
increasing energy. This makes QCD a valuable
tool to investigate the strong interaction: In the weak coupling
regime of QCD the agreement of perturbative calculations with an
enormous number of measurements available is truly impressive. This is
the case despite the fact that QCD violates an essential basis of a
perturbative description, namely field-particle duality. This duality
assumes that each field in a quantum field theory is associated with a
physical elementary particle.

It is evident that hadrons are not elementary particles. The partonic
substructure of the nucleon has been determined to an enormous
precision leaving no doubt that the parton picture emerges from quarks
and gluons, the elementary fields of QCD. It is a well-known fact that
these quarks and gluons have not been detected outside hadrons, which
is known as confinement. Although this
hypothesis was formulated decades ago, the understanding of the
confinement mechanism(s) is still not satisfactory, see, e.g.,
\cite{Greensite:2011zz} for a recent discussion of the different
aspects of the confinement problem. 

Noting that the Kinoshita--Lee--Nauenberg theorem on infrared divergences 
\cite{Kinoshita:1962ur,Lee:1964is} applies to non-Abelian gauge
theories in four dimensions, order by order in perturbation theory
\cite{Kinoshita:1975ie}, a description of confinement in terms of
perturbation theory (at least in any na\"ive sense) is excluded. This
finding corroborates the simple argument that, since confinement 
arises at small momentum scales, the relevant values of the strong
coupling $\alpha_s$ are too large to justify a perturbative
treatment. Therefore nonperturbative methods are required to study the
dynamics of confinement. Furthermore, the quest for the confining
gluonic field configuration(s) has led to the anticipation that a
possible picture of confinement is directly related to the vacuum
structure of QCD.

In the first section of this chapter we will comment on our current
understanding of the QCD vacuum as it is obtained from lattice
gauge theory and the duality to string theory. In the second section we
briefly review some aspects of confinement and dynamical breaking of
chiral symmetry from the perspective of functional methods. In the
third section, additional aspects of chiral symmetry breaking as
inferred from lattice calculations are revisited.

\subsection{Confinement}\label{sec:secA1}

Confinement is a fascinating phenomenon which precludes observation of
free quarks in our world.  Mathematically, the property of confinement
is usually formulated in terms of the potential $V_{\bar{Q}Q}(R)$
between external heavy quarks.  In case of pure gluodynamics
(i.e.,\ without dynamical quarks) this potential grows at large
distances $R$, not allowing for the separation of the quarks. Lattice
simulations indicate that
\begin{equation}\label{potential}
\lim_{R\to\infty}{V_{\bar{Q}Q}(R)} = \sigma\cdot R+\frac{\text{const}}{R},
\end{equation}
where $\sigma$ is a constant and $\text{const}/R$ is the leading
correction.  Equation~(\ref{potential}) can be interpreted in terms of a
string, with tension $\sigma$ stretched between the heavy quarks.

Despite many years of intense efforts, there is no analytic solution
yet to the problem of confinement in case of non-Abelian gauge
interactions in four dimensions, i.e.,\ in the real world. There are
examples, however, of Abelian theories where confinement is
demonstrated analytically
\cite{Polyakov:1975rs,Polyakov:1976fu,Polyakov:1987ez}. What is common
to all these models is that confinement is associated with particular
vacuum field configurations, or with the structure of the vacuum.

Moreover, there is a strong correlation between confinement of some
charges and condensation of the corresponding magnetic degrees of
freedom \cite{Polyakov:1975rs,Polyakov:1976fu,Polyakov:1987ez}. For
example, in case of an Abelian charge, confinement of charged
particles is due to condensation of magnetic monopoles, and vice
versa.  An important example is provided by superconductors: it is a
charged field which is condensed and (external, heavy) magnetic
monopoles which are confined.  Thus, observation of confinement
probably indicates a kind of duality between electric and magnetic
degrees of freedom. According to the modern theoretical views, the
vacuum of non-Abelian theories is populated by condensed, or
percolating, magnetic degrees of freedom.  By studying the vacuum
structure we expect to observe the dual world of the magnetic degrees
of freedom.

Studies using lattice gauge theory have produced strong support of this
idea; for a review see, e.g., \cite{Greensite:2003bk}. The fact
that one can observe and make measurements on vacuum fluctuations is
far from trivial. Indeed, in the continuum-theory language one
usually subtracts vacuum expectations of various operators,
concentrating on the physical excitations.  In this respect, the
vacuum of the latticized space-time is rather similar to the ``vacuum''
of percolation theory.
\footnote{In its simplest form, the percolation theory introduces a
   probability $p$ of a link or of a plaquette to be ``occupied'' and
   studies the properties of the congregate of the occupied links or
   plaquettes. More generally, quantum geometry provides alternative
   formulations of field theories and of string theories in terms of
   trajectories and surfaces, respectively, for an introduction see,
   e.g., \cite{Ambjorn:1994yv}.} In the latter case the properties of
the vacuum condensates, in the so-called overheated phase, are
subject of theoretical predictions and measurements.  For example,
the phase transition is signaled by emergence of an infinite cluster
of closed trajectories at some critical value $p_\text{cr}$ . At
$p>p_\text{cr}, |p-p_\text{cr}|\ll 1$ the probability of a given link to belong
to the infinite cluster is still small:
\begin{equation}
    \theta_\text{inf.cluster} \sim (p-p_\text{cr})^{\alpha},
\end{equation}
where $\alpha> 0$.

Lattice studies of the vacuum of the Yang-Mills theories revealed the
existence of infinite clusters of trajectories and surfaces with
remarkable scaling properties:
\begin{eqnarray}\begin{aligned}\label{defects}
    &\theta_\text{link} \approx (\text{const})(\Lambda_\text{QCD} a)^3,\\
    &\theta_\text{plaquette} \approx (\Lambda_\text{QCD} a)^2~,
\end{aligned}\end{eqnarray}
where $\theta_\text{link}$ and $\theta_\text{plaquette}$ are the
probabilities of a given link, respectively, plaquette, to belong to
the infinite clusters, $a$ is the lattice spacing,
$\Lambda_\text{QCD}$ is the hadronic scale, $\Lambda_\text{QCD}\sim
100~{\rm MeV}$. Moreover, the trajectories are contained in the
surfaces~\cite{Ambjorn:1999ym,deForcrand:2000pg,Gubarev:2002ek}. These
lines and surfaces can be called defects of lower dimension. Indeed,
the trajectories represent $D=1$ defects in the Euclidean $D=4$ space
and surfaces represent $D=2$ defects. Removal of the defects, which
occupy a vanishing fraction of the lattice in the continuum limit
$a\to 0$ results in the loss of confinement (and of the spontaneous
breaking of chiral symmetry). In terms of the non-Abelian fields, the
defects are associated with an excess of action and topological
charge.

A theory of confinement in terms of field-theoretic defects is an
unfinished chapter. Although truly remarkable observations were
made in lattice studies and illuminating theoretical insights were
suggested there is no concise picture yet. Elaborating such a picture
would be of great importance for the field theory in general. It is
like going from the Hooke's law for continuum media to a theory of
dislocations where the same law arises only after averaging over many
defects. Whether a similar step can indeed be made in the case of
confinement remains an open question, to be addressed in future.

So far we discussed temperature $T=0$ and Euclidean space-time.  Thus,
the defects percolate in all four dimensions. A remarkable phenomenon
occurs at the temperature of the deconfining phase transition, $T_c$:
the defects become predominantly parallel to the time direction, while
still percolating in the three spatial dimensions, see, in
particular~\cite{Greensite:2003bk,Engelhardt:1999fd}. On the lattice,
one studies geometrically defined asymmetries like
\begin{equation}
    A = \frac{N_{\tau}-\frac{1}{3}N_{x,y,z}}{N_{\tau}+\frac{1}{3}N_{x,y,z}},
\end{equation}
where $N_{\tau}$ is the number of links (belonging to the 1D defects
above) looking in the Euclidean time direction and $N_{x,y,z}$ is the
number of links looking in one of the spatial directions.  The
asymmetry $A\approx 0$ below $T_c$ and $A\approx 1$ at temperatures
above $T_c$.

It is worth mentioning that lower-dimensional defects in field
theories have been discussed in many papers. One of the best known and
early examples is \cite{Callan:1984sa}.  Moreover, the quantized
vortices in rotating superfluids, known for about seventy years, can be
thought of as 1D defects. Indeed, within the hydrodynamic
approximation vortices introduce a singular flow of the liquid, with
the singularity occupying a line, the axis of the vortex.

However, in the particular case of Yang-Mills (YM) theories the only
example of nonperturbative fluctuations which can be studied in
quasiclassical approximation is provided by instantons, and this
example does not help to interpret the lattice data mentioned above.
Probably, this is one of the reasons why the observations
(Eq.~(\ref{defects})) and their extensions did not have much feedback to the
continuum theory. Also, the algorithm for the search of defects is
formulated in a specific lattice language and this makes the
interpretation of data difficult. Actually, in case of temperature
$T>T_c$ a well known example of a field-theoretic
operator exists, which might serve as a field theoretic image of the
(Euclidean) time-independent defects. We have in mind the Polyakov
line, or path-ordered exponent:
\begin{equation}\label{polyakov}
L=\mathrm{tr} P\big(\exp{i\int_0^{1/T}A_0({\bf x},\tau)d\tau}\big)~,
\end{equation}
where $A_0$ is the gauge potential and $\tau$ is the Euclidean time,
$1/T \ge \tau\ge 0$. The loop is an extended object defined in four
dimensions. However, since it is parallel to the Euclidean time, we
have in fact a 3D object. Note that condensation of the Polyakov lines
at $T>T_c$ has been discussed in many papers, for a review see, e.g.,
\cite{Dumitru:2010mj}. If this is true, then the Polyakov lines could
be considered as an example of lower-dimensional defects in the
language used here.

 It is worth mentioning that in case of supersymmetric gauge theories,
 with elementary scalar fields, the theory of defects is developed
 much further, for a review see, e.g., \cite{Tong:2008qd}. Moreover,
 some features of the defects present in SUSY YM theories are in
 striking accord with the lattice observations concerning pure
 Yang-Mills theories (with no elementary scalar fields). In
 particular, in both cases the fields of monopoles are locked onto the
 magnetic surfaces (defined independently):
\begin{equation}
\epsilon_{ijk}H^i\Sigma^{jk} = 0~,
\end{equation}
where $H^i$ is the magnetic field of monopoles and $\Sigma_{jk}$ are
surface elements, constructed from the tangent vectors.  However,
these two approaches -- lattice studies of the defects in pure
Yang-Mills case and theoretical studies of defects in the
supersymmetric case -- have been developing independently, with almost
no interaction between the corresponding mini-communities.

A new chapter in the theory of gauge interactions with strong coupling
was opened with the formulation of the Maldacena duality, for a review
see, e.g., \cite{Polchinski:2010hw}. It was forcefully argued that the
infrared completion of gauge theories is provided by string theories
with extra dimensions and nontrivial geometry. The ordinary $4D$ space,
where the gauge theories are defined, is assumed to constitute a
boundary of the multi-dimensional space.  There are certain rules to
relate the stringy physics in the extra dimensions to the physics of
gauge theories in ordinary four dimensions. For this reason one talks
about the ``holographic'' approach to gauge theories. Exact results
apply, however, only to supersymmetric gauge theories (with elementary
scalars) in the limit of a large number of colors, $N_c\to \infty$.

In case of pure Yang-Mills theories, with no elementary scalar fields,
the strongest claim was made quite some time
ago~\cite{Witten:1998zw,Sakai:2004cn} and not much progress has been
made since then. Namely, it was shown that in the {\it far infrared
  limit}, i.e.,\ formally in the limit
$$R \gg \Lambda_\text{QCD}^{-1},$$ the pure, large-$N_c$ Yang-Mills theory
 belongs to the same universality class as a particular string
theory, specified in~\cite{Witten:1998zw,Sakai:2004cn}. However, this
very string theory in the ultraviolet is dual to a supersymmetric {\it
  five-dimensional} Yang-Mills (YM) theory which is radically different
from the YM theory in $4D$ which we are interested in. Thus, only
large-distance, or nonperturbative physics of the gauge theories can
be captured within this model. On the other hand, the separation
between the perturbative and nonperturbative contributions is actually
not uniquely defined at any distances, large distances included.

Nevertheless, it is just in the case of vacuum defects that the
holographic approach can be tested. Indeed, from the lattice
simulations we know that there are percolating defects which survive,
therefore, in the far infrared. Remarkably, the holographic
approach based on~\cite{Witten:1998zw,Sakai:2004cn} is able to explain the
basic observations concerning the vacuum structure of pure Yang-Mills
theories.

A nonexhaustive list of theoretical predictions looks as follows:
\begin{itemize}
\item{The model incorporates, without any tuning, the confinement
  phenomenon~\cite{Witten:1998zw,Sakai:2004cn}, i.e.,\ reproduces the
  large-distance behavior of the heavy-quark potential
  (Eq.~(\ref{potential})) with $\sigma \sim \Lambda_\text{QCD}^2$.

Geometrically, confinement is related to the properties of a fifth,
$z$-direction. The physical meaning of the coordinate $z$ is that it
is conjugate to the resolution of measurements. In more detail, $z\to
0$ corresponds to the ultraviolet limit, or to measurement with fine
resolution. Larger values of $z$ correspond to momentum transfer of
order $\Delta p \sim 1/z$. One of the basic geometric properties of
the theory considered is the existence of a horizon,
$$z \le z_H \sim \Lambda_\text{QCD}.$$
One can show that the existence of
this horizon in the $z$-direction implies confinement. Moreover, there
is indeed a string stretched between the heavy quarks. } \item{The
  stringy completion of gauge theories drastically extends the number
  of topologically stable classical solutions, see, e.g.,
  \cite{Gorsky:2009me,Chernodub:2009bv} and references therein. The
  geometric reason is that the model~\cite{Witten:1998zw,Sakai:2004cn}
  has two compact dimensions, Euclidean time (as usual) and one extra,
  sixth dimension $\theta$. Lower-dimensional defects correspond to
  D0-, D2-, D4-branes.\footnote{``D'' for ``Dirichlet'': points (D0), surfaces (D2),
  and $4D$ hypersurfaces (D4) on which strings end.}
  If the branes wrap around at least one
  of the compact dimensions, the corresponding solutions are 
  stable. The ordinary instantons correspond to D0-branes wrapped
  around the $\theta$-direction~\cite{Bergman:2006xn}. Moreover, it is
  a general property that wrapping around the $\theta$-coordinate
  implies a nontrivial topological charge of the defect in terms of the
  Yang-Mills fields.

In particular, there are D2 branes wrapped around $\theta$ which would
match topologically charged strings in the vacuum of YM
theories~\cite{Gorsky:2009me,Chernodub:2009bv}.}
\item{At low temperatures, the D2 branes just discussed are expected
  to percolate.  The geometrical reason is that the radius of the
  sixth dimension, $R_{\theta}$ depends in fact on the
  $z$-coordinate. Moreover, the crucial observation is that
$$R_{\theta}(z_H) = 0.$$ Since the action associated with any defect
  is proportional to its (D+1) dimensional volume, the action of
  defects wrapped around the $\theta$-coordinate vanishes at $z=z_H$,
  or in the infrared. This implies vanishing of the action of the
  D2-branes and this makes plausible their
  percolation. } \item{Holography predicts a phase transition to
  deconfinement at some $T_c$~\cite{Witten:1998zw,Sakai:2004cn}. In
  the geometric language this is a so-called Hawking-Page transition
  \cite{Hawking:1982dh}, i.e.,\ a transition between two geometries in
  general relativity.  In the case considered, there are two similar
  compact directions, $\theta$- and $\tau$-directions.  Below $T_c$,
\begin{eqnarray}\begin{aligned}\label{theta}
&R_{\tau}=(2\pi T)^{-1},\\ &R_{\theta}(z=0)=\text{const}_{\theta},R_{\theta}(z_H)=0,
\end{aligned}\end{eqnarray}
while above $T_c$ the roles of the two compact dimensions are
interchanged:
\begin{eqnarray}\begin{aligned}\label{tau}
&R_{\theta}=\text{const}_{\theta}, \\ &R_{\tau}(z=0)=(2\pi T)^{-1},R_{\tau}(z_H)~=0,
\end{aligned}\end{eqnarray}
and the phase transition occurs at $(2\pi T_c)^{-1} = \text{const}_{\theta}.$
} \item{According to the holographic picture, the deconfining phase
  transition can be viewed as dimensional reduction at finite
  temperature, $T=T_c$ \cite{Gorsky:2009me,Chernodub:2009bv}:
\begin{eqnarray}\begin{aligned}\label{3d}
(4D ~\text{percolation}, T<T_c)~\rightarrow \\ \rightarrow (3D~\text{percolation},
    T>T_c).
\end{aligned}\end{eqnarray}
Indeed, because of the vanishing of the radius of the time circle at
the horizon, $R_{\tau}(z_H)=0$, see Eq.~(\ref{tau}), percolating defects
are those which are wrapped around the compact $\tau$-direction at
$T>T_c$. The wrapping around the $\tau$-direction implies in turn that
the nonperturbative physics in the infrared becomes three-dimensional,
see discussion around Eq.~(\ref{polyakov}) above.  }
\item{Generically, the holographic models predict a low value of the
  shear viscosity $\eta$~\cite{Kovtun:2004de}:
\begin{equation}\label{viscosity}
\eta/s = 1/4\pi.
\end{equation}
This prediction is shared by the model considered, see, e.g.,
\cite{Chernodub:2010mm}. }
\end{itemize}

Thus, we can summarize that the holographic model based
on~\cite{Witten:1998zw,Sakai:2004cn} does, in fact, reproduce all the
basic observations concerning defects in pure Yang-Mills theories.
However, the predictions are mostly qualitative in nature. Since the
holographic model does not work in the ultraviolet, it is not
possible to fix scales, such as the tensions associated with the
defects.

Also, there is no established one-to-one correspondence between
defects inherent to the holography and defects observed on the
lattice. The reason is the proliferation of the defects in the
holographic model. At the moment, there are a few possibilities open
to accommodate the defects known from the lattice studies and new,
not-yet-observed defects are predicted.  Let us mention in this
connection that it is only recently that it was observed that the
defects called thermal monopoles in the lattice nomenclature are in
fact dyons~\cite{Bornyakov:2011th}:
\begin{equation}
|{\bf E}^a| = |{\bf B}^a|~,
\end{equation}
where ${\bf E}^a,{\bf B}^a$ are the color electric and magnetic fields
associated with the thermal monopoles.

In recent years it was recognized that the phenomenon of confinement
has much in common with superfluidity and superconductivity. This
similarity is most explicit in holographic models; for reviews of
applications of holography to condensed matter systems see, e.g.,
\cite{Sachdev:2010ch,Hartnoll:2009sz}.  Namely, basically similar
holographic models describe confinement in four dimensions and
superfluidity (superconductivity) in three dimensions.

From the technical point of view, however, this change in
dimensionality of the space considered is quite crucial. Namely, the
difficulty in solving the confinement problem is that it is reduced to
string theory in terms of defects, and the string theory in $4D$ is
poorly developed. Reduction by one dimension transforms 2D defects
into 1D defects. The one-dimensional defects, in turn, correspond to
field theory in the language of the quantum geometry, and  field
theory is much better understood than  string theory. This is the
reason why in case of superfluidity and superconductivity the
holographic approach allows to get more detailed predictions than in
case of the confinement in $4D$. Another implication of this simple
counting of dimensions is that in the deconfining phase one can expect
to find superfluidity. Indeed, in the deconfining phase the $4D$
nonperturbative physics becomes 3D physics, see Eq.~(\ref{3d}). And,
indeed, the holographic models predict relation (Eq.~(\ref{viscosity}))
which is, according to the modern views~\cite{Kovtun:2004de}, the
lowest possible value of the shear viscosity.

In the deconfining phase one expects to find a
dissipation-free electric current as well.  We have in mind the so
called chiral magnetic effect
(CME)~\cite{Kharzeev:2007jp,Fukushima:2008xe,Son:2009tf}. The effect
is the induction of electric current flowing along the external
magnetic field in the presence of a nonvanishing chiral chemical
potential $\mu_5$:
\begin{equation}\label{current}
\bm{j}_\text{el} = \frac{\mu_5}{2\pi^2}\bm{B}_\text{ext}.
\end{equation}
There is an exciting possibility that the effect of charge separation
with respect to the collision plane observed in experiments at
RHIC~\cite{Voloshin:2008jx,Wang:2012qs} and at
ALICE~\cite{Abelev:2012pa,Hori:2012hi} is a
manifestation of a (fluctuating) chiral chemical potential.
For further discussion, see Chapter~\ref{sec:chapd}.

From the theoretical point of view, it is most exciting that the
current (Eq.~(\ref{current})) is dissipation free and can exist in
equilibrium, provided that the chiral limit is
granted~\cite{Kharzeev:2011ds,Alekseev:1998ds}.  In this respect, the
CME effect is similar to superconductivity.  On the other hand, the
current (Eq.~(\ref{current})) is carried by fermionic degrees of freedom and
there is, unlike the superfluidity case, no coherent many-particle
state. In this sense Eq.~(\ref{current}) rather describes ballistic
transport, i.e., collisionless transport along the external magnetic
field, and without any driving force. Unlike the ordinary ballistic
transport (which refers simply to propagation at distances less than
the mean free path), Eq.~(\ref{current}) is to be quantum and
topological in nature. An explicit quantum state responsible for the
dissipation-free flow (Eq.~(\ref{current})) has not been constructed yet.

Discussion of the dissipationless nature of the CME brings us to
mention the, probably, most dramatic shift of direction of our studies
which is taking place nowadays. We mean exploration of
condensed-matter systems which are similar in their properties to
relativistic chiral-invariant field theories. The spectrum of 
fermionic excitations in these systems is linear in the momentum
$p$:
\begin{equation}
\label{spectrum}
\epsilon \approx v_s\cdot p,
\end{equation}
where $v_s$ is the fermionic speed of sound. The spectrum
(Eq.~(\ref{spectrum})) is similar to the spectrum of a superfluid. However,
now it refers to fermions. The implication is that in such materials
there should exist a kind of chiral superconductivity, exhibited by
Eq.~(\ref{current}). Moreover, for the condensed-matter systems the
condition of validity of the chiral limit can be satisfied to a much
better accuracy than in case of QCD. This point could be crucial for
applications. The best known example of such ``chiral materials'' is
graphene. The analog of the chiral magnetic effect is expected to be
observed in semi-metals which are also chiral materials; for details
and references see, in particular, \cite{Kharzeev:2012dc}.

Apart from the CME, there are other interesting phenomena which are
expected to happen in strongly interacting gauge theories in external
magnetic fields.  In particular, it was argued in \cite{Chernodub:2010qx} that at some critical value of the
external magnetic field there is a phase transition of the ordinary
vacuum of QCD to a superconducting state. The estimate of the
critical field is
\begin{equation}
(B_\text{ext}^\text{crit})^2 \approx (\text{0.6~GeV})^2.
\end{equation}
 Moreover, one expects that the new superconducting state represents a
 lattice-like structure of superconducting vortices,
 see~\cite{Braguta:2013uc} and references therein.

To summarize, the most intuitive model of confinement, the so-called
dual superconductor, appealed to the analogy with
superconductivity~\cite{Polyakov:1975rs,Polyakov:1976fu,Polyakov:1987ez,
Nambu:1974zg,'tHooft:1975pu,Mandelstam:1974pi,Kronfeld:1987vd,Kronfeld:1987ri}.
However, there is no complete implementation of this analogy so far because
confinement in $4D$ gauge theories is rather related to string theory
which is not developed enough yet. This relation to strings is
manifested especially clear once one turns to study defects
responsible for the
confinement~\cite{Greensite:2003bk,Ambjorn:1999ym,deForcrand:2000pg,Gubarev:2002ek}.
However, above the critical temperature, $T>T_c$, the nonperturbative
physics of the gauge theories comes much closer to the physics of
superconductivity~\cite{Chernodub:2010mm,Gorsky:2009me,Chernodub:2009bv}. This
time it is a relativistic, or chiral
superconductivity~\cite{Kharzeev:2007jp,Fukushima:2008xe,Son:2009tf}
which is a new chapter in theoretical physics. The
phenomenon of the chiral superconductivity seems inherent not only to
relativistic field theories but to some condensed-matter systems, like
graphene and semimetals, as well.

\subsection{Functional methods}\label{sec:secA2}

As we have seen above, the confining field configurations are, at
least to our current understanding, given by lower dimensional
defects. Furthermore, the long-range correlations in between these
defects are of crucial importance. This makes confinement a
phenomenon based on the behavior of glue in the deep infrared.

It is evident that in such a situation, at least to complement
lattice gauge theory, continuum methods are highly
desirable. As stated in the introduction to this chapter, a
perturbative description of confinement is excluded, leaving us with
a need for nonperturbative tools in quantum field theory. One of
the very few such approaches is given by the one of functional
methods. We use here this term summarizing all those 
nonperturbative methods which are based on generating functionals
and/or Green functions. The basic idea is to rewrite exact
identities in between functionals such that they become amenable to an
exact treatment in certain kinematical limits and controlled truncation
schemes for general kinematics. The truncated set of equations is then
subsequently solved numerically. Typically the cost of obtaining
numerical solutions is then orders of magnitude less than for a
lattice Monte Carlo calculation.  In the last decade several of these methods have
been used to study the infrared behavior of QCD, amongst them most
prominently Dyson-Schwinger equations (see, e.g.,
\cite{Alkofer:2000wg,Binosi:2008qk}), exact renormalization group
equations (see, e.g., \cite{Pawlowski:2005xe}), $n$-particle
irreducible actions (see, e.g., \cite{Berges:2004pu}), and the
so-called pinch technique (see, e.g., \cite{Binosi:2009qm}). Several
recent investigations exploit possible synergies and use different
functional methods in a sophisticatedly combined way. Before going
into details a few general remarks are in order.

The challenge to describe confinement adequately is given by the fact
that the physical Hilbert space of asymptotic (hadron) states does not
contain any states with particles corresponding to the elementary
fields in QCD, {i.e.}, quarks and gluons.  For a satisfactory
description of color confinement within local quantum field theory,
the elementary fields have to be disentangled completely from a
particle interpretation. Within (nonperturbative) gauge field theories
the elementary fields implement locality. Those fields are chosen
according to the underlying symmetries and charge structure and
reflect only indirectly the empirical spectrum of
particles. Furthermore, to circumvent the production of colored states
from hadrons, strong infrared singularities are anticipated.  This
expectation is supported by the fact that the absence of unphysical
infrared divergences in Green functions of elementary fields would
imply colored quark and gluon states in the spectrum of QCD to every
order in perturbation theory \cite{Poggio:1976qr}. And, even more
directly, the linearly rising static potential, discussed in the last
section, indicates a strong ``$1/k^4$-type'' infrared singularity in
four-point functions of heavy colored fields.

It will be useful for the following discussion to revisit the formal
argument for the nonperturbative nature of the confinement scale in
four-dimensional gauge field theories: In the chiral limit QCD is
classically scale invariant. It therefore needs to dynamically
generate the physical mass scale related to confinement.  Furthermore,
it is an asymptotically free theory with a Gaussian ultraviolet
fixed point, and its renormalization group (RG) equations, in the
presence of such a mass scale, imply (at least in expressions for
physical quantities) an essential singularity in the coupling at $g
=0$. The dependence of the RG invariant confinement scale on the
coupling and the renormalization scale $\mu$ near the ultraviolet
fixed point is determined by
\begin{equation}
  \Lambda = \mu \exp \left( - \int ^g \frac {dg'}{\beta (g')} \right)
  \stackrel{g\to 0}{\rightarrow } \mu \exp \left( - \frac 1
           {2\beta_0g^2} \right)\,,
  \label{Lambda}
\end{equation}
where with asymptotic freedom $\beta_0>0$. Since all RG invariant
mass scales in QCD at the chiral limit will exhibit the behavior
(Eq.~(\ref{Lambda})) up to a multiplicative constant, this has, besides the
inadequacy of a perturbative expansion for the problem at hand,
another important consequence: in the chiral limit the ratios of all
bound state masses do not depend on any parameter.

The objectives of the application of functional methods to QCD and
hadron physics can be typically separated into two issues: One is the
description of hadrons and their properties from elementary Green
functions. This is described in Chapter~\ref{sec:chapb}. The other
is understanding of fundamental implications of QCD as, e.g.,
dynamical breaking of chiral symmetry or the axial anomaly. One
should note that the formation of bound states with highly
relativistic constituents provides hinge between the two types of
investigations. But most prominently, a possible relation of the
phenomenon of confinement to the infrared behavior of QCD amplitudes
has been the focus of many studies.

Although one aims at the calculation of physical and therefore
gauge-invariant quantities, functional methods (based on the Green 
functions of elementary fields) are required by mere definition to fix
the gauge. Most investigations have been performed in Landau, Coulomb
or maximally Abelian gauge. The reasons for the respective choices are
quite distinct. As studies in Landau gauge are in the majority, we
discuss them first.

Some relations between different confinement scenarios become most
transparent in a covariant formulation which includes the choice of a
covariant gauge. First, we note that covariant quantum theories of
gauge fields require indefinite metric spaces. Abandoning the
positivity of the representation space already implies to give up one
of the axioms of standard quantum field theory. Maintaining the much
stronger principle of locality, gluon confinement then naturally
relates to the violation of positivity in the gauge field
sector. Therefore one of the main goals of corresponding lattice and
functional studies of the Landau gauge gluon propagator was to test
them for violation of positivity. As a matter of fact, convincing
evidence has been found for this property, see, e.g.,
\cite{Maas:2011se} for a recent review.

Noting that positivity violation beyond the usual perturbative
Gupta-Bleuler or Becchi-Rouet-Stora--Tyutin (BRST) quartet
mechanism\footnote{See, e.g, Chapter~16 of \cite{Peskin:1995ev}} has
been verified, the question arises how a physical positive-definite
Hilbert space can be defined. If, in Landau  gauge QCD, BRST symmetry is 
softly broken  (as some recent investigations indicate, see below) the 
answer is unknown. On the other hand, for an unbroken BRST
symmetry the cohomology of the BRST charge provides a physical Hilbert
space as has been shown more than three decades ago
\cite{Kugo:1979gm}. Given some well-defined assumptions, known as
Kugo-Ojima confinement criteria, it has been subsequently proven that
in this scenario the color charge of any physical state must
vanish. As a corollary to these confinement criteria, it is shown that
then the ghost propagator diverges more than a massless pole
\cite{Kugo:1995km}. Such a behavior is exactly the one found in one
type of solutions of Dyson-Schwinger and Exact Renormalization Group
studies. It now goes under the name of the scaling solution and is
characterized by an infrared enhanced ghost and an infrared vanishing
gluon propagator with correlated infrared exponents, see, e.g.,
\cite{Maas:2011se} and references therein.

Lattice calculations of Landau gauge propagators lead, however,
seemingly to another conclusion, namely an infrared finite gluon
propagator and a simple massless ghost propagator,\footnote{Here we
  consider only the case of four spacetime dimensions. Note that for
  two spacetime dimensions there is only the scaling solution found in
  the continuum as well as on the lattice
  \cite{Cucchieri:2012cb,Huber:2012zj,Zwanziger:2012xg}.}  see, e.g.,
\cite{Sternbeck:2005tk,Bogolubsky:2009dc,Oliveira:2012eh,Bogolubsky:2013rla}.
Functional equations, on the other hand, also have such a type of
solution, and as matter of fact, it turns out that these are actually
a whole one-parameter family of solutions depending on the chosen
renormalization constant for one of the propagators
\cite{Aguilar:2008xm,Fischer:2008uz,Boucaud:2011ug}. These are called
either decoupling or massive solutions. The latter name should,
however, be understood with some care. Of course, in Landau gauge the
gluon propagator, although infrared finite, stays transverse. No
degenerate longitudinal component of the gluon develops as it is the
case in the Higgs phase of Yang-Mills theory with a massive gauge
boson: Also for the decoupling or ``massive'' solution the gluon stays
in the massless representation of the Poincar{\'e} group with only two
polarizations attributed, and as already true on the perturbative
level the timelike and the longitudinal gluon stay in the fundamental
BRST quartet together with the Faddeev-Popov ghost and the antighost.

The relation between the one scaling and many decoupling solutions
can be understood most easily if one chooses to renormalize the ghost
propagator at vanishing four-momentum: Denoting by $G(p^2)$ the
dressed ghost renormalization function, a nonvanishing choice for
$G^{-1}(0)$ leads to one of the decoupling solutions, choosing
$G^{-1}(0)=0$ to the scaling solution which then identifies itself as
one of the two endpoints of the one-parameter family of solutions.

Recently, a verification of the multitude of propagator solutions has
been obtained within a lattice calculation \cite{Sternbeck:2012mf}: On
the lattice it turns out that the choice of eigenvalues of the
Faddeev-Popov operator in between different Gribov copies of the same
configurations (and all of them fixed to Landau gauge!)\ provides the
discrimination in between different members of the decoupling solution
family. Therefore one can conclude that the existence of several
solutions of functional equations is related to the difficulties of
fixing nonperturbatively the gauge in the presence of the Gribov
ambiguity as has been already speculated in \cite{Maas:2009se} based
on lattice calculations, where Gribov copies have been chosen on the
basis of the infrared behavior of the ghost propagator.

Another well-investigated topic for the realization of confinement is
the so-called Gribov-Zwanziger scenario. The generic idea is hereby to
take into account only one gauge copy per gauge orbit. Within the
state of all gauge field configurations the ones fulfilling the
na\"ive Landau gauge, {i.e.\/}, the transverse gauge fields, form a
``hyperplane'' $\Gamma = \{A:\partial \cdot A=0\}$. A gauge orbit
intersects $\Gamma$ several times and therefore gauge fixing is not
unique. The so-called minimal Landau gauge, obtained by minimizing
$||A||^2$ along the gauge orbit, is usually employed in corresponding
lattice calculations. It restricts the gauge fields to the Gribov
region
\begin{eqnarray}
    \Omega & = & \{A: ||A||^2 \, \text{minimal}\} \nonumber \\
        & = & \{A: \partial \cdot A=0, -\partial \cdot D (A) \ge 0 \},
\end{eqnarray}
where the Faddeev operator $-\partial \cdot D (A)$ is strictly
positive definite.  Phrased otherwise, on the boundary of the Gribov
region, the Gribov horizon, the Faddeev operator possesses at least
one zero mode.  Unfortunately, this is not the whole story. There are
still Gribov copies contained in $\Omega$, therefore one needs to
restrict the gauge field configuration space even further to the
region of global minima of $||A||^2$, which is called the fundamental
modular region.  Usually, a restriction to the fundamental modular
region can be obtained in neither lattice calculations nor functional
methods.  Note, however, that the restriction to the first Gribov
region $\Omega$ is fulfilled when using functional equations as long
as the ghost propagator does not change sign. To include contributions
from field configurations which are exactly the ones being in $\Omega$
leads to the requirement that the ghost propagator is more singular in
the infrared than a simple pole, i.e., one obtains the same condition
as in the Kugo-Ojima approach.

In \cite{Dudal:2009xh} the relation of the Kugo-Ojima to the
Gribov-Zwanziger scenario has been investigated showing that the
occurrence of the same condition is not at all accidental but points to
a deep connection in between these scenarios. Besides this positive
result the authors of \cite{Dudal:2009xh} obtained that conventional
BRST symmetry is softly broken by the introduced boundary
terms. Unfortunately, it is not clear yet whether some modified
symmetry might be left unbroken. If not, one has to face the
disturbing fact that an analysis of the Kugo-Ojima picture leads to a
contradiction to one of its basic prerequisites.

To allow within the Gribov-Zwanziger scenario for a less divergent
ghost and an infrared nonvanishing gluon propagator the so-called
refined Gribov-Zwanziger picture has been developed. Some details can
be found in the recent review \cite{Vandersickel:2012tz} and
references therein. Although the refined Gribov-Zwanziger scheme
yields propagators in qualitative agreement with lattice results, it
has not contributed to the question whether and, if so, how, the
infrared behavior of Green functions is related to confinement. It is
probably fair to say that with respect to the Kugo-Ojima and
Gribov-Zwanziger pictures of confinement in linear covariant gauges
the current understanding is inconclusive.  In order to make progress
several questions need to be answered: First, is BRST softly or
dynamically broken in Landau gauge QCD? Second, are there other
symmetries similar to BRST which need or should be considered? Third,
is the multitude of possible infrared behaviors of QCD Green functions
a failure of the employed methods, or are all these solutions correct
ones in the sense that their existence is an issue of complete
nonperturbative gauge fixing and all of them lead to identical
gauge-invariant observables?

Much work on functional approaches to Coulomb gauge QCD has been
performed over the last decades, see, e.g., \cite{Watson:2011kv} and
references therein.  On the one hand, there is no confinement without
Coulomb confinement \cite{Zwanziger:2002sh}, and the strong infrared
divergence of the time component of the gluon propagator seems to
offer a relatively easy understanding of confinement. On the other
hand, functional methods for Coulomb gauge QCD have proven to be
utterly complicated and no definite conclusion can be reached yet.
Given the fact that lattice results leave room for (but also do not
show) the analogue of the Gribov-Zwanziger scenario, it seems
worthwhile to continue the corresponding efforts.

As explained in detail in the previous section, an intriguing scenario for
confinement is the dual-superconductor picture. Intimately related to
this picture is the use of the so-called maximally Abelian gauge. The
corresponding gauge condition is such that it maximizes the diagonal
part of the gluon field.\footnote{In mathematical terms, it maximizes
  the elements of the gluon field being in the Cartan subalgebra which
  then also gave the name to this gauge.} This gauge keeps
Poincar{\'e} invariance but breaks the covariance under gauge
transformations.  Quite general arguments allow to establish a
connection between confinement, on the one hand, and the dominance of
the Abelian gluon field components in the deep infrared on the other
hand. Therefore it is encouraging that this picture has been verified
in lattice calculations \cite{Gongyo:2013sha} and in an exact infrared
analysis of combined functional equations \cite{Huber:2009wh}.
Nevertheless, the provided evidence is not (yet) compelling. Progress
has been made, however, with respect to the understanding of the
Kugo-Ojima scenario in the maximally Abelian gauge: Whereas a na\"ive
implementation of the Kugo-Ojima criteria fails
\cite{Suzuki:1983cg,Hata:1992dn}, a generalization of this confinement
scenario to Coulomb and the maximally Abelian gauge has been
constructed recently \cite{Mader:2013}.

All these studies described so far in this subsection concentrated on the
Yang-Mills sector of QCD. They provide essential insights into color
confinement (and hereby especially gluon confinement) but put the
question of quark confinement aside. In several recent studies---see
\cite{Fister:2013bh} and references therein---the question of quark
confinement has been addressed by computing the Polyakov loop  potential
from the fully dressed primitively divergent  correlation functions. For
static quarks with infinite masses the free energy of a single quark will
become infinite as the following Gedanken experiment shows: Removing  the
antiquark in a colorless quark-antiquark pair to infinity requires an
infinite amount of energy for a confined system. The gauge field part of
the related operator is the Polyakov loop (Eq.~(\ref{polyakov})), and the free
energy of the ``single'' quark state, $F_q$, can be expressed with help of
the expectation value of $L$
\begin{equation}
\langle L \rangle \propto \exp (-F_q/T).
\end{equation} 
Therefore, $\langle L \rangle$ is strictly vanishing in the confined phase
(but will be nonzero in the deconfined phase). For gauge groups $SU(N_c)$ 
this relates the question of confinement to the center symmetry $Z_{N_c}$:
In the center-symmetric phase the only possible value for  $\langle L
\rangle$ is zero, and one necessarily has confinement. Exploiting (i)
$L[\langle A_0 \rangle]\ge \langle L[A_0] \rangle$, and (ii) the fact that 
the full effective potential related to $L[\langle A_0 \rangle]$ can be 
calculated in terms of propagators in constant $A_0$ background
within functional methods, allows to derive a criterion for quark
confinement in terms of the infrared behavior of the ghost and gluon 
propagators, see \cite{Fister:2013bh} and references therein. Corresponding
studies have been performed in the Landau gauge, the Polyakov gauge and in
the Coulomb gauge hereby confirming the gauge independence of the formal
results.\footnote{For recent lattice calculations of the Polyakov loop
potential see \cite{Diakonov:2012dx,Greensite:2012dy}.} The main result of
these studies can be summarized as follows: infrared suppression of
gluons but nonsuppression of ghosts is sufficient to confine static
quarks.

A similar link of confinement to the infrared behavior of gauge-fixed
correlation functions has been established in the last years with the help
of so-called dual order parameters, see e.g.,\ \cite{Fischer:2012vc} and
references therein. These order parameters are, on the one hand, related to
the spectral properties of the Dirac operator \cite{Gattringer:2006ci} and
therefore tightly linked to the quark correlation functions, on the other
hand, they represent ``dressed'' Polyakov loops.  Corresponding
calculations have been extended to fully dynamical 2- and 2+1-flavor
QCD at nonvanishing temperatures and densities.  It turns out that 
the different classes of here discussed order
parameters are closely related to each other, for a discussion see e.g.,\ 
\cite{Braun:2009gm}.

Summarizing recent work on this topic one can conclude that, both on a
quantitative and a qualitative level, confinement criteria have been
developed further and one has gained more insight into the relation of 
confinement to the infrared properties of QCD with the help of functional
approaches. These studies, however, provide only a basis to tackle the
problem of the dynamical origin of confinement.

In this context one should note that 
the above discussion does not touch on the origin  
of the linearly rising potential in between static
quarks. First of all, one has to realize that the question
whether and how such a linearly rising static potential can be encoded
in the $n$-point Green functions of quenched QCD is a highly
nontrivial one. In lattice gauge theory, this potential is extracted
from the behavior of large Wilson loops. Due to the exponentiation of
the gluon field the Wilson loop depends on infinitely many $n$-point
functions.  Therefore, the observed area law of the Wilson loop does
not provide a compelling reason why a finite set of $n$-point
functions should already lead to confinement in the sense of a
linearly rising potential. On the other hand, one can show that an
infrared singular quark interaction can provide such a linearly rising
potential. The typical starting point for such an investigation is the
hypothesis that some tensor components of the quark four-point
function diverge like $1/k^4$ for small exchanged momentum $k$. If
such an infrared divergence is properly
regularized~\cite{Gromes:1981cb} and then Fourier transformed, it
leads, in the nonrelativistic limit, to a heavy quark potential with
a term linear in the distance $r$, {i.e.}, to the anticipated linearly
rising potential. This provides an example how confinement can be
encoded already in a single $n$-point function.

With the Landau gauge gluon being confined (instead of being
confining) it is immediately clear that in Landau gauge QCD the
quark-gluon vertex function needs to have some special properties if
quark confinement is realized in the quark four-point function as
described above. In this respect it is interesting to note that in the
scaling solution of Dyson-Schwinger and Functional Renormalization
Group Equations the quark-gluon vertex can be infrared singular such
that the four-point function assumes the $1/k^4$ singularity
\cite{Alkofer:2008tt}.  Furthermore, such an infrared singularity
provides a possibility of a Witten-Venezanio realization of the
$U_A(1)$ anomaly within a Green function approach
\cite{Alkofer:2008et}. As the origin of the pseudoscalar flavor
singlet mass in the Witten-Veneziano relations is the topological
susceptibility, these findings verify the deep connection of the
infrared behavior of QCD Green functions to the topologically
nontrivial properties of the QCD vacuum. Note that such a relation
between Green functions and vacuum comes naturally in the
Gribov-Zwanziger picture of confinement: In the deep infrared Green
functions are dominated by the field configurations on the Gribov
horizon which, on the other hand, are mostly (or maybe even
completely) of a topologically nontrivial type.

Taken together all this motivates the idea that in Landau gauge QCD the
quark-gluon vertex is of utter importance, and therefore it is in the
focus of several recent studies, see, e.g.,
\cite{Aguilar:2013ac,Hopfer:2013np} and references therein. These are
not only interesting with respect to confinement but show also some
very important result for the understanding of dynamical chiral
symmetry breaking. Usually one considers the generation of quark
masses as the most important effect of chiral symmetry breaking. The
recent studies of the quark-gluon vertex prove unambiguously that the
dynamical generation of scalar and tensor components in this vertex
takes place. In the deep infrared the chiral symmetry violating scalar
and tensor interactions are as strong (if not even stronger) as the
chiral symmetry respecting vector interactions: even in the light
  quark sector QCD generates due to chiral symmetry breaking scalar
  confinement, in addition to vector confinement, dynamically.

\subsection{Mechanism of chiral symmetry breaking}\label{sec:secA3}

Already in 1960 Nambu~\cite{Nambu:1960xd} concluded from the low value
of the pion mass that the pion is a collective excitation
(Nambu-Goldstone boson) of a spontaneously broken symmetry. He suggested
that the breaking of chiral symmetry gives origin to a pseudoscalar
zero-mass state, an idealized pion. After formulation of the
QCD Lagrangian it turned out that for massless quark fields $\psi$
(the chiral limit) left and right-handed species
\begin{equation}\label{eq:lrDecomp}
\psi_r=\frac{1}{2}(1+\gamma_5)\psi,\quad
\psi_l=\frac{1}{2}(1-\gamma_5)\psi,
\end{equation}
are not coupled, they have independent $SU(N_f)$ symmetries,
$SU(N_f)_l\times SU(N_f)_r$, where $N_f$ is the number of
favors. These symmetries can be decomposed into vector and axial vector
symmetries, $SU(N_f)_V\times SU(N_f)_A$. The small pion mass
$m_\pi=140~{\rm MeV}$ is an indication that in the groundstate of QCD
the axial vector symmetry is broken. In the chiral limit it is only
broken by the dynamics of QCD and not by the Lagrangian. This
spontaneous breaking of chiral symmetry (SB$\chi$S) is an effect which
is strongly related to the structure of the nonperturbative vacuum of
QCD. The only method at present available to tackle this
nonperturbative problem is lattice QCD. As discussed in detail in
Sec.~\ref{sec:secA1}, lattice studies of the vacuum of Yang-Mills
theories revealed the existence of infinite clusters of surfaces with
quantized magnetic fluxes (vortices) and of trajectories of magnetic
monopoles localized on vortices. If monopoles or vortices are removed
from the vacuum, both confinement and chiral symmetry
breaking~\cite{deForcrand:1999ms} are gone. Whereas the vortex and
monopole pictures give a consistent picture of quark confinement, the
mechanism for chiral symmetry breaking is not yet clarified and
therefore is under intensive discussion. There are several recent
investigations possibly indicating where to search for this
mechanism. The main questions to be addressed are:
\begin{itemize}
\item What are the configurations/degrees of freedom responsible for
  chiral symmetry breaking?
\item How do we study them and what are the quantitative general
  results so far?
\item Are quark confinement and chiral symmetry breaking related, and
  if yes, how?
\end{itemize}

\begin{figure}[btp]
   \includegraphics*[width=\figwid]{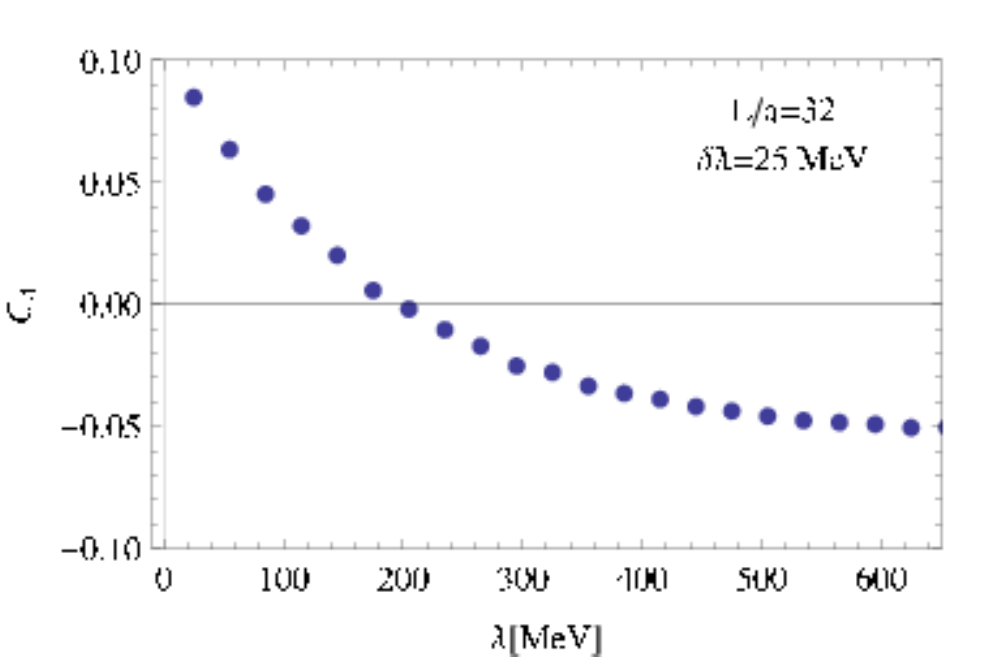}
   \caption{$C_A(\lambda)$ for an $L=32a$ lattice with $a=0.085~$fm of
     quenched QCD. From~\cite{Alexandru:2012xj}.}
\label{fig:Fig31}
\end{figure}

\begin{figure}[btp]
   \includegraphics*[width=\figwid]{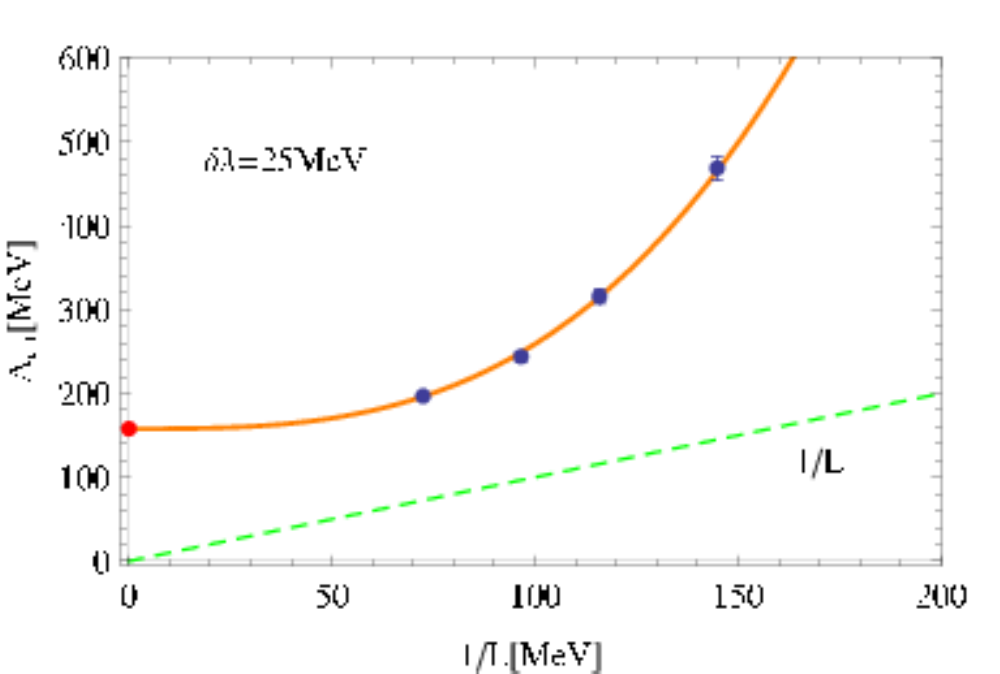}
   \caption{Infinite volume extrapolation of
     $\Lambda_\mathrm{ch}$. From~\cite{Alexandru:2012xj}.}
\label{fig:Fig32}
\end{figure}

The origin of chiral symmetry breaking may be described as an analog
to magnetization, its strength is measured by the fermion condensate
\begin{equation}\label{eq:qqbar}
\bar\psi\psi=\bar\psi_l\psi_r+\bar\psi_r\psi_l,
\end{equation}
which is an order parameter for chiral symmetry breaking. It is a
vacuum condensate of bilinear expressions involving the quarks in the
QCD vacuum, with an expectation value
$\langle0|\bar\psi\psi|0\rangle\approx -(250~{\rm MeV})^3$ given by
phenomenology and confirmed by direct lattice evaluations (see, e.g.,
\cite{Chiu:2003iw}). The Banks-Casher equation~\cite{Banks:1979yr}
\begin{equation}\label{BanksCasher}
\langle0|\bar\psi\psi|0\rangle=-\pi\rho(0),
\end{equation}
relates this expectation value to the density $\rho(0)$ of near-zero
Dirac eigenmodes, i.e.,\ low-lying nonzero eigenmodes $\psi_\lambda$ of
the Dirac equation $D\psi_\lambda=\lambda\psi_\lambda$, distributed
around $\lambda=0$. Hence, the breaking of chiral symmetry should be
imprinted in the chiral properties of the near-zero modes. Since the
Dirac eigenmodes appear in pairs with eigenvalues $\pm\lambda$ and
have opposite chiralities, there can be no preference for left or
right modes, hence the modes have to have specific chiral properties
locally. Reference~\cite{Alexandru:2010sv,Alexandru:2012xj} considers the
left-right decomposition (Eq.~(\ref{eq:lrDecomp})) of the local value
$\psi_{\lambda}(x)$ of Dirac modes. For an ensemble of gauge
configurations they analyze a probability distribution
$\mathcal{P}_\lambda(|\psi_r|,|\psi_l|)$ of these local values in some
surrounding $\delta\lambda$. In order to determine whether the
dynamics of QCD enhances or suppresses the polarization, they define
an uncorrelated distribution
$\mathcal{P}_\lambda^\text{u}(|\psi_r|,|\psi_l|)=
P_\lambda(|\psi_r|)P_\lambda(|\psi_l|)$ from $P_\lambda(|\psi_r|)=\int
d\psi_l\mathcal{P}_\lambda(|\psi_r|,|\psi_l|)$. Then, they determine
whether the correlation $C_A$ for a sample chosen from $\mathcal
P_\lambda$ is more polarized than a sample chosen from $\mathcal
P_\lambda^\text{u}$, indicating enhanced polarization for $C_A>0$ and
anticorrelation for $C_A<0$. The values of $C_A(\lambda)$ for an
$L=32a$ lattice with $a=0.085~$fm of quenched QCD in
Fig.~\ref{fig:Fig31} show that the lowest modes exhibit a dynamical
tendency for chirality, while the higher modes dynamically suppress
it.
Chirally polarized low-energy modes condense and are thus carriers of
the symmetry breaking. The width $\Lambda_\mathrm{ch}$ of the band of
condensing modes provides a new dynamical scale as the dependence on
the infrared cutoff in Fig.~\ref{fig:Fig32} indicates, where the
numerical data are compared with a fit of the form $\Lambda_\text{ch}(1/L)
= \Lambda_\text{ch}(0) + b\, (1/L)^3$ and the cutoff $1/L$ itself. This fit
yields an infinite volume limit of $\Lambda_\text{ch}\approx 160~{\rm
  MeV}$.
Further,~\cite{Alexandru:2012xj} presents evidence that $\Lambda_\text{ch}$
is nonzero in the chiral limit of $N_f=2+1$ QCD and spontaneous
breaking of chiral symmetry thus proceeds via chirally
polarized modes, and $\Lambda_\text{ch}$ vanishes simultaneously with the
density of near-zero modes when temperature is turned on.

This leads to the question about the origin of the near-zero modes. A
first indication about the origin of the near-zero modes came from the
instanton liquid
model~\cite{Ilgenfritz:1980vj,Diakonov:1984vw,Hutter:1995sc}. There is
no unique perturbative vacuum of QCD, different vacua are
characterized by a winding number. Instantons and anti-instantons are
transitions between neighboring winding numbers. They have topological
charge $Q=\pm1$ and give rise to a single zero mode $\psi_0$ with
eigenvalue $\lambda=0$ and definite chirality, i.e.,\ they exhibit
either $\psi_l$ or $\psi_r$. For field configurations with instantons
and anti-instantons these (would-be) zero modes get small shifts of
their eigenvalues and distribute around zero along the imaginary axis
as the Dirac operator is anti-Hermitian, they become near-zero
modes. Hence, overlapping would-be zero modes belonging to single
instantons or anti-instantons split into low-lying nonzero modes and
contribute to the above density of near-zero modes. The instanton
liquid model provides a physical picture of chiral symmetry breaking
by the idea of quarks ``hopping'' between random instantons and
anti-instantons, changing their helicity each time. This process can
be described by quarks propagating between quark-instanton
vertices. In the random instanton ensemble one finds the value of the
chiral condensate $\langle0|\bar\psi\psi|0\rangle\approx -(253~{\rm
  MeV})^3$~\cite{Diakonov:1979nj}, which is quite close to the
phenomenological value\footnote{The above value depends on a scale
  given by the average instanton size.}. Despite their striking
success providing a mechanism for chiral symmetry breaking, instantons
are not able to explain quark confinement. There are models where
instantons may split into merons~\cite{Lenz:2003jp},
bions~\cite{Poppitz:2008hr} or at finite temperature into
calorons~\cite{Bruckmann:2009pa} which may provide a monopole-like
confinement mechanism. Since the QCD-vacuum is strongly
nonperturbative, it does not contain semiclassical instantons but is
crowded with topologically charged objects that, after smooth
reduction of the action (also known as cooling), may become
instantons.

Reference~\cite{Buividovich:2011cv} demonstrates that the above mentioned
smoothing procedures affect the dimensionality of the regions where
the topological charge density $q(x)$ is localized. They measure the
local density $q(x)$ of the topological charge with the trace of the
zero-mass overlap operator $D(x,x)$~\cite{Neuberger:1997fp,Neuberger:1998wv}
\begin{equation}
q(x)=-\mathrm{Tr}\left[\gamma_5\left(1-\frac{a}{2}D(x,x)\right)\right],
\end{equation}
where the trace is taken over spinor and color indices. These
investigations demonstrate that topological charge and zero modes are
localized on low-dimensional fractal structures and tend to occupy a
vanishing volume in the continuum limit. With the inverse
participation ratio
\begin{equation}
IPR=N\sum_x^N\alpha^2(x),\quad\textrm{for}\quad\sum_x\alpha(x)=1,
\end{equation}
for arbitrary normalized distributions $\alpha(x)$ they derive a
fractal dimension of fermionic zero modes. Distributions localized on
a single site get $IPR=N$ and constant distributions $IPR=1$. With the
eigenfunctions $\psi_\lambda$ of the overlap Dirac operator to the
eigenvalues $\lambda$ they measure the average over all zero modes and
all measured gauge field configurations of the local chiral condensate
\begin{equation}\label{CondDens}
\rho_\lambda(x)=\psi_\lambda^\dagger(x)\psi_\lambda(x).
\end{equation}
\begin{figure}[btp]
   \includegraphics*[width=\figwid]{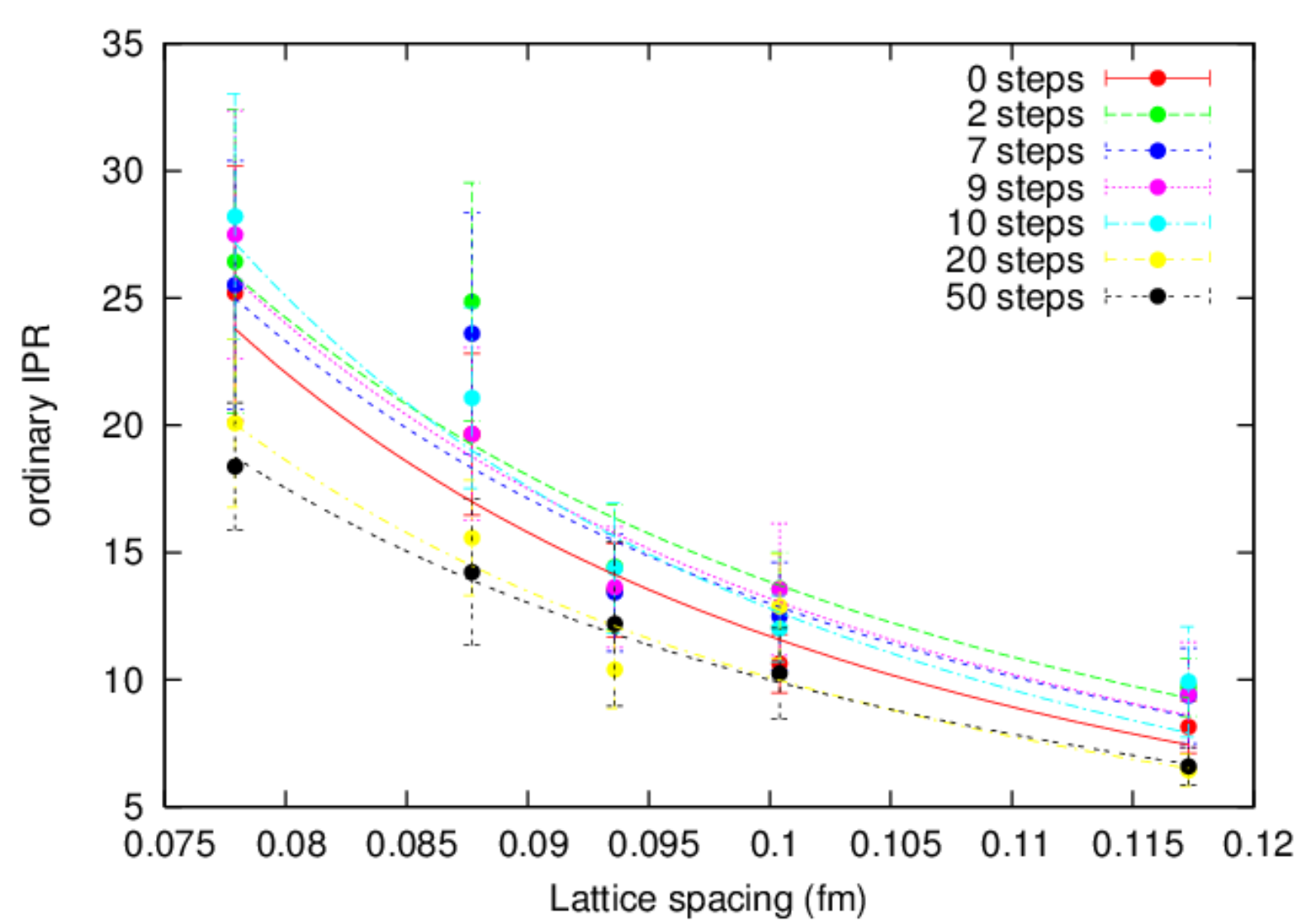}
   \caption{Ordinary IPR for zero modes,
     Eq.~(\ref{CondDens}). From~\cite{Buividovich:2011cv}.}
\label{fig:Fig33}
\end{figure}
Figure~\ref{fig:Fig33} shows how the localization depends on the
lattice spacing $a$ and the number of cooling steps. The finer the
lattice is the larger gets the IPR. This agrees very well with the
idea that the volume occupied by the fermionic zero modes in the
continuum limit approaches zero~\cite{Zakharov:2006vt}. Since zero
modes, $\lambda=0$, have definite chirality the results for the local
chirality agree with the local chiral condensate.

\begin{figure}[btp]
   \includegraphics*[width=\figwid]{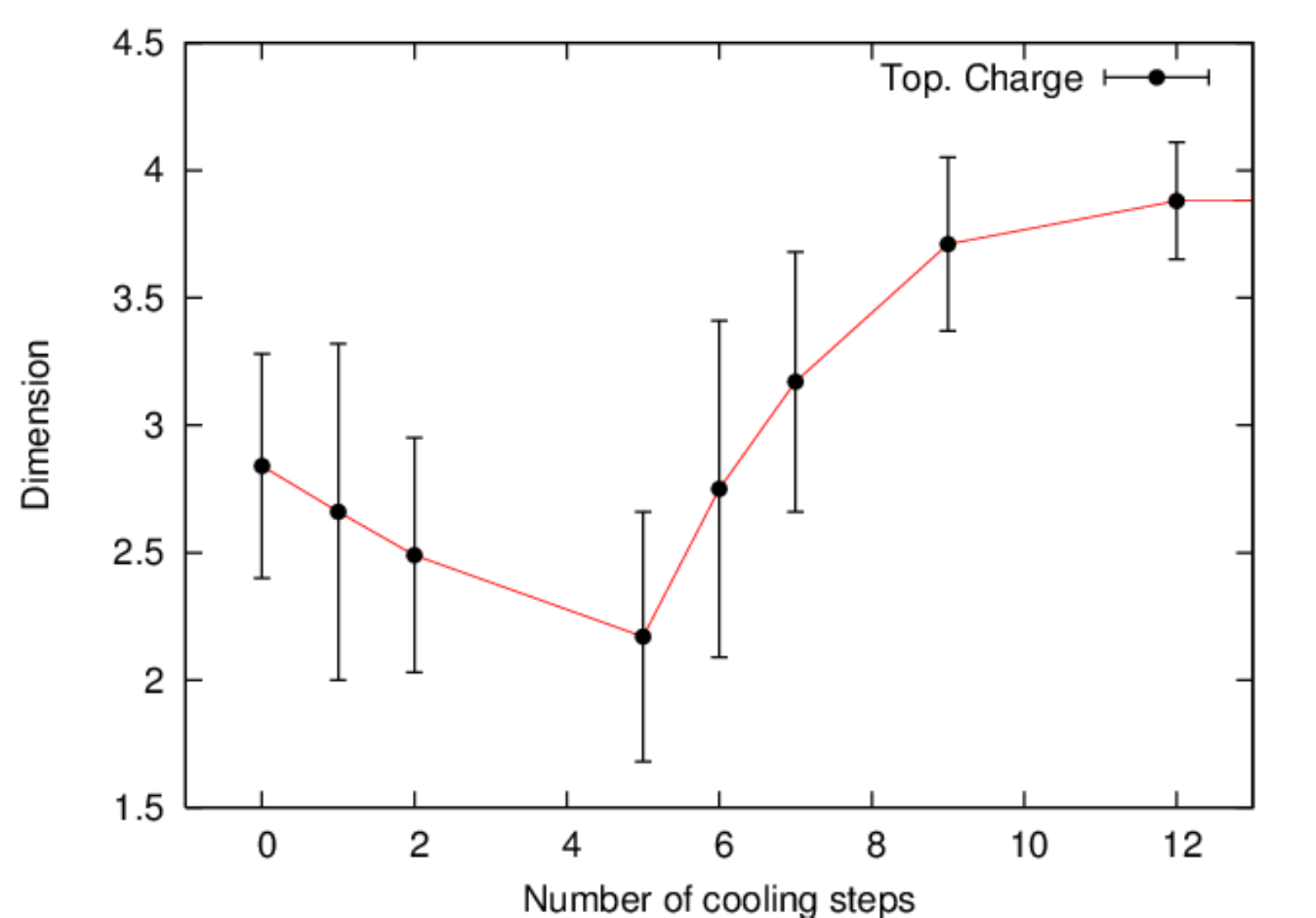}
   \caption{Fractal dimensions at various cooling stages. The solid
     line is shown to guide the eye. From~\cite{Buividovich:2011cv}.}
\label{fig:Fig34}
\end{figure}

Performing a number of measurements with various lattice spacings $a$~\cite{Buividovich:2011cv} is able to define a fractal dimension
$d$ by
\begin{equation}
IPR(a)=\frac{\mathrm{const}}{a^d},
\end{equation}
see Fig.~\ref{fig:Fig34}.
These results show that fermionic zero modes and chirality are
localized on structures with fractal dimension $2\le D\le3$, favoring
the vortex/domain-wall nature of the
localization~\cite{Engelhardt:1999xw,Polikarpov:2004iv,Kovalenko:2004xm}. The
fractal dimension of these structures depends on the number of cooling
steps. A long sequence of cooling iterations destroys the
low-dimensional structures leading to gauge fields close to classical
minima of the action where instantons dominate the properties of field
configurations.

In~\cite{Engelhardt:1999xw,Reinhardt:2001kf} it was shown that center
vortices, quantized magnetic fluxes in the QCD vacuum, contribute to
the topological charge by intersections with $Q_U=\pm1/2$ and writhing
points with a value of $\pm1/16$.

\begin{figure}[btp]
   \includegraphics*[width=\figwid]{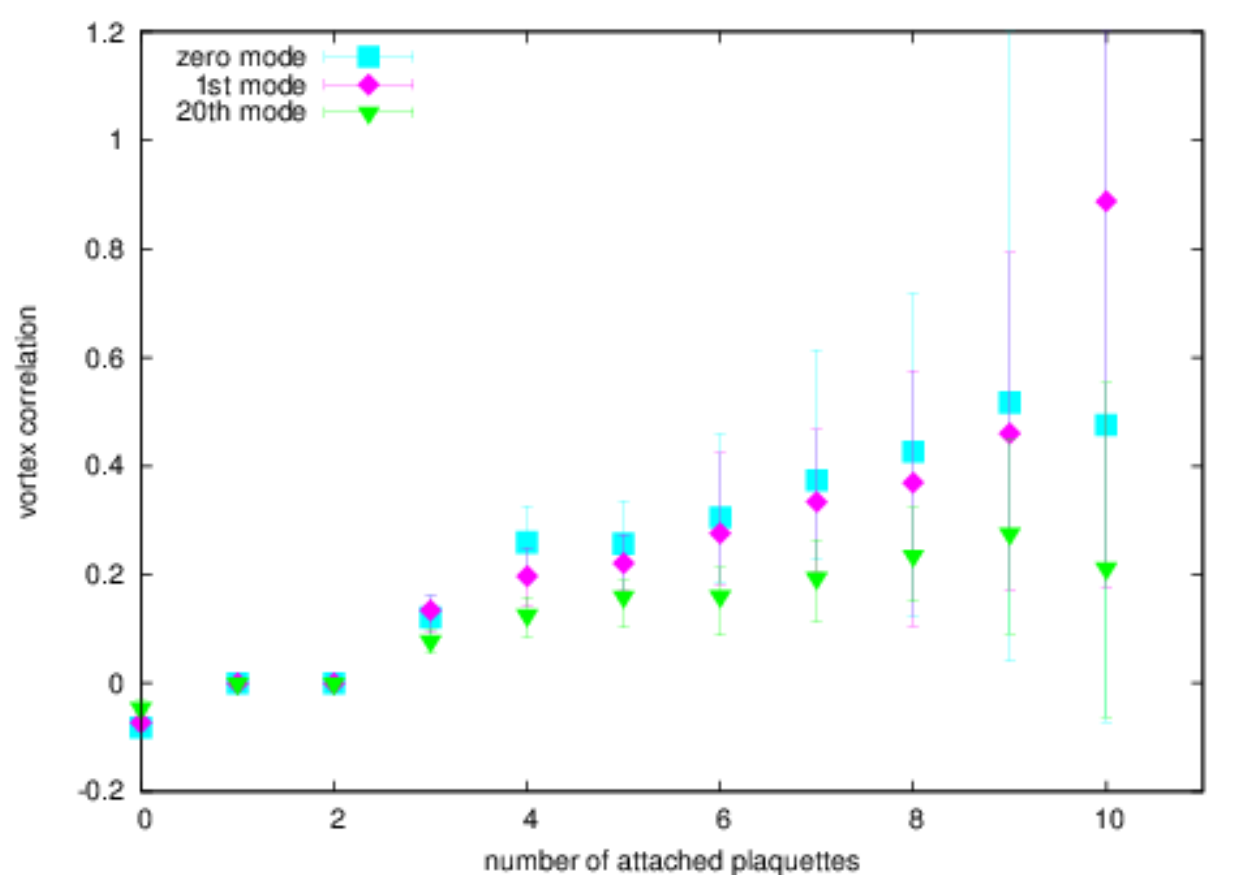}
   \caption{Vortex correlation $C_\lambda(N_v)$ for overlap eigenmodes
     on a $16^4$ lattice at
     $\beta_\mathrm{LW}=3.3$. From~\cite{Hollwieser:2008tq}.}
\label{fig:Fig35}
\end{figure}

Since it is expected that zero modes of the Dirac operator concentrate
in regions of large topological charge density, a correlation between
the location of vortex intersections and writhing points and the
density $\rho_\lambda(x)=|\psi_\lambda(x)|^2$ of eigenmodes of the
Dirac operator $D$, where $D\psi_\lambda=\lambda\psi_\lambda$ with
$\lambda=0$ in the overlap formulation and $\lambda\approx0$ in the
asqtad formulation supports this
picture~\cite{Hollwieser:2008tq}. Reference~\cite{Kovalenko:2005rz}
proposed the observable
\begin{equation}
C_\lambda(N_v)=\frac{\sum_{p_i}\sum_{x\in
    H}(V\rho_\lambda(x)-1)}{\sum_{p_i}\sum_{x\in H}1},
\end{equation}
as a measure for the vortex-eigenmode correlation. To explain this
formula we have to recall that center vortices are located by center
projection in maximal center gauge~\cite{DelDebbio:1996mh}. Plaquettes
on the projected lattice, ``P-plaquettes'', are either $+1$ or $-1$;
they form closed surfaces on the dual lattice. Each point on the
vortex surface belongs to $N_v$ P-plaquettes. $N_v=0$ we get for
points which do not belong to a vortex surface, $N_v=1$ or $2$ is
impossible since vortex surfaces are closed, for corner points or
points where the surface is flat we get $N_v=3, 4$ or $5$, when the
surface twists around a point $N_v=6$ or $7$ and at points where
surfaces intersect $N_v\ge 8$. In Fig.~\ref{fig:Fig35}, we display the
data for $C_\lambda(N_v)$ versus $N_v$ computed for eigenmodes of the
overlap Dirac operator. The lattice configurations are generated by
Monte Carlo simulations of the L\"uscher-Weisz action at
$\beta_\mathrm{LW}=3.3$. The correlations for the first eigenmode and
the twentieth Dirac eigenmode are shown. Since the correlator
increases steadily with increasing $N_v$, we conclude that the Dirac
eigenmode density is significantly enhanced in regions of large $N_v$.

By the Atiyah-Singer index
theorem~\cite{Atiyah:1971rm,Schwarz:1977az,Brown:1977bj,Narayanan:1994gw}
zero modes are related to one unit of topological charge. Therefore,
the question emerges, how vortex intersections and writhing points are
related to these zero modes. Reference~\cite{Hollwieser:2011uj} compares
vortex intersections with the distribution of zero modes of the Dirac
operator in the fundamental and adjoint representation using both the
overlap and asqtad staggered fermion formulations in SU(2) lattice
gauge theory. By forming arbitrary linear combinations of zero modes
they prove that their scalar density peaks at least at two
intersection points~\cite{Hollwieser:2011uj}.

In recent investigations a further source of topological charge was
discovered. A contribution with one unit of topological charge comes
from colorful center vortices~\cite{Schweigler:2012ae}. Vortices may
have a color structure with a winding number and contribute to the
topological charge. Covering of the full SU(2) color group leads to
actions of a few instanton actions only and indicates that such
configurations are possibly appearing in Monte-Carlo
configurations. According to the index theorem and the Banks-Casher
relation, interacting colorful vortices contribute to the density of
near-zero modes.

These observations lead to a picture similar to the instanton liquid
model. The lumps of topological charge appearing in Monte-Carlo
configurations interact in the QCD-vacuum and determine the density of
near-zero modes. Therefore, it is not the true zero modes deciding on
the value of the topological charge of a field configuration which
lead to the breaking of chiral symmetry. The number of these modes is
small in the continuum limit. It is the density of interacting
topological objects which leads to the density of modes around zero
and according to the Banks-Casher relation (Eq.~(\ref{BanksCasher}))
determines the strength of chiral symmetry breaking.

Due to the color screening by gluons the string tension of pairs of
static color charges in $SU(N)$ gauge theories depends on their
$N$-ality. From the field perspective this $N$-ality dependence has
its origin in the gauge field configurations which dominate the path
integrals in the infrared. Center vortices are the only known
configurations with appropriate properties. Concerning chiral symmetry
breaking a remarkable result was found in~\cite{deForcrand:1999ms},
namely removing vortices from lattice configurations leads to
restoration of chiral symmetry. If one considers that a phase
transition of the gauge field influences both gluons and fermions,
then one would expect that deconfinement and chiral phase transition
are directly related, as indicated by lattice
calculations~\cite{Karsch:2001cy}.

\begin{figure}[btp]
\includegraphics*[width=\figwid]{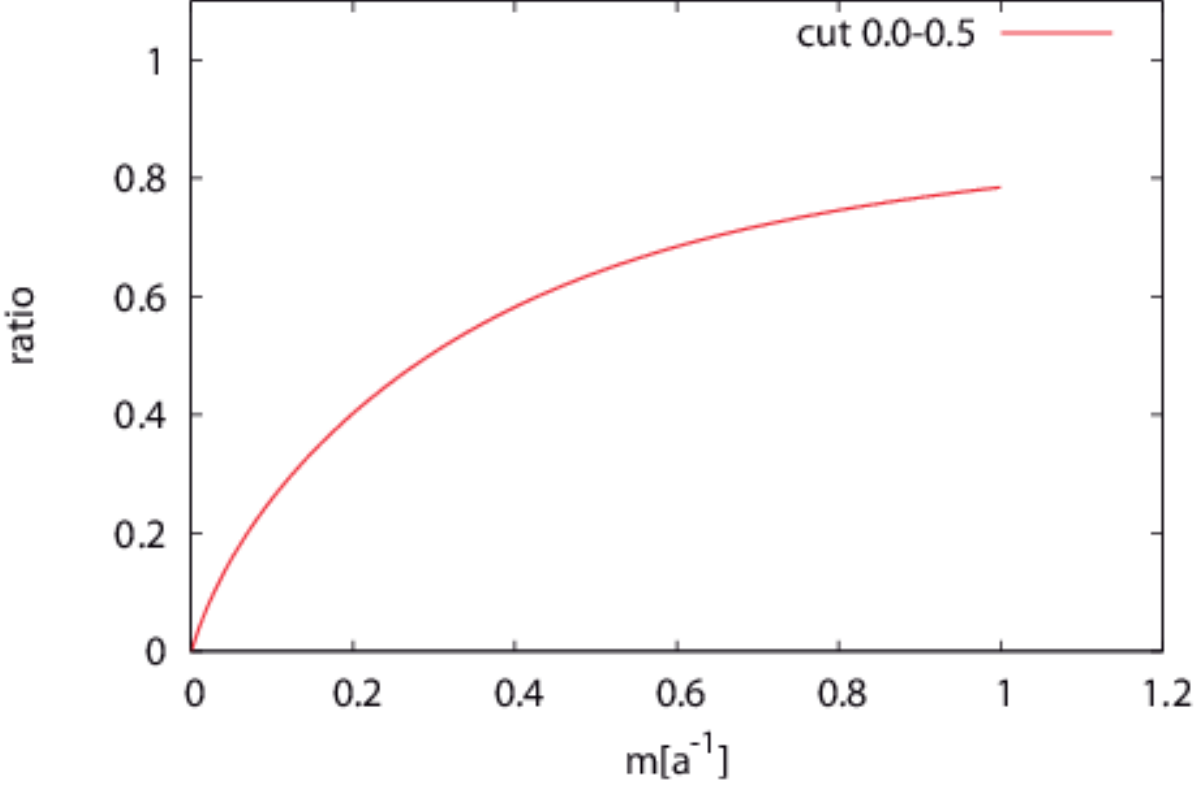}
\put(-180,130){\Large{$\frac{\langle\bar\psi\psi\rangle_{\Lambda_{\rm
          IR}}} {\langle\bar\psi\psi\rangle}$}}
\caption{$\langle\bar\psi\psi\rangle_{\Lambda_{\rm
      IR}}/\langle\bar\psi\psi\rangle$ for an IR cut of $\Lambda_{\rm
    IR}=0.5a^{-1}$, plotted against the current quark mass $m$.  A
  large reduction of $\langle\bar\psi\psi\rangle_{\Lambda_{\rm
      IR}}/\langle\bar\psi\psi\rangle\simeq0.02$ is found in the
  physical case of $m\simeq0.006a^{-1}\simeq5~{\rm
    MeV}$. From~\cite{Suganuma:2013tx}.}
\label{fig:Fig36}
\end{figure}

It is an interesting check of this picture whether field
configurations with restorations of chiral symmetry still have
confinement. This problem was attacked recently from two different
sides. Using the completeness of the Dirac mode basis and restricting
the Dirac mode space by a transition to the corresponding projection
operator
\begin{equation}
\sum_\lambda|\lambda\rangle\langle\lambda|=1\quad\to\quad \hat
P_A=\sum_{\lambda>k}|\lambda\rangle\langle\lambda|.
\end{equation}
A manifestly gauge covariant Dirac-mode expansion and projection
method was developed in~\cite{Suganuma:2012sp,Gongyo:2012vx}. They had
to deal with the technical difficulty to find all eigenvalues and
eigenfunctions of huge matrices and used therefore the Dirac operator
for staggered fermions~\cite{Susskind:1976jm,Sharatchandra:1981si} in
SU(3)-QCD and rather small $6^4$ lattices. After removing the lowest
$k$ Dirac modes they got a strong reduction of the chiral condensate
to $2\%$ in the physical case of $m\simeq0.006a^{-1}\simeq5~{\rm
  MeV}$, see Fig.~\ref{fig:Fig36}.
\begin{figure}[btp]
   \includegraphics*[width=\figwid]{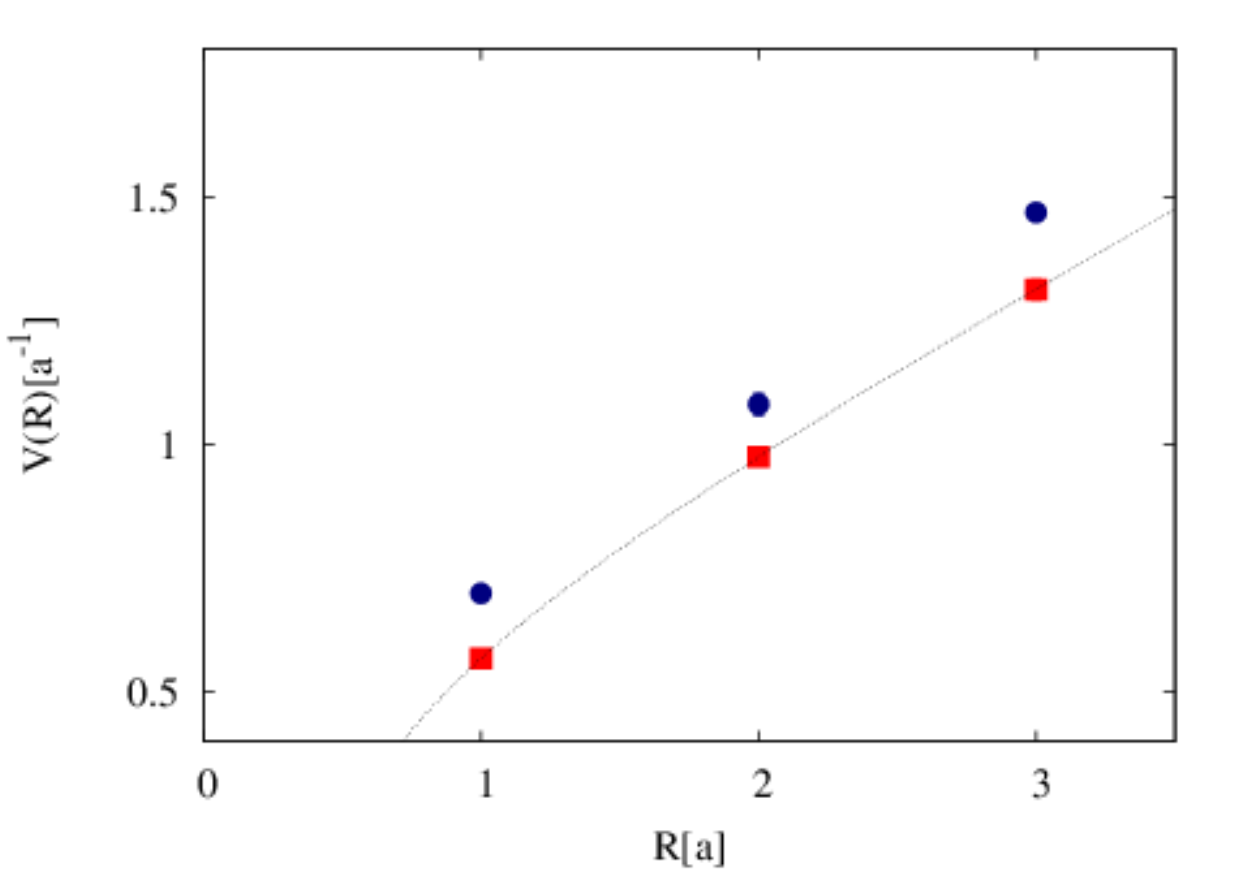}
   \caption{Inter-quark potential (circles) after removal of low-lying
     Dirac modes with the IR-cutoff $\Lambda_{\rm IR}=0.5a^{-1} \simeq
     0.4~{\rm GeV}$ and original potential (squares), apart from an
     irrelevant constant. From~\cite{Suganuma:2013tx}.}
\label{fig:Fig37}
\end{figure}
This removal conserved the area law behavior of Wilson loops without
modifying the slope. Besides an irrelevant constant the inter-quark
potential is almost the same, see Fig.~\ref{fig:Fig37}. The Polyakov
loop remains almost zero indicating that the center symmetry is still
unbroken~\cite{Iritani:2012tc,Iritani:2013tq,Suganuma:2013tx}.

\begin{figure}[btp]
   \includegraphics*[width=\figwid]{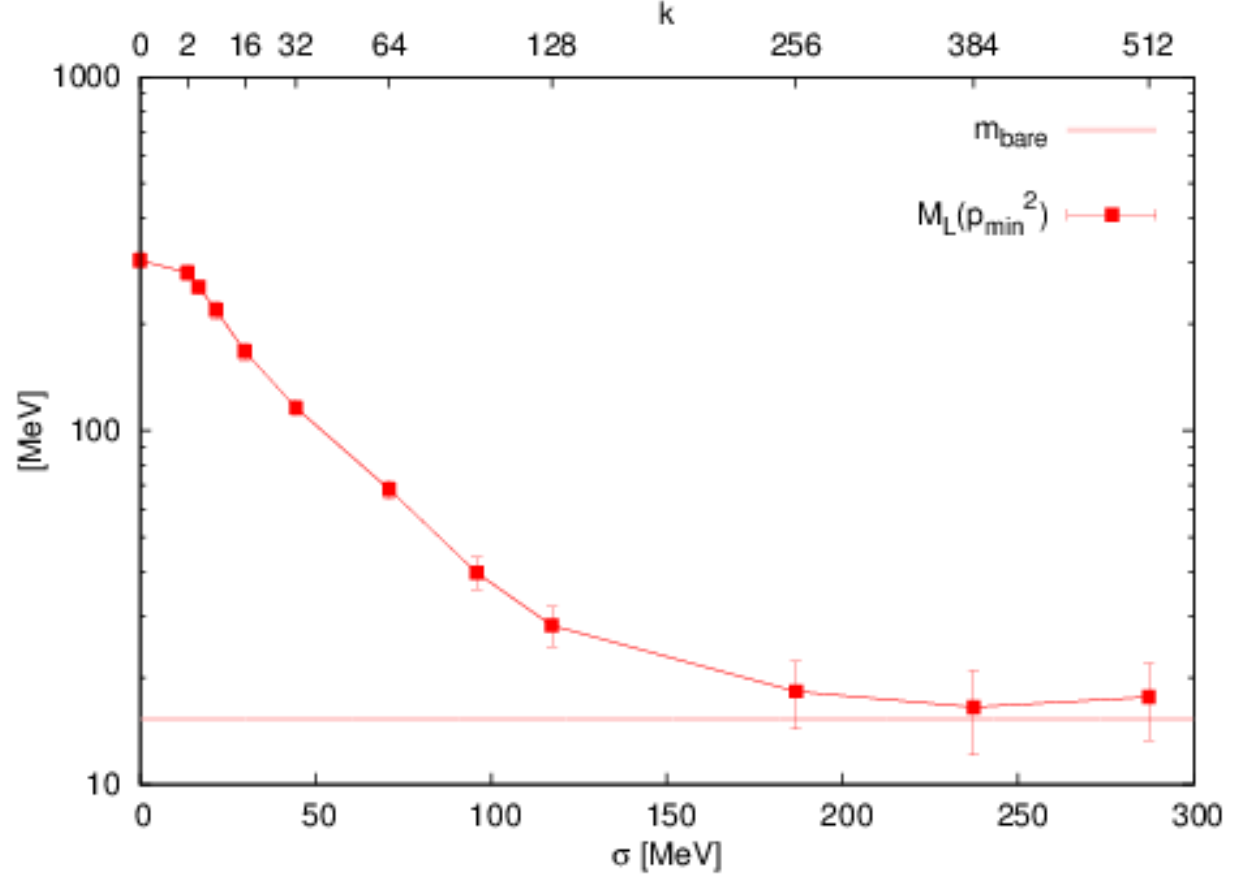}
   \caption{Lattice mass function $M_L(p_\mathrm{min}^2)$ for the
     smallest available momentum $p_\mathrm{min}=0.13~{\rm GeV}$ as a
     function of the truncation level. On the lower axis the level $k$
     is translated to an energy scale. For comparison, the bare quark mass
     is plotted as a horizontal line. From~\cite{Schrock:2013xf}.}
\label{fig:Fig38}
\end{figure}

\begin{figure}[btp]
   \includegraphics*[width=\figwid]{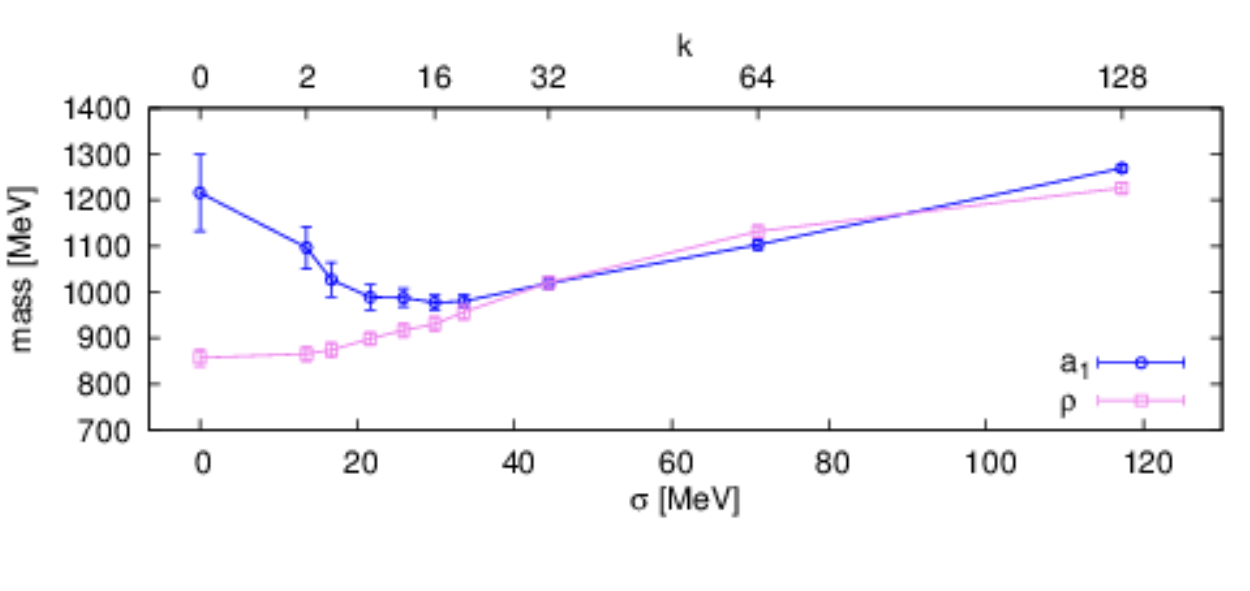}
   \caption{Influence of the removal of the lowest $k$ modes of the
     Dirac operator on the masses of chiral partner mesons, the vector
     meson $\rho$ and axial vector meson
     $a_1$. From~\cite{Schrock:2013xf}.}
\label{fig:Fig39}
\end{figure}

The Graz
group~\cite{Lang:2011vw,Lang:2011ai,Glozman:2012fj,Glozman:2012hw,Glozman:2012iu}
studied hadron spectra after cutting low-lying Dirac modes from the
valence quark sector in a dynamical lattice QCD calculation. They
expressed the valence quark propagators $S$ directly by the
eigenfunctions of the Dirac operator and removed an increasing number
$k$ of lowest Dirac modes $|\lambda\rangle$
\begin{equation}\label{RedQuaProp}
S_{\mathrm{red}(k)}=S-\sum_{\lambda\le
  k}\mu_\lambda^{-1}|\lambda\rangle\langle\lambda|\gamma_5,
\end{equation}
with $\mu_\lambda$ the (real) eigenvalues of the Hermitian Dirac
operator $D_5=\gamma_5D$. They extracted the mass function $M_L(p^2)$
from the reduced quark propagator (Eq.~(\ref{RedQuaProp})) for chirally
improved fermions. In Fig.~\ref{fig:Fig38} the dynamical generated
mass $M_L(p_\mathrm{min}^2)$ for the smallest available momentum
$p_\mathrm{min}=0.13~{\rm GeV}$ is plotted as a function of the
truncation level $k$. Removing the low energy modes the dynamic mass
generation ceases and the bare quark mass is approached successively.
Except for the pion, the hadrons survived this artificial restoration
of chiral symmetry by this truncation. The quality of the exponential
decay of the correlators increases by this procedure indicating a
state with the given quantum numbers. In Fig.~\ref{fig:Fig39} the
influence of the truncation of the masses of two mesons which can be
transformed into each other by a chiral rotation, the vector meson
$\rho$ and the axial vector meson $a_1$. These would-be chiral
partners become degenerate after restoration of chiral
symmetry. Interestingly these meson masses increase with increasing
truncation level $k$.
These results demonstrate that even without a chiral symmetry breaking
vacuum confined hadrons can exist, at least with rather large mass.

We got some picture for spontaneous breaking of chiral symmetry
(SB$\chi$S). We could call it a ``kinematical'' picture: Interacting
lumps of topological charge lead to low-lying Dirac modes which via
the Banks-Casher relation (Eq.~(\ref{BanksCasher})) induce SB$\chi$S. But
there is still no answer to the question about the dynamics of
SB$\chi$S. We could try to conjecture a mechanism: It is the low
momentum modes changing chirality, when they enter a combination of
electric and magnetic fields present in regions of nonvanishing
topological charge density. In such fields slow color charges would
move along spiraling paths changing their momentum and conserving
their spin. Fast moving charges would be less influenced by such field
combinations. This could explain the importance of low-lying Dirac
modes for SB$\chi$S and clarify why Goldstone bosons do not survive
the removal of low-lying Dirac modes and heavy hadrons with increasing
removal rather tend to increase their masses, see
Fig.~\ref{fig:Fig39}. Hence field configurations with lumps of
topological charge contributions increase the density of low-lying
Dirac eigenmodes with pronounced local chiral properties producing a
finite chiral condensate.

\subsection{Future Directions}\label{sec:secA4}
As described above there are many unsolved interesting problems
concerning the vacuum structure of QCD, confinement and chiral
symmetry breaking.  Most important, there is still no satisfactory
solution to the confinement problem in non-Abelian gauge theories and
therefore no proof that continuum QCD confines.  Some of the
most interesting questions for future work to generate progress in
this field are as follows:

\begin{itemize}
\item Can one treat confinement as a string theory of lower dimensional
  topological defects in four dimensions? Is confinement related to percolation 
  of these defects?
\item Is the BRST symmetry of gauge-fixed non-Abelian Yang-Mills theory softly 
  broken in the nonperturbative domain? If so, what is implied then for
  the Kugo-Ojima and Gribov-Zwanziger confinement scenarios? 
\item Is the existence of families of solutions for Green functions of
  elementary fields related to some yet not fully understood gauge degree of
  freedom?
\item In the Coulomb gauge, what balances  the ``over-confinement'' (i.e., the to
  large string tension) due the   time-component of the gluon field?
\item Is there Abelian dominance in the maximally-Abelian gauge? Do
  chromomagnetic monopoles correlate along the lower dimensional
  topological defects mentioned above? If so, is the picture of a dual 
  superconductor still valid? 
\item Does the quark four-point function display the anticipated ``$1/k^4$''
 infrared singularity? If so, in all gauges?  Is then the cluster
 decomposition property violated? 
\item Can one construct an explicit quantum state 
  responsible for a dissipation-free flow of an electric current along an
  external magnetic field (chiral magnetic effect)?
\item Does there exist a kind of chiral superconductivity, inherent
  not only to relativistic field theories but to some condensed-matter
  systems, like graphene and semimetals, as well?
\item Do chirally polarized low-energy modes condense? 
  What is the physical origin of the band width
  $\Lambda_\mathrm{ch}$ of condensing modes?
\item Does the result that fermionic zero modes and chirality are
  localized on structures with fractal dimension $D=2-3$ favor the
  vortex/domain-wall nature of the localization?
\item Which kind of effective quark-gluon interactions are generated 
by dynamically breaking chiral symmetry? Will this include a scalar confining 
force?
\newpage
\item Why do Goldstone bosons not survive the removal of low-lying
  Dirac modes?
\item What is the relative contribution of the various interacting
  topological objects to the Dirac operator's density of modes around 
  zero virtuality?
\item Do low-momentum modes change chirality in regions of nonvanishing
topological charge density with electric {\bf and} magnetic fields 
present and thus dynamically break chiral symmetry?
\end{itemize}
An answer to these questions may hold the key to understand infrared QCD and the 
related phenomena, most prominently, confinement and dynamical chiral symmetry
breaking. Especially the confinement problem is one of the truly fundamental
problems in contemporary physics. Until it is well understood something
essential is lacking in our comprehension of particle and nuclear physics.
Although the problems described in this section have proven to be very hard
their solution is important, and they are certainly worth pursuing. 

\clearpage
\section[StronglyCoupled]{Strongly coupled theories and conformal symmetry \protect\footnotemark}
\footnotetext{Contributing authors:
J.~Erdmenger$^{\dagger}$, E.~Pallante$^{\dagger}$, K.~Papadodimas, A.~Pich, R.~Pittau}
\label{sec:chapg}



Most of the multifaceted physics of QCD and a good part of the
theories of fundamental interactions beyond the standard model (BSM)
are or have sectors that are strongly coupled.
It is therefore of relevance to devise new and increasingly sophisticated theoretical approaches to study strongly coupled physics. The same methods often provide significant guidance in other branches of physics, from cosmology to material science.
This chapter provides a short review of recent progress and present challenges in the theoretical formulation of gauge theories at strong coupling and their applications to particle and 
condensed matter physics.

Conformal symmetry has recently emerged as a key ingredient in this context and as a guide in the study of the many aspects of the phase diagram of non-Abelian gauge theories in four spacetime dimensions, as well as in the phenomenological search for models of new physics (NP) beyond the electroweak symmetry breaking (EWSB) scale and the standard model (SM) of particle interactions. The physics output has progressed jointly with the refinement of theoretical and computational approaches.
Among the first, gauge-gravity duality and, in particular, the AdS/CFT correspondence \cite{Maldacena:1997re} between higher dimensional string theories living in anti-de Sitter spacetime and conformal field theories (CFTs) living at their boundaries have introduced new classes of strong/weak coupling dualities and allowed predictions for (near)conformal strongly coupled systems that complement 
other effective field theory approaches, such as the large-$N$ expansion, the functional renormalization group, or methods to solve Schwinger-Dyson equations; examples of these approaches can be found in the rest of this document. Computational approaches essentially amount to lattice field theory, to date the only method we know that should provide the full nonperturbative solution of QCD, once the continuum limit is reached.
Lattice field theory investigations have recently benefited from algorithmic advances and a huge step forward in supercomputer technology and performance, see also Chapter~\ref{sec:chapz}.

The interplay of conformal symmetry and the strongly coupled regime of quantum field theories has led to new paradigms and has highlighted the existence of families of gauge theories and regions of their phase diagram that might be relevant in describing high energy particle physics between the electroweak symmetry breaking scale and the Planck scale.
The same theoretical advances have motivated the development of
a number of methods for describing strongly coupled systems in condensed matter physics. Interesting examples in this context are the phase structure and transport properties of materials of the latest generation such as graphene, non-Fermi liquids and high-$T_c$ superconductors.

This chapter is organized in four sections.
Section~\ref{sec:G_QFT} provides an overview of the most recent formal developments in quantum field theories with and without supersymmetry, with emphasis on conformal field theories and the ways they connect to QCD at strong coupling.
Section~\ref{sec:G_ConfBSM} discusses in more detail how conformal symmetry can be restored in non-Abelian gauge theories with matter content and outlines the theory of the conformal window.   One interesting possibility  is that theories close to the conformal window could be realized in nature and play a relevant role for new physics at the weak scale. Section~\ref{sec:G_EWSB} discusses electroweak symmetry breaking and BSM scenarios for its realization that involve strongly coupled dynamics and/or spontaneously broken conformal symmetry.  In particular, we discuss the theoretical premises for a wide class of strongly coupled models, composite-Higgs or dilaton-Higgs, using a general effective field theory approach to constrain them with electroweak precision measurements and the recent discovery of a Higgs-like boson of about 126 GeV at the Large Hadron Collider (LHC).
As an alternative to strongly coupled new dynamics,
we revisit the appealing scenario of a minimally extended SM, where an underlying  conformal symmetry would govern the dynamics from the Planck scale all the way down to the weak scale.
Finally, Sec.~\ref{sec:G_CM} is devoted to recent advances in the study of condensed matter systems using lattice gauge theory and gauge-gravity duality. We discuss future prospects in Sec.~\ref{sec:G_Conclusions}.

\subsection{New exact results in quantum field theory}
\label{sec:G_QFT}

In this section we review recent developments in exact methods in quantum field theory (QFT), some of which were inspired by string theory and/or the AdS/CFT correspondence.

Many of these developments refer  to quantum field theories in the {\it large $N$ limit}. As is well known \cite{'tHooft:1973jz},
gauge theories simplify  by scaling the number of colors $N$ to infinity  while at the same time sending the coupling constant $g$ to zero, keeping the
combination $\lambda = g^2 N$, called the 't Hooft coupling, fixed. In this limit Feynman diagrams acquire a topological
classification, with {\it planar} diagrams providing the leading contribution to any given process, while the contribution of
{\it non-planar} diagrams is suppressed by powers of $1/N$. Although QCD, whose gauge group is $SU(3)$, corresponds to the value $N=3$, certain aspects are captured by the large-$N$ approximation.

The large $N$ limit plays a central role in the AdS/CFT correspondence (gauge-gravity duality).
The latter asserts that large $N$ gauge theories admit a holographic description in terms of
higher-dimensional string theories \cite{Maldacena:1997re,Gubser:1998bc,Witten:1998qj}.
In certain limits of the parameter space the higher-dimensional string theory can be well approximated by
semi-classical supergravity, which allows computations to be performed in the strongly-coupled regime of the
gauge theory.
The best studied example is the duality between the ${\cal N}=4$ Super-Yang-Mills (SYM) and type IIB string
theory in AdS$_5\times$S$^5$.
Several generalizations of the AdS/CFT correspondence have been developed, which attempt to describe gauge
theories closer to QCD.
For an entr\'ee to the vast literature see the classic review \cite{Aharony:1999ti} and
Secs.~\ref{sec:G_integrability}, \ref{sec:G_3dCFT}, \ref{sec:G_ConfBSM} and \ref{sec:G_CM} of this chapter.

A large part of this section is focused on CFTs. While QCD is not conformal, the study of CFTs is important for several reasons. CFTs are the ultraviolet (UV) and infrared (IR) limits of renormalization group (RG) flows of other quantum field theories; so, any other well-defined
quantum field theory can be understood as a UV CFT perturbed by relevant operators. Moreover, CFTs can be studied with more general methods (some of which are described below) than the usual perturbative expansion. This allows us to probe them in the strong coupling regime. 
Finally, CFTs have applications in string theory and condensed matter physics.

\subsubsection{Integrability of planar ${\cal N}=4$ SYM}
\label{sec:G_integrability}

One of the most remarkable recent achievements in QFT is the proposed exact solution of planar ${\cal N}=4$ SYM using methods of integrability and input from the AdS/CFT correspondence. See Ref.~\cite{Beisert:2010jr} for an extensive review and list of references. Unlike QCD, the ${\cal N}=4$ SYM is conformal and does not have asymptotic multiparticle states. Instead, the ``spectrum'' of the theory is encoded in the conformal dimensions of local, single trace operators. At small values of the 't Hooft coupling $\lambda=g^2 N$ the conformal dimensions can be computed perturbatively by usual Feynman diagrams. As $\lambda$ is increased the computations quickly become intractable. Nevertheless, it is believed that for any value of the 't Hooft coupling $\lambda$ the spectrum of the ${\cal N}=4$ SYM at large $N$ is governed by a 1+1 dimensional integrable system. The exact S-matrix of this integrable system has been
determined.
Using this exact worldsheet S-matrix, the conformal dimensions of single trace operators can be determined by the solutions of complicated algebraic equations derived by the thermodynamic Bethe ansatz or Y-system. For instance, the anomalous dimension of the Konishi operator (a particular single trace operator) has been evaluated for all values of $\lambda$ by solving these equations numerically. As expected, the anomalous dimensions smoothly interpolate between the perturbative values at small $\lambda$ and the AdS/CFT predictions of type IIB string theory on AdS$_5\times$S$^5$ at large $\lambda$. The results from integrability constitute a notable nontrivial verification of the AdS/CFT correspondence. More recently, there have been promising attempts to extend the methods of integrability to the computation of correlation functions and to investigate the connections with scattering amplitudes in the ${\cal N}=4$ SYM. It would of course be exciting if integrability persists, in some form, in theories closer to QCD.

\subsubsection{Scattering amplitudes}

The computation of scattering amplitudes in perturbative QCD is of central importance both for theoretical and practical reasons\,---\,for instance, the analysis of backgrounds at the LHC. While straightforward in principle, the evaluation of scattering amplitudes using QCD Feynman diagrams grows very quickly in complexity as the number of external lines and/or number of loops is increased.

In the last decades we have seen remarkable progress in developing alternative methods to compute scattering amplitudes in QCD as well as in more general gauge theories, most prominently for the ${\cal N}=4$ SYM. These methods are based on ``on-shell'' techniques\,---\,generalized unitarity as well as input from the AdS/CFT correspondence. For a summary of these developments see Ref.~\cite{Dixon:2011xs}. In the 1980s Parke and Taylor presented a compact formula for the tree-level maximally helicity violating (MHV) amplitudes of gluons in QCD \cite{Parke:1986gb}, which is vastly simpler than what appears in the intermediate steps of the computation via Feynman diagrams. More recently, a relation was conjectured between tree-level scattering amplitudes and a string theory in twistor space \cite{Witten:2003nn}, which eventually led to generalizations and the Cachazo-Svrcek-Witten rules \cite{Cachazo:2004kj}. Another important step was the development of the Britto-Cachazo-Feng-Witten on-shell recursion relations \cite{Britto:2005fq}. By considering a
particular analytic continuation of tree-level amplitudes and exploiting the fact that, in certain theories, the resulting meromorphic function has simple behavior at infinity of the complex plane, higher-point amplitudes can be reconstructed by gluing together lower-point amplitudes. This technique simplifies the computation of tree level and, to some extent, higher-loop amplitudes. For outcomes of these developments we refer to Chapter~\ref{sec:chape} of this document.

Further insights are provided by the AdS/CFT correspondence and the work  \cite{Alday:2007hr}, which relates scattering amplitudes of gluons at strong coupling in the ${\cal N}=4$ SYM to minimal area surfaces in AdS. The AdS/CFT computation of scattering amplitudes led to the discovery of a hidden symmetry of amplitudes called {\it dual conformal invariance}, which was also independently noticed in perturbative field theory computations at weak coupling. It also led to uncovering the relation between Wilson loops and scattering amplitudes, see Ref.~\cite{Dixon:2011xs} for further discussions and references to the original literature.

These developments suggest that gluon scattering amplitudes, especially those for planar ${\cal N}=4$ SYM, may be governed by
additional symmetries, such as the dual conformal invariance which together with ordinary conformal invariance closes into a larger
``Yangian'' symmetry, which may not be manifest in the Lagrangian formulation of the theory. This has led to an ambitious attempt of describing the all-loop scattering amplitudes of ${\cal N}=4$ SYM in terms of new mathematical structures; see Ref.~\cite{ArkaniHamed:2012nw} for latest developments in this direction.

\subsubsection{Generalized unitarity and its consequences}

The main inspiring idea behind generalized unitarity is that the only information needed to compute one-loop
amplitudes, independently of the number of external legs, are the coefficients of a very well-known and
tabulated set of 1-, 2-, 3- and 4-point scalar integrals \cite{Hahn:1998yk,Ellis:2007qk,vanHameren:2010cp},
plus rational parts which have to be added separately \cite{Ossola:2008xq}.
Each coefficient is sitting in front of a unique combination of polydromic functions (logarithms and
di-logarithms) which can be identified by looking at the discontinuities of the amplitude
\cite{Anastasiou:2006jv}, while the rational parts are not-polydromic in four dimensions.
In the pioneering work of Refs.~\cite{Bern:1994zx,Bern:1994cg}, the discontinuities are determined
analytically by combining different ways of putting on-shell two internal particles in the loop (two-particle
cuts) and the rational parts are reconstructed from the soft/collinear limits of the full amplitude.
In Ref.~\cite{Britto:2004nc} Britto, Cachazo and Feng generalized, for the first time, this procedure by
introducing the concept of quadruple cut: all possible ways of putting four loop particles on-shell
completely determine the coefficients of the contributing 4-point scalar integrals (boxes), fully solving, at
one-loop, theories so symmetric that only boxes are present, such as ${\cal N}=4$ SYM.
Since a quadruple cut factorizes the amplitude in four tree amplitudes, the box coefficients are simply
computed in terms of the product of four tree amplitudes evaluated at values of the loop momenta for which
the internal particles are on-shell.
The solution for general theories\,---\,where also lower point functions contribute, such as triangles,
bubbles and tadpoles\,---\,is provided by the Ossola-Papadopoulos-Pittau (OPP) approach of
Refs.~\cite{Ossola:2006us,Ossola:2007ax}, in which the coefficients are directly inferred from the one-loop
{\it integrand}.
The advantage of this method is that, once the coefficients of the box functions are determined, a simple
subtraction from the original integrand generates an expression from which the coefficients of the 3-point
scalar functions can be computed by means of triple cuts \cite{Forde:2007mi}, and so on.
The one-loop integrand can be either determined by gluing together tree-level amplitudes, as in the
generalized unitarity methods \cite{Berger:2008sj,Ellis:2007br}, or computed numerically
\cite{vanHameren:2009dr,Cascioli:2011va}, the only relevant information being the value of the integrand at
certain values of the would-be loop momentum.
As for the missing rational parts, OPP uses special tree-level vertices (involving up to four fields and
determined once for all for the theory at hand \cite{Draggiotis:2009yb,Garzelli:2009is,Page:2013xla}) to
included them, while they can be computed via their d-dimensional cuts in generalized unitarity
\cite{Ellis:2008ir}.
Alternatively, it is possible to construct the rational parts recursively in the number of legs
\cite{Berger:2008sj}.

Since the integrand of a one-loop amplitude is a tree-level like object, tree-level Feynman-diagrams-free recursion techniques \cite{Berends:1987me,Caravaglios:1995cd,Draggiotis:1998gr,Mangano:2002ea} can be applied also in generalized unitarity and OPP. This has been dubbed {\it ``NLO revolution''} and made possible to calculate numerically twenty-gluon amplitudes at NLO in QCD \cite{Giele:2008bc} or six-photon amplitudes \cite{Ossola:2007bb} in QED, and to attack the NLO computation of complicated  processes needed in the LHC phenomenology, such as $t\bar t b \bar b$ production \cite{Bevilacqua:2009zn} (as an irreducible QCD background to the $Ht\bar t$ signal), $pp \to 4$ leptons \cite{Frederix:2011ss} (as a background to the Higgsstrahlung production mechanism),  $H$ + 3 jets \cite{Cullen:2013saa}\,---\,using the effective $ggH$ coupling\,---\,, $W$ + 5 jets \cite{Bern:2013gka} and $pp \to 5$ jets \cite{Badger:2013vpa}.
On the basis of the above computational progress, NLO Monte Carlo codes have been constructed in the last few years allowing the LHC experimental collaborations to analyze their data at the NLO accuracy. Among them BlackHat \cite{Berger:2008sj}, GoSam \cite{Cullen:2011ac}, HELAC-NLO \cite{Bevilacqua:2011xh} and Madgraph5-aMC@NLO~\cite{Alwall:2011uj,Frixione:2002ik,Hirschi:2011pa}. The last two Monte Carlo programs are general purpose ones: the user inputs the process to be simulated and the programs provide the complete NLO answer by combining virtual and real contributions, including merging with parton shower and hadronization effects. For instance, realistic NLO simulations of $W$ + 2 jets \cite{Frederix:2011ig} production and  $pp \to H t \bar t$ \cite{Frederix:2011zi} can be obtained in a completely automated fashion within the Madgraph5-aMC@NLO framework\footnote{See {\tt http://amcatnlo.cern.ch} for more examples.}.

The idea of getting the loop amplitude from its integrand (or equivalently from its cuts) can be generalized beyond one-loop \cite{Mastrolia:2011pr,Kleiss:2012yv,Badger:2012dp,Johansson:2012zv}, with the important difference that no minimal basis for multi-loop integrals is known. A particularly interesting approach is the multivariate polynomial division \cite{Mastrolia:2012an,vanDeurzen:2013vga}, which generalized OPP to multi-loop integrands, although the field is still in its infancy compared with the full automation achieved at one-loop.

\subsubsection{Supersymmetric gauge theories}

Several new results about strongly-coupled supersymmetric field theories have been developed in the last several years. The work in \cite{Pestun:2007rz} showed how  supersymmetric localization can be used to derive  exact results in four-dimensional ${\cal N}=2$ and ${\cal N}=4$ supersymmetric gauge theories. The main point of this important result is that, under certain conditions, supersymmetric field theories can be placed on compact spheres while preserving the action of a supercharge.  One can then demonstrate that the full path integral of the theory\,---\,even with the insertion of certain supersymmetric operators\,---\,reduces to a finite dimensional integral (matrix model) over configurations preserving the unbroken supercharge. This makes possible the exact, nonperturbative computation of partition functions, Wilson and 't Hooft loop expectation values, and other observables in several supersymmetric theories in two, three, and four spacetime dimensions.

In parallel, a large class of four-dimensional superconformal field theories with ${\cal N}=2$ supersymmetry was discovered \cite{Gaiotto:2009we}, which do not always have a weakly coupled Lagrangian description. These theories can be engineered in string theory by wrapping multiple M5 branes\footnote{{M5 branes are membranes of 5+1 spacetime dimensions in 11 dimensional M-theory.}} on Riemann surfaces and they have interesting mathematical structure and dualities between them. This led to the discovery of the Alday-Galotto-Tachikawa correspondence \cite{Alday:2009aq}, which relates partition functions (and other observables) in four-dimensional
supersymmetric field theory, to correlation functions in certain two-dimensional CFTs.

\subsubsection{Conformal field theories}

CFTs constitute an important class of quantum field theories. An ambitious long-standing goal is to study conformal field theories by the method of {\it conformal bootstrap}. CFTs have the property that all correlation
functions can be computed given the spectrum (dimensions and spins of local operators) and operator product expansion (OPE) coefficients. By performing successive OPEs,  any correlation function can be computed in terms of these basic CFT data. Requiring the consistency of the OPE expansion in all possible channels leads to an infinite set of equations for the spectrum and OPE coefficients, which are known as the {\it conformal bootstrap} or {\it crossing symmetry} equations. These equations are exact and hold beyond perturbation theory. It is, however, difficult to extract useful data from them, as they are an infinite set of equations for an infinite number of variables.

In recent years progress has been made in extracting concrete, rigorous and universal constraints for higher-dimensional CFTs from the conformal bootstrap equations. This work began  with \cite{Rattazzi:2008pe}, which demonstrated how\,---\,in certain CFTs\,---\,the conformal bootstrap can provide bounds for the conformal dimensions of certain operators. For this analysis, the explicit expressions for 4d conformal blocks, first discovered in \cite{Dolan:2000ut, Dolan:2003hv}, played a crucial role. Subsequently, similar methods have been applied to derive bounds to OPE coefficients, central charges, and other aspects of the spectra of higher-dimensional CFTs. More recently, the conformal bootstrap has been applied towards solving the 3d Ising model \cite{ElShowk:2012ht}. Interesting new constraints for the spectrum of CFTs can be found by considering the bootstrap equations in the Lorentzian regime \cite{Fitzpatrick:2012yx,Komargodski:2012ek}.

In a different direction, a remarkable new result has been the proof of the so-called $a$-theorem \cite{Komargodski:2011vj}. By design, RG transformations integrate out degrees of freedom.  Therefore, if two QFTs are connected by an RG flow, one expects the UV theory to contain more degrees of freedom than the IR theory. In two-dimensional theories this feature is expressed by the Zamolodchikov c-theorem \cite{Zamolodchikov:1986gt}. In 4d CFTs Cardy \cite{Cardy:1988cwa} proposed that 
 a certain coefficient $a$ in the trace anomaly be used to count the degrees of freedom, and he conjectured that $a$ would decrease into the IR. Over twenty years later, Komargodski and Schwimmer \cite{Komargodski:2011vj}  proved that $a$ (appropriately defined away from the conformal point) does indeed decrease under RG flow. The now established $a$-theorem strongly suggests the irreversibility of the RG flow and can be used to verify the consistency of conjectured RG-flow relations between different QFTs.

Another interesting development  \cite{Maldacena:2011jn} proved the analogue of the Coleman-Mandula theorem for conformal field theories. It has been  demonstrated that if a CFT contains a single higher spin conserved charge, then it necessarily has to contain an infinite tower of higher spin conserved currents and additionally, the correlators of those currents have the form of free-field correlators. In subsequent work \cite{Maldacena:2012sf}, it was further proven that weakly broken higher spin symmetry is sufficient to constrain the leading order three-point functions.

\subsubsection{3d CFTs and higher spin symmetry}
\label{sec:G_3dCFT}

Significant progress has been made in the study of three-dimensional CFTs. A large class of such theories can be constructed by coupling Chern-Simons gauge theory to matter in various representations of the gauge group. Among them, of special importance are the Aharony-Bergman-Jafferis-Maldacena theories \cite{Aharony:2008ug} with matter in the bifundamental of the gauge group $U(N)_k\times U(N)_{-k}$, where $k$ is the Chern-Simons level. These  theories describe the low energy excitations of coincident M2 branes in M-theory. In the large $N$ limit they are holographically dual to M-theory on AdS$_4\times$S$^7/Z_k$ (or type IIA string theory on AdS$_4\times \mathbb{CP}^3$).

Three-dimensional Chern-Simons theory coupled to matter in the fundamental representation has also attracted attention recently. In the large $N$ limit these theories exhibit (slightly broken) higher-spin symmetry and interesting dualities between theories with bosons and fermions, a 3d version of ``bosonization" \cite{Giombi:2011kc, Aharony:2011jz, Aharony:2012nh, Aharony:2012ns}. These CFTs are important from a theoretical point of view, because they are believed to be  holographically dual to higher-spin gravity (of Vasiliev type) in AdS$_4$ \cite{Klebanov:2002ja, Sezgin:2003pt, Giombi:2012ms}. Vasiliev-type gravitational theories \cite{Vasiliev:1992av}, while more complicated than ordinary Einstein gravity in AdS, have vastly fewer fields than string theory and hence provide an example of AdS/CFT of intermediate complexity. Moreover, Chern-Simons fundamental matter theories provide examples of QFT-gravity (and QFT-QFT) dualities without any amount of supersymmetry. Finally, it has been proposed \cite{
Chang:2012kt} that there is a triality between a supersymmetric ${\cal N}=6$ version of Vasiliev gravity in AdS, the ABJ Chern-Simons-matter theory with gauge group $U(N)_k\times U(M)_{-k}$ and IIA string theory on AdS$_4\times \mathbb{CP}^3$ which might provide an understanding of closed strings in AdS as the flux tubes of (non-Abelian) Vasiliev theory.

Similar relations between higher spin CFTs and higher spin gravity have been discovered in lower dimensions. In \cite{Gaberdiel:2010pz} an AdS$_3$/CFT$_2$ type of duality has been proposed between the two-dimensional ${\cal W}_N$ minimal model CFTs and Vasiliev gravity in AdS$_3$. This duality is interesting because the boundary theory is exactly solvable and can serve as a useful toy-model for AdS/CFT. Further related developments are reviewed in \cite{Gaberdiel:2012uj}.

\subsection{Conformal symmetry, strongly coupled theories and new physics}
\label{sec:G_ConfBSM}

In this section we focus on non-Abelian gauge theories in four dimensions, and discuss the emergence of conformal symmetry when varying the matter content.
In the realm of four dimensions, the amount of exact results based on duality arguments is still limited if compared with theories in lower dimensions. It is also generally true that most of the theoretical arguments require exact supersymmetry. 
 
The existence of a conformal window, i.e., a family of theories that develop an attractive infrared fixed point (IRFP) at nonzero coupling and are deconfined with exact chiral symmetry at all couplings, has been long advocated for QCD with many flavors \cite{miransky_conformal_1997,appelquist_1996} and for supersymmetric QCD (SQCD) \cite{Seiberg:1994pq}.
The  structure of  the perturbative beta function \cite{caswell_asymptotic_1974,banks_phase_1982} for non-Abelian gauge theories without supersymmetry and the Novikov-Shifman-Vainshtein-Zakharov \cite{Novikov:1983uc} beta function of SQCD  suggest that a conformal window is a general feature of non-Abelian gauge theories with matter content, while its extent and location depend on the gauge group, the number of colors $N$, the number of flavors $N_f$ and the representation of the gauge group to which they belong.
The IRFP moves towards stronger coupling if the number of flavors is decreased, approaching the lower end of the conformal window.
This is the reason why only a genuinely nonperturbative study, possibly complemented by the existence of duality relations,  can establish the mechanism underlying its emergence or disappearance, its properties, and the differences between realizations with and without supersymmetry. The theory of the conformal window is further discussed in Sec.~\ref{sec:G_cw}.

Interestingly, theories just below the conformal window may develop a precursor near-conformal behavior, characterized by a slower change of the running coupling with the energy scale (``walking")  and provide a potentially  interesting class of candidates for BSM physics and the EWSB mechanism.
The interplay of lattice field theory and AdS/CFT in this context will be considered in Sec.~\ref{sec:CFT_EWSB_LAT_models}, while strongly coupled BSM candidates and LHC constraints will be more extensively discussed in Sec.~\ref{sec:G_EWSB}. There, we also review the appealing possibility that conformal symmetry and its spontaneous breaking may play a role up to  the Planck scale.

\subsubsection{ Theory of the conformal window}
\label{sec:G_cw}

\begin{figure}[tb]
  \includegraphics*[width=\figwid]{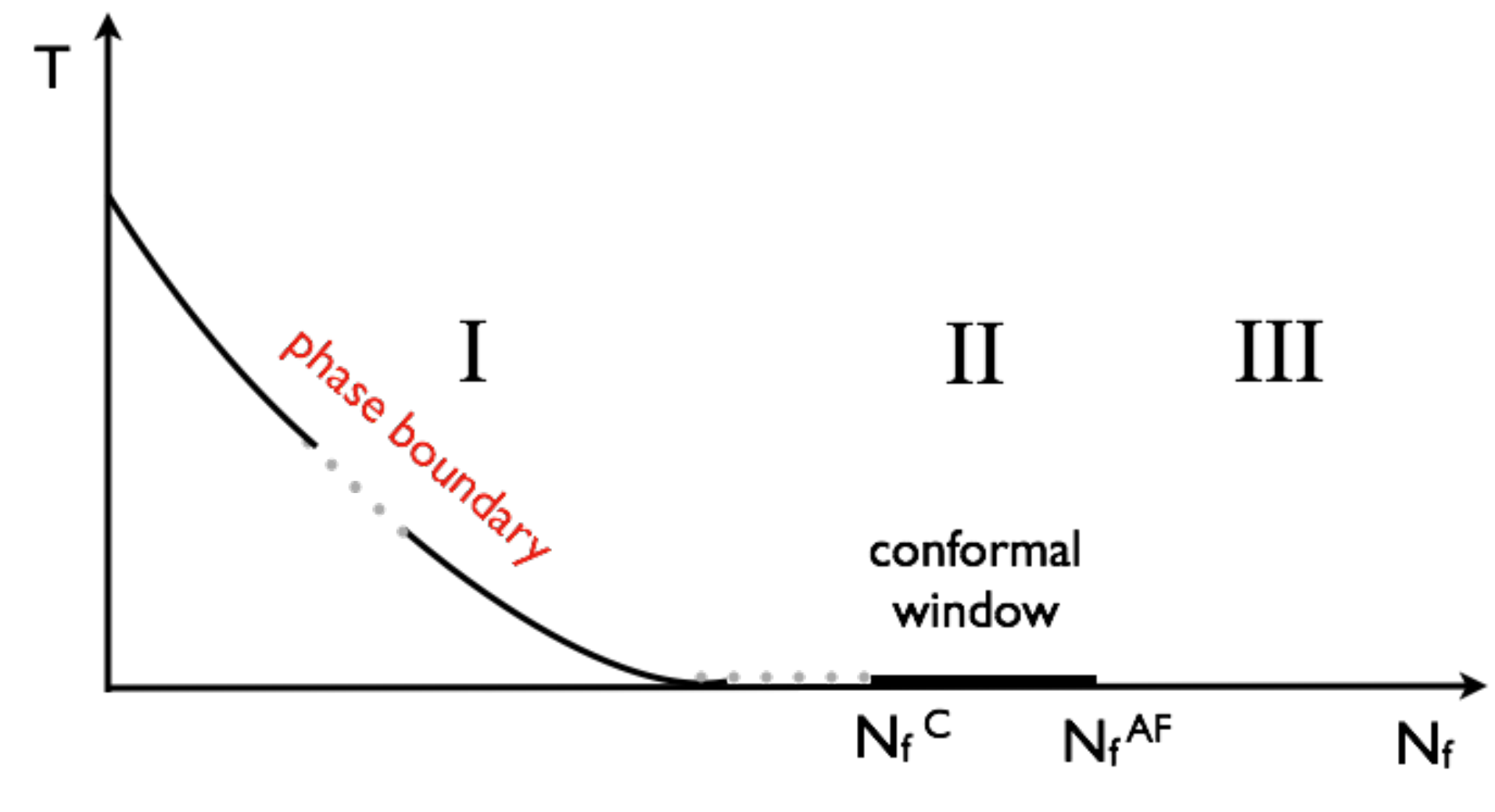}
   \caption{Temperature ($T$) and number of flavors ($N_f$) phase diagram for a generic non-Abelian gauge theory at zero density. In region I, one or more phase boundaries separate a low-temperature region from a high-temperature region. The nature of the phase boundary and which symmetries identify the two phases, in particular the interplay of confinement and chiral symmetry breaking, depend on the fermion representation and the presence or absence of supersymmetry. Region II identifies the conformal window at zero temperature, for $N_f^c<N_f<N_f^{AF}$, while region III is where the theory is no longer asymptotically free.
            }
\label{fig:Phase_TNf}
\end{figure}
Figure~\ref{fig:Phase_TNf} summarizes the salient features of the phase diagram of non-Abelian gauge theories with massless fermions in the temperature--flavor-number ($T$-$N_f$) plane.
 In particular, we identify three regions from left to right: region I  describes one or more families of theories below the conformal window, region II identifies the conformal window above a critical flavor number and before the loss of asymptotic freedom $N_f^c<N_f<N_f^{AF}$, while region III is where theories have lost asymptotic (UV) freedom for $N_f>N_f^{AF}$. Details of region I depend on  the way in which deconfinement and chiral symmetry restoration are realized at finite temperature. This realization in turn depends upon the transformation of the fermionic matter representations under the gauge group and the presence or absence of supersymmetry.   The simplest realization of region I is provided by QCD, i.e., the case of fermions in the fundamental representation of the $SU(N)$ gauge group with $N$ colors. In this case, chiral symmetry is broken at zero temperature for any $N_f<N_f^c$, and a chiral phase boundary\,---\,a line of thermal chiral phase transitions (or crossovers)\,---\,separates the low-temperature chirally-broken phase from the high-temperature chirally-restored phase.

The region above the chiral phase boundary, at low $N_f$, describes the strongly coupled quark-gluon plasma (QGP). 
Just above the phase boundary, QCD and ${\cal N}=4$ SYM carry similar features according to the AdS/CFT correspondence. Both predict the QGP to be a strongly coupled, nearly ideal fluid. 
 The two descriptions should depart from each other at higher temperatures, where QCD becomes weakly coupled while ${\cal N}=4$ SYM remains strongly coupled. Properties of the QGP are reviewed in Chapter~\ref{sec:chapd} of this document.

In QCD, the presence of a single true order parameter, i.e., the chiral condensate associated with chiral symmetry, suggests that the end point of the finite temperature chiral phase boundary in Fig.~\ref{fig:Phase_TNf}  should be identified with the lower end of the conformal window.  A phase transition would signal its opening  at some $N_f^c$ (region II in Fig.~\ref{fig:Phase_TNf}), and chiral symmetry is restored for theories inside the conformal window. Lattice studies \cite{Deuzeman2010} support this scenario, where theories inside the conformal window appear to be chirally symmetric also away from the IRFP. Eventually, chiral symmetry is expected to be broken again at sufficiently strong coupling in the lattice theory. An interesting observation is that chiral symmetry could  also be broken by the emergence of a UVFP at strong coupling, signaling the appearance of a new continuum field theory. 

While the nature of the phase transition at $N_f^c$ is yet to be uncovered, it is natural to expect that the chiral dynamics plays a role in its appearance. In fact, it has been suggested \cite{miransky_conformal_1997,appelquist_1996} that a phase transition of the Berezinskii-Kosterlitz-Thouless (BKT) type (conformal phase transition) should be expected when the chiral dynamics is taken into account. Such a phase transition would be  signaled  by a preconformal scaling  of chiral observables \cite{miransky_conformal_1997,appelquist_1996, braun_chiral_2006, Kaplan:2009kr}  just below $N_f^c$, known as BKT or Miransky scaling.

Moving to different fermion representations, lattice results \cite{Kogut:1984sb} suggest that QCD with fermions in the adjoint representation develops an intermediate phase at finite temperature, for a given $N_f$ in region I, where the theory is deconfined with broken chiral symmetry. In other words the restoration of chiral symmetry would occur at temperatures higher than the deconfinement temperature, i.e., $T_{ch} > T_{dec}$. In this scenario, it is plausible to expect that the two phase boundaries should merge at the lower end of the conformal window, thus $T_{ch}=T_{dec}=0$ for $N_f=N_f^c$.

Supersymmetric QCD offers yet another realization of region I, where the dual, free magnetic phase for $N+2<N_f<N_f^c=3N/2$ \cite{Seiberg:1994pq} implies a confined electric phase, where chiral symmetry is not yet broken \cite{Strassler:2001px}.  A better understanding of the interplay of chiral symmetry and confinement in the presence of supersymmetry might also shed light into some aspects of the nonsupersymmetric case \cite{Kogut:1985pp, Abel:2012un}.
Region I of nonsupersymmetric theories is being currently explored on the lattice \cite{Liao:2012tw, Miura:2011mc, Appelquist:2012nz}.

As said before, the conformal window in region II identifies theories with $N_f^c<N_f<N_f^{AF}$; they have a conformal IRFP and  are deconfined with exact chiral symmetry at zero temperature. The existence of a  conformal window and its properties can thus be established not only by directly probing the IRFP\,---\,a delicate task for lattice simulations\,---\,but also indirectly through the inspection of chiral observables, confinement indicators, the spectrum of low-lying states, and more generally by  identifying the symmetry properties of the weak and strong coupling sides of the IRFP.   The latter strategy was advocated in \cite{Deuzeman2010}, while the running gauge coupling has also been studied in \cite{appelquist_lattice_2008, appelquist_lattice_2009, Petropoulos:2012mg}, and strategies to directly probe conformality at the IRFP have been explored in \cite{DelDebbio:2010ze,Patella:2012da,DeGrand:2011cu}.
Region III is where theories are no longer asymptotically free.
The weak coupling beta function is now positive and the theory is free in the infrared. The emergence of a UVFP at stronger coupling would make these theories interesting.

There are still many questions to be answered.
What is the detailed nature of the finite temperature phase boundary, and what is the interplay  of confinement and chiral symmetry breaking for theories in region I?
What is the nature of the phase transition that opens the conformal window in region II? And what is the fate of the IRFP at $N_f^c$?  The IRFP coupling can (i) flow to zero, (ii) flow to infinity, (iii) flow to a finite value at which a discontinuity occurs, see, e.g., Ref.~\cite{Antipin:2012sm}, (iv) merge with a UVFP \cite{Kaplan:2009kr}.
The latter can only be realized if the UVFP is developed at strong coupling for theories inside region II, or simply at its lower-end.
The AdS/CFT correspondence can in principle be a useful and complementary tool to explore these scenarios, for CFTs in four or lower dimensions. Preliminary attempts can be found in \cite{Maldacena:2000yy,Casero:2007jj, Casero:2006pt, Conde:2011rg, Barranco:2011vt}.

\subsubsection{Lattice, AdS/CFT, and the electroweak symmetry breaking}
\label{sec:CFT_EWSB_LAT_models}
Lattice studies of non-Abelian gauge theories just below the conformal window  aim to establish or exclude the existence of a preconformal behavior, characterized by an almost zero beta function (to which we have referred previously as the walking regime) and a preconformal scaling of the finite temperature phase boundary and the chiral observables.
These theories are expected to be rather strongly coupled, confining in a broad sense  and have chiral symmetry spontaneously broken at zero temperature. Depending on their specific matter content (Goldstone bosons and resonances), they may be viable candidates for EWSB and BSM physics at the (multi) TeV scale.

The pattern of color $N$ and flavor $N_f$ dependence of their beta functions is sufficient to infer that, for fixed $N$, the conformal window shifts to lower $N_f$ and shrinks when increasing the Casimir of the fermion representation. Furthermore, lowering $N$ is qualitatively equivalent to increasing $N_f$.
Hence, a preconformal behavior with minimal fermionic content could be realized by gauge groups with $N=2$ or 3 and Dirac fermions in representations higher than the fundamental (adjoint, two-index symmetric and two-index antisymmetric), or mixed Weyl and Dirac fermions in the fundamental and nonfundamental representations; for example, the conformal window of $SU(2)$ with adjoint fermions is expected to open at about $N_f=2$. This theory and other variations are  extensively studied on the lattice, see, e.g., Ref.~\cite{catterall_phase_2008, DeGrand:2013uha, Fodor:2012ty, DeGrand:2012qa}.
The conformal window for $SU(3)$ with fermions in the fundamental representation is, in contrast, expected to open in the surroundings of $N_f=12$, and most results suggest the range between $N_f=8$ and $N_f=12$ \cite{Deuzeman2010,deuzeman_physics_2008, appelquist_lattice_2009, Fodor:2011tu, Aoki:2012eq}. This theory offers an optimal playground for the theoretical understanding of the emergence of conformality and its connection with QCD and QGP physics.
Notice also that a preconformal regime with a lower fermionic content in the fundamental representation can be obtained by lowering the color content to $N=2$.

It is also important to observe that the most traditional lattice strategies, well tested and optimized in the context of QCD, can be far from optimal when studying the theory inside or close to the conformal window and at strong coupling. This is due to the different symmetry patterns and structure of the beta function for QCD as compared to theories inside the conformal window, and the fact that many optimization methods for lattice QCD have been devised to work close to the continuum limit, at rather weak coupling. 
It has recently been shown \cite{Deuzeman2010a,Deuzeman:2011pa,Deuzeman:2012ee,Cheng:2011ic} how the Symanzik improvement program and its generalizations inherited from QCD can lead to exotic phases, genuine lattice artifacts, when used in the study of these systems at strong coupling. The same conclusions may be generalized to the lattice study of strongly coupled condensed matter systems such as graphene \cite{Deuzeman:2012ee}; the latter is a QED system with a chiral symmetry breaking transition at strong coupling, analogous in many respects to theories inside the conformal window in the QED-like region at the strong-coupling side of the IRFP. 
For reviews and a more complete list of references to recent work see, e.g., Refs.~\cite{GiedtLAT,Neil:2012cb,DelDebbio:2010zz,Pallante:2009hu}.

The genuinely nonperturbative nature of the lattice formulation for theories inside or just below the conformal window allows,  in principle, exploring all salient aspects of their dynamics, in particular the mass ratio of the vector and scalar low-lying states, their first excitations, the pseudo-Goldstone boson decay constant, and the anomalous dimension of the fermion mass operator at the would-be IRFP.
The relevance of higher dimensional operators, such as four-fermion operators, can also be explored,  as well as the Yukawa interaction with a scalar field and/or the addition of a dilaton.
By varying the details of the interaction Lagrangian and of a Higgs-dilaton potential, one can  explore  the nonperturbative regime of an entire class of models, from Higgsless, to composite Higgs and dilaton-Higgs models. Also, the lattice study of Yukawa-Higgs models provides a genuinely nonperturbative information on the stability of the Higgs potential and the UV safety of the SM \cite{Bulava:2012rb,Fodor:1994sj, Gerhold:2009ub, Gerhold:2007yb}.

Exploiting the AdS/CFT correspondence  beyond ${\cal N}=4$ SYM in four spacetime dimensions is not a straightforward task.
The first steps in this direction aimed to find the AdS realizations that are approximately dual to ${\cal N}=1$ SYM  \cite{Maldacena:2000yy}, or to SQCD with $N_f$ dynamical flavors in the fundamental representation \cite{Casero:2007jj, Casero:2006pt, Conde:2011rg}.  The beta functions of the approximately dual gauge theories can also be studied \cite{Barranco:2011vt}, in order to explore possible realizations of the conformal window and the relevance of Kaluza-Klein excitations. The latter do not decouple in general,  and give rise to  higher-dimensional operators in the dual gauge theory.  While still in their infancy, these studies may provide useful insights into the role of supersymmetry for the emergence of conformality and the interplay of chiral symmetry and confinement.  Leaving aside AdS/CFT and supersymmetry, a recent attempt to derive the large-$N$ Yang-Mills beta function and the glueball spectrum from first principles \cite{Bochicchio:2013eda,Bochicchio:2008vt} may finally help to clarify the relevant differences between supersymmetric and nonsupersymmetric theories, and eventually suggest a new class of dualities for nonsupersymmetric gauge theories.

\subsection{Electroweak symmetry breaking}
\label{sec:G_EWSB}

A new Higgs-like boson with mass $M_H = 125.64 \pm 0.35$~GeV has been discovered at the LHC \cite{Aad:2012tfa,ATLAS:2013sla,Chatrchyan:2012ufa,Chatrchyan:2013lba}, with a spin/parity consistent with the SM assignment $J^P=0^+$ \cite{Aad:2013xqa,Chatrchyan:2012jja}. Although its properties are not yet precisely measured, it complies with the expected behavior and therefore it is a very compelling candidate to be the SM Higgs \cite{Pich:2013vta}.
An obvious question to address is the extent to which alternative scenarios of EWSB remain viable. In particular, what are the implications for strongly-coupled models
in which the electroweak symmetry is broken dynamically? Alternatively, can a minimally extended SM be a valid theory up to the Planck scale?

\subsubsection{Strongly coupled scenarios for EWSB }
\label{sec:G_BSM_SC}

Usually, strongly-coupled theories do not contain a fundamental Higgs field, bringing instead resonances of different types as in QCD. For instance, Technicolor \cite{Weinberg:1979bn,Weinberg:1975gm,Susskind:1978ms},
the most studied strongly-coupled model, introduces an asymptotically-free QCD replica at TeV energies which breaks the electroweak symmetry in the infrared,
in a similar way as chiral symmetry is broken in QCD.
This gives rise to the appearance of a tower of heavy resonances
in the scattering amplitudes.
Other models consider the possibility that the ultraviolet theory remains close to a
strongly-interacting conformal fixed point over a wide range of energies
(Walking Technicolor) \cite{Holdom:1984sk,Holdom:1981rm,Appelquist:1986an,Yamawaki:1985zg}; recent work in this direction incorporates conformal field theory techniques (Conformal Technicolor)
\cite{Luty:2004ye,Galloway:2010bp,Rattazzi:2008pe}.
Strongly-coupled models in warped \cite{Randall:1999ee} or deconstructed \cite{ArkaniHamed:2001ca,Georgi:2004iy,Hill:2000mu} extra dimensions~\cite{Csaki:2003zu,Csaki:2003dt,Cacciapaglia:2004jz,Cacciapaglia:2004rb,SekharChivukula:2001hz,Chivukula:2002ej,Chivukula:2003kq,Chivukula:2004pk,Chivukula:2004mu,Agashe:2003zs,Burdman:2003ya,Agashe:2004rs,Contino:2006qr,Foadi:2003xa,Hirn:2004ze,Casalbuoni:2004id,Perelstein:2004sc}
have been also investigated.

The recently discovered scalar boson could indeed be a first experimental signal of a new strongly-interacting sector: the lightest state of a large variety of new resonances of different types. Among the many possibilities, the relatively light mass of the discovered Higgs candidate has boosted the interest \cite{Espinosa:2010vn,Contino:2010mh,Azatov:2012bz}
in strongly-coupled scenarios with a composite pseudo-Goldstone Higgs boson
\cite{Kaplan:1983fs,Kaplan:1983sm,Georgi:1984ef,Georgi:1984af,Dugan:1984hq,Dimopoulos:1981xc}, where the Higgs mass is protected by an approximate global symmetry and is only
generated via quantum effects. Another possibility would be to interpret the  Higgs-like scalar as a dilaton, the pseudo-Goldstone boson associated with the spontaneous breaking of scale (conformal) invariance
\cite{Goldberger:2007zk,Matsuzaki:2012mk,Matsuzaki:2012xx,Bellazzini:2012vz,Chacko:2012vm,Elander:2012fk}.

In the absence of direct evidence of a particular ultraviolet completion, one should investigate the present phenomenological constraints, independently of any
specific implementation of the EWSB.
The precision electroweak data confirm the
$SU(2)_L\times SU(2)_R\rightarrow SU(2)_{L+R}$
pattern of symmetry breaking, giving rise to three Goldstone bosons which, in the
unitary gauge, become the longitudinal polarizations of the gauge bosons.
When the $U(1)_Y$ coupling $g'$ is neglected, the electroweak Goldstone dynamics
is described at low energies by the same Lagrangian as
the QCD pions, replacing the pion decay constant by the
EWSB scale $v=(\sqrt{2}\, G_F)^{-1/2} = 246\,$GeV~\cite{Appelquist:1980vg}.
Contrary to the SM, in strongly-coupled scenarios the symmetry is nonlinearly realized.

The dynamics of Goldstones and massive resonance states can be analyzed in a generic way by using an effective Lagrangian based on a $SU(2)_L\times SU(2)_R$ symmetry, spontaneously broken to the diagonal subgroup $SU(2)_{L+R}$.
The theoretical framework is analogous to the Resonance Chiral Theory description of QCD at GeV energies~\cite{Ecker:1988te,Ecker:1989yg,Cirigliano:2006hb}.
Let us consider a low-energy effective theory containing the SM gauge bosons coupled
to the electroweak Goldstone bosons and the light scalar state $S_1$ with mass $m_{S_1} = 126$~GeV, discovered at the LHC, which is assumed to be an $SU(2)_{L+R}$ singlet. We also include the lightest vector and axial-vector triplet multiplets, $V_{\mu\nu}$ and $A_{\mu\nu}$, with masses $M_V$ and $M_A$, respectively. To lowest order in derivatives and number of resonance fields \cite{Pich:2012jv,Pich:2012dv,Pich:2013fea},
\begin{eqnarray}\label{eq:Lagrangian}
\mathcal{L}\; &=\; &
\frac{v^2}{4}\,\langle u_\mu u^\mu \rangle\,\left( 1 + \frac{2\,\omega}{v}\, S_1\right)
 + \frac{F_A}{2\sqrt{2}}\, \langle A_{\mu\nu} f^{\mu\nu}_- \rangle
\nonumber \\ &&\mbox{}
+ \frac{F_V}{2\sqrt{2}}\, \langle V_{\mu\nu} f^{\mu\nu}_+ \rangle
+ \frac{i\, G_V}{2\sqrt{2}}\, \langle V_{\mu\nu} [u^\mu, u^\nu] \rangle
\nonumber \\ &&\mbox{}
+ \sqrt{2}\, \lambda_1^{SA}\,  \partial_\mu S_1 \, \langle A^{\mu \nu} u_\nu \rangle\, ,
\end{eqnarray}
plus the gauge boson and resonance kinetic terms.
The electroweak Goldstone fields $\vec\varphi(x)$ are parameterized through the matrix
$U=u^2= \exp{\left\{ i \vec{\sigma}\cdot \vec{\varphi} / v \right\} }$,
$u^\mu = -i\, u^\dagger  D^\mu U\, u^\dagger$ with $D^\mu$
the appropriate gauge-covariant derivative, and
$\langle A\rangle$ stands for the trace of the $2\times 2$ matrix $A$.
The first term in Eq.~(\ref{eq:Lagrangian}) gives the Goldstone Lagrangian, present in the SM, plus the scalar-Goldstone interactions.
For $\omega=1$ one recovers the $S_1\to\varphi\varphi$ vertex of the SM.
The $F_V$ and $F_A$ terms incorporate direct couplings of the vector and axial-vector resonances with the gauge fields through
$f^{\mu\nu}_\pm = -\frac{g}{2}\, u^\dagger \vec\sigma\,\vec{W}^{\mu\nu} u
\mp \frac{g'}{2} u \sigma_3 B^{\mu\nu} u^\dagger$.

The presence of massive states coupled to the gauge bosons modifies the $Z$ and $W^\pm$ self-energies,
which are characterized by the so-called oblique parameters $S$ and $T$ \cite{Peskin:1991sw,Peskin:1990zt}.
$S$ measures the difference between the off-diagonal $W^3B$ correlator and its SM value, while $T$ parameterizes the
difference between the $W^3$ and $W^\pm$ self-energies, after
subtracting the SM contribution. To define the SM correlators, one needs a reference value for the SM Higgs mass; taking it at $m_{S_1}= 126$ GeV,
the global fit to electroweak precision data gives the constraints
$S = 0.03\pm 0.10$ and $T=0.05\pm0.12$~\cite{Baak:2012kk}.

The oblique parameter $S$ receives tree-level contributions from vector and axial-vector exchanges \cite{Peskin:1991sw,Peskin:1990zt}, while
$T$ is identically zero at lowest-order (LO):
\begin{equation}
S_{\mathrm{LO}} = 4\pi \left( \frac{F_V^2}{M_V^2}\! -\! \frac{F_A^2}{M_A^2} \right)  \,,
\qquad\quad
T_{\mathrm{LO}}=0 \,.
\label{eq:LO}
\end{equation}

Assuming that weak isospin and parity are good symmetries of the strong dynamics, the $W^3 B$ correlator
is proportional to the difference of the vector and axial-vector two-point Green's functions.
In asymptotically-free gauge theories this difference vanishes as $1/s^3$ at $s\to\infty$  \cite{Bernard:1975cd}, implying two super-convergent sum rules,
known as the first and second Weinberg sum rules (WSRs)~\cite{Weinberg:1967kj}. At LO they give the identities
\begin{equation}
F_{V}^2 - F_{A}^2  = v^2\, ,
\qquad\quad
F_{V}^2  \,M_{V}^2 - F_{A}^2 \, M_{A}^2  = 0\, ,
\end{equation}
which relate  $F_V$ and $F_A$ to the resonance masses, leading to
\begin{equation}
S_{\mathrm{LO}}\; =\; \frac{4\pi v^2}{M_V^2}\,  \left( 1 + \frac{M_V^2}{M_A^2} \right) \, .
\end{equation}
Since the WSRs also imply $M_A>M_V$, this prediction turns out to be bounded by~\cite{Pich:2012jv}
\begin{equation}
\frac{4\pi v^2}{M_V^2} \; < \; S_{\rm   LO} \; < \;   \frac{8 \pi v^2}{M_V^2} \, . \label{eq:SLOtwoWSR}
\end{equation}
It is likely that the first WSR is also true in gauge theories
with nontrivial ultraviolet fixed points~\cite{Orgogozo:2011kq,Appelquist:1998xf},  while
the second WSR is questionable in some scenarios.
If only the first WSR is considered,
but still assuming the hierarchy  $M_A>M_V$,
the lower bound in Eq.~(\ref{eq:SLOtwoWSR}) remains~\cite{Pich:2012jv}.

The allowed experimental range for $S$ implies that $M_V$ is larger than 1.5 (2.4) TeV at 95\% (68\%) CL. Thus, strongly-coupled models of EWSB should have a quite high dynamical mass scale. While this was often considered as an undesirable property, it fits very well with the LHC findings, which are pushing the scale of new physics beyond the TeV region. It also justifies our approximation of only considering the lightest resonance multiplets.

The experimental constraints on $S$ and $T$ depend on the chosen reference value for the SM Higgs mass, which we have taken at $m_{S_1}$. Since the SM Higgs contribution only appears at the one-loop level,
there is a scale ambiguity when comparing a LO theoretical result with
the experimental measurements, making necessary to consider NLO corrections~\cite{Pich:2012jv,Pich:2012dv,Pich:2013fea,Orgogozo:2011kq,Matsuzaki:2006wn,Barbieri:2008cc,Cata:2010bv,Foadi:2012ga,Orgogozo:2012ct}.
Imposing proper short-distance conditions on the vector and axial-vector correlators, the NLO contributions to $S$ from $\varphi\varphi$, $V\varphi$ and $A\varphi$ loops have been evaluated in \cite{Pich:2012jv}. These corrections are small and strengthen  the lower bound on the resonance mass scale slightly.

Much more important is the presence of a light scalar resonance with $m_{S_1}= 126$ GeV. Although it does not contribute at LO, there exist sizable $S_1 B$ ($S_1\varphi$) loop contributions to $T$ ($S$). Neglecting the mass-suppressed loop corrections from vector and axial-vector resonances and terms of $\mathcal{O}(m_{S_1}^2/M_{V,A}^2)$,
one finds \cite{Pich:2012dv,Pich:2013fea}
\begin{equation}
 T =  \frac{3}{16\pi \cos^2 \theta_W} \bigg[ 1 + \log \frac{m_{S_1}^2}{M_V^2}
 - \omega^2 \left( 1 + \log \frac{m_{S_1}^2}{M_A^2} \right)  \bigg]  \, .
\label{eq:T}
\end{equation}
Enforcing the second WSR, one obtains the additional constraint
$\omega = M_V^2/M_A^2$, which requires this coupling to be in the range $0\leq \omega \leq 1$, and \cite{Pich:2012dv,Pich:2013fea}
\begin{eqnarray}
S  &= &   \frac{4 \pi v^2}{M_{V}^2} \left(1+\frac{M_V^2}{M_{A}^2}\right) + \frac{1}{12\pi}
\bigg[ \log\frac{M_V^2}{m_{S_1}^2}  -\frac{11}{6}
\nonumber  \\ &&
+\;\frac{M_V^2}{M_A^2}\log\frac{M_A^2}{M_V^2}
 - \frac{M_V^4}{M_A^4}\, \bigg(\log\frac{M_A^2}{m_{S_1}^2}-\frac{11}{6}\bigg) \bigg] \, .\quad
\label{eq:1+2WSR}
\end{eqnarray}

\begin{figure}[tb]
\includegraphics*[width=\figwid]{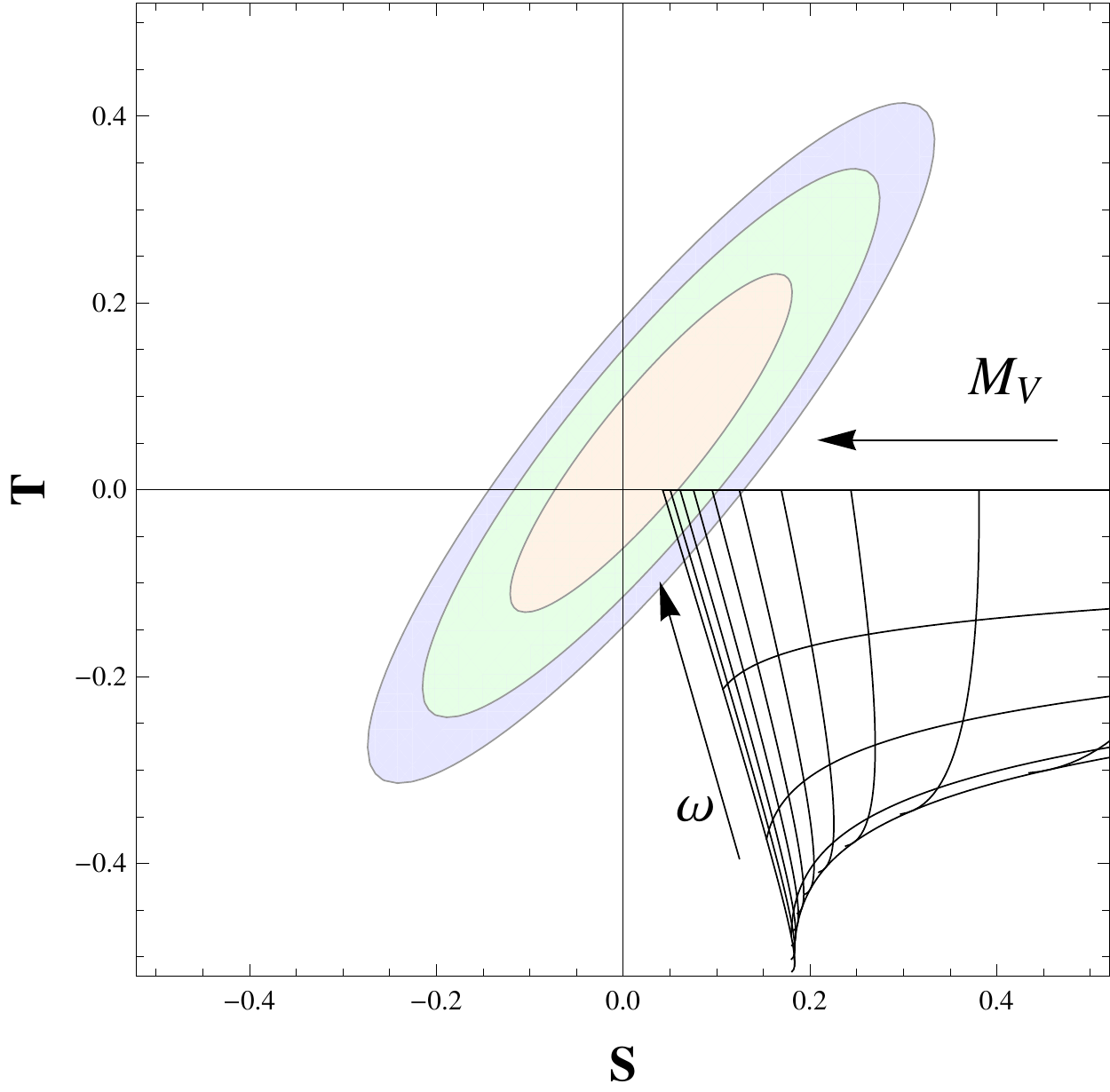}
\caption{\small{NLO determinations of $S$ and $T$, imposing the two WSRs.
The approximately vertical (horizontal) lines correspond to values
of $M_V$, from $1.5$ to $6.0$~TeV at intervals of $0.5$~TeV
($\omega= M_V^2/M_A^2$: $0.00, \, 0.25, 0.50, 0.75, 1.00$).
The arrows indicate the directions of growing  $M_V$ and $\omega$.
The ellipses give the experimentally allowed regions at 68\% (orange), 95\% (green) and 99\% (blue) CL \cite{Pich:2012dv}.}}
\label{fig:2WSR}
\end{figure}

These NLO predictions are compared with the experimental bounds
in Fig.~\ref{fig:2WSR}, for different values of $M_V$ and $\omega= M_V^2/M_A^2$. The line with $\omega= 1$ ($T=0$) coincides with the LO upper bound in Eq.~(\ref{eq:SLOtwoWSR}).
This figure demonstrates a very important result in the two-WSR scenario: the precision electroweak data require that the Higgs-like scalar should have a $WW$ coupling very close to the SM one. At 68\% (95\%) CL, one gets
$\omega\in [0.97,1]$  ($[0.94,1]$) \cite{Pich:2012dv,Pich:2013fea},
in nice agreement with the present LHC evidence \cite{Aad:2012tfa,ATLAS:2013sla,Chatrchyan:2012ufa,Chatrchyan:2013lba}, but much more restrictive.
Moreover, the vector and axial-vector states should be very heavy (and quite degenerate);
one finds $M_A \approx M_V> 5$~TeV ($4$~TeV) at 68\% (95\%) CL \cite{Pich:2012dv,Pich:2013fea}.

\begin{figure}[tb]
\includegraphics*[width=\figwid]{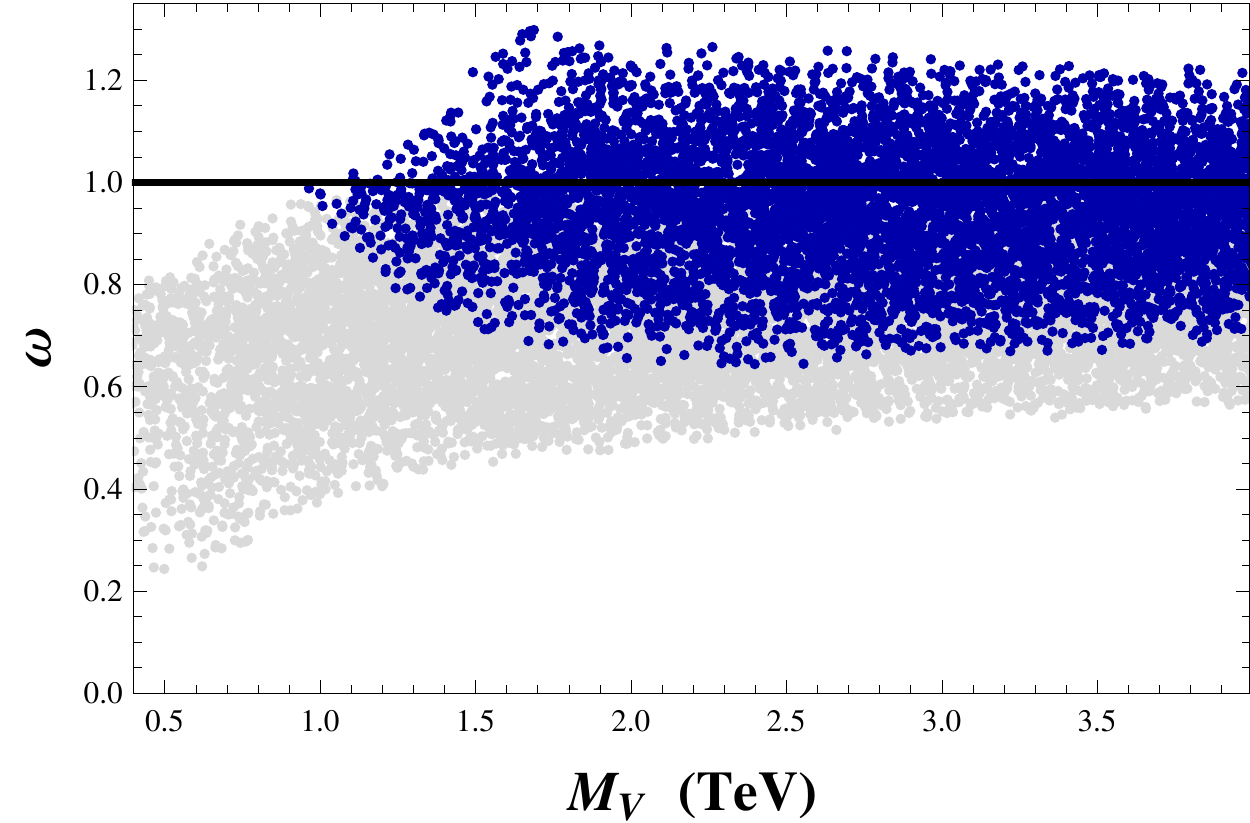}
\caption{\small Scatter plot for the 68\% CL region, in the case
when only the first WSR is assumed.
The dark blue and light gray regions
correspond, respectively,  to
$0.2<M_V/M_A<1$ and $0.02<M_V/M_A<0.2$ \cite{Pich:2012dv}.}
\label{fig:1WSR}
\end{figure}

If the second WSR is dropped, one can still obtain a lower bound at NLO
(assuming $M_V<M_A$):
\begin{equation}
S \geq   \frac{4 \pi v^2}{M_{V}^2} + \frac{1}{12\pi}  \bigg[ \log\frac{M_V^2}{m_{S_1}^2} -\frac{11}{6}
- \omega^2 \bigg(\!\log\frac{M_A^2}{m_{S_1}^2}-\frac{17}{6}
 + \frac{M_A^2}{M_V^2}\!\bigg) \bigg]  .
\label{eq:lower-bound-1WSR}
\end{equation}
In the limit $\omega \to 0$, this lower bound reproduces the corresponding result in Eq.~(\ref{eq:1+2WSR}), which is excluded by Fig.~\ref{fig:2WSR}. Thus, a vanishing scalar-Goldstone coupling would be incompatible with the data, independently of whether the second WSR is assumed.
Figure~\ref{fig:1WSR} shows the allowed 68\% CL region in the space of parameters $M_V$ and $\omega$, varying $M_V/M_A$ between 0 and 1. Values of $\omega$
very different from the SM and/or vector masses below the TeV scale
can only be obtained with a large splitting of the vector and axial-vector masses, which looks quite unnatural. In general there is no solution for $\omega >1.3$.
Requiring $0.2\, (0.5) <M_V/M_A<1$, leads to $1-\omega<0.4\, (0.16)$
and $M_V > 1\, (1.5)$~TeV \cite{Pich:2012dv,Pich:2013fea}.

The tree-level exchanges of the light Higgs-like boson regulate quite well the high-energy behavior of the longitudinal gauge-boson scattering amplitudes:
\begin{equation}\label{eq:WW}
\mathcal{M}(W_L^+W_L^-\to W_L^+W_L^-)\,\sim\, (1-\omega^2)\, u/v^2\, ,
\end{equation}
where $u$ is the usual Mandelstam variable.
With $\omega\approx 1$, the perturbative unitarity bounds can only be approached at very high energies, where the strongly-coupled dynamics will restore the right behavior
\cite{Espriu:2012ih,Delgado:2013loa,Delgado:2013hxa}.

These conclusions are quite generic, only using mild assumptions about the ultraviolet behavior of the underlying strongly-coupled theory, and they can be easily adapted to more specific models obeying the $SU(2)_L\times SU(2)_R\rightarrow SU(2)_{L+R}$ pattern of EWSB. For instance, in the $SO(5)/SO(4)$ minimal composite Higgs model \cite{Agashe:2004rs,Contino:2006qr}, the $S$ and $T$ constraints
are directly given by Fig.~\ref{fig:2WSR} with the identification
$\omega=\cos\theta\leq 1$, where $\theta$ is the $SO(4)$ vacuum angle~\cite{Giudice:2007fh,Contino:2010rs,Contino:2011np}.
A Higgs-like dilaton, associated with the spontaneous breaking of scale (conformal) invariance at the scale $f_\phi$ \cite{Goldberger:2007zk,Matsuzaki:2012mk,Matsuzaki:2012xx,Bellazzini:2012vz,Chacko:2012vm,Elander:2012fk}, would correspond to $\omega = v/f_\phi$.
The experimental constraints on $\omega$ discussed above require $f_\phi\sim v$, making unlikely this light-dilaton scenario.

Thus, strongly-coupled electroweak models are allowed by current data provided the resonance mass scale stays above the TeV scale and the light Higgs-like boson has a gauge coupling close to that of the SM. This has obvious implications for future LHC studies, since it leads to a SM-like scenario. A possible way out would be the existence of new light scalar degrees of freedom, sharing the strength of the SM gauge coupling; at available energies, this possibility could result in phenomenological signals similar to perturbative two-Higgs-doublet models \cite{Celis:2013rcs}.

Future progress requires a thorough investigation of the fermionic sector. The couplings of the Higgs-like scalar with ordinary fermions are not well known yet and could show deviations from the SM Yukawa interactions. Generally, a proper understanding of the pattern of fermion masses and mixings  is also missing; in particular, the huge difference between the top mass scale and the small masses of the light quarks or the tiny neutrino ones remains to be explained.

\subsubsection{Conformal symmetry, the Planck scale, and naturalness}
\label{sec:G_Planck}

Should the LHC experiments ultimately discover no new particles, beyond the Higgs-like boson at about 126 GeV, then entire families of BSM theories  would be excluded or would have to depart from naturalness \cite{Weinberg:1978ym,'tHooft:1979bh,Susskind:1978ms,Veltman:1977fy,Veltman:1976rt,Wilson:1970ag} in a substantial way; it would be true for all scenarios that invoke a relevant new energy scale $\Lambda_{EW}\lesssim \Lambda\ll \Lambda_{Planck}$, such as most versions of weakly-coupled supersymmetry or strongly-coupled compositeness.
One should, instead, aim to formulate a theoretically viable completion of the SM that does not imply a proliferation of new particles up to scales $\Lambda\lesssim\Lambda_{Planck}$, possibly embedding gravity.
In other words, to which extent is it possible to enhance the symmetries of the SM without enlarging its particle content?

It seems not accidental that a Higgs boson with a mass of about 126 GeV allows for a SM vacuum that is at least metastable, or perhaps stable \cite{Ellis:2009tp, Bezrukov:2012sa, Degrassi:2012ry, Alekhin:2012py, Antipin:2013sga, Bulava:2013ep, Jegerlehner:2013cta}, with the SM ultraviolet cutoff as high as the Planck scale. 
A precise determination of the boundary between the metastable and stable vacuum solution for the SM has become especially relevant after the discovery of the Higgs-like boson at the LHC. The full knowledge of the RG coefficients for all the SM parameters (gauge couplings, Yukawa couplings, masses and Higgs sector parameters), from the weak scale to the Planck scale, is necessary to establish the  fate of the SM vacuum. Most of the current predictions \cite{Bezrukov:2012sa, Degrassi:2012ry, Alekhin:2012py, Antipin:2013sga} suggest that the SM vacuum is at least metastable. Interestingly, 
the work in \cite{Jegerlehner:2013cta} concludes for a stable solution, accompanied by a first order phase transition at about ~$7\times 10^{16}$ GeV, above which the system is in the unbroken phase, i.e., the Higgs VEV vanishes; in this analysis the phase transition is induced by the zero in the coefficient of the quadratic divergence of the Higgs mass counterterm. It is interesting to explore further the implications of this scenario for inflation and baryogenesis. 
 
All the above mentioned results should be considered as work in progress, since predictions and their accuracy are still affected by theoretical uncertainties (such as higher order contributions in the perturbative expansion, or inclusion of operators with dimension higher than four), and by the experimental uncertainty on the top quark mass, the running strong coupling $\alpha_s$ and the Higgs mass itself. Calculations are currently done in perturbation theory, see, e.g., Refs.~\cite{Bezrukov:2012sa, Degrassi:2012ry, Alekhin:2012py, Antipin:2013sga, Jegerlehner:2013dpa, Jegerlehner:2013cta}, and on the lattice \cite{Bulava:2013ep}.

A vacuum that is at least metastable, with a lifetime longer than the age of the Universe, or stable would imply that the SM may be a valid effective field theory up to the Planck scale. It would not contradict the stringent bounds coming from flavor physics, on the contrary it would avoid the long-standing difficulties of most BSM models to produce tiny deviations from the SM predictions in all flavor sectors and for all relevant observables; among the latter are flavor changing neutral current  processes, radiative decays such as $b\to s\gamma$, and CP-violating observables, such as permanent electric dipole moments, see Chapter~\ref{sec:chape}.
Ultimately, we would like to answer the first question of all: what is the symmetry, if any, that protects the Higgs mass from running all the way to the Planck mass, or equivalently, what is the source of the large hierarchy $\Lambda_{EW}/\Lambda_{Planck}\simeq 10^{-16}$?
\begin{figure}[tb]
 \includegraphics*[width=\figwid]{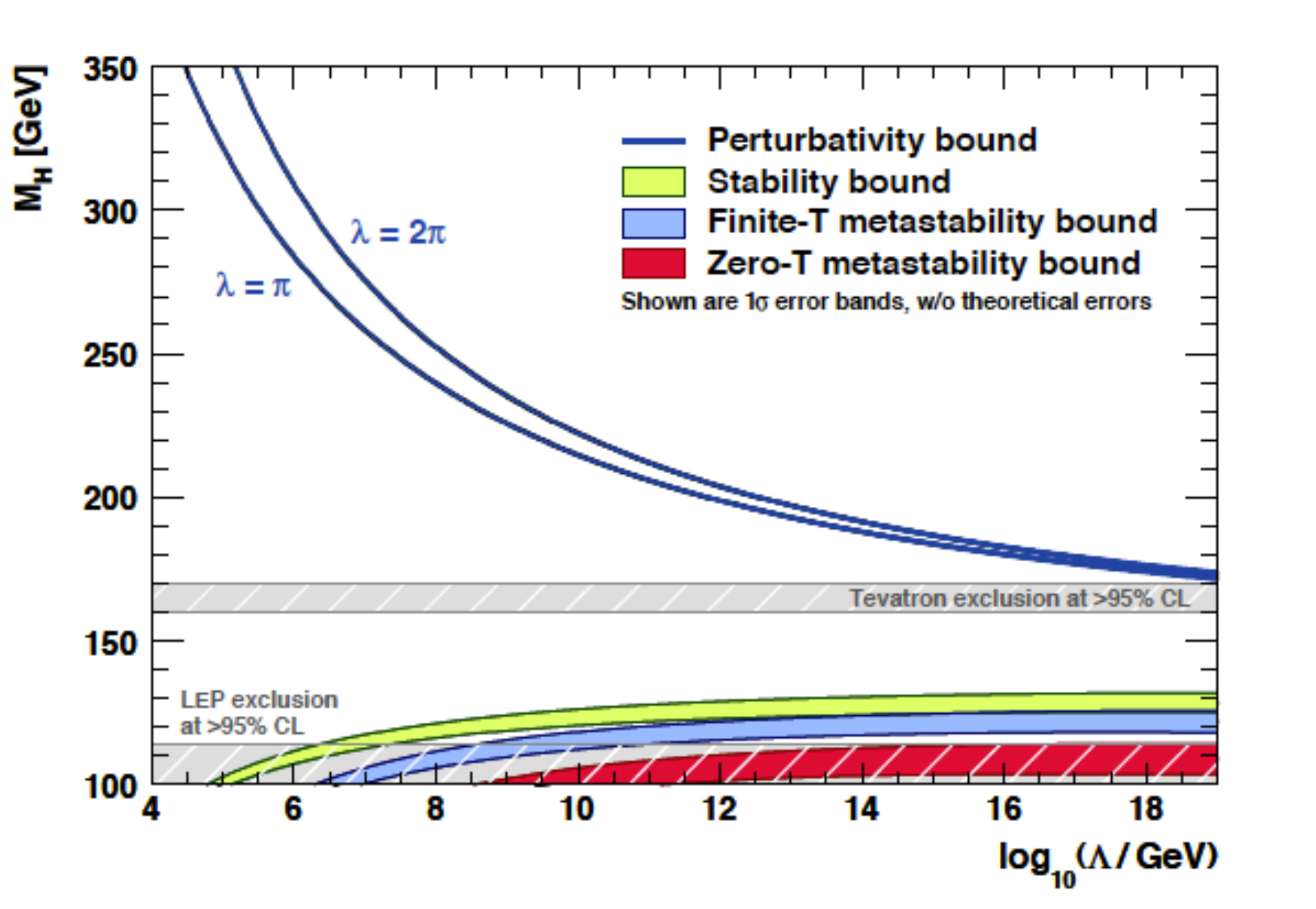}
   \caption{ An illustration of the trend of the stability bounds (lower curves) and perturbativity bounds (upper curves) for the SM vacuum from \cite{Ellis:2009tp} as function of the quartic self-coupling of the Higgs field and the Higgs boson mass. 
The determination of the boundary between the metastable and stable vacuum solution is  work in progress, and depends on the experimental uncertainty on the top quark mass, the running strong coupling $\alpha_s$, and the Higgs mass itself.
            }
\label{fig:Stability}
\end{figure}
The line of thought in \cite{Jegerlehner:2013cta} would answer this question without invoking an  underlying symmetry beyond the SM gauge group. Instead, it is the RG evolution of the quadratic divergences (treated as physical, for the theory with a finite UV cutoff) to protect the SM from instabilities \cite{Jegerlehner:2013cta}. 
An alternative line of thought is to invoke an underlying symmetry beyond the SM gauge group. 
In line with ideas put forward more than a decade ago \cite{Bardeen:1995kv} and ideas that inspired walking technicolor models \cite{Holdom:1984sk,Holdom:1981rm,Appelquist:1986an,Yamawaki:1985zg}, one could conceive  that scale invariance (and the invariance under the full conformal group) is the symmetry underlying the RG evolution of the SM well above the TeV scale and up to the Planck scale.
At the classical level, the SM Lagrangian is scale and conformally invariant, with the exception of the Higgs mass term. To  maintain full conformal symmetry at the Lagrangian level, one can generate the Higgs mass through a Higgs-dilaton coupling \cite{AoSColeman} and the spontaneous breaking of conformal symmetry \cite{Shaposhnikov:2008xi, Englert:1976ep}. The dilaton, which is the Goldstone boson of the spontaneously broken symmetry, remains massless or may acquire a nonzero mass through terms that explicitly break conformal symmetry.

At the quantum level, the SM scale and conformal invariance is  explicitly  broken by the logarithmic running of the coupling constants, so that the divergence of the dilatation current, which is equal to the trace of the energy-momentum tensor, has the general form 
$$\partial_\mu s^\mu = T_\mu^\mu = \sum_i\, \beta_i (\{g\},\{\lambda\} )\cdot O_i^{(d=4)} + {\mbox{mass terms}}\, ,$$
with $\beta_i$ the beta function of the SM coupling for the operator $O_i$,  $\{g\}$ the set of gauge couplings, and $\{\lambda\}$ the set of scalar couplings. In the absence of mass terms, scale (and conformal) invariance will be restored at RG fixed points, where $\beta_i =0$. Approximate scale invariance, with $\beta_i\simeq 0$, might also be sufficient for the viability of the SM beyond the EWSB scale.
Work dating before the discovery of the top quark \cite{Pendleton:1980as} pointed at the appealing possibility that  SM physics at the weak scale is driven by the presence of infrared pseudo-fixed points for the SM couplings. The relevant observations can be summarized as follows: i) the top Yukawa coupling $g_t$ and the self-interaction Higgs coupling $\lambda$ develop a IRFP in the limit where the electroweak couplings $g,g^\prime$ are neglected with respect to the strong coupling $g_c$, ii) the running of the light quark masses and charged leptons is unaffected by $g_t$, while the light down quarks receive small corrections from it, iii) the RG running of the ratio $m_b/m_\tau$ is dominated by $g_t$, (iv) the gauge couplings are unaffected by $g_t$ at one-loop order, and v) the CKM mixing angles and phase seem to have a IRFP at zero that is approached very slowly. Detailed studies of the RG equations of the SM and its supersymmetric extensions have followed during the years, essentially without changing early conclusions. For a review and analysis of the fixed point and manifold structure of the SM see Ref.~\cite{Schrempp:1996fb}. 

The ultraviolet fate of the SM is not yet established, and the LHC has not yet provided hints of a specific BSM completion close to the TeV scale. In the context of RG studies, the reduction of parameters program introduced in \cite{Zimmermann:1984sx}  may be resurrected and provide insights into possible ultraviolet behaviors in light of the most recent experimental findings.
Recall that in the matter sector with $g=g^\prime =0$, the top and Higgs couplings in the top-Higgs-$g_c$ subsector share asymptotic freedom and have a IRFP.  Even if the SM cannot be taken to the limit $\mu\to\infty$ due to the Landau pole of $g^\prime$, nor to the limit $\mu\to 0$ due to confinement of strong interactions, it might well be that the underlying conformal symmetry in one or more sectors of the SM is enough to drive its evolution from the electroweak to the Planck scale.

Conformal symmetry would also be able to avoid 
the source of the gauge hierarchy problem, since it can protect the mass of the Higgs boson from additive quantum corrections of the order of the ultraviolet cutoff of the theory. As an alternative to the most familiar SM extensions with and without supersymmetry, one can invoke conformal invariance at the quantum level and its spontaneous breaking at the Planck scale, see, e.g., Ref.~\cite{Shaposhnikov:2008xi}. 
In the context of quantum gravity, the authors of \cite{Englert:1976ep} conjectured that it is always possible  to render a theory  conformally invariant at the quantum level (at least perturbatively), if its action is  conformally invariant in any $d$ spacetime dimensions\,---\, obtained via dilaton couplings\,---\,and if conformal symmetry is only spontaneously broken. In other words, there would exist conformally invariant counterterms to all orders in perturbation theory.  Within the scalar sector of the SM, it was recently shown  \cite{Shaposhnikov:2008xi} that a ``Scale-Invariant" (SI) prescription does exist for which i) the theory is conformally invariant at the quantum level to all orders in perturbation theory,  ii) it reproduces the low-energy running of the coupling constants, iii) it embeds unimodular gravity, and (iv) it protects the mass of the Higgs boson from additive ultraviolet corrections, i.e., $\delta m_H^2 \propto m_H^2$ and not $\Lambda_{Planck}^2$. 

It seems worthwhile to explore further the consequences of this program for the Yukawa and gauge sectors of the SM, taking as a reference starting point the spontaneously broken conformal symmetry at the Planck scale.
It remains to be seen if the resulting theory is renormalizable and, most importantly,  unitary,  and to be established how unique is the prescription that both ensures conformality at the quantum level and reproduces the low-energy running of the SM couplings.
In view of the most recent LHC findings, the scenario of a minimally extended SM up to the Planck scale with conformal invariance as an underlying symmetry, remains an appealing possibility. 
The next round of LHC data will hopefully provide further hints into a preferred high-energy completion of the SM.

\subsection{Methods from high-energy physics for strongly coupled, condensed matter systems}
\label{sec:G_CM}

The investigation of QCD at low energies, a prototypical example of a
strongly-coupled quantum field theory, has lead to the development of
a number of methods for describing strongly-coupled theories also in
other areas of physics. The example
studied here is condensed matter
physics, where methods developed for QCD are applied to strongly
coupled theories of relevance for the study of systems such as graphene
and high-$T_c$ superconductors.
 Both lattice gauge theory and gauge-gravity duality methods
have been applied to condensed matter systems. While lattice gauge
theory is an established method for studying QCD at low energies,
gauge-gravity duality was developed more recently as a generalization
of the AdS/CFT correspondence of string theory. It has proved very
useful in studies of transport processes and the
calculation of spectral functions of the quark-gluon plasma and for
QCD-like theories at high density, reviewed in Chapter \ref{sec:chapd} of
this document.

As examples of lattice gauge
theory and gauge-gravity duality applications to condensed matter physics, we review lattice gauge theory results for the
conductivity in graphene as function of the coupling strength, as well as gauge-gravity duality results for the Green's functions and
conductivities in non-Fermi liquids and superconductors. These methods may be applied more generally for the description of strongly coupled systems in condensed matter physics, for which traditional methods are scarce. They may also be used to predict new phases of matter.

\subsubsection{Lattice gauge theory results}
\label{sec:CM_LGT}

Graphene is a material which displays a relativistic dispersion
relation. Near the Fermi-Dirac points, the charge carriers display an
energy spectrum similar to the one of free 2+1-dimensional massless
Dirac fermions. This leads to unusual transport properties which have
recently been investigated using lattice gauge theory
\cite{Buividovich:2012uk,Buividovich:2012nx,Ulybyshev:2013swa}.
The lattice study of graphene was initiated in
\cite{Drut:2008rg}, where evidence for a second order
semimetal-insulator  transition was found, which is associated to
spontaneous chiral symmetry breaking and the opening of a gap in the
energy spectrum.

\begin{figure}[tb]
  \includegraphics*[width=\figwid]{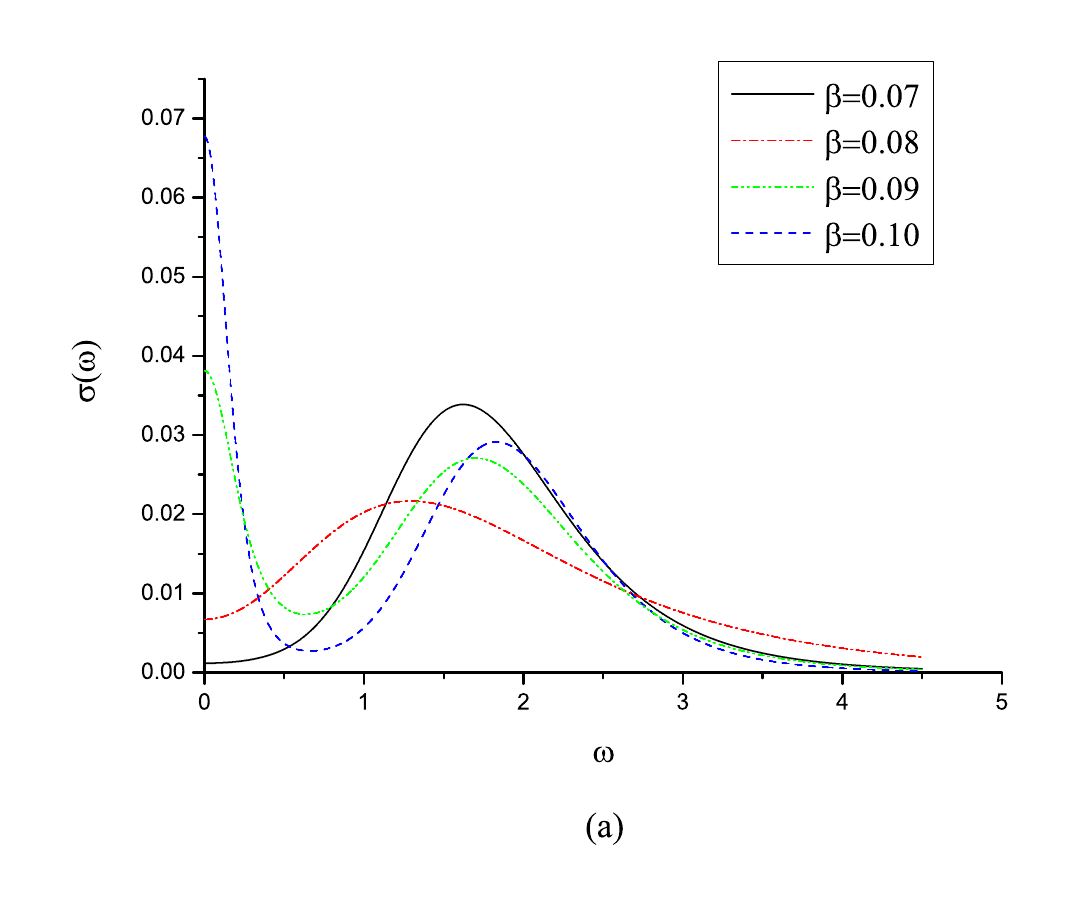}
        \caption[Conductivity]{AC conductivity by varying the frequency $\omega$ from the lattice study in \cite{Buividovich:2012uk}  for
          different values of the inverse coupling $\beta$ in the strong coupling regime.
           \label{fig3a} }
\end{figure}

As in  \cite{Drut:2008rg}, the starting
point of \cite{Buividovich:2012uk} is a 3+1-dimensional Abelian lattice gauge field coupled to
2+1-dimensional staggered lattice fermions. The conductivity is
calculated as a function of the inverse lattice gauge coupling $\beta$ given by
\begin{equation}
\beta \equiv \frac{1}{g^2} = \frac{v_F}{4 \pi e^2} \frac{\epsilon
  +1}{2} \, ,
\end{equation}
where $\epsilon$ is the dielectric permittivity and $v_F$ the Fermi velocity. It is found that at large values of
the coupling $g$, a fermion condensate $\langle \bar \psi \psi \rangle$ forms.
Simultaneously, the
DC conductivity is smaller in the strong coupling regime ($g=4.5$) as compared to
the weak coupling regime ($g < 3.5$) by three orders of magnitude.
At small values of $\beta$, the AC conductivity
as calculated from linear response theory shows the behavior
displayed in Fig.~\ref{fig3a}. In the opposite limit of vanishing
interaction (large $\beta$), the AC conductivity should develop a  $\delta (\omega)$ contribution
due to translational invariance from the absence of scattering. When the
interaction is increased, thus for decreasing $\beta$ in Fig.~\ref{fig3a}, the peak becomes broader. The second peak in Fig.~\ref{fig3a}
is expected to correspond to the optical frequency range for
graphene. These results have been obtained using the maximum entropy
method, while a more refined analysis based on a tight-binding model can be found
in \cite{Buividovich:2012nx}.
An alternative QCD-inspired strong-coupling approach to study graphene  is to use the
Schwinger-Dyson equations. The dynamical gap generation by
long-range Coulomb interactions in suspended graphene has been
investigated with this approach in \cite{Popovici:2013fwp}.

\subsubsection{Gauge-gravity duality results}
\label{sec:CM_GG}

Generalizations of the AdS/CFT correspondence \cite{Maldacena:1997re}, referred to as {\it
  gauge-gravity duality}, are naturally suited for describing strongly
coupled systems. Gauge-gravity duality is a conjecture which states
that strongly coupled $SU(N)$ field theories with $N \rightarrow
\infty$ in $d$ dimensions are mapped to weakly coupled
gravity theories in $d+1$ dimensions. The two theories share the
global symmetries and the number of degrees of freedom. Supersymmetry as well as conformal symmetry may be completely broken within gauge-gravity duality by considering more complicated metrics than the original anti-de Sitter space, and RG flows may be described in which the additional coordinate corresponds to the energy scale. Several nontrivial examples within QCD support the gauge-gravity duality
conjecture, such as the result for the shear viscosity over entropy
ratio \cite{Kovtun:2004de}, results for jet quenching \cite{Liu:2006ug}, as well as for chiral symmetry
breaking and the $\rho$ meson mass as function of the $\pi$ meson mass squared at
large $N$ \cite{Erdmenger:2007cm} (see Chapter~\ref{sec:chapd}).

Of course, the microscopic
degrees of freedom in a condensed matter system are very different
from those described by a non-Abelian gauge theory at large $N$.
Nevertheless, the idea is to make use of {\it universality}  and
to consider systems at second order phase transitions or, more  generally, at renormalization group
fixed points, where the microscopic details may not be important.
A prototype example for this scenario are {\it quantum phase
  transitions}, i.e., phase transitions at zero temperature which are
induced by quantum rather than thermal fluctuations \cite{SachdevBook,Sachdev:2011wg,Hartnoll:2007ih}. These transitions
generically appear when varying a parameter or coupling which is not
necessarily small.

In many cases, the study of models relevant to condensed matter physics
involves the introduction of a finite charge density in addition to
finite temperature. This applies for instance to Fermi surfaces
or condensation processes. In the
gauge-gravity duality context, this is obtained in a natural way by
considering charged black holes, the Reissner-Nordstr\"om black holes. Their
gravity action involves additional gauge fields,
\begin{gather} \label{eq:EinsteinMaxwell}
{\cal S} = \int \! \mathrm{d}^{d+1} x \, \sqrt{-g} \left[ \frac{1}{2 \kappa^2}
\left(R  - 2 \Lambda\right) - \frac{1}{4 e^2} F^{m
  n}F_{m n} \right] \, .
\end{gather}
Here, $\kappa^2$ is the gravitational constant in $d+1$ dimensions, $R$
is the Ricci scalar for the metric $g_{m n}$ with determinant $g$,
$\Lambda$ is  the negative cosmological constant
associated with anti-de Sitter space, and $F_{mn}$ is the field
strength for a $U(1)$ gauge field $A_m$ on the gravity side.
According to the prescriptions of the AdS/CFT correspondence, this
gauge field $A_m$  couples
to a conserved global $U(1)$ current in the dual $SU(N)$ gauge theory,
for which it acts as a source,
\begin{gather}
\langle J_\mu \rangle = \frac{\delta W}{\delta A^\mu} \, ,
\end{gather}
with $W$ the generating functional  of connected Green's functions.
Similarly, the metric of the curved space is the source for the
energy-momentum tensor in the dual field theory.
A chemical potential and finite charge density are obtained
from a nontrivial profile for the time component of the gauge field
in Eq. \eqref{eq:EinsteinMaxwell}.
Within this approach,
standard thermodynamic quantities such as the free energy and the
entropy may be calculated.  An important observable characterizing the properties of condensed
matter systems, and already discussed in Sec.~\ref{sec:CM_LGT} in the context of lattice studies, is the frequency-dependent conductivity. This can be
calculated in a straightforward way using gauge-gravity duality
techniques.  Below, we discuss examples for results obtained using this approach.

In several holographic models, instabilities may lead to new
ground states with lower free energy. This includes models with
properties of superfluids and superconductors
\cite{Hartnoll:2008kx,Hartnoll:2008vx}. In addition to
condensed matter physics, such new ground states occur also in models
describing the quark-gluon plasma at finite isospin density and
predict the frictionless motion of mesons through the plasma
\cite{Ammon:2008fc,Ammon:2009fe}.
In some cases, the new
ground state is characterized by a spatially modulated condensate \cite{Domokos:2007kt,Nakamura:2009tf,Donos:2012wi,Bu:2012mq}.
These findings have analogues
also within QCD itself. For instance, an external magnetic field leads
to a spatially modulated $\rho$ meson condensate \cite{Chernodub:2011mc,Chernodub:2011tv}, similar to
earlier results for Yang-Mills and electroweak fields
\cite{Nielsen:1978rm,Ambjorn:1988tm}.

\begin{figure}[tb]
  \includegraphics*[width=\figwid]{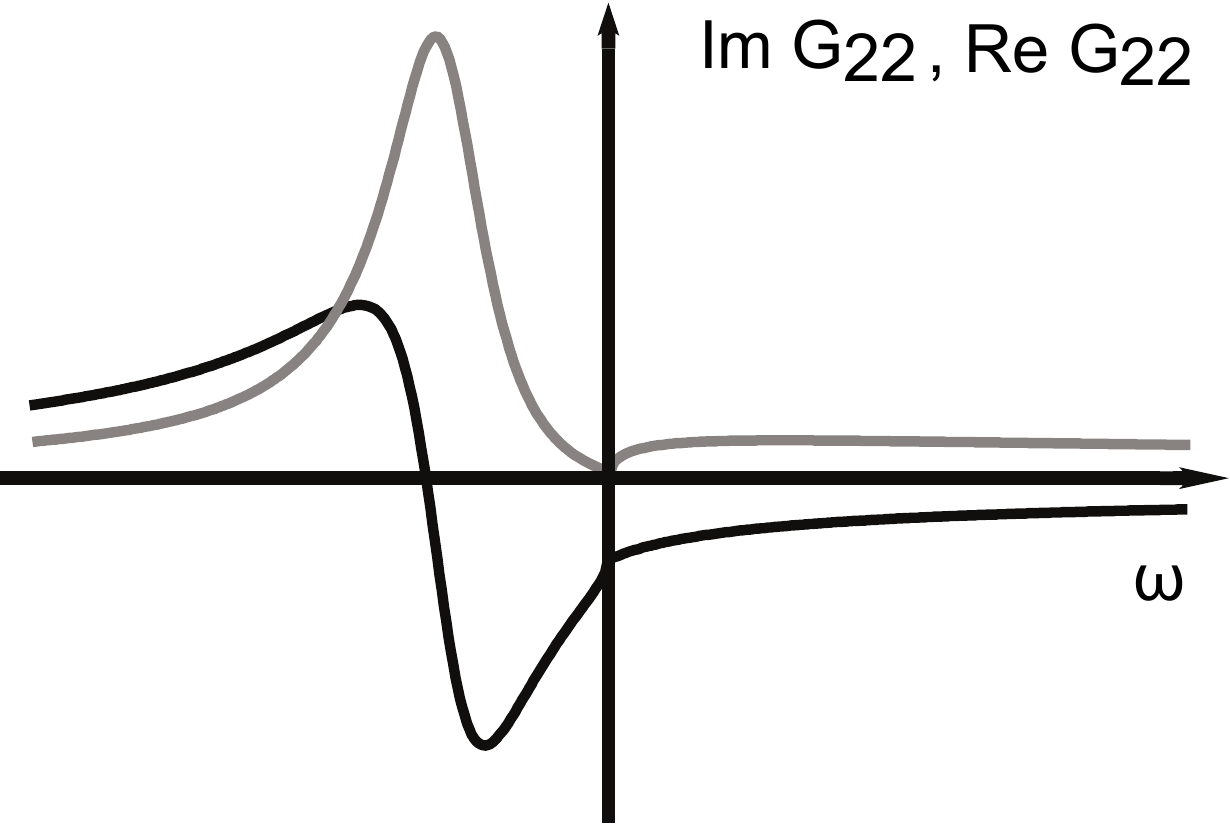}
        \caption[Conductivity2]{Typical frequency dependence of the real part (black) and imaginary part (gray) of the fermionic
          retarded Green's function calculated from gauge-gravity
          duality. We display the 2-2-component in spin space.
           \label{fig1} }
\end{figure}

\begin{figure}[tb]
  \includegraphics*[width=\figwid]{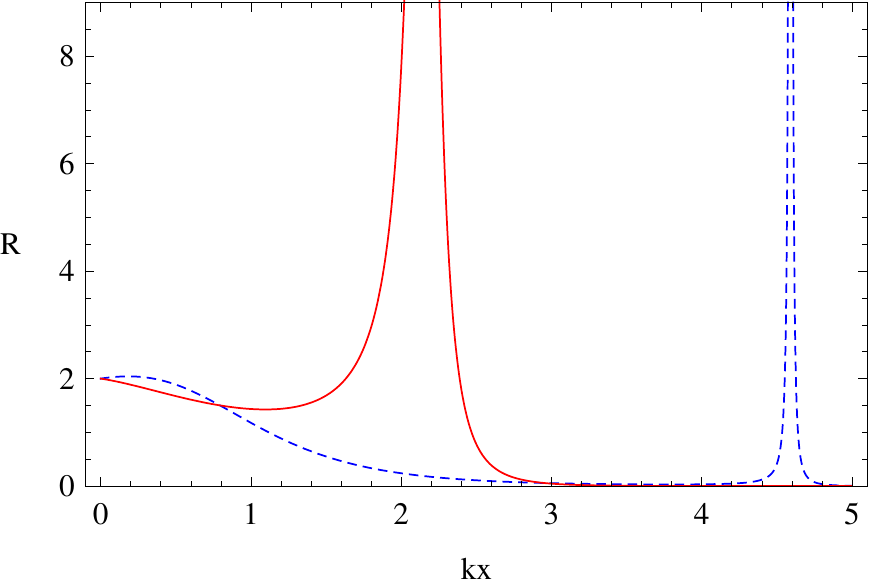}
        \caption[Conductivity3]{Spectral function from \cite{Ammon:2010pg} as function of the
          momentum $k_x/ \pi T$ from a gauge-gravity
          dual model showing non-Fermi liquid behavior. The two
          curves correspond to different components  ${R}_{11} $ (solid red) and ${R}_{22} $  (dashed blue) of the fermion
          matrix in spin space, related by ${R}_{11} (k_x) = {R}_{22} (-k_x)$.
           \label{fig2} }
\end{figure}

\begin{figure}[tb]
  \includegraphics*[width=\figwid]{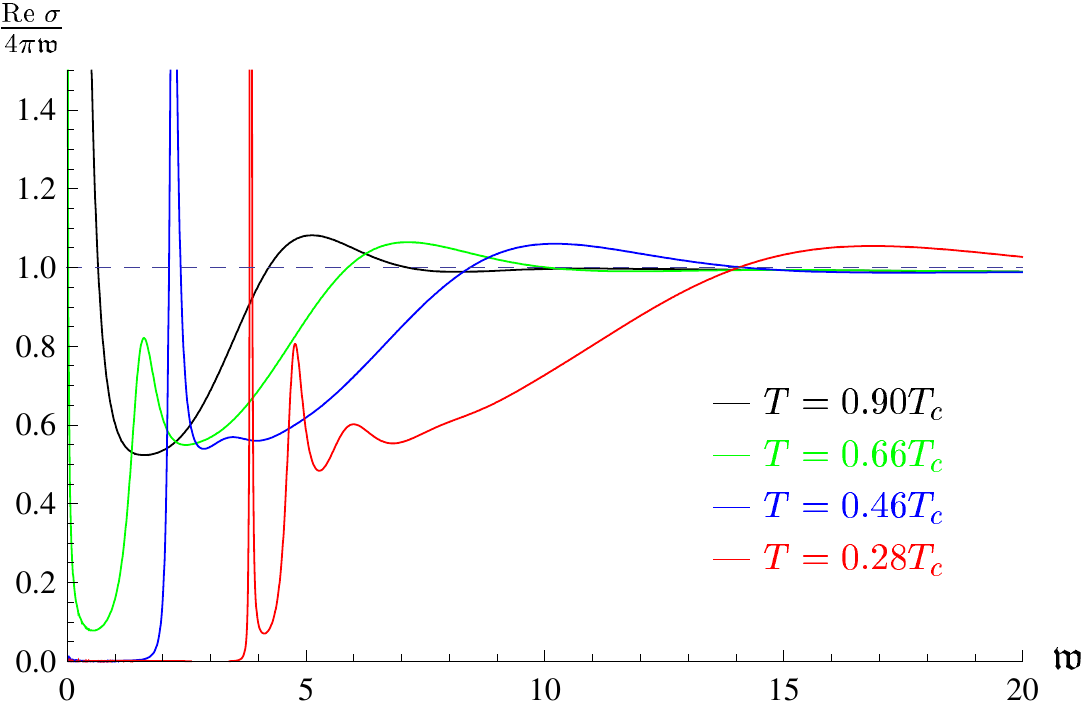}
        \caption[Conductivity4]{Frequency-dependent conductivity for the gauge-gravity superfluid from \cite{Ammon:2008fc}. The horizontal axis corresponds to the reduced frequency $\omega/(2 \pi T)$.
 This relativistic model involves a finite isospin density and the new
 ground state corresponds to a $\rho$ meson condensate. At low
 frequencies, a gap develops when lowering the temperature.
 The peaks at higher frequencies above the gap  correspond to higher excited modes (similar to
 the $\rho^*$) in this
 strongly coupled system.
           \label{fig4} }
\end{figure}

A further important aspect of condensed matter applications is
the study of fermions in strongly coupled systems using gauge-gravity
duality \cite{Liu:2009dm,Cubrovic:2009ye}. The standard
well-understood approach for describing fermions in weakly coupled
systems is Landau
Fermi liquid theory. These systems have a Fermi surface, and the
low-energy degrees of freedom are quasi-particle excitations around
the Fermi surface. However, many systems have been observed in
experiments which do not exhibit Landau Fermi liquid
behavior. Although they have a Fermi surface, their low-energy
degrees of freedom do not correspond to weakly coupled quasi-particles.
Nevertheless, the Fermi surface contains essential information about the physical
properties also of strongly
coupled systems. For instance for high-$T_c$ superconductors,
it reveals the $d$-wave symmetry structure.
Gauge-gravity duality
provides means for calculating Fermi surfaces and spectral functions
for strongly coupled systems \cite{Faulkner:2009wj}. An example for
the real and imaginary parts of the retarded Green's function is shown
in Fig.~\ref{fig1}. This result is obtained from the Reissner-Nordstr\"om black hole discussed above and corresponds to the Fermi surface of a strongly coupled non-Fermi liquid which is difficult to obtain using standard approaches. Note that for fermions, the Green's function is a $2 \times 2$ matrix in spin space. 
Within the gauge-gravity duality framework, an approach giving control over the microscopic degrees of freedom
 of the quantum field theory involved
 is to calculate Fermi surfaces for fermionic supergravity
fields dual to composite gauge-invariant fermionic operators in the
dual field theory \cite{Ammon:2010pg}. This requires starting from a ten-dimensional gravity action involving an internal manifold in addition to the asymptotically anti-de Sitter space.
Due to the strong coupling, the resulting systems may be of marginal or non-Fermi
liquid type. An example is shown in Fig.~\ref{fig2}.
The predicted dispersion relation  and momentum dependence of the spectral
function read \cite{Liu:2009dm,Ammon:2010pg}
\begin{align}
\omega - \omega_f & \sim \left( k - k_f\right)^z \, \\
{R_{ii}} & \sim \left( k - k_f \right)^{-\alpha}, \quad i=1,2,
\end{align}
with
\begin{gather}
z = 1.00 \pm 0.01 \, , \qquad
\alpha = 2.0 \pm 0.1 \, .
\end{gather}
This result deviates substantially from the Landau Fermi liquid theory, where $z = \alpha =1$.
More recently, progress has been made towards
holographically calculating the Fermi surfaces for the elementary
fermions present in the dual field theory \cite{Huijse:2011ef}.
For gauge-gravity dual models of superconductivity and superfluidity,
the conductivity as obtained from the current-current correlator
\begin{gather}
\sigma_{ij} (\omega) = \frac{i}{\omega} G^R_{ij} (\omega) \, , \qquad i,j \in \{1,2 \} \, ,
\end{gather}
displays a gap as function of the frequency as expected. For the model
of  \cite{Ammon:2008fc,Ammon:2009fe}, which corresponds to a relativistic superfluid at finite isospin
density, this is shown in Fig.~\ref{fig4}.
Since condensed matter systems are generically nonrelativistic, it is
useful to consider extensions of gauge-gravity duality to spaces which
share nonrelativistic symmetries \cite{Adams:2008wt,Kachru:2008yh}. Some of these spaces have the
additional advantage of naturally providing a zero ground state
entropy. Moreover, in addition to the thermodynamic entropy, the
quantum mechanical entanglement entropy may also be realized within the
gauge-gravity duality \cite{Ryu:2006bv}, with significant consequences for the models considered.
Generally, the entanglement entropy provides an order parameter, for
instance for topologically ordered states.

The examples given show that both lattice gauge theory and gauge-gravity duality have useful applications to strongly coupled systems also within condensed matter physics. Further new developments along this recent line of research are expected in the near future.

\subsection{Summary and future prospects}
\label{sec:G_Conclusions}

The study of strongly coupled systems, from particle to condensed matter physics, has recently acquired new
prospects and directions.
One important aspect of these developments, and partly a revival of ideas sketched in the past, is the
realization of the utility of conformal symmetry because many properties of systems in Nature can be
described in terms of small deviations from the conformal point.
This happens to be particularly useful when trying to solve field theories in their nonperturbative regime,
i.e., strongly coupled systems.

As reviewed in Sec.~\ref{sec:G_QFT}, many exact methods have been recently refined to describe QFTs in the
large $N$ limit, often inspired by string theory and its lower dimensional realizations.
The AdS/CFT correspondence, and more generally gauge-gravity dualities have become an inspiring tool for
effective field theory realizations of strongly coupled systems in four or lower spacetime dimensions, by
identifying their dual string theory realization.
However, it will remain difficult to extend the use of duality arguments beyond the large $N$ limit, and to
predict from first principles the size of deviations from the large $N$ and the conformal limits in this
framework.
Conformal bootstrap methods offer a powerful alternative, fully based on field theory arguments.
Still remains true that in some cases\,---\,properties of QGP, conductivities in graphene and
superconductors\,---\,the predictions based on the gauge-gravity duality turn out to be promisingly close to
the experimental results, as discussed in Sec.~\ref{sec:G_CM}.
Lattice field theory computations remain, as of today, the only genuinely nonperturbative description of
these systems that is a priori able to provide the complete answer, from strong to weak coupling, once the
continuum limit is reached.

As reviewed in Sec.~\ref{sec:G_ConfBSM}, the combination of lattice computations and analytic field
theoretical methods is being especially successful in uncovering the physics of the conformal window and,
more generally, the approach to conformal symmetry in non-Abelian gauge theories with matter content, with
and without supersymmetry.
Strongly coupled theories close to the conformal window provide interesting candidates for BSM physics.
A~wide class of viable BSM theories have been discussed in Sec.~\ref{sec:G_ConfBSM} and \ref{sec:G_EWSB}.
In particular, we have considered theories with a strongly coupled new sector, invoking compositeness at the
multi-TeV scale with or without conformality, and minimal SM extensions where conformal symmetry is invoked
at the Planck scale.
Weakly coupled supersymmetric extensions of the SM have not been discussed here, and, except for the
maximally constrained minimal supersymmetric SM (cMSSM), they remain a viable alternative to the mentioned
scenarios.

\newpage

All these attempts aim to accommodate a Higgs boson of 126 GeV and the absence of significant deviations from
the SM, as in accordance with the LHC observations collected up to now.
The next run of the LHC experiments will either confirm the validity of the SM by pushing all exclusion
bounds to higher energies, or, in the most striking case, will find direct evidence of a new sector\,---\,be
it, perhaps, a resonance of a composite strongly coupled extension, additional scalar(s), or supersymmetric
partner(s).
While awaiting further experimental signatures, the task of particle theorists is to rethink their models in
light of the recent Higgs boson discovery and broaden their scope by exploring implications for cosmology and
a possible unification with gravity.

Ultimately, the high energy completion of the SM ought account for anything that is not yet embedded in it,
i.e., neutrino masses and oscillations, baryo- and leptogenesis, dark matter and dark energy\,---\,or
anything that identifies with that $27\%$ and $68\%$, respectively, of our Universe beyond a tiny $5\%$
\cite{Ade:2013ktc, Ade:2013zuv} of visible baryonic matter.
In other words, it should account for the evolution of the Universe once gravity is embedded in the theory,
and it ought to explain any possible deviation from the SM eventually observed at the LHC.

\clearpage
\section*{Appendix: Acronyms}
\begin{longtable*}[t]{rll}
& & \\
\hline
& {\bf Accelerators} & \\
\hline
AGS & Alternating Gradient Synchrotron \\&  \protect{\url{http://www.bnl.gov/rhic/ags.asp}} \\
DA$\Phi$NE	&Double Annular $\Phi$ Factory for Nice Experiments	\\&\protect{\url{http://www.lnf.infn.it/acceleratori/}}	\\
ELSA & ELectron Stretcher Accelerator \\&\protect{\url{http://www-elsa.physik.uni-bonn.de/elsa-facility\_en.html}} \\ 
HERA & Hadron-Electron Ring Accelerator \\& \protect{\url{http://adweb.desy.de/mpy/hera/}} \\
LEP	&Large Electron-Positron collider	\\&\protect{\url{http://home.web.cern.ch/about/accelerators/large-electron-positron-collider}}	\\
LHC	&Large Hadron Collider	\\&\protect{\url{http://home.web.cern.ch/topics/large-hadron-collider}}	\\
MESA	&Mainz Energy-Recovering Superconducting Accelerator	\\&\protect{\url{http://www.prisma.uni-mainz.de/mesa.php}}	\\
NICA	&Nuclotron-based Ion Collider fAcility	\\&\protect{\url{http://nica.jinr.ru/}}	\\
RHIC	&Relativistic Heavy Ion Collider	\\&\protect{\url{http://www.bnl.gov/rhic/}}	\\
SPS	&Super Proton Synchrotron	\\&\protect{\url{http://home.web.cern.ch/about/accelerators/super-proton-synchrotron}}	\\
Tevatron	&	\\&\protect{\url{http://www.fnal.gov/pub/tevatron/}}	\\

& & \\
\hline
& {\bf Experiments} & \\
\hline
ACME	&Advanced Cold Molecule Electron EDM	\\&\protect{\url{http://laserstorm.harvard.edu/edm/}}	\\
ACORN 	&A CORrelation in Neutron decay	\\&\protect{\url{http://www.ncnr.nist.gov/expansion/individual\_instruments/aCORN041911.html}}	\\
ALEPH	&Apparatus for LEP Physics at CERN	\\&\protect{\url{http://home.web.cern.ch/about/experiments/aleph}}	\\
ALICE	&A Large Ion Collider Experiment	\\&\protect{\url{http://aliceinfo.cern.ch/}}	\\
ATLAS	&A Toroidal LHC ApparatuS	\\&\protect{\url{http://atlas.ch}}	\\
BaBar	&	\\&\protect{\url{http://www-public.slac.stanford.edu/babar/}}	\\
Belle	&	\\&\protect{\url{http://belle.kek.jp/}}	\\
BES~III  & Beijing Spectrometer \\& \protect{\url{bes3.ihep.ac.cn/}} \\
BRAHMS & Broad RAnge Hadron Magnetic Spectrometers Experiment at RHIC \\& \protect{\url{http://www4.rcf.bnl.gov/brahms/WWW/brahms.html }}\\
CB & Crystal Ball (MAMI) \\&\protect{\url{http://wwwa2.kph.uni-mainz.de/internalpages/detectors-and-setup/cb-mami.html}} \\
CB & Crystal Barrel (ELSA) 4$\pi$ photom spectrometer) \\& \protect{\url{http://www1.cb.uni-bonn.de/index.php?id=4&L=1}} \\
CDF 	&Collider Detector at Fermilab	\\&\protect{\url{http://www-cdf.fnal.gov/}}	\\
CLAS	&CEBAF Large Acceptance Spectrometer	\\&\protect{\url{http://www.jlab.org/Hall-B/clas-web/}}	\\
CLEO	&	\\&\protect{\url{http://www.lepp.cornell.edu/Research/EPP/CLEO/WebHome.html}}	\\
CMD-3   & Cryogenic Magnetic Detector \\& \protect{\url{cmd.inp.nsk.su/cmd3 }}\\
CMS	&Compact Muon Solenoid	\\&\protect{\url{http://cern.ch/cms}}	\\
COMPASS	&COmmon Muon and Proton Apparatus for Structure and Spectroscopy	\\&\protect{\url{www.compass.cern.ch/}}	\\
CPLEAR	&	\\&\protect{\url{http://cplear.web.cern.ch/cplear/welcome.html}}	\\
D\O\	& after its location on the Tevatron ring	\\&\protect{\url{http://www-d0.fnal.gov/}}	\\
DELPHI & DEtector with Lepton, Photon and Hadron Identification \\& \protect{\url{www.cern.ch/delphi}} \\
DIRAC & DImeson Relativistic Atom Complex \\&  \protect{\url{dirac.web.cern.ch/DIRAC/}}  \\
DISTO	&Dubna-Indiana-Saclay-TOrino	\\&\protect{\url{http://oldsite.to.infn.it/activities/experiments/disto/disto\_overview.html}}	\\

E158 & E158 experiment (SLAC) A precision measurement of the Weak Mixing Angle in Moeller Scattering \\& \protect{\url{www.slac.stanford.edu/exp/e158/}} \\

E246 & E246 experiment (KEK)  \\& \\

E549 & E549 experiment (KEK)  \\& \\

E609 & E609 experiment (Fermilab) \\& \\

E653 & E653 experiment (Fermilab) \\& \\

E665 & E665 experiment (Fermilab) Muon spectrometer \\& \protect{\url{http://www.nuhep.northwestern.edu/~schellma/e665/}}\\

E791 & E791 experiment (Fermilab) \\& \protect{\url{http://ppd.fnal.gov/experiments/e791/welcome.html}}  \\

E852 &  A Search for Exotic Mesons \\&\protect{\url{http://hadron.physics.fsu.edu/~e852/}} \\ 
E862  & E862 experiment, Antihydrogen at Fermilab \\& \protect{\url{http://ppd.fnal.gov/experiments/hbar/}} \\

E989 & E989 experiment (Fermilab) \\&  \\

emiT & A search for Time-reversal Symmetry Violation in Polarized Neutron Beta Decay \\& \protect{\url{http://ewiserver.npl.washington.edu/emit}} \\
FINUDA	&Fisica NUcleare a DAFNE	\\&\protect{\url{http://www.lnf.infn.it/esperimenti/finuda/finuda.html}}	\\
FOCUS  & \\& \protect{\url{http://www-focus.fnal.gov}} \\
FOPI	&4$\pi$	\\&\protect{\url{https://www.gsi.de/en/work/research/cbmnqm/fopi.htm}}	\\
GRAAL & GRenoble Anneau Accelerateur Laser \\& \protect{\url{http://graal.ens-lyon.fr/}} \\
HADES	&High Acceptance DiElectron Spectrometer	\\&\protect{\url{http://www-hades.gsi.de/}}	\\
HAPPEX	&Hall A Precision Parity EXperiment	\\&\protect{\url{http://hallaweb.jlab.org/experiment/HAPPEX}}	\\
HERMES	&	\\&\protect{\url{http://www-hermes.desy.de/}}	\\
H1 & H1 detector (HERA) \\& \protect{\url{www-h1.desy.de/}} \\
JADE & JApan, Deutschland, and England \\& \protect{\url{https://wwwjade.mpp.mpg.de/}} \\
KaoS	& Kaon Spectrometer	\\& \protect{\url{http://www-aix.gsi.de/~kaos/html/kaoshome.html}}	\\
KEDR & \\& \protect{\url{kedr.inp.nsk.su/}} \\
KLOE	&K LOng Experiment	\\&\protect{\url{http://www.lnf.infn.it/kloe/}}	\\
LHCb	&Large Hadron Collider beauty	\\&\protect{\url{http://lhcb.web.cern.ch/lhcb/	}}\\
MINER$\nu$A	&Main Injector Experiment for $\nu$-A	\\&\protect{\url{http://minerva.fnal.gov/	}}\\
MiniBoone	&BOOster Neutrino Experiment	\\&\protect{\url{http://www-boone.fnal.gov/}}	\\
MOLLER	&	after M\o ller scattering \\&\protect{\url{http://hallaweb.jlab.org/12GeV/Moller/}}	\\
MuLan & MUON Lifetime ANalysis \\& \protect{\url{https://www.npl.uiuc.edu/exp/mulan//}} \\
MUSE & The MUon proton Scattering Experiment \\& \protect{\url{www.physics.rutgers.edu/~rgilman/elasticmup/}}
Muon $g-2$ & \\ & \protect{\url{http://www.g-2.bnl.gov/}} \\
NA10  & NA10 experiment (CERN) \\& \protect{\url{http://greybook.cern.ch/programmes/experiments/NA10.html/}} \\
NA45	& NA45 experiment (CERN)	\\&\protect{\url{http://greybook.cern.ch/programmes/experiments/NA45.html}}	\\
NA48   & NA48 experiment (CERN) CP violation \\& \protect{\url{http://greybook.cern.ch/programmes/experiments/NA48.html/}} \\
NA49	& NA49 experiment (CERN)	\\&\protect{\url{http://na49info.web.cern.ch/na49info/}}	\\
NA57	& NA57 experiment (CERN)	\\&\protect{\url{http://wa97.web.cern.ch/WA97/}}	\\
NA60	&  NA60 experiment (CERN)	\\&	\\
NA61/SHINE	& SPS Heavy Ion and Neutrino Experiment	(CERN) \\&\protect{\url{http://home.web.cern.ch/about/experiments/na61shine}}	\\
NA62  & NA62 experiment (CERN) \\& \protect{\url{http://na62.web.cern.ch/na62/Home/Aim.html/}} \\
Nab at ORNL	&	\\&\protect{\url{http://www.phy.ornl.gov/groups/neutrons/beta.html}}	\\
nEDM-SNS & neutron EDM experiment at the Spallation Neutron Source \\&  \protect{\url{www.phy.ornl.gov/nedm/}}  \\
New Muon $g-2$ & \\&  \protect{\url{http://muon-g-2.fnal.gov/}} \\
NuSea & \\& \protect{\url{http://p25ext.lanl.gov/e866/e866.html}} \\
NuTeV	&	\\&\protect{\url{http://www-e815.fnal.gov/}}	\\
OBELIX &  \\&\protect{\url{http://www.fisica.uniud.it/~santi/OBELIX/OBELIX.html}} \\
OPAL	&Omni-Purpose Apparatus at LEP	\\&\protect{\url{www.cern.ch/opal}}	\\
 PANDA & anti-Proton ANnihilation at DArmstadt \\& \protect{\url{http://www-panda.gsi.de/}} \\
PAX	&Polarized Antiproton eXperiment	\\& \protect{\url{http://collaborations.fz-juelich.de/ikp/pax/index.shtml}}	\\
PEN & \\& \protect{\url{http://pen.phys.virginia.edu/}} \\
PERC	&Proton and Electron Radiation Channel	\\&	\\
PERKEOIII &	\\&\protect{\url{http://www.physi.uni-heidelberg.de/Forschung/ANP/Perkeo/perkeo3.php}}	\\
PHENIX	&Pioneering High Energy Nuclear Interaction eXperiment	\\&\protect{\url{http://www.bnl.gov/rhic/PHENIX.asp}}	\\
PHOBOS & \\& \protect{\url{http://www.phobos.bnl.gov/}} \\
PIBETA & PI BETA experiment \\& \protect{\url{http://pibeta.phys.virginia.edu}} \\
PLANCK	&	\\&\protect{\url{http://www.rssd.esa.int/index.php?project=planck}}	\\
PNDME	&Precision Calculation of Neutron-Decay Matrix Elements	\\&\protect{\url{http://www.phys.washington.edu/users/hwlin/pndme/index.xhtml}}	\\
PREX	&208Pb Radius Experiment	\\&\protect{\url{http://hallaweb.jlab.org/parity/prex/}}	\\
PRIMEX	&Primakoff experiment	\\& \protect{\url{http://www.jlab.org/primex/}}	\\
QWEAK	&	\\&\protect{\url{http://www.jlab.org/qweak/}}	\\
REX-ISOLDE	&Radioactive Beam EXperiment at ISOLDE	\\&\protect{\url{http://isolde.web.cern.ch/rex-isolde}}	\\
SIDDHARTA	&SIlicon Drift Detectors for Hadronic Atom Research by Timing Application\\&\protect{\url{http://www.lnf.infn.it/esperimenti/siddharta/}}	\\
& & \\
SND & Spherical Neutral Detector \\& \protect{\url{wwwsnd.inp.nsk.su}} \\
SNO	&Sudbury Neutrino Observatory	\\&\protect{\url{http://www.sno.phy.queensu.ca/}}	\\
STAR & Solenoid Tracker At RHIC\\& \protect{\url{http://www.star.bnl.gov/}} \\
ThO	&ACME Thorium Oxide Electron EDM Experiment	\\&\protect{\url{http://www.doylegroup.harvard.edu/wiki/index.php/ThO}}	\\
TREK & \\& \protect{\url{http://trek.kek.jp}} \\
UCNA & Ultra Cold Neutron Apparatus \\& \protect{\url{http://www.krl.caltech.edu/research/ucn}} \\
UCNB	&UltraCold Neutron source	\\&\protect{\url{http://www.ne.ncsu.edu/nrp/ucns.html}}	\\
UCNb	&UltraCold Neutron source	\\&\protect{\url{http://www.ne.ncsu.edu/nrp/ucns.html}}	\\
VES & VErtex Spectrometer (IHEP) \\&\protect{\url{http://pcbench.ihep.su/ves/index2.shtml}} \\ 
WA102 & WA102 experiment (CERN) \\&\protect{\url{http://www.ep.ph.bham.ac.uk/exp/WA102/}} \\ 
WASA & Wide Angle Shower Apparatus \\&\protect{\url{http://collaborations.fz-juelich.de/ikp/wasa/index.shtml}}  \\ 
WMAP	&Wilkinson Microwave Anisotropy Probe	\\&\protect{\url{http://map.gsfc.nasa.gov/}}	\\
ZEUS & ZEUS experiment (HERA) \\& \protect{\url{www-zeus.desy.de}} \\

\hline
& {\bf Laboratories} & \\
\hline
BNL     &Brookhaven National Laboratory \\&\protect{\url{www.bnl.gov}} \\
CERN	&European Organization for Nuclear Research	\\&\protect{\url{http://home.web.cern.ch/}}	\\
CSSM	& Centre for the Subatomic Structure of Matter \\& \protect{\url{http://www.physics.adelaide.edu.au/cssm/}} \\
DESY	& Deutsches Elektronen-SYnchrotron  \\& \protect{\url{http://www.desy.de/}} \\
FAIR	&Facility for Antiproton and Ion Research	\\&\protect{\url{http://www.fair-center.eu/}}	\\
Fermilab	&Fermi National Accelerator Laboratory	\\&\protect{\url{https://www.fnal.gov/}}	\\
FNAL	&Fermi National Accelerator Laboratory	\\&\protect{\url{https://www.fnal.gov/}}	\\
FRM-II	&Heinz Maier-Leibnitz research neutron source	\\&\protect{\url{http://www.frm2.tum.de/}}	\\
GSI	&GSI Helmholtz Centre for Heavy Ion Research	\\&\protect{\url{http://www.gsi.de/}}	\\
ILL	&Institut Laue-Langevin	\\&\protect{\url{http://www.ill.eu/}} \\	
JINR	&Joint Institute for Nuclear Research	\\&\protect{\url{http://www.jinr.ru/}}	\\
JLab	&Jefferson Lab	\\&\protect{\url{https://www.jlab.org/}}	\\
JPARC	&Japan Proton Accelerator Research Complex	\\&\protect{\url{http://j-parc.jp/index-e.html}}\\
KEK & High Energy Accelerator Research Organization \\& \protect{\url{http://legacy.kek.jp/}} \\
LANL	& Los Alamos National Laboratory \\& \protect{\url{http://www.lanl.gov/}} \\			
LBNL	&Lawrence Berkeley National Laboratory	\\&\protect{\url{http://www.lbl.gov/}}	\\
MAMI	&MAinz MIcrotron \\&\protect{\url{http://wwwkph.uni-mainz.de/B1/}} \\
NIST	& National Institute of Standards and Technology \\&  \protect{\url{www.nist.gov}} \\
ORNL	&Oak Ridge National Laboratory	\\&\protect{\url{http://www.ornl.gov/}}	\\
RBRC	& RIKEN BNL Research Center \\& \protect{\url{www.bnl.gov/riken}} \\
SLAC	& SLAC National Accelerator Laboratory (originally Stanford Linear Accelerator Center)	\\& \protect{\url{https://www6.slac.stanford.edu/}}	\\

& & \\
\hline
& {\bf Lattice-QCD Collaborations} & \\
\hline
ALPHA & ALPHA collaboration  \\& \protect{\url{http://www-zeuthen.desy.de/alpha/}} \\
APE &  Array Processor Experiment \\&  \protect{\url{http://apegate.roma1.infn.it/mediawiki/index.php/Main_Page}} \\
BGR & Bern-Graz-Regensburg &\\ \\
BMW & Budapest-Marseille-Wuppertal \\& \protect{\url{http://www.bmw.uni-wuppertal.de/Home.html}} \\
CLS & Coordinated Lattice Simulations \\& \protect{\url{https://twiki.cern.ch/twiki/bin/view/CLS/WebHome}} \\
CSSM Lattice & Centre for the Subatomic Structure of Matter Lattice \\& http://www.physics.adelaide.edu.au/cssm/lattice/ \\
CP-PACS & Computational Physics by Parallel Array Computer Systems \\& \protect{\url{http://www.rccp.tsukuba.ac.jp/cppacs/project-e.html}} \\
DiRAC & Distributed Research utilising Advanced Computing \\& \protect{\url{http://ukqcd.swan.ac.uk/dirac/}} \\
ETM & European Twisted Mass \\& \protect{\url{http://www-zeuthen.desy.de/~kjansen/etmc/}} \\
Fermilab Lattice \\& \protect{\url{http://inspirehep.net/search?ln=en\&ln=en\&p=find+cn+fermilab+lattice}} \\
FLAG & Flavour Lattice Averaging Group \\& \protect{\url{http://itpwiki.unibe.ch/flag/index.php}}, \protect{\url{http://www.latticeaverages.org}}\\
Hadron Spectrum \\& \protect{\url{http://usqcd.jlab.org/projects/Spectrum/}} \\
HotQCD \\& \protect{\url{http://quark.phy.bnl.gov/~hotqcd/}} \\
HPQCD & High-precision QCD \\& \protect{\url{http://www.physics.gla.ac.uk/HPQCD/}} \\
JLQCD & Japanese Lattice QCD \\& \protect{\url{http://jlqcd.kek.jp}} \\
LHP & Lattice Hadron Physics \\& \\
MILC & MIMD Lattice Computations \\& \protect{\url{http://www.physics.utah.edu/~detar/milc/}} \\
PACS-CS & Parallel Array Computer System for Computational Sciences \\& \protect{\url{http://www2.ccs.tsukuba.ac.jp/PACS-CS/}} \\
PNDME & Precision Calculation of Neutron-Decay Matrix Elements \\& \protect{\url{http://www.phys.washington.edu/users/hwlin/pndme/index.xhtml}} \\
QCD-Taro & \\& \\
QCDSF & QCD Structure Function \\& \protect{\url{http://inspirehep.net/search?p=find+cn+qcdsf}} \\
RBC & RBRC-BNL-Columbia \\& \protect{\url{http://rbc.phys.columbia.edu}} \\
SPQcdR & Southampton-Paris-Rome QCD &\\ \\
SWME & Seoul-Washington Matrix Element \\& \protect{\url{http://lgt.snu.ac.kr/}} \\
UKQCD & United Kingdom QCD \\& \protect{\url{http://www.ukqcd.ac.uk}} \\
USQCD & United States QCD \\& \protect{\url{http://www.usqcd.org}} \\
WHOT-QCD & \\& \\

& & \\
\hline
& {\bf Other} & \\
\hline

CKMfitter & Global analysis of CKM matrix \\& \protect{\url{http://ckmfitter.in2p3.fr/}} \\

CODATA &Committee on Data for Science and Technology	\\&\protect{\url{http://www.codata.org/}}	\\
CTEQ	& The Coordinated Theoretical-Experimental Project on QCD \\&\protect{\url{http://users.phys.psu.edu/~cteq/}}	\\
HFAG & Heavy Flavor Averaging Group \\& \protect{\url{http://www.slac.stanford.edu/xorg/hfag/}} \\

NNPDF & Neural Network Parton Distribution Functions \\& \protect{\url{https://nnpdf.hepforge.org}} \\

PDG & Particle Data Group \\& \protect{\url{http://pdg.lbl.gov}} \\
PYTHIA & after an ancient Greek priestess \\& \protect{\url{http://home.thep.lu.se/~torbjorn/Pythia.html}} \\
QWG & Quarkonium Working Group \\& \protect{\url{http://www.qwg.to.infn.it}} \\

UTfit & Unitarity Triangle fits \\& \protect{\url{http://www.utfit.org}} \\


\end{longtable*}

\clearpage
\section*{Acknowledgements}


We dedicate this document to the memory of Mikhail Polikarpov, who passed away in July 2013.
Misha worked with us for decades as a convener of the ``Confinement'' section of the Quark Confinement and
Hadron Spectrum Series.
He guided and expanded the scientific discussion of that topic, inspiring and undertaking new research
avenues.
From its initial conception, he supported the enterprise of this document and organized
Chapter~\ref{sec:chapa}, writing the part on confinement himself.
He attracted the XI$^{\mathrm{th}}$ Conference on Quark Confinement and the Hadron Spectrum to St.\
Petersburg (September 8-12, 2014; see \href{http://phys.spbu.ru/confxi.html}{\tt
http://phys.spbu.ru/confxi.html}).
His warm and kind personality, his high sense of humor, his ideas in physics and his special energy in
imagining and realizing new projects will be always a loss and an example for us.
We also miss four other physicists who made lasting contributions to the field of strong interactions: Dmitri
Diakonov, Nikolai Uraltsev, Pierre van Baal, and Kenneth Wilson.
We remember Misha, Dima, Ken, Kolya, and Pierre with fondness and gratitude.
In addition, S.G.\ would like to dedicate her work on this document to the memory of her mother and mentor,
Gladys Strom Gardner.

The authors thank Mikko Laine for collaboration during initial 
stages of the project.
In addition,
V\'eronique Bernard,
Alex Bondar,
Geoffrey T. Bodwin, 
Davide Caffarri,
Gilberto Colangelo,
Lance Dixon,
Gernot Eichmann,
Christoph Hanhart,
Ulrich Heinz,
Aleksi Kurkela,
Vittorio Lubicz,
Alexander Milov,
Bachir Moussallam,
Antonio Pineda,
Brad Plaster,
Massimiliano Procura,
Joan Soto,
Reinhard Stock,
Ubirajara van~Kolck,
Julia Velkovska,
and
Richard Williams 
provided helpful correspondance and contributions 
during the preparation of this review.
Finally, the Editors thank Ma{\l}gorzata Janik for maintaining document 
and bibliography updates and for
coordinating an editorial team of Jeremi Niedziela and Anna Zaborowska.

{\small
The authors appreciate and acknowledge support for work on this
document provided, in part or in whole, by
\begin{itemize}
\item U.S.~Department of Energy under contracts \#DE-FG02-05ER41375 and \#DE-FG02-91ER40628
(M.~Alford and K.~Schwenzer)
\item the EU I3HP project EPOS ``Exciting Physics of Strong Interactions'' (WP4 of HadronPhysics3)
(R. Alkofer)
\item U.S. Department of Energy grant no. DE-SC0007984,
(P.~Arnold)
\item Deutsche Forschungsgemeinschaft DFG EClust 153 ``Origin and Structure of the Universe''
(N. Brambilla, L. Fabbietti, B. Ketzer, A. Vairo)
\item Deutsche Forschungsgemeinschaft DFG grants BR 4058/2-1 and BR 4058/1-1; work 
supported in part by DFG and NSFC (CRC 110)
(N. Brambilla, A. Vairo)
\item U.S. Department of Energy
(T.~Cohen)
\item Istituto Nazionale di Fisica Nucleare (INFN) of Italy
(P. Di Nezza)
\item the Ministry of Education and Science of the Russian Federation;
the Russian Foundation for Basic Research under grants 12-02-01032
and 12-02-01296 and the German Research Foundation (DFG) under grant
HA 1457/9-1.
(S. ~Eidelman)
\item Bundesministerium f\"ur Bildung und Forschung BMBF Grant No.\ 05P12WOGHH;
Helmholtz Young Investigator University Group, VH-NG-330
(L.~Fabbietti)
\item Swiss National Science Foundation (SNF) under the Sinergia grant number
CRSII2\underline{ }141847\underline{ }1
(X.~Garcia i Tormo)
\item U.S. Department of Energy Office of Nuclear Physics under contract DE-FG02-96ER40989
(S.~Gardner)
\item DOE Contract No.\ DE-AC05-06OR23177 under which JSA operates the
Thomas Jefferson National Accelerator Facility, and by the National Science
Foundation through grants PHY-0855789 and PHY-1307413
(J. Goity)
\item Austrian Science Foundation FWF under project No.\ P20016-N16
(R.~H\"ollwieser)
\item Polish National Science Centre reaserch grant UMO-2012/05/N/ST2/02757
(M.~A.~Janik)
\item European Union in the framework of European Social Fund through the Warsaw University of Technology Development Programme
(M.~A.~Janik)
\item U.S. Department of Energy, Office of Nuclear Physics, under grant DE-FG02-89ER40531
(D. Keane)
\item Bundesministerium f\"ur Bildung und Forschung BMBF Grant No.\ 05P12WOCC1
(B. Ketzer)
\item European Union program ``Thales'' ESF/NSRF 2007-2013 
(E.~Kiritsis)
\item European Union (European Social Fund, ESF) and Greek national funds through the Operational Program
``Education and Lifelong Learning'' of the National Strategic Reference Framework (NSRF) under ``Funding of
proposals that have received a positive evaluation in the 3rd and 4th Call of ERC Grant Schemes''
(E.~Kiritsis)
\item European Union's Seventh Framework Programme under grant agreements (FP7-REGPOT-2012-2013-1) No.\ 
316165, and PIF-GA-2011-300984 and by the European Commission under the ERC Advanced Grant BSMOXFORD~228169
(E. Kiritsis) 
\item Fermilab is operated by Fermi Research Alliance, LLC, under Contract No.~DE-AC02-07CH11359 with  the 
U.S. Department of Energy  
(A.~S.~Kronfeld)
\item U.S. Department of Energy Office of Nuclear Physics under contract DE-FG02-97ER4014 
(H.-W.~Lin)
\item Spanish grant FPA2011-27853-C02-01
(Felipe J.~Llanes-Estrada)
\item Center for Computational Sciences as part of the Rhineland-Palatinate Research Initiative  
(H.~B.~Meyer)
\item the DFG grant ME 3622/2-1 
(H.~B.~Meyer)
\item the European Research Council under the European Community's Seventh Framework Programme 
(FP7/2007-2013) / ERC grant agreement No.\ 210223
(A.~Mischke)
\item Netherlands Organisation for Scientific Research (project number 680-47-232)
(A.~Mischke)
\item Dutch Foundation for Fundamental Research (project number 10PR2884)
(A.~Mischke)
\item the Russian Foundation for Basic Research (grant 14-02-01220)
(R.~Mizuk)
\item U.S. Department of Energy by Lawrence Berkeley National Laboratory, Nuclear Science Division, under 
Contract No. DE-AC02-05CH11231
(G.~Odyniec) 
\item Spanish Government and EU funds for regional development grants FPA2011-23778 and CSD2007-00042
(A.~Pich)
\item Generalitat Valenciana grant PrometeoII/2013/007 
(A.~Pich)
\item the European Research Council under the European Community's Seventh Framework Programme 
(FP7/2007-2013) / ERC grant agreement No.\ 291377
(R. Pittau) 
\item Spanish grant FPA2011-22398
(R. Pittau)
\item U. S. Department of Energy under contract No.~DE-AC02-98CH10886 and the National Science Foundation 
under grant Nos.~PHY-0969739 and -1316617 
(J.-W. Qiu)
\item Italian MIUR under project 2010YJ2NYW and INFN under specific initiative QNP
(G.~Ricciardi)
\item  The European Research Council grant HotLHC ERC-2011-StG-279579;
by Ministerio de Ciencia e Innovacion of Spain (FPA2009-11951) and by Xunta de Galicia.
(C.~A.~Salgado)
\item the EU I3HP project ``Study of Strongly Interacting Matter'' (acronym HadronPhysics3),
Grant Agreement No.\ 283286
(H.~Sazdjian)
\item Austrian Science Foundation FWF under project No.~P23536-N16
(A.~Schmitt)
\item NSF PHY-0969490 and Indiana University Center for Spacetime Symmetries
(W.~M.~Snow)
\item Lawrence Livermore National Laboratory is operated by Lawrence Livermore National Security, LLC, for 
the U.S. Department of Energy, National Nuclear Security Administration, under Contract DE-AC52-07NA27344.
The work is also supported in part by the US DOE JET topical collaboration. 
(R.~Vogt)  
\item Academy of Finland, grant \# 266185 
(A.~Vuorinen)
\end{itemize}
}

\vfill
{\bf\small \hspace{0.7em}Preprint numbers}
\begin{table}[h]
    \centering
    \begin{tabular}{ll}
        CCQCN-2014-24 &
        CCTP-2014-5 \\ 
        CERN-PH-TH/2014-033 &
        DF-1-2014 \\
        FERMILAB-PUB-14-024/T &
        HIP-2014-03/TH \\
        ITEP-LAT-2014-1 &
        JLAB-THY-14-1865 \\
        LLNL-JRNL-651216 &
        MITP/14-016 \\
        NT@UW 14-04 &
        RUB-TPII-01/2014 \\
        TUM-EFT 46/14 &
        UWThPh-2014-006        
    \end{tabular}
\end{table}

\bibliography{strongtemplateb}
\end{document}